\documentclass[english]{elsarticle}
\usepackage[T1]{fontenc}
\usepackage[latin9]{inputenc}
\usepackage{geometry}
\usepackage{url}
\usepackage{xcolor}
\usepackage{array}
\usepackage{textcomp}
\usepackage{bm}
\usepackage{amsmath}
\usepackage{amssymb}
\usepackage{graphicx}
\usepackage{multirow}
\PassOptionsToPackage{normalem}{ulem}
\usepackage{ulem}

\makeatletter

\DeclareFontEncoding{LGR}{}{}
\DeclareTextSymbol{\~}{LGR}{126}
\providecommand{\tabularnewline}{\\}
\providecolor{lyxadded}{rgb}{0,0,1}
\providecolor{lyxdeleted}{rgb}{1,0,0}

\usepackage{bm}\usepackage{epstopdf}

\makeatother

\usepackage{babel}
\begin{document}
\begin{frontmatter}
\title{Growth of Epitaxial Graphene: Theory and Experiment}

\author{H. Tetlow$^{\dagger}$, J. Posthuma de Boer$^{\ddagger}$, I. J.
Ford$^{\sharp}$, D. D. Vvedensky$^{\ddagger}$, J. Coraux$^{\natural}$
and L. Kantorovich$^{\dagger}$\footnote{The corresponding author}}


\address{$^{\dagger}$Physics Department, King's College London, Strand, London
WC2R 2LS, United Kingdom\\$^{\ddagger}$The Blackett Laboratory,
Imperial College London, London SW7 2AZ, United Kingdom\\
$ $$^{\natural}$Universit{\'e}
Grenoble Alpes, Institut NEEL, F-38042, Grenoble, France, and CNRS,
Institut NEEL, F-38042, Grenoble, France\\$^{\sharp}$Department
of Physics and Astronomy and London Centre for Nanotechnology, University
College London, Gower Street, London WC1E 6BT, United Kingdom.}

\begin{abstract}
A detailed review of the literature for the last 5-10 years on epitaxial
growth of graphene is presented. Both experimental and theoretical
aspects related to growth on transition metals and on silicon carbide
are thoroughly reviewed. Thermodynamic and kinetic aspects of growth
on all these materials, where possible, are discussed. To make this
text useful for a wider audience, a range of important experimental
techniques that have been used over the last decade to grow (e.g.
CVD, TPG and segregation) and characterize (STM, LEEM, etc.) graphene
are reviewed, and a critical survey of the most important theoretical
techniques is given. Finally, we critically discuss various unsolved
problems related to growth and its mechanism which we believe require
proper attention in future research.
\end{abstract}

\end{frontmatter}

\tableofcontents{}

\section{Introduction }

\begin{figure}
\begin{centering}
\includegraphics[height=6cm]{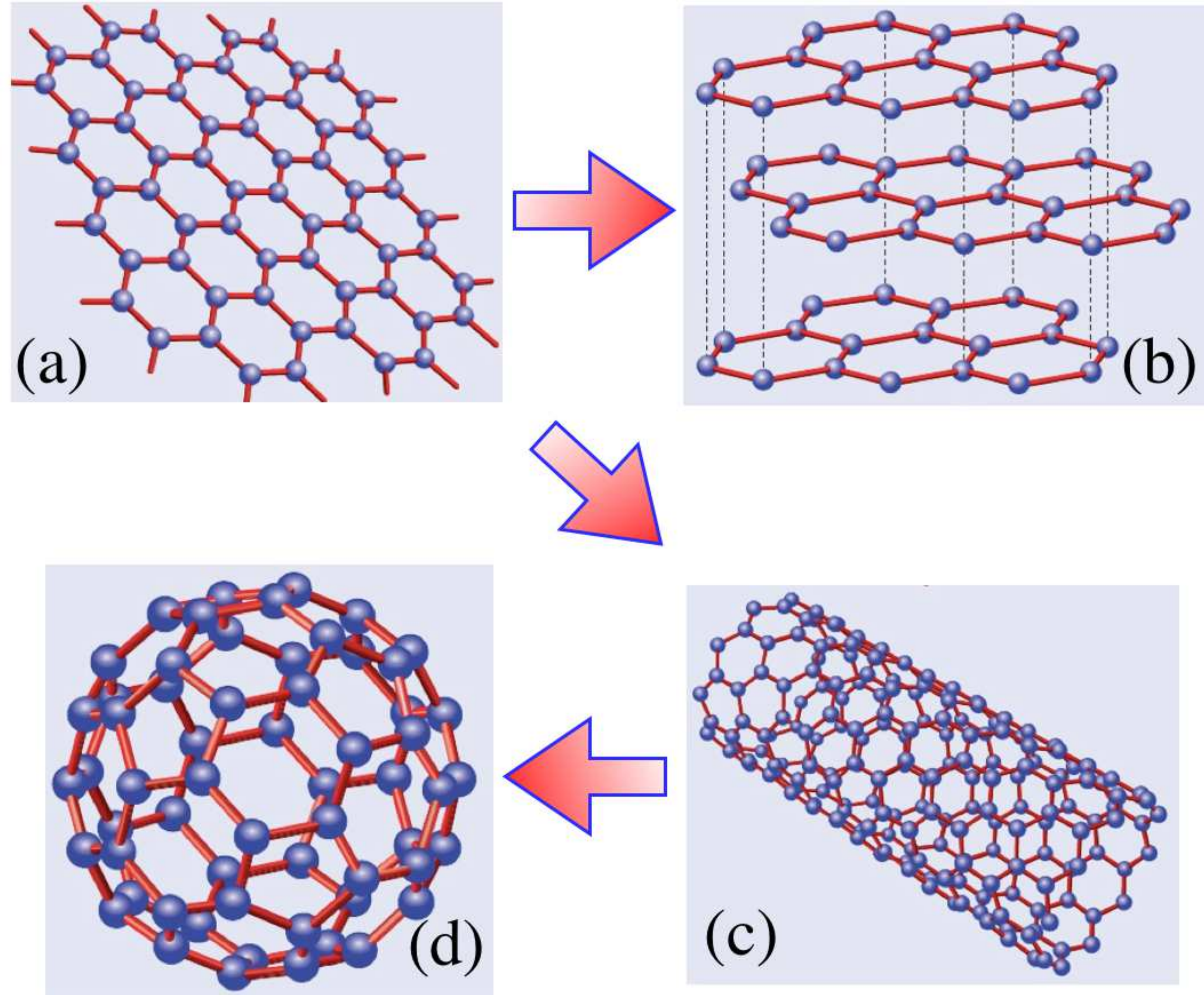}
\par\end{centering}

\caption{Various form of carbon: (a) graphene; (b) graphite; (c) nanotube;
(d) buckyball. {[}Adapted with permission from \cite{C-transformation-source}.{]}
\label{fig:Various-form-of-C}}
\end{figure}

Graphene, one of the allotropic forms of elemental carbon, is a planar
monolayer of carbon atoms arranged on two-dimensional hexagonal lattice
with a carbon-carbon bond length of 0.142 nm \cite{weiss58}. Graphene
is also the conceptual building block for some of these allotropes,
including graphite, which is layered graphene, as well as nanotubes,
which are graphene sheets that are seamlessly rolled up into cylinders
with nanometre-scale diameters, and buckminsterfullerene (``buckyballs''),
which are graphitic molecular cages (Fig. \ref{fig:Various-form-of-C}).

While the existence of graphene has been known for some time, 
the name ``graphene'' was formally adopted by IUPAC only in 1997
as a replacement for the term ``graphitic layers'' in their \textit{Compendium
of Chemical Technology}. The explosive growth in graphene research
began in 2004-2005 with the publication of a series of papers by several
groups, pioneered by those at Manchester in the United Kingdom \cite{novoselov04,novoselov05a,novoselov05b}
and the Georgia Institute of Technology \cite{berger04} and Columbia
\cite{zhang05} in the United States. There are several reasons for
this worldwide interest in graphene, above all the remarkable electronic,
mechanical, optical, and transport properties \cite{neto09} that
have generated many avenues for applications \cite{soldano10,avouris10,bonaccorso10,dassarma11,sun11,Review_applications_2012}.
For example, graphene is optically transparent (between 70\%--97\%
transparency, depending on thickness), and has outstandingly high
electronic and thermal transport and thermal conductivity. The pace
of progress has been so frantic that, in May 2011, the British Broadcasting
Corporation (BBC) dubbed graphene the ``miracle material'' \cite{bbc}.
Additionally, there was already a community of scientists working
on carbon materials, most notably nanotubes and fullerenes, that was
quick to recognize the potential of graphene. Finally, small samples
of graphene could be produced by mechanical exfoliation (the ``scotch
tape'' method) \cite{novoselov04}. Though simple and cost-effective
and ideally suited for laboratory work, this method is unsuitable
for the amount of material that would be required for mass production.
The search for alternative methods for producing large amounts of
graphene has been attempted in many research groups worldwide over
the last decade or so and will be discussed below. Government agencies
have been quick to seize the potential of graphene by funding several
large-scale programs:~the European Union is funding a 10-year 1.35
billion euros flagship program on graphene, South Korea is spending \$350
million on commercialization initiatives, and the United Kingdom is
investing £50 million in a commercialization hub.

The remarkable electronic properties of graphene are due ultimately
to its crystal structure of a honeycomb lattice of carbon atoms, which
can be regarded as two interpenetrating triangular lattices. The $s^{2}p^{2}$
configuration of atomic carbon hybridizes in graphene into a configuration
in which the $2s$, $2p_{x}$, and $2p_{y}$ orbitals of each carbon
are $sp^{2}$-hybridized to form in-plane $\sigma$ bonds with its
three nearest neighbors. The remaining $p_{z}$ orbital is oriented
along the $z$-direction, perpendicular to the graphene plane, and
forms $\pi$-bonds which merge with neighboring $2p_{z}$ orbitals
to form delocalized states across the graphene plane. The ease of
movement of electrons in these $\pi$ states is responsible for the
extraordinary electrical conductivity of graphene. The graphene unit
cell has two $\pi$ orbitals forming two bands that may be thought
of as bonding (the lower energy valence band) and anti-bonding (the
higher energy conduction band) in nature. These are referred to as
$\pi$ and $\pi^{\ast}$ bands, respectively. The gap between the
$\pi$ and $\pi^{\ast}$ bands closes at the $K$-points of the Brillouin
zone, resulting in an energy-momentum dispersion which is approximately
\textit{linear} around the $K$ points. This behavior is one of the
many fascinating properties of graphene and is largely responsible
for much of the excitement about this material. Interestingly, this
basic picture of the electronic bands of graphene was known as far
back as 1947, when Wallace \cite{wallace47} used a two-band tight-binding
model incorporating the bonding and anti-bonding $\pi$-bands to obtain
these results for a single graphite layer.

But there are other noteworthy properties of graphene. The $\sigma$
bonds between the relatively light carbon atoms are quite strong,
creating a bonding situation similar to that in diamond: the speed
of sound in graphene is very high, and so is its thermal conductivity,
being mainly governed by phonons. The conductivity is in fact comparable
to that of diamond and an order of magnitude larger than in conventional
semiconductors, which can be harnessed for applications. In particular,
graphene could be useful for applications in electronic devices, since
a high thermal conductivity facilitates the diffusion of heat to the
contacts and allows for more compact circuits. Another consequence
of the in-plane bonding is that graphene has a very high spring constant
$K\sim50~\mbox{eV $\textrm{\AA}$}^{-2}$ \cite{viola09}, elastic modulus $Y\sim1$
TPa \cite{lee08}, and tensile strength of 130~GPa \cite{lee08},
making graphene the strongest material ever measured. Various combinations
of these properties can be used in applications either to replace
existing materials, or to create new device concepts. However, a major
obstacle in the transfer of proof-of-concept from the laboratory to
the commercial sector is the production of high-quality graphene in
sufficient quantities.

The initial studies which provided the first glimpses of the intriguing
properties of graphene were made on relatively small samples produced
by a mechanical exfoliation method developed by Geim and Novoselov
\cite{novoselov04}. A highly oriented pyrolytic graphite (HOPG) precursor
was subjected to oxygen plasma etching to create 5~$\mu$m-deep mesas
which were then pressed into a layer of photoresist. The photoresist
was baked and the HOPG cleaved from the resist. Scotch tape was used
to repeatedly peel flakes of graphite from the mesas. These thin flakes
were then released in acetone and captured on the surface of Si/SiO$_{2}$.
Mechanical exfoliation can produce graphene samples with areas up
to 1~mm$^{2}$ with good electrical properties. The ease and cost-effectiveness
of mechanical exfoliation is a key factor in the rapid expansion of
fundamental studies of graphene. However, this is time-consuming and
suited only to small-scale production. The realization of the potential
of graphene, for example, as a post-silicon electronic material, will
require the development of techniques for the production of high-quality
large-area sheets of graphene.

An attractive alternative to mechanical exfoliation is the epitaxial
growth of graphene on a hexagonal substrate. Considerable progress
has been achieved in this direction, using both silicon carbide and
close-packed metals as the substrate. The thermal decomposition of
SiC consists of heating the sample in ultra-high vacuum (UHV) to temperatures
between 1273 K and 1773 K, which causes Si to
sublimate, leaving a carbon-rich surface. This technique is capable
of generating wafer-scale graphene layers and is potentially of interest
to the micro-electronics industry. In contrast to the thermal decomposition
of SiC, where carbon is already present in the substrate, in chemical
vapor deposition (CVD), carbon is supplied in gas form and a metal
is used as both catalyst and substrate to grow the graphene layer.
CVD more closely resembles the epitaxial methods that have been developed
since the 1960s to produce films of semiconductors, metals, and other
materials, and some considerable work has been done to date to understand
the epitaxial growth of graphene. Still, there have been few systematic
studies of the atomistic formation kinetics of epitaxial graphene
by any method on any substrate, so no coherent picture of the growth
mechanisms is available. It is the purpose of this review to assemble
the information available from theory and experiment on the epitaxial
kinetics of graphene in the light of the vast work that has been done
on other materials and to highlight where additional work is required
to provide a more complete understanding of graphene growth kinetics.

The review covers both experimental and theoretical aspects of the growth of graphene.
The idea behind this
was that we wanted to present first
the existing experimental picture, limiting ourselves mainly to the
results relevant for understanding of growth and  kinetics; then
we review existing literature with the idea to rationalize, as much
as possible, the substantial volume of various experimental results.
Therefore, in Section \ref{sub:Experimental-techniques} we briefly review
the main experimental techniques that have been employed to study
and grow graphene and then we consider in some considerable detail
those which became popular over the last decade. These include CVD,
temperature programmed growth (TPG) and segregation based methods, as well as growth of graphene by
sublimation from SiC. We believe that these sections should be useful
mostly for non-experimentalists wishing to understand better how theory
can be compared with experiment. Then in Section \ref{sec:Structure-of-Graphene}
we discuss mostly structural properties of graphene on a number of
metal surfaces; in particular, we consider such surfaces as Ru(0001),
Rh(111), Ir(111), Pt(111), Ni(111), Cu(111) and Cu(100).
In Section \ref{sec:Simulation-methods} we briefly review some of the
existing simulation techniques which have proven extremely useful
in understanding graphene growth. These include density functional
theory (DFT), tight binding (TB) and empirical (classical) potential
methods, which have been found useful in obtaining relaxed structures,
formation energies and thermodynamic properties. These are the subject
of Section \ref{sub:Atomistic-modelling-methods}. In Section \ref{sub:Simulations-of-dynamics-methods}
the main techniques that are useful in studying the dynamics of growth
are briefly reviewed, such as molecular dynamics (MD), kinetic Monte
Carlo (KMC) and continuum methods. We believe this introductory section
should be useful for experimentalists wishing to appreciate the capabilities
of theory, and how theory may assist their experiments in providing
the necessary background for understanding growth and hence devising
and optimizing the growth methods they use. This is followed by our
review of various theoretical results relevant to graphene growth.
We first discuss available results relevant for CVD and TPG growth.
In Section \ref{sub:Rate-equation-analysis} rate equation based methods
are considered. Then in Section \ref{sub:Atomistic-approaches} we
discuss various atomistic approaches, mostly based on first principles
DFT simulations, that have been used to date for studying various
stages of graphene growth. These include dehydrogenation of hydrocarbon
molecules, early stages of the nucleation and growth of carbon clusters,
attachment of carbon species to existing graphene islands and the
role of defects in free-standing as well as metal-supported graphene.
Then in Section \ref{sub:kMC} we review several dynamic approaches
that have been explored so far to simulate the process of growth itself.
A special Section \ref{sec:SiC} is devoted to simulations of growth
of graphene by the sublimation of SiC where we present a variety of
results from structure studies up to phenomenological growth models.
Finally, a general discussion and conclusions are contained in Section
\ref{sec:Discussion} where we summarize the main current achievements
of both theory and experiment in understanding of graphene growth,
and discuss unsolved problems and possible directions of further studies.{\small{} }{\small \par}

There have been many reviews written on all aspects of graphene \cite{neto09,park09,shao09,bonaccorso10,choi10,dreyer10,first10,riedl10,schwierz10,balandin11,dassarma11,neto11,pumera11,singh11,young11,batzill12,bonaccorso12,shen12,young12,zhang13},
but this is the first review devoted exclusively to growth.

\section{Fundamentals of nucleation and growth}

\subsection{Main processes}

\textcolor{black}{}
\begin{figure}[t!]
\centering{}\textcolor{black}{\includegraphics[height=6cm]{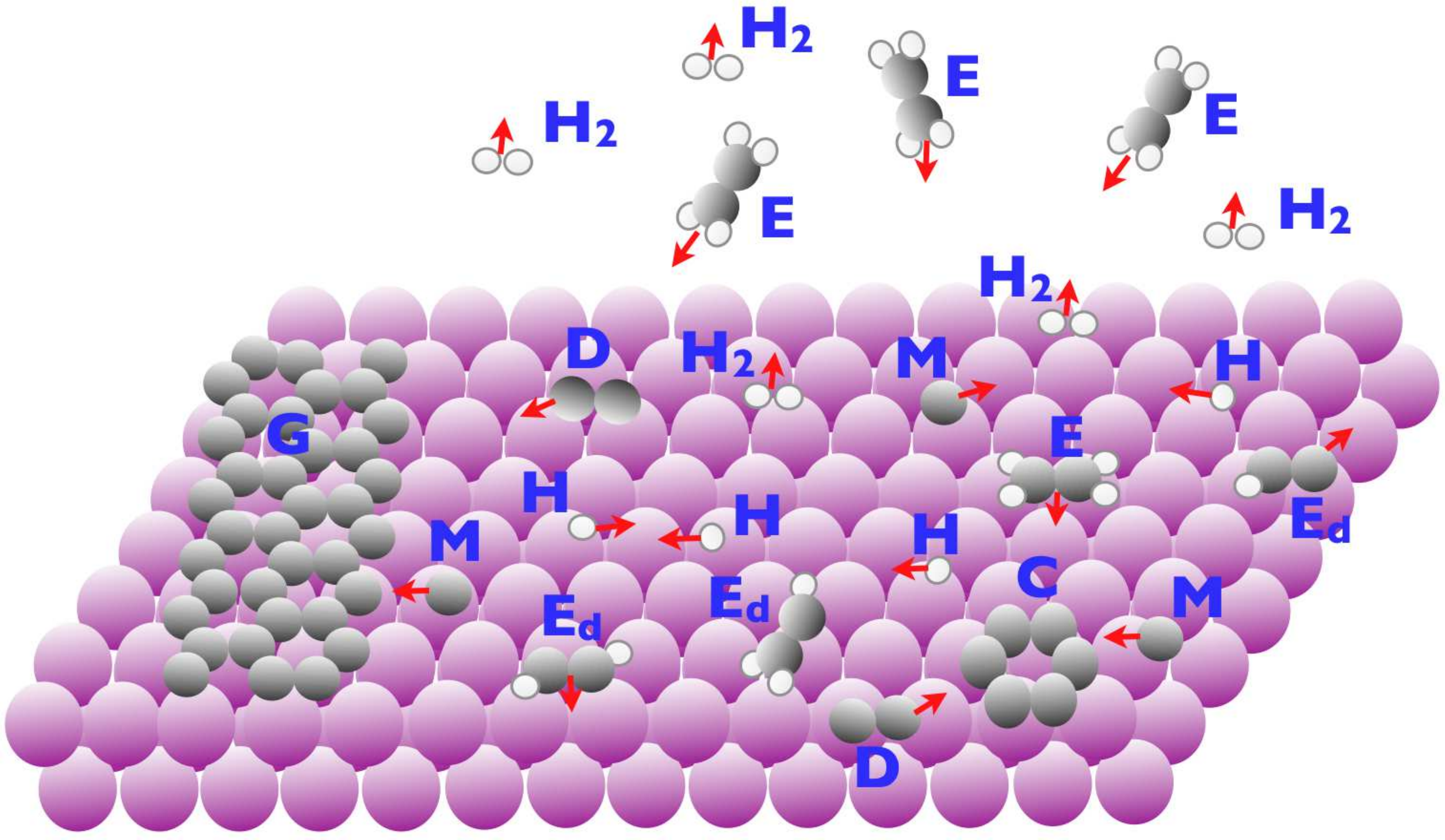}
\caption{Schematic representation of fundamental processes during epitaxy:
hydrocarbon
molecules E are deposited on the surface, undergo decomposition
via a series of dehydrogenation reactions giving rise to various C$_x$H$_y$ species shown as E$_{d}$,
and H atoms.
The new species
are all able to diffuse across the surface.
Smaller carbon species M and D form and diffuse on the surface, aggregating
into larger clusters C.
H atoms of the original molecule
migrate
on the surface, and form H$_{2}$ molecules
that evaporate from the
surface. Finally, some of the species like M and D, or even their
bigger clusters C, may attach to the island G at its edge. Other processes
(not shown) are also possible based on e.g. diffusion of atoms along
the edge of an island,
nucleation of second- and higher layers on the  islands,
downward movement of atoms
adsorbed on top of  islands  to
lower layers,
and the break up,
or dissolution, of islands. \textcolor{black}{\label{fig:processes} }}
}
\end{figure}

\textcolor{black}{Figure~\ref{fig:processes} illustrates typical
processes that occur during epitaxy. Depending on the material and
the growth method, some of these processes may need to be amended.
We begin with deposition. This an experimentally controllable process
both in terms of the deposition rate and the species being deposited.
In the simplest case, the deposited species is the atomic (or molecular)
constituent of the growing material; we assumed in the Figure that
these are polyatomic precursors E since this is what is frequently
used in growing graphene. In the latter case, the reaction kinetics
become important, as certain surface sites may dominate catalytic
activity, and the reaction products (shown as
hydrogen molecules H$_{2}$)
must desorb
from the surface. Even the atoms of the growing material may desorb
from the surface at high temperatures (not shown). After deposition
and, possibly, decomposition kinetics, the atomic species M diffuse
on the surface. Even such an ostensibly simple process might have
complex steps, such as exchange processes, or even involve the formation
of dimers D or larger
carbon clusters C
that migrate along the surface.
The concentration of diffusing species increases as deposition proceeds
and eventually attains a level where islands G can nucleate. There
are several processes by which this might occur. In }\textit{\textcolor{black}{homogeneous
nucleation}}\textcolor{black}{, islands are formed by the collision
of two or more migrating species which then bind to form an island.
Depending on the temperature, this island may be stable, or may break up.
Eventually, the concentration of the migrating species is sufficient
to induce island formation, however many individual units are required.
In }\textit{\textcolor{black}{heterogeneous nucleation}}\textcolor{black}{,
the rate of nucleation is increased by the presence of a surface defect
such as a vacancy, a screw dislocation, or a pre-existing step edge.
Island size distributions can be used to study nucleation mechanisms
and, in favorable circumstances, to identify the nucleation mechanism.}

\textcolor{black}{Once the islands are formed, they can grow by the
capture of migrating units. This attachment process may have additional
barriers. Similarly, the reverse of this process, detachment, is the
process by which atoms
are separated
from an island. The shapes
of islands can be strongly influenced by edge diffusion, that is,
the motion of atoms around the edge of an island. At low temperatures,
some systems exhibit fractal islands which, despite having a large
surface energy, cannot reduce it by any type of rearrangement because
surface migration is too slow. At higher temperatures, edge diffusion
becomes effective at reducing the edge lengths of islands, which are
now more compact. As the islands grow, they become large enough such
that deposition on the islands becomes appreciable. Two processes
then become important. The first is downward hops when species (e.g.
atoms) diffuse down from the upper layers of the islands. If there
is no barrier at step edges, then the atoms can hop down from an island
to the terrace below with no additional impediment. If there is an
additional barrier, the atoms have a longer residence time on the
island, which builds up their concentration and makes second-layer
nucleation more likely.}

\subsection{The nucleation process\label{sub:Nucleation-Theory}}

Beginnings are often complicated and messy, and difficult to understand
after the event, irrespective of whether they relate to historical,
social or scientific processes. The formation, or nucleation, of new
materials or phases of matter lies firmly within this category. What
makes nucleation hard to pin down is the relative inaccessibility
of the process to experimental investigation. Nucleation events involve
the coming together of precursor species on a molecular scale, and
in a quasi-random fashion, with a successful event often lasting a
very short time from start to finish. It is akin to a chemical reaction
scheme with many intermediate steps, some of which are very rare.
Added to this is the relative difficulty of placing the process within
a fully developed theoretical framework, together with the sensitivity
of calculations to the details of the assumed intermolecular interactions.
Yet theoretical progress can be made in tandem with experiment and
the nucleation of phase transformations in materials science is now
becoming better understood as a result of the new richness of experimental
methods on the nanoscale.

Our objective is to review current understanding of the nucleation
and growth of islands of graphene on various substrates, driven by
a range of thermal and chemical processing schemes, but it is necessary
to start with a general survey of the basic ideas that are used in
discussions of phase transformations \cite{Ford2004,Kalikmanov-book2012}.
Our allusion to a sequence of chemical reactions is a perfectly valid
kinetic description, but on top of this a thermodynamic framework
has been built that provides a simplifying viewpoint. We shall start,
as is often the case, with an appeal to the second law of thermodynamics,
an imperative placed on the universe to increase its entropy, or equivalently,
a requirement that systems in weak contact with their surroundings
should evolve to reduce a free energy or grand potential of some kind
\cite{Ford-book2013}. Of course, according to a strict interpretation
in equilibrium thermodynamics, free energies properly characterize
equilibrium states, and not states that are evolving towards equilibrium,
but we take the point of view that these concepts can transfer across
to nonequilibrium processes as long as the states of the system in
question exist for long enough to achieve an internal thermalization
and equilibration.

Thus the phase of a material with the lowest chemical potential under
the prevailing conditions is naturally selected, and this approach
underpins the construction of phase diagrams. But such considerations
are seriously flawed, since they will not account for supercooling
of a liquid below its freezing point or metastability of any kind.
The reason that materials can disregard the mandate of a phase diagram
is that the conversion from liquid into solid, or indeed carbon adatoms
into graphene sheets, must go through an intermediary structure that
is thermodynamically less stable than the initial phase, even though
the final phase is ultimately more stable. The process of transformation
has to proceed through a bottleneck: viewed thermodynamically, a system
will apparently need to increase its free energy in order ultimately
to decrease it. Conspiracies against the second law such as this are
not illegal, of course, since the law is only to be obeyed on the
average. But in order to quantify the process of nucleation, we need
to understand the statistical physics of molecular attachment and
detachment from relatively short-lived molecular clusters.

\begin{figure}
\begin{centering}
\includegraphics[height=4cm]{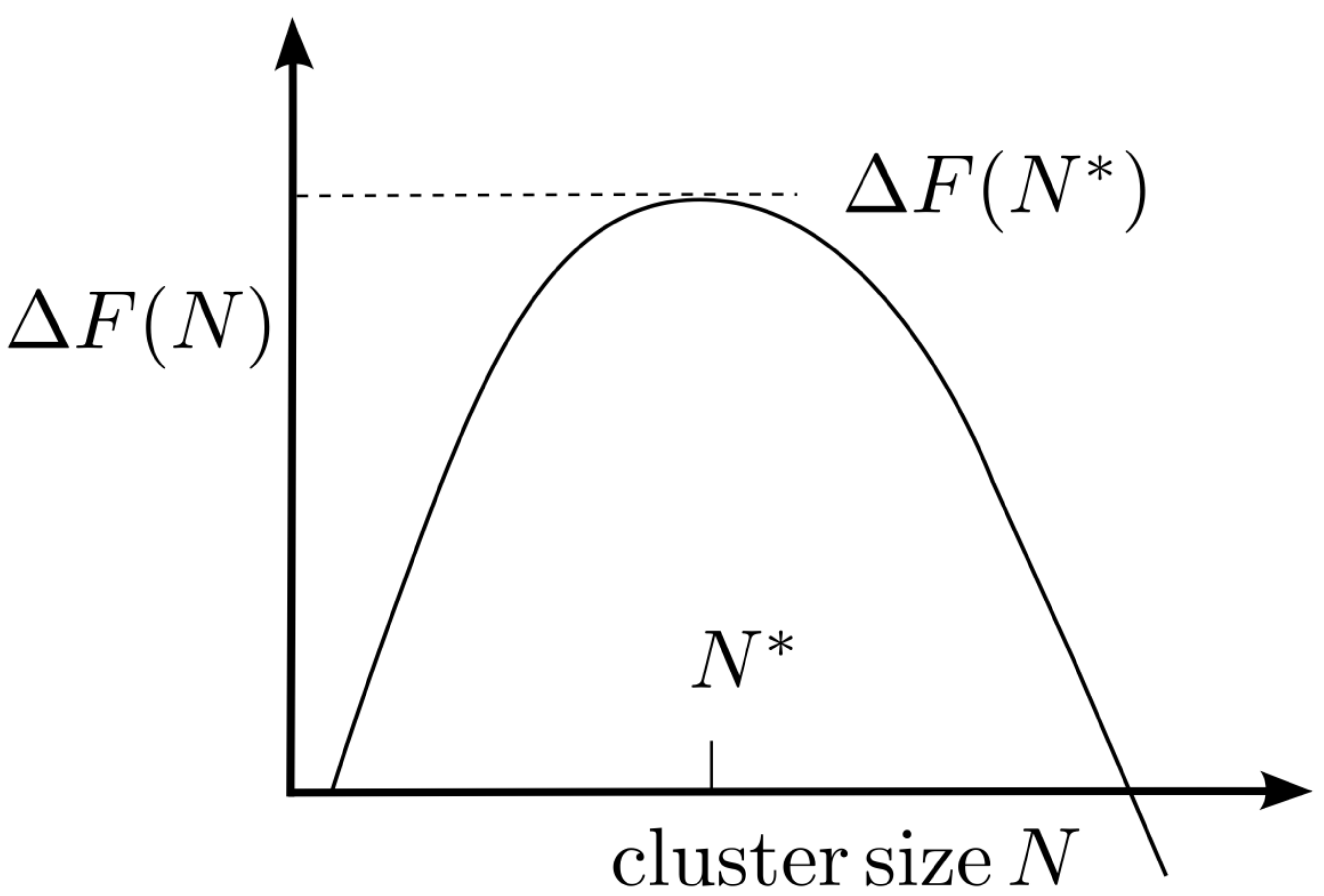}
\par\end{centering}

\caption{Nucleation is often described using a free energy $\Delta F(N)$ defined
for clusters of size $N$. The peak in free energy at the critical
size $N^{*}$ is related to the nucleation rate.\label{fig:Nucleation-1}}
\end{figure}

The classic diagram of nucleation theory is the so-called free energy
barrier sketched in Figure \ref{fig:Nucleation-1}. This represents
a difference in thermodynamic potential $\Delta F$ of a cluster of
$N$ atoms or molecules (referred to generically as monomers) calculated
with respect to that of a single monomer, and specific to the prevailing
conditions of temperature and monomer chemical potential in the environment.
It is commonly referred to as a cluster free energy but more properly
it is a difference in grand potential \cite{ford97}. The size of
the cluster is defined in some fashion that usually involves a minimum
separation between monomers, or a requirement that the cluster energy
should lie below some monomer emission threshold \cite{harris03}.
$\Delta F$ is employed to estimate the probability of generating
a cluster of a given size from the complex dynamics operating between
the cluster and its environment. The probability is estimated to be
$P(N)=P(1)\exp(-\Delta F(N)/k_{B}T)$ where $P(1)$ is the probability
of generating a cluster consisting of a single monomer and $T$ is
the temperature of the environment. This formulation is consistent
with Einstein's theory of fluctuations.

The free energy difference may be written
\begin{equation}
\Delta F(N)=F_{{\rm ex}}(N)-N\Delta\mu\;,\label{eq:nuc}
\end{equation}
where $\Delta\mu$ is the difference in chemical potential between
the metastable and stable thermodynamic phase (for example the chemical
potential of C atoms on the substrate minus that in the graphene flake)
under the prevailing environmental conditions. The negative sign of
the second term therefore reflects the tendency for the new, or stable
phase to grow in accordance with the second law. However, the first
term, known as the excess free energy, is a positive contribution
that impedes the transformation to the new phase. It represents the
thermodynamic cost of forming a fragment of the new phase, rather
than the new phase in bulk. It is often characterized as an interfacial
free energy cost, though it can arise from more profound features
of the fragment, related to structural or density differences with
respect to the bulk. In the very simplest approach, known as the classical
theory of nucleation, the excess free energy for the fragment, or
nucleus, is taken to be proportional to the surface area, for growth
of a three dimensional structure, with a proportionality constant
given by the surface tension $\gamma$ of the bulk planar interface.
The free energy barrier may then be written
\[
\Delta F(N)=\gamma A(N)-N\Delta\mu=\theta N^{\alpha}-N\Delta\mu,
\]
where $A(N)$ is the surface area of the cluster, and the proportionality
to cluster size $N$ raised to the power $\alpha=2/3$ follows on
the grounds of geometry; $\theta$ is a constant related to bulk density
of the new phase. For a fragment that is two-dimensional, like a sheet
or flake, the excess free energy might be taken to depend on cluster
size raised to the power $\alpha=1/2$. In either case, the free energy
barrier in Figure \ref{fig:Nucleation-1} will possess a peak at the
so-called critical cluster size $N^{*}$. We interpret this to mean
that a nucleus with this size is thermodynamically unstable with respect
to both the addition or loss of monomers. The probability of a nucleation
event might then be represented as the probability of forming the
critical cluster (per unit time and system volume) multiplied by the
probability of its subsequent growth, which is presumed to be of order
one half. The rate of nucleation is therefore proportional to $\exp(-\Delta F(N^{*})/kT)$
and the height of the free energy barrier $\Delta F(N^{*})$ is then
clearly a measure of the likelihood that a nucleation event will take
place quickly or slowly.

The classical theory can be criticized for its use of a bulk, planar
interfacial tension to represent the excess free energy, but more
realistic models can be developed using specific intermonomer interactions.
There are other difficulties with the theoretical framework, but they
can be removed by developing a set of kinetic or master equations
to represent the attachment and detachment of monomers from the nucleus,
and indeed to bring into consideration processes involving dimers,
trimers etc. Thus the intuitive picture offered by Figure \ref{fig:Nucleation-1}
remains central to nucleation theory. The critical free energy barrier
is a principal ingredient of many approaches, supplemented by a variety
of kinetic models to account for growth by attachment of clusters
as well as monomers. Such models would then describe coalescence between
clusters, known as Smoluchowski ripening \cite{Smoluchowski06}.

Should it be doubted that the free energy can really be employed to
describe small and short-lived nuclei, then nonequilibrium corrections
to the thermodynamics and hence the kinetics can be introduced, allowing
them, for example, to relax towards a quasiequilibrium state in a
finite time. Additional features of the theory can be included to
allow attachment of more than one monomer species \cite{Vehkamaki-book2006},
and to explore more complicated definitions of cluster size .

The framework relies on specifying an environmental monomer chemical
potential, and this is not always a constant. Furthermore, the phenomenon
of Ostwald ripening follows from a coupling between clusters through
their common environment. There is a tendency for clusters larger
than the critical size to grow, leading to the depletion of monomers
in the environment, which then has the effect of making clusters smaller
than the critical size become more unstable. The overall effect is
the increase in size of large clusters at the expense of small clusters,
and this is accompanied by an increase in the critical size, destabilizing
small clusters further. Of course this is not a nucleation mechanism
but rather a process of redistribution of material and the coarsening
of a size distribution of clusters. It tends to be a delicate process
since the relative differences in stability between clusters of different
sizes tends to be small and the process is slow rather than fast.

\subsection{Thermodynamics and kinetics of growth\label{sub:Thermodynamics-and-kinetics}}

The growth of a system such as graphene using material from a specific
source is reasonably well understood from a thermodynamic perspective.
It is necessary that there should be species (single carbon atoms
or clusters of atoms), available on or beneath the surface of the
substrate, that have a higher chemical potential than that of the
same species when incorporated into the growing graphene sheet. Such
precursors might be generated from other species. For example adsorbed
carbon dimers might form from carbon monomers, if it is thermodynamically
favorable to do so, before they are then incorporated into the graphene
phase. This is an example of Ostwald's rule of stages, whereby a thermodynamically
favored transformation might take place through intermediaries that
are less metastable than the original feedstock.

The important point, therefore, is to give careful consideration to
the nature of the attaching species and to calculate the free energy
change associated with such a process. According to phenomenological
nonequilibrium thermodynamics, the rate of growth is proportional
to this change. For example, if the attaching species were adsorbed
carbon monomers then the appropriate free energy change would be $\Delta\mu=\mu_{C}^{{\rm ad}}-\mu_{C}^{{\rm gr}}$,
where the chemical potentials on the right hand side refer to carbon
atoms in the metastable and stable situations of surface adsorption
and within graphene, respectively. There is, of course, a situation
where the chemical potentials are equal, where no growth occurs and
the flake is in equilibrium with the adsorbed monomers. Since we can
often write $\mu_{C}^{{\rm ad}}\propto\ln c$ where $c$ is the density
(or concentration) of adsorbed monomers, we obtain $\Delta\mu\propto\ln(c/c^{eq})$
where $c^{eq}$ is the concentration of adsorbed monomers when in
equilibrium with graphene. If we define an adsorbed monomer supersaturation
by $S=c/c^{eq}$ then this becomes $\Delta\mu\propto\ln S\approx S-1$
for small supersaturations. This formulation when written as $\Delta\mu\propto c-c^{eq}$
allows growth to be interpreted kinetically as a balance between the
attachment and detachment of monomers at rates proportional to their
(meta)stable concentrations. Furthermore, for edges of graphene flakes
that have a significant radius of curvature, the detachment rate will
be modified by a Kelvin factor that represents the thermodynamic destabilization
of carbon in that phase, owing to the undercoordination of edge sites
with respect to a planar interface.

But if the primary attaching species were carbon dimers, then the
relevant free energy change associated with growth would be $\Delta\mu_{2C}=\mu_{2C}^{{\rm ad}}-\mu_{2C}^{{\rm gr}}$.
Following the above argument, this would reduce to $\Delta\mu_{2C}\propto c_{2}-c_{2}^{eq}$,
for small deviations from equilibrium, in terms of concentrations
of dimers. We might wish to relate this to the density of adsorbed
carbon monomers, requiring a relationship between the chemical potentials
of the dimer and the monomer, or equivalently between the (meta)stable
dimer and monomer populations. We would expect $\mu_{2C}^{{\rm ad}}\approx2\mu_{C}^{{\rm ad}}-\epsilon$
where $\epsilon$ is the adsorbed dimer binding energy with respect
to the monomer, or equivalently we might appeal to a law of mass action
that states that $c_{2}\propto c^{2}$, etc. The attachment rate would
then become proportional to $c^{2}$. In this way, nonlinear growth
kinetics emerges from fundamentally linear attachment processes, as
is well-known in chemical kinetics.

In practice, graphene growth laws appear to be nonlinear, a feature
that allows us insight into the nature of the primary attaching species.
But it is not simply the thermodynamics that determines the growth
process; the kinetics also plays a role. In addition to the condition
that attachment should be associated with a reduction in species chemical
potential, the species in question needs to be mobile enough to make
its way to the edge of the growing flake, and this is usually size
and structure dependent. An understanding of the kinetics of growth
can often provide the proportionality factors that accompany the thermodynamic
driving forces considered above, and conversely, experimental data
can provide clues as to the nature of the attaching species. A combination
of the two is necessary in order to reveal the nature of graphene
growth.

\section{Experimental techniques\label{sub:Experimental-techniques}}

Experiments addressing the growth of epitaxial graphene are in essence
closely related to the study of graphene structure, a field covering
eight orders of magnitudes of length-scales in the case of graphene
growth, from sub-$\textrm{\AA}$ngstr\"{o}m to millimeters. The picture of the snake
biting its tail here largely applies, as often in materials science:
one usually tracks the formation of graphene with structural observations,
for instance of the hexagonal shape of growing islands, or of a typical
reciprocal-space lattice revealed by diffraction; but one usually
needs graphene growth to be optimized prior to such high resolution
structural characterizations. The two-dimensional nature of graphene,
with all of its atoms being surface atoms, naturally suits it for
surface science studies, either scanning probe microscopies, or other
microscopy or diffraction techniques in geometries and/or with energies
yielding shallow penetration of electrons/photons into the supporting
substrate used for graphene growth. There are however two examples
of non-surface science techniques also being used: transmission electron
microscopy (TEM) and Raman spectroscopy.
The latter technique is
well suited for
studying such low amounts of matter as that contained in a single
(or less than that) layer of graphene thanks to a resonance scattering
process. Various techniques which have been applied thus far to
the study of graphene growth and structure are summarized in Table
\ref{tab:Experimental-techniques}.


\begin{table}
\caption{Experimental techniques which have been used for studying the growth and structure of graphene, including their acronyms, physical mechanism, and operating conditions. \label{tab:Experimental-techniques}}
{\small
\begin{tabular}{|c|c|c|c|c|c|c|}\hline
\multirow{2}{*}{Acronym}&\multirow{2}{*}{Full name}&\multirow{2}{*}{Type}&Spatial&Pressure&Temperature&\multirow{2}{*}{\emph{in operando}}\\
&&&resolution&range (mbar)& range (K)&\\
\hline\hline
\multirow{2}{*}{AFM}&Atomic force&Scanning&$\lesssim$~\AA\ in UHV,&\multirow{2}{*}{$10^{-10}$--1}&\multirow{2}{*}{4--300}&\multirow{2}{*}{No}\\
&microscopy&probe&$\sim$~nm otherwise&&&\\
\hline
\multirow{4}{*}{LEED,}&Low-energy&&\multirow{6}{*}{$\sim$~100~$\mu$m}&\multirow{6}{*}{$10^{-10}$}&\multirow{6}{*}{300--1500}&\multirow{6}{*}{Yes}\\
\multirow{4}{*}{SPALEED,}&electron&&&&&\\
\multirow{4}{*}{$\mu$LEED}&diffraction&Electron&&&&\\
&spot-profile-&scattering&&&&\\
&analysis-LEED,&&&&&\\
&micro-LEED&&&&&\\
\hline
\multirow{3}{*}{LEEM}&Low-energy&Full-field micro-&\multirow{3}{*}{$\sim$~nm}&\multirow{3}{*}{$10^{-10}$--$10^{-7}$}&\multirow{3}{*}{300--1500}&\multirow{4}{*}{Yes}\\
&electron&scopy based on&&&&\\
&microscopy&electron scattering&&&&\\
\hline
\multirow{3}{*}{PEEM}&Photoemission&Full-field micro-&\multirow{3}{*}{10~nm}&\multirow{3}{*}{$10^{-10}$--$10^{-7}$}&\multirow{3}{*}{300--1500}&\multirow{3}{*}{Yes}\\
&electron&scopy based&&&&\\
&microscopy&on photoelectrons&&&&\\
\hline
\multirow{3}{*}{Raman}&\multirow{4}{*}{--}&Ensemble average&\multirow{2}{*}{$\sim 100$~nm}&\multirow{4}{*}{1}&4-700&\multirow{4}{*}{No}\\
\multirow{3}{*}{spectroscopy}&&or microscopy in&\multirow{2}{*}{(diffraction}&&(isolated&\\
&&a confocal&\multirow{2}{*}{limit)}&&from&\\
&&microscope&&&substrate)&\\
\hline
\multirow{4}{*}{RHEED}&Reflection&\multirow{2}{*}{Ensemble average}&\multirow{4}{*}{$\sim$~100~$\mu$m}&\multirow{4}{*}{$10^{-10}-10^{-7}$}&\multirow{4}{*}{300-1500}&\multirow{4}{*}{Yes}\\
&high-energy&\multirow{2}{*}{based on electron}&&&&\\
&electron&\multirow{2}{*}{scattering}&&&&\\
&diffraction&&&&&\\
\hline
\multirow{3}{*}{SEM}&Scanning&Scanning micro-&\multirow{3}{*}{$\sim$~nm}&\multirow{3}{*}{$10^{-6}$}&\multirow{3}{*}{300}&\multirow{3}{*}{No}\\
&electron&scopy based on&&&&\\
&microscopy&electron scattering&&&&\\
\hline
\multirow{3}{*}{STM}&Scanning&\multirow{3}{*}{Scanning probe}&\multirow{3}{*}{$\lesssim$~\AA}&\multirow{3}{*}{$10^{-10}$}&\multirow{3}{*}{$<\,$1--1100}&Yes\\
&tunneling&&&&&(with special\\
&microscopy&&&&&instruments)\\
\hline
\multirow{3}{*}{SXRD}&\multirow{2}{*}{Surface X-ray}&Ensemble average&\multirow{3}{*}{$\sim$~100~$\mu$m}&\multirow{3}{*}{$10^{-10}$}&\multirow{3}{*}{300--1100}&\multirow{3}{*}{No}\\
&\multirow{2}{*}{diffraction}&based on photon&&&&\\
&&scattering&&&&\\
\hline
\multirow{3}{*}{TEM,}&Transmission&\multirow{3}{*}{Full-field}&\multirow{4}{*}{$\lesssim$~\AA}&\multirow{4}{*}{$<10^{-7}$}&\multirow{4}{*}{300--900}&\multirow{2}{*}{Yes}\\
\multirow{3}{*}{HRTEM}&electron micro-&\multirow{3}{*}{microscopy}&&&&\multirow{2}{*}{(with special}\\
&scopy, high-&&&&&\multirow{2}{*}{instruments)}\\
&resolution TEM&&&&&\\
\hline
\multirow{4}{*}{STEM}&Scanning&\multirow{3}{*}{Scanning}&\multirow{4}{*}{$\lesssim$~\AA}&\multirow{4}{*}{$<10^{-7}$}&\multirow{4}{*}{300}&\multirow{4}{*}{No}\\
&transmission&\multirow{3}{*}{electron beam}&&&&\\
&electron&&&&&\\
&microscopy&&&&&\\
\hline
\multirow{3}{*}{XSW}&\multirow{2}{*}{X-ray}&Ensemble average&\multirow{3}{*}{$\sim100$~$\mu$m}&\multirow{3}{*}{$10^{-10}$}&\multirow{3}{*}{300}&\multirow{3}{*}{No}\\
&\multirow{2}{*}{standing waves}&based on photon&&&&\\
&&scattering&&&&\\
\hline
\multirow{3}{*}{XPS/UPS}&X-ray/UV&Ensemble average&\multirow{3}{*}{$\sim100$~$\mu$m}&\multirow{3}{*}{$10^{-10}$--$10^{-7}$}&\multirow{3}{*}{4--1100}&\multirow{3}{*}{Yes}\\
&photoelectron&based on electron&&&&\\
&spectroscopy&excitation&&&&\\
\hline
\end{tabular}}
\end{table}


Basically two types of approaches are employed for studying graphene
growth: \emph{in operando} analysis over time-scales down to (typically) a
second while graphene is growing, and post-growth studies, which
rely on the study of a collection of samples for which growth has
been interrupted at various stages. The former requires ultra-high
vacuum (UHV) conditions in all experiments performed thus far, while
the later can involve ex situ studies, i.e. outside the growth environment,
but UHV is necessary for the highest resolution characterizations.
\emph{In operando} studies offer a wealth of advantages, especially
the absence of reproducibility issues and fast optimization of growth,
but remain rather rare, presumably because they are only possible
in a few instruments worldwide and because they are not suitable in
certain growth conditions. For this reason, they are limited practically
to the growth of graphene on metals, single crystals and SiC under
UHV, but graphene growth in atmospheric conditions and/or very high
temperatures (above 1500\textdegree{}C) retains many of its secrets.

None of the techniques which we will now briefly describe is able
alone to provide a full picture of the structure or growth in a given
system. Usually a combination of experimental techniques, sometimes
confronted by structural calculations, is required instead. Some quantities
remain difficult to determine, a typical example being the distance
between graphene and a metal, and its nanoscale variations.

\subsection{Scanning tunneling microscopy (STM)}

The scanning tunneling microscope was invented in 1982 \cite{Binnig1982}.
It is a powerful tool for studying the structural and electronic properties
of \emph{conducting} surfaces, which can be employed in a variety
of environments, from UHV to air and liquids \cite{Hansma1987}. It
consists of a sharp tip scanning the surface of a sample, in a regime
of such tip-sample distances that electrons can tunnel to (from) the
tip from (to) the sample surface. In practice this implies very short
distances between the surface and the tip apex, of the order of 1-10
$\textrm{\AA}$. Due to the exponential decay of the tunneling current
as a function of the tip-sample distance, STM is intrinsically well-suited
to atomic resolution imaging \cite{Tersoff1985}. Since the conductance
(first derivative of the current with respect to the voltage) is approximately
proportional to the local density of states \cite{Tersoff1985}, the
technique is also capable of local electronic spectroscopy in a mode
in which the current is measured as a function of the bias applied
between tip and sample.

STM can be performed in the constant current or constant height mode.
In the first mode, an electronic feedback loop modifies the distance
between tip and sample on the fly during the scan, thereby ensuring
a constant current during the image acquisition. In this case the
STM image consists of spatial variations of the sample-tip apparent
distance. Note that what is measured is only the \emph{change} of
the tip-sample distance during the scan, not the actual \emph{absolute}
distance which remains unknown. Also, even the change of the distance
measured during the scan may not necessarily correspond to the actual
change of the surface height (e.g. when scanning over a molecule)
as what is basically measured is the local density of states. In the
second mode, the feedback loop is opened (or set with a response time
smaller than the scan rate), leaving the average sample-tip distance
constant, and then the STM image consists of spatial variations of
the tunneling current. The latter mode allows faster scanning, not
being limited by the mechanical response time of the tip and its holder.
Very roughly speaking, this is the basis of high speed STM imaging,
which goes beyond video rate \cite{Rost2005}. Most STM studies are
performed from room temperature down to low or very low temperatures.
The low temperature, when operating at liquid helium or nitrogen temperatures,
provides thermal stability which is beneficial to high resolution
distortion-free imaging, and is also a requirement for high energy
resolution spectroscopic analysis. The operation of STM at temperatures
above room temperature is more problematic, firstly due to appreciable
thermal drifts, and secondly due to the tolerance of the instrument
parts to temperatures of a few 100 K above room temperature. The so-called
environmental STM is possible at temperatures approaching 1300 K,
however, when specially designed instruments are used \cite{Hoogeman1998}.

\begin{figure}
\begin{centering}
\includegraphics[height=5cm]{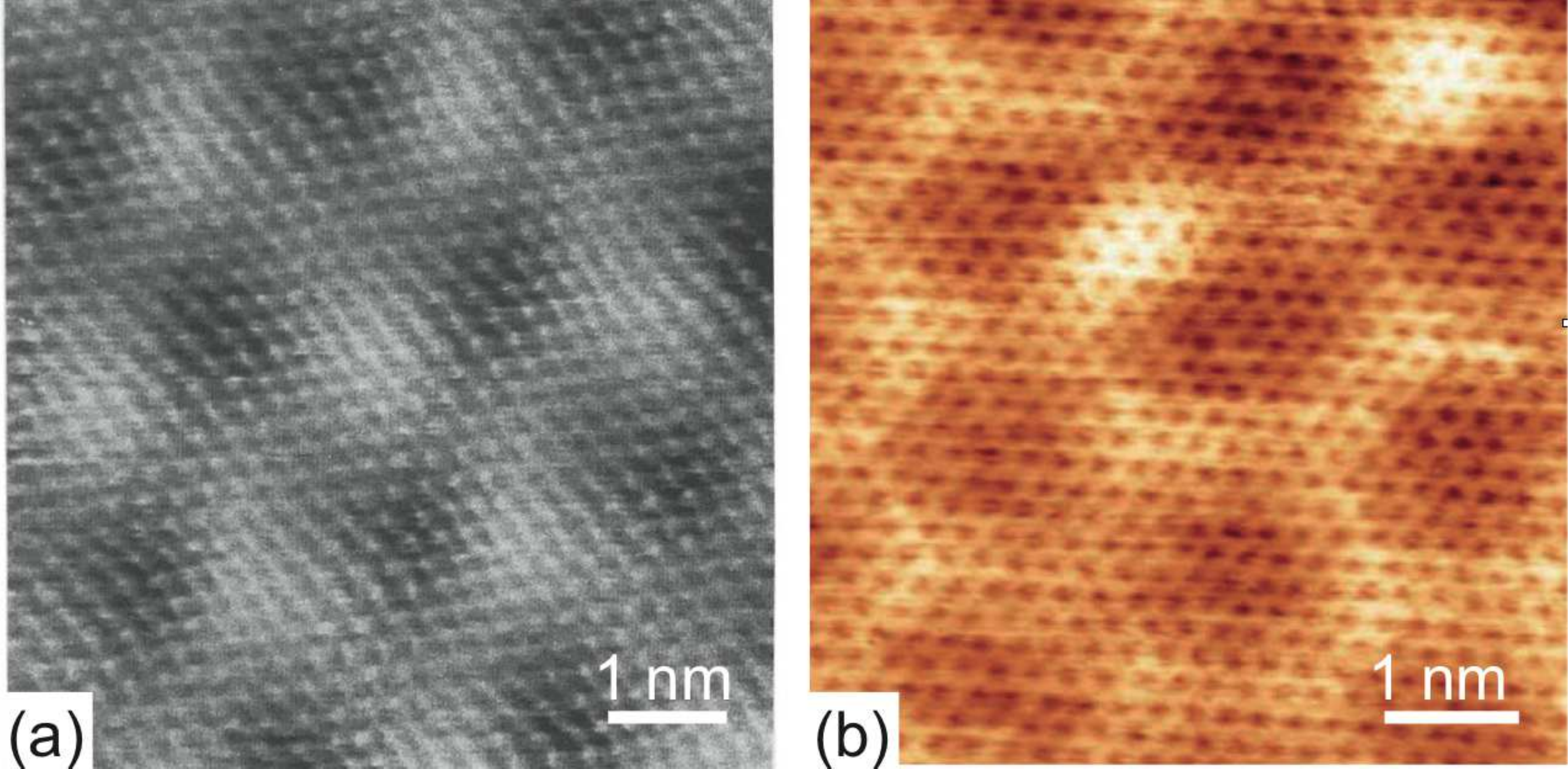}
\par\end{centering}

\caption{\label{fig:experimental techniques STM}UHV atomic resolution STM
topographs of graphene on (a) Pt(111) and (b) SiC(0001). The centers
of carbon rings appear dark, while atoms appear bright. Only three
out of six atoms of each ring are seen in (a) due to the effect of
the tip, while in (b) one infers the positions of atoms to lie at
the edges of the small bright hexagons. In addition to the atomic
contrast, a larger scale contrast also appears: a moir\'{e} pattern between
graphene and Pt(111) in (a) and a $\left(6\sqrt{3}\times6\sqrt{3}\right)R30\text{\textdegree}$
pattern of the carbon buffer layer between SiC(0001) and graphene
in (b). {[}Reprinted (a) from  \cite{Land1992} with permission from Elsevier and
(b) with permission from \cite{Mallet2007}. Copyright (2007) by the American Physical Society{]}.}
\end{figure}

UHV STM has been the instrument of choice in a number of early and
recent advanced surface science studies. The first atomic resolution
images of graphene were actually obtained with this technique \cite{Land1992}
in the 1990's for graphene on Pt(111), as shown in Fig. \ref{fig:experimental techniques STM}(a).
The technique gave a strong push to studies of the structure and growth
of epitaxial graphene. It allowed the resolution of superstructures
in graphene on SiC, see Fig. \ref{fig:experimental techniques STM}(b),
and metals \cite{Diaye2006,Marchini2007,Riedl2007,Varchon08,Loginova08,Hiebel2008,Otero2010},
revealed the presence of defects in graphene on substrates \cite{Coroux2008b,Sutter2009,Biedermann2009,Levy2010},
made it possible to study growth at different stages \cite{Coraux},
including at  elevated growth temperatures \cite{Kwon_NL_2009} and with
time resolution at growth temperatures   \cite{Dong}.
Determination
of the structure around defects, as well as of the height or corrugation
of graphene, remain difficult due to the mixed topography/electronic
sensitivity of STM.

\subsection{Atomic force microscopy (AFM)}

The force microscope was invented in 1986 \cite{Binnig1986} shortly
after the STM. Since then it has become a standard tool for characterizing
surfaces in atmospheric conditions, and is capable of atomic resolution
when employed in UHV in well-chosen detection modes \cite{Giessibl1992}.
It relies on the scanning of a sample surface with a sharp tip sufficiently
close to it so that short-range atomic forces between tip apex and
surface vary substantially during the scan. The tip force is transduced
into an electrical signal \cite{Meyer1992}, for instance by means
of the optical detection of the movement of the cantilever holding
the tip. The detailed analysis of the total force sensed by the tip
is rather complex: the tip-surface interaction includes many contributions
which have different dependence on the tip-surface distance and can
be repulsive or attractive: van der Waals, electrostatic, ionic, frictional,
chemical, and capillary forces might be of importance. Some of these
contributions are more prominent in UHV (e.g. those of a chemical
nature), others more in atmospheric conditions (e.g. capillary forces
linked to the presence of a water meniscus between the tip and sample
surface). Atomic resolution, if achieved, is solely due to short-ranged
chemical and/or electrostatic forces and an appropriate apex structure
of the tip (which is almost impossible to control); the long range
interactions, such as van der Waals forces between the surface and
the macroscopic part of the tip, provide the background contribution
which may be dominant. This is why achieving atomic resolution with
AFM can be difficult. Still, it has been achieved in many cases over
the last decade, for different surfaces in UHV and using the dynamic
mode of operation (see below). It is probably fair to say that achieving
atomic resolution with AFM is more difficult than with an STM. Under
atmospheric conditions, the operation of AFM is, however, usually
much more convenient than that of an STM, as it can operate at larger
working distances and accordingly cope with the presence of adsorbates
on the surface. Besides, unlike STM, the usage of AFM is not restricted
to conductive samples, i.e. the AFM is in principle more versatile.

Depending on the tip-surface working distance AFM is characterized
by different mechanical behaviors, which define different AFM operation
modes \cite{Giessibl2003}. In the static mode, no vibration is imposed
to the tip, which is merely dragged across the surface during the
scan. This mode relies on soft cantilevers, which can be deflected
by the weak atomic forces, and on suitable eigenfrequencies lying
away from the resonance frequency of the cantilever. These requirements
make this apparently simple mode of operation difficult to implement
in many situations; in addition, it results in both the tip and surface
being damaged during the scan. Dynamic modes, in which the tip is
oscillating perpendicular to the surface while scanned, are more widely
employed. Making the tip oscillate with an amplitude that may change
between sub-$\textrm{\AA}$ for stiff cantilevers to hundreds of $\textrm{\AA}$ for soft ones
avoids the problem where the tip jumps into contact with the surface
and hence maintains both the tip and the surface intact during the
scan. Two major scanning modes are in use: (i) in the amplitude modulation
regime the oscillation frequency is kept constant and out of resonance
and the oscillation amplitude and the phase are measured, while (ii)
in the frequency modulation regime the oscillation frequency is maintained
at resonance during the scan and the oscillation amplitude is kept
constant. The latter mode, which for UHV is the most suited for achieving
true atomic resolution, can be run either by keeping the resonance frequency
constant and measuring the corresponding vertical displacement of
the sample (similar to the constant current mode of STM), or by keeping
the sample height unchanged and hence measuring changes in the resonance
frequency (detuning), which is similar to the constant height mode
of STM. The latter mode is also called the non-contact mode (NC-AFM),
when the average working tip-surface distance is around 2-4 $\textrm{\AA}$
which is within the region of attractive forces; in the tapping mode
the working distance is smaller, and mostly repulsive forces are probed.
Dynamically probing the surface in AFM provides information about
the topography of the surface through the measurement of the shift
between the excitation and the actual tip oscillation needed to ensure
a constant tip amplitude vibration in frequency modulation, or the
variation of the amplitude of excitation needed to ensure a constant
tip frequency vibration in amplitude modulation. AFM also provides
information about energy dissipation at surfaces; this information
is obtained either by monitoring the phase shift between excitation
and tip vibration frequencies in the amplitude modulation mode, or
by measuring the amplitude change during the scan in the frequency
modulation.

\begin{figure}
\begin{centering}
\includegraphics[width=1\textwidth]{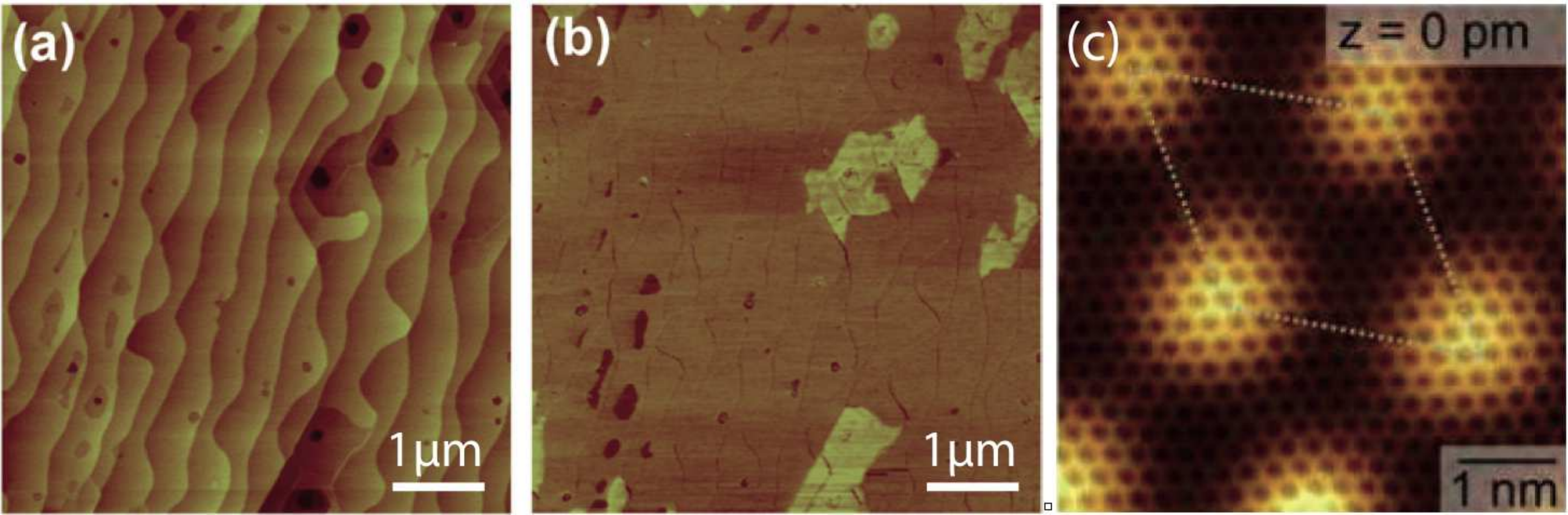}
\par\end{centering}

\caption{\label{fig:experimental techniques AFM} AFM (a) height and (b) phase
images measured in tapping mode in air for graphene with various numbers
of layers on SiC(0001): single, bi-, and trilayer graphene regions
show up as dark, medium, and bright shades in (b). (c) Atomic resolution
AFM map of the frequency shift, measured under UHV in dynamic non-contact
frequency modulation mode with a tip made inert by termination with
a CO molecule, of graphene on Ir(111). {[}Reproduced
from \cite{Hibino2010} by permission of IOP Publishing. All rights reserved (a) and by permission of \cite{Boneschanscher2012}. Copyright (2012) American Chemical Society (b).{]}}
\end{figure}

AFM is widely used under atmospheric conditions for rapid and rough
characterization of graphene on surfaces, usually at room temperature
\cite{Novoselov2004,Berger2006,Reina2008}. In this capacity it is
mostly employed to check the uniformity of graphene. It was also realized
that the phase signal, which is sensitive to the dissipation properties
of materials under mechanical excitation of the tip, shows strong
contrast from graphene-free regions to graphene-covered ones \cite{Bolen2009,Hannon2008},
and even from a single layer graphene to few-layer graphene \cite{Hibino2010},
see Fig. \ref{fig:experimental techniques AFM}. Though the in-depth
understanding of this effect is missing, it allows efficient determination
of the number of layers and of the size of graphene flakes or multilayer
patches.

Advanced AFM studies of graphene at high-resolution are rather scarce
\cite{Sun2011,Boneschanscher2012,Voloshina2013}, especially in the
case of epitaxial graphene. It was shown that for quantitative height
determinations care should be taken considering the tip-graphene chemical
interaction \cite{Boneschanscher2012}, Fig. \ref{fig:experimental techniques AFM}(c),
and concerning the working distance and frequency shift (i.e. the
resonance frequency change when bringing the oscillating tip closer
to the surface from infinity) in constant frequency shift imaging
\cite{Voloshina2013}. With such care, the height modulations in the
moir\'{e} pattern between graphene and Ir(111) surface (see Section \ref{sub:Ir(111)})
could be estimated.

\subsection{Transmission electron microscopy (TEM)}

TEM development dates back to the 1930's \cite{Ruska1980}. The electrons
generated by an electron source are injected into a column by means
of electromagnetic lenses that shape a high energy electron beam (from
several 10 keV to few 100 keV). The electrons then pass through a
sample, and the sample plane is imaged through another set of electromagnetic
lenses. The sample must be sufficiently small that it only absorbs
a limited part of the electron beam. The high energy of the electrons
allows high-resolution imaging of the sample; however, aberrations
in the electron optics restricted the ultimate resolution achievable
with TEM for a long time. Aberration-corrected instruments are now
available and they allow resolutions close to 1 $\textrm{\AA}$ \cite{Haider1998}
and below. The electron optics column allows us to observe, instead
of the image plane, the Fourier plane, i.e. to perform a diffraction
experiment. The electron optics instrumentation in a conventional
TEM is designed to make full-field images of the samples. Adding scanning
coils and forming a small electron spot allows performance of STEM,
i.e. imaging of the sample by scanning it. This technique enables
energy dispersive X-ray spectroscopy, electron energy loss spectroscopy,
or annular dark field imaging (the latter allowing atomic number contrast
imaging) with ultra-high spatial resolution. The aberration corrected
STEM offers spatial resolution below 1 $\textrm{\AA}$ \cite{Batson2002}. Transmission
electron microscopes capable of high resolution imaging of samples
heated to high temperatures were demonstrated in the early 2000's
\cite{Hansen2002}. Due to the cluttered environment around the sample
in a TEM or STEM, there is however little space for performing \emph{in
operando} growth experiments.

\begin{figure}
\begin{centering}
\includegraphics[height=5cm]{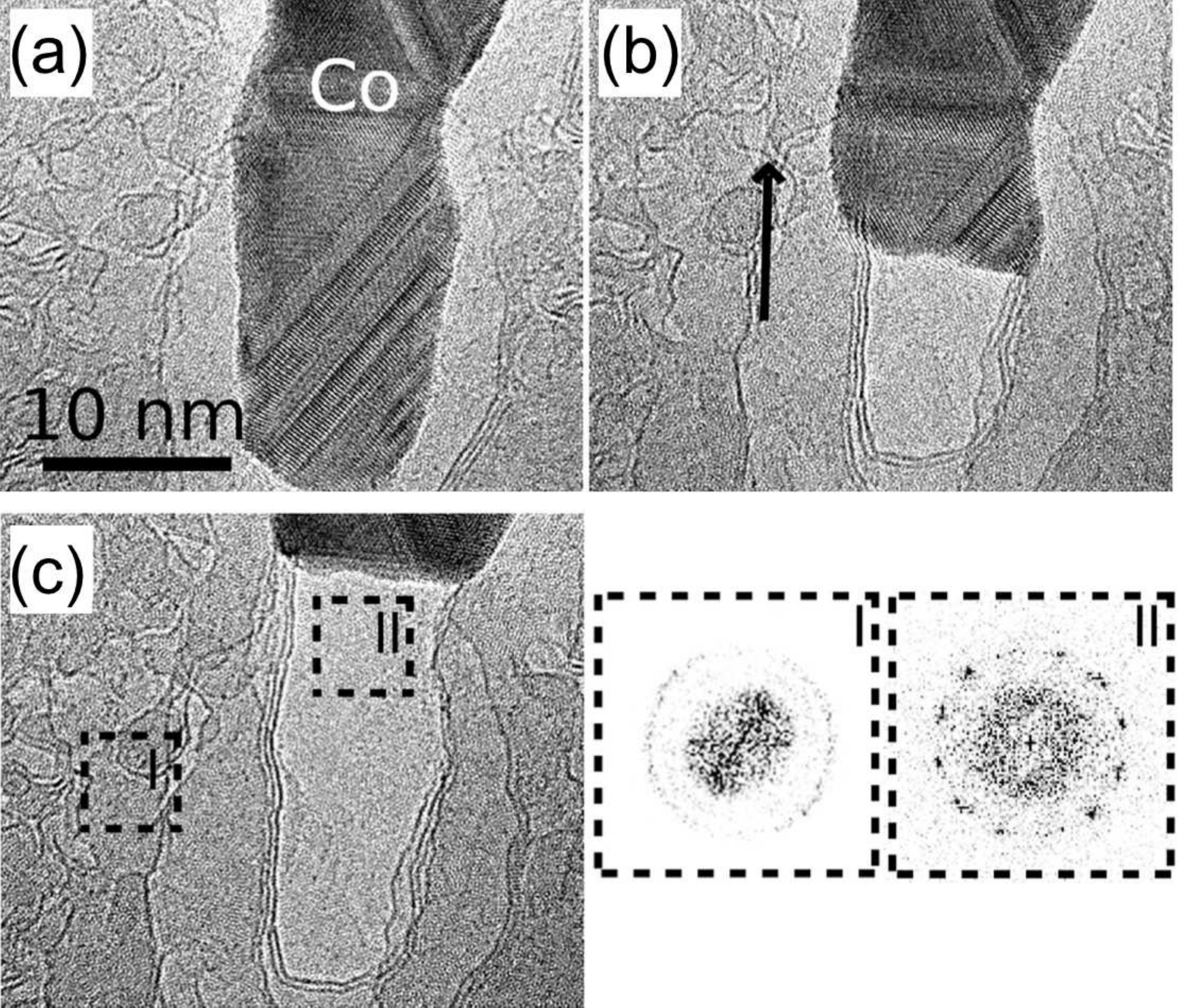}
\par\end{centering}

\caption{\label{fig:experimental techniques TEM} 10 nm thick lamella of Co
on amorphous carbon, at 913 K, growing ((a) to (b)) and
retracting ((b) to (c)) under an energetic (200 keV) electron beam,
causing a transformation from amorphous carbon to graphene. The presence
of graphene after the retraction of Co is confirmed in the Fourier
transform of the images (II). {[}Reprinted with permission from \cite{RodriguezManzo2011}. Copyright (2011) American Chemical Society.{]}}
\end{figure}

The study of fragile samples, such as biological matter, and of materials
composed of light (low atomic number) elements (e.g. graphene, boron
nitride), required the development of high resolution TEM with the
electron energy below 100 keV in order to prevent the knock-on of
atoms by electrons. For carbon the knock-on occurs for electrons with
about 80 keV energy \cite{Smith2001}. Imaging graphene without inducing
a substantial amount of defects thus is not possible using beams of
above this energy. The purely two-dimensional nature of graphene is
an additional difficulty in studies of its growth. Indeed, in a plane-view
geometry (graphene observed from above), the presence of a substrate
composed of strong electron scatterers (the metal atoms), and thicker
than several nm, makes it impossible to detect the presence of graphene.
In a cross-section geometry, detecting a single layer is a tedious
task which requires simulation of TEM images, and observing the projection
of a two-dimensional phenomenon (growth on a planar substrate) in
one dimension (perpendicular to the cross-section) greatly complicates
understanding. In practice very few studies of the growth of graphene
with TEM exist. One noticeable exception is the plane-view \emph{in
operando} study of the transformation of amorphous carbon into graphene
in the presence of a thin (few nanometres) metal catalyst \cite{RodriguezManzo2011}
(Fig. \ref{fig:experimental techniques TEM}). Noteworthy also are studies in
the 1960s of the structure of graphene prepared by evaporation
of carbon onto metal foils: after growth, graphene was transferred
to TEM grids by chemically etching their metallic substrate, and was
identified in TEM images with the help of electron diffraction \cite{Irving1967},
an approach which was used again years later \cite{Reina2008}.

\subsection{Low-energy and photoemission microscopy (LEEM and PEEM)}

The first images of surfaces obtained with PEEM and LEEM were published
in 1966 \cite{Turner1966} and 1985 \cite{Telieps1985a,Telieps1985b}.
The two techniques rely, just like TEM, on an electron optics imaging
column composed of electromagnetic lenses and operating using high
energy (typically 20 keV) electrons. PEEM relies on the extraction
of photoelectrons from the sample, which is usually achieved with
an ultraviolet source (laser, lamp or X-rays). LEEM relies on secondary
or reflected electrons, created by a low-energy electron beam. This
beam is shaped from electrons created by an electron source, accelerated
to high energy electron beam inside an electron column (like in TEM),
and decelerated before reaching the sample surface. In both techniques
the low energy (0-100 eV, typically) of the electrons used for imaging
ensures extreme surface sensitivity to a few topmost layers of the
sample.

Since no actual contact occurs between the electron optics elements
and the sample surface, both techniques are well suited for temperature-dependent
studies, and, as long as the pressure remains reasonably low (typically
below 10$^{-7}$mbar), they can be run \emph{in operando},
i.e. during the evaporation of species onto the sample surface or
in the presence of gases reacting at the sample surface. Thanks to
the strong interaction between low-energy electrons and matter and
due to the fact that PEEM and LEEM are full-field microscopies, high
measurement rates are possible, typically from 1 to 100 Hz, which
allows true real time monitoring of the growth on surfaces, as anticipated
early on \cite{Bauer2011}. These studies are routinely performed
with lateral resolution of about 10 nm. Recent developments, using
aberration-corrected electron optics inspired by TEM instrumentation,
allow the attainment of lateral resolutions of a few nm \cite{Tromp2010}.

In PEEM, images reveal work function contrasts. In LEEM, the electron
reflectivity is imaged, which is linked to the density of states and
the sample structure perpendicular to the surface. As in a standard
optical or transmission electron microscope, instead of observing
the image plane, one can observe the Fourier plane. In LEEM, this
allows the performance of LEED experiments (see next subsection),
and micro-LEED experiments by using an aperture placed before the
sample in order to select one region of the sample. In PEEM, provided
that an energy analyzer is added after the imaging column, this allows
performance of angle-resolved photoemission spectroscopy to obtain
constant energy cuts of the electron band structure of the surface
\cite{Schmidt1998}.

\begin{figure}
\begin{centering}
\includegraphics[height=4cm]{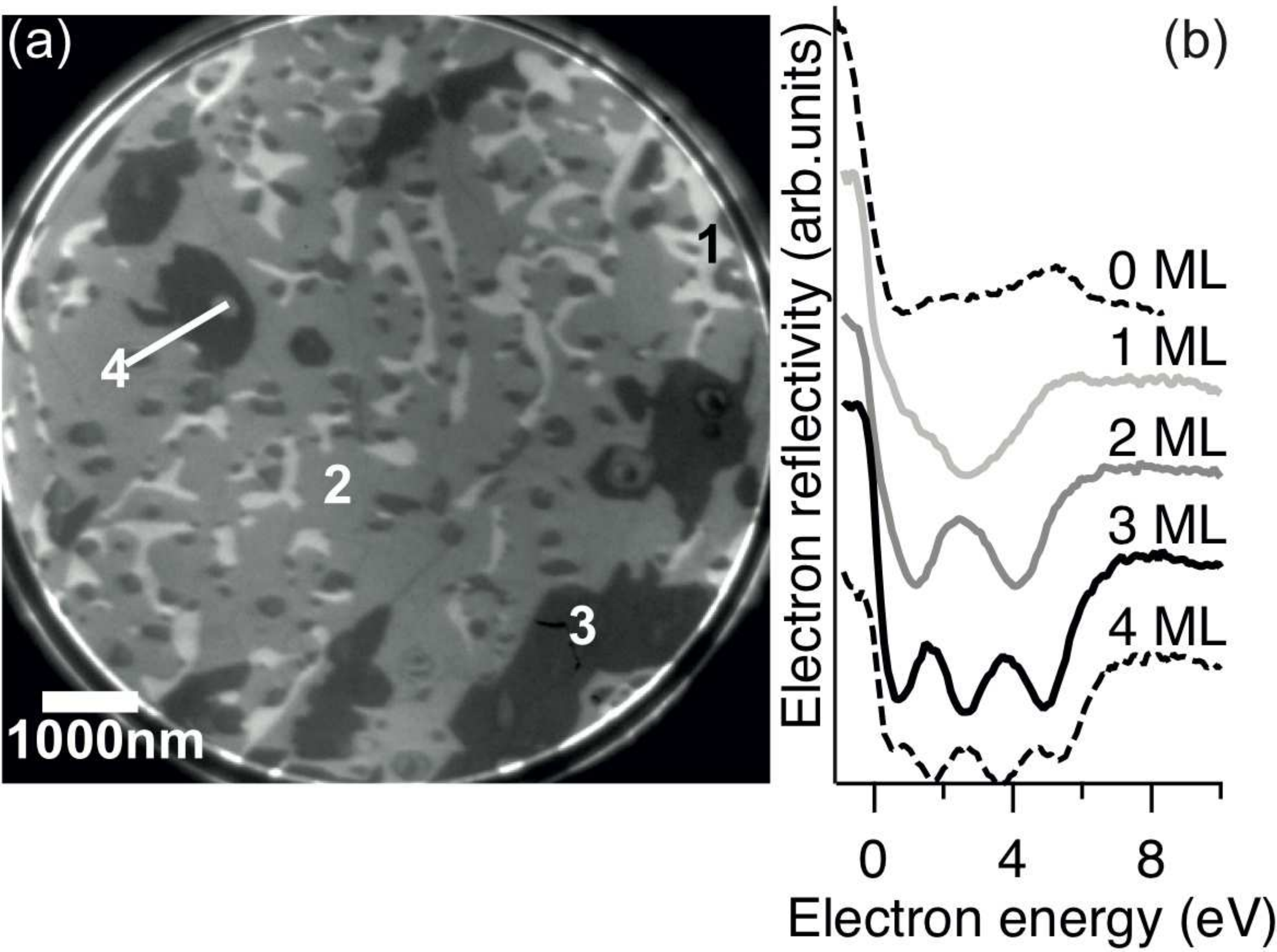}
\par\end{centering}

\caption{\label{fig:experimental techniques LEEM} (a) LEEM image of graphene
on SiC(0001) and (b) electron reflectivity at different locations
in (a), allowing a count of the number of layers (the number of low
energy oscillations plus one). {[}Reproduced from \cite{Ohta2008} by permission of IOP Publishing. All rights reserved.{]}. }
\end{figure}

The first published LEEM studies of graphene \cite{Ohta2008} allowed
determination of the extent of few-layer graphene regions and a count
of the number of graphene layers in graphene/SiC samples by taking
advantage of the electron reflectivity sensitivity to the out-of-plane
structure, see Fig. \ref{fig:experimental techniques LEEM}. LEEM
has been employed for an \emph{in operando} analysis of the growth
of graphene on Ru(0001) \cite{Sutter08}, which revealed for instance
that graphene islands grow down the atomic step edge staircase of
the substrate (see Section \ref{sub:Ru(0001)-in-methods}). In the
same year it was established that electron reflectivity measurements,
possible in a spatially-resolved manner in LEEM, allowed \emph{in
operando} measurements of the C adatom concentration on metal surfaces
and a connection to be made to the elementary processes occurring
during growth \cite{Loginova08} (Sections \ref{sub:CVD-Nucleation}
and \ref{sub:Rate-equation-analysis}). LEEM, coupled to micro-LEED
measurements, was then employed to conveniently identify rotational
variations in graphene on metals \cite{Loginova09}. PEEM also has
been employed to study the growth of graphene on metals \cite{vanGastel2009},
though it provides only limited structural information (no diffraction
capability like in a LEEM).

\subsection{Surface-sensitive electron diffraction (LEED and RHEED)}

LEED and RHEED are two techniques whose use spread in the 1960s,
as they were accompanying the development of UHV instrumentation and
other techniques designed for studying the growth of highly crystalline
surfaces. The first technique uses low-energy electrons (below 100
eV typically). Such electrons experience scattering events whereby
they rapidly lose their energy, which makes the technique very sensitive
to the few topmost layers of the sample. In the typical scattering
geometry, LEED provides snapshots of in-plane cuts of the reciprocal
space lattice of the topmost layer of the sample. RHEED uses high-energy
electrons (typically a few 10 keV). Such electrons penetrate matter
much deeper than the low energy electrons used in LEED. In RHEED however,
this extended penetration is avoided by the scattering geometry, which
corresponds to grazing incidence and exit. A RHEED image captures
part of the scattering contribution which, due to the finite thickness
of the sample (as seen by electrons), extends perpendicular to the
surface. The aspect of the RHEED image is determined by the length-scale
of the surface roughness as compared to the coherence length of the
electron beam, which is typically of the order of 10 nm: electrons
penetrate much deeper in samples with roughness length-scale well
below this value, which yields spotty patterns, while in the opposite
situation the sample is actually seen as two-dimensional which yields
streaky patterns.

The analysis of electron diffraction diagrams can only be performed
quantitatively in the rather complex framework of dynamical theory
of diffraction explicitly taking into account multiple diffraction
events \cite{Lagally1985,vanHove1986}. This is noticeably required
for interpreting the variations of LEED intensity as a function of
electron energy, from which one can derive valuable information about
the structure of the sample. The location of the diffraction contributions
in reciprocal space are, however, little affected by multiple diffraction
effects, which allows straightforward determination of the lattice
parameter in the plane of the surface. Best accuracy is usually achieved
with the help of RHEED or of spot-profile analysis LEED \cite{Sheithauer1986}.
Moreover, the strong interaction between electrons and matter allows
fast acquisition rates, which is essential for the \emph{in operando}
monitoring of surface growth; RHEED, with its open geometry, is ideally
suited for this purpose \cite{Blanc-Coraux-2013}.

\begin{figure}
\begin{centering}
\includegraphics[height=5cm]{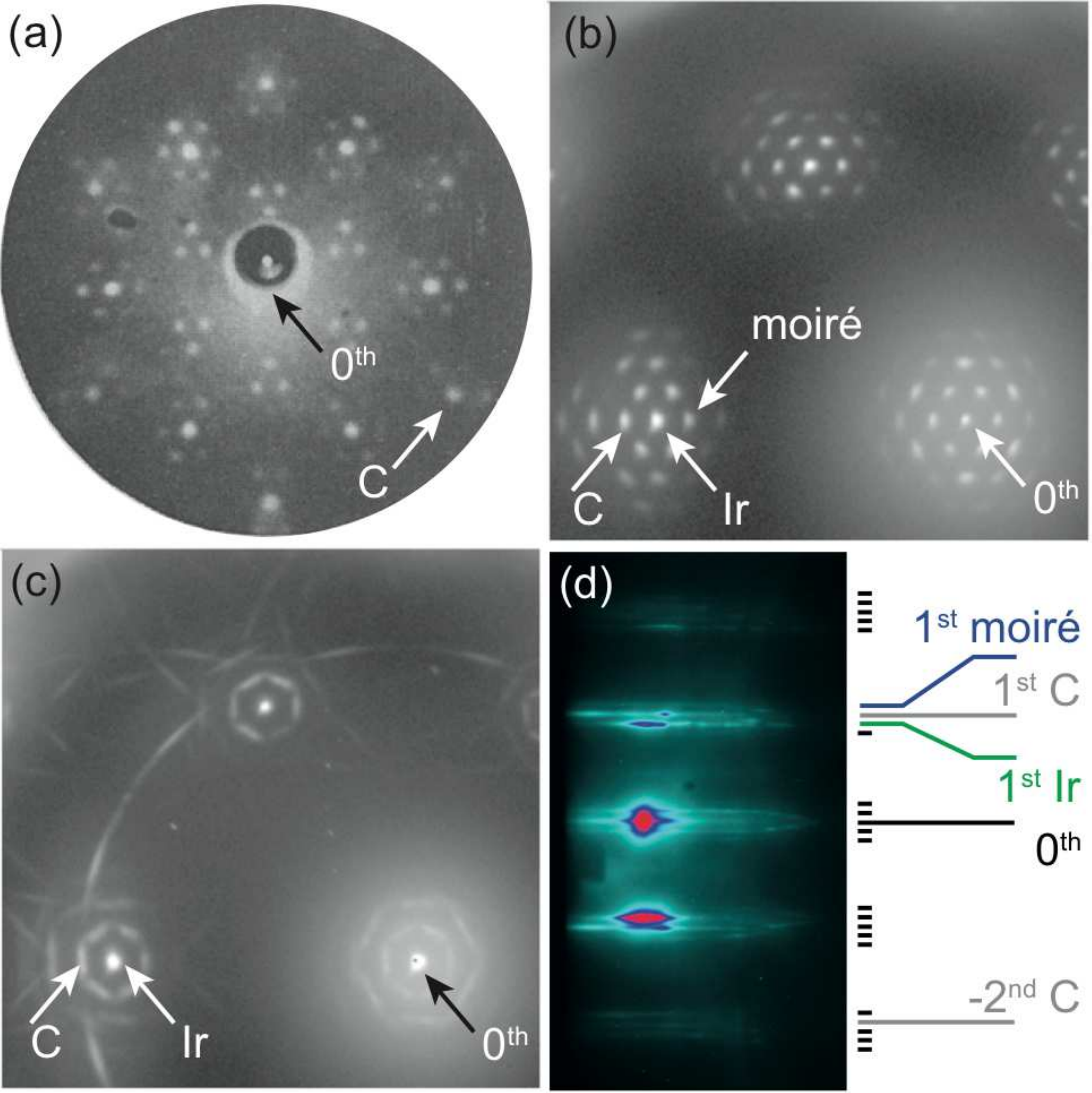}
\par\end{centering}

\caption{\label{fig:experimental techniques ED} (a) LEED pattern of graphene
on SiC(0001), and SPALEED pattern on Ir(111) covered by (b) an ill-
and (c) a well-ordered graphene layer. The reflected beam (labeled
0$^{{\rm th}}$) is marked by an arrow. The presence of diffusion
rings around the reflected beam (c) signals a scatter in the in-plane
orientation of graphene with respect to its substrate; on the contrary,
well-defined spots track well-defined epitaxial relationships. (d)
RHEED pattern of graphene on Ir(111) around the specularly reflected
beam (labeled 0$^{{\rm th}}$). All patterns were obtained under UHV
conditions. Both LEED and RHEED reveal the presence of superstructures
(e.g. moir\'{e} in (a) and (d)). {[}Reprinted from \cite{vanBommel1975}, with permission from Elsevier
(a), from \cite{Hattab2011} by permission of IOP Publishing. All rights reserved (b,c) and with permission from \cite{VoVan2011}. Copyright (2011), AIP Publishing LCC. (d).{]}}
\end{figure}

LEED structural studies are amongst the earliest characterizations
of epitaxial graphene. They provided understanding of the first steps
of graphene formation on SiC \cite{vanBommel1975,Forbeaux1998} (Fig.
\ref{fig:experimental techniques ED}(a)) and unveiled the moir\'{e} superstructure
arising between graphene and a metallic substrate \cite{Nieuwenhuys-TPD-1976}.
LEED was employed to understand the detailed structure of graphene
on metals, not only in the plane of the surface (Fig. \ref{fig:experimental techniques ED}(b,c))
but also perpendicular to it, with the help of quantitative dynamical
theory analysis \cite{Gamo1997,Moritz,Hamalaninen-2014}.
 More
recently, when performed with micro-sized electron beams inside a
LEEM, LEED revealed the presence of rotational variants in graphene
on metals \cite{Loginova2009b,Sutter2009}. RHEED has been used much
less than LEED thus far for studying graphene, and mostly as a way
to determine the graphene/substrate epitaxy \cite{Moreau2010,VoVan2010}
(Fig. \ref{fig:experimental techniques ED}(d)). Clearly, the potential
of this technique for studying graphene growth is far from being fully
exploited.

\subsection{Scanning electron microscopy (SEM)}

The early development of the SEM instrumentation dates back, similarly
to TEM, to the 1930s \cite{vonArdenne1938}. As in a TEM, electrons
produced by an electron gun are shaped into a beam through a column
composed of electromagnetic lenses, which is focussed onto the sample
surface and scanned through it with the help of coils. The energy
of the electron beam ranges from a few 100 eV to a few keV. Both secondary
electrons (having experienced energy losses due to scattering events
inside the sample) and back-scattered electrons (elastically scattered)
can be analyzed independently using specific detectors. The secondary
electrons, having low energy, can only escape the sample if they are
produced sufficiently close to the impact point of the electron beam
onto the surface, which allows high spatial resolution of the order
of 1 nm. The back-scattered electrons are those reflected elastically
from the sample. Since they have a rather high energy, they can scatter
from rather deep within the sample (typically, hundreds of nm) from
the surface. Back-scattering, being more efficient when the atomic
number of the scatterer is high, provides chemical information about
the sample. The back-scattered electrons may also be employed for
performing diffraction experiments (electron back-scatter diffraction,
EBSD) which is valuable for determining the bulk structure of the
sample. Other detection modes are available in SEM, such as those
based on exploiting X-rays, but we will not discuss them here. Environmental
SEM was developed for a large variety of \emph{in operando} studies
of materials, especially at high temperature \cite{Biswas1983,Erhart1984}
(note, however, that the time resolution is limited by the scanning
time).

\begin{figure}
\begin{centering}
\includegraphics[height=5cm]{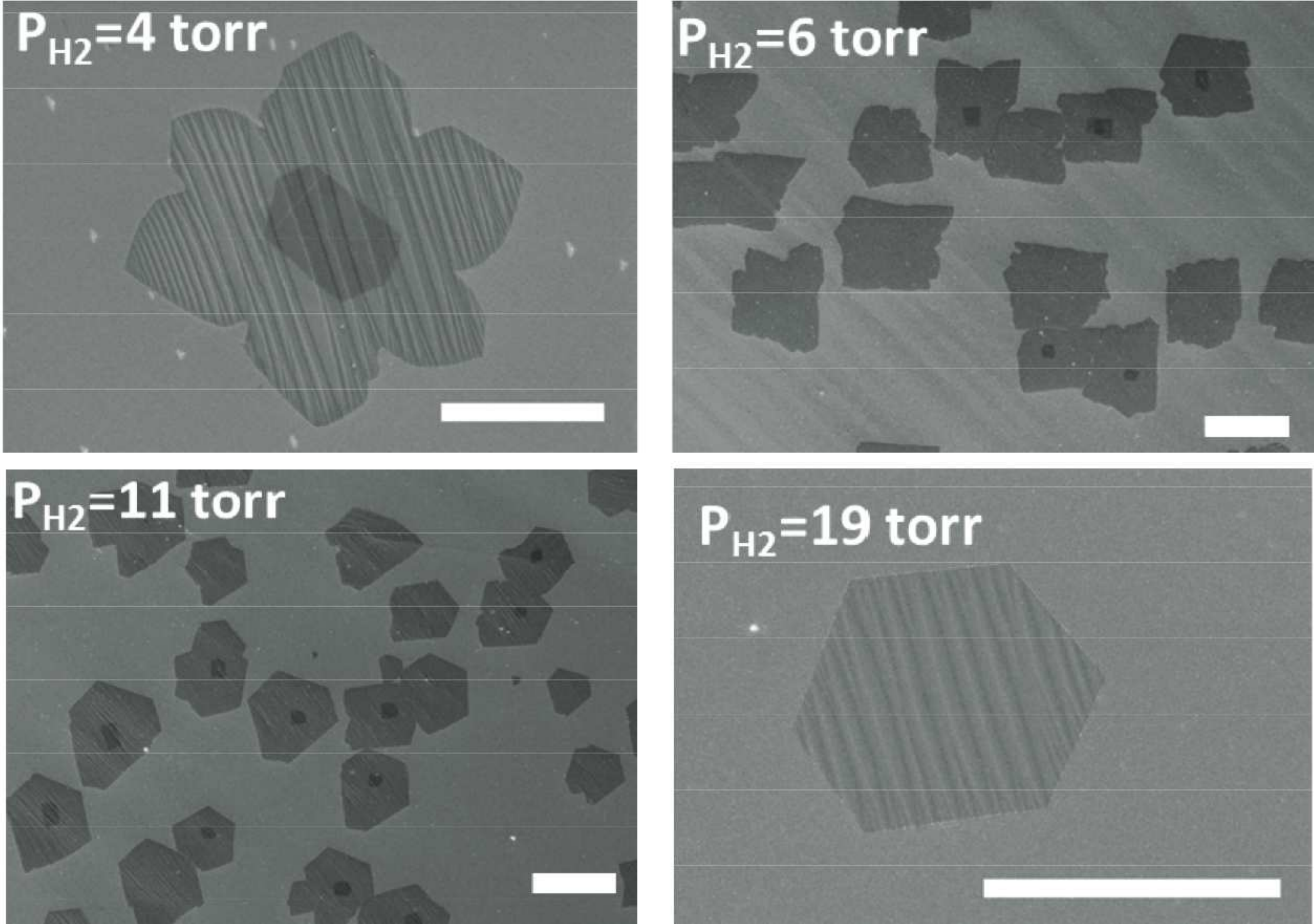}
\par\end{centering}

\caption{\label{fig:experimental techniques SEM} SEM images with secondary
electrons of graphene prepared by CVD on Cu foils with various partial
pressures of hydrogen. Copper appears bright. The graphene islands
often contain a bilayer (or multilayer) region at their center. {[}Reprinted
with permission from \cite{Vlassiouk2011}. Copyright (2011) American Chemical Society.{]}}
\end{figure}

SEM is widely employed for \emph{ex situ} characterizations of the
growth of graphene \cite{Kim2009,Li2009}. Because of its atomic thickness
graphene is usually detected with secondary electrons (usually unspecified
in publications) which probe only a small topmost fraction of the
sample. It allows the study of defects such as wrinkles \cite{Chae2009}
and provided valuable insight into the growth of graphene on Cu foils,
which was beneficial, e.g., to understand the dependence of the shape
of graphene islands on growth conditions \cite{Vlassiouk2011} (Fig.
\ref{fig:experimental techniques SEM}). Using back-scattered electrons,
EBSD was also successful employed to relate the shape of graphene
islands, as observed with secondary electrons, to the surface termination
of the individual metal grains supporting the growth \cite{Wood2011}.
The full power of SEM has not yet been fully exploited, especially
in the view of \emph{in operando} studies, which are now possible
at temperatures above 1300 K \cite{Subramanian2006}.

\subsection{Raman spectroscopy and microscopy}

Raman spectroscopy was used in materials science as early as the 1940's.
The technique relies on the Raman effect, whereby an electromagnetic
wave gains or loses part of its energy by exciting phonons in a material.
The energy lost or gained measured in a Raman spectroscopy experiment
characterizes, among other things, the bonds (e.g. interatomic, van
der Waals) responsible for cohesion in the material of interest. A
Raman spectrum thus provides a signature of a material, and contains
information about processes that weaken or strengthen bonds (by charge
transfers or strains, for example), and concerning interactions between
phonons and other excitations (such as electronic waves, plasmons,
etc). The typical energy losses (discussion of energy gains, yielding
smaller Raman signals, is not discussed here) of photons in matter
range between few 10 to few 1000 cm$^{-1}$, i.e. 1 to
100 meV.

In practice, the light source used for a Raman spectroscopy experiment
is most often a visible light or infrared laser. After interaction
with the sample, the beam is passed through a filter rejecting the
strongest scattering contribution, which is elastic (Rayleigh, zero
energy loss), and then injected into a spectrometer coupled to a high
sensitivity detector. In an alternative setup Raman spectroscopy may
be implemented in a confocal microscope, allowing for high spatial
resolution (down to the diffraction limit) of samples.

\begin{figure}
\begin{centering}
\includegraphics[height=6cm]{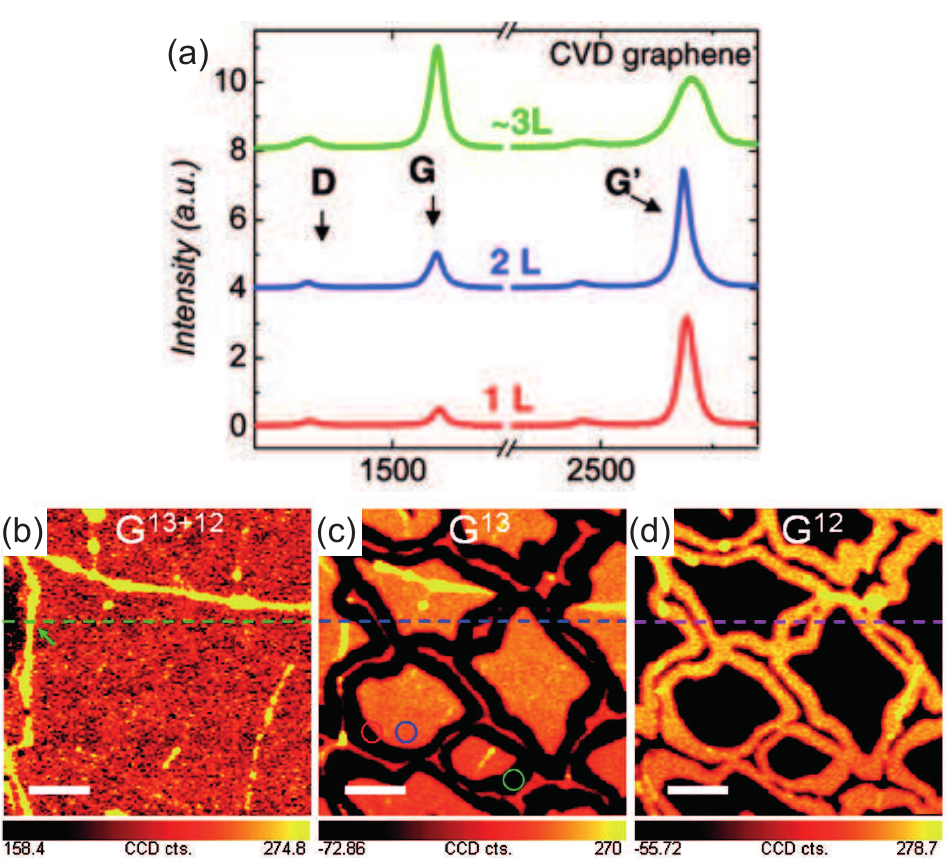}
\par\end{centering}

\caption{\label{fig:experimental techniques Raman} (a) Raman spectrum measured
at various locations in a graphene film, prepared by CVD on Ni thin
films and transferred to SiO$_{2}$/Si, for single, bi-, and trilayer
regions, showing the presence of D, G, and G' (2D) peaks. (b-d) Raman
maps of the G band intensity of a single layer of graphene prepared
by CVD with $^{12}$CH$_{4}$ and $^{13}$CH$_{4}$ on Cu foils and
transferred to SiO$_{2}$/Si. The growth has been performed in two
steps, one with $^{12}$CH$_4$ and the other with $^{13}$CH$_{4}$.
(b) is the sum of the intensities around the G bands of the two isotopes,
(c) is the intensity around the G band of $^{12}$C, and (d) the intensity
around the G band of $^{13}$C (the two bands are shifted by about
60 cm$^{-1}$). {[}Reprinted with permission from \cite{Reina2008}. Copyright (2008) American Chemical Society
(a) and \cite{Li2009b}. Copyright (2009) American Chemical Society (b-d).{]}.}
\end{figure}

Graphene has three main features in its Raman spectrum, the so-called
D, G, and 2D (also called G') modes, appearing respectively at about
1350, 1580 and 2700 cm$^{-1}$ \cite{Ferrari2006}, see
Fig. \ref{fig:experimental techniques Raman}(a). The first mode is
only present in samples with defects. The G mode is a first order
process, doubly degenerate, involving in-plane transverse optical
and longitudinal optical excitations at the Brillouin zone center
\cite{Ferrari2007,Malard2009}. The two others are second order modes,
but doubly resonant (thus readily observable, like the G mode), involving
one in-plane transverse optical phonon at the K point of the Brillouin
zone and one defect for the D mode, and two in-plane transverse optical
phonons at the K point for the 2D (G') mode \cite{Ferrari2007,Malard2009}.
The relative intensity and width of the 2D mode \cite{Ferrari2006},
as well as the occurrence of a shear mode for small energy losses
in polarized light experiments \cite{Tan2012}, provide valuable information
about the number of layers in graphene stacks. Polarized light is
useful in identifying the nature of the edges of graphene flakes \cite{Casiraghi2009}.
The relative orientation of graphene layers in graphene stacks can
be tracked by specific Raman signatures \cite{Righi2011}. The high
sensitivity of Raman scattering to the mass of the atomic nuclei makes
the study of experiments performed with different carbon isotopes
(Fig. \ref{fig:experimental techniques Raman}(b-d)) a rich area for
understanding growth, for instance the competition between surface
and volume processes \cite{Li2009b}. Raman modes are also strongly
affected by various perturbations, including the strains of C-C bonds
occurring under stress \cite{Mohiuddin-2009}, which effectively modify
the stiffness of the C-C bond and charge transfer \cite{Das-Pisana-2008}
through the strong electron-phonon coupling in graphene. This sensitivity
may be exploited, for instance, to characterize the nature of the
interaction between graphene and its substrate.

\subsection{Surface-sensitive X-ray diffraction (SXRD)}

In the energy range corresponding to high resolution structural characterizations,
typically 10 keV, X-rays interact only weakly with the electronic
clouds of atoms. This results in large penetration depth of X-rays
in matter. A grazing incidence geometry limits this penetration depth
to a around 10 nm, which allows the study of interfaces close to the
surface \cite{Marra1979}. The study of surfaces with SXRD usually
requires UHV and samples prepared \emph{in situ} \cite{Robinson1986,Feidenhans'l1989}
in order to avoid contaminations which would alter the surface. In
an SXRD experiment, a well-focussed X-ray beam is directed onto the
sample surface with an incidence angle of a few tenths of a degree.
The requirement for a well-defined incidence angle makes low-divergence
beams, such as those available from synchrotron sources, preferable.
Due to the weak interaction between hard X-rays and matter, the high
intensity available at such a source is another definitive advantage.
An SXRD experiment consists of measuring the scattering from the sample
perpendicular to its surface (by varying the exit angle) and along
lines (crystal truncation rods) passing through the in-plane cut of
the reciprocal space. The distance between these lines characterizes
in-plane strains, the width of these lines is related to strain fields
and the size of the structurally coherent domains, and the intensity
modulation along the lines relates to the out-of-plane structure of
the sample.

\begin{figure}
\begin{centering}
\includegraphics[height=6cm]{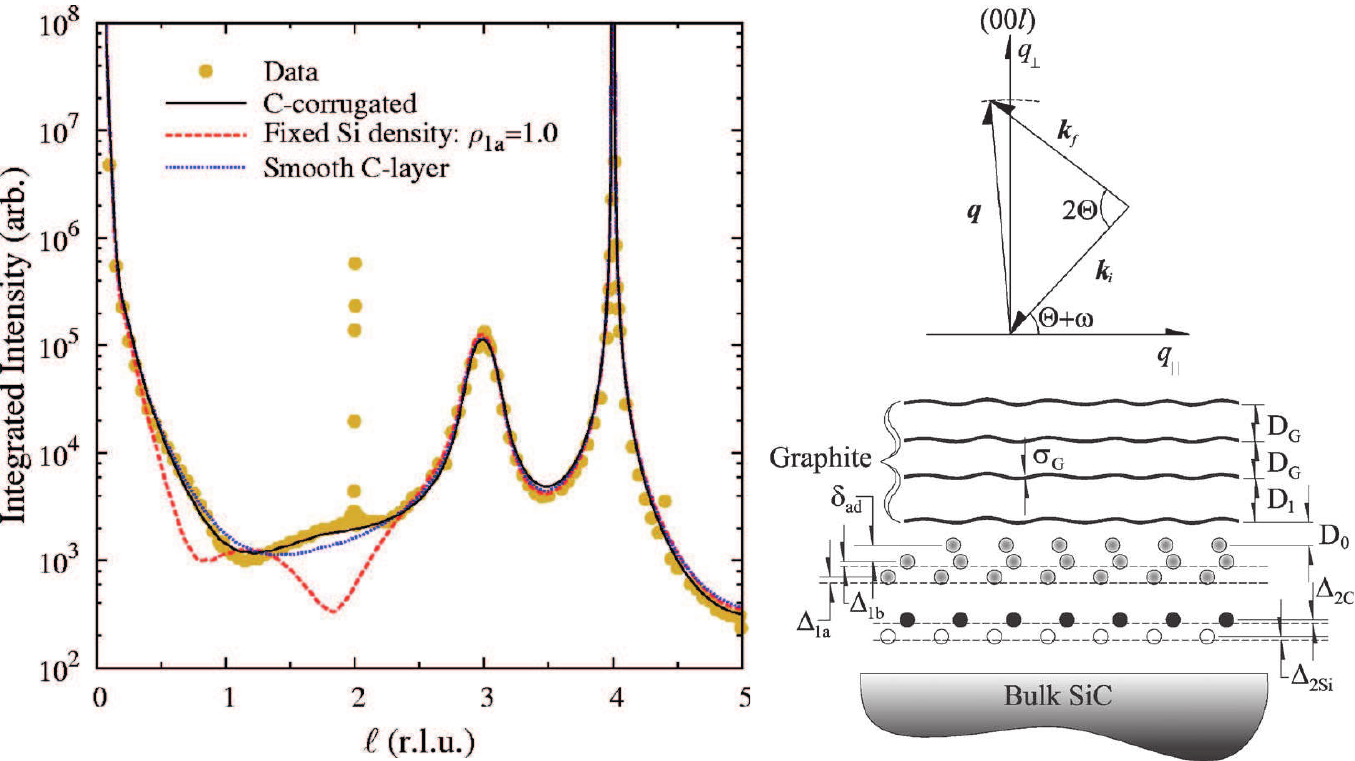}
\par\end{centering}

\caption{\label{fig:experimental techniques SXRD} Specular reflectivity (X-ray
scattered intensity along the reciprocal space coordinate $l$ perpendicular
to the surface) of nine-layer graphene on SiC(000$\overline{1}$),
and fittings of the data (disk symbols) based on different structural
models. The scattering geometry is shown in the top right cartoon;
the various parameters employed for parametrizing the model employed
for the fittings is shown at the bottom right. {[}Reprinted with
permission from \cite{Hass2007}. Copyright (2007) by the American Physical Society.{]}}
\end{figure}

A study of graphene on a substrate with SXRD is in principle a difficult
one: carbon is one of the lightest scatterers, and is only present
in very low amounts. The first SXRD study was performed in a rather
favorable situation of multilayer graphene on a substrate also composed
of rather light scatterers, SiC. The study yielded the out-of-plane
structure of multilayer graphene, revealing for instance rotational
disorder \cite{Hass2007}, see Fig. \ref{fig:experimental techniques SXRD}.
The first study on a single layer graphene on Ru(0001) revealed a
surprisingly large superstructure and allowed determination of the
graphene-metal distance \cite{Martoccia}. SXRD was later employed
for studying the growth of graphene on a metal, highlighting the role
of preparation conditions in the formation of defects in graphene
\cite{Blanc2012,Jean-Zhou-2013}, and the positive thermal expansion
of graphene on a metal down to low temperature \cite{Jean-Zhou-2013}.
Many of the features observed in these studies are almost undetectable
by other techniques. This high resolution however comes at the expense
of demanding experiments and relatively heavy data analysis.

\subsection{X-ray and ultraviolet photoelectron spectroscopy (XPS/UPS)}

The development of X-ray photoelectron spectroscopy (XPS) and ultraviolet photoelectron spectroscopy (UPS),
for
the the investigation of the properties of materials, dates back to the 1950s \cite{Siegbahn1956}
and 1960s \cite{Turner1962}, respectively. These techniques are named after the photoelectron effect, whereby
an energetic photon (X-ray, UV) excites a core-level electron above the Fermi level. If
the photon energy is sufficient for the photoelectron to exceed the sample work-function, a free electron
with sufficient kinetic energy is created which escapes from the sample.
By measuring
the energy threshold for photoemission, one may extract the work-function, and then
subtracting it away from the photoelectron kinetic
energy, measured with the help of an electron energy analyzer, an electron binding energy for each of
the photo-excited core levels is obtained.

In the laboratory, metal anodes (e.g. Al) and discharge sources (e.g. He) are used as X-ray and UV sources.
Relying on electrons, XPS and UPS are usually operated under high or ultra-high vacuum. Synchrotron radiation
sources are valuable alternatives which offer high photon flux (thus allowing one to reduce considerably the
counting times) and energy tunability (thus surface sensitive tunability). XPS and UPS are surface sensitive
techniques, whose probing depth is set by the strong interaction between electrons and matter: depending on the
sample atomic number, a photoelectron will only escape the sample, in the photon excitation energy range usually
employed, from depths of between a few to a few tens of  $\textrm{\AA}$ngstr\"{o}ms.
The precise value of the core level energy
not only depends on the atomic element of interest, but also on the kind and nature of bonds these atoms form with
their environment: chemical identification, hybridization state, and charge transfer/oxidation states are
accordingly readily identified with XPS and UPS.

XPS and UPS are routinely used to characterize the degree of purity and presence of oxygen groups in graphene
samples, which are obtained via chemical routes, by means of the analysis of the carbon core levels.
XPS and UPS have also  been
employed since the 1990s to characterize graphene on metals \cite{Nagashima1994}, and more recently graphene
on SiC \cite{Rollings2006}. In these studies the interaction between graphene and its support, as well as the
number of graphene layers, could be analyzed by studying the carbon core levels.
The study of metal core level
changes upon graphene growth also made it possible to investigate the graphene-metal interaction \cite{Preobrajenski2008,Lacovig}.
Noteworthy is also the use of XPS at synchrotron sources as a tool for {\em in situ} monitoring of the growth of graphene
and its evolution upon, e.g. heating \cite{Lizzit201068,Miniussi2011}.

\subsection{Near-edge X-ray absorption fine structure (NEXAFS)}

In X-ray absorption experiments in a transmission geometry, one measures elastically-scattered photons: above an
absorption edge, corresponding to the excitation of photoelectrons from core levels, elastic scattering is strongly
reduced (due to the increase of inelastic scattering corresponding to photoelectron excitation). One may also measure
the electron yield, especially in a non-transmission geometry, which is linked to the absorption process. Synchrotron
sources, owing to their energy tunability, high brilliance, and control over the photon polarization, have given a strong
push to X-ray absorption spectroscopies, making it possible to extract polarization-dependent absorption {\it vs.}
energy spectra
with high signal to noise ratios, and accordingly, to detect low amounts of absorbing atoms in matter and to derive valuable
information about the local environment of the absorbing atoms. In the following we will address specifically the
10 eV-region above the absorption edge, the so-called near-edge X-ray absorption fine structure (NEXAFS) or X-ray absorption
near edge structure (XANES) region. In this energy range, electronic transitions towards empty states close to the Fermi Level
or low-energy continuum states are involved \cite{koningsberger1988}.

NEXAFS studies of graphene (on substrates) have focussed on the C 1s spectra. Addressing the spectra of the topmost
substrate layer, for instance in order to explore the effects of the graphene-substrate interaction, is a challenging
issue. Indeed, the absorption edges of Si and metals are generally above 1 keV, an energy at which a very grazing incidence
is required to make the technique actually surface sensitive; even then, the surface sensitivity is no better than roughly
10 atomic layers from the substrate, which makes it difficult to extract information from the very surface of a sample. In
the case of graphene on SiC, C is in both graphene and the substrate. In order to probe specifically C from graphene, a
retarding potential of a few 100 V allows one to discriminate those low-energy electrons, detected in electron yield modes,
which originate from layers deeper in the sample \cite{aristov2010}. A typical C 1s NEXAFS spectrum exhibit several features, mainly
three, associated each with electronic transition from the core level to the $\pi$* and $\sigma$* orbitals, like in
graphite \cite{bruhwiler1995}. Adjusting the polarization vector of the X-ray beam with respect to the graphene surface
makes it possible to probe selectively some of these specific transitions \cite{fischer1991}, and to detect the effects of
the graphene-substrate interaction \cite{lee2010}. Other features make it possible to detect charge transfers and hybridization between
graphene and the substrate \cite{Preobrajenski2008}.

\subsection{X-ray standing waves (XSW)}

X-ray standing waves is a technique which combines diffraction and
a chemically sensitive process, such as X-ray fluorescence of photoemission
\cite{Batteran1964}. It gives the positions of atoms in the standing
wave field generated by a crystalline substrate, while providing information
about the interaction between the atoms at the surface. The technique
proved especially powerful in determining the structure of interfaces
and atomic and molecular adsorbates in the sub-monolayer regime \cite{Zegenhagen1993,Woodruff2005}.
A typical XSW experiment consists of measuring photoemission spectra
from an adsorbate for various energies of the X-ray beam close to
the Bragg condition. While the wavelength of the standing wave varies
only slightly, the spatial phase shift of the standing wave varies
strongly, and thus strong photoemission occurs when the adsorbate
position coincides with the nodes of the standing wave. The tunability
of the photon wavelength and the availability of intense beams at
synchrotron sources allows high resolution and high sensitivity in
XSW studies.

Only two studies of graphene have been performed with XSW \cite{Busse2011,Runte2014},
though studying graphene does not impose more stringent conditions
than those needed for studying molecules. This seems however to be
an ideal method for the accurate determination of the graphene-metal
distance, which is characteristic of the graphene-metal interaction.

\section{Simulation methods\label{sec:Simulation-methods}}

\begin{figure}
\begin{centering}
\includegraphics[height=8cm]{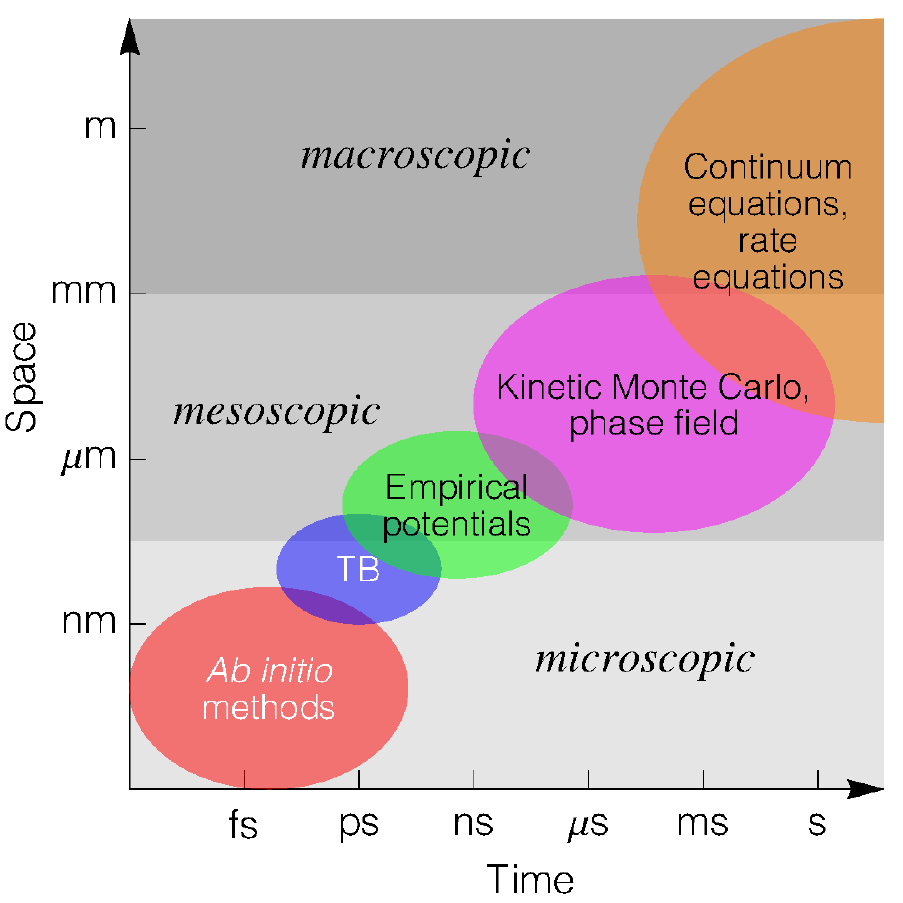}
\par\end{centering}

\caption{Comparison of various theoretical techniques
with respect to the characteristic spatial extent and the time scale associated with them.
Here TB stands for tight-binding and semi-empirical methods.
\label{fig:Compar-of-theoretical-methods} }
\end{figure}

In this section we shall give a brief overview of theoretical techniques
(see Fig. \ref{fig:Compar-of-theoretical-methods}) and methods which
we believe must be useful in graphene related modeling. In fact, each
of these methods alone or together in various combinations have already
proved extremely beneficial in existing theoretical simulations of
graphene growth. The overview is intended not solely for consistency
and convenience of our discussion; it is believed that a quick run
over a modern toolkit in use for modeling surface processes should
be useful to readers who are less familiar with them. Only a brief
outline will be given as there are many good reviews (e.g. \cite{Payne-DFT,Stich:2007vt,LK-book-2004})
and books (e.g. \cite{Allen-Tildesley-simul-book,Parr-Yang-DFT-book-1989,LK-book-2004,Martin-DFT-book-2008,Frenkel-Smit-simul-book})
on the subject.

\subsection{Atomistic modeling methods\label{sub:Atomistic-modelling-methods}}

The observable properties of all solids are governed by quantum mechanics,
as expressed by solutions of a Schr\"{o}dinger equation for the motion
of the electrons and the nuclei. However, because of the inherent
difficulty of obtaining even grossly approximate solutions of the
full many-body Schr\"{o}dinger equation, one typically focuses on reduced
descriptions that are believed to capture the essential features of
the problem of interest.

\subsubsection{Density Functional Theory \label{DFT}}

Most of the computational techniques applied at the atomic scale to
the graphene growth are based on Density Functional Theory (DFT) which
is a reasonable compromise between efficiency and precision. In its
standard formulation this method is only applicable to the ground
state of a quantum system at zero temperature, and in most cases of
crystal surfaces and carbon species on them, which is the main interest
for us here, this is sufficient. Modern DFT methods allow us to calculate
the total energy of a system of nuclei and electrons, as well as the
forces acting on atoms, from first principles, i.e. without using
any empirical information (such as experimentally measured interatomic
distances, elastic moduli, optical and vibrational spectra, etc.).
An ability to calculate the total energy in turn allows us to relax
structures to their mechanical equilibrium geometries (corresponding
to zero atomic forces), or to run molecular dynamics (MD) simulations
to take account of temperature effects. Although no empirical parameters
enter the calculation,
it is essential to remember that
the results (e.g. the total energy, atomic
geometry and vibrational frequencies) are not ``exact'' since DFT
methods employ density functionals which serve only as an approximation.
We shall briefly discuss how precise DFT may be in regard to the problem of graphene on metals
in sections \ref{metallic-substrates}.

The DFT is based on an exact statement \cite{Hohenberg-Kohn} that
the total energy of an arbitrary system of electrons in the ground
state can be written as a functional of the electron density $\rho(\mathbf{r})$:
\begin{equation}
E\left[\rho\right]=F\left[\rho\right]+\int V_{n}(\mathbf{r})\rho(\mathbf{r})d\mathbf{r}\;,\label{eq:HK-energy}
\end{equation}
where $V_{n}(\mathbf{r})$ is an external potential and $F[\rho]$
is a \emph{universal functional} of the electron density which does
not depend on the nature of the system, i.e. the number of electrons
$N_{e}$, positions of atoms, their chemical identity, etc. The external
potential in our case corresponds to the potential created by atomic
nuclei and is the only term in the energy which explicitly depends
on the atomic positions. However, since the electron density is obtained
by varying the total energy with respect to it subject to the normalization
condition
\[
\int\rho(\mathbf{r})d\mathbf{r}=N_{e}\;,
\]
the electron density in the energy minimum (the ``true'' density)
depends on the atomic positions as well. This fact is essential when
calculating the forces on atoms.

The functional $F[\rho]$ is universal, and this is the indisputable
strength of the DFT; however, its weakness is that this functional
is not known. It is clear that it should contain the kinetic energy
of the electrons and their interactions between themselves, i.e. their
Coulomb (Hartree) interaction, plus exchange and correlation, however,
it is a very difficult (maybe impossible) task to find exactly all
these contributions as density dependent functionals. Kohn and Sham
\cite{Kohn-Sham} proposed a computational scheme which somehow remedies
this problem by mapping the real system of interacting electrons to
an artificial system of non-interacting electrons of the same density
\begin{equation}
\rho(\mathbf{r})=2\sum_{i=1}^{N_{e}/2}\left|\psi_{i}(\mathbf{r})\right|^{2}\;,\label{eq:HK-density}
\end{equation}
moving in an one-electron effective potential $V^{eff}(\mathbf{r})=V_{H}(\mathbf{r})+V_{xc}(\mathbf{r})+V_{n}(\mathbf{r})$.
This consist explicitly of the Hartree term, $V_{H}(\mathbf{r})=\int\rho\left(\mathbf{r}^{\prime}\right)/\left|\mathbf{r}-\mathbf{r}^{\prime}\right|d\mathbf{r}^{\prime}$,
exchange-correlation contributions $V_{xc}(\mathbf{r})$, , as well
as the potential due to atomic cores. $V_{xc}$ should non-trivially
depend on the electron density. In Eq. (\ref{eq:HK-density}) $\psi_{i}(\mathbf{r})$
is the one-electron wavefunction for an electron occupying the state
$i$, and in the ground state it is assumed that the first $N_{e}/2$
states with the lowest one-electron energies, $\epsilon_{i}$, are
occupied (for simplicity we assume an even number of electrons, and
that a pair of electrons with opposite spins occupy each one-electron
energy level). The mapping to a non-interacting electron gas resolves
the problem of the unknown kinetic energy term in the total energy
since for the non-interacting electron gas the kinetic energy can
be trivially written in terms of the one-electron orbitals $\psi_{i}$.
However, this will not be the actual kinetic energy of the interacting
gas; the introduced error is transferred into the exchange and correlation
energy which now includes this discrepancy as well. The one-electron
orbitals are obtained by varying the total energy with respect to
them, subject to their orthonormality condition, yielding the Kohn-Sham
equations \cite{Kohn-Sham}:
\begin{equation}
\left[-\frac{1}{2}\Delta+V_{H}(\mathbf{r})+V_{xc}(\mathbf{r})+V_{n}(\mathbf{r})\right]\psi_{i}=\epsilon_{i}\psi_{i}\;,\label{eq:KS-equs}
\end{equation}
where the Hartree, $V_{H}$, and exchange-correlation, $V_{xc}$,
potentials are functionals of the density, and we adopted atomic units
for simplicity. Since the density itself is composed of the orbitals,
the equations (\ref{eq:KS-equs}) for all orbitals are highly non-linear with respect
to them and their solution has to be sought self-consistently starting
from some initial guess for the orbitals. Knowledge of the orbitals
provides the electron density (\ref{eq:HK-density}) and the total
energy. This calculation can be conducted for various atomic positions
(giving a different potential $V_{n}$ due to the nuclei) which yields
the dependence of the total energy of the electron-nuclei system on
atomic positions in the electronic ground state. Therefore, as was
mentioned above, this enables one to determine the whole potential
energy surface (PES) of the system. In particular, one can find potential
energy minima by relaxing the atomic geometry from different initial
geometries \cite{Payne-DFT}, calculate the phonon spectrum at any
geometry \cite{LK-book-2004,Baroni:2001tn} and investigate energy
barriers along the minimum energy paths (MEP) connecting different
minima. In addition, using atomic forces calculated entirely quantum-mechanically,
one can run MD simulations by solving numerically Newton's equations
of motion for atoms and hence attempt to investigate the temporal
evolution of the system when in the electronic ground state. These
tasks are greatly facilitated by the Hellmann-Feynman theorem \cite{Payne-DFT}
which allows atomic forces for the given geometry to be calculated
from the orbitals directly; differentiating the orbitals with respect
to atomic positions (which is not straightforward) is not required.

In practice, calculations are performed using periodic boundary conditions
\cite{Payne-DFT,LK-book-2004} whereby the system of interest (e.g.
a carbon cluster adsorbed on the Ir(111) surface) is periodically
repeated along the surface; in turn the latter consists of a finite
number of atomic layers, which together with the periodic array of
molecules on them form a slab. Then the slabs are periodically repeated
in the perpendicular direction to make the whole system 3D periodic.
It is essential to have sufficient distance between the clusters on
the slab (across the surface) to avoid artificial interaction between
the images of the clusters; similarly, sufficient vacuum gap is to
be selected between slabs to avoid interaction between them. Two other
technical points also need to be mentioned. The first one is related
to the fact that normally only valence electrons are explicitly represented
in the calculation and this is accomplished by means of the method
of pseudopotentials \cite{Payne-DFT,LK-book-2004}. Finally, the second
point concerns the basis set used in expanding the orbitals $\psi_{i}$:
these could be either plane waves $\sim e^{i\mathbf{K}\cdot\mathbf{r}}$
(with the reciprocal vector $\mathbf{K}$) or atomic orbitals centered
on atoms. The former set is easy to control and bring to convergence,
however, very many plane waves are normally needed for large systems
and hence the calculations can be very expensive. The atomic basis
set is localized and has a built-in quality of the constituent atoms;
therefore, the behavior of the electrons around atoms can be well
reproduced with a rather limited number of orbitals, while adding
a small number of additional diffused orbitals may be sufficient to
reproduce the behavior of the electrons in the spaces between atoms.
As a result, calculations in a localized basis set could be much cheaper
than the plane wave calculations. In particular, in calculations of
surface properties the vacuum gap can be taken as large as necessary
with no detrimental effect on the performance; this is not true for
the plane wave basis set since a bigger vacuum gap would result in
a finer reciprocal lattice and, as a consequence, more plane waves
need to be used to achieve the same precision. At the same time, calculations
with the localized basis set are not so straightforward: it is much
more difficult to converge the results of the calculations with respect
to the basis set, and on top of that there is a specific error in
calculating energy differences along a reaction path (e.g. an adsorption
energy) as different parts of the system are calculated using basis
sets which are not compatible. For instance, in the reaction $A+B\rightarrow AB$
different bases are used for calculating each reactant and the product
(remember that only orbitals centered on atoms are used!), and hence
the reaction energy $\Delta E=E_{AB}-E_{A}-E_{B}$ contains an error
due to different levels of convergence of each energy component with
respect to the basis set. This so-called basis set superposition error
(BSSE) can be compensated to some extend by means of the counterpoise
correction method \cite{Boys:1970BSSE}.

Although DFT is a ground state theory, it can still be quite useful
in trying to understand the mechanism of growth of graphene. To start
with, one may study the stability of various carbon species such as
monomers, dimers, trimers, etc. on the surfaces and their preferential
adsorption sites via geometry optimization calculations. This information
in turn provides understanding about which carbon species are in abundance
on the surface under current experimental conditions and hence must
play a dominant role in the growth. In this respect the problem of locating
the global energy minimum on the PES is of special importance \cite{Wales-PES-book}.
Normally the calculations start from a guessed initial geometry, and
the forces on atoms are used to move the system across the PES to
a lower energy geometry. This so-called geometry optimization procedure,
however, may easily arrive at a local energy minimum; finding the
lowest one (the global minimum on the PES) requires performing such
simulations starting from many different initial geometries. Physical
intuition may help here, however, and various techniques discussed
in e.g. Refs. \cite{Wales-PES-book,Wales:2006tn,Goedecker:2005vj}
may do a much better job. In particular, methods like \emph{simulated
annealing} are frequently used. In this method MD simulations are
run at initially very high temperature allowing the system to explore
the PES over a rather extensive portion of the phase space; then the
temperature is gradually reduced to zero, trapping the system in the
global minimum.

Although DFT based methods are in most cases highly reliable, they
are computationally expensive, especially when large clusters of atoms
need to be modeled on a surface. This is because periodic boundary
conditions are normally employed, and a significant chunk of the surface
needs to be included in the unit cell in order to ensure reasonable
distances between cluster images such that there is no spurious interaction
between them. Also, surfaces of crystals are modeled using slabs consisting
of a number of surface layers, and in many cases many of those layers
need to be explicitly included for the slab to be representative enough
of the proper bulk termination \cite{Zhao:2010gq,Henkelman:2001fg}.
Finally, first-principles DFT simulations scale very unfavorably with
the system size and become prohibitively expensive for systems containing
more than a few hundred atoms, although attempts to develop more efficient
methods which scale ``linearly'' with the system size are ongoing
\cite{Goedecker:2005vj,Goedecker-1999,Bowler-Gillan-2002,O-N-workshop-JPCM-2008,ONETEP,SIESTA}.
The implementation of this strategy places many important phenomena
within the capabilities of DFT.

\subsubsection{Dispersion interaction within DFT methods \label{sec:vdW}}

Over the years DFT has earned a lot of respect in the material science community due to its numerous
successes in describing structure and a variety of ground state properties of a wide class of materials.
The local density approximation (LDA) \cite{LDA} and different flavours of
the generalised gradient approximations (GGA), such as for instance the PBE density functional
\cite{Perdew:1996PBE,Perdew:1998PBE}, are the most popular exchange-correlation  functionals
being used today. The LDA density functional depends only on the electron density alone (the functional
is local), while the GGA functionals also depend on the density gradient and are hence semi-local.
This means that the dispersion interaction which is  non-local in nature is not taken
into account by LDA or GGA methods, and hence weakly bound layered compounds where the dispersion
interaction must be mainly responsible for their stability, are to be described incorrectly
by these methods. Indeed, if LDA overbinds the layers (which is due to the wrong reasons as
the binding comes from the exchange
while the missing dispersion interaction is the correlation effect by nature),
GGA results in significant underbinding \cite{Bjorkman-JPCM-2012,Bjorkman-PRL-2012}.

Therefore, over the last decade a significant effort has been made to incorporate the
dispersion interaction into the DFT methodology. A number of approaches is now
available:
(i) A semi-empirical DFT-D method due to Grimme \cite{Grimme-JComptChem-2004,Grimme-JCC-2006,Grimme-vdW}
in which a pair-wise force-field expression proportional to $C_6^{AB}/R^6$ (where $R$ is the distance between
atoms A and B and $C_6^{AB}$ a fitting parameter) is
added to the DFT energy for all atomic pairs to account for the dispersion forces;
the $C_6^{AB}$ coefficients are  fitted to high-quality data for a large number of molecules.
(ii) van der Waals density functional (vdW-DF) methods
\cite{Dion-Rydberg-Schroder-Langreth-Lundqvist-PRL-2004,vdW-DF2,Langreth-review,VV10-PRB-2012,
Vydrov-VV10-2010,VV10-PRB-2012}
in which the correlation energy is represented by a special functional which is non-local in electron density;
as it vanishes for a uniform density, it is supplemented by the LDA correlation functional; various
exchange functionals can then be added leading to a variety of different techniques (see e.g.
\cite{Langreth-review,Klimes-JPCM-2010,Bjorkman-JPCM-2012});
(iii) Tkachenko-Scheffler many-body self-consistent screened theory (TS-SCT)
\cite{Tkatchenko:2012fm,Tkatchenko-JCP-2013,Bucko-PRB-2013} in which a Grimme-like energy contribution
is added to the DFT energy with the $C_6^{AB}$ coefficients
derived from an adiabatic-connection fluctuation-dissipation theorem (ACFDT) \cite{ACFDT-Langreth-Perdew-1977}.

In fact, the latter theorem allows us to formally
obtain the exact exchange-correlation energy within the DFT framework.
The calculation requires knowledge of the frequency-dependent density-response function for the interacting
electrons, which in turn needs an object called the exchange-correlation kernel.
Neglecting the latter altogether results in the so-called random-phase approximation (RPA)
for the exchange-correlation functional \cite{ACFDT-Langreth-Perdew-1977,Harl-Kresse-PRL-2009}.
This is one of the most sophisticated many-body treatments of the electron correlations available today
and is expected  to be highly accurate for describing dispersion interactions. It is thought to be
less accurate for treating short-range covalent interactions \cite{Bjorkman-PRL-2012}.

The quality of vdW-DF based approaches was recently  \cite{Bjorkman-JPCM-2012,Bjorkman-PRL-2012}
 compared with that of RPA for a wide class of layered materials.
It was found that the results very much depend on which exchange functional is used. A general
conclusion was reached that neither of the existing vdW-DF approaches
is capable of reproducing correctly both geometry and binding energies (as compared to RPA)
of all studied solids. At the same time, since the RPA calculations are
about 200-300 times more expensive than GGA, it is currently still computationally
prohibitively expensive to use them for complex systems requiring geometry optimisation;
the number of atoms in the unit cell is also to be limited to a few.
Therefore, RPA still
cannot be used in realistic calculations of complex systems and one has to
resort to simpler methods such as DFT-D, vdW-DF or TS-SCT. However, care
is currently needed in applying these dispersion corrected DFT approaches as results may depend
on the method used. Application of several techniques, such as e.g. vdW-DF and TS-SCT,
to the same system (see e.g. \cite{Bamidele-Stich-JCTC-2013}) may be required to make meaningful conclusions.


\subsubsection{Tight Binding and Empirical Potentials based methods}

The expense of the DFT approach is why atomistic simulations based
on empirical potentials (EPs) have also been used, and we shall review
some of the work which has been done using this type of theories.
In these methods the total energy of the whole system is written using
simplified analytical expressions (either pairwise or many-body) containing
atomic positions explicitly. This allows a quick calculation of atomic
forces, so that large systems can be considered and MD simulations
evolved over long time scales. The energy expressions also contain
fitting parameters, and these are determined by comparing the results
of EP calculations with those done by DFT-based methods on the same
systems, and/or from experimental data (e.g. bulk lattice and elastic
constants, etc.). Because of the nature of EPs and the way in which
their parametrization is derived (based on a limited set of trial
systems and empirical information), one has to apply EP-based methods
with care. Usually, these methods are able to produce reliable predictions
of energies of various systems (e.g. comparison of cluster formation
energies on a surface), but may fail to give correct energy barriers
or predict new adsorption geometries or structures as these may deviate
substantially from the trial set used to fit the parameters.

The choice of potential is determined by factors such as the bond
type, the desired accuracy, transferability, and the available computational
resources. Potentials can be categorized broadly as: (i) pair potentials
and (ii) empirical many-body potentials. Two-body, or pair, potentials,
such as the Lennard--Jones \cite{jones24} and Morse \cite{morse29}
potentials, are used for large-scale simulations where computational
efficiency is paramount, but where a generic description is sufficient,
rather than detailed comparisons with a particular materials system.
For systems where multi-body interactions are important \cite{carlsson90},
the Stillinger--Weber \cite{stillinger85}, Tersoff \cite{tersoff86,tersoff88},
and Brenner \cite{brenner90} potentials are often used for covalent
carbon-like materials, and embedded-atom \cite{daw83,daw84}, effective
medium \cite{jacobsen87}, and Finnis--Sinclair potentials \cite{finnis84}
are common choices for metals. Such potentials are empirical in that
they are parametrized by fitting either to a set of experimental measurements
or to quantum mechanical calculations of representative atomic configurations.

There are also semi-empirical electronic structure methods available,
such as the tight-binding (TB) technique, which have also been used
for studying graphene growth related problems. These methods are intermediate
between more rigorous DFT and EP methods \cite{Haghi,Amara:2009hn}.
TB methods are based on a model Hamiltonian which is solved for each
geometry giving the total energy and a set of one-electron wavefunctions.
The Hamiltonian is parametrized to reproduce results of more refined
(e.g. DFT) calculations and some experimental data, and hence may
be more universal when compared with the EP approach. In spite of
that, this is still a semi-empirical method, and hence the results
should be carefully verified against more precise calculations from
time to time.

\subsubsection{Formation energies}

As mentioned in Sections \ref{sub:Nucleation-Theory} and \ref{sub:Thermodynamics-and-kinetics},
calculations of the stability of various carbon species on metal surfaces
play a central role in understanding which species are of special importance
for  graphene growth phenomena. Several definitions of the formation
(or stabilization) energy, $E_{f}(N)$, of a C-cluster with $N$ carbon
atoms on a metal surface can be found in the literature. They are
broadly based on the following formula:
\begin{equation}
E_{f}(N)=E_{tot}-(E_{metal}+N\mu_{C})\;,\label{eq:cluster-form-energy}
\end{equation}
where $E_{tot}$ is the total (e.g. DFT) energy of the cluster relaxed
on the metal, $E_{metal}$ is the energy of a bare metal surface,
and $\mu_{C}$ is the energy of a single carbon atom, but sometimes
referred to as its chemical potential (but this would only be appropriate
terminology if the temperature were zero). The difference between
various approaches adopted in the literature lies in the definition
of $\mu_{C}$. For instance, Gao \emph{et al. }\cite{Gao11}\emph{
}define $\mu_{C}$ as being the formation energy per carbon atom of
a free-standing graphene sheet. This is in contrast to Wesep \emph{et
al. }\cite{wesep:171105}\emph{ }who define $\mu_{C}$ to be the energy
of an isolated carbon atom. In a different study by Riikonen \emph{et
al. }\cite{Riikonen} both definitions were used and it was shown
that there are differences, of course, in the values of the formation
energies obtained using each definition.\emph{ }Although this matter
might seem to be unimportant, it does affect our conclusions concerning
the relative stabilities of clusters with different numbers $N$ of
atoms on the metal surface (as it is e.g. done in \cite{Zhangdoi:10.1021/jp2006827}
as discussed in Section \ref{sub:Early-Stages-of}).

If $\mu_{C}$ is taken to be the energy of an individual carbon atom
\emph{on a terrace} then equation (\ref{eq:cluster-form-energy})
can be used to compare binding energies of various carbon clusters
on the terrace or at steps. In this case the expression would correspond
to the formation energy of a cluster from individual adsorbed carbon
atoms. In order to determine whether an existing diffusing cluster
of $N$ carbon atoms on a terrace might preferentially stick to a
metal step, a slightly different definition is to be used:
\begin{equation}
E_{f}^{s}(N)=E_{tot}-E_{step}-E_{f}(N)\;,\label{eq:cluster-at-step-form-E}
\end{equation}
where $E_{tot}$ is the total energy of the cluster attached to the
step, $E_{step}$ is the energy of an bare metal step and $E_{f}(N)$
the formation energy (\ref{eq:cluster-form-energy}) of the $N-$atom
carbon cluster on the surface. A slightly more complex expression
for the formation energy is required when assessing the stability
of an $N-$atom carbon cluster at the edge of a graphene island (graphene
ribbons are normally used in practical calculations relying on periodic
boundary conditions implemented in widely used DFT codes) as detailed
in Section \ref{sub:Atomistic}.

Thermodynamic equilibrium between graphene islands and single carbon
atoms and clusters requires the chemical potential $\mu_{C}$ of the
C atoms to be the same in each phase. If the calculated chemical potential
of C atoms in clusters is higher than that in islands, the clusters
are thermodynamically unstable and should eventually attach to graphene;
the same can be said for single carbon atoms roaming the surface.
If however the chemical potential of C atoms in graphene flakes on
the surface is higher that in clusters, graphene formation is not
sustainable and large flakes will not grow; instead, there will be
many C clusters on the surface. Using this kind of argument one might
attempt to speculate on which cluster size is critical for the growth
of graphene. This will be the size $N$ for which the calculated energy
per C atom of the cluster C$_{N}$ on the surface, used as an estimate
of its chemical potential, becomes lower (more negative) than the
C atom energy (i.e. the chemical potential) in graphene \emph{on the
surface}. Note that the latter might be quite different from the C
atom energy in the free standing graphene, especially in the cases
of transition metals such as Ru(0001) and Rh(111) for which the interaction
of the metal with graphene on top of it is the strongest. In the situation
of a real experiment, however, this kind of analysis may only serve
as a guide and hence should be used with care. It is more relevant
for the nucleation stage (see Section \ref{sub:Nucleation-Theory})
when the difference of chemical potentials (supersaturation) drives
the phase change. During growth the system is far from equilibrium,
and the total energies of structures may be of less importance compared
to the energy barriers which mostly determine the kinetics. In addition,
there is a much richer variety of carbon containing species on the
surface during growth, not just C atoms and clusters (see Section
\ref{sec:Carbon-Feedstock-Theory-1}).

\subsubsection{Nudged Elastic Band Method}

Static calculations can provide a lot of information about the \emph{dynamics
of a} system as well, which is absolutely essential for understanding
the growth phenomenon. For instance, ground state DFT calculations
are well suited for studying carbon species mobility (i.e. diffusion
barriers) by calculating the minimum energy paths (MEP) connecting
two stable minima on the PES. This problem is related to that associated
with finding a transition state (TS), or a saddle point, between two
energy minima and hence a calculation of the energy barrier to cross
from one minimum to the other. This task can be solved by so-called
constrained minimization whereby the system is taken between initial
and final states along a trajectory realized using a set of some selected
atomic coordinate(s). For instance, a diffusion barrier can be calculated
by taking a molecule between two adsorption sites by way of fixing
the lateral (reaction) coordinates of a single atom of the molecule
at each position along the diffusion path; all other coordinates of
the system including the position of the atom in a plane orthogonal
to the reaction coordinates are allowed to relax. Moving the atom
between the initial and final states in small steps and relaxing the
system in this manner can allow the system to be taken over the barrier.

The problem with this approach however is that in real situations
it might be very difficult to guess what
reaction coordinate one should choose that
would allow passing exactly
over the actual saddle point on the PES corresponding
to the top point on the MEP.
The wrong choice will result in the energy barrier being overestimated.
The Nudged Elastic Band (NEB) method \cite{Jonsson:1999tf,Henkelman:2000tl,Henkelman:2000wb}
eliminates this problem. In this method the MEP is found numerically
by means of a finite number of ``images'' of the system which correspond
to system geometries along the MEP. The image(s) in the middle of
the sequence would correspond to the system geometry at or near the
transition state on the MEP. Assuming there is only one TS along the
MEP connecting the initial and the final states, the energy difference
between the TS and the initial (final) state corresponds to the energy
barrier for going from the initial (final) to the final (initial)
state over the TS. The images are ``connected'' by springs to keep
the images apart from each other and hence map the MEP uniformly.
To avoid the band slipping off the actual MEP during the optimization
process, only projections of forces on atoms due to springs which
are parallel to the band are considered, while projections of DFT
forces in the same direction are removed, i.e. only the perpendicular
part of the DFT forces on atoms with respect to the band direction
are kept. Using NEB one can also calculate attachment (and detachment)
energies and associated barriers for carbon atoms and clusters to
attach to (coalescence with) or detach from each other or step edges.
This wealth of information can be compared with the corresponding
results on terraces, and hence conclusions can be drawn concerning
the role played by extended surface defects in the growth of graphene
islands. Finally, detailed energetic information can be used in building
up kinetic models of growth either at phenomenological (section \ref{sub:Rate-equation-analysis})
or atomic-scale (section \ref{sub:kMC}) levels. We shall briefly
explain the main ideas of kinetic simulations based on atomic-scale
theories in the next section.

\subsection{Simulations of dynamics of growth processes\label{sub:Simulations-of-dynamics-methods}}

\subsubsection{Molecular dynamics simulations}

During growth atoms move around to assume geometries and configurations
which correspond to the lowest free energy under the given thermodynamic
conditions (temperature $T$ and pressure $P$). In this sense it
seems natural in the atomistic simulations to allow the atoms to move
according to the forces acting on them as this basically repeats what
nature is doing. If $\mathbf{F}_{i}$ is the force acting on atom
$i$, then atomic positions change according to Newton's equations
of motion:
\begin{equation}
m_{i}\dot{\mathbf{v}}_{i}=\mathbf{F}_{i}\;,\quad\dot{\mathbf{r}}_{i}=\mathbf{v}_{i}\;\label{eq:MD-eqs}
\end{equation}
Here $\mathbf{r}_{i}(t)$ and $\mathbf{v}_{i}(t)$ are the position
and velocity of atom $i$ at time $t$. These equations are the essence
of MD simulations. In practice the calculations are solved numerically
using various discretization schemes \cite{Frenkel-Smit-simul-book}.
Special algorithms have been also developed to simulate in these numerical
experiments particular thermodynamic ensembles such as $NVT$ (constant
number of particles $N$, volume $V$ and temperature $T$) and $NPT$
(constant $N$, $T$ and pressure $P$) \cite{Frenkel-Smit-simul-book}.
The latter ensemble is more relevant to the case of simulating the
growth phenomenon as the pressure rather than the volume is kept constant
in the experiments.

Although MD simulations at first glance seem to be the method of choice
in studying growth processes, they are not strictly speaking designed
for them. This is because growth is associated with many elementary
(local) events requiring the surmounting of (sometimes large) energy
barriers when the system jumps between different minima on the PES.
Indeed, the basic limitation of the MD method that has prevented long
simulation times is that the integration time step must be small enough
to capture the dynamics of the vibration modes of the system, with
frequencies of the order of $10^{13}\:{\rm s}^{-1}$. This requires
time steps in the \textit{femtosecond} range. But the residence time
of, say, an adatom between hops can extend to \textit{microseconds}
because of the energy barriers that separate different energy minima
on the PES, and the interactions responsible for aggregation phenomena
occur over a time scale of \textit{milliseconds} to \textit{minutes}.
If one attempts to run an MD simulation of a growth process in which
numerous activation events are present, the system would spend most
of the time sitting in the potential energy wells with extremely rare
jumps between them; this is evident from the trajectories of mobile
atoms, which are complex paths localized around their initial sites
with only rare excursions to neighboring sites. In fact, if the energy
barriers are much larger than  $k_{B}T$ (here $k_{B}$ is Boltzmann's
constant), no transitions will be observed at all during the course
of a typical MD run, even if we run them over a very long time exceeding
any available resources many times over. In spite of the fact that
several methods have been developed for accelerating the MD treatment
of such rare events, based on stimulating the transitions to occur
faster than in an ordinary simulation \cite{voter02}, these methods
are still very expensive and hence can only be applied in classical
MD simulations. The other reason why MD simulations may be of more
limited use in studying growth processes is that normally we are not
really interested in the detailed knowledge on the atomic level of
what the system is doing when spending most of its time sitting in
the energy well; we are more concerned with the transitions themselves
between the wells as the system attempts to reduce its free energy
during growth of the new phase.

\subsubsection{Kinetic Monte Carlo simulations}

A necessary advance can be made by going beyond the atomistic detail
and considering kinetic processes in a time (and space) coarse-grained
manner. In Monte Carlo methods \cite{landau00} the \textit{deterministic}
equations (\ref{eq:MD-eqs}) of MD are replaced by\emph{ }\textit{\emph{equations
based on }}\textit{stochastic} transitions for the slow processes
in the system. In their most general form, Monte Carlo methods are
stochastic algorithms for exploring phase space, but their implementation
for equilibrium and nonequilibrium calculations is somewhat different
\cite{kang89,fichthorn91}. Here, as we are interested in the simulation
of growth driven by an external source of material, we will focus
on the kinetic Monte Carlo method (KMC).

KMC simulations are capable of providing a realistic platform for
studying system dynamics (kinetics) related to performing activation
events, and this is modeled as a sequence of consecutive jumps between
free energy minima, in real time \cite{Gillespie-1976,Reuter-Stampfl-Scheffler-2005,Gillespie-KMC,Voter-review-2005}.
Suppose that the probability of finding a system in state $\sigma$
at time $t$ is $P(\sigma,t)$ and that the transition rate per unit
time from $\sigma$ to $\sigma^{\prime}$ is $W(\sigma,\sigma^{\prime})$.
The equation of motion for $P$ is the master equation \cite{vankampen81}:
\begin{equation}
\frac{\partial P}{\partial t}=\sum_{\sigma^{\prime}}P(\sigma^{\prime},t)W(\sigma^{\prime},\sigma)-\sum_{\sigma^{\prime}}P(\sigma,t)W(\sigma,\sigma^{\prime})\,.\label{eq:master-eq}
\end{equation}
KMC methods are algorithms that solve the master equation by accepting
or rejecting transitions with probabilities that yield the correct
evolution of a nonequilibrium system. The effect of fast dynamical
events is taken into account phenomenologically by stochastic transition
rates for slower events. Hence in KMC simulations only jumps between
states of the systems are considered, ignoring completely residence
events, i.e. the time the system spends within a given state (potential
well). Suppose at a given time $t$ the system occupies a state $i$
associated with a certain arrangement of atoms, clusters, molecules,
etc. on the surface. We are interested in the evolution of the system
in time as various surface species diffuse across the surface, desorb,
decompose (e.g. hydrocarbon molecules lose their H atoms), and coalesce
(e.g. C atoms attach to clusters, clusters attach to islands and so
on). At each time step one has to specify then all the states $j$
the system may propagate into from the given state $i$ according
to the various local processes mentioned above. For each of those
processes one calculates the transition rate according to transition
state theory (TST) \cite{Nitzan-book}
\begin{equation}
r_{i\rightarrow j}=\nu_{i}\exp\left(-\frac{\Delta E_{i\rightarrow j}}{k_{B}T}\right)\;,\label{eq:KMC-rate}
\end{equation}
where $\nu_{i}$ is an attempt frequency which is of the order of
the frequency of atomic vibrations (around $10^{12}-10^{13}$ s$^{-1}$)
\cite{zangwill88,wang01} and $\Delta E_{i\rightarrow j}$ is the
energy barrier required for making the transition $i\rightarrow j$
which can be calculated, for example, using NEB as was discussed earlier.
The rate $r_{i\rightarrow j}$ corresponds to a single transition
which takes the system from its current state $i$; the sum of the
rates
\[
R_{i}=\sum_{j\,(\neq i)}r_{i\rightarrow j}
\]
corresponds to the total rate of leaving the current state, i.e. it
is an \emph{escape rate}. The probability $P_{i}(t)=\exp\left(-R_{i}\Delta t\right)$
to \emph{remain} in the current state over time $\Delta t$ is exponentially
decaying with time since eventually the system would leave the current
state and hence the probability to remain in it must decay to zero
(because of transitions into other states). Consequently, the time
of a successful transition can be determined at random from
\[
\Delta t=-\left(1/R_{i}\right)\ln x_{1}\;,
\]
where $x_{1}$ is a random number between 0 and 1. To determine explicitly
which state $j$ is to be selected for the successful transition,
we assume that the probability of each such event is given by the
ratio $P_{i\rightarrow j}=r_{i\rightarrow j}/R_{i}$. In practice,
the state $j$ is calculated also at random by picking up the second
random number $x_{2}\in(0,1)$ and then picking the event $j$ from
the condition
\[
\sum_{k=1}^{j-1}r_{i\rightarrow k}<x_{2}<\sum_{k=1}^{j}r_{i\rightarrow k}\;.
\]
Once the state $j$ is chosen, all possible transitions from $j$
to other states are analyzed again, the list of all of them is constructed
and the total escape rate $R_{j}$ calculated. Then two new random
numbers are taken again to determine the time for the next transition
$\Delta t$ and which transition from the new list the system would
jump into. This process is repeated many times until the required
state of the system is reached, i.e. the new phase has been fully
formed.

It is seen that during the evolution the system indeed jumps from
one potential energy well to another. The states after each transition
and the transition times are chosen at random, i.e. the system propagates
in time via a particular trajectory which depends very much on the
available elementary processes and their energy barriers. If at a
given state $i$ there are processes $j$ with small and large barriers,
the rates of these $r_{i\rightarrow j}$ will be strongly disparate
and the transition with the smallest barrier (the largest rate) is
most likely to be selected. As a result, it is probable that during
the course of the evolution various events with the smallest barriers
will play the dominant role; at the same time, it is worth remembering
that least probable transitions can (and will) be selected as well
from time to time. In fact, the system would progress quickly in time
until all fast processes are realized (large rates, small time steps
$\Delta t$); then further progress would require a less probable
process to be chosen which is associated with a larger barrier, smaller
rate and longer time $\Delta t$.

Although the details of the underlying mechanism for kinetic processes
are lost, the explicit calculation of atomic trajectories is avoided,
so KMC simulations can be performed over real times, running into
seconds, hours, or days, as required. The KMC method offers considerable
advantages over the MD method both in terms of the real time over
which the simulation evolves, as well as in the number of atoms included
in the simulation, because much of the computational overhead in MD
used to evolve the system between rare events is avoided.

The main challenge for KMC simulations is to be able to identify
all the essential states the system can jump into from a given state.
Usually, to simplify the problem an intelligent selection of the moves
performed on ``the lattice'' of system configurations is made,
and the corresponding transition rates are obtained from the energy
barriers calculated for the chosen ``moves'' before the KMC simulations.
In some cases this approach may lead to oversimplified kinetics in
which essential processes are missing. Therefore, care as well as
detailed study of the possible elementary processes is required prior
to devising the KMC simulation schemes. Thus KMC models can often
benefit from a related classical or quantum MD simulation to identify
the important physical process, and NEB simulations to estimate the
prefactors and kinetic barriers. Quite often, experiments themselves
can suggest a particular mechanism. The feasibility of performing
detailed simulations over experimental time scales enables various
parametrizations to be tested and models of kinetic phenomena to be
validated.

\subsubsection{Grand Canonical Monte Carlo simulations}

Sometimes real dynamical simulations are replaced by an intuitive
approach based on the Grand Canonical Monte Carlo (GCMC) method \cite{Frenkel-Smit-simul-book}.
In this method the system of atoms is propagated by a sequence of
``moves''; at each move the system is changed in a certain way,
e.g. an atom is added from an atomic reservoir (the source), removed
from the system into the reservoir (sink), or displaced on the surface,
and then all atoms are relaxed to mechanical equilibrium. However,
not all the moves are automatically accepted as the method has a stochastic
element in it. The decision on whether to accept the move or reject
it is made depending on the temperature and a comparison of the energy
difference between the final and initial states with the chemical
potential $\mu$ of the species of interest (e.g. of C atoms), which
is the characteristic thermodynamic feature of the source-sink reservoir
(assumed to be very large, i.e. macroscopic, e.g. carbon feedstock).
This is done by employing a Metropolis-like algorithm: if the energy
after adding/removing an atom is reduced, the probability of this
event is taken as unity, however, if the energy is higher, then the
step is accepted at random with a certain probability. In can be shown
\cite{Frenkel-Smit-simul-book} that this method generates a sequence
of system states that eventually converges to the correct grand canonical
distribution of species (e.g. C atoms and C-clusters on the surface)
in a sense that after some number of moves, depending on the nature
of the system and the starting configuration, the system ``equilibrates'',
i.e. new moves would sample the phase space according to the grand
canonical distribution. However, care is needed here. Even though
the moves may resemble real processes happening during growth, it
is essential to realize that this kind of simulation cannot be considered
as real kinetics, i.e. as a realistic propagation of the system in
time, as the algorithm is not designed for this purpose. The method
simply drives the simulated system to thermal equilibrium with the
source-sink system (e.g. C gas above the surface), and the path along
which this neighbourhood in the phase space is reached has nothing
to do with the real time propagation of the system during growth.
Moreover, as was already mentioned, during the growth process the
system may be far from equilibrium, while the GCMC simulations is
a genuinely equilibrium approach. And therefore there is a problem
with this technique which is that it is not straightforward to connect
the results of the simulations with experiment \cite{Amara:2006hd}
as experimentally the chemical potential of carbon is not maintained
at a fixed value.

\subsection{Continuum equations}

In models of nonequilibrium systems based on continuum equations,
typically in the form of deterministic or stochastic partial differential
equations, the underlying atomic structure of matter is neglected
altogether and is replaced by a spatially continuous and differentiable
mass density. Analogous replacements are made for other physical quantities
such as energy and momentum. Differential equations are then formulated
either from basic physical principles, such as the conservation of
energy or momentum, or by invoking approximations within a particular
regime, using e.g. a coarse-graining by smearing out fast degrees
of freedom as compared to the characteristic time of interest (e.g.
growth kinetics).

There are many benefits of a continuum representation of kinetic phenomena.
Foremost among these is the ability to examine macroscopic regions
in space over extended periods of time. This is facilitated by extensive
libraries of numerical methods for integrating deterministic and stochastic
differential equations. Complementing the numerical solution of partial
differential equations is the vast analytic methodology for identifying
asymptotic scaling regimes and performing stability analyses. Additionally,
if a continuum equation can be systematically derived from atomistic
principles, there is the possibility of discriminating between inherently
atomistic effects and those that find a natural expression in a coarse-grained
framework. Continuum equations also provide the opportunity for examining
the effect of apparently minor modifications to the description of
atomistic processes on the coarse-grained evolution of a system which,
in turn, facilitates the systematic reduction of full models to their
essential components. 
Because continuum theories based on methods such as the phase field
for graphene are only beginning to emerge (see Section 7.4), these methods will be given a quick mention here, as they
have the potential to uncover the growth kinetics of graphene at the mesoscopic scale.
We hope
that this would stimulate theorists in applying these techniques for
studying this fascinating growth problem.

\subsubsection{Burton-Cabrera-Frank Theory and Phase-field Method}

The Burton--Cabrera--Frank (BCF) theory \cite{burton51} describes
growth on a stepped surface of a monatomic crystal in terms of the
deposition and migration of single adatoms. The central quantity in
this theory is, therefore, the adatom concentration $c({\bf x},t)$
at position ${\bf x}$ and time $t$. The processes which cause this
quantity to change are the surface migration of adatoms, which have
diffusion constant $D$, the flux $J$ of adatoms onto the surface
from an incident flux, and desorption of the atoms from the surface
at an average rate $\tau_{s}^{-1}$. The equation determining $c({\bf x},t)$
on a terrace is a two-dimension diffusion equation with source and
sink terms:
\begin{equation}
\frac{\partial c}{\partial t}=D\bm{\nabla}^{2}c+J-\frac{c}{\tau_{s}}\,.\label{eq3}
\end{equation}
This equation is supplemented by boundary conditions which give the
normal velocity of a step edge,
\begin{equation}
v_{n}=\bigl(D\bm{\nabla}c\big|_{+}-D\bm{\nabla}c\big|_{-}\bigr)\cdot\bm{n}\,,
\end{equation}
where the unit normal $\bm{n}$ points from the upper (denoted by
``+'') to the lower (by ``-'') terrace. The scale of the adatom
concentration is determined by the competition between the deposition
flux, which drives the surface away from equilibrium and \textit{increases}
the adatom density, and the relaxation of the surface towards equilibrium
through adatom diffusion, which \textit{decreases} the adatom density.
Since the BCF theory neglects interactions between adatoms, the growth
conditions must be chosen to ensure that the adatom concentration
is maintained low enough to render their interactions unimportant.
Thus, this theory is valid only for high temperatures and/or low fluxes,
where growth is expected to occur by the advancement of steps.

The phase-field method provides a mathematical description to free-boundary
problems for phase transformations, such as solidification and the
assembly of structures on a surface, in which the interface has a
finite, but small, thickness. The central quantity in this method
is an auxiliary function, called the \textbf{\textit{phase-field}},
whose value identifies the phase at every point in space and time.
The phase field model of the solid-liquid phase transition was first
proposed by Langer \cite{langer86}, and has developed into a widely-used
method of computing realistic growth structures in a variety of settings
\cite{sekerka04}.

The phase field method has been applied to stress-induced instabilities
\cite{kassner01}, the motion of steps \cite{liu94,karma98,ratz04},
island nucleation and growth \cite{castro03,yu04,ming10}, and self-organized
nanostructures \cite{kim04}.

\subsubsection{Rate Equations}

With increasing temperature or decreasing deposition rate, growth
by the nucleation, aggregation and coalescence of islands on the terraces
of a substrate becomes increasingly likely and a description of growth
by the advancement of steps is no longer appropriate. One way of providing
a theoretical description of this regime within an analytic framework
is with equations of motion for the densities of adatoms and islands.
These are called rate equations \cite{venables84}.

We will consider the simplest rate equation description of growth,
the most relevant to the issues being considered here, where adatoms
are the only mobile surface species and the nucleation and growth
of islands proceeds by the \textit{irreversible} attachment of adatoms,
i.e.~once an adatom attaches to an island or another adatom, subsequent
detachment of that adatom cannot occur. We will signify the density
of surface atoms by $c\equiv c_{1}(t)$ and the density of $n$-atom
islands by $c_{n}(t)$, where $n>1$. Thus, the rate equation for
$c_{1}$ is
\begin{equation}
\frac{dc}{dt}=J-2D\sigma_{1}c^{2}-Dc\sum_{n=2}^{\infty}\sigma_{n}c_{n}\label{eq6}
\end{equation}
In common with most formulations of rate equations, the adatom and
island densities are taken to be spatially homogeneous. In particular,
there is no diffusion term, $D\nabla^{2}c$, despite the fact that
adatoms are mobile. This description is most suitable for flat surfaces,
where there are no pre-existing steps to break the translational symmetry
of the system and induce a spatial dependence in the adatom and island
densities. But even in this case, spatial effects cannot be neglected
altogether. This will be discussed below.

The first term on the right-hand side of (\ref{eq6}) is the deposition
of atoms onto the substrate (with rate $J$), which increases the
adatom density, and so has a positive sign. The next term describes
the nucleation of a two-atom island by the irreversible attachment
of two migrating adatoms. This term decreases the number of adatoms
(by two) and thus has a negative sign. The rate for this process is
proportional to the \textit{square} of the adatom density because
two adatoms are required to form a two-atom island, and to $D$, the
adatom diffusion constant, because these adatoms are mobile. The third
term accounts for the depletion rate of adatoms due to their capture
by islands with all possible numbers of atoms. This term is proportional
to the product of the adatom and total island densities and must also
have a negative sign. The quantities $\sigma_{i}$ ($i\geq1$), called
''capture numbers'', account for the diffusional flow of atoms into
the islands \cite{venables84,bales94,bartelt96,amar01}.

The rate equation for the density of $n$-atom islands $c_{n}(t)$
is
\begin{equation}
\frac{dc_{n}}{dt}=Dc\sigma_{n-1}c_{n-1}-Dc\sigma_{n}c_{n}\label{eq7}
\end{equation}
The first term on the right-hand side is the creation rate of $n$-atom
islands due to the capture of adatoms by ($n-1$)-atom islands. Similarly,
the second term is the depletion rate of $n$-atom islands caused
by their capture of adatoms to become $(n+1$)-atom islands. There
is an equation of this form for every island comprised of two or more
atoms, so Eqs. (\ref{eq6}) and (\ref{eq7}) represent an infinite
set of coupled ordinary differential equations. However, since the
density of large (compared to the average size) islands decreases
with their size, in practice the hierarchy in (\ref{eq7}) is truncated
to obtain solutions for $c(t)$ and the remaining $c_{n}(t)$ to any
required accuracy. Notice that in writing (\ref{eq7}) we have omitted
any direct interactions between islands. This restricts us to a regime
where there is no appreciable coalescence of these islands.

Rate equations produce correct scaling of island densities with $D/J$
in the\textbf{ }\textit{aggregation regime} of island growth, i.e.~where
the island density has saturated and existing islands capture all
deposited atoms. However, the scaling of $c$ and $c_{n}$ with coverage
is not correctly accounted for. This can be traced to the approximation
of constant capture numbers, which misses important aspects of island
kinetics. The next level of approximation is to include the spatial
extent of the islands in an average way by assuming that the local
environment of each island is independent of its size and shape \cite{bales94}.
This produces the correct scaling of $c_{n}$ with both $D/J$ and
coverage, but still not the correct distribution of island sizes.
The calculation of the latter quantity requires proceeding one step
further by including explicit spatial information in the capture numbers
to account for the correlations between neighboring nucleation centers
and the different local environments of individual islands \cite{bartelt96,bartelt99}.
In practice, only total island densities are analyzed with rate equations;~simulations
or semi-analytic methods, such as the phase-field and level-set methods,
are used for the quantitative analysis of measured island-size distribution
functions.

\section{Growth Methods\label{sec:Growth-Methods}}

It is now known that graphene can be grown via several different methods.
Of these some involve producing graphene epitaxially on transition
metal surfaces \cite{Wintterlin}. In the major method of production
called \emph{chemical vapor deposition} (CVD) carbon or carbon based
molecules are deposited onto a surface under conditions for which
absorption into the bulk can be neglected. In CVD the species are
deposited onto a surface that is originally at some high temperature.
Though not discussed here, there has also been recent interest in
depositing onto transition metal films supported on a separate substrate
\cite{Ago2010,Iwasaki2010,Reddy201,Sutter2010,VoVan2011,
Tonnoir2013}.
As well
as this method, \emph{temperature programmed growth} (TPG) has also
been used. In this approach temperature control is employed to stimulate
graphene growth on a surface using as the carbon source hydrocarbons
that are deposited on the surface initially at room temperature. Another
kind of surface-confined growth on metals involves the formation of
a surface carbide as a transient state. In another method, to be called
the \emph{segregation method} hereafter, carbon is deposited on a
surface maintained at such a high temperature that a proportion of
C atoms are absorbed into the bulk of the metal. Upon cooling, the
absorbed carbon segregates from the bulk to the surface and becomes
mobile forming graphene flakes, i.e. the segregated carbon in the
bulk is used as a source. In addition to these methods where graphene
is formed on metal surfaces, graphene can be grown on SiC. The most
widespread approach consists in heating SiC to a high enough temperature
for silicon to sublimate from the surface, leaving behind carbon atoms
that may then form graphene. A less studied approach (especially from
the point of view of the processes at play during growth) involves
molecular beam epitaxy \cite{AlTemimy2009,Moreau2010} and CVD \cite{Michon2010,Hwang2010}
on SiC. In this section we review experimental observations of graphene
growth using methods mentioned above. Since our main interest is understanding
the growth of graphene, to which both experiment and theory have been
contributing over the years, we start in this section discussing the
main experimental techniques that have proven extremely useful in
understanding the growth phenomenon.

\subsection{Chemical Vapor Deposition (CVD)\label{sec:Chemical-Vapour-Deposition}}

We consider here the deposition of molecules containing carbon onto
a transition metal surface. Typical carbon precursors are ethene or
methane \cite{Loginova09,Coraux} but it is also possible to grow
graphene using a variety of different larger carbon based molecules
\cite{Wang}. Prior to the deposition of hydrocarbons the surface
of the transition metal is cleaned. The sample is placed in a UHV
chamber, with a base pressure of approximately $10{}^{-11}$ mbar
and
is subjected to cycles of room temperature sputtering with energetic
(typically 1 keV) ions (e.g. Ar$^+$) and
flash annealing between 1200 and 1500 K.
Then the sample is kept
at some specific temperature, usually greater than 700 K \cite{Coraux,Hwang11}.
Gaseous molecules are introduced onto the hot surface.
The pressures and flow rates of vapors in the dosing tube are varied
to produce different fluxes of molecules onto the transition metal
surface.

There are a number of different substrates that have been used to
grow graphene. These include crystalline substrates such as
Cu(111) \cite{Gao10},
Ni(111) \cite{Yamamoto1992,Lahiri2010},
Co(0001) \cite{Eom2009,Rader2009},
Fe(110) \cite{Vinogradov2012},
Au(111) \cite{Nie2012},
Pd(111) \cite{Kwon_NL_2009},
Pt(111) \cite{Sutter2009},
Re(10$\bar{1}$0) \cite{Gall1987},
Ru(0001) \cite{Loginova2009b},
Rh(111) \cite{Voloshina}, Ir(111) \cite{Coraux,Lacovig,Loginova08,Loginova09},
but also polycrystalline surfaces \cite{Wofford,Li10,Hwang11} and
thin films evaporated onto a different bulk material \cite{Chi,Kongara}.
Here we only mention metals used for CVD; other substrates will be mentioned later on in Sections \ref{sec:Temperature-Programmed-Growth},
\ref{sec:Carbon-Feedstock-Experimental} and \ref{sec:Segregation}.
We now summarize some of the key observations concerning nucleation
and growth of graphene for a number of transition metal surfaces.
Unless stated, the experiments are carried out in UHV and it is ethene
which is deposited onto a hot transition metal surface.

\subsubsection{Nucleation\label{sub:CVD-Nucleation}}

Nucleation of graphene on a surface can occur homogeneously on terraces
or at defect sites on the surface (e.g. at steps). It is of particular
interest to understand graphene nucleation for applications requiring
extremely high quality material that is free of defects. This is because
when graphene nucleates at many places on the surface the eventual
coalescence of individual graphene islands can create grain boundaries
or edge defects that can have detrimental effects on some of graphene's
electronic properties. We thus consider some of the experimental observations
on different surfaces.

\begin{figure}
\begin{centering}
\includegraphics[height=6cm]{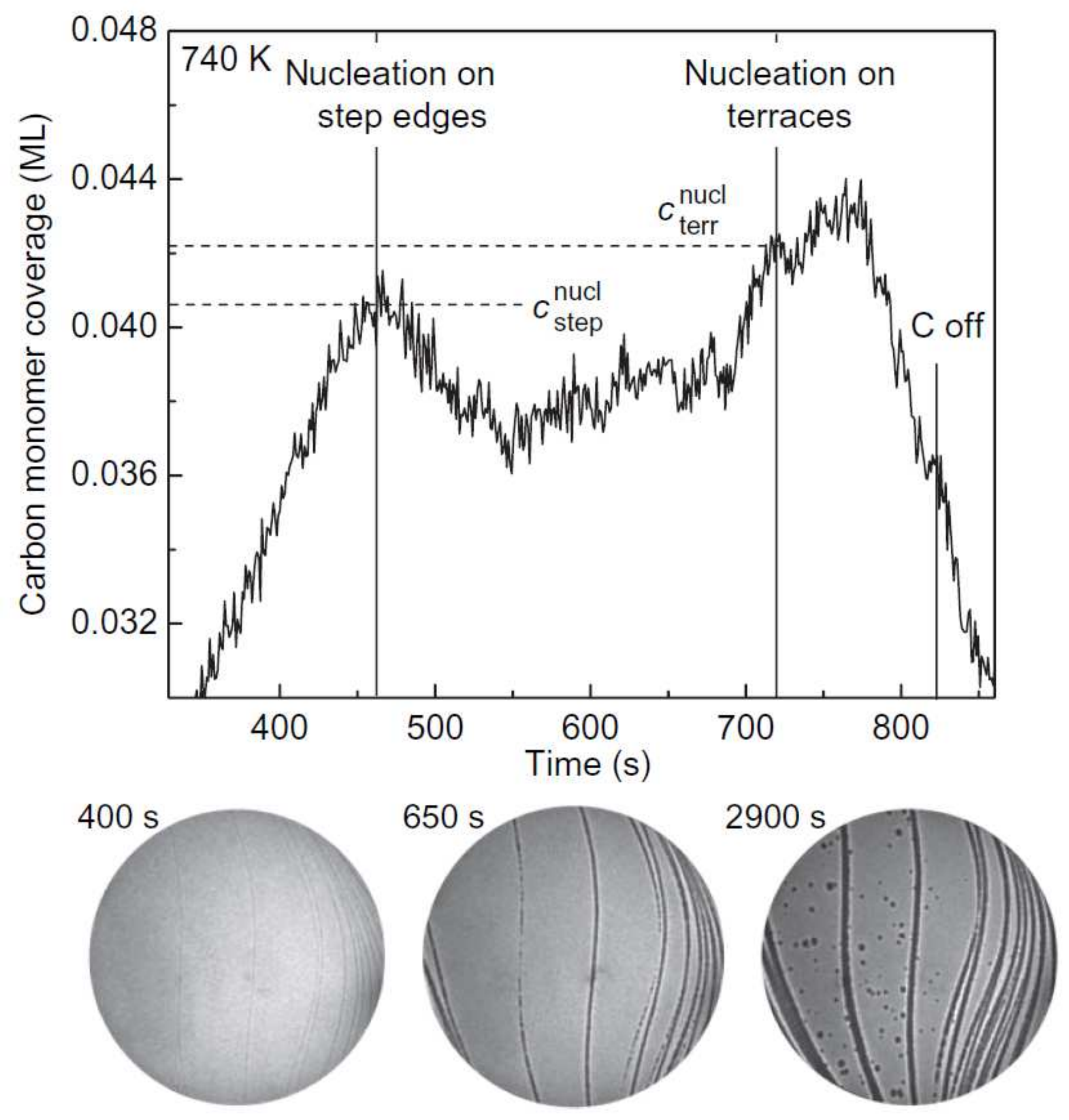}
\par\end{centering}

\caption{Time evolution of adatom coverage during and after C deposition (marked
``C off'') on Ru(0001) surface kept at 740 K (top panel), shown
together with LEEM images (at the bottom) at various times during
growth. The first decrease in adatom concentration corresponds to
nucleation at step edges. This is clear from the LEEM image at 650
s which clearly shows Ru steps decorated with graphene. The second
decrease in adatom concentration occurs at a higher coverage and corresponds
to nucleation on terraces. {[}Reproduced from \cite{Loginova08} by permission of IOP Publishing. All rights reserved.{]}\label{fig:Longinova - nucleation}}
\end{figure}

On Ir(111) at an ethene partial pressure of $5\times10^{-10}$ mbar,
graphene islands nucleate almost exclusively at step edges for growth
temperatures between 790-1320 K \cite{Coraux}, although for temperatures
greater than 1120 K there is an extremely small fraction of islands
found to nucleate at terraces as well. This is similar to growth on
Ru(0001) where nucleation also begins at step edges and specifically
on lower terraces at low C atom concentrations, but also on terraces
for high C atom concentration \cite{Loginova08}, see Fig. \ref{fig:Longinova - nucleation}.
This is in contrast to growth on Rh(111) and Pt(111) where nucleation
has been observed to occur homogeneously on terraces \cite{Hwang11,Wang,Gao11}
while on Cu(111) and polycrystalline copper surfaces, graphene has
been observed to nucleate heterogeneously at steps, defects and impurities
\cite{Wofford,Nie,Celebi:2013cc}.

\begin{figure}
\begin{centering}
\includegraphics[height=7cm]{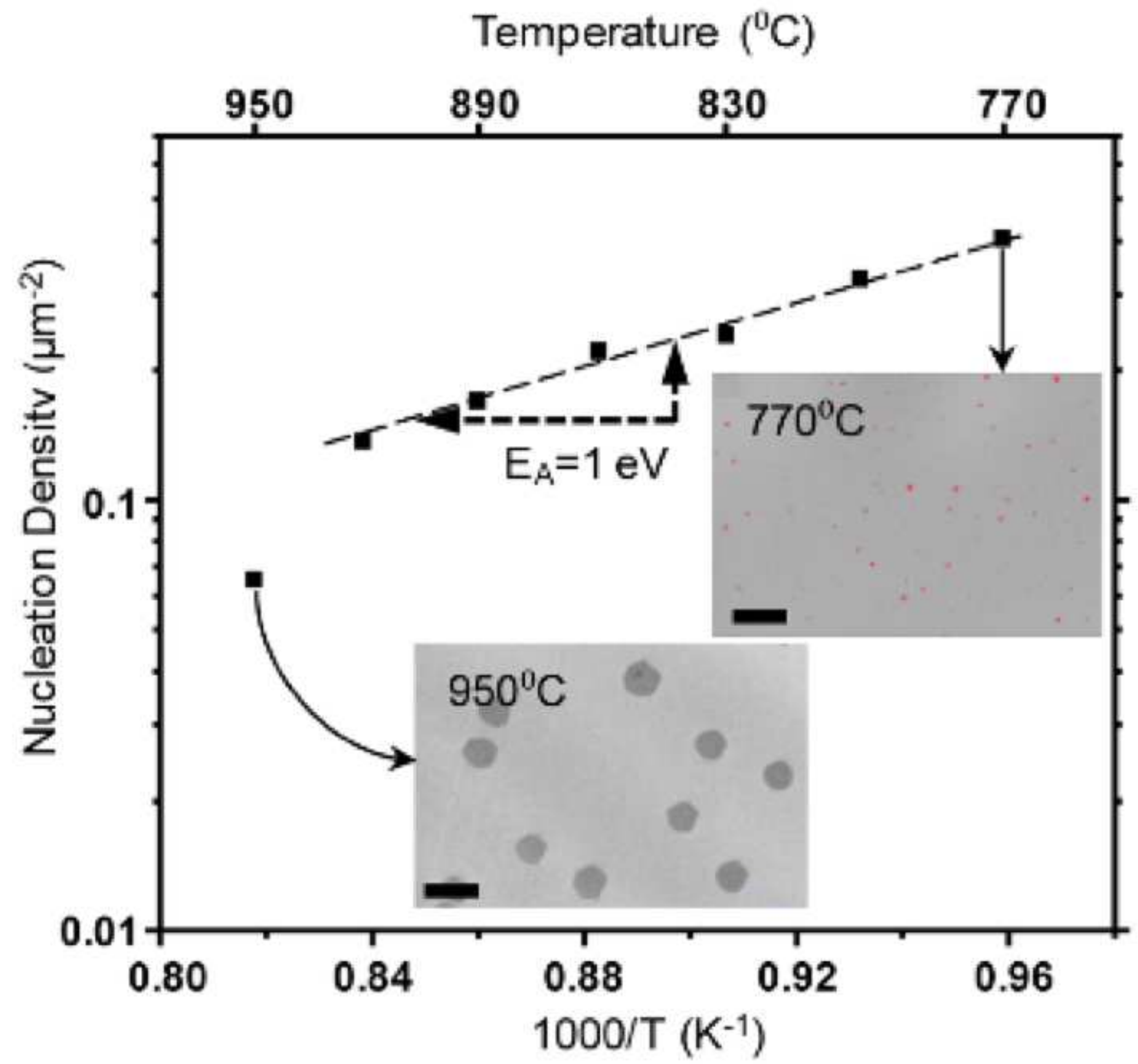}
\par\end{centering}

\caption{Measured nucleation density on copper fitted to an Arrhenius type
of behavior across a temperature range from 1043 K (770$^{\circ}$) to 1223 K (950$^{\circ}$C).
Insets: typical SEM images of the copper surface at the initial stages
of graphene growth for 1043 K (770$^{\circ}$) and 1223 K (950$^{\circ}$C) (scale bar:
1 $\mu$m). {[}Reprinted with permission from \cite{Celebi:2013cc}. Copyright (2013) American Chemical Society.{]}
\label{fig:Measured-nucleation-density-Celebi}}
\end{figure}

It was also observed in \cite{Celebi:2013cc} that the number of nucleation
events on copper is reduced with temperature as shown in Fig. \ref{fig:Measured-nucleation-density-Celebi}.
At not very high temperatures the dependence of the nucleation density
is fitted well by an Arrhenius dependence with an effective activation
energy of $E_{A}=$1 eV. This regime is attributed to an attachment-limited
nucleation. However, at higher temperature of 1223 K the
nucleation density drops significantly. This data is corroborated
by the SEM images shown in the insets for two temperatures. Although
no specific explanation for this effect was suggested, it may be due
to enhanced diffusion of active species during nucleation processes
at higher temperatures. Indeed, the growth of fluctuations as the
temperature is increased should lead to an increase of homogeneous
nucleation events; however, this is not observed in experiment as
small clusters formed in this way would quickly merge with other ones
due to ripening processes leading to a smaller number of clusters
at larger sizes.

\subsubsection{Island morphology and growth\label{sub:CVD-Island-morphology-and-growth}}

We expect the strength of the interaction between carbon species and
the transition metal to be an important factor in influencing the
shape of growing islands, and indeed, for different transition metals,
which are thought to have different magnitudes of interaction with
graphene, we see different island morphologies.

\paragraph{Ruthenium\label{sub:CVD-Ruthenium}}

\begin{figure}
\centering{}\includegraphics[height=6cm]{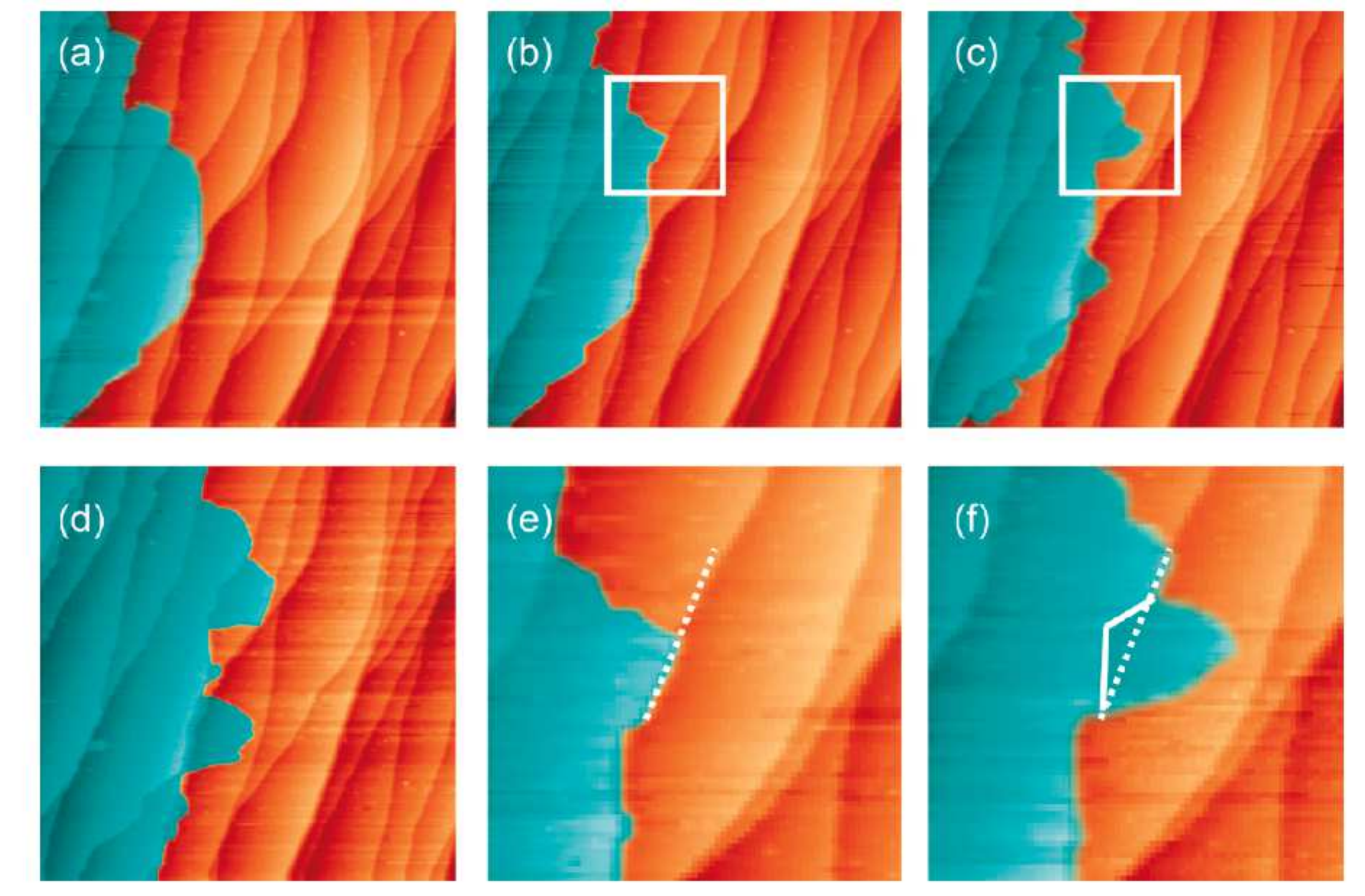}\caption{(a-d) A series of \emph{in situ} STM images at 938 K showing
the growth of graphene (blue) across the steps of the Ru(0001) surface
(orange). (e,f) Details from panels (b) and (c); the dashed lines
mark the changed step position between the consecutive panels (e)
and (f). {[}Reprinted with permission from \cite{Gunther:2011jq}. Copyright (2011) American Chemical Society.{]}\label{fig:Gunther - finger-like growth}}
\end{figure}

A number of groups have demonstrated the step-down growth of graphene
on Ru(0001), whereby islands are only able to grow from upper terraces
to lower terraces over steps and not the other way round \cite{Loginova09,Loginova08,Sutter08,Gunther,Gunther:2011jq}.
This particular scenario is seen to occur both when carbon is provided
by carbon segregating from the bulk (Section \ref{sec:Segregation})
and also when carbon is deposited onto a hot transition metal surface.
The processes occurring during the step down growth have been observed
\emph{in situ} using high temperature STM by dosing the surface held
at 938 K with ethene at $2\times10{}^{-8}$ mbar to induce nucleation,
and then as soon as islands were observed the ethene pressure was
reduced to $5\times10^{-9}$ mbar \cite{Gunther:2011jq}. In this
study, whose aim was to explore this step down growth process in detail,
the edges of these islands were observed to be ``finger-like'',
see Fig. \ref{fig:Gunther - finger-like growth}, meaning that the
growth front did not move coherently over a step \cite{Gunther:2011jq}.
Instead a mechanism was observed whereby part of the growth front,
through some fluctuation, was able to attach to a lower terrace. This
point on the island then presumably acts as a seed for the growth
to continue onto the lower terrace.

\begin{figure}
\centering{}\includegraphics[height=6cm]{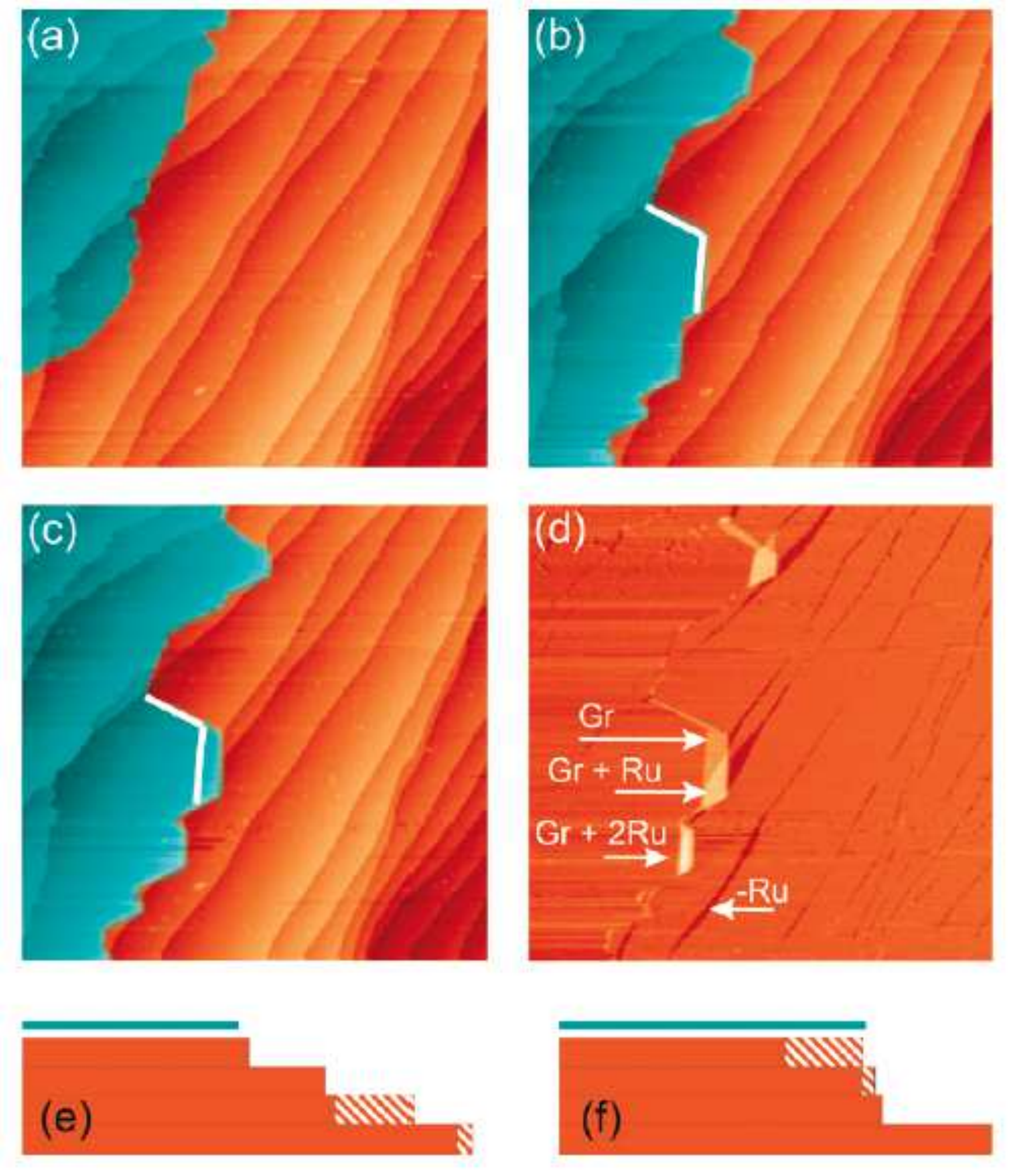}\caption{\emph{In situ} STM images at 938 K, showing the single terrace
mode. (a) After dosing 4.5 L of $\mbox{C\ensuremath{_{2}\mbox{H\ensuremath{_{4}}}}}$
at $2\times10^{-8}$ mbar the ethene valve was closed. (b,c) Ru(0001)
terraces grow and steps are no longer overgrown. (d) Differential
image of (b) and (c), showing areas where Ru atoms are removed (dark)
and areas where graphene has grown and one or two layers of Ru atoms
deposited (bright). Colors as in Fig. \ref{fig:Gunther - finger-like growth}.
(e,f) Proposed growth mechanism with the shading indicating reshuffled
Ru layers. {[}Reprinted with permission from \cite{Gunther:2011jq}. Copyright (2011) American Chemical Society.{]}
\label{fig:Gunther}}
\end{figure}

In the same study a different growth pattern was observed at high
temperatures and extremely low pressures that resulted in yet another
growth process \cite{Gunther:2011jq}. Ethene was deposited at 938
K at $2\times10^{-8}$ mbar. The pressure was then reduced by shutting
the ethene entry valve so that the only deposition was due to the
molecules remaining in the dosing tube. Graphene islands under these
conditions were observed not to overgrow the Ru step but instead to
grow at the same rate as the Ru step as described in Fig. \ref{fig:Gunther}.
The authors speculated that this growth process requires Ru atoms
to be transported to sites underneath the graphene layer and they
further suggested the source of Ru atoms are etched terraces, observed
as the dark lines labelled ``-Ru'' in Fig. \ref{fig:Gunther}. Although
graphene grown in this way was observed to be of extremely high quality,
the low ethene pressures would inhibit the growth rate and, as the
authors note, could limit the applicability of this procedure to the
production of large graphene sheets.

\paragraph{Iridium\label{sub:CVD-Iridium}}

\begin{figure}
\centering{}\includegraphics[height=6cm]{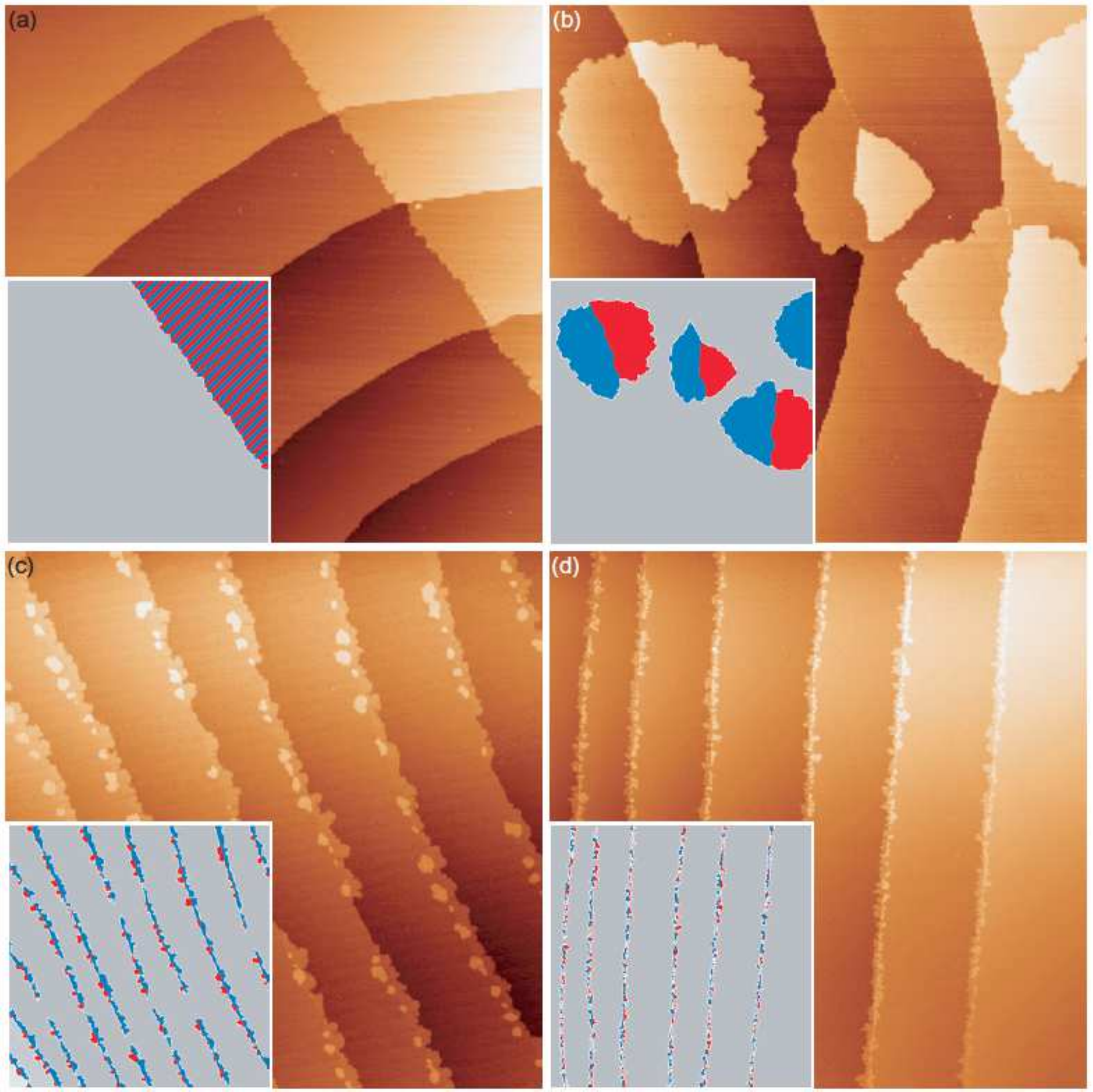}\caption{STM images of graphene on Ir(111) during CVD growth. Images taken
at 300 K after growth has been allowed to continue for: (a) 80 s and
(b)-(d) 40 s. Surface temperatures are: (a) 1320 K, (b) 1120 K, (c)
970 K, (d) 870 K, and in all cases the pressure of ethene during deposition
was $5\times10{}^{-10}$ mbar. Insets show graphene attached to ascending
step edges (blue) and descending steps (red). {[}Reproduced
from \cite{Coraux} by permission of IOP Publishing. All rights reserved.{]} \label{fig:STM-images-Ir111}}
\end{figure}

Unlike the growth of graphene on Ru, on iridium islands growing on
lower terraces are not impeded by steps and so can grow both up and
down iridium steps \cite{Coroux2008b}, although they still demonstrate
a preference for growing down steps \cite{Loginova09}. STM images
taken at 300 K of islands grown at 1120 K and $5\times10^{-10}$ mbar
ethene pressure show a number of islands of uniform size after 20
s of growth \cite{Coraux}, see Fig. \ref{fig:STM-images-Ir111}.
After 40 s of growth a drop in island density is seen indicating the
coalescence of graphene islands. The growth rate of immobile islands
on iridium has been shown to proceed by exactly the same cluster attachment
mechanism as seen on Ru \cite{Loginova08}. Whether only a limited
number of rotational domain types exist on Ir(111) \cite{Loginova2009b}
or a scatter of orientations is present \cite{Hattab2011} remains
an open question; this probably depends on the growth conditions.
It has also been shown that graphene on Ir(111) exists in different
rotational domains with respect to the Ir substrate \cite{Loginova09}.
This is discussed further in Section \ref{sub:Ir(111)}. Formation
of different rotational domains has been shown to be temperature dependent
and further that certain rotational domains seem to grow at a different
rate to others \cite{Hattab2011,Loginova09}.

\paragraph{Copper\label{sub:CVD-Copper}}

Graphene has been grown on both crystalline and polycrystalline surfaces
of copper using CVD with different feed molecules and under different
conditions. We consider here the growth on Cu(111), Cu(001) and on
polycrystalline copper foils. Islands growing on these surfaces demonstrate
a number of different morphologies which may indicate that growth
proceeds by different processes depending on the experimental conditions.
Copper and graphene interact only very weakly so that (i) graphene
can easily ascend steps and (ii) graphene islands grow with many different
rotational domains, like on Ir(111) (see the previous section) and
on Pt(111) \cite{Land1992,Sutter2009}.

\begin{figure}
\begin{centering}
\includegraphics[height=4cm]{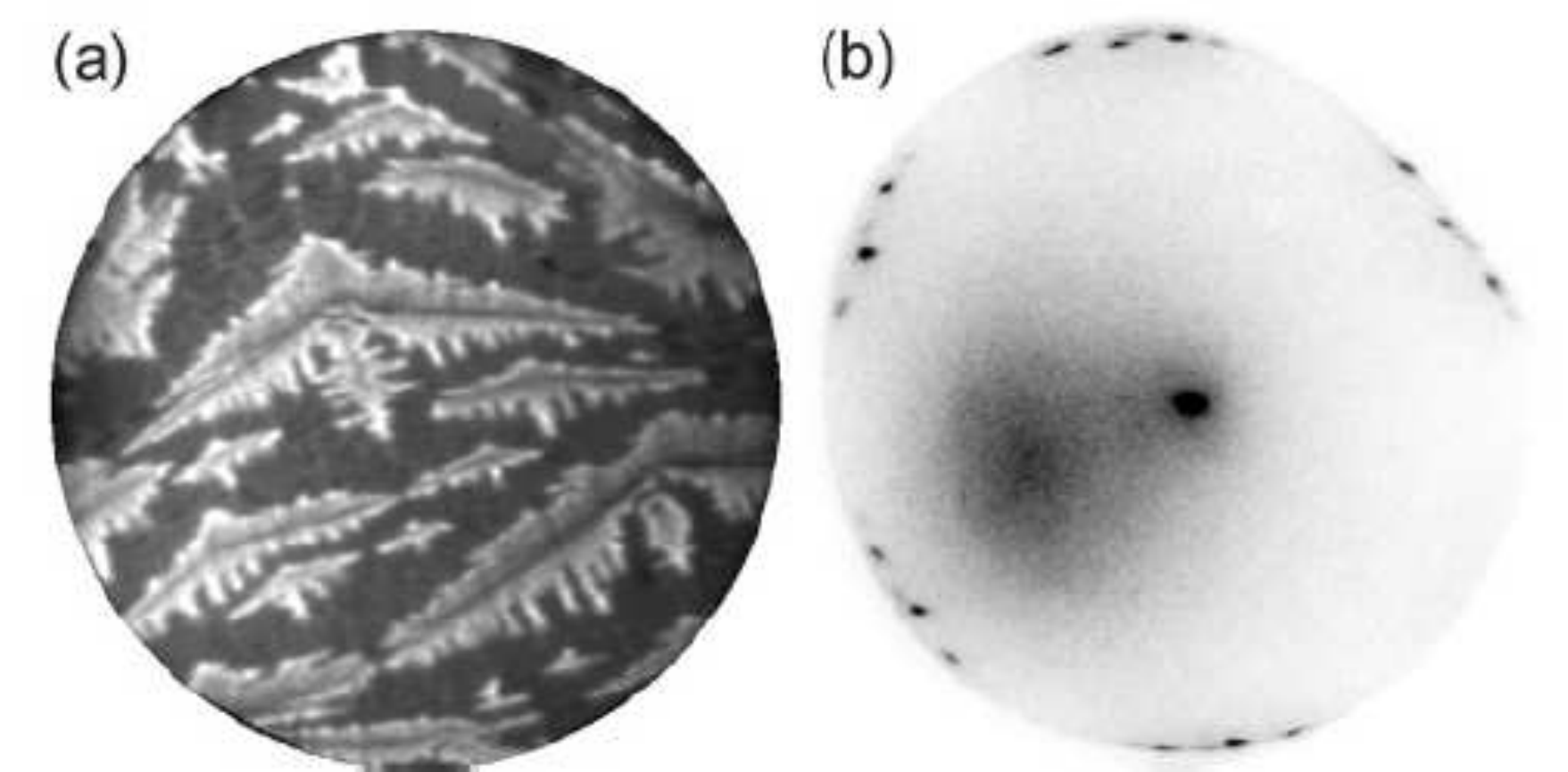}
\par\end{centering}

\caption{(a) Dendritic graphene growing on Cu(111) at 963 K. Field of view
is 7$\mu$m. (b) LEED pattern (44 eV) from a 0.5-$\mu$m-diameter
region of an island, showing that it is polycrystalline. {[}Reprinted
with permission from \cite{Nie}. Copyright (2011) by the American Physical Society.{]}\label{fig:Dendritic-graphene} }
\end{figure}

\begin{figure}
\begin{centering}
\includegraphics[height=4cm]{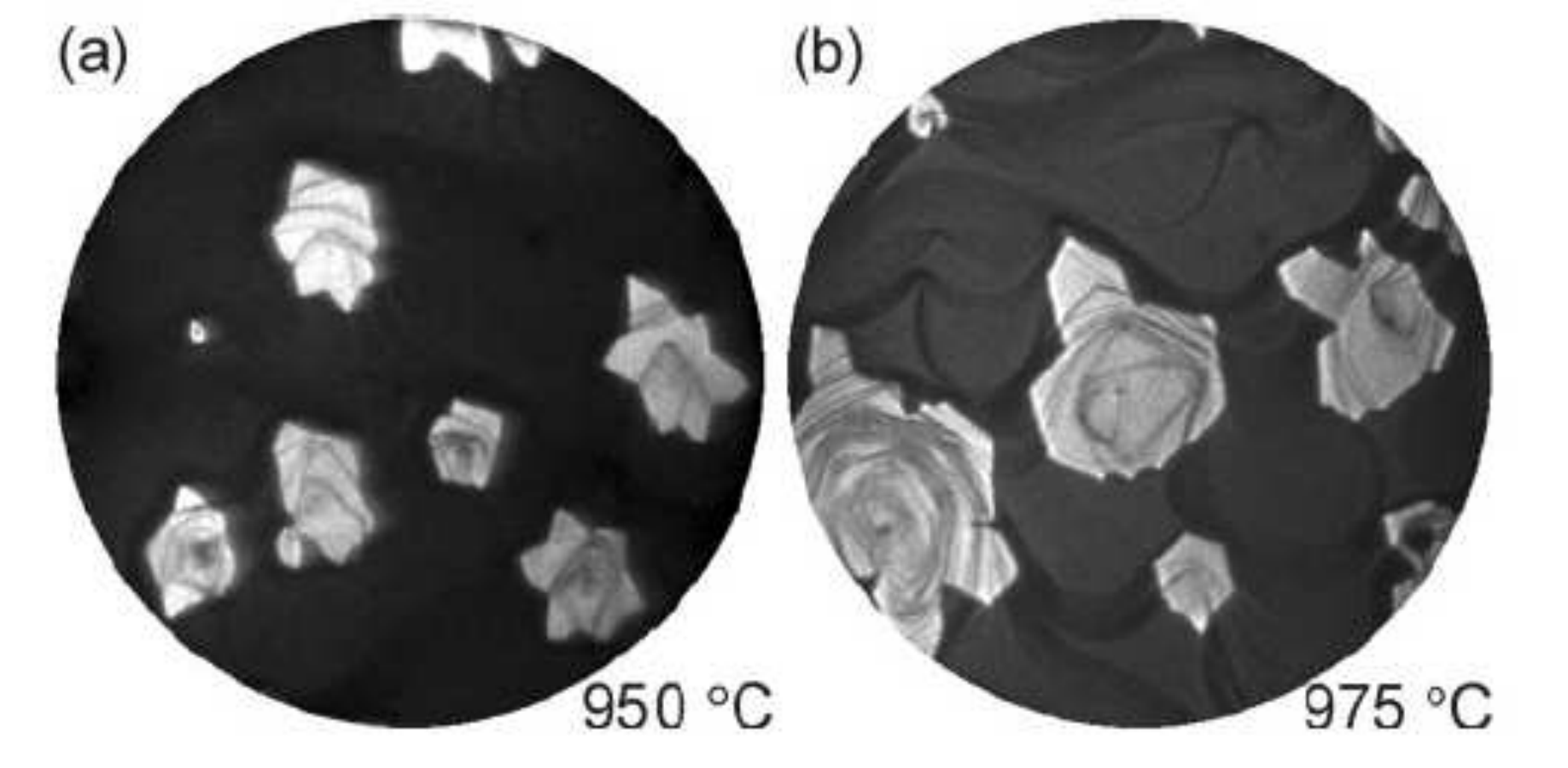}
\par\end{centering}

\caption{LEEM images of faceted islands grown at (a) 1223 K (950$^{\circ}$C) (20-$\mu$m
field of view) and (b) 1248 K (975$^{\circ}$C) (14.5-$\mu$m field
of view). {[}Reprinted with permission from \cite{Nie}. Copyright (2011) by the American Physical Society.{]} \label{fig:faceted graphene}}
\end{figure}

It has been shown that it is possible to grow large single layer graphene
sheets on Cu(111). In 2010 Gao \emph{et}\textcolor{black}{\emph{ al.
}}\textcolor{black}{\cite{Gao10} deposite}d ethene at a partial pressure
of 10$^{-5}$ mbar onto Cu(111) while repeatedly cooling and flash
heating the surface to 1273 K. For 0.35 \% of the Cu(111)
covered with graphene a number of hexagonal domains were observed
with some of them bonded at substrate step edges. At larger coverages
continuous graphene sheets were observed suggesting that smaller graphene
domains coalesce. The sheets were shown to consist of different rotational
domains although only two rotational domains were seen in high concentrations.
Further, hexagonal domains were observed to form two different rotational
domains with respect to the Cu lattice. Nie \emph{et al. }\cite{Nie}
examined in detail the effect of Cu defects and growth temperatures
on graphene morphology. Instead of exposing the surface to ethene
and cooling, they deposited carbon obtained from heating a carbon
rod onto a Cu(111) surface at different temperatures. The morphologies
of graphene islands growing on Cu(111) during deposition of carbon
from a graphitic rod have been shown to be strongly dependent on the
temperature of the Cu surface as well as whether islands nucleate
at step bunches or defect free regions. At lower growth temperatures
($\sim963$ K) islands are considerably more dendritic than those
grown at higher temperatures, see Fig. \ref{fig:Dendritic-graphene},
while at temperatures exceeding 1173 K graphene islands are more compact
and are obviously faceted as demonstrated in Fig.\ref{fig:faceted graphene}.

There is also considerable interest in the growth of graphene on polycrystalline
copper, to some extent due to the low cost of Cu compared to other
transition metal surfaces but also because of the low solubility of
carbon in Cu. Early attempts to grow graphene on Cu by Li \emph{et
al. }\cite{Li10} resulted in graphene grains up to tens of micrometres
in size. Here the graphene produced can itself be considered ``polycrystalline''
in that it contains regions with different rotations with respect
to the substrate. As discussed this can have detrimental effects on
the transport properties of electrons in graphene. Recently efforts
have been made in decreasing the nucleation density, especially by
improving the surface of Cu with careful polishing and by lowering
the proportion of carbon precursor with respect to that of the inert
gas and etching gas (H$_{2}$) \cite{Li2012,Yan2012}. This allows
production of single-crystal graphene grains with sizes of up to a
few millimetres \cite{Yan2012}.

\begin{figure}
\begin{centering}
\includegraphics[height=4cm]{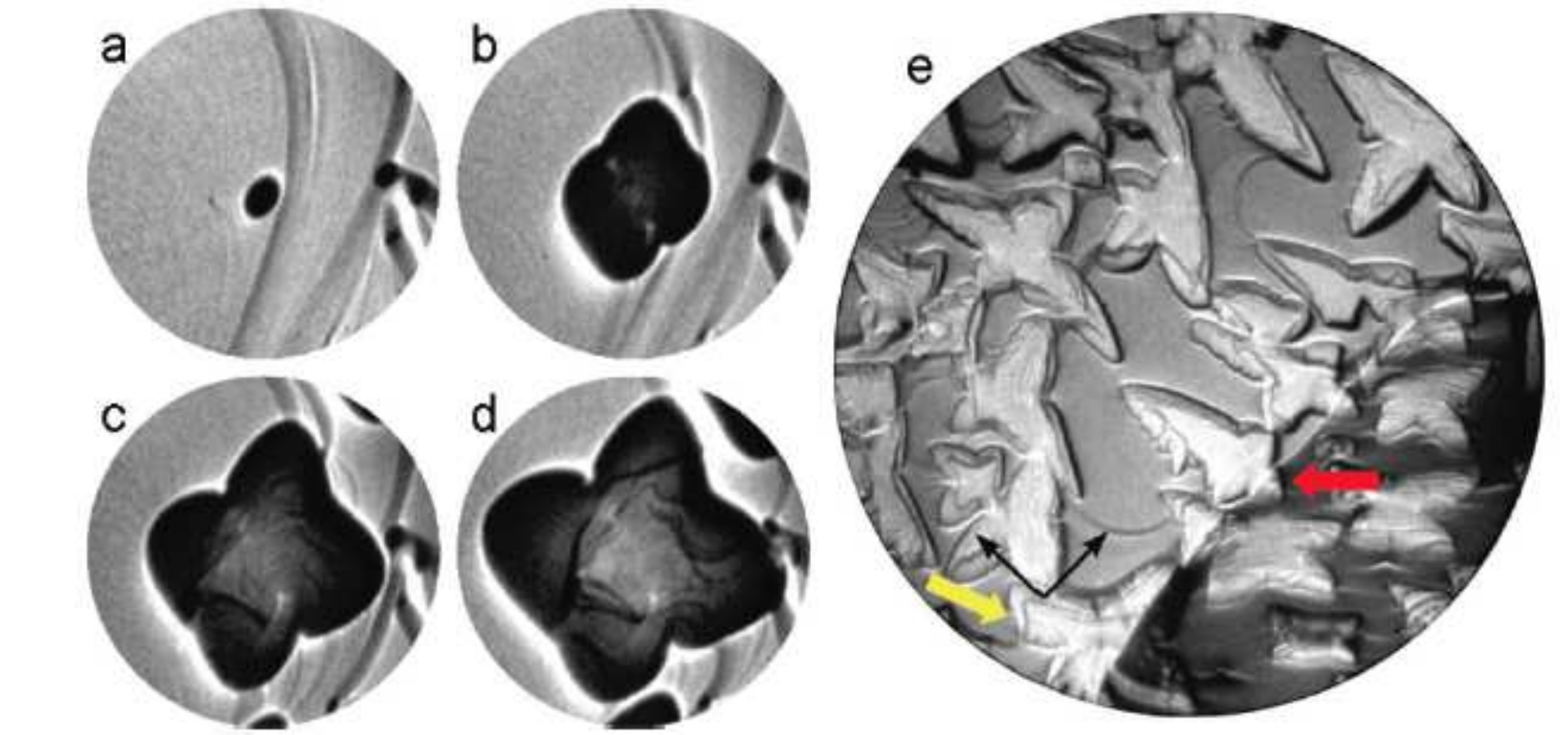}
\par\end{centering}

\caption{LEEM images of the evolution of a graphene island on Cu(100) at 1115 K
after (a) 15 s of C deposition, (b) 90 s, (c) 240 s, and (d) 390 s
(field of view is 10 $\mu$m). This process results in distinctly
four-lobed islands (e). The axes of the graphene lobes tend to align
along the Cu$\left<001\right>$ directions (black arrows). Graphene
lobes are able to grow across Cu grain boundaries (red arrow) and
can be distorted by large Cu step bunches (yellow arrow, field of
view is 46 $\mu$m, grown at 1063 K). {[}Reprinted
with permission from \cite{Wofford}. Copyright (2011) American Chemical Society.{]} \label{fig:Cu(001)}}
\end{figure}

One of the common surface terminations of Cu grains in polycrystalline
foils and films is the (100) facet. On this surface graphene forms
a distinct four lobed structure, Fig. \ref{fig:Cu(001)}, with each
lobe having a different crystallographic orientation with respect
to the copper lattice \cite{Wofford,Li10}. Wofford \emph{et al. \cite{Wofford}
}used LEEM images to observe\emph{ in situ }the formation of graphene
islands growing on the (100) surface, Fig. \ref{fig:Cu(001)}. The
nucleation of graphene islands was only observed for growth temperatures
greater than 1063 K and occurred heterogeneously at defect
sites and imperfections. The 4-lobed structure was attributed to the
simultaneous nucleation of graphene crystals with different orientations
at a single nucleation site. Eventually these graphene islands coalesce
to form a single polycrystalline film.

\begin{figure}
\begin{centering}
\includegraphics[height=6cm]{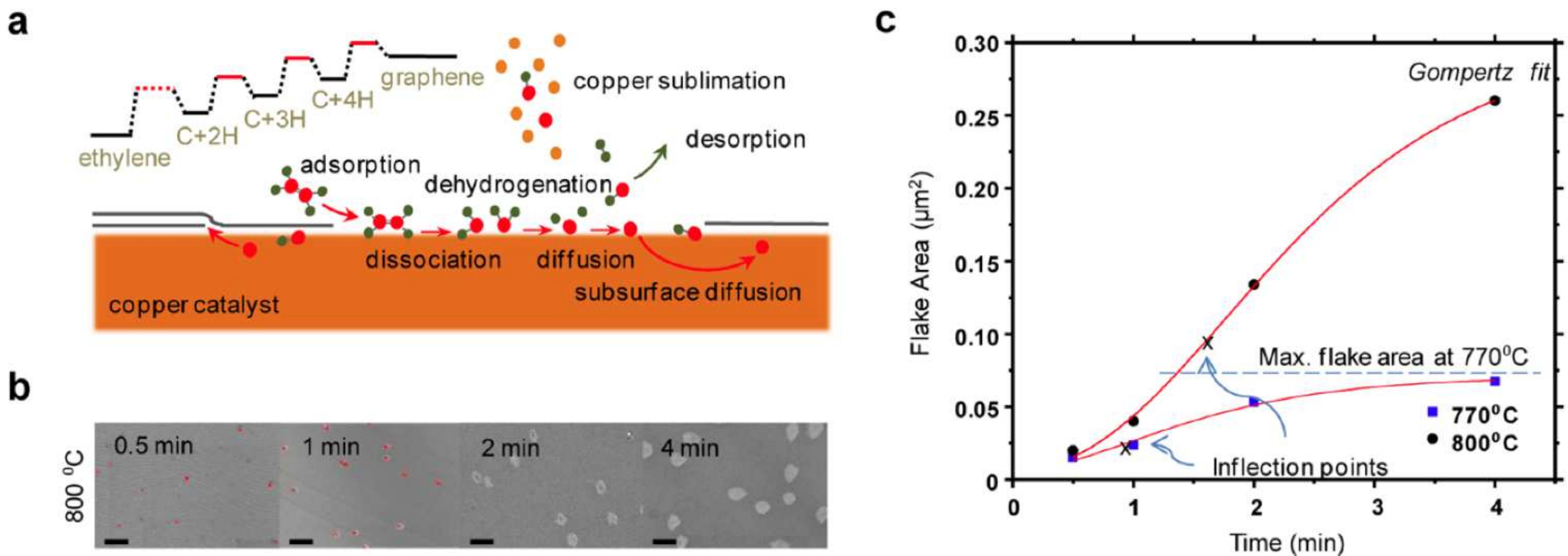}
\par\end{centering}

\caption{(a) General picture of processes taking place during CVD growth of
graphene on copper; (b) typical SEM micrographs showing evolution
of graphene islands with time (800$^{\circ}$C (1073 K), scale bar is 1 $\mu$m);
(c) evolution of the measured mean area of graphene flakes with time
at two temperatures demonstrating Gompertzian kinetics. Inflection
points in both cases are also indicated. {[}Reprinted with permission
from \cite{Celebi:2013cc}. Copyright (2013) American Chemical Society.{]} \label{fig:Celebi-kinetics}}
\end{figure}

A detailed study of graphene formation kinetics has been attempted
by Celebi \emph{et al.} \cite{Celebi:2013cc}. In these CVD experiments
the copper surface was exposed to ethene and a small amount of hydrogen
gas at the same time. It was found that the growth is sustained by
continuous hydrocarbon input, but is hampered by copper sublimation.
The latter has a complex effect: on the one hand, sublimation produces
surface defects which may facilitate dehydrogenation reactions of
the feedstock molecules and likely serve as nucleation centers, but
on the other hand the evaporated Cu atoms facilitate desorption of
hydrocarbon molecules thereby reducing the density of active species
on the surface, see the schematics in Fig. \ref{fig:Celebi-kinetics}(a).
It was also found that maintaining a high temperature after the supply
of ethene was switched off did not increase further the mean flake
size. The authors also studied the effect of hydrogen gas on graphene
CVD growth and found little difference with and without it. It was
also observed that secondary graphene flakes grow underneath existing
primary flakes, and their growth is stopped only when the primary
graphene layer is complete. Interestingly, the shapes of secondary
flakes are clearly hexagonal, while the primary flakes are observed
to be mostly circular. It is suggested that the circular shape of
the primary flakes is due to Cu sublimation; it is suppressed when
secondary flakes are formed underneath the primary ones resulting
in hexagonal shapes.

Using regimes for which individual graphene islands could be resolved,
their mean area $A(t)$ as a function of time was measured from the
SEM images as shown in Fig. \ref{fig:Celebi-kinetics}(c) for two
temperatures; typical SEM micrographs over a 4 minute time interval
are shown in Fig. \ref{fig:Celebi-kinetics}(b). The island's growth
shows a typical sigmoidal (``S-shaped'') time dependence \cite{Skrdla:2012jt}
: initially when the density of active carbon species is low, the
growth is limited by the small number of nucleated seeds (incubation);
with time growth progresses faster due to attachment of active carbon
species to existing nuclei and new nuclei are formed; finally, as
the supply of active carbon species becomes scarce due to their fast
previous consumption, growth slows down again. Assuming that the flakes'
areal enlargement rate, $dA/dt$, is proportional to the their area,
$A$, with a proportionality constant which decays exponentially with
time (dispersive kinetics), i.e.
\begin{equation}
\frac{dA}{dt}\sim Ae^{-kt}\;,\label{eq:Gompertzian1}
\end{equation}
the Gompertz type of the area time dependence is obtained:
\begin{equation}
A(t)=A_{max}\exp\left\{ -\exp\left[\frac{e\mu_{m}}{A_{max}}\left(t-\lambda\right)+1\right]\right\} \;,\label{eq:Gompertzian}
\end{equation}
where $A_{max}$ is the maximum possible flake size ($t\rightarrow\infty$),
and the other constants ($\lambda$ and $\mu_{m}$, the latter being
the growth rate at the inflection point where $dA/dt$ has a maximum)
are fitting parameters ($e$ is the base of the natural logarithm).
This Gompertzian kinetics was found to fit extremely well the observed
time dependence of the area of the flakes measured at different temperatures,
see examples in Fig. \ref{fig:Celebi-kinetics}(c), which shows that
underlying assumptions made when writing Eq. (\ref{eq:Gompertzian1})
are probably reasonable. Finally, this type of kinetic analysis of
flake size was performed at different temperatures, and effective
activation energies (time dependent) were analyzed. It was argued
that the rate limiting step in graphene growth on copper can be related
to the dissociation of hydrocarbon species, which involves large energy
barriers. Different processes essential for an understanding of graphene
growth are depicted schematically in Fig. \ref{fig:Celebi-kinetics}(a).

\subsubsection{CVD with carbide as a transient state}

\begin{figure}
\begin{centering}
\includegraphics[height=4cm]{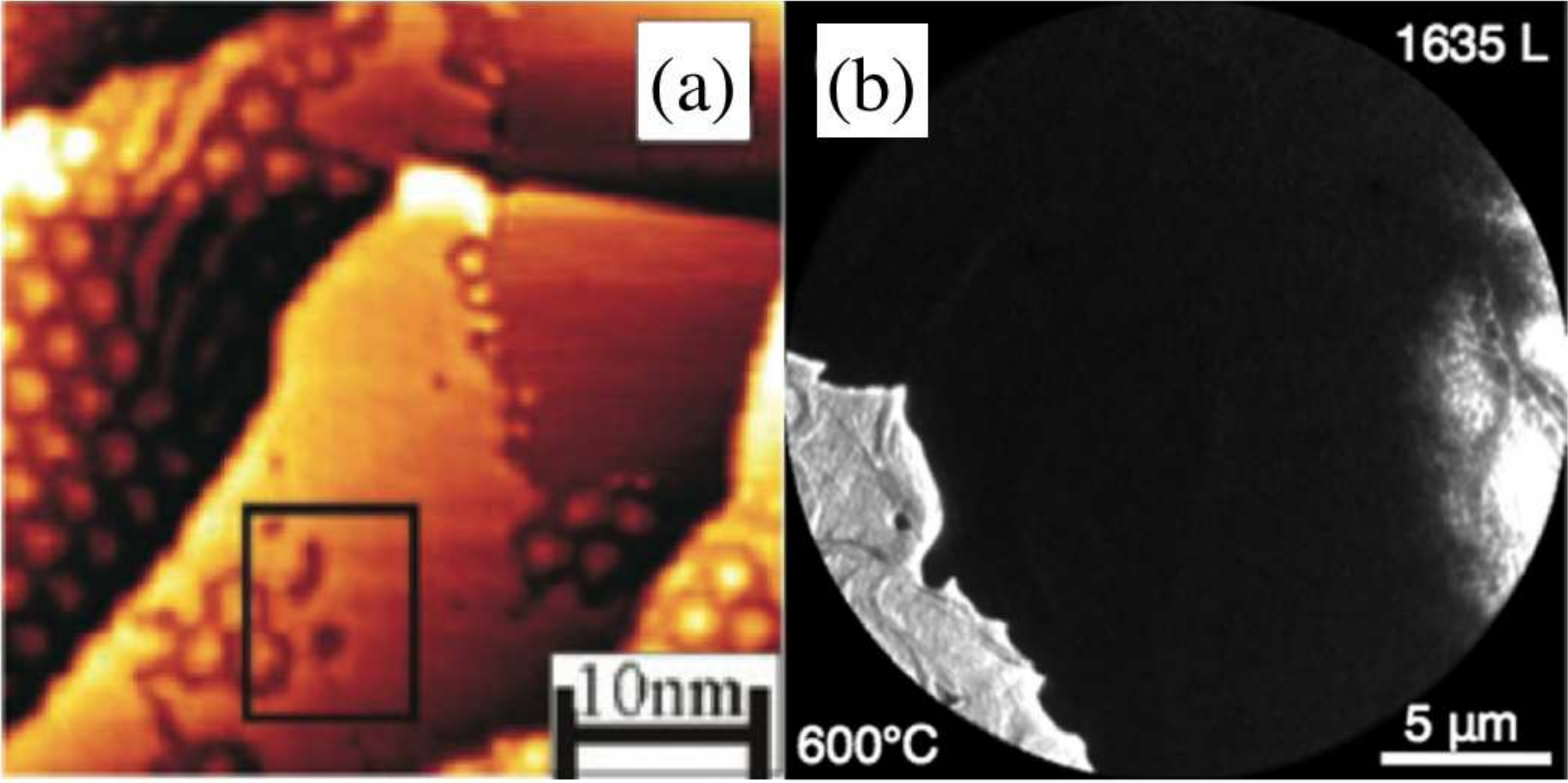}
\par\end{centering}

\caption{(a) Coexistence of a surface carbide and graphene on Ni(111), as shown
by STM at room temperature of a sample prepared at 730 K. (b) LEEM
image showing the carbon-free Ni surface between graphene and the
carbide, in growth conditions at 870 K. {[}Reprinted with permission
from \cite{Lahiri2010}. Copyright (2010) American Chemical Society (a) and \cite{Addou}. Copyright (2012), AIP Publishing LCC (b).{]}\label{fig:Coexistence-of-carbide-graphene-Ni}}
\end{figure}

The phase diagram of the metal-carbon system is complex in some cases.
Not only graphene on a metal, carbon dissolved in the bulk, but also
carbon-metal alloys
may exist within
different ranges of composition and temperature. This is especially
true on metals such as Fe, Ni, Rh, or Re. While in some cases the
formation of
a
surface carbide state must be avoided because it
is more thermodynamically stable than the graphene phase itself, as
was found on Fe(110) \cite{Vinogradov2012}, in other cases the
carbide can be an intermediate state during CVD process during the
growth of graphene. This is what was found on Ni(111), a surface on
which a well-known surface carbide, Ni$_{2}$C, readily
forms at a few 100 K above room temperature. Around 770 K, it was
found that the carbide phase, which corresponds to a high density
carbon phase, transforms into a graphene layer whose zigzag rows are
3\textdegree{}-rotated with respect to the Ni dense-packed rows \cite{Lahiri2010},
see Fig. \ref{fig:Coexistence-of-carbide-graphene-Ni}(a). If the
system is quenched to room temperature the transformation is stopped
and the coexistence of neighboring (touching) carbide and graphene
regions is observed. How this transformation occurs remains unknown.
At slightly higher temperature (50 K higher), a different transformation
is observed: graphene islands nucleate after the carbide fully covers
the substrate surface, and progressively grow by depleting the Ni
surface from the carbide, as observed by \emph{in operando} LEEM \cite{Addou},
see Fig. \ref{fig:Coexistence-of-carbide-graphene-Ni}(b).
In the case of Re(10$\bar{1}$0) it was found that saturating the bulk at high
 temperature (1700 K) with carbon yields an $\alpha$-carbide. After
formation of this carbide, graphene forms on the surface of the crystal
at high temperature \cite{Gall1987}.

\subsection{Temperature Programmed Growth (TPG)\label{sec:Temperature-Programmed-Growth}}

\subsubsection{Method\label{sub:Method}}

Another method of growing graphene on transition metals is
 to first adsorb carbon-containing species at room temperature on the
surface, and then to progressively heat the sample. In this method, the
reaction between the adsorbed species, which drives the formation of graphene,
 is triggered by the temperature increase. Thus the method has been coined
"temperature programmed growth" (TPG) \cite{Coraux}.
The main difference with the CVD method is
that in TPG the carbon containing species are added to the
'cold' surface with subsequent annealing to stimulate the growth,
while in CVD the species are added to the surface which is already 'hot'.

TPG experiments are performed in an ultra high vacuum (UHV) to prevent
contamination. As
in CVD,
the surface of the transition
metal is initially cleaned. The metal is maintained at room temperature
and exposed to the hydrocarbon source, and the molecules become adsorbed
onto the surface. The system is then heated to a fixed temperature
(TPG temperature).
Upon heating, the adsorbed molecules experience progressive dehydrogenation.
After full dehydrogenation, the resulting carbon species readily assemble
to form carbidic species, whose structure gets better ordered as the
temperature increases, eventually yielding graphene islands.
Depending
on the heating temperature the graphene growth will progress and the
properties of the islands change.
STM, performed both after cool-down \cite{Coraux} and during heating
\cite{Dong}, have provided valuable insights into graphene growth by CVD.

\subsubsection{Observations\label{sub:Observations}}

\begin{figure}
\centering{}\includegraphics[scale=0.5]{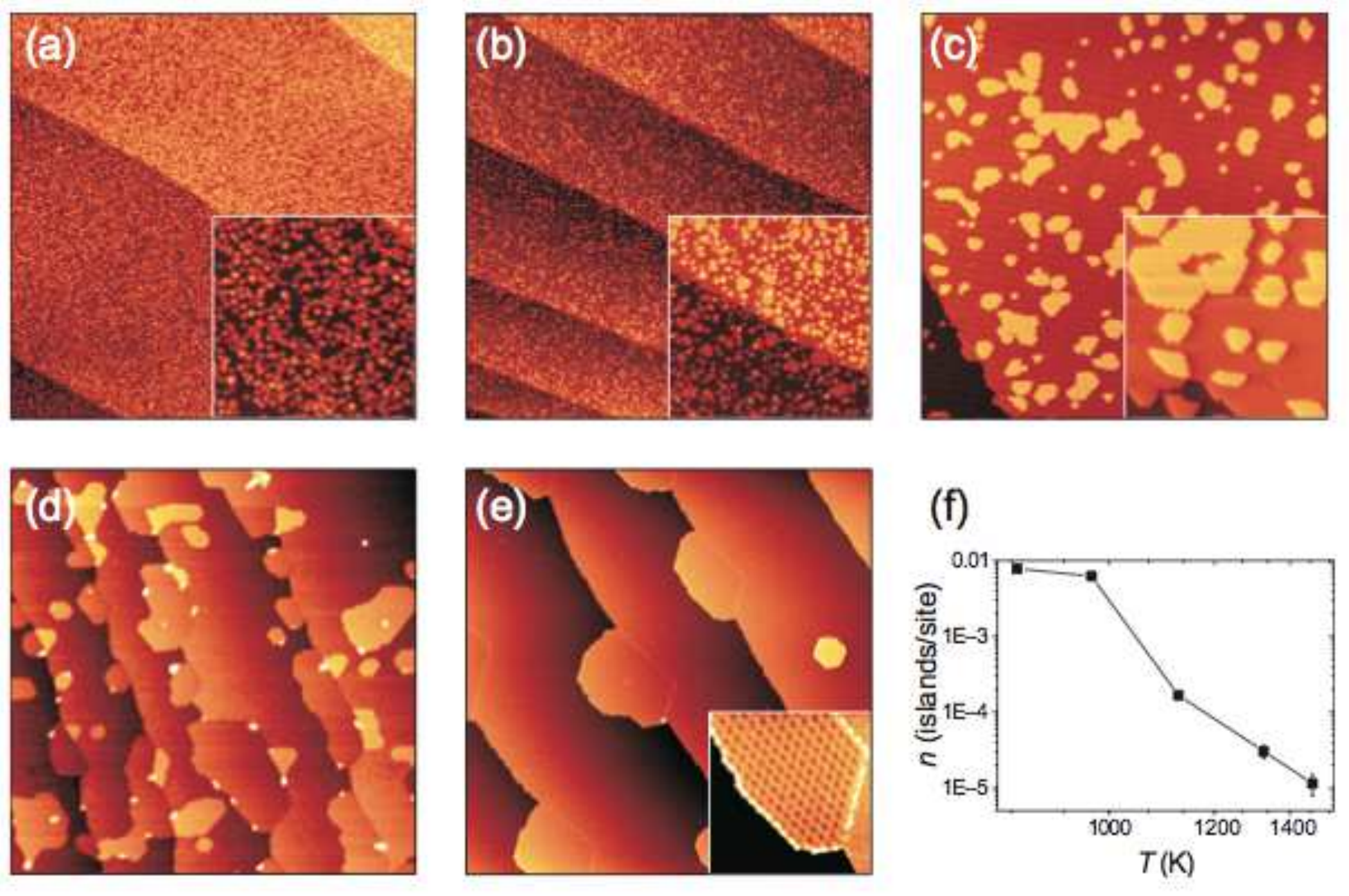}\caption{STM images of graphene grown on the Ir(111) surface with TPG after
heating to (a) 870 K, (b) 970 K, (c) 1120 K, (d) 1320 K, and (e) 1470
K for 20 s. (f) Graphene island density $n$ for the different annealing
temperatures $T$. {[}Reproduced from \cite{Coraux} by permission of IOP Publishing. All rights reserved.{]}
\label{fig:Coroux-TPG1}}
\end{figure}

The growth of graphene has been imaged on Ir(111) \cite{Coraux},
Rh(111) \cite{Dong,Wang} and Ru(0001) surfaces for a range of TPG
temperatures varying from around 770 K to 1470 K. On Ir(111) at temperatures
below 870 K carbon species are found randomly distributed on the surfaces.
Islands formed from these structures have non-uniform height and a diameter
less than 2 nm. Also the moir\'{e} structure that is characteristic of
graphene (see Section \ref{sec:Structure-of-Graphene}) is not observed.
At these temperatures hydrocarbons have decomposed but graphene formation
has not yet begun. At TPG temperatures between 870-970 K an increase
in the graphene island density is noticed and small graphene islands
with a moir\'{e} structure are identified. Therefore graphene starts to
form on these surfaces in this temperature range. At temperatures
larger than 970 K a decrease in graphene island density is observed
because of an increase of the mean island size, Fig. \ref{fig:Coroux-TPG1}(f).
The formation temperature also varies depending on the surface. The
lowest temperatures at which graphene formation is initiated during
TPG experiments are summarized in Table \ref{tab:table1}. At higher
TPG temperatures the graphene islands are found to be larger in size
and are located primarily at step edges. It was also found that at
higher temperatures the difference in orientation between the islands
becomes \textcolor{black}{smaller (the direction distribution is narrower)
as some orientations are more energetically favorable. The orientations
found are influenced by the underlying substrate \cite{Coraux,Dong}.
The dynamics of graphene growth on the Ir(111) surface for various
annealing temperatures is shown in Fig}. \ref{fig:Coroux-TPG1}.

\begin{table}
\begin{centering}
\begin{tabular}{|c|c|c|c|}
\hline
Surface & Carbon species & Graphene & Ref.\tabularnewline
\hline
\hline
Rh(111) & 770 K & 870 - 973 K & \cite{Wang}\tabularnewline
\hline
Rh(111) & < 870 K & %
\begin{tabular}{c}
808 - 969 K\tabularnewline
(dissolves at 1053 K)\tabularnewline
\end{tabular} & \cite{Dong}\tabularnewline
\hline
Ir(111) & 870 K & 970 - 1470 K & \cite{Coraux}\tabularnewline
\hline
Ru(0001) & No results & 973 - 1173 K & \cite{Huang2012}\tabularnewline
\hline
Ru(0001)  & 900 K & 1000 - 1100 K & \cite{Cui2011}\tabularnewline
\hline
Pt(111) & 900 K & 1000 - 1100 K & \cite{Cui2011}\tabularnewline
\hline
Re(0001) & No results & 1100 K & \cite{Miniussi2011}\tabularnewline
\hline
Co(0001) & <410 K & 600 K & \cite{Eom2009}\tabularnewline
\hline
\end{tabular}
\par\end{centering}

\caption{The temperature ranges where carbon species and graphene were observed
on different surfaces.\label{tab:table1}}
\end{table}

\subsubsection{Ripening\label{sub:Ripening}}

\begin{figure}
\centering{}\includegraphics[width=15cm]{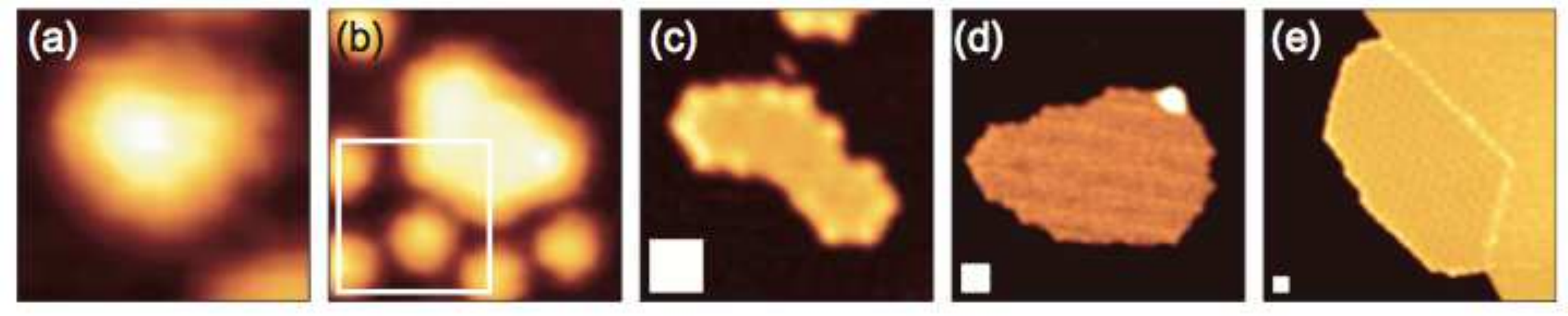}\caption{STM topographs of graphene islands grown with TPG on Ir(111) after
heating to (a) 870 K, (b) 970 K, (c) 1120 K, (d) 1320 K, and (e) 1470
K for 20 s. At low temperatures, (a) and (b), the islands are compact,
then at intermediate temperatures (c) they become non-compact. They
return to being compact at higher temperatures (d), (e). {[}Reproduced from \cite{Coraux} by permission of IOP Publishing. All rights reserved.{]} \label{fig:Coroux_TPG2}}
\end{figure}

By imaging the growth at a range of TPG temperatures different stages
in the growth of graphene can be identified. From this useful conclusions
about how the growth progresses have been drawn. Coraux\emph{ et al.}
\cite{Coraux} observed how the size and shape of the graphene islands
changes at various TPG temperatures. They annealed the sample to a
particular temperature and then cooled it so that the surface can
be imaged with STM. At higher annealing temperatures the graphene
growth is noticed to be more advanced. At 970 K when graphene starts
to form, the islands are small and have a compact shape, shown in
Figure \ref{fig:Coroux_TPG2}(a) and (b). As the temperature is increased
to 1120 K the islands appear larger, with a non-compact irregular
shape, see Figure \ref{fig:Coroux_TPG2}(c). At even higher temperatures
of 1320 K and 1470 K the islands continue to grow in size, however
their shape is compact again as demonstrated by Figure \ref{fig:Coroux_TPG2}(d)
and (e).

These observations suggest that the islands grow by ripening processes.
Two possible mechanisms for this are Ostwald ripening and Smoluchowski
ripening. In Ostwald ripening large islands grow by incorporating
carbon adatoms arising from the dissolution of smaller islands \cite{Ostwaldripening},
whereas in Smoluchowski ripening entire mobile islands coalesce \cite{Schmoluchowskiripening}.
For Ostwald ripening adatoms need to be able to detach from the smaller
islands, which is unlikely at temperatures below 1500 K \cite{Coraux}.
Since the temperature range of the TPG experiment is below this, it
is concluded \cite{Coraux} that the main mechanism for graphene growth
must be Smoluchowski ripening. This causes the reduction in observed
particle density as the growth progresses due to island migration
and subsequent joining together. Their high mobility may be facilitated
by increased distance of the graphene islands from the transition
metal surface \cite{Amara:2009hn}, see also Section \ref{sec:Structure-of-Graphene},
as this effectively reduces the interaction between a graphene flake
and the metal and decreases the diffusion energy barrier. The Smoluchowski
ripening process is also suggested by the changes in the compactness
of the islands. As temperature increases the islands become more mobile.
At 1120 K the islands have enough mobility to make contact with each
other and coalesce. However, it is argued \cite{Coraux}, that at
intermediate temperatures there is not enough energy for the recently
coalesced islands to reshape themselves in the time period during
which the heat is applied. This causes non-compact islands to form.
At 1320 K and above, large islands are located mainly at the step
edges of the substrate where they have become attached. As the islands
are less mobile at the step edges the chances of further coalescence
are rarer and their shape stabilizes over time. Smoluchowski ripening
has also been reported for Rh(111) and Ru(0001) surfaces \cite{Wang,Huang2012}.

\subsubsection{Cluster Formation\label{sub:Cluster-Formation}}

The initial shape of clusters growing on Ir(111) has been investigated
using \emph{in situ} high-resolution photoelectron spectroscopy \cite{Lacovig}.
It was discovered that in the early stages of growth a number of carbon
clusters are formed. These clusters vary in size but all were observed
to be dome-like, with the interaction between cluster and surface
occurring only at edge sites.

\begin{figure}
\centering{}\includegraphics[height=5cm]{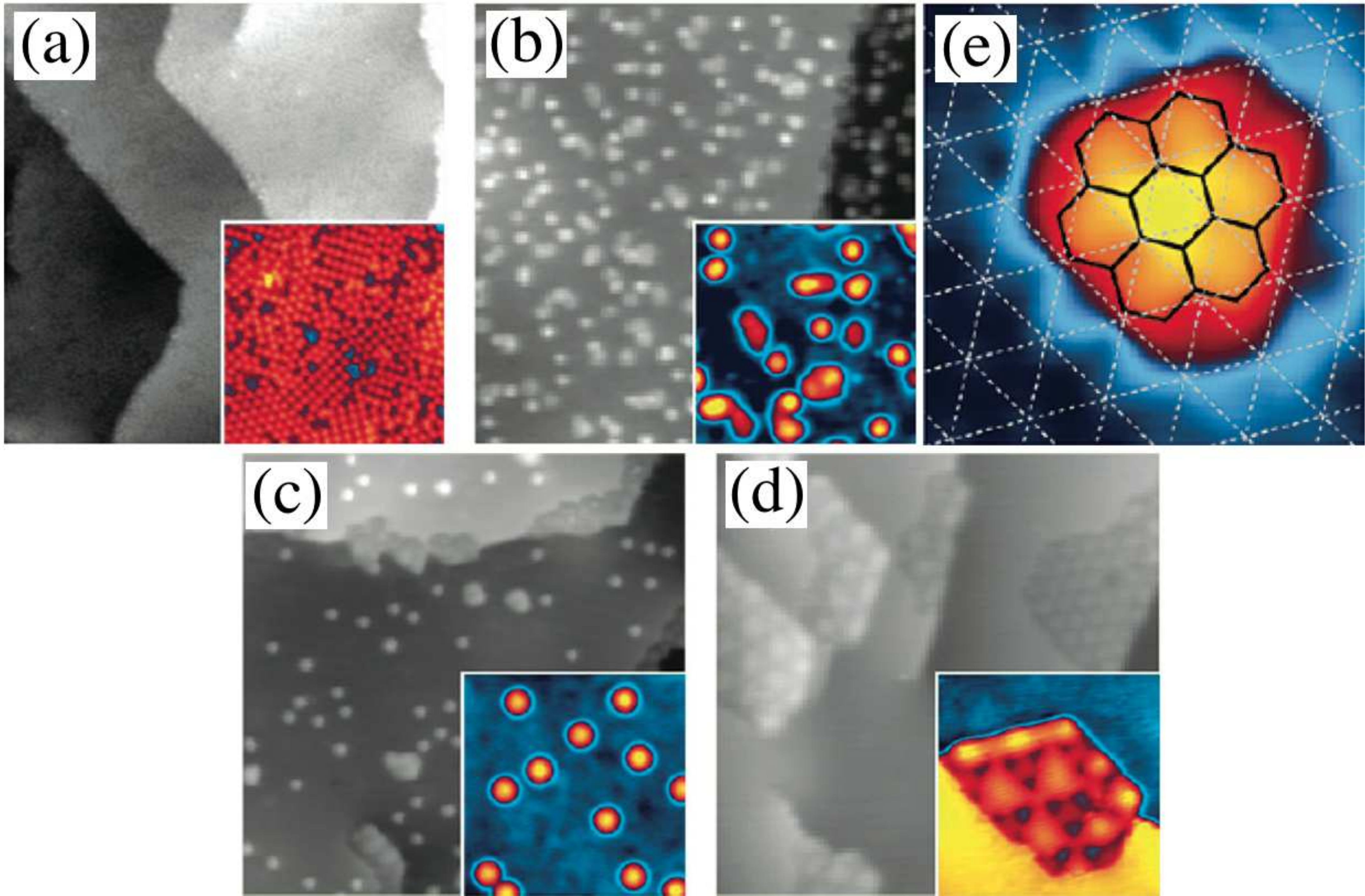} \caption{STM images of the Rh(111) surface annealed to (a) 470 K, (b) 770 K,
(c) 870 K, and (d) 973 K after ethene deposition. (e) A STM topograph
of clusters seen in (b) and (c). \label{fig:Wang-TPG} {[}Reprinted
with permission from \cite{Wang}. Copyright (2011) American Chemical Society.{]}}
\end{figure}

In a TPG study of graphene growth on Rh(111) initiated by the ethene
deposition, the presence of randomly distributed small clusters was
found to coincide with the formation of graphene islands \cite{Wang}.
STM images show that just prior to graphene formation the size distribution
of the clusters is dramatically reduced and the cluster particle density
decreases. At this point carbon clusters were identified with a distinct
size of 1 nm across. With high STM magnification these were identified
as a carbon honeycomb of 7 hexagons joined together (known as 7C$_{6})$,
see Fig. \ref{fig:Wang-TPG}(e). DFT calculations determined that
the most stable structure for the clusters is a dome shape, with the
outer atoms interacting more strongly with the surface. The relative
concentration of these clusters with respect to all carbon species
increases with TPG temperature and eventually they are the only species
present other than graphene islands. This led the authors to conclude
that these uniformly sized clusters are the only precursors for graphene
growth.

Once the temperature is high enough for ethene to decompose, the clusters
start to form. At 770 K and 870 K, as is demonstrated by Fig. \ref{fig:Wang-TPG}(b)
and (c), the clusters are observed exclusively on terraces. Since
there are no clusters at step edges, then they cannot have formed
by using step sites to facilitate their growth e.g. via creating dimers
from C monomers as was previously proposed \cite{Chen}. They are
suggested to have formed instead from the decomposition of heavy hydrocarbon
species created from the polymerization of ethylidyne (cf. Sections
\ref{sec:Carbon-Feedstock-Experimental} and \ref{sec:Carbon-Feedstock-Theory-1}).
At 870 K clusters were observed to start diffusing on terraces towards
steps (Smoluchowski ripening) where they coalesce into graphene flakes.

At 973 K the diffusion is greatly enhanced, and graphene islands that
have been formed from the small clusters are found predominately at
step edge sites, Fig. \ref{fig:Wang-TPG}(d). This suggests that most
of the mobile 1 nm clusters eventually settle at step edges, where
they grow to form graphene by Smoluchowski ripening. However a few
graphene islands are found on large terraces, and therefore it was
deduced that it is the formation of the clusters rather than their
attachment to step edges that limits the graphene growth rate. This
is unlike graphene formation by other methods. For instance (see also
discussion in Section \ref{sub:CVD-Iridium}), in the cases of hot
C deposition onto Ir(111) it was determined that the formation of
5 atom carbon clusters (which are precursors for graphene growth in
this case) occurs only as they attach to the substrate step edges
\cite{Loginova08}. The activation energy for this process is dependent
on the step edge structure.

\begin{figure}
\centering{}\includegraphics[scale=0.4]{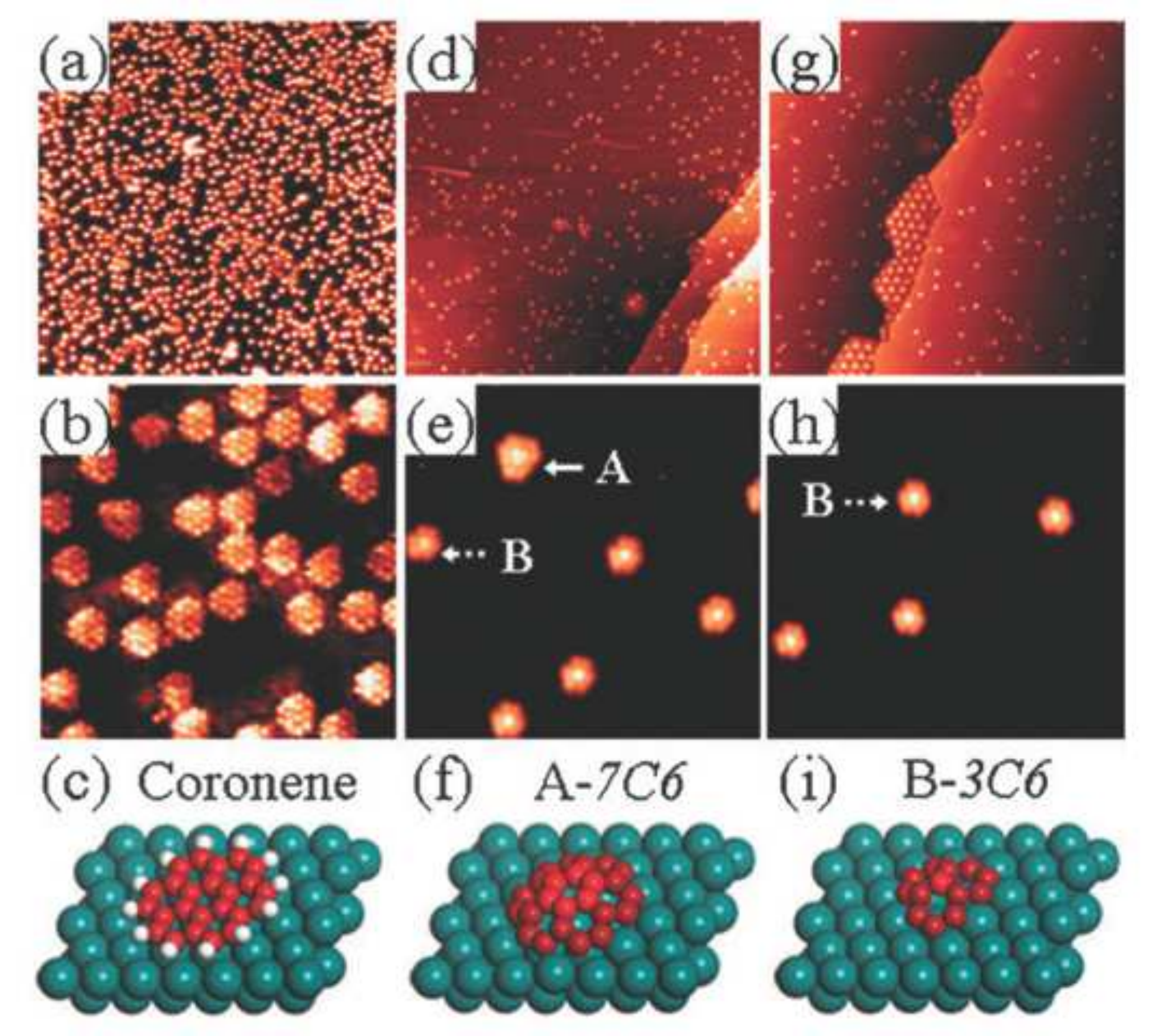}\caption{STM images of the Ru(0001) surface annealed to various temperatures:
(a) 100 $\times$ 100 nm$^{2}$, 500 K; (b) 10 $\times$ 10 nm$^{2}$,
500 K; (d) 100 $\times$ 100 nm$^{2}$, 900 K; (e) 10 $\times$ 10
nm$^{2}$, 900 K; (g) 100 $\times$ 100 nm$^{2}$, 1000 K; (h) 10
$\times$ 10 nm$^{2}$, 1000 K. In (e) and (h) the labels A and B
correspond to the 7C$_{6}$ and the 3C$_{6}$ clusters, respectively.
The structures of (c) coronene, (f) the 7C$_{6}$ cluster and (i)
the 3C$_{6}$ cluster on the Ru(0001) surface are also shown. The
colored balls represent: cyan = Ru, red = C, white = H. \label{fig:Cui-TPG}
{[}Reproduced from \cite{Cui2011} with permission of The Royal Society of Chemistry (RSC) on behalf of the European Society for
Photobiology, the European Photochemistry Association and the RSC.{]}}
\end{figure}

Cluster precursors to graphene growth using TPG have also been reported
on Ir(111) \cite{Lacovig} and Ru(0001) \cite{Cui2011} surfaces.
For coronene and ethene deposition onto a Ru(0001) surface two cluster
types were found at 900 K, see Fig. \ref{fig:Cui-TPG}(e): one the
dome-shaped 7C$_{6}$ type seen also on the Rh(111) surface \cite{Wang}
and the other a 3 hexagon carbon structure 3C$_{6}$, also dome-shaped
\cite{Cui2011}. These clusters were inferred to have the structures
shown in Fig. \ref{fig:Cui-TPG}(f) and (i), respectively. At 1000
K the 3C$_{6}$ clusters became dominant and with further heating
to 1100 K eventually only graphene islands were present.

Studies on the Pt(111) surface found that clusters are not formed
\cite{Cui2011}. This is explained in terms of interactions between
graphene and the different surfaces. For Ru(0001) and Rh(111) surfaces
a strong interaction is assumed, whereas for Ir(111) and Pt(111) the
interaction is thought to be weaker \cite{Preobrajenski2008}. The
interaction between graphene and the surface drives the formation
of the clusters. The clusters are metastable on the surface\textcolor{black}{{}
and occur only at a particular temperature range. For larger clusters
the stability increases} \cite{Wang}.

\subsubsection{Using TPG to Optimize Further Growth\label{sub:Using-TPG-to-further-growth}}

In the work of Dong \emph{et al.} \cite{Dong} the growth of graphene
was studied on Rh(111). To form the graphene the TPG method of adding
ethene at room temperature before heating was compared with depositing
ethene directly on the surface held at high temperatures (which is
basically the CVD procedure). In both cases the growth of graphene
was imaged simultaneously with the heating using a high temperature
STM. In the case of the TPG method, graphene was found over a wide
range of temperatures, from 808 to 1053 K, and therefore there is
a large temperature window for graphene nucleation. The graphene formed
had partial surface coverage and different regions had multiple orientations
with respect to the substrate. This is undesirable as it leads to
domain boundaries where graphene islands meet each other when growing
further and coalescing.

\begin{figure}
\centering{}\includegraphics[scale=0.4]{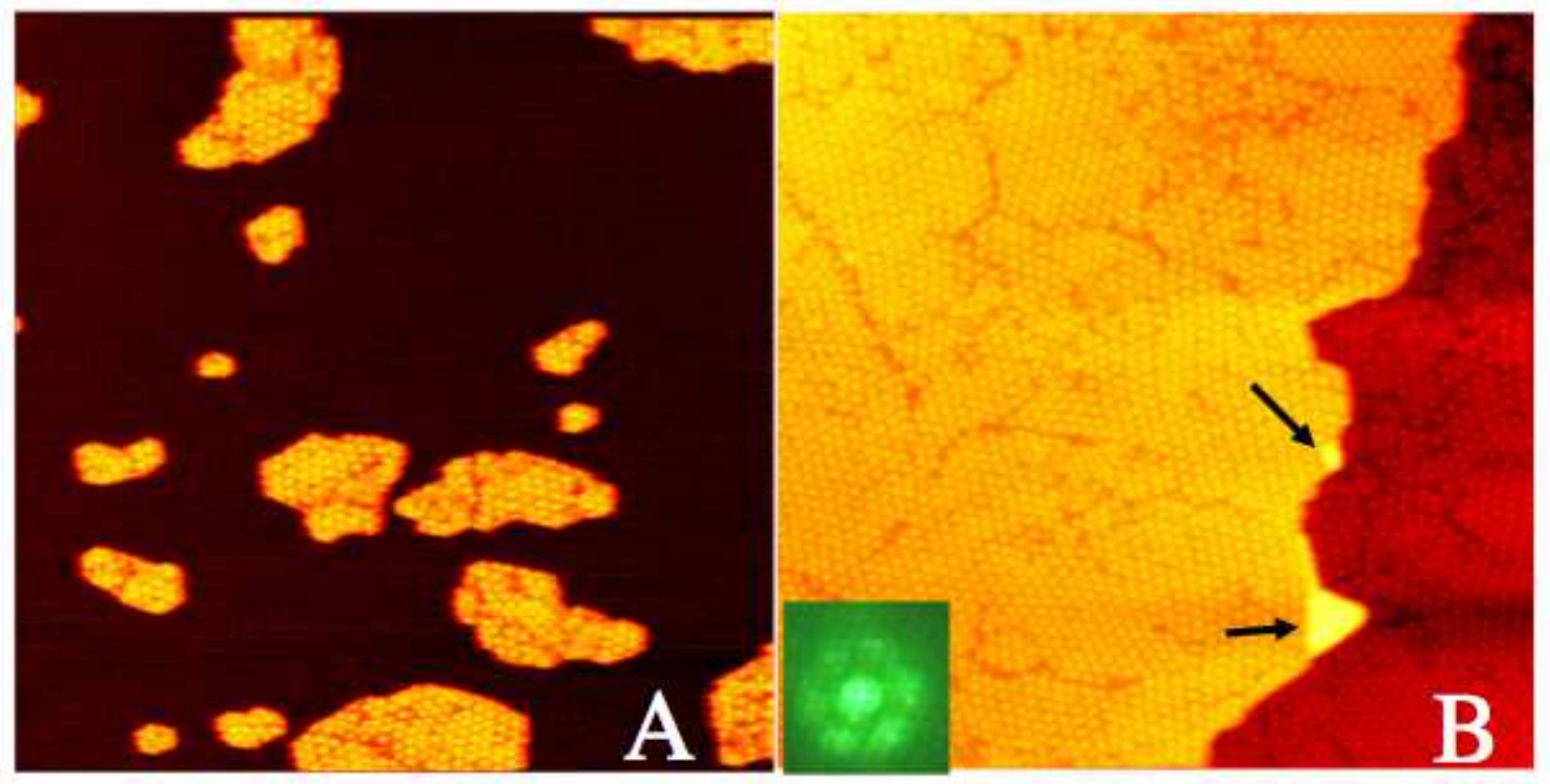}\caption{(A) The starting Rh(111) surface for graphene growth prepared using
TPG. Ethene is deposited onto the surface at room temperature, which
is then heated to 975 K. (B) The surface after ethene has been deposited
onto the surface in (A) and then kept for 76 min at 975 K. Complete
graphene coverage has been achieved with a nearly uniform orientation.
\label{fig:Dong-TPG} {[}Reproduced from \cite{Dong} by permission of IOP Publishing. All rights reserved.{]}}
\end{figure}

Fo\textcolor{black}{r ethene} deposition at high substrate temperature
(as in Section \ref{sec:Chemical-Vapour-Deposition}), the temperature
range which ensured graphene growth was much narrower: graphene was
only found when the ethene was deposited between 1016 and 1053 K.
At lower deposition temperatures a rhodium carbide structure forms
instead. Moreover, this structure was observed not to transform into
graphene upon increasing the temperature and at 1016 K it dissolves
into the substrate. When ethene was deposited at 1028 K, graphene
was produced with complete coverage and good alignment with the substrate.
However upon cooling the graphene moir\'{e} pattern distorted. This is
believed to be due to segregation of C adatoms that can dissolve into
the bulk at high temperatures. The segregated C adatoms could cause
the nucleation of new graphene and carbide islands on the surface,
which may distort the previously formed graphene layer above.

To optimize graphene growth the nucleation and growth stages were
proposed to be separated using the following method. First graphene
islands are nucleated using TPG. Then the growth of these is maintained
by deposi\textcolor{black}{ting ethene} onto the hot surface kept
at 975 K. As graphene has already nucleated it continues to grow and
if the ethene flux is sufficiently low the formation of carbides is
prevented. With this method the entire surface was eventually covered
with graphene\textcolor{red}{{} }as shown in Fig. \ref{fig:Dong-TPG}(b).
From direct observation of the growth with \emph{in situ }STM it was
found that when two domains with different orientations meet the growth
continues in the direction closest to the substrate orientation. As
a result the density of domain boundaries in the final structure was
lower than in graphene grown with regular TPG which is indeed highly
desirable for practical applications. Upon cooling the moir\'{e} pattern
remained unchanged.

\subsection{Carbon Feedstock\label{sec:Carbon-Feedstock-Experimental}}

The crucial step in graphene formation in CVD and TPG processes is
the dehydrogenation of hydrocarbon molecules (the feedstock). Unraveling
the involved mechanisms is essential for understanding the initial
stages of graphene growth just before (or maybe during) the nucleation
of carbon species into graphene islands. There is an essential difference
though between the CVD and TPG procedures. In CVD experiments the
impinging molecules possess non-zero kinetic energy (which can be
controlled in experiments) and may even be vibrationally excited (if
evaporated at sufficiently high temperatures), and these may facilitate
their sticking to the surface and initial partial dehydrogenation
\cite{Fuhrmann:2005jr}. In TPG experiments the molecules are adsorbed
on the surface prior to temperature treatment and hence the surface
temperature is the only factor that controls their decomposition.

Temperature programmed high-energy XPS (TP-XPS) is an especially powerful
technique that can be used to determine temperature windows in which
various carbon species appear and disappear on the surface of a metal
as its temperature is gradually increased. This is done by monitoring
the position of the $1s$ core level of the carbon atoms, which is
very sensitive to the immediate chemical environment. At the same
time, one can monitor the core levels of the transition metal atoms,
and these can give invaluable information, for example on how the
metal is affected at different temperatures by the adsorbed hydrocarbon
species, and may even indicate specifically when graphene starts forming.
Combined with the temperature programmed desorption (TPD) measurements which provide precise information
on the species
desorbing
from the surface during the experiments,
it is possible to build a detailed picture of the thermal evolution
of carbon species on the given surface and hence determine the processes
that occur during the initial stages of graphene growth.

The thermal evolution of feedstock on transition metal surfaces has
been investigated by several authors with the aim of determining the
carbon species that are actively involved in these early stages of
growth, and whether the choice of carbon feedstock and growth conditions
can affect how graphene is produced. Lizzit and Baraldi \cite{Lizzit201068}
studied thermal evolution of ethene on the Ir(111) surface. Throughout
the experiment the temperature was raised from 170 to 1120 K. The
TP-XPS spectra of the C 1s and the Ir 4$f{}_{7/2}$ levels were studied
in order to monitor changes in their core electron binding energies
and thereby determine the different species present during the thermal
treatment. Also the photoemission intensity thermal evolution, S factor
and vibrational splitting were used to identify the species through
comparison with previous XPS results.

\begin{figure}
\centering{}\includegraphics[height=6cm]{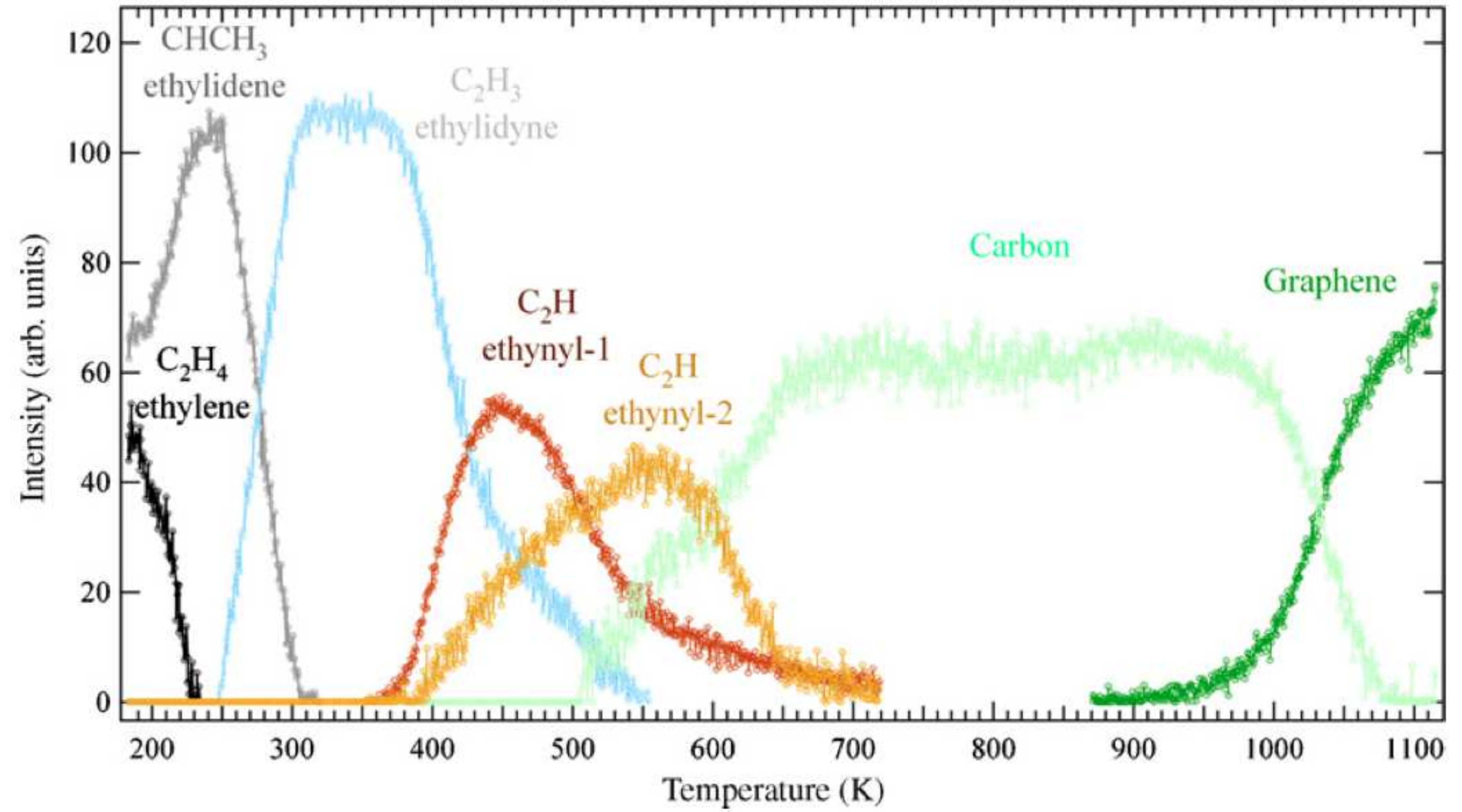}\caption{The thermal evolution of ethene into different surface species on
the Ir(111) surface during the course of the TP-XPS experiments. {[}Reprinted from \cite{Lizzit201068}, with permission from Elsevier.{]} \label{fig:lizzit-thermal-evolution}}
\end{figure}

The evolution of ethene C$_{2}$H$_{4}$ as the temperature is increased
is shown in Fig. \ref{fig:lizzit-thermal-evolution}. Initially it
is assumed that both ethene and ethylidene CHCH$_{3}$ are chemisorbed
on the surface. At 170 K the ethene molecules are suggested to begin
to transform to ethylidene, which is the only species found at temperatures
between 235 and 250 K, i.e. complete conversion (isomerization) C$_{2}$H$_{4}$
$\rightarrow$ CHCH$_{3}$ is believed to have taken place by 235
K. Upon increasing the temperature, dehydrogenation of ethylidene
occurs; the ethylidene feature disappears at around 310 K and subsequently
C$_{2}$H$_{3}$ (ethylidyne) and later on C$_{2}$H (ethynyl) species
are found on the surface. The dehydrogenation processes are accompanied
by a clear H$_{2}$ TPD signal \cite{Nieuwenhuys-TPD-1976}. Two types
of signals associated with the ethynyl are attributed to two possible
configurations of the species on the surface. At 500 K complete dehydrogenation
of the existing carbon species starts which correlates well with the
second hydrogen TPD feature between 630 and 730 K reported in \cite{Nieuwenhuys-TPD-1976};
pure adsorbed carbon species start forming eventually leading to the
appearance of graphene at around 900 K.

\begin{figure}
\centering{}\includegraphics[height=5cm]{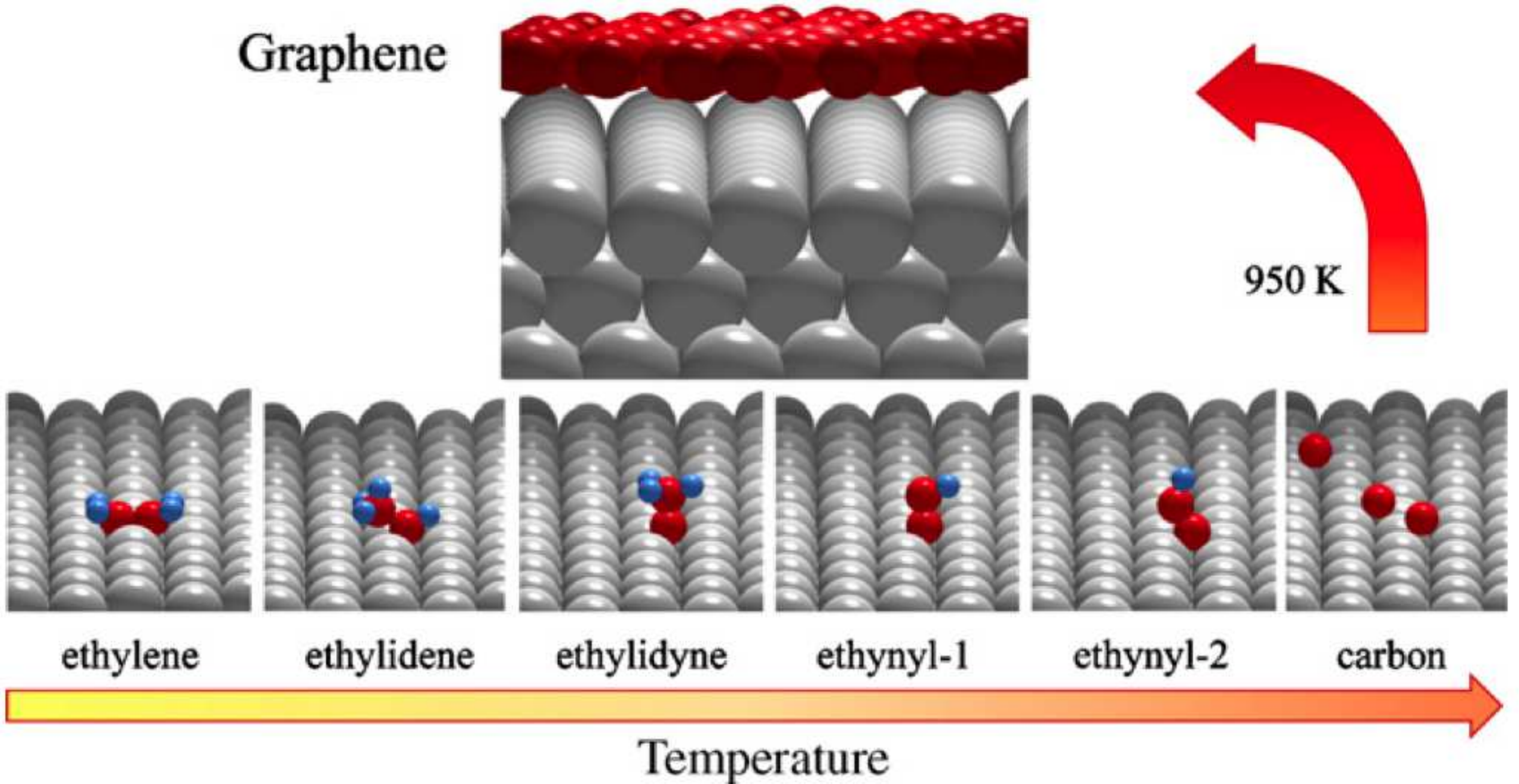}\caption{Schematic of the thermal evolution of ethene (ethylene) into graphene
on the Ir(111) surface. {[}Reprinted from \cite{Lizzit201068}, with permission from Elsevier.{]}
\label{fig:lizzit-thermal-evolution2}}
\end{figure}

The onset of graphene formation is also supported by the evolution
of the Ir 4$f_{7/2}$ core level: the corresponding XPS feature approaches
its position for the clean Ir surface indicating graphene formation,
as it is known that graphene interacts weakly with the Ir surface.
It is concluded in these experiments that on the Ir(111) surface ethene
undergoes complete dehydrogenation. Moreover, it is clear that the
evolution is quite complex and involves many stages as depicted in
Fig. \ref{fig:lizzit-thermal-evolution2}, characterized by dehydrogenation
and isomerization processes leading to coexistence of various species
on the surface at the same conditions. Interestingly, as follows from
the schematics in the Figure, the authors assumed that only atomic
carbon is present at the very last stages on the surface prior and
during graphene formation. This may however not be the case as it
is well known that much higher activation energies (over 2 eV \cite{Andersin:2009kd})
are required for breaking the C-C bonds, and hence there must be a
certain distribution of C$_{2}$ species alongside atomic carbon on
the surface prior to the formation of larger carbon clusters and graphene
nucleation.

Similar work has been conducted for the Pt(111) surface \cite{Fuhrmann:2005jr}
albeit for a much limited temperature range corresponding to the very
initial stages of ethene decomposition. It was found that ethene completes
its transformation into ethylidyne C$_{2}$H$_{3}$ on this surface
at the higher temperature of 290 K (230 K on Ir(111)). Therefore the
barrier for converting ethene into C$_{2}$H$_{3}$ must be higher
on Pt(111) and it is a poorer catalyst for C-H breaking than Ir(111).
As with the case of the Ir(111) surface, the authors also observed
ethylidene CHCH$_{3}$ as an intermediate product for temperatures
of up to 290 K. At the same time, during the conversion C$_{2}$H$_{4}$
$\rightarrow$ C$_{2}$H$_{3}$ a hydrogenation reaction takes place
as well whereby ethene C$_{2}$H$_{4}$ forms again and desorbs from
the surface (40\%). This effect was not observed on Ir(111) \cite{Lizzit201068}
though.

Decomposition of the simplest hydrocarbon, methane CH$_{4}$, on the
Pt(111) surface was also considered in \cite{Fuhrmann:2005jr}. At
120 K methane immediately dehydrogenates into methyl CH$_{3}$ which
from 260 K starts partially transforming into CH species. The dehydrogenation
is confirmed by the corresponding TPD signal. At the same time, the
backward hydrogenation competes with the dehydrogenation reaction
resulting in formation of methane again which desorbs from the surface
at these temperatures. At temperatures above and around 300 K, CH
is established as a stable species.

The comparison of these two studies, on Ir(111) and Pt(111) surfaces,
clearly shows that mechanisms governing decomposition of the carbon
feedstock, although involving the same intermediates, may however
differ in detail, i.e. each particular surface requires individual
consideration.

In order to produce graphene, various hydrocarbon feedstocks have
been successfully used. However, very little is known about the effect
of feedstock on the growth mechanism and graphene quality. To our
knowledge, comprehensive comparative study of various feedstocks has
not yet been done; however, single studies indicate the fundamental
effect of using various molecules as a feedstock in growing graphene.
Methane (in the gaseous form) and benzene C$_{6}$H$_{6}$ (liquid)
precursors were compared as the carbon source for growing graphene
on Cu foils in \cite{Zhanchengdoi:10.1021/nn200854p}. It was found
that when using methane the graphene structure could not be identified
below 873 K. For benzene SEM images showed that even at temperatures
as low as 573 K high quality graphene monolayer flakes can be formed.
This was explained by two considerations illustrated by DFT calculations:
(i) dehydrogenation of benzene requires lower barriers than for methane;
moreover, in the latter case the required reactions may be quite complex
and multistage, resulting in a mixture of various CH$_{x}$ species
on the surface prior to nucleation in the case of methane; (ii) benzene
after dehydrogenation already has a carbon ring, the main building
block of graphene, and hence the nucleation step may effectively require
a smaller barrier than in the case of methane where coalescence of
single carbon atoms needs to take place. Therefore nucleation and
coalescence are expected to require more energy when methane is used
as a precursor. The higher energy profile for methane compared with
benzene throughout all stages of growth (adsorption, dehydrogenation,
nucleation) explains the difference in growth temperatures required
for producing graphene.

\subsection{Segregation\label{sec:Segregation}}

A less frequently used method to produce graphene layers is to make
use of carbon solubility in different substrates. Here we will consider
carbon precipitation from both nickel \cite{Zhang10} and ruthenium
\cite{McCarty,Sutter08}, but the method also applies to most of the
metals that are usually employed for CVD and TPG, provided that they
are heated to sufficiently high temperature so that their carbon solubility
becomes non-negligible \cite{Sutter2009,Nie2011}; note that as a
consequence of the presence of grain boundaries in the metal \cite{Su2011},
the carbon solubility is also increased. The method begins with heating
the material to high enough temperatures to allow carbon to dissolve
into the bulk. This temperature is strongly dependent on the specific
material. The carbon may be added in a usual way, e.g. using hydrocarbon
molecules. The whole sample is then cooled slowly, leading to a reduced
solubility of C in the material and hence causing C atoms to segregate
from the bulk onto the surface and form graphene islands there.

\subsubsection{Ru(0001)\label{sub:Ru(0001)-in-methods}}

\begin{figure}
\begin{centering}
\includegraphics[height=5cm]{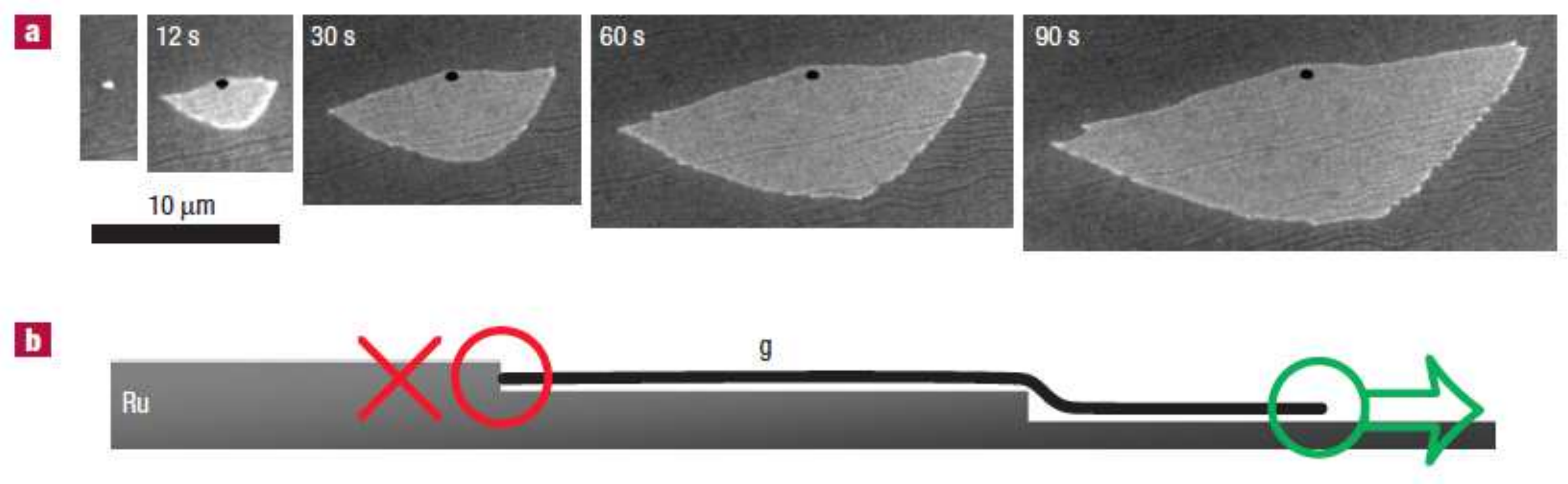}
\par\end{centering}

\caption{(a) LEEM images of a first-layer graphene island growing on Ru(0001)
at 1123 K in real time. (b) Schematics illustrating the predominant
growth down the steps. {[}Reprinted by permission from Macmillan Publishers Ltd: Nature Materials \cite{Sutter08}, copyright (2008).{]}
\label{fig:Sutter-LEEM-images-of-growing-island}}
\end{figure}

Sutter \emph{et al.} \cite{Sutter08} used electron scattering and
microscopy as well as micro-Raman spectroscopy to observe and characterise
graphene grown on Ru(0001) using the segregation method. Here carbon
was deposited at high temperatures (>1423 K) onto the Ru surface and
subsequently absorbed into the bulk. The sample was then cooled to
1098 K driving carbon atoms to the surface. Graphene islands were
observed to nucleate sparsely and grow by incorporation of carbon
species at edges. Interestingly, it was also found that growth only
occurs down Ru steps and not up, producing a number of ``lens''
shaped graphene islands, see Fig. \ref{fig:Sutter-LEEM-images-of-growing-island}.

To investigate how the segregation of carbon atoms from the bulk affects
graphene growth on Ru(0001), McCarty \emph{et al. \cite{McCarty}}
performed experiments under two different conditions required to grow
graphene by segregation and deposition, respectively, and used LEEM
measurements to measure the adatom concentration during growth. The
segregation of bulk carbon was examined by first heating the surface
to a high enough temperature for absorption to occur. While the authors
do not explicitly state what temperature they heated the surface to
(only briefly mentioning 1404 K (1131$^{\circ}$C) in Fig. 31), presumably
it is similar to that of Sutter \emph{et al.} \cite{Sutter08}. The
surface is then cooled to 973-1223 K (700-950$^{\circ}$C) (a rate of cooling was
not stated). To make quantified comparisons they deposit carbon onto
a different Ru(0001) surface heated to a lower temperature and measured
the concentration of adatoms in equilibrium with graphene islands
\cite{Loginova08}. This method of production is described in Section
\ref{sec:Temperature-Programmed-Growth}. Both experiments were performed
in UHV.

\begin{figure}
\begin{centering}
\includegraphics[height=4cm]{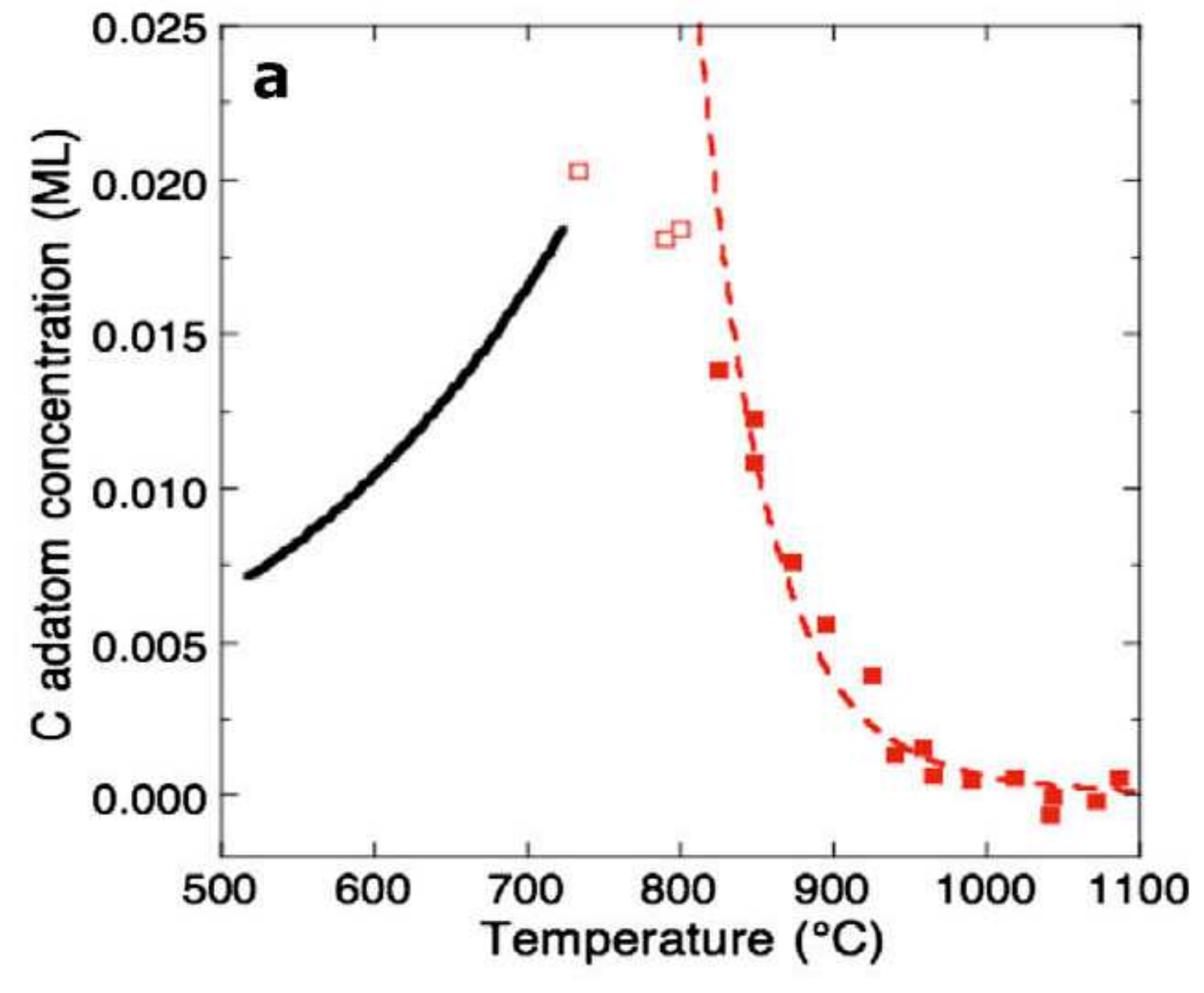}
\par\end{centering}

\caption{The concentration of adatoms in equilibrium with carbon dissolved
in the Ru bulk. The initial temperature before cooling was 1404 K (1131$^{\circ}$C).
Open red squares show the concentration of metastable carbon just
before nucleation of graphene and red-filled squares indicate the
concentration of adatoms on the surface. The black curve shows the
adatom concentration in equilibrium with graphene \cite{Loginova08}.
The red dotted line is a fit to the Langmuir-McLean model of segregation.
{[}Reprinted from \cite{McCarty}, with permission from Elsevier.{]} \label{fig:McCarty-equi.-with-bulk}}
\end{figure}

\begin{figure}
\centering{}\includegraphics[height=5cm]{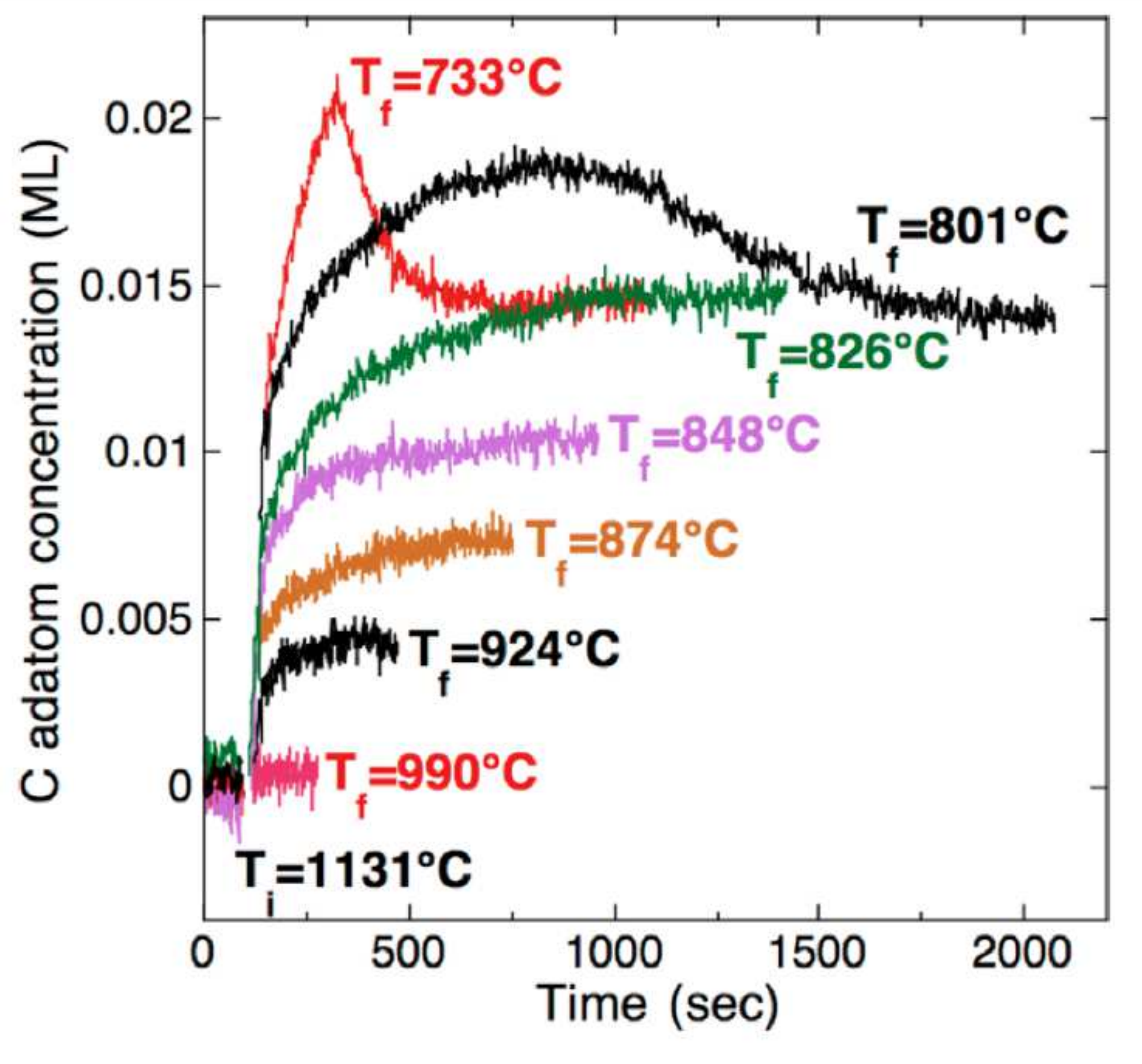}\caption{The adatom concentration on the surface of Ru(0001) as a function
of time during a temperature decrease from 1401 K (1131$^{\circ}$C). After
an initial increase, the plots for both 733 and 801$^{\circ}$C (1006 and 1074 K) show
an eventual decrease in concentration which the authors connect to
the nucleation of graphene. {[}Reprinted from \cite{McCarty}, with permission from Elsevier.{]}\label{fig:mccarty adatom}}
\end{figure}

By measuring the change in reflectivity of the surface at 1404 K
and at lower temperatures the authors were able to measure the amount
of carbon that has segregated from the bulk, see Fig. \ref{fig:McCarty-equi.-with-bulk}.
As can be seen in the Figure the increase in adatom concentration
at 1223 K is an indication of segregation from the
bulk. At 1046 and 1074 K the adatom concentration begins to
increase before decreasing sharply as is seen in Fig. \ref{fig:mccarty adatom},
and the authors connect this decrease in concentration with the appearance
of a graphene island at $\sim$450 s.

In exactly the same way as described in \cite{Loginova08} the authors
measured the velocity of the island growth front and found that the
results obtained for segregation and deposition agreed. Therefore,
the nucleation of islands and early growth is due to the segregated
carbon atoms, not by direct attachment of bulk carbon, and thus the
mechanisms involved in the nucleation and growth of graphene on Ru(0001)
seem to be independent of how carbon atoms reach the surface. In other
words, the way in which the carbon atoms are delivered to the surface
(either from adsorbed molecules, as in CVD and TPG, or from the bulk
as in segregation methods) is not the main rate-determining process
for the growth, as the formation of graphene islands happens on a
much longer timescale. A slow enough cooling procedure would limit
the influx of C atoms from the bulk and would leave enough time for
the C atoms on the surface to diffuse around and eventually nucleate
with a low density, whereby a higher quality graphene, with a low
density of defects such as grain boundaries, would result.

\subsubsection{Nickel\label{sub:Nickel}}

The experimental procedure of growing graphene on Ni(111) and polycrystalline
Ni surfaces usually begins with heating the surfaces to above 1170
K \cite{Odahara,Zhang10}. In a number of experiments the formation
of monolayer, bilayer and trilayer graphene has been observed with
the number of monolayers gr\textcolor{black}{own being strongly dependent
on temperature \cite{Zhang10,Odahara}. However, for CVD temperatures
between 733-923 K growth on Ni(111) is self-terminating
after one monolayer (growth is limited to one layer) \cite{Addou}
and for growth below 673 K a surface carbide phase has been
observed \cite{Lahiri11,Addou}. }

Multilayer graphene has been observed on both Ni(111) and polycrystalline
substrates. The difference between growth on these surfaces has been
examined by Zhang \emph{et al. }\cite{Zhang10}. They grew graphene
by initially depos\textcolor{black}{iting methane at atmospheric pressure
onto the Ni surface at 1173 K. The sample was then cooled to 773 K
and examined using Raman spectroscopy and AFM. They found that while
on Ni(111) monolayer and bilayer graphene is formed, on polycrystalline
surfaces the number of layers increases. }

\textcolor{black}{Similar behavior on Ni(111) has been observed using
LEEM \cite{Odahara}. Here carbon is absorbed into the bulk after
the surface is exposed to benzene at 1200 K for ten minutes. The surface
is then cooled to 1125K whereb}y graphene begins to nucleate and the
coverage eventually reached a full layer. Cooling the temperature
further to 1050 K results in a second graphene layer and the initial
nucleation of a third monolayer. The edge structure on the first and
third layers is hexagonal while the second layer demonstrates dendrite
structures. It should also be mentioned that growth of the first layer
was observed to be almost ten times faster than that of the second
or third layers.

There appear to be much fewer nucleation sites on Ni(111) compared
to polycrystalline Ni \cite{Zhang10,Odahara}, where it has been suggested
that nucleation occurs at grain boundaries between different crystalline
surfaces \cite{Zhang10}. This leads to a large graphene sheet with
a single rotational domain.

\subsection{SiC Sublimation\label{sec:SiC-Sublimation}}

\subsubsection{Graphene on SiC\label{sub:Graphene-on-SiC}}

An alternative approach to producing epitaxial graphene is to grow
it on silicon carbide. Heating to high temperatures causes sublimation
of silicon from the surface of SiC. The remaining carbon atoms form
a carbonic surface, and by controlling growth conditions they can
arrange themselves to form a graphene layer. It is believed (although
not yet fully established) that as the growth progresses graphene
layers are formed between current graphene layers and the SiC surface,
i.e. secondary layers grow underneath the primary one. This is a bottom-up
process, which is unlike other growth methods discussed above where
new layers are formed on top of the previous layers (top-down processes).

While the initial studies involving growth under UHV required heating
to about 1600 K for control of the thickness of the graphene film
\cite{Forbeaux1998}, now samples of better controlled quality are
grown under rougher conditions, but require higher heating temperatures.
Under a Si-rich atmosphere \cite{deHeer2007} or by using a buffer
gas \cite{Virojanadar2008,Tromp2009,Emtsev:2009wd}, high quality
samples are indeed prepared at temperatures of about 1900 K.

\begin{figure}
\centering{}\includegraphics[height=4cm]{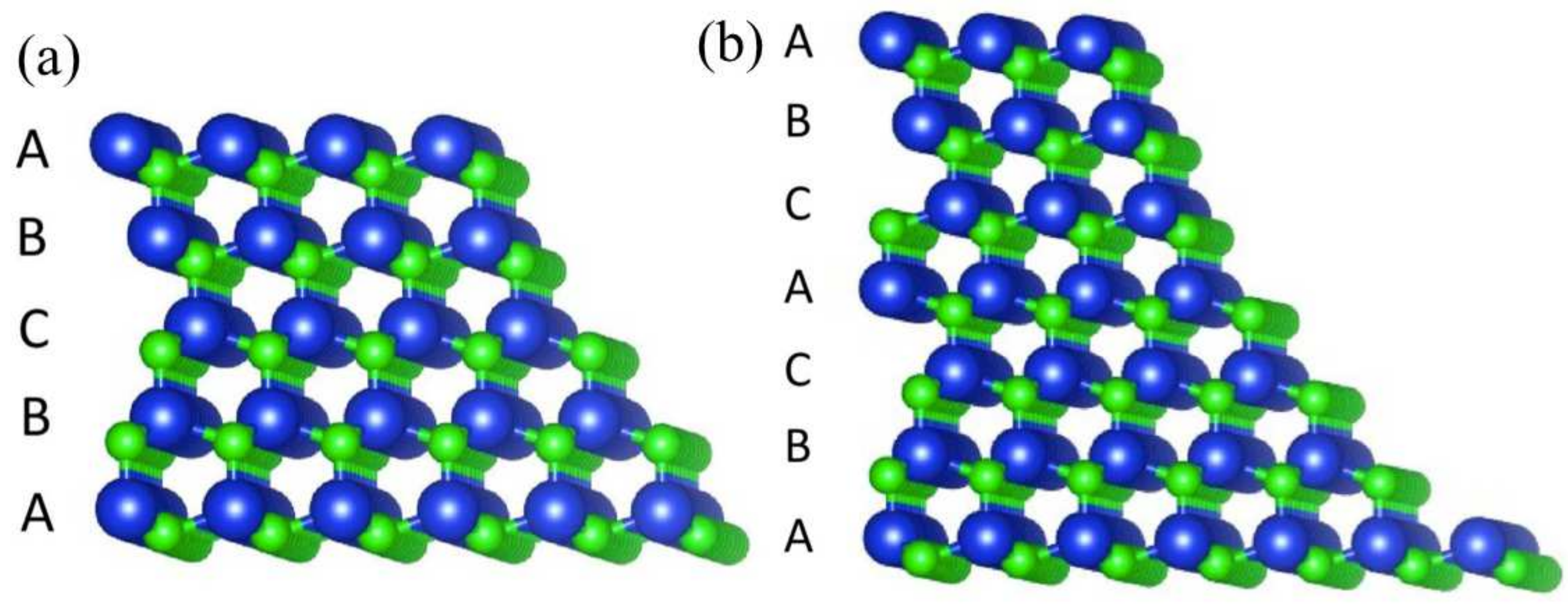}\caption{The lattice structure for (a) 4H-SiC and (b) 6H-SiC. The letters indicate
the stacking sequences. C and Si atoms are shown with green and blue
colors, respectively. {[}Reproduced with permission from\textcolor{red}{{}
}\cite{ming2011thesis}.{]} \label{fig:Ming-lattices}}
\end{figure}

Graphene formation is dependent on the structure of the SiC surface.
The most commonly used SiC polytype structures for growing epitaxial
graphene are 4H-SiC and 6H-SiC. The unit cells of these structures
are both hexagonal and contain 4 and 6 SiC bilayers, respectively,
with different stacking sequences. Each bilayer is composed of a plane
of C atoms and a plane of Si atoms. The structure of these lattices
is shown in Fig. \ref{fig:Ming-lattices}. The occupation site positions
are denoted with the letters A, B and C for each distinct SiC layer.
For 4H-SiC and 6H-SiC the stacking sequences are ...(ABCB)... and
...(ABCACB)..., corresponding to 4 and 6 distinct layers, respectively,
which are then periodically repeated.

Both of these lattices have either a S\textcolor{black}{i- or C-terminated
surface. These are the (0001) surface for Si termination and the $(000\overline{1})$
surface for C termination. The growth of the graphene and its final
structure are dependent on which of these surfaces is initially exposed.
On the Si-terminated surface the graphene layer produced has an orientation
of 30\textdegree{} with respect to the SiC plane \cite{Forbeaux2000,vanBommel1975}.
For the C-surface the growth is faster \cite{muehlhoff:2842} but
the graphene has a variety of different orientations \cite{Forbeaux2000,vanBommel1975}.
Because of this most studies of graphene growth on SiC have focused
on the Si-terminated surface.}

\textcolor{black}{Pre-patterned SiC surfaces, for instanc}e, with
arrays of trenches with abrupt walls created by lithography, were
found to spontaneously transform into inclined facets with a well-defined
surface normal other than (0001) or $(000\overline{1})$ under annealing.
Graphene growth on such patterned surfaces was found to be preferential
on the inclined facets, so that extended graphene ribbons, with size
determined by the depth of the initial trenches, can be obtained in
a rather versatile manner \cite{Sprinkle2010}.

\subsubsection{The Growth Process\label{sub:The-Growth-Process}}

To produce graphene using Si sublimation, the SiC surface must first
be treated. Scratches are removed by etching with hydrogen, and then
the SiC is rapidly heated to the growth temperature, typically between
1473 and 1933 K, for a specific time. During the heating process a
Si flux is applied to remove any produced SiO gas and maintain the
concentration of Si on the surface \cite{Forbeaux2000,Li1996141,Kaplan1989111}.
Alternatively, growth can be performed under an Ar atmosphere. The
SiC is then cooled at a slower rate until reaching room temperature.
The etching process forms regular steps along the surface of the SiC,
which are a unit cell in height.

\begin{figure}
\begin{centering}
\includegraphics[height=4cm]{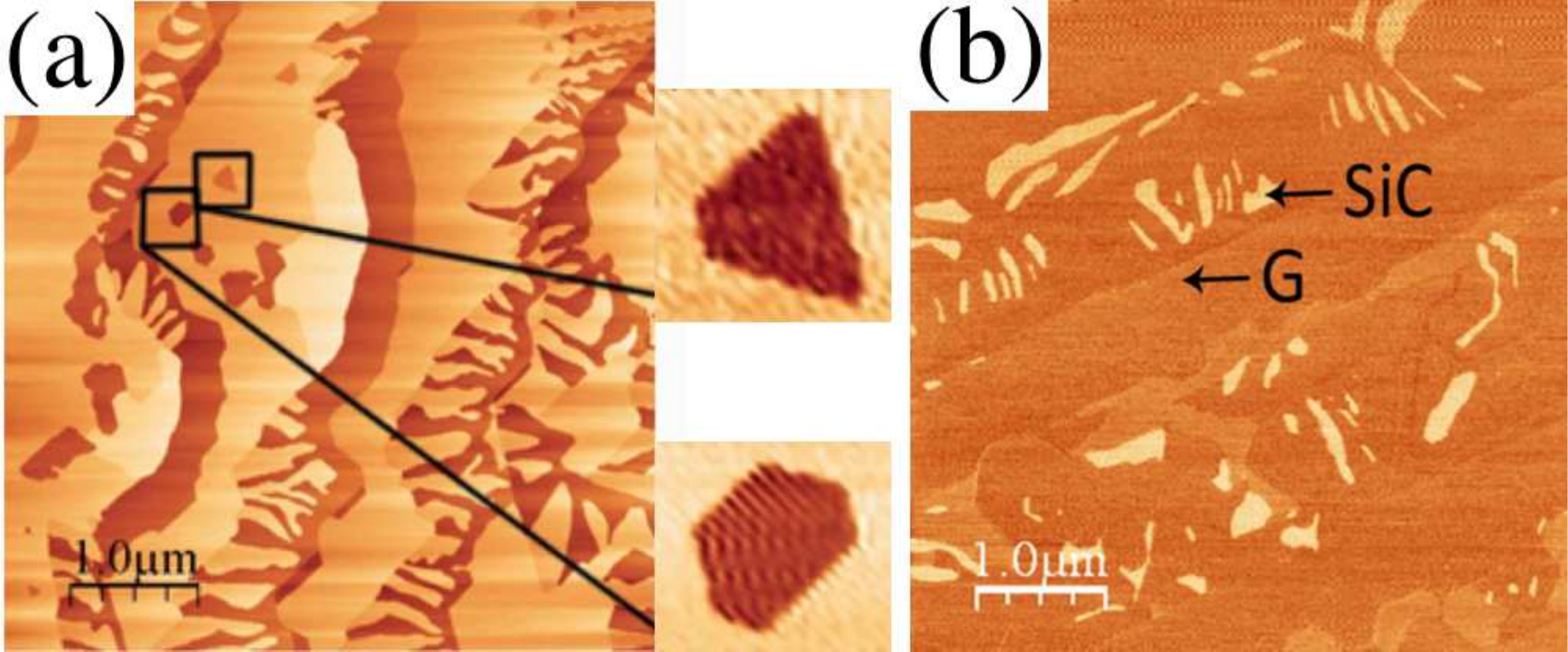}
\par\end{centering}

\centering{}\caption{AFM topography of the SiC surface after heating to (a) 1748 K and
(b) 1823 K. At 1673 K no graphene is present (shown using AFM phase
measurements). At 1748 K fingers are formed with graphene in between
them (a) (two terrace pits highlighted in the AFM image are shown
separately at high resolution) and these eventually erode to become
islands (b). At 1873 K the islands have completely eroded and graphene
covers the whole surface. \label{fig:SiC-AFM-with-T} {[}Reprinted
with permission from \cite{Bolen2009}. Copyright (2009) by the American Physical Society.{]}}
\end{figure}

To understand how graphene grows from this process, the 4H-SiC(0001)
surface was monitored as the growth temperature and heating time were
separately varied \cite{Bolen2009}. In each case AFM (tapping mode)
and STM were used to observe the changes on the surface and the progression
of graphene growth. Scans were performed in a nitrogen environment
at atmospheric pressure. To determine the temperature dependence of
the growth the SiC was heated to a range of different temperatures
between 1673 and 1873 K for 10 minutes. At 1673 K no graphene was
found on the surface, however pits in the SiC surface had started
to develop on the terraces. Increasing the temperature increased the
fraction of the surface covered by pits. At 1748 K, see Fig. \ref{fig:SiC-AFM-with-T}(a),
the bottom of a pit was analyzed with high STM magnification. Taking
a 2D Fast Fourier transform of the STM data revealed the hexagonal
structure of graphene. From this it was assumed that there are two
phases present on the surface: reconstructed SiC on the surface terrace
and graphene further down. From this the contrast in the phase data
from AFM images was interpreted to determine the surface composition
at each temperature.

\begin{figure}
\begin{centering}
\includegraphics[height=4cm]{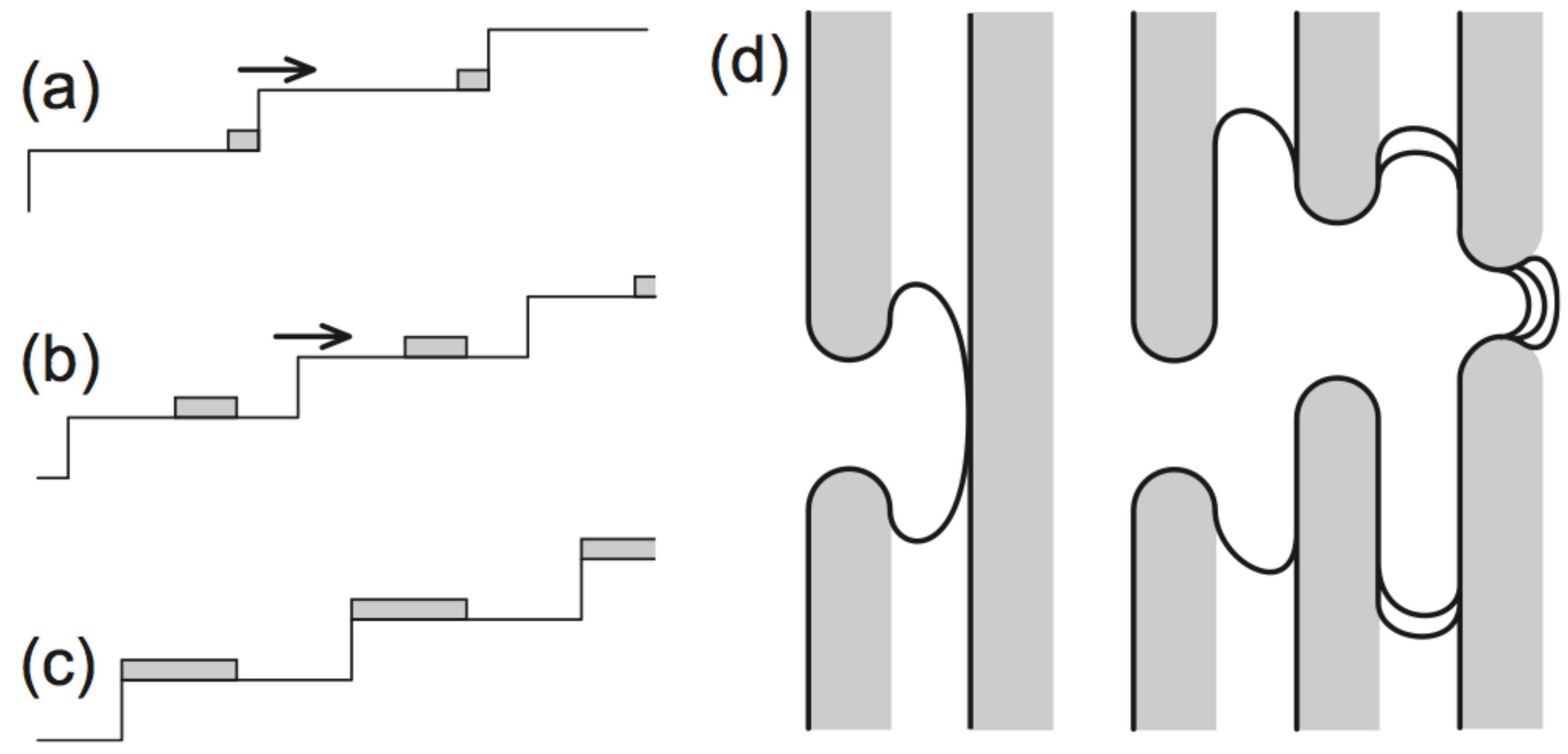}
\par\end{centering}

\caption{\label{fig:pit formation Hannon} (a) The buffer layer (gray) nucleates
at the lower sides of $\left(\sqrt{3}\times\sqrt{3}\right)$-reconstructed
SiC steps. (b) $\left(\sqrt{3}\times\sqrt{3}\right)$-reconstructed
SiC steps retract uphill as the buffer layer grows. (c) Eventually
a state is reached in which all $\left(\sqrt{3}\times\sqrt{3}\right)$-reconstructed
SiC steps have been consumed. (d) Pits form when the domains of the
buffer layer on each terrace are not continuous and $\left(\sqrt{3}\times\sqrt{3}\right)$-reconstructed
SiC steps (thick black lines) advance to the next terrace. {[}Reprinted
with permission from \cite{Hannon2008}. Copyright (2008) by the American Physical Society.{]}}
\end{figure}

The origin of a pit formation has been tracked to the processes governing
the nucleation and growth of the buffer layer \cite{Hannon2008},
an intermediate layer formed before graphene is grown (at lower temperature)
and eventually sandwiched between graphene and the substrate (see
\ref{sub:Multilayer-Growth}). This buffer layer nucleates on SiC
terraces upon consumption of the Si atoms expelled from the SiC(0001)
step edges (actually $\left(\sqrt{3}\times\sqrt{3}\right)$-reconstructed
ones) above 1300 K, leaving regions elongated along the SiC step edges,
which are three SiC bilayers-deep, see Fig. \ref{fig:pit formation Hannon}(a,b).
Continued growth of the buffer layer proceeds upon retraction of $\left(\sqrt{3}\times\sqrt{3}\right)$-reconstructed
SiC steps until these encounter a region covered by a buffer layer
uphill of the step (Fig. \ref{fig:pit formation Hannon}(c)). In practice
this leads to stripes of the buffer layer parallel to the initial
substrate step edge pattern, except at some locations where the buffer
layer stripes are not continuous, where the retraction of the $\left(\sqrt{3}\times\sqrt{3}\right)$-reconstructed
SiC steps can continue, which leads to the formation of pits (Fig.
\ref{fig:pit formation Hannon}(d)).

At the same time, finger-like structures associated with pits can
be found on the surface at 1748 K as illustrated in Fig. \ref{fig:SiC-AFM-with-T}(a).
These fingers are formed perpendicular to a step edge and correspond
to reconstructed SiC with graphene between them. At higher temperatures
as the step edges diminish, the fingers lengthen and eventually become
islands after an entire step front has eroded. These islands are found
at 1823 K, Fig. \ref{fig:SiC-AFM-with-T}(b). Above this temperature
the SiC islands shrink further and finally the surface is completely
covered by graphene.

\begin{figure}
\centering{}\includegraphics[height=5cm]{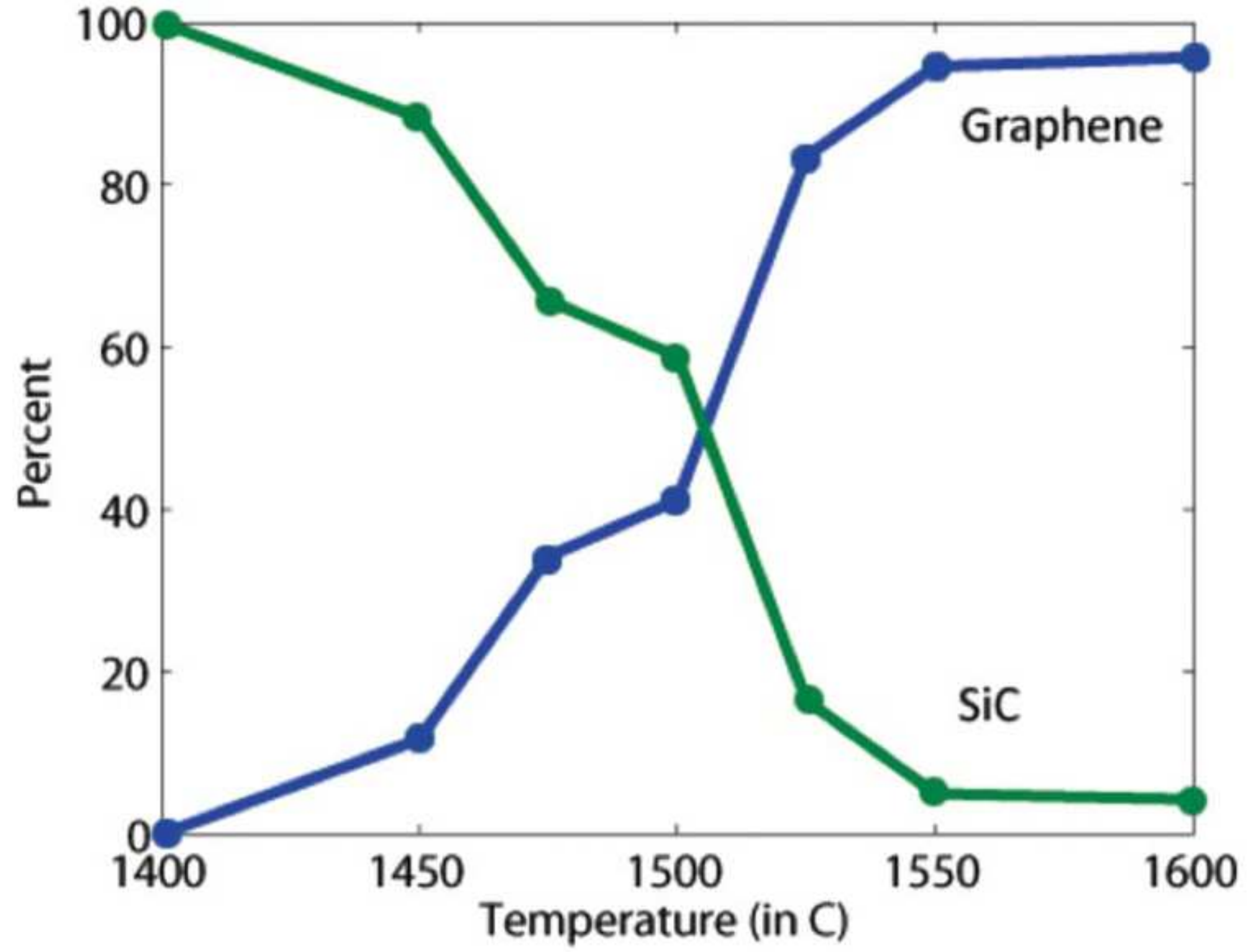}\caption{The surface coverage of both graphene (blue) and SiC (green) as a
function of annealing temperature. \label{fig:SiC-growth} {[}Reprinted
with permission from \cite{Bolen2009}. Copyright (2009) by the American Physical Society.{]}}
\end{figure}

The surface coverage of SiC and graphene at different temperatures
is shown in Fig. \ref{fig:SiC-growth}. The increase in graphene coverage
above 1773 K is accompanied by the reduction of surface features such
as pits and fingers, as a graphene layer begins to cover them. At
1873 K graphene completely covers the surface.

The time dependence of the growth was studied by keeping the growth
temperature fixed at 1748 K and varying the growth time between 0
and 60 minutes. AFM images taken at different time intervals show
that the surface progresses in almost the same way as when the temperature
dependence was studied. Fingers are formed after 2-14 minutes of growth,
which become islands at 16 minutes. After 60 minutes the graphene
layer is nearly entirely complete. This shows that the growth mechanism
for graphene from SiC is the same whether the temperature or heating
time is varied.

\begin{figure}
\begin{centering}
\includegraphics[height=6cm]{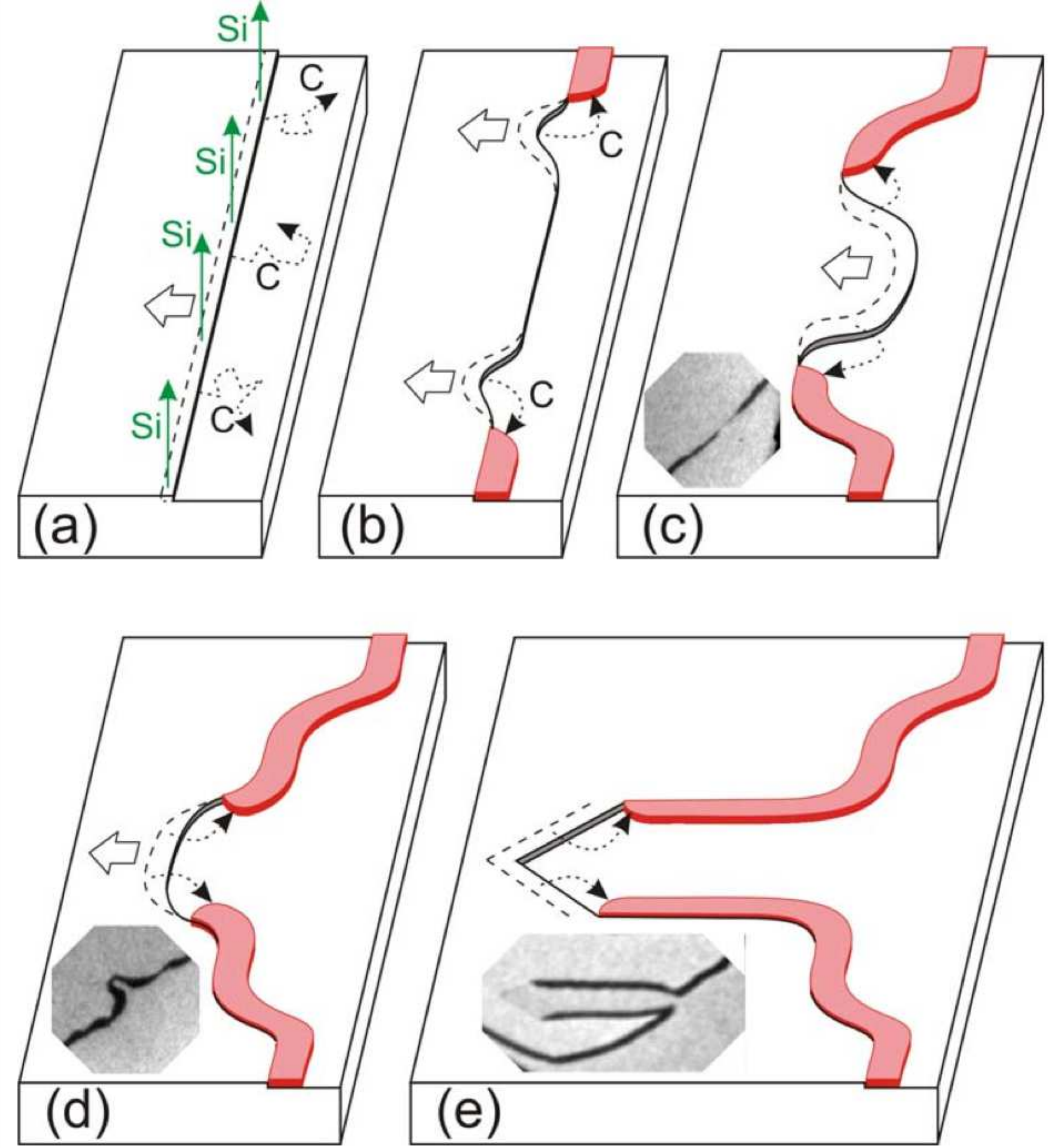}
\par\end{centering}

\caption{\label{fig:Arrows Ohta} Schematic of the formation of an arrow feature.
The insets show static LEEM images of SiC steps at similar stages
of evolution. LEEM images are 2$\times$2 $\mu$m$^{2}$
(c,d) and 4$\times$2 $\mu$m$^{2}$ (e). {[}Reprinted
with permission from \cite{Ohta2010}. Copyright (2010) by the American Physical Society{]}}
\end{figure}

A better quality of graphene prepared on SiC far from UHV conditions
motivated a few studies aiming at unraveling growth processes which
could be specific to such conditions. For a SiC(0001) surface almost
fully covered by the buffer layer after a 1820 K annealing procedure
under Ar atmosphere, further UHV growth yields high quality graphene
\cite{Ohta2010} as obtained by Emstev \emph{et al.} \cite{Emtsev:2009wd}
who used only non-UHV conditions throughout growth. These findings
question a qualitative understanding that the reduction of Si evaporation
rate due to the presence of the Ar atmosphere is crucial for slowing
down graphene growth close to thermodynamic equilibrium. Instead,
the formation of a high quality buffer layer, with the help, for instance,
of a high temperature treatment in Ar atmosphere, seems to be the
prerequisite for growing high quality graphene. Such a high quality
buffer layer should consist of straight triple SiC step edges, separated
by large reconstructed (buffer layer) terraces. Such terraces indeed
contain about the right amount of carbon necessary for building a
full graphene layer on top of the buffer layer in a step flow fashion
\cite{Ohta2010}. The presence of single SiC steps should be avoided
as they lead to the formation of graphene on their side, which, unlike
in the case of triple SiC steps, inhibits Si evaporation due to strong
C-Si bonds at the SiC-graphene interface. In practice, this leads
to the formation of arrow-like growth fronts, as explained in Fig.
\ref{fig:Arrows Ohta}, instead of smooth ones, thanks to which high
quality graphene can be obtained.

In a UHV study, graphene of very high quality has been obtained using
evaporation of Si (actually disilane, which readily transforms into
Si upon reaching the substrate surface at high temperature, H atoms
resulting from its decomposition being efficiently desorbed into vacuum)
\cite{Tromp2009}. In this study, different surface transformations
of SiC(0001) upon heating could be slowed down considerably, and even
reversed by applying an appropriate Si partial pressure. This produced
a series of surface transitions in conditions much closer to thermodynamic
equilibrium, yielding graphene with a much more uniform number of
layers (1-2) and larger-scaled features (few microns) as compared
to the situation without Si evaporation (few 100 nm features and 1-4
layers) at the same growth temperature \cite{Tromp2009}. Further
increase in the growth temperature and Si pressure (in ranges not
suitable for the utilization of the LEEM instrument used in Ref. \cite{Tromp2009})
is expected to yield further improved morphology and structure.

\subsubsection{Multilayer Growth\label{sub:Multilayer-Growth}}

SiC sublimation often produces multiple graphene layers on
top of one another. They can have different properties depending on
their interaction with each other and the SiC substrate.

To determine the thickness of graphene layers STM observations of
the surface have been used \cite{Lauffer2008}. It was found that
the surface roughness can be related to the layer thickness with an
exponential relationship. By measuring the surface roughness, steps
between upper and lower terraces were identified with a height of
0.4 nm. This height does not correspond to the usual gap between two
graphene layers of 0.335 nm. The difference is attributed to the existence
of a buffer layer in between the SiC(0001) surface and the graphene
layers. This layer has a honeycomb lattice structure with a ($6\sqrt{3}\times6\sqrt{3}$)R30\textdegree{}
reconstruction \cite{Riedl2007} involving the formation of strong
C-Si bonds \cite{Varchon08}, thus not exhibiting the electronic properties
typical of graphene \cite{Mattausch2007}.

\begin{figure}
\centering{}\includegraphics[height=4cm]{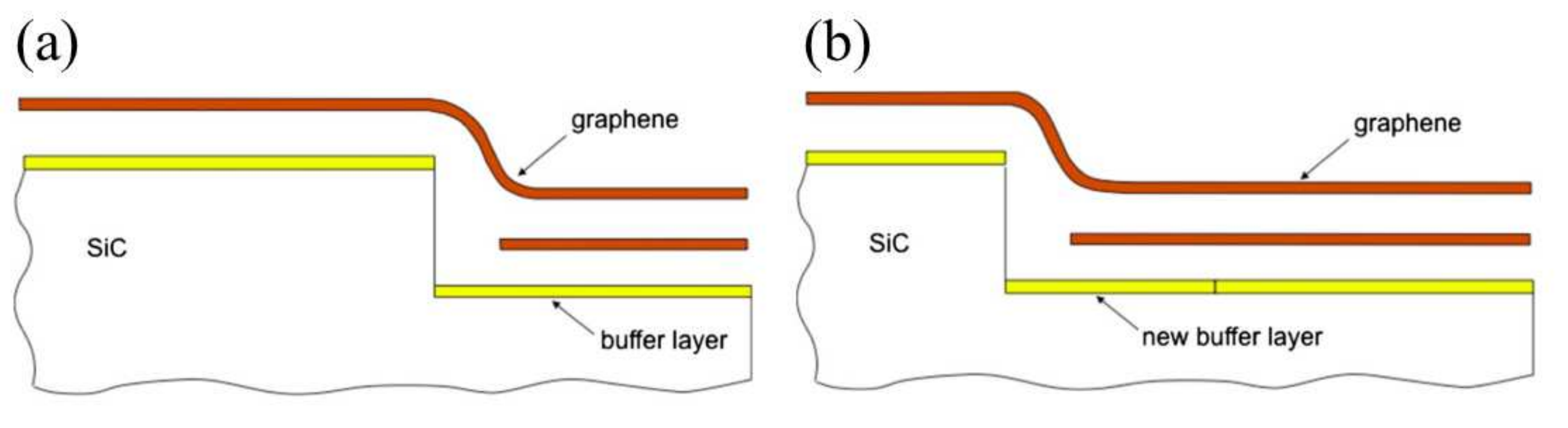} \caption{Growth of a second graphene layer at a step edge from the state in
(a) to the state in (b). \label{fig:Borovikov_growth} {[}Reprinted
with permission from \cite{Borovikov2009}. Copyright (2009) by the American Physical Society.{]}}
\end{figure}

As a second graphene layer grows it is calculated from the C density
that 3 layers of SiC must be consumed \cite{Lauffer2008}. Because
the initial graphene layer continues to cover the surface, it is suggested
that the additional layer grows underneath it. Above 1473 K the SiC
underneath the buffer layer on the upper terrace decomposes, Si atoms
desorb and the liberated C atoms create a graphene layer and a new
buffer layer underneath \cite{Borovikov2009}. This bottom-up growth
process is depicted in Fig. \ref{fig:Borovikov_growth}.

Evolution of the growth of graphene multilayers has been observed
using LEEM \cite{Ohta2008}. In order to achieve this, the contrast
in the LEEM images was interpreted as the number of graphene layers
on the SiC surface. As a check, angle-resolved photoemission spectroscopy
(ARPES) was used to determine the populations of the layers from their
different photoemission intensities. This was compared with the layer
populations found from the LEEM image contrast. As these populations
were the same it shows that the LEEM contrast indeed corresponds to
the different layer thicknesses.

\begin{figure}
\begin{centering}
\includegraphics[height=6cm]{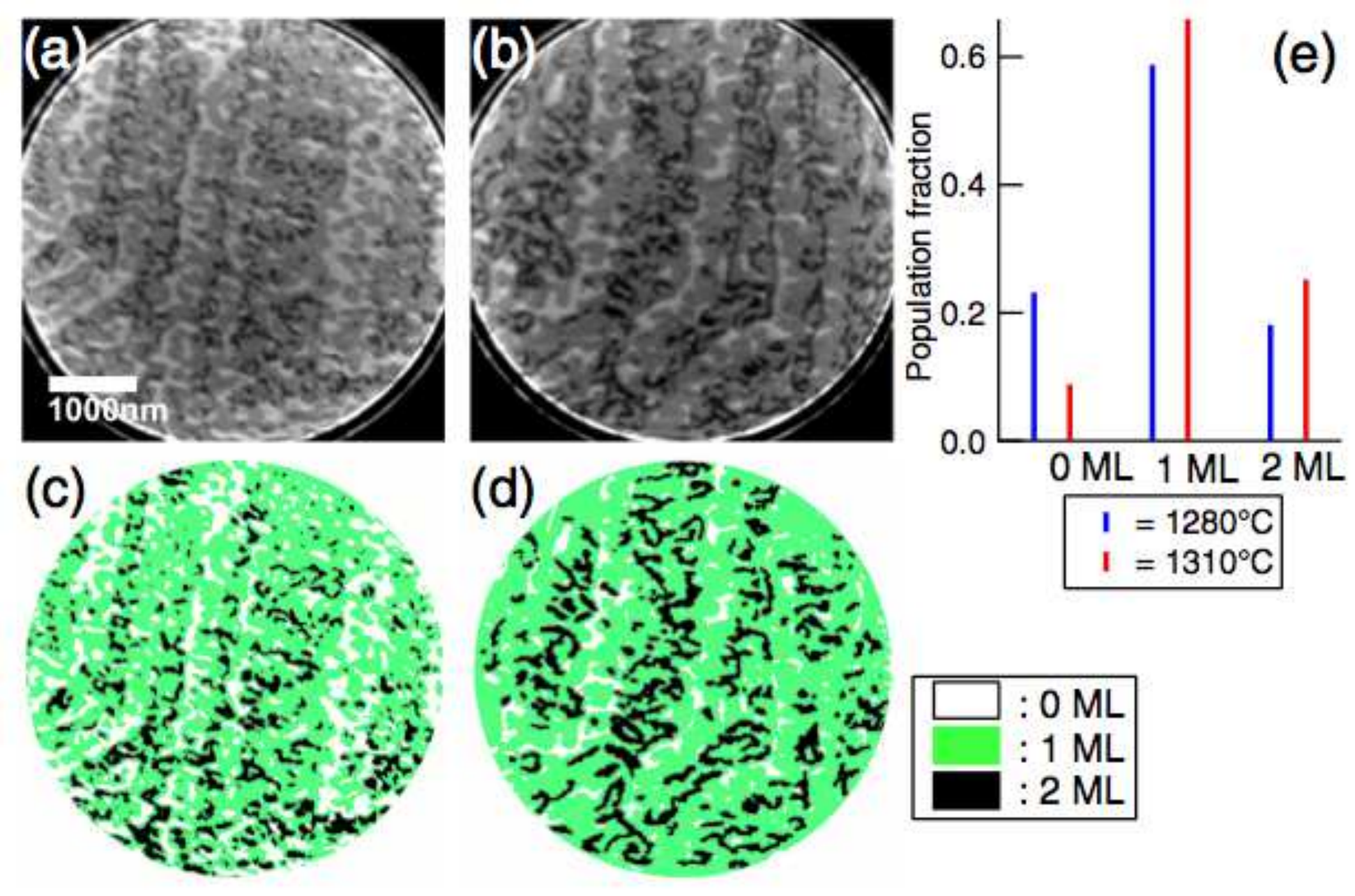}
\par\end{centering}

\centering{}\caption{LEEM images of graphene on SiC after heating to (a) 1553 K and (b)
1583 K. (c) and (d) show the regions of different layer thicknesses
for (a) and (b) after an object extraction process has been applied.
(e) The population fraction of each of these layers at both of the
temperatures. {[}Reproduced from \cite{Ohta2008} by permission of IOP Publishing. All rights reserved.{]}\label{fig:Ohta-LEEM} }
\end{figure}

Using LEEM images of the SiC surface the population fraction of the
different graphene layers was determined at 1553 K and 1583 K, see
Fig. \ref{fig:Ohta-LEEM}(e). The light, medium and dark gray regions
of the image show graphene-free, single layer and bilayer graphene,
respectively, Fig. \ref{fig:Ohta-LEEM}(a) and (b). Heating from 1553
to 1583 K causes the graphene-free population to decrease and the
single layer and bilayer populations to increase. This shows how multilayer
graphene on SiC develops, with new layers growing underneath previous
layers, causing the increase in thickness as the temperature is raised.

\section{Structure of Graphene on Substrates\label{sec:Structure-of-Graphene}}

\subsection{Metallic Substrates\label{metallic-substrates}}

\begin{figure}
\begin{centering}
\includegraphics[height=5cm]{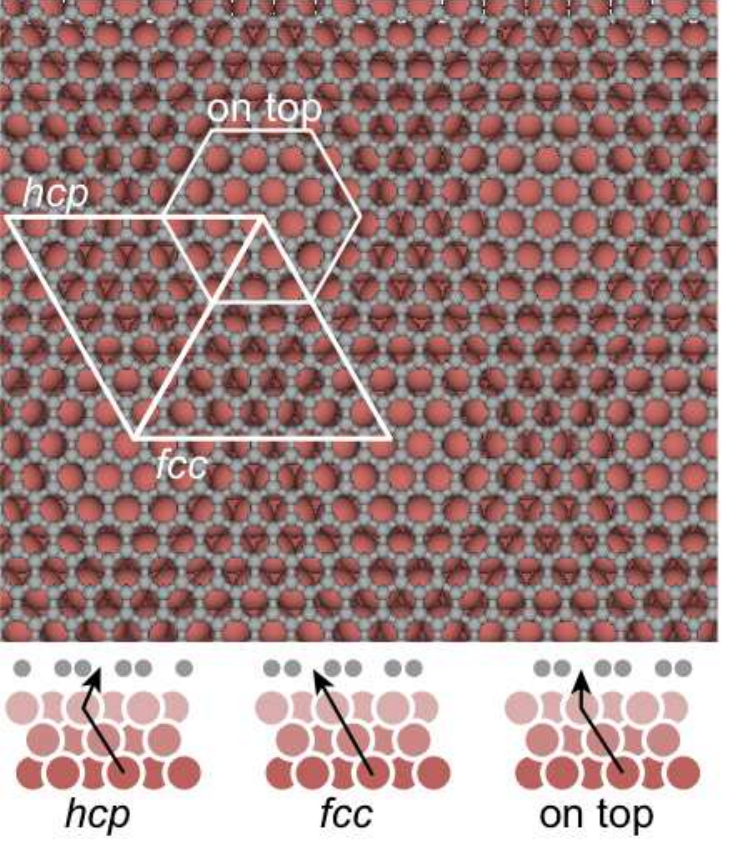}
\par\end{centering}

\caption{\label{fig:Ball-model-moire}Ball-model of the distinctive moir\'{e} sites,
top view (upper panel) and side view (bottom panels); the so-called
hcp, fcc, and on top sites are also indicated.}
\end{figure}

When graphene is grown on transition metal surfaces, various properties
arise due to the lattice mismatch between the metal and graphene.
The C-C bond is extremely strong, much stronger for instance than
C-metal bonds. As already pointed out, this translates into a superior
stiffness of graphene. Biaxial strain in graphene, which may arise
from epitaxial stress during graphene growth on a substrate, is consequently
especially energetically costly. More than about 1\% strain is actually
prohibitively costly, and even such strain cannot be achieved in the
graphene metal systems where C-metal bonds are weak. With the exception
of Co(0001) and Ni(111), graphene cannot have its lattice sufficiently
strained to form a commensurate structure on the substrate. The so-called
moir\'{e}s result, which may be described by the number of graphene hexagon
units ($m\times m$) matching $(n\times n)$ surface cells of the
substrate, in the case of commensurate structures. In a moir\'{e} structure,
carbon atoms are located at different positions with respect to the
atoms of the substrate, and it is usually named after the position
of the center of the carbon ring with respect to the underlying metal
stacking (see Fig. \ref{fig:Ball-model-moire} for the definition
of ``hcp'', ``fcc'', and ``on top'' sites). Their position affects
the strength of the interaction between the surface and the carbon
atoms. Across the moir\'{e} superstructure there are bonding and nonbonding
regions where the C atoms are mostly found on top of substrate atoms
or in between them. Periodic corrugations as well as changes in the
separation between graphene and the substrate are observed. For Ni,
however, the lattice mismatch is small and there is no moir\'{e} structure.

Graphene can also orientate itself at different angles with respect
to the substrate lattice. For one particular surface there may be
many graphene rotational domains that can form. A graphene lattice
whose zigzag rows align to the dense-packed rows of the metal is denoted
as the R0 domain. Other domains are defined in terms of the angle
between them and the R0 phase, such as R30 where the graphene lattice
is rotated by 30\textdegree{} with respect to the substrate lattice.
Each of the domains have an individual moir\'{e} structure, and are therefore
corrugated differently too. Here we report experimental and theoretical
results of the details of the graphene structure on the Ru(0001),
Rh(111), Ir(111), Pt(111), Ni(111), Re(0001), Au(111), Co(0001) and Cu surfaces. The main results
for the different domains found on each of these surfaces are listed
in Table \ref{tab:structure} along with details of their moir\'{e} structure
and the corrugation. We note that the occurrence of the different
moir\'{e}s, as well as the lattice parameter of graphene (thus the type
of commensurability or incommensurability with the metal underneath)
is expected to depend on the growth temperature. Indeed, roughly speaking,
one would expect the graphene to lock in on the substrate at the given
growth temperature, with a structure which is influenced by the lattice
parameter of the substrate at that temperature. This might explain
the observation of different structures in graphene/Ru(0001) \cite{Marchini2007,Vazquez2008}
and graphene/Ir(111) \cite{Blanc2012} systems.

\begin{table}
\caption{Structure of graphene on various surfaces.  The surfaces are listed in order from the strongest interaction with graphene (Ru(0001)) to the weakest (Pt(111)) [185]. The corrugation is described in terms of the peak-to-peak height, with the data in parentheses obtained from DFT (where available). \label{tab:structure}}
\smallskip
\begin{tabular}{|c|c|c|c|c|}\hline
\multirow{3}{*}{Surface}&Metal&Angle between graphene&\multirow{2}{*}{Graphene Moir\'e}&\multirow{3}{*}{Corrugation (\AA)}\\
&lattice&zig-zag rows and metal&\multirow{2}{*}{superstructure}&\\
&constant (\AA)&densely-packed rows&&\\
\hline\hline
\multirow{4}{*}{Ru(0001)}&\multirow{4}{*}{2.71}&\multirow{4}{*}{$0^\circ$}&$(25\times25)$ \cite{Martoccia10,Moritz}
&0.82 \cite{Martoccia10},
1.53 \cite{Moritz},\\
&&&&0.15 \cite{Borca2010}, (1.5 \cite{Martoccia})\\
&&&$(13\times 13)$ \cite{Moritz}&(1.59 \cite{Moritz})\\
&&&$(12\times12)$ \cite{WangRuB801785A}&(1.5 \cite{WangRuB801785A})\\
\hline
Rh(111)&2.69&$0^\circ$&$(12\times12)$ \cite{Voloshina} &0.5--1.5 \cite{Voloshina}\\
\hline
\multirow{7}{*}{Ir(111)}&\multirow{7}{*}{2.72}&$0^\circ$ \cite{Loginova2009b,Meng2012} &incommensurate  \cite{Blanc2012}
&$\sim0.3$ \cite{Diaye2006,Loginova2009b}\\
&&&$(10\times10)$  \cite{Meng2012}&0.423  \cite{Meng2012}\\
&&$\sim14^\circ$ \cite{Loginova2009b,Meng2012} &$(4\times4)$ \cite{Meng2012}&(0.101 \cite{Meng2012})\\
&&$19^\circ$ \cite{Meng2012}&$(3\times3)$ \cite{Meng2012}&(0.051\cite{Meng2012} )\\
&&$23^\circ$ \cite{Meng2012}&$(\sqrt{19}\times\sqrt{19})$ \cite{Meng2012}&(0.022\cite{Meng2012}) \\
&&$26^\circ$ \cite{Meng2012}& $(\sqrt{37}\times\sqrt{37})$ \cite{Meng2012} &(0.015\cite{Meng2012} )\\
&&$30^\circ$ \cite{Loginova2009b,Meng2012}& $(2\times2)$ \cite{Meng2012}&0.04 \cite{Loginova2009b} (0.014 \cite{Meng2012})\\
\hline

\multirow{6}{*}{Pt(111)}&\multirow{6}{*}{2.77}&$30^\circ$&$(2\times2)$&\\
&&$19^\circ$&$(3\times3)$&$<0.3$\\
&&$\hskip29.5pt 14^\circ$\hskip6pt\multirow{2}{*}{\cite{Gao2011}}&\hskip40pt$(4\times4)$\hskip17pt\cite{Gao2011},&\hskip28pt\multirow{2}{*}{\line(1,0){50}\hskip6pt\cite{Gao2011}}\\
&&$6^\circ$&\hskip26pt$(\sqrt{37}\times\sqrt{37})$\hskip6pt\cite{Merino2011}&\\
&&$3^\circ$&$(\sqrt{61}\times\sqrt{61})$&0.5--0.8\\
&&$2^\circ$&$(\sqrt{67}\times\sqrt{67})$&\\

\hline
\multirow{2}{*}{Ni(111)}&\multirow{2}{*}{2.49}&$0^\circ$&\multirow{2}{*}{$(1\times1)$ \cite{Gamo1997,rosei1983,Zhao2011}}&\\
&&$17^\circ\pm7^\circ$ \cite{Dahal}&&\\
\hline
\multirow{2}{*}{Cu(111)}&\multirow{2}{*}{2.56}&$0^\circ$ \multirow{2}{*}{\cite{Gao10}} &&\\
&&\hskip-25pt$7^\circ$ &&\\
\hline

\multirow{6}{*}{Pd(111)}&\multirow{6}{*}{2.75}&\multirow{2}{*}{$-2^\circ$}&$(3\sqrt{7}\times3\sqrt{7})$--$R\,19^\circ$&\\
&&\multirow{2}{*}{$-5^\circ$}&$(\sqrt{39}\times\sqrt{39})$--$R\,16^\circ$&\\
&&\multirow{2}{*}{\hskip30pt$17^\circ$}\hskip6pt\multirow{2}{*}{\cite{Murata2010}}&\hskip30pt$(\sqrt{21}\times\sqrt{21})$--$R\,11^\circ$\hskip6pt\multirow{2}{*}{\cite{Murata2010}}&\\
&&\multirow{2}{*}{$22^\circ$}&$(\sqrt{13}\times\sqrt{13})$--$R\,14^\circ$&\\
&&\multirow{2}{*}{$26^\circ$}&$(\sqrt{7}\times\sqrt{7})$--$R\,19^\circ$&\\
&&&$(7\times7)$&\\

\hline
\multirow{4}{*}{Au(111)} &  \multirow{4}{*}{2.88}  &  $0^\circ$  &   &\\

            & & \hskip30pt$11^\circ$\hskip6pt\multirow{2}{*}{\cite{Nie2012}} & &\\

&&$14^\circ$&&\\
&&$26^\circ$&&\\
\hline
Re(0001)&2.76&$0^\circ$\hskip6pt \cite{Miniussi2011,Tonnoir2013}&$(8\times8)$ \hskip6pt\cite{Tonnoir2013}&1.7\hskip6pt\cite{Tonnoir2013}\\
\hline
Fe(110)&2.87/4.06&$0^\circ$\hskip6pt\cite{Vinogradov2012}&$(6\times18)$\hskip6pt\cite{Vinogradov2012}&0.9\hskip6pt\cite{Vinogradov2012}\\
\hline
\end{tabular}
\end{table}

Before we move on to discuss each particular metallic substrate, it is worthwhile
to say a few words here on the role of the dispersion interaction (Section \ref{sec:vdW}) in stabilising
adsorption of a single layer of graphene on metals.
As this is a layered geometry, one would expect that the role of
dispersion interaction is fundamental in binding graphene. Various DFT based methods
were recently compared for the graphene adsorption on Ni(111)
\cite{Mittendorfer2011,Olsen-PRL-2011}, Cu(111) and Co(0001) \cite{Olsen-PRL-2011}.
The calculations were performed using LDA, PBE, various flavours of vdW-DF and RPA methods.
In the latter case exact exchange was also applied. Because of
the fact that RPA is extremely expensive computationally, only primitive cells
were considered in these calculations
with the graphene lattice being appropriately scaled to that of the underlying
metal.
In all cases LDA calculations resulted in a strong binding of graphene with the distance of around
$d=$ 2  $\textrm{\AA}$ to the metal.
PBE simulations yielded binding only in the case of Co(0001) which was found rather weak; no minimum
was obtained in the cases of Cu(111) and Ni(111) substrates. Calculations using the vdW-DF methods
resulted in all three cases in a minimum at rather large distance of  $d=$ 3.7  $\textrm{\AA}$ to the metal
indicating physisorption with the adsorption energy (per C atom) of around 40 meV \cite{Olsen-PRL-2011}.
However, RPA simulations showed that the situation is much more complex as there is a subtle balance
between the chemical (short range) and dispersion (long range) interactions in these systems.
In particular, while for Cu(111) only a single minimum with $d=$ 3.25  $\textrm{\AA}$ was found
(which is between the values given by LDA and vdW-DF), for Ni(111) and Co(0001) two minima were
found corresponding to chemisorption and physisorption. For instance,
in the case of Ni(111) these were at the distances
of 2.3 and 3.25 $\textrm{\AA}$, respectively, with the physisorption state being by 4-8 meV per C atom
(depending on the k-point sampling used) more favourable. Interestingly, two minima were also
found in a similar study \cite{Mittendorfer2011}, however, the chemisorption state was found to be
more energetically favourable by 7 meV per C atom. These energy differences are however too small
to conclude decisively at which distance graphene is preferentially adsorbed on these metals.
The most likely conclusion from these studies is
 that the potential energy surface for graphene adsorbed on a metal surface as a function of $d$
 between 2.3 and 3.3 $\textrm{\AA}$ is rather
flat. It is clear however that there is
subtle balance between the chemical and dispersive interactions in these systems which RPA method
captures successfully, while other methods  do not, so great care is indeed needed when
applying DFT based technique to this type of system.

\subsubsection{Ru(0001)\label{sub:Ru(0001)}}

Graphene can be grown on the Ru(0001) surface using any of the methods
described in Sections \ref{sec:Chemical-Vapour-Deposition}, \ref{sec:Temperature-Programmed-Growth}
and \ref{sec:Segregation}. The strong interaction between Ru and
carbon results in the graphene growth having\textcolor{black}{{} only
the R0 rotational domain. When graphene is grown on the Ru(0001) surface,
a moir\'{e} superstructure is formed due to the lattice mismatch between
Ru and graphene. This structure h}as a hexagonal lattice with a measured
repeat distance of around 30 $\textrm{\AA}$ \cite{WangRuB801785A}. A variety of
methods such as LEED, STM and surface X-ray diffraction (SXRD) studies
have been used to identify the moir\'{e} superstructure. STM measurements
have suggested periodicities of graphene/metal as (11 $\times$ 11)C/(10
$\times$ 10)Ru \cite{Vazquez2008}, and (12 $\times$ 12)C/(11 $\times$
11)Ru \cite{Marchini2007,Zhang2009}. SXRD studies, which allowed
unprecedented resolution, have shown that the moir\'{e} structure has
a periodicity of (25 $\times$ 25)C/(23 $\times$ 23)Ru \cite{Martoccia10,Moritz}.
Using LEEM coupled to micro-LEED experiments, it was found that the
actual structure of the samples varies slightly about this second
order R0 commensurability, with small deviations from R0 \cite{Man2011}.

\begin{figure}
\centering{}\includegraphics[scale=0.3]{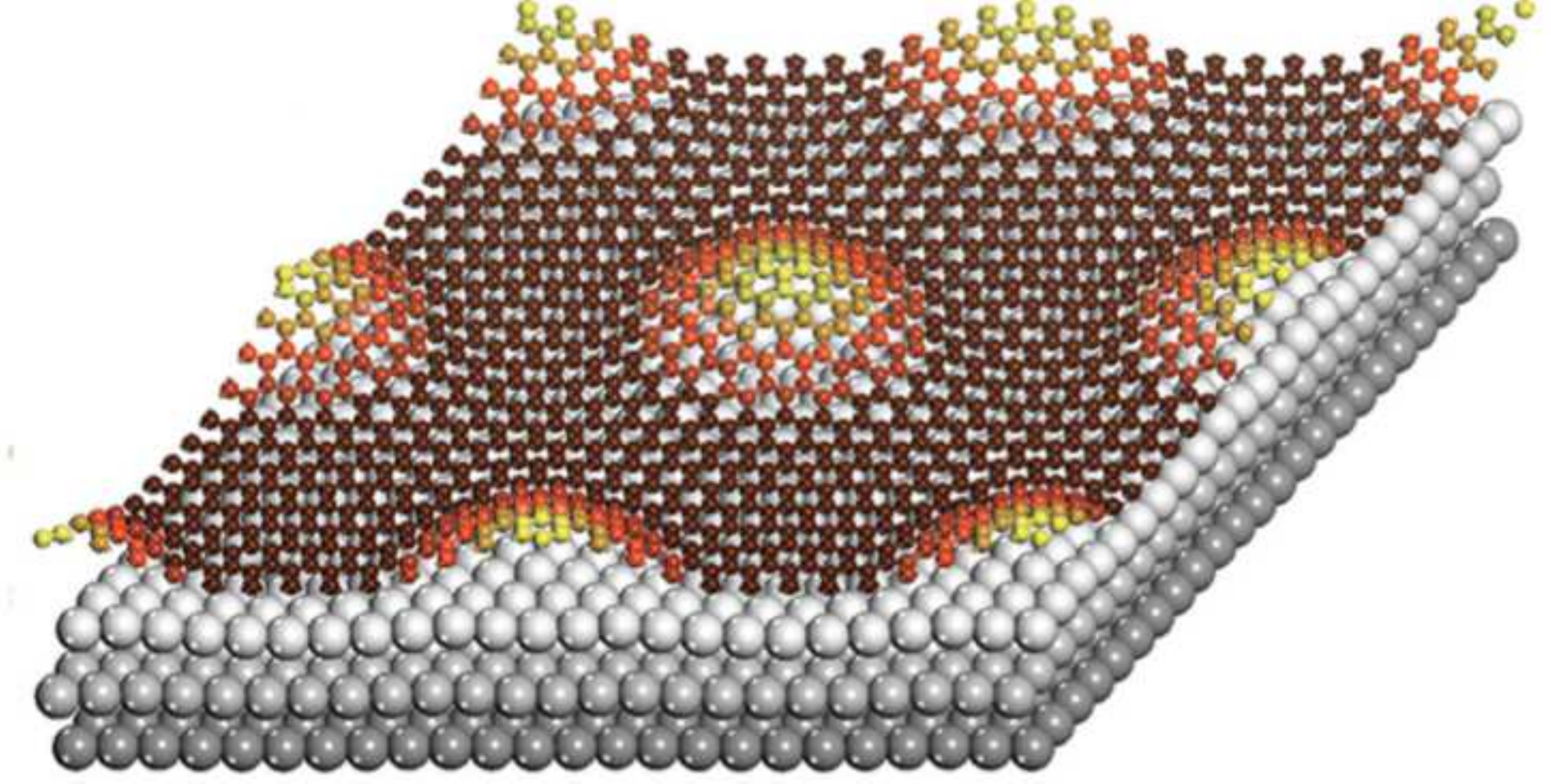}\caption{The 3D surface structure model of graphene on Ru(0001) showing the
corrugation across the lattice, as determined from LEED.
For better visibility, the vertical displacements of the C and Ru atoms are enhanced by factors
of 3 and 7, respectively.
{[}Reprinted
with permission from \cite{Moritz}. Copyright (2010) by the American Physical Society.{]}\label{fig:The-3D-surface-Moritz}}
\end{figure}

Due to the strong interaction between graphene and Ru(0001) the graphene
becomes highly corrugated. Measurements of the peak to peak height
of these corrugations have been made with LEED, from helium atom scattering
(HAS) data and using X-ray diffraction. Each method observes different
corrugation heights with LEED, HAS and XRD measuring the height as
1.5 $\textrm{\AA}$ \cite{Moritz}, 0.15 $\textrm{\AA}$ \cite{Borca2010} and 0.82$\pm$0.15
$\textrm{\AA}$ \cite{Martoccia10}, respectively. DFT calculations determine a
corrugation of 1.5 $\textrm{\AA}$ \cite{Martoccia} and 1.59 $\textrm{\AA}$ \cite{Moritz}.
As a result of the corrugation the separation between the graphene
and the Ru surface also varies. LEED studies have found that the separation
changes from 2.1 to 3.64 $\textrm{\AA}$ across the periodicity of the cell \cite{Moritz}.
A similar result of 2 to 3 $\textrm{\AA}$ has been found with X-ray diffraction
\cite{Martoccia10}. The model of graphene on the Ru(0001) surface
is shown in Fig. \ref{fig:The-3D-surface-Moritz}.

The graphene-Ru interaction has also been found to cause corrugation
in the Ru substrate. For the first Ru layer the corrugation was measured
as 0.23 $\textrm{\AA}$ with LEED \cite{Moritz} and 0.1$\pm$0.02 $\textrm{\AA}$ with SXRD \cite{Martoccia10}.
For the layers further below this the corrugation is reduced and it
eventually decays into the bulk substrate \cite{Martoccia}. This
is unlike the graphene layers where additional layers on top of the
first layer are unaffected by the Ru substrate.

\begin{figure}
\centering{}\includegraphics[scale=0.5]{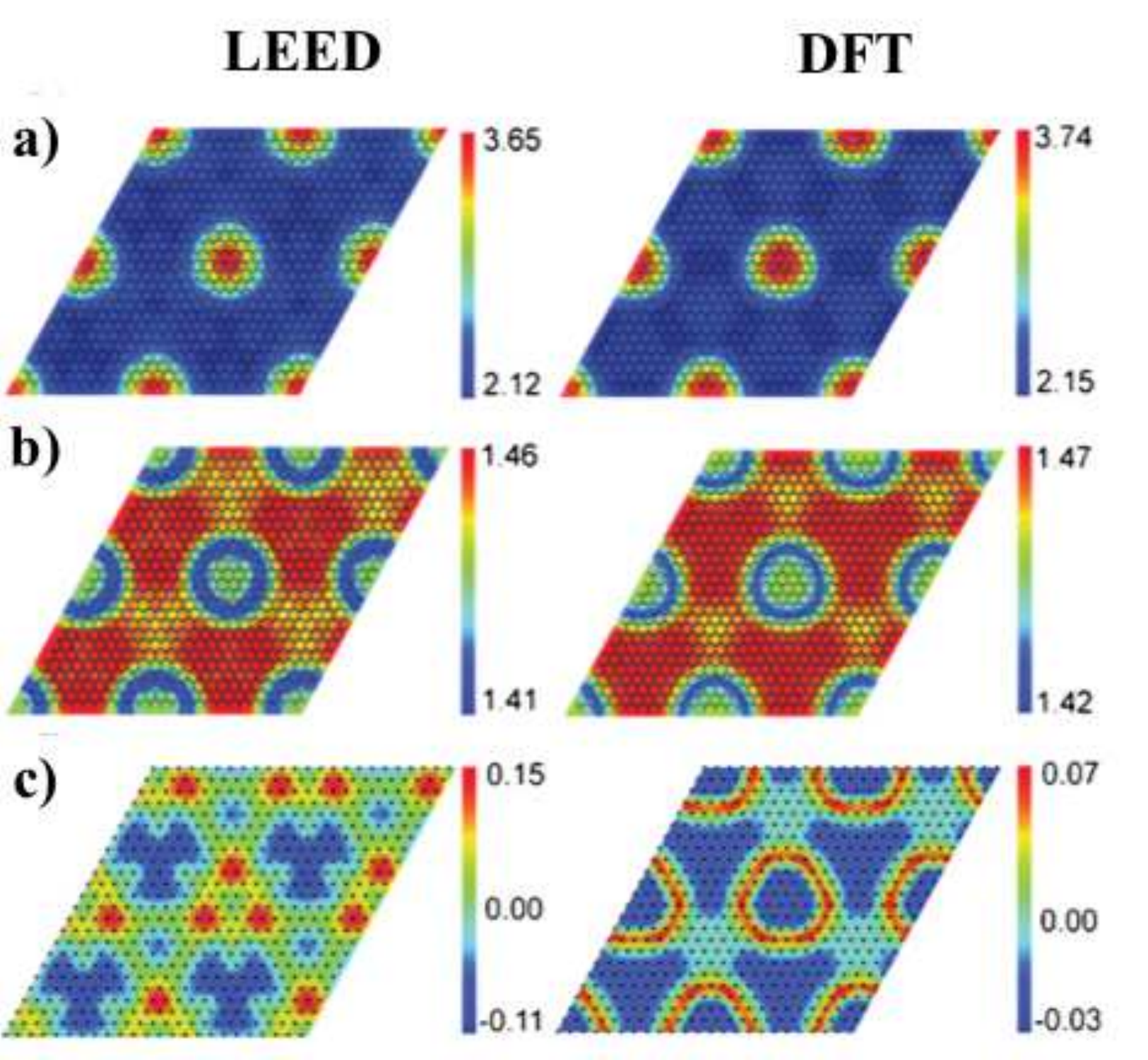}\caption{LEED and DFT results for (a) the corrugation of the graphene layer,
(b) the C-C bond length in the graphene layer, (c) the corrugation
of the upper Ru layer.\label{fig:LEED-and-DFT}{[}Reprinted wi\textcolor{black}{th
permission from \cite{Moritz}. Copyright (2010) by the American Chemical Society.{]}}}
\end{figure}

In the work of Mor\textcolor{black}{itz }\textcolor{black}{\emph{et
al.}}\textcolor{black}{{} \cite{Moritz} DFT was u}sed to study the
graphene corrugation, the bond length of the graphene layer and the
corrugation of the Ru underlayer for a simulated (13 $\times$ 13)C/(12
$\times$ 12)Ru moir\'{e} lattice. The results are shown in Fig. \ref{fig:LEED-and-DFT}
along with the results for the same properties as determined by LEED
experiments. The variation in the C-C bond length throughout the moir\'{e}
superstructure can be related to the strength of the binding in different
regions across the lattice. Where the C-C binding is strong the C
atoms have shorter bonds to each other and are close to the Ru substrate,
whereas in the weakly bound regions they are found further from the
substrate. This gives rise to the corrugations shown in Fig. \ref{fig:LEED-and-DFT}(a).
At the corrugation maxima the bond lengths match that of freestanding
graphene, while in the minima the bonds are stretched due to the interaction
with the Ru substrate. In comparison to the LEED data the DFT calculated
results are similar, but the graphene corrugation height is overestimated
with DFT and the corrugation of the Ru surface is underestimated.
The difference in the Ru corrugation is suggested to be due to the
fact that only two Ru layers were relaxed in the calculation, whereas
experimental results have shown that the relaxation occurs for up
to 7 Ru layers below the surface.

\subsubsection{Rh(111)}

\begin{figure}
\centering{}\includegraphics[height=8cm]{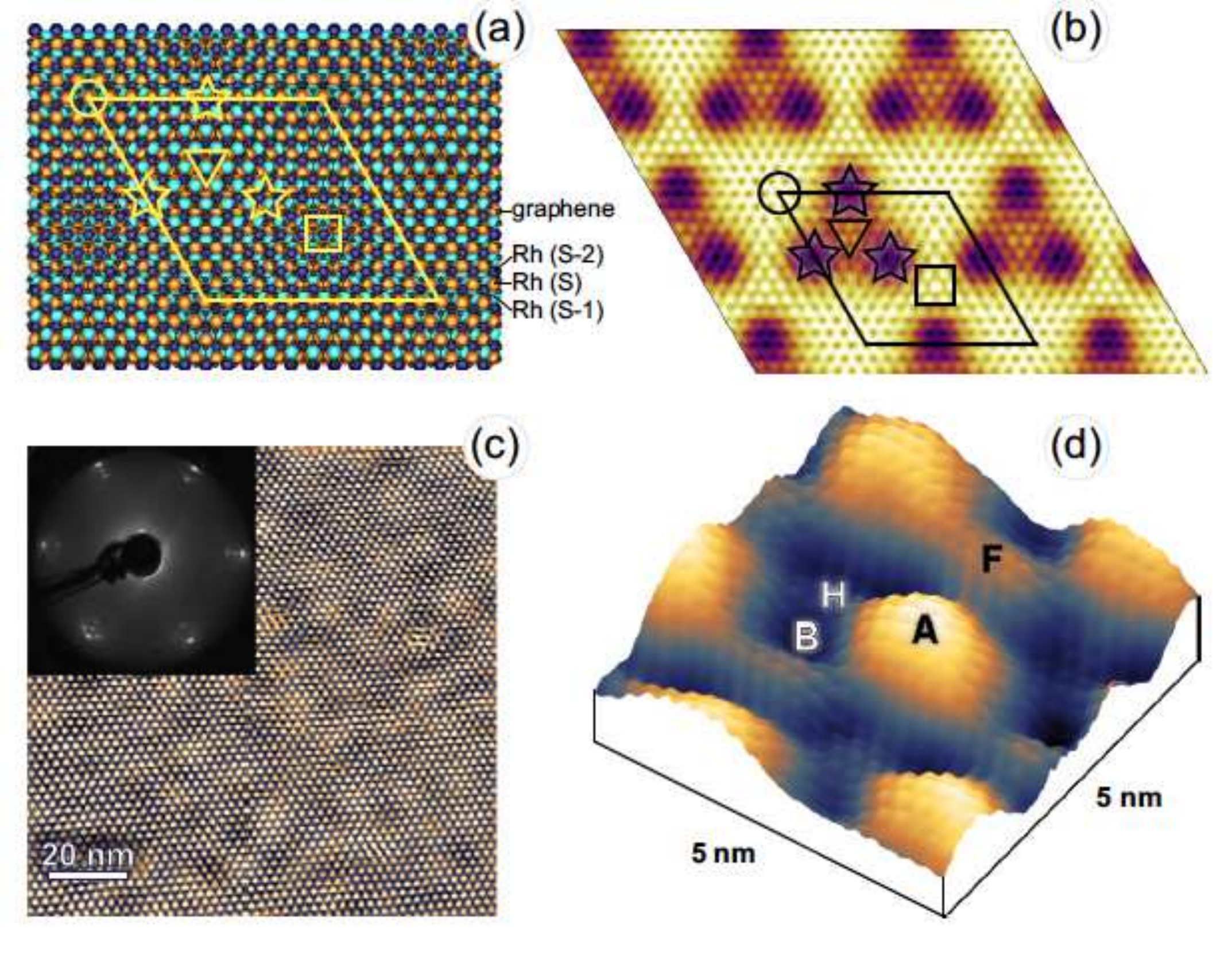}\caption{(a) Crystallographic structure s\textcolor{black}{howing the 12$\times$12
supercell of graphene positioned on top of a 11$\times$11 cell of
Rh(111); (b) the corresponding calculated STM image of graphene/Rh(111).
Here the symbols refer to positions of the centers of hexagonal units
of graphene relative to the Rh lattice in high symmetry configurations
\cite{Wang-Schaub-NanoLett-2010} (labelled Rh(S), Rh(S-1) and Rh(S-2)
in (a)). These are A sites whereby carbon atoms lie on top of Rh(S-2)
and Rh(S-1) atoms, (circles); H sites where atoms lie on top of Rh(S)
and Rh(S-2) atoms (down-triangle); F sites where atoms lie on top
of Rh(S) and Rh(S-1) (square), and finally B sites which bridge (star).
(c) Large scale STM image of the graphene layer on Rh(111). Tunneling
conditions: U$_{T}$=+1 V, I$_{T}$=1 nA. (d) 3D view of the 5$\times$5
nm$^{2}$ region from (c) showing the structure of the graphene layer
on Rh(111). Tunneling conditions: U$_{T}$=0.55 V, I$_{T}$=10 nA.
{[}Reprinted wit}h permission from \cite{Voloshina}. Copyright (2012), AIP Publishing LLC.{]}.\label{fig:(a)-Crystallographic-structure}}
\end{figure}

Graphene on Rh(1\textcolor{black}{11) has similarities with that on
Ru(0001) and in terms of the strength of interaction between the surface
and graphene Rh lies in between Ir and Ru \cite{Voloshina}. Further,
graphene forms just a single rotational domain on Rh(111) with a mismatch
corresponding to a 12$\times$12 graphene cell occupying 11$\times$11
Rh supercell \cite{Voloshina,Wang-Schaub-NanoLett-2010}. This has
been observed experimentally using both AFM and STM, see Fig. \ref{fig:(a)-Crystallographic-structure}.
These images demonstrate a moir\'{e} structure with a measured corrugation
of between 0.5 - 1.5 $\textrm{\AA}$. In this study, Voloshina }\textcolor{black}{\emph{et
al }}\textcolor{black}{\cite{Voloshina} used AFM to examine the differences
in bonding between graphene and Rh at the different sites that carbon
atoms in graphene occupy on the Rh surface. These regions correspond
to the different arrangements of carbon atoms on top of Rh(111) as
explained in the caption of Fig. \ref{fig:(a)-Crystallographic-structure}(a).
They found that the strongest bonded region corresponds to the bridge
region and the most weakly bound region is the A region or the T region
in the notation of } Wang \textcolor{black}{\emph{ et al. }}\textcolor{black}{\cite{Wang-Schaub-NanoLett-2010}.
Unsurprisingly, due to the changes in the interaction strength the
different regions vary in height above the Rh(111) surface. In the
A region the C atoms are 3.15 $\textrm{\AA}$ above the surface, whereas
for the strongly bound bridge region the gap between graphene and
the substrate is much lower, about 2.08 $\textrm{\AA}$. Thus, just
like Ru(0001), graphene on Rh(111) forms a strongly corrugated sheet
with regions that interact much more strongly with the underlying
substrate than others. }

\subsubsection{Re(0001)\label{sub:Re(0001)}}

The preparation of graphene on Re(0001) is achieved either by repeated TPG
cycles \cite{Miniussi2011} or by
segregation from thin films \cite{Tonnoir2013}.
 Graphene on Re(0001) exhibits many similarities with that on Ru(0001) and
Rh(111), given the relatively strong interaction between graphene and Re.
The moir\'{e} of graphene on Re may consist of different coincidence lattices
\cite{Miniussi2011,Tonnoir2013}, presumably depending on the preparation
conditions. These coincidences  correspond preferentially to carbon zig-zag
rows aligning Re dense packed rows, for instance, with 8 carbon rings matching
 7 ones. In this configuration, DFT calculations including van der Waals
interaction predict a 1.7 $\textrm{\AA}$ corrugation along the moir\'{e} sites,
 with the lowest-lying sites being 2.2 $\textrm{\AA}$ and
the highest-lying ones 4 $\textrm{\AA}$ above Re atoms \cite{Tonnoir2013}.

\subsubsection{Ir(111)\label{sub:Ir(111)}}

Preparation of graphene on Ir(111) is usually achieved using hydrocarbon
decomposition either with TPG (Section \ref{sec:Temperature-Programmed-Growth})
or CVD (Section \ref{sec:Chemical-Vapour-Deposition}). In these methods
the surface acts as a catalyst for the growth of graphene from hydrocarbons.
Due to this and the fact that carbon has low solubility in Ir, graphene
growing on Ir(111) is usually limited to just one layer. The graphene
formed on the Ir(111) surface has a moir\'{e} structure, whose geometry
is more complex than that of graphene/Ru(0001), as shown by a recent
SXRD study revealing the incommensurate nature of the moir\'{e} between
graphene and Ir(111) \cite{Blanc2012}. The repeat distance of the
R0 domain moir\'{e} structure (where the graphene and Ir lattices are
aligned) was found from LEED and STM measurements to be 25.8$\pm$2
$\textrm{\AA}$ and 25.2$\pm$0.4 $\textrm{\AA}$, respectively \cite{Diaye2008}, and depends
on preparation temperature \cite{Blanc2012}.

\begin{figure}
\begin{centering}
\includegraphics[height=5cm]{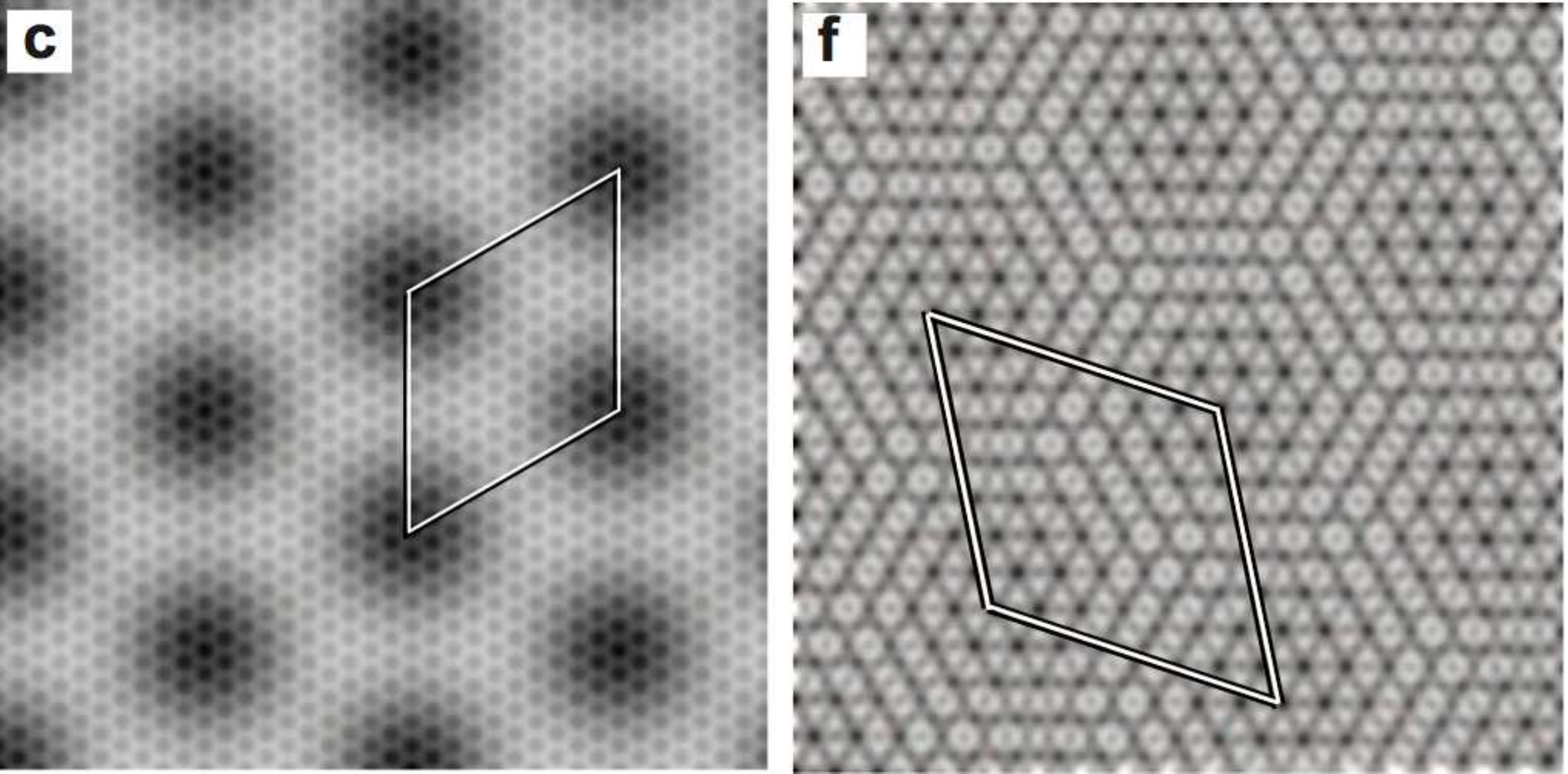}
\par\end{centering}

\caption{Schematic illustrations from STM images of the graphene on Ir(111)
rotated by (a) 0\textdegree{} (R0 phase) and (b) 30\textdegree{} (the
R30 phase) with respect to the Ir(111) lattice. {[}Reprinted with
permission from \cite{Loginova2009b}. Copyright (2009) by the American Physical Society.{]}\label{fig:Loginova-Ir-moire}}
\end{figure}

Unlike graphene on Ru(0001), the Ir(111) surface gives rise to the
growth of graphene with various orientations, see Fig. \ref{fig:Loginova-Ir-moire}.
As well as the commonly found R0 phase \cite{Diaye2008,Coraux,Loginova2009b},
multiple domain phases which are rotated by various angles with respect
to the R0 phase have been identified on the Ir(111) surface \cite{Loginova2009b,Meng2012},
the most frequently found are listed in Table \ref{tab:structure}.
These domains each have different structural properties such as moir\'{e}
repeat distance, G-Ir separation and corrugation. The cause of these
multiple domains is suggested to be due to the interaction strength
between the graphene and the substrate surface. For Ru(0001) the interaction
is strong, and hence there is large corrugation and strong orientational
dependence on the surface that allows only one R0 orientation. For
Ir(111) the interaction with graphene is weaker \cite{Preobrajenski2008}
and it is possible then that multiple orientations exist due to the
limited influence of the surface. The various growth alignments found
on Ir(111) have also been shown to be temperature dependent \cite{Hattab2011}.
In a TPG study of the growth from ethene deposition and then annealing
to temperatures between 1000 and 1530 K the preferred orientation
was determined at different temperatures. Below 1200 K the
orientation distribution of the domains is rather broad.
At 1255, 1350
and 1460 K there is evidence of the R0 and R30 phases as well as randomly
orientated domains. At 1530 K the graphene becomes well-ordered with
only the R0 phase present.

The corrugation of graphene on Ir(111) has been estimated with STM
for the 0$^{\circ}$ and 30$^{\circ}$ orientations. For the R0 phase
the corrugation is measured as approximately 0.3 $\textrm{\AA}$, whereas for R30
the corrugation is lower at 0.04 $\textrm{\AA}$, see Fig. \ref{fig:Loginova-Ir-moire}.
This suggests that the R0 domains interact more strongly with the
surface than the R30 domains \cite{Loginova2009b}. The corrugation
of the R0 phase was later confirmed in a reliable manner with XSW
\cite{Busse2011} and AFM \cite{Boneschanscher2012,Voloshina2013}
experiments (STM, in principle, only provides information about apparent
heights). It was found that the corrugation depends on the internal
strain in graphene islands \cite{Busse2011}.

Additional graphene rotational domains were found on the Ir(111) surface
with LEED and STM \cite{Meng2012}. R14, R19, R23 and R26 domains
have been reported, but also in some preparation conditions a continuous
spread of orientations \cite{Hattab2011}, like on Pt(111) \cite{Land1992},
is observed. The moir\'{e} periodicity of these domains is listed in Table
\ref{tab:structure} along with the corrugation determined from DFT
calculations which included van der Waals forces as well. The size
of the corrugations is claimed to be due to the positions of the C
atoms within the moir\'{e} structure. For the R0 domain C atoms are found
in distinct regions which are either on top of Ir atoms in the first
(top region), second (fcc region) and third (h\textcolor{black}{cp
region) layers \cite{Diaye2006}. Therefore the corrugation is large.
For R23 and R26 no fcc or hcp regions can be noticed with STM and
accordingly the graphene is less corrugated. In R30 graphene the fcc
and hcp regions are similar due to its small size and the corrugation
is also small. The separation between graphene and the Ir(111) surface
was also calculated in the same manner. The size of the gap between
t}he lowest carbon atom and the first layer of Ir substrate was shown
to vary by about 0.1 $\textrm{\AA}$ across all of the orientations.
The smallest gap found was for the R0 domain (3.159 $\textrm{\AA}$),
which was also the most corrugated structure of the different domains.
Those with the smallest corrugations also had the largest graphene-Ir
separations \cite{Meng2012}.

\subsubsection{Au(111)\label{sub:Au(111)}}

The structure of graphene on Au(111) was investigated by STM, which
revealed that different orientations, like on Ir(111), coexist,
corresponding to 0$^{\circ}$, 11$^{\circ}$, 14$^{\circ}$ and 26$^{\circ}$
rotations between carbon zig-zag rows
and Au dense packed ones  \cite{Nie2012}. Interestingly, the interaction
between graphene and Au(111) is sufficiently weak that the herringbone
reconstruction is not lifted after graphene growth, and can readily be
observed in STM. This interaction is the weakest in the case of non-0$^{\circ}$
rotational domains, which translates into a different (non determined)
corrugation along the moir\'{e} sites. Based on this observation, an upper
limit for the graphene-metal interaction energy was given by trying to
reproduce, in the framework of a  Frenkel-Kontorova model, the features
of the Au reconstruction in the presence of graphene. This upper limit
was determined to be 13 meV per C atom \cite{Nie2012}.

\subsubsection{Pt(111)\label{sub:Pt(111)}}

\begin{figure}
\centering{}\includegraphics[height=5cm]{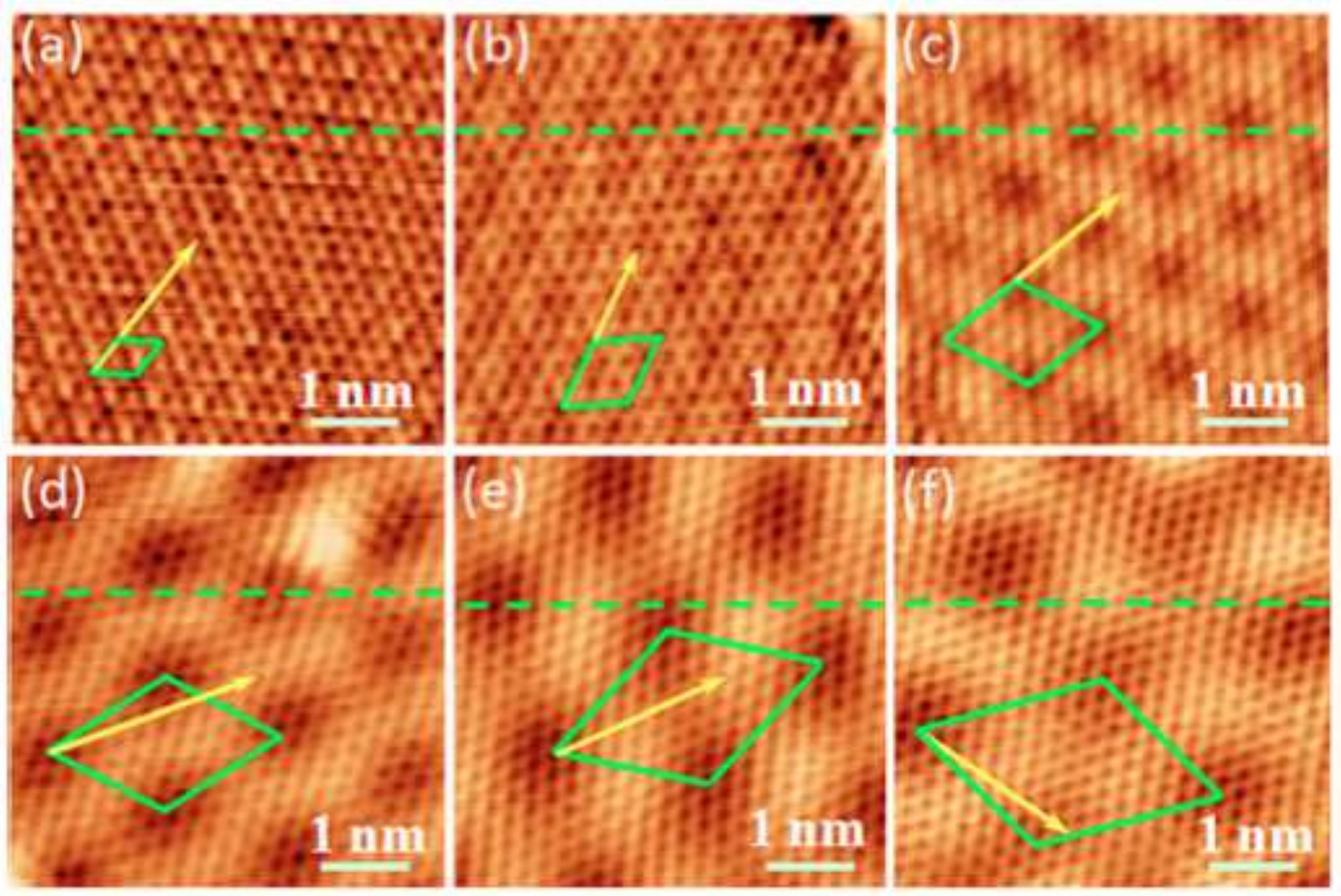}\caption{STM images of the graphene lattice on the Pt(111) surface rotated
by (a) 30\textdegree{}, (b) 19\textdegree{}, (c) 14\textdegree{},
(d) 6\textdegree{}, (e) 3\textdegree{}, and (f) 2\textdegree{} with
respect to the Pt lattice. The moir\'{e} supercells are highlighted in
each case. {[}Reprinted with permission from \cite{Gao2011}. Copyright (2011), AIP Publishing LLC.{]}\label{fig:Gao2011-moire}}
\end{figure}

Compared to the other surfaces the interaction between the Pt(111)
surface and the graphene is considerably weaker \cite{Preobrajenski2008}.
As a result of this a main feature of graphene grown on Pt(111) is
that there are many rotational domains \cite{Gao2011,Merino2011,Sutter2009}.
Some which have been reported are listed in Table \ref{tab:structure}.
Each of the domains has a moir\'{e} type structure with unit cells that
are each sized differently \cite{Gao2011} as shown in Fig. \ref{fig:Gao2011-moire}.
The structure of the domains is also corrugated across the unit cell
with the amount of corrugation dependent on the cell size \cite{Gao2011}.
For the R30 and R19 domains the moir\'{e} unit cells are small (5 and
7.38 $\textrm{\AA}$ in size, respectively) and in both cases corrugation observed
with STM has an apparent height less than 0.3 $\textrm{\AA}$. For the domains with
a larger repeating unit the measured corrugation is slightly more
significant and ranges within 0.5 - 0.8 $\textrm{\AA}$ for the 14\textdegree{},
6\textdegree{} and 2\textdegree{} rotations. As discussed previously,
more corrugated structures are the result of a stronger interaction
between the graphene and the surface. For the 7 $\times$ 7 and 2
$\times$ 2 moir\'{e} structures the separation between graphene and Pt
was calculated to be above 3.1 $\textrm{\AA}$. The large size of
this gap is due to the weak graphene-Pt interaction \cite{Gao2011}.

In the experimental study \cite{Sutter2009} graphene was grown from
segregation of C atoms from the Pt bulk. From a LEEM study it was
found that smaller
rotational-angle
domains grow more quickly than larger
domains. The suggested reason for this is that it is due to kinetic
factors and that for the smaller domains there is a lower attachment
barrier for C \cite{Sutter2009} (assuming of course that the growth
progresses via atomic, not cluster, attachments). This could relate
to the fact that the interaction with the surface may be weaker for
smaller
rotational-angle
domains
(as seen in the corrugation data). A similar phenomena
has been noticed on the Ir(111) surface, where the R30 domains grow
faster than the R0 domains \cite{Loginova09}. Of the two it is probable
that the R0 domain is more corrugated and is assumed to interact more
strongly with the surface \cite{Loginova2009b}. This may cause the
attachment barrier for C monomers or clusters to be greater.

In the work by Merino \emph{et al. }\cite{Merino2011} the composition
of the graphene grown on Pt(111) was analyzed in order to find the
fractional coverage of the different rotational domains. The study
was based on hundreds of STM images. From this it was noticed that
only some particular domains are present,
which
make up a larger fraction of the total
graphene coverage. To understand these results a model was proposed
where the mismatch between the graphene and the surface for each of
the domains was related to how frequently the structures form. The
model suggests that rotational domains with smaller mismatch should
be found more frequently on the surface, which agrees with the experimental
observations. This explains why graphene grows along only some particular
angles with respect to the Pt lattice.

\subsubsection{Ni(111) and Co(0001)\label{sub:Ni(111)}}

The discussion in this subsection is mainly concerned with the Ni(111) surface.
However, it is worth keeping in mind that the Co(0001) surface has
a lattice very
similar to that of Ni(111), and therefore its mismatch and hence
interaction with graphene are expected to be similar.

The Ni(111) surface has a lattice constant that is very similar to
the free-standing graphene lattice and so under most growth conditions
graphene forms a commensurate structure on Ni(111) with a very small
lattice mismatch. Thus, unlike graphene growth on many transition
metal surfaces, usually no moir\'{e} pattern is observed during high temperature
growth.

As a result usually (1 $\times$ 1) commensurate structures are formed, which are
characterized by a short graphene-metal distance \cite{rosei1983}. Among the
many possible (1 $\times$ 1) structures \cite{Zhao2011}, corresponding to hcp,
fcc, ontop configurations of Fig. \ref{fig:Ball-model-moire},
other kinds of configurations,
e.g. with the centre of carbon rings atop Ni bridge sites of the different Ni
layers, have been identified in experiments. High resolution XPS
recently showed that, besides the hcp configuration inferred from LEED
measurements \cite{Gamo1997},  a configuration with carbon ring centres atop
the bridge sites of the topmost Ni layers was formed \cite{Zhao2011}.
It has
also been shown that under some specific growth conditions it is possible
to form graphene which exists in two different rotational domains
with respect to the Ni(111) lattice \cite{Dahal,Lahiri11},
and that in the case when a surface carbide forms as an intermediate during
graphene preparation, a graphene phase rotated by 17 $\pm$ 7$^{\circ}$ with
respect to the Ni lattice forms \cite{Dahal}.
DFT works \cite{Bertoni2005,Kalibaeva2006,Giovanetti2008,Fuentes2008,
Vanin2010,Mittendorfer2011,Zhao2011} have focused on determining which of
the (1 $\times$ 1) commensurate structures is most favorable.
Most of these works
rely on DFT calculations
which do not include the dispersion interaction leading,
as was already mentioned in Section \ref{metallic-substrates}, to artificial binding
with LDA and no binding at all with PBE.
Thus, as follows from the above mentioned studies, a conclusion
concerning which structure
is the most stable, should be taken with care. Whether DFT calculations
which account for dispersion interaction \cite{Vanin2010,Zhao2011} can make reliable predictions
about the stability of the graphene/Ni(111) system
seems questionable at the moment \cite{Mittendorfer2011} (see also discussion
in Section \ref{sec:vdW}).
Correspondingly, the question
of whether graphene is weakly adsorbed on Ni(111) or not is still under debate.

\subsubsection{Cu(111) and Cu(100)}

\begin{figure}
\begin{centering}
\includegraphics[height=8cm]{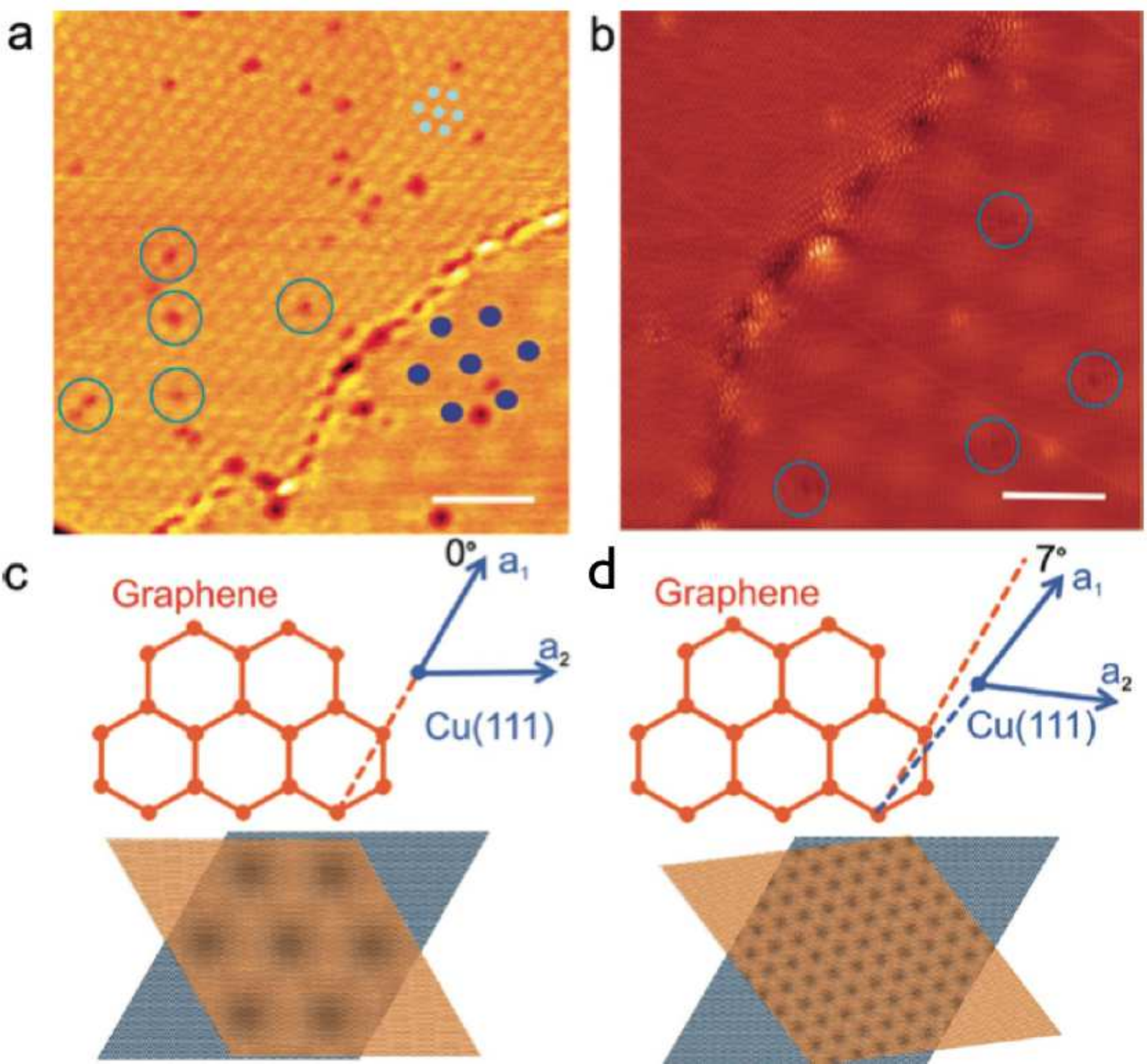}
\par\end{centering}

\caption{STM topography images at the domain boundaries of graphene on Cu(111).
(a) STM image at a domain boundary showing two different moir\'{e} patterns
in the two domains. The periodicity of the moir\'{e} pattern for the upper
left domain is $\sim$2.0 nm, and that for the lower right domain
is around 5.8 nm. Scale bar: 10 nm. (b) Atomic resolution STM image
at a domain boundary, showing the honeycomb structure of graphene.
The moir\'{e} pattern in the upper left domain cannot be observed under
these scanning conditions. The periodicity of the moir\'{e} pattern in
the lower right domain is $\sim$3.0 nm. Scale bar: 4 nm. Blue circles
indicate some of the adsorbates on the graphene surface in (a) and
(b). (c) The most observed ($\sim$30\%) moir\'{e} pattern of graphene
on Cu(111) with around 6.6 nm periodicity. The lattice orientation
of graphene is strictly aligned with that of Cu(111) surface. (d)
Another typical moir\'{e} pattern of graphene on Cu(111) with a periodicity
of $\sim$2.0 nm and the misorientation angle of $\sim$7$^{\circ}$.
{[}Reprinted with permission from \cite{Gao10}. Copyright (2010) American Chemical Society.{]}\label{fig:STM-topography-images-Gao10}}
\end{figure}

As described by Gao \emph{et al. }\cite{Gao10} STM images of graphene
grown on Cu(111) using CVD have been observed to form two different
rotational domains which are rotated with respect to the Cu(111) lattice
by $0^{\circ}$ and $7^{\circ}$, see Fig. \ref{fig:STM-topography-images-Gao10}.
These correspond to a periodicity of $\sim$6.6 nm and $2$ nm, respectively.
Note that other periodicities have been observed as well, albeit in
a lower concentration. On Cu(100) graphene shows moir\'{e} patterns which
are not triangular \cite{Rasool2010}; this is due to the different
symmetries of graphene and Cu(100).

\subsection{SiC\label{sub:buffer-ripples-SiC}}

\begin{figure}
\begin{centering}
\includegraphics[width=10cm]{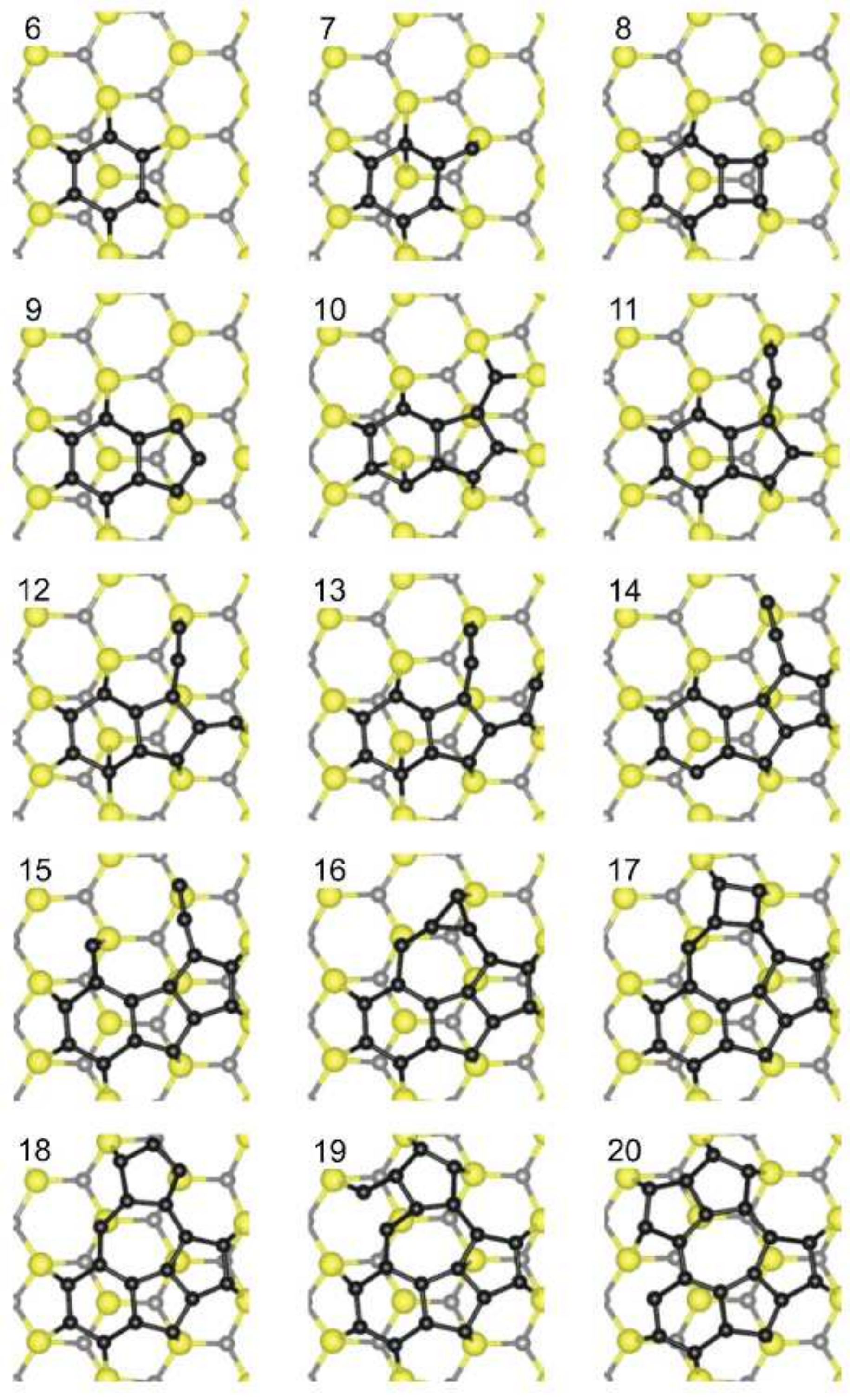}
\par\end{centering}

\caption{Structural models of clusters obtained by adding single C atoms one-by-one
to the initial seed cluster C$_{6}$ shown in the upper left corner.
Numbers correspond to the total number of C atoms (shown as dark small
circles) in the clusters. Si and C atoms of the SiC surface are shown
as large yellow and smaller grey circles, respectively. {[}Reprinted
with permission from \cite{Inoue:2012ih}. Copyright (2012) by the American Physical Society.{]} \label{fig:Inoue-clusters}}
\end{figure}

To investigate the importance of C clusters on the SiC surface at
the nucleation stage of formation of the buffer layer, Inoue\emph{
et al.} \cite{Inoue:2012ih} studied clusters of up to 20 atoms on
the surface employing a DFT method and geometry relaxation. Their
approach was based on starting from a hexagonal cluster C$_{6}$,
see Fig. \ref{fig:Inoue-clusters}(a), and then adding one-by-one
single C atoms to various available sites of the cluster at each step;
the geometry of all clusters built in that way were optimized and
formation energies calculated. This process was continued until a
C$_{20}$ cluster was finally constructed. The formation energy was
calculated using essentially Eq. (\ref{eq:cluster-form-energy}),
with the chemical potential $\mu_{C}$ taken to be that of C in graphene
on the surface; however, the energy per C atom was actually presented.
Analyzing the bond lengths formed by the initial seed C$_{6}$ with
the SiC surface, it was concluded that C atoms of the hexagonal cluster
form covalent bonds with the nearest Si atoms of the surface which
explains why this particular cluster was found to be the most appropriate
choice for the seed. Once C atoms are added, however, hexagonal structures
were found to be less favorable in bigger clusters; instead, penta-heptagonal
clusters were found to have lower formation energies. This was explained
by a greater flexibility of their cages to distortion caused by the
mismatch with the SiC surface, as compared to the hexagonal cages.
These results suggest that non-hexagonal clusters must play important
role in nucleating graphene on SiC. It was also explicitly shown in
\cite{Inoue:2012ih} that the chemical potential of individual C atoms
on the surface becomes greater at larger coverages than the average
chemical potential of C atoms in clusters which indicates that at
some coverage the formation of clusters becomes indeed energetically
favorable, and hence clusters provide a driving force for growth.

\begin{figure}
\begin{centering}
\includegraphics[height=6cm]{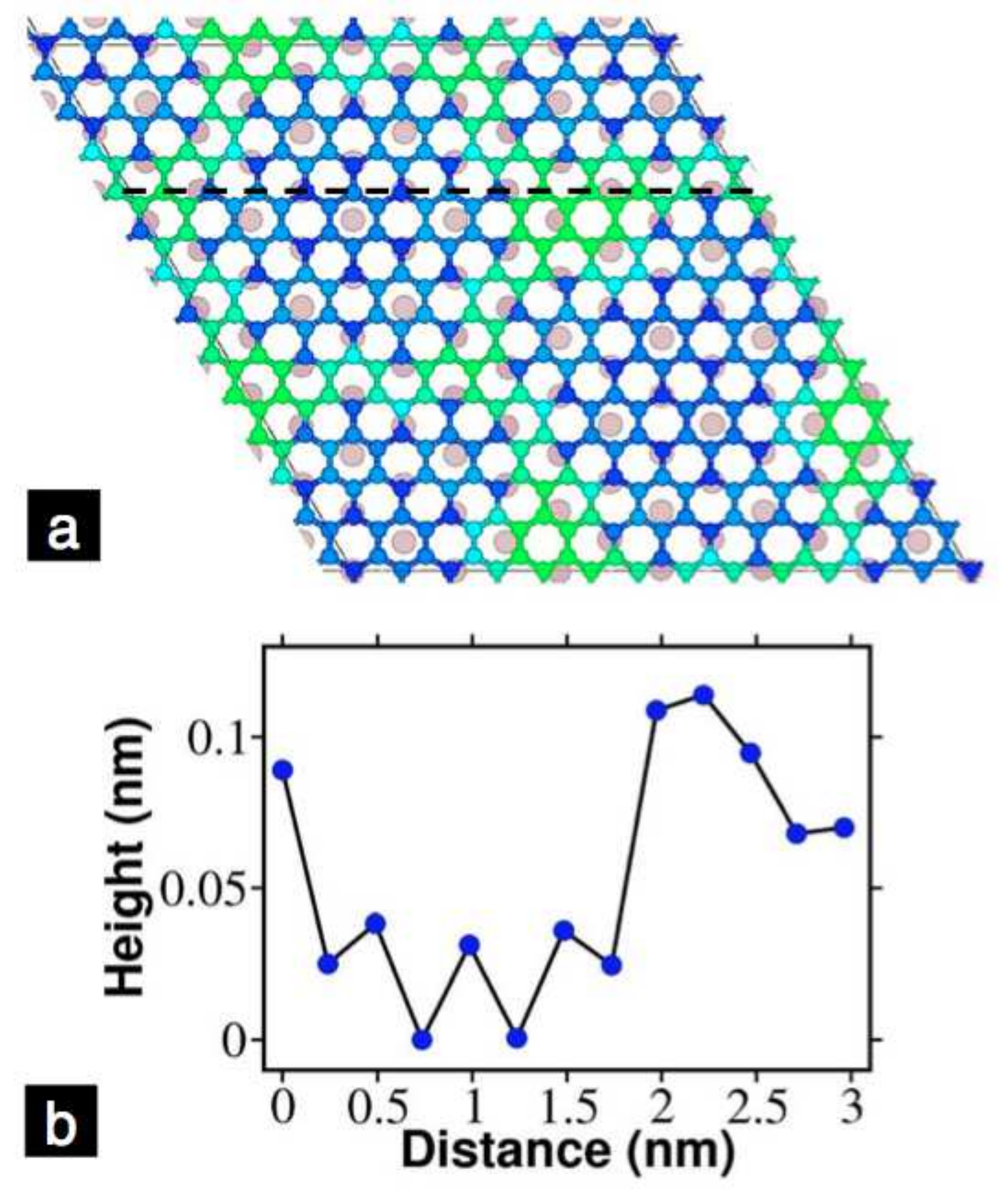}
\par\end{centering}

\caption{(a) Shown are atomic positions in the buffer layer on top of the Si-terminated
SiC surface: gray circles represents the top layer Si atoms of the
last SiC plane. C atoms of the buffer layer are shown with the color
reflecting their height, ranging from blue close the substrate to
green for those which are further away. (b) Heights of the buffer
layer atoms along the black dashed line defined in (a). {[}Reprinted
with permission from \cite{Varchon08}. Copyright (2008) by the American Physical Society.{]} \label{fig:Varchon}}
\end{figure}

\begin{figure}
\begin{centering}
\includegraphics[height=6cm]{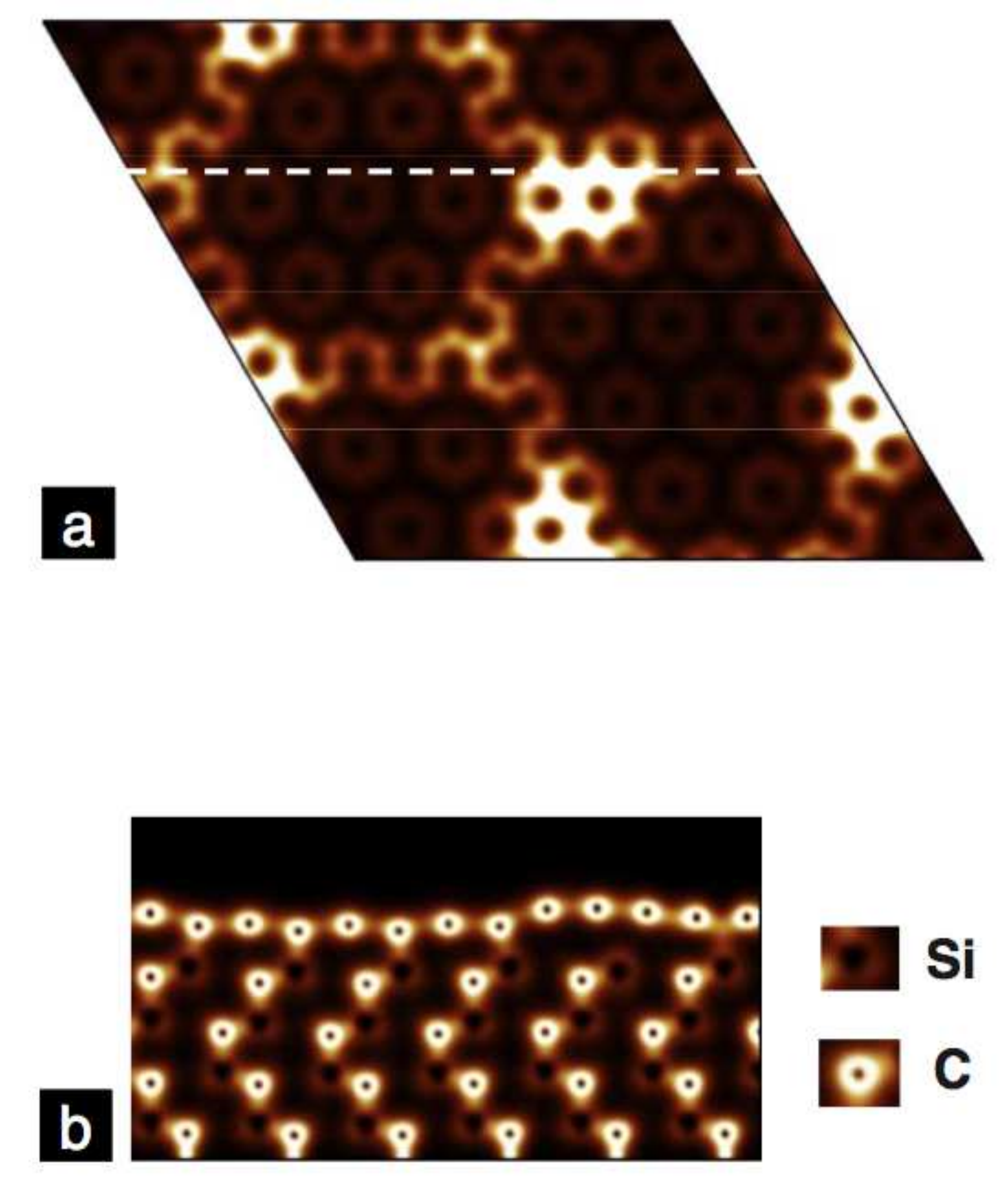}
\par\end{centering}

\caption{Calculated charge density of the graphene layer placed on top of a
C buffer layer resting on the Si-terminated SiC surface. (a) Top and
(b) side views. A cut through the system perpendicular to the surface
and running along the dashed line indicated in (a) is shown in (b).
{[}Reprinted with permission from \cite{Varchon08}. Copyright (2008) by the American Physical Society.{]} \label{fig:varchon-charge-dens}}
\end{figure}

Varchon \emph{et al.} \cite{Varchon08} examined the structure of
both the buffer layer and the graphene layer on top of the buffer
layer for the Si and C terminations of the 6H-SiC surface using DFT.
In the case of the Si-terminated SiC surface, the first (buffer) layer
was found to be strongly corrugated, with a height variation of 1.20
$\textrm{\AA}$. These corrugations are induced by strong covalent bonding between
carbon and silicon at parts of their hexagonal lattices where they
are commensurate with each other, Fig. \ref{fig:Varchon}, in a similar
way to what was found in graphene on transition metal surfaces (Section
\ref{sec:Structure-of-Graphene}). The corrugation of the buffer
layer has a noticeable effect on the second (graphene) layer as well,
where a rather small calculated corrugation of 0.4 $\textrm{\AA}$ (ripples) running
with periodicity of over 19 $\textrm{\AA}$ was found, see Fig. \ref{fig:varchon-charge-dens}.
In the case of the C-terminated SiC surface, much stronger interaction
between the buffer layer and the C atoms underneath was found, and
the ripple effect characteristic for the Si-terminated SiC surface
was not observed. The average height of the buffer layer from the
SiC surface was also calculated using DFT in \cite{Mattausch2007},
but the value they found (2.58 $\textrm{\AA}$) was considerably larger.

As discussed in Section \ref{sec:SiC} the growth of a carbon buffer
layer precedes that of quasi-free standing graphene (QFSG) during
annealing on SiC. DFT calculations of the buffer layer of graphene
on SiC \cite{Kim,Varchon08} show that it does not display a number
of graphene characteristics, most likely due to its strong covalent
bonding to the SiC surface. An example of this would be the clear
difference between the band structures of the buffer layer compared
to free standing graphene demonstrated in \cite{Varchon07,Mattausch2007}.

Other than first principle examinations, the first and second layers
of graphene on SiC have also been examined using empirical potentials
(EP) \cite{Lampin,Tang,Tang08,Jakse}. These studies not only examined
some aspects of the buffer layer on the Si face of SiC; they also
have shown that EP in conjunction with molecular dynamics (MD) simulations
can be used to obtain results that agree quite well with \emph{ab
initio }calculations where available \cite{Lampin}. This is of some
interest because large scale and longer time simulations of growth
mechanisms using EP could be performed that would not be possible
for first principles based methods \cite{Lampin}. We shall now briefly
review a selection of the results obtained using EP.

\textcolor{black}{Lampin }\textcolor{black}{\emph{et al. }}
\textcolor{black}{\cite{Lampin}}
used the many-body empirical potentials EDIP \cite{Lucas} to perform
simulations of a layer of carbon atoms on top of the Si face of SiC.
EDIP potentials were fitted to be able to describe C and Si based
crystals and their defects; they do not account for the dispersion
interaction as is the case for the standard density functionals commonly
used in DFT calculations. After relaxing a number of different starting
configurations they obtained the lowest energy structure, shown in
Fig. \ref{fig:Lampin-relaxation}, which broadly agrees with corresponding
DFT calculations discussed above. The separation between the Si surface
and the lowest carbon atom in graphene was found to be 1.68 $\textrm{\AA}$ with
the graphene sheet itself being strongly corrugated with an amplitude
of 1.50 $\textrm{\AA}$ (1.20 $\textrm{\AA}$ with DFT \cite{Varchon08,Mattausch2007}). Similar
rippled structures of the buffer layer were obtained by Tang
\emph{et al.} {\cite{Tang} who used modified Tersoff potentials
\cite{Tersoff-potentials-1994} and simulated annealing to find the
structure of graphene nanoribbons (GNR) of different widths on top
of a Si-terminated SiC bulk. They reported corrugation of the buffer
layer C atoms of the order of 1 $\textrm{\AA}$ and ripples with wavelength
of around 18 $\textrm{\AA}$. They also find that the ripples remain
periodic up to some temperature (around 1500-2000 K depending on the
GNR width), after which thermal fluctuations dominate the GNR structure.

\begin{figure}
\begin{centering}
\includegraphics[height=3cm]{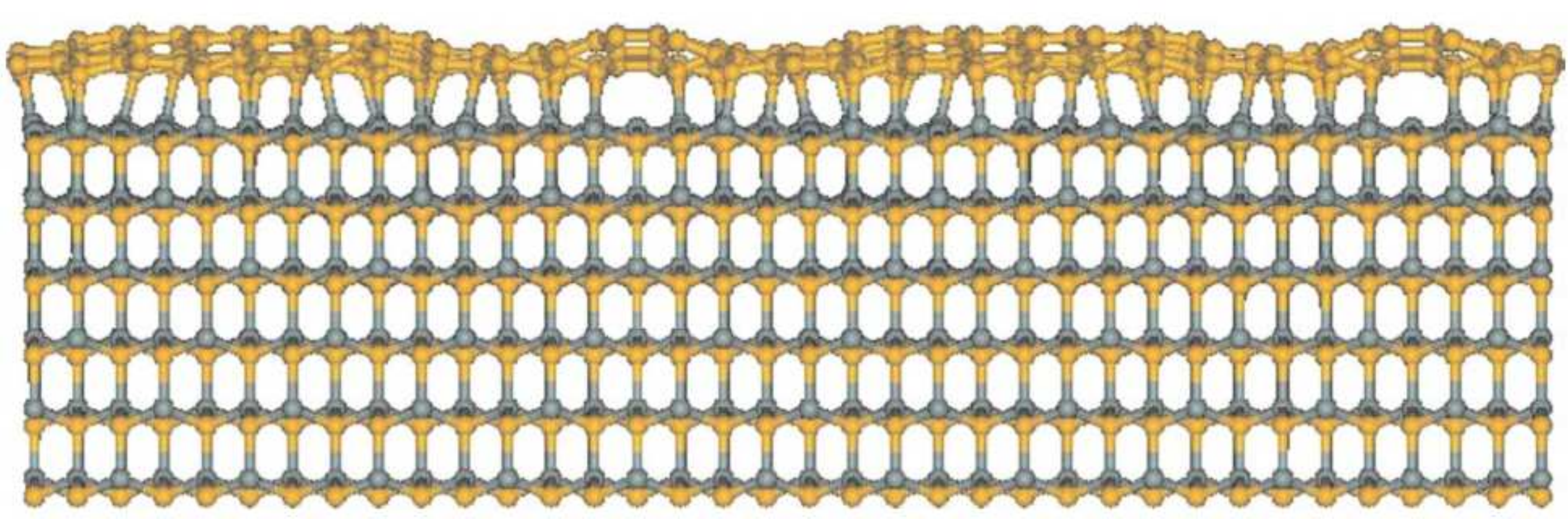}
\par\end{centering}

\caption{Side view of the EDIP relaxed C atoms buffer layer on top of the Si-terminated
SiC surface. {[}Reprinted with permission from \cite{Lampin}. Copyright (2012), AIP Publishing LLC.{]}
\label{fig:Lampin-relaxation}}
\end{figure}

\begin{figure}
\begin{centering}
\includegraphics[height=6cm]{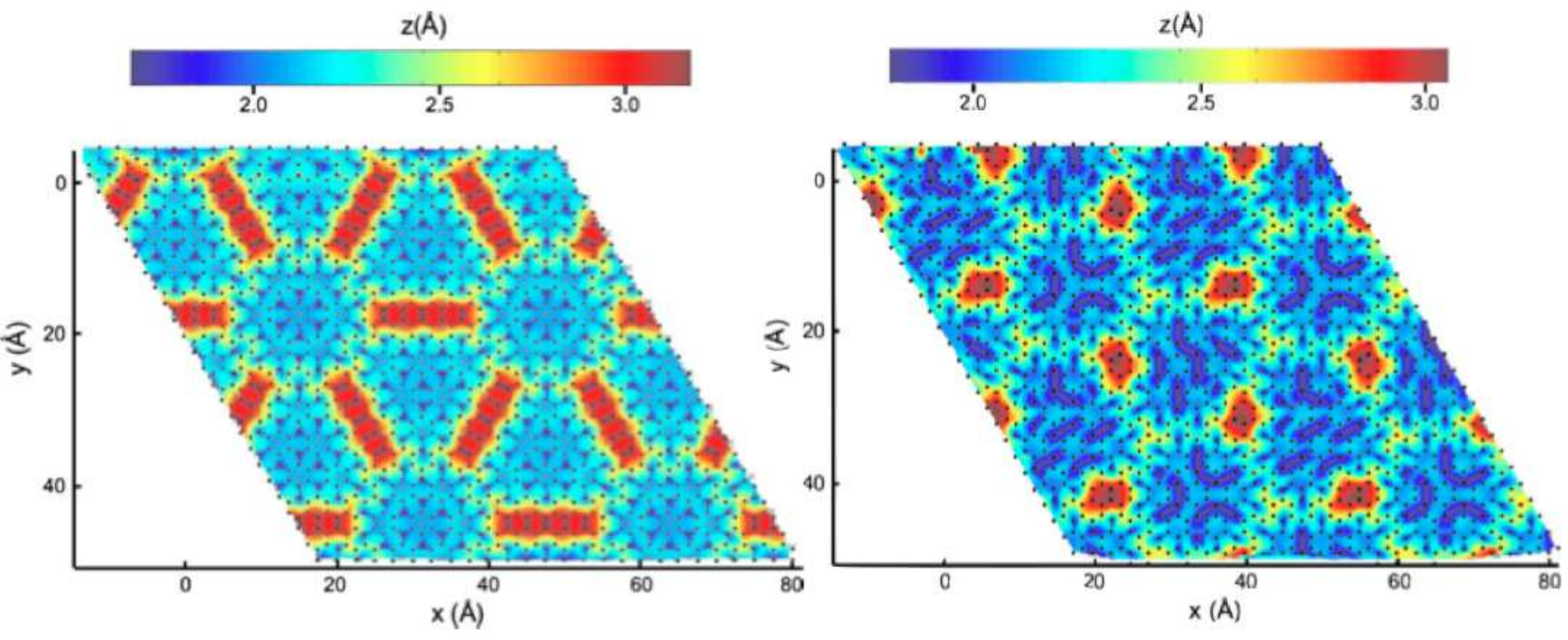}
\par\end{centering}

\caption{Representation of the heights of the carbon atoms in the buffer layer
placed on top of the Si-terminated 6H-SiC surface calculated using
EDIP (left) and DFT (right, from \cite{Varchon08}) shown using different
colors. Blue represents the closest atoms to the surface while red
represents atoms further away from the surface. Black dots show C
atoms in the buffer layer, while light crosses indicate the topmost
Si atoms of SiC. The height $z$ (in $\textrm{\AA}$) corresponds
to the vertical distance between the C atoms in the buffer layer and
a reference plane associated with the topmost Si atoms of SiC prior
to its relaxation. {[}Reprinted with permission from \cite{Lampin}. Copyright (2012), AIP Publishing LLC.{]}
\label{fig:Lampin}}
\end{figure}

The structure that both groups \cite{Tang,Lampin} obtained have clear
similarities with that found using DFT calculations\textcolor{red}{{}
}\cite{Varchon08}, although in \cite{Lampin} there are some differences
indicated in the heights from the surface of a number of carbon atoms
of which the buffer layer comprises, see Fig. \ref{fig:Lampin}. Thus,
the results obtained from EP calculations do agree globally with those
obtained using \emph{ab initio} methods, although there are discrepancies
in some of the details in the obtained geometries. The results also
seem to depend on the particular EP used in each case.


\section{Simulation of graphene growth on metals\label{sec:Simulation-of-graphene}}

As should have become clear from the previous sections, the growth
of graphene is a rather complicated process which happens across large
length and time scales. Below we shall try to reflect on the recent
progress in understanding the complexity of this phenomenon based
on theoretical techniques.

We shall start by discussing phenomenological theories in section
\ref{sub:Rate-equation-analysis}. The usefulness of these approaches
stems from the fact that these may be directly compared to some of
the experimental results, for instance, with time dependent measurements
of the concentration of various carbon species on the surface; moreover,
cumbersome experimental procedures (e.g. temperature ramp) can be
easily incorporated as well. The main drawback of these methods is
in their phenomenological character which lacks atomistic detail.
In addition, spatial information (e.g. distribution of clusters and
adatoms on the surface) is also not contained in these methods.

At the other end of the spectrum are atomistic approaches, mainly
based on density functional theory (DFT) simulations; these will be
covered in section \ref{sub:Atomistic-approaches}. These methods
enable one to uncover various elementary processes responsible for
the growth phenomenon in great detail. These include the formation
energies and mobilities of different carbon species on the given surface,
including terraces and surface defects such as steps, the rate of
attachment (detachment) of the species to (from) clusters and islands,
and the favorable alignment of nuclei, and hence the growth of large
clusters and islands, with respect to the substrate symmetry directions,
etc. Moreover, rates of these elementary processes can be calculated
and hence their importance for the graphene growth established. The
ability to provide a detailed information on the nature of the binding
of the carbon species to the substrate is also a particular strength
of these techniques. However, only relatively small systems can be
considered at this level of theory, and dynamical simulations may
only be run over very short times, many orders of magnitude shorter
than the growth process itself.

However, the atomistic information on the elementary processes and
their rates can be incorporated into kinetic approaches, e.g. based
on Kinetic Monte Carlo (KMC) simulations, which could provide a wealth
of information on growth, such as e.g. the time dependent distribution
of relevant carbon species on the surface, and this kind of information
can be directly compared with experiment. Moreover, KMC rates can
be fed back into the phenomenological rate equations and hence provide
a platform for developing more refined phenomenological models. To
our knowledge, just a few KMC simulations have been done to date on
graphene growth, and we shall discuss them in detail in section \ref{sub:kMC}.

But first we review the fundamental processes that are widely believed
to occur during epitaxial growth of graphene.

\subsection{Rate Equation Analysis of Growth\label{sub:Rate-equation-analysis}}

The detailed mechanism by which graphene growth occurs is not fully
understood; moreover, the processes are also shown to be dependent
on the type of surface and its orientation. From experimental observations
of the graphene coverage using LEEM it was suggested that the rate
of graphene growth on Ru(0001) \cite{Loginova08} and Ir(111) \cite{Loginova09}
surfaces has a non-linear dependence on the C adatom concentration.
In the experiment it was noticed that as C adatoms are deposited onto
the surface, graphene islands start to nucleate at a critical concentration
$c^{nucl}$ at which point the growth of islands leads to a sudden
drop in the concentration of carbon adatoms, see Fig. \ref{fig:Rate-equations}.
When the C deposition is turned off the C adatom concentration, $c(t)$,
decreases further until it reaches a concentration $c^{eq}=c(t=\infty)$,
where an equilibrium is established between the two-dimensional ``gas''
of carbon adatoms on the surface and the formed graphene islands.
It was assumed that $c^{eq}$ has the Arrhenius form,

\begin{equation}
c^{eq}\propto e^{-E_{form}/k_{B}T}\;,\label{eq:(1)}
\end{equation}
where the formation energy $E_{form}$ can be found from its measured
temperature dependence. Here $k_{B}$ is Boltzmann's constant and
$T$ is the temperature. The change in the excess concentration (proportional
to adatom supersaturation) $c(t)-c^{eq}$, over time was determined
by observing the graphene coverage from LEEM images. The growth velocity
$v$ was also found; this is defined via

\begin{equation}
v=\frac{1}{P}\frac{dA}{dt}\label{eq:(2)}
\end{equation}
where $P$ is the island perimeter and $dA/dt$ is the rate of change
of the island area. Note that this growth velocity can only be considered
as an order-of-magnitude estimate, as this expression is only valid
for circular islands. By assuming that the velocity is proportional
to the supersaturation, and that the barrier for a single C atom to
attach to an island is much larger than for an attachment of a cluster
containing $n$ atoms, the authors obtained an equation for the dependence
of the growth on the C adatom concentration:

\begin{equation}
v=m_{n}(c_{n}-c^{eq})=m_{n}e^{-E_{n}/k_{B}T}\left[\left(\frac{c}{c^{eq}}\right)^{n}-1\right]=B\left[\left(\frac{c}{c^{eq}}\right)^{n}-1\right]\;,\label{eq:rate}
\end{equation}
where

\begin{equation}
c_{n}=e^{n\Delta\mu/k_{B}T-E_{n}/k_{B}T}=\left(\frac{c}{c^{eq}}\right)^{n}e^{-E_{n}/k_{B}T}\label{eq:(4)}
\end{equation}
is the concentration of clusters containing $n$ carbon atoms, and
$\Delta\mu=k_{B}T\ln(c/c^{eq})$ is the excess chemical potential
of the adsorbed C atoms. $E_{n}$ in this expression is an excess
free energy of cluster formation, denoted $F_{{\rm ex}}$ in Section
\ref{sub:Nucleation-Theory}. By fitting the experimental data to
the form of equation (\ref{eq:rate}), $n$ was estimated as 4.8$\pm$0.5.

The value found is atypical of normal crystal growth where $n=1$
(corresponding to growth via attachments of individual atoms). This
suggests that intermediate states such as C clusters are indeed involved,
which led to the conclusion that in this case the growth of graphene
proceeds by the addition of 5 atom clusters, rather than C monomers,
to graphene flakes.

Motivated by this result, Zangwill and Vvedensky formulated rate equations
for the epitaxial growth of graphene on metal surfaces \cite{Vvedensky}.
Based on the experiments by Loginova \emph{et al}. \cite{Loginova08,Loginova09}
the flux of the carbon atoms arriving at the surface, $F$, and the
diffusion related constant for their motion across it, $D$, were
used in rate equations for the homogeneous densities of carbon monomers
$c(t)$, five atom clusters $c_{5}(t)$ and graphene islands $G$.
It is assumed that $n=5$ monomers collide to form a 5 atom cluster
and then these clusters move across the surface in correspondence
with a diffusion related constant $D'$. Then it is assumed that $j$
of these clusters collide to form an island of $nj$ carbon atoms.
The value of $j$ is decided from the temperature dependence of the
carbon adatom concentration, which suggests that $j=6$. Using these
constants the authors determined the rate equations:

\begin{equation}
\frac{dc}{dt}=F-nDc^{n}+nKc_{n}-DcG+K'G\;,
\end{equation}

\begin{equation}
\frac{dc_{n}}{dt}=Dc^{n}-Kc_{n}-D'c_{n}G-jD'c_{n}^{j}\;,
\end{equation}

\begin{equation}
\frac{dG}{dt}=D'c_{n}^{j}\;.
\end{equation}
Here $K$ and $K^{\prime}$ are the cluster dissociation rate and
the detachment rate of adatoms from the islands, respectively, while
$D$ and $D^{\prime}$ are the corresponding diffusion related constants;
$c_{n}$ is the concentration of $n-$atom C clusters. $D$, $D^{\prime}$,
$K$ and $K^{\prime}$ each have the Arrhenius from of $\nu e{}^{-E/k_{B}T}$,
where $E$ is the energy barrier of the process and $\nu=2k_{B}T/h$
is a prefactor assumed to be the same for each of the rate parameters
($h$ is Planck's constant). $D$, $D^{\prime}$, $K$ and $K^{\prime}$
were determined by optimization of the energy barriers while taking
into account some natural constraints on their values.

Using $n=5$ and $j=6$ the time dependence of $c$, $c_{n}$ and
$G$ were calculated. The change in adatom concentration $c(t)$ over
time determined from the rate equations, and shown in Fig. \ref{fig:Rate-equations},
compares well with the experimental results; however there are discrepancies.
The rate equations underestimate the experimental value of the adatom
density when graphene nucleation occurs and overestimate the steady
adatom concentration after the onset of nucleation. This is thought
to be due to the assumption that the densities are homogeneous and
there is no spatial information about the locations of the clusters
and islands on the surface built into the rate equations. 
In a recent study \cite{Posthuma14} a refined rate equation theory was presented
which correctly accounts for the temperature dependence of the graphene island density. It was also shown that the values
of $j$ between 5 and 7 are almost equally likely.

\begin{figure}
\centering{}\includegraphics[height=5cm]{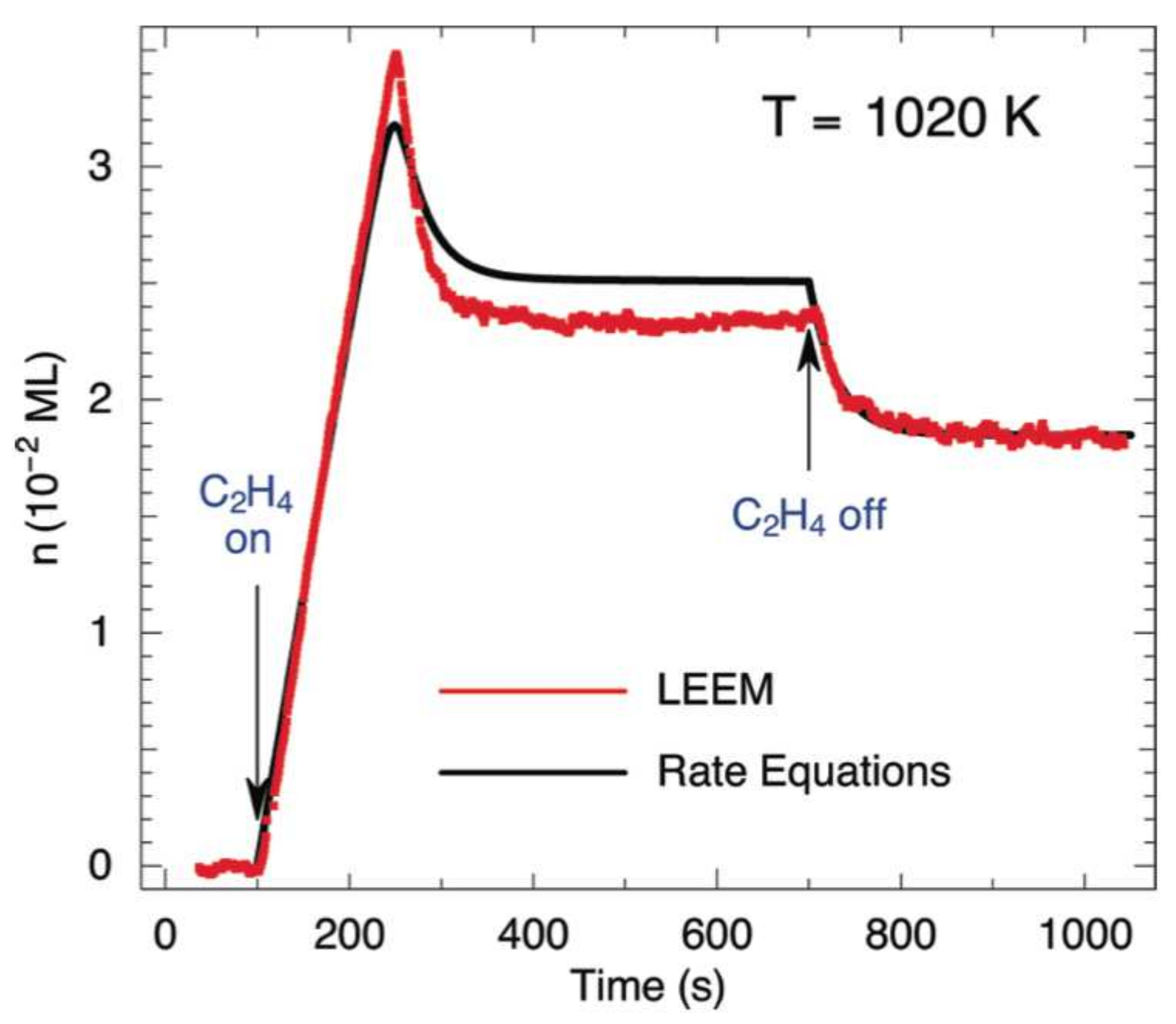}\caption{The change in the C adatom density over time as calculated from the
rate equations (black line) and determined from LEEM data (red line).
{[}Reprinted with permission from \cite{Vvedensky}. Copyright (2011) American Chemical Society.{]} \label{fig:Rate-equations}}
\end{figure}

In order to provide an inhomogeneous description of the attachment
processes that accounts for the growth, multi-scale KMC approaches
are necessary (section \ref{sub:kMC}). These would be based on rates
determined from first principles calculations (section \ref{sub:Early-Stages-of}
and \ref{sub:Atomistic}). But before we enquire into these, let us
first discuss available results on atomistic modeling of the early
stages of growth which shed light on the relevant nucleation processes
preceding the actual growth of graphene islands.

\subsection{Atomistic approaches from first principles\label{sub:Atomistic-approaches}}

\subsubsection{Dehydrogenation \label{sec:Carbon-Feedstock-Theory-1}}

As was discussed in Section \ref{sec:Carbon-Feedstock-Experimental},
initial stages of graphene growth rely on the dehydrogenation of hydrocarbon
molecules. H atoms presumably detach from the molecules with the transition
metal surface acting as catalyst to facilitate the reaction. The adsorbed
atomic hydrogens would not desorb from the surface due to strong binding
to it. Instead, it is most likely that they would diffuse across the
surface until they collide with another adsorbed H atom to form the
H$_{2}$ molecule which then is easily desorbed. However, it is also
clear that the dehydrogenation must be associated not with one but
with many competing processes, and that various intermediate species
C$_{x}$H$_{y}$ must be present on the surface until the dehydrogenation
and desorption of hydrogen is complete and graphene nucleation starts.
Although the existing literature related to dehydrogenation of hydrocarbons
in the early stages of CVD (TPG) processes is rather scarce, DFT simulations
which have been done  are reviewed below in this Section, do illustrate
the complexity of the processes leading to graphene nucleation.

We shall start with the simplest hydrocarbon which is methane. Its
dehydrogenation on the Cu(111) surface was considered in \cite{gajewski:064707}
using plane wave spin-polarized DFT calculations in a supercell geometry.
All possible adsorption sites on the surface for both the removed
H atom and the CH$_{x}$ species ($x$ = 0 - 4) were considered to
find the minimum energy pathway for dehydrogenation. The most favorable
sites were used as the initial and final configurations in NEB calculations
of the dehydrogenation process. The saddle points along the minimum
energy path (MEP) were precisely located, which was confirmed by the
existence of a single imaginary vibrational frequency. The energy
profile of the chain of dehydrogenation reactions CH$_{4}$ $\rightarrow$
CH$_{3}+$H $\rightarrow$ CH$_{2}+$2H $\rightarrow$ CH$+$3H $\rightarrow$
C+4H is shown in Fig. \ref{fig:gajewski-energy-profile}. Also in
the Figure one can see the relative energies of the initial, final
and all intermediate configurations and the corresponding energy barriers.
The dehydrogenation reactions are all endothermic with the calculated
energy barriers lying between 0.94 and 1.84 eV. The energy of the
final state corresponding to the adsorbed C and four H atoms is 3.2
eV less favorable than the gas-phase CH$_{4}$ molecule and the Cu(111)
surface, so at first glance the complete dehydrogenation of the methane
molecule seems to be highly unfavorable on this surface. However,
when the final state was compared with the energy of an adsorbed C$_{2}$
molecule, a reduction of 1.5 eV was found, see Fig. \ref{fig:gajewski-energy-profile}.
This result led the authors to argue that forming carbon dimers on
the surface after complete methane dehydrogenation must be an essential
step in graphene nucleation; moreover, it was also speculated that
formation of larger C$_{n}$ ($n>2$) carbon clusters on the surface
would result in even larger energy gains and hence would be essential
in understanding the kinetics of graphene growth.

\begin{figure}
\centering{}\includegraphics[height=6cm]{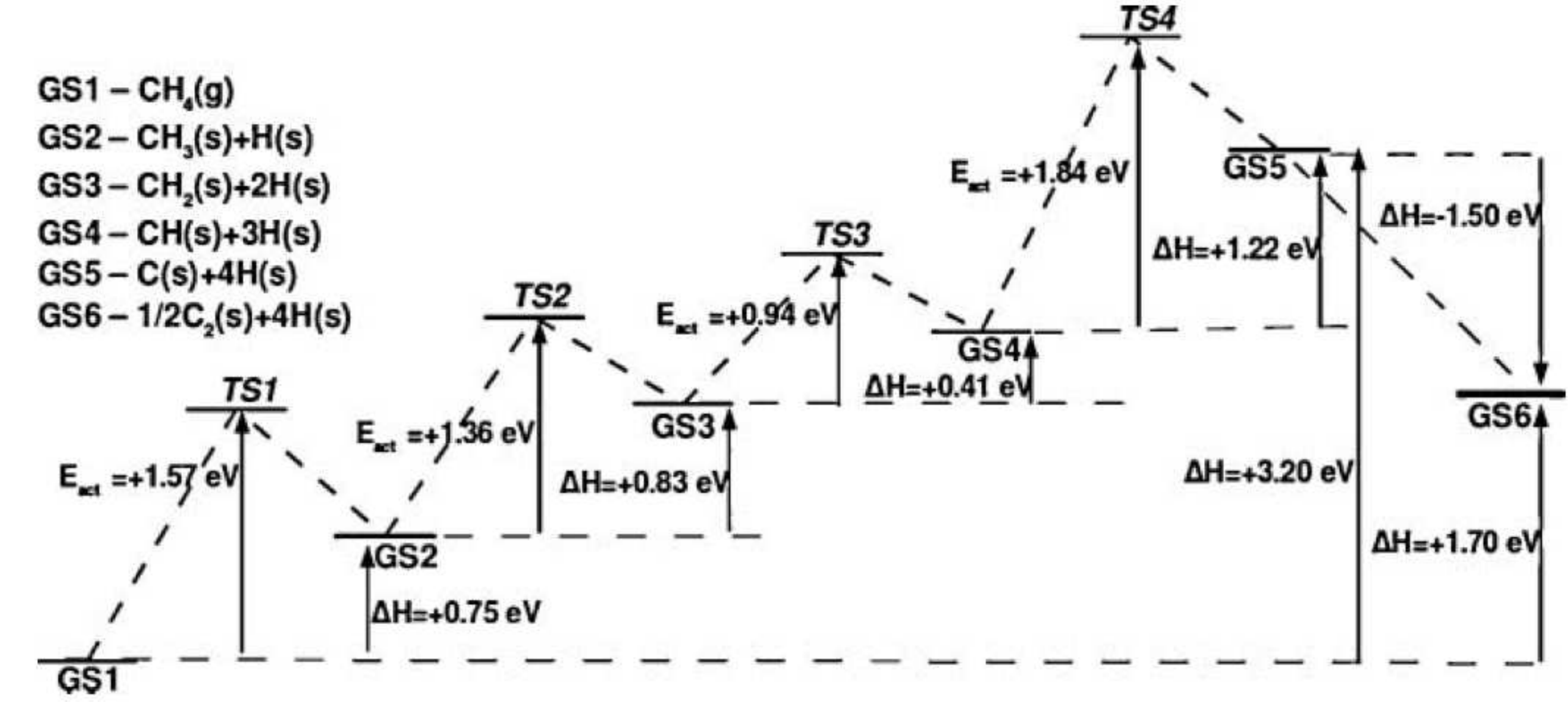}\caption{The energy profile of the complete dehydrogenation process of methane
on the Cu(111) surface with subsequent formation of C$_{2}$ species
(C dimers). {[}Reprinted with permission from \cite{gajewski:064707}. Copyright (2011), AIP Publishing LLC.{]}
\label{fig:gajewski-energy-profile}}
\end{figure}

The same system was independently considered also in \cite{Zhangdoi:10.1021/jp2006827}.
The initial, final and transition states for each stage of the dehydrogenation
reaction are shown in Fig. \ref{fig:Zhang_dehydro}(a), and the complete
energy profile in Fig. \ref{fig:Zhang-energy-profile} (black curves).
As can be seen by comparing this energy profile with the one shown
in Fig. \ref{fig:gajewski-energy-profile}, the general picture is
basically confirmed: the process of dehydrogenation alone leads to
a highly energetically unfavorable state; each H dissociation step
is endothermic and has an energy barrier between 1.0-2.0 eV, and hence
complete dehydrogenation of methane is unfavorable on the Cu(111)
surface. It was concluded from these calculations, and quite rightly,
that various C$_{x}$H$_{y}$ species must be present on the surface
at the same time: since dehydrogenation is not favored, coalescence
reactions involving various species must be favored, leading to a
rich variety of carbon-containing species on the surface. Indeed,
it was found that the reaction CH+CH $\rightarrow$ C$_{2}$H$_{2}$
is exothermic by 1.94 eV with the energy barrier of only 0.3 eV. The
attachment of additional CH in the reaction C$_{2}$H$_{2}$+CH $\rightarrow$
C$_{3}$H$_{3}$, although leading to 0.98 eV energy gain (i.e. is
also exothermic), required a higher energy barrier of 1.1 eV. The
authors also considered the Cu(100) surface, see again Fig. \ref{fig:Zhang-energy-profile}
(red curves), and arrived at the same conclusions. Complete dehydrogenation
of ethene was studied as well in \cite{Zhangdoi:10.1021/jp2006827}
leading to qualitatively similar conclusions: during the four dehydrogenation
steps required to fully decompose the molecule into a C$_{2}$ (dimer)
and four H atoms, the system energy increases by 0.48, 0.09, 0.67,
and 0.62 eV, respectively.

\begin{figure}
\begin{centering}
\includegraphics[height=6cm]{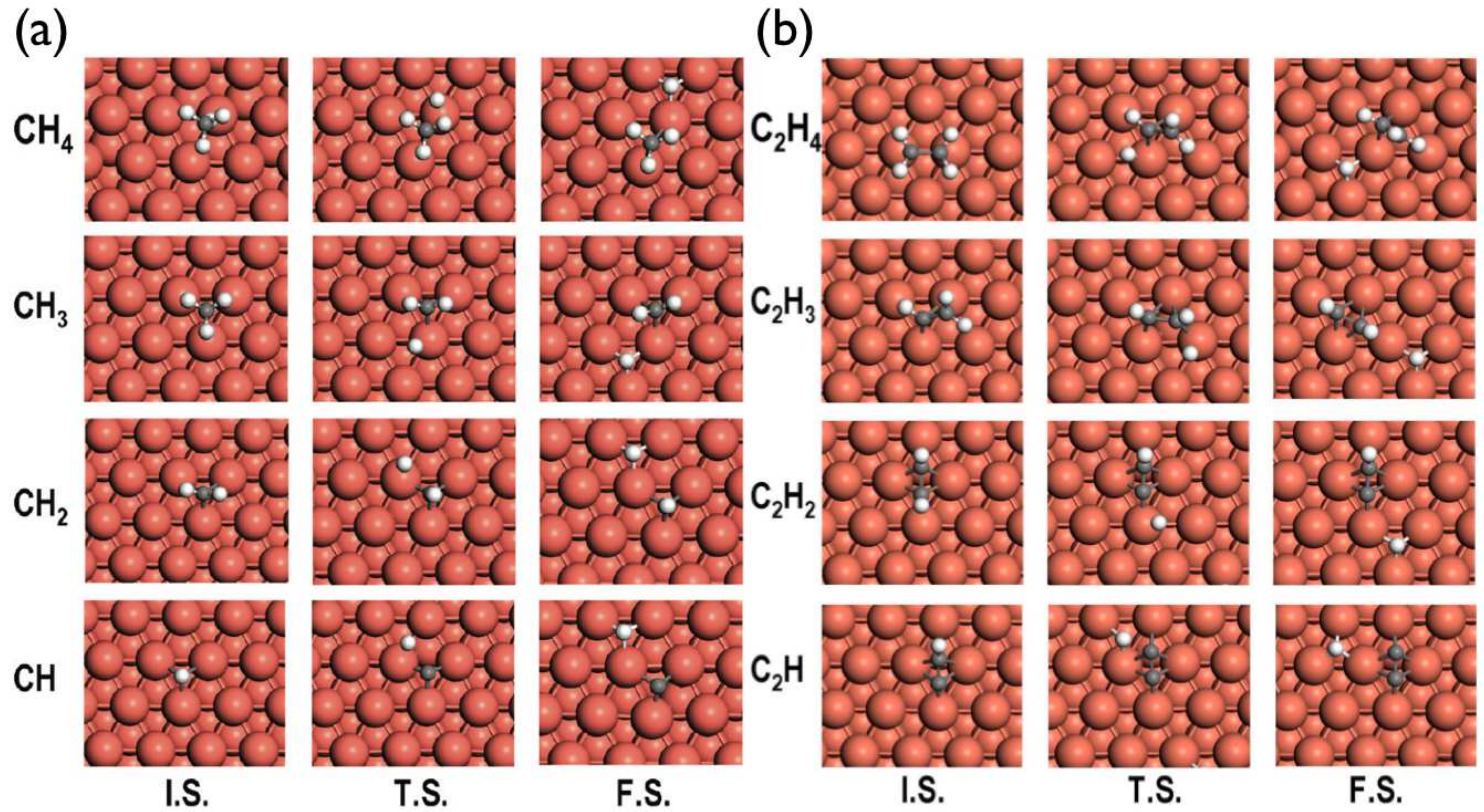}
\par\end{centering}

\centering{}\caption{The structures of the initial (I.S.), transition (T.S.) and final
(F.S.) states corresponding to the processes leading to complete dehydrogenation
of (a) methane and (b) ethene on the Cu(111) surface. The red, white
and grey spheres represent Cu, H and C atoms, respectively. {[}Reprinted
with permission from \cite{Zhangdoi:10.1021/jp2006827}. Copyright (2011) American Chemical Society.{]}\label{fig:Zhang_dehydro}}
\end{figure}

\begin{figure}
\centering{}\includegraphics[height=5cm]{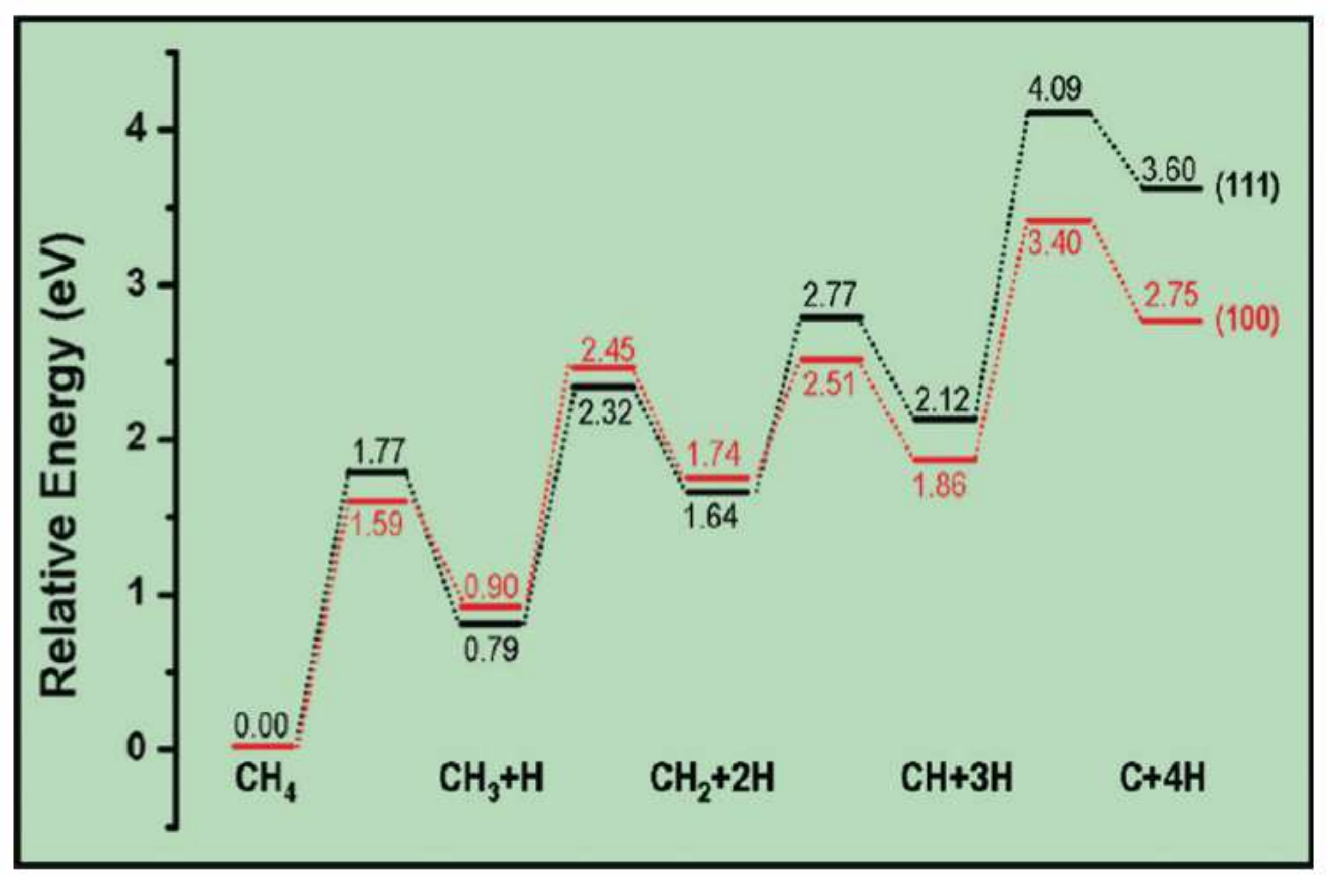}\caption{The energy profile of the complete dehydrogenation process of methane
on the Cu(111) (black) and Cu(100) (red) surfaces. {[}Reprinted with
permission from \cite{Zhangdoi:10.1021/jp2006827}. Copyright (2011) American Chemical Society.{]}\label{fig:Zhang-energy-profile}}
\end{figure}

We have seen above that direct dehydrogenation, at least on the Cu(111)
surface, may be highly unfavorable, and that other reactions such
as hydrogenation, may take place leading to a high likelihood of having
a complicated mixture of various carbon-containing species on the
surface during nucleation and growth. In fact, one can say with certainty
that the actual kinetics of these reactions is quite complicated,
and the models considered so far must be highly oversimplified. This
conclusion can be made from the analysis of existing literature \cite{Aleksandrov2012187,Li:2010cm,Zhao:2010gq,Chen:2010ez}
on the initial stages (Section \ref{sec:Carbon-Feedstock-Experimental})
of ethene decomposition into ethylidyne CH$_{3}$C (or C$_{2}$H$_{3}$).
The complete scheme of all reaction paths is shown in Fig. \ref{fig:Aleksandrov-different-reaction}.
The conversion of ethene into ethylidyne was suggested to occur by
several different and competing mechanisms, composed of a sequence
of hydrogenation, dehydrogenation and (possibly) isomerization steps
in various orders. For all elementary reactions the corresponding
energy barriers have been calculated using DFT \cite{Aleksandrov2012187,Li:2010cm,Zhao:2010gq,Chen:2010ez}
which gave the complete set of all required transition rates.

\begin{figure}
\centering{}\includegraphics[height=8cm]{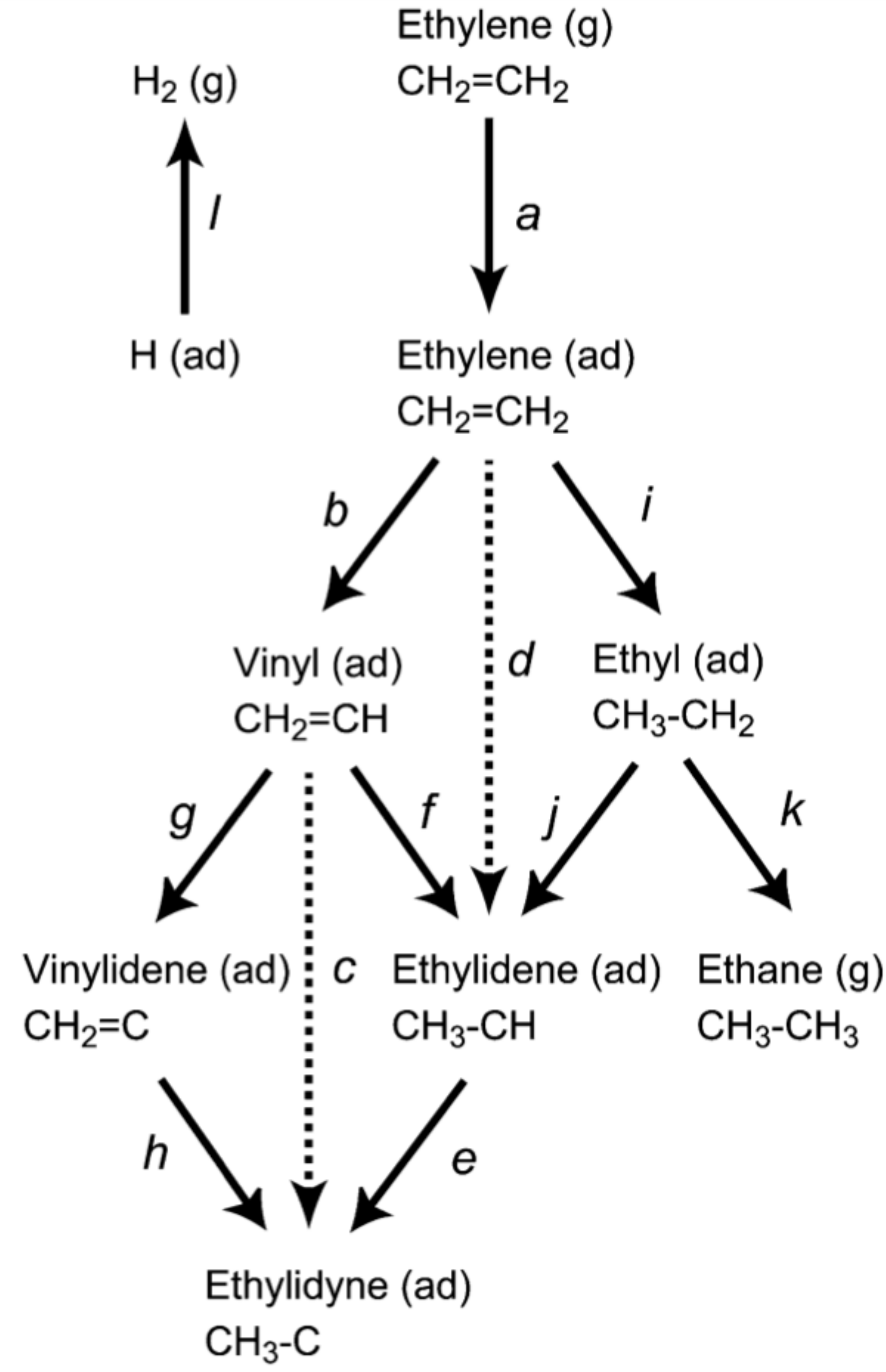}\caption{The different reaction mechanisms of ethene conversion to ethylidyne
C$_{2}$H$_{3}$ on the Pt(111) and Pd(111) surfaces. Arrows pointing
to the left and right indicate dehydrogenation and hydrogenation steps,
respectively, while the dashed line arrows show possible isomerization
routes. Down and up arrows represent adsorption and desorption, respectively.
{[}Reprinted from \cite{Aleksandrov2012187}, with permission from Elsevier.{]}
\label{fig:Aleksandrov-different-reaction}}
\end{figure}

However, as is stressed in \cite{Aleksandrov2012187}, one cannot
assess the preferred reaction path just on the basis of energy barriers
(rates) as other factors such as availability of intermediate species
(such as hydrogen) should be taken into account as well. This has
been done in \cite{Aleksandrov2012187} using a KMC approach. It is
based on considering ethene molecules on the surface with subsequent
conversion reactions taking place according to the scheme of Fig.
\ref{fig:Aleksandrov-different-reaction}. In the case of hydrogenation
reactions the reactant should be located in neighboring positions.
Diffusion of molecules on the surface between reaction events was
considered approximately by assuming a force field describing inter-molecular
interactions and a local equilibrium which was accomplished by moving
the molecules using a set of MC steps.

Simulations of the conversion of ethene to ethylidyne on Pt(111),
Pd(111) \cite{Aleksandrov2012187,Zhao:2010gq} and Rh(111) \cite{Li:2010cm}
surfaces demonstrated that the dominant mechanism is through vinyl
and vinylidene, i.e. via a sequence of direct dehydrogenation reactions.
For the Pt(111) surface the rate limiting step was the hydrogenation
of the vinylidene to C$_{2}$H$_{3}$. Despite hydrogenation reactions
having lower energy barriers than dehydrogenation reactions, the lack
of atomic H adsorbed on the surface results in a low probability for
this process to occur. On the Pt(111) surface it was found that the
adsorbed H from dehydrogenation of ethene is mainly lost through H$_{2}$
production. In the case of Pd(111) the rate limiting process is the
initial dehydrogenation of ethene. Due to these differences in the
limiting process, the conversion of ethene into C$_{2}$H$_{3}$ was
calculated to begin at a temperature of around 230 K on Pt(111), but
is not complete until reaching 325 K.

It is clear that full consideration of the conversion of a given hydrocarbon
feedstock into graphene, as suggested by TP-XPS experiments described
in Section \ref{sec:Carbon-Feedstock-Experimental}, would contain
many more reaction paths and intermediates, and this would require
a rather sophisticated and detailed modeling. This kind of modeling
must not only rely on the calculations of energy barriers, but must
also be based on the time evolution of all species involved which
must take into account their diffusion across the surface. This kind
of modeling can only be achieved using KMC type of approaches which
are based on the spatial distribution of reactants and products over
the surface, the stochastic nature of their diffusion and reactions
between them requiring  surmounting  potential energy barriers.
Such future modeling should also take into account a finite probability
of C-C bond scission which would be possible at the high temperatures
at which graphene is normally grown using TPG or CVD methods. As a
result, one would be able to consider the actual distribution of carbon
clusters C$_{n}$ with $n\geq1$ on the surface as a function of the
given feedstock used in the given experiment.

\subsubsection{Carbon monomers and dimers on metal surfaces }

\begin{figure}
\begin{centering}
\includegraphics[scale=0.5]{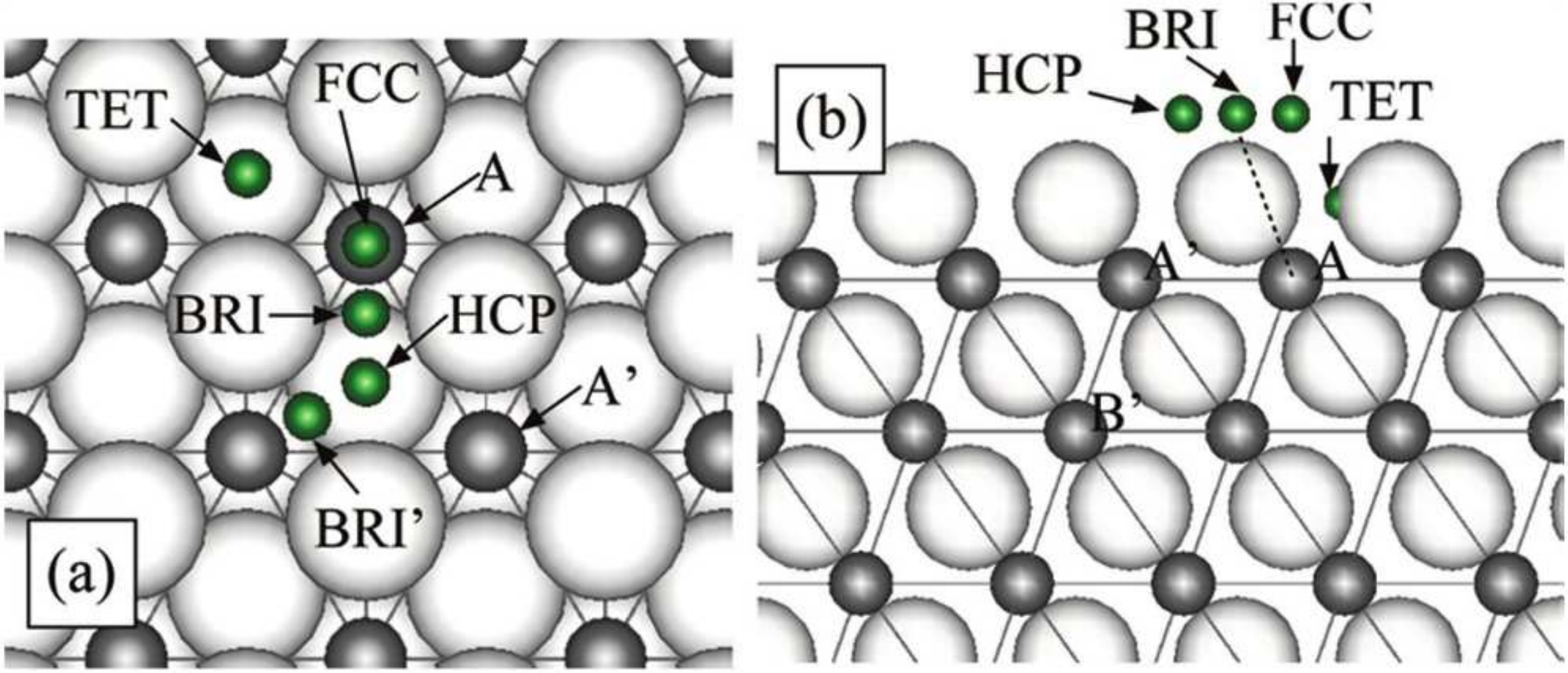}
\par\end{centering}

\caption{\textcolor{black}{Various adsorption sites on the Cu(111) surface:
(a) top and (b) side views. {[}Reprinte}d with permission from \cite{Riikonen}. Copyright (2012) American Chemical Society.{]}
\label{fig:Riikonen-absorption}}
\end{figure}

Before discussing stability and other properties of carbon clusters
of different sizes adsorbed on metal surfaces, which is obviously
related to the first stages of graphene growth, we shall first review
theoretical studies in which monomers and dimers at different sites
on the metal surfaces were investigated \cite{Chen,Riikonen,Saadi}.
This may give us some essential information regarding the nucleation
of graphene on a surface.

Recently Riikonen \emph{et al. }\cite{Riikonen}\emph{ }examined the
adsorption energy of monomers and dimers at different lattice sites,
both on and below the surface of Cu(111), see Fig. \ref{fig:Riikonen-absorption}.
They also calculated the diffusion barriers of monomers between different
equilibrium surface and bulk sites. Before discussing a number of
interesting conclusions regarding the initial stages of nucleation
on Cu(111) made in their study, it is interesting to mention an important
technical point. The authors noted that energies may depend strongly
on the simulation cell size: for instance, the C atom migration barrier
obtained using a larger supercell was found to be almost 0.5 eV smaller
than the previously reported value obtained using a smaller sized
cell.

The most stable position found in\emph{ }\cite{Riikonen} for a carbon
monomer on Cu is the bulk interstitial site, referred to as A and
A$^{\prime}$ in Fig. \ref{fig:Riikonen-absorption}. The surface
sites labelled BRI, FCC and HCP are found to be metastable with respect
to adsorption of a single C atom as it would quickly diffuse (with
the barrier of the order of 0.1 eV) from these into either the A or
A$^{\prime}$ site. The diffusion deeper into the Cu bulk (A$^{\prime}\rightarrow$B$^{\prime}$)
was found to require the crossing of an energy barrier of at least
1.5 eV in height. The formation of dimers from two carbon atoms lying
on several nearby surface sites was found to require no activation
energy, i.e. dimerisation is spontaneous. Further, the diffusion barrier
they calculated for a dimer was found to be very small (0.27 eV).
Thus, the energy required to dissociate a dimer is found to be actually
larger than the diffusion barrier between different adsorption sites.
This suggests that dimers are a persistent species and could possibly
dominate the nucleation of larger carbon structures that eventually
form graphene.

In a similar manner to the above study, Chen \emph{et al. }\cite{Chen}\emph{
}investigated the stability of monomers at different sites on the
surfaces of Ru(0001), Ir(111) and Cu(111). The most stable sites on
Ru and Ir were both found to be the HCP sites, with adsorption energies
of -7.66 and -7.44 eV, respectively. In agreement with the results
of Riikonen \emph{et al.} \cite{Riikonen}, the most stable absorption
site on Cu(111) was found to be the subsurface octahedral (A and A$^{\prime}$
in the notation of Riikonen \emph{et al.}. On both Ir and Ru similar
values were obtained for the adsorption energy at step edges and it
was also found that while the barrier for dimer formation is relatively
high, 1.37 and 1.49 eV respectively, on Cu(111) the barrier for dimerization
is much lower, 0.32 eV. The authors also found that the energy required
to form a dimer is greatly reduced on Ru and Ir when dimerization
occurs at a step edge. They further suggest that these results could
possibly explain the observations of heterogeneous nucleation at step
edges on Ir and Ru. Since the formation energy of dimers on the Cu(111)
surface is so low, it was also suggested that nucleation on weakly
interacting metals (like Cu) should occur homogeneously. On metals
with a stronger C-metal bond the above results for dimer formation
at step edges would mean that nucleation is more likely to occur at
steps, i.e. heterogeneously. However, as noted by Wang \emph{et al.
}\cite{Wang}\emph{,} while these results would predict that nucleation
on Rh(111) would occur predominantly at steps, this was found not
to be the case. Further, islands with different morphologies growing
on Cu(111) have been found to exist predominantly at step edges and
other defect sites \cite{Nie}.

\begin{figure}
\begin{centering}
\includegraphics[scale=0.5]{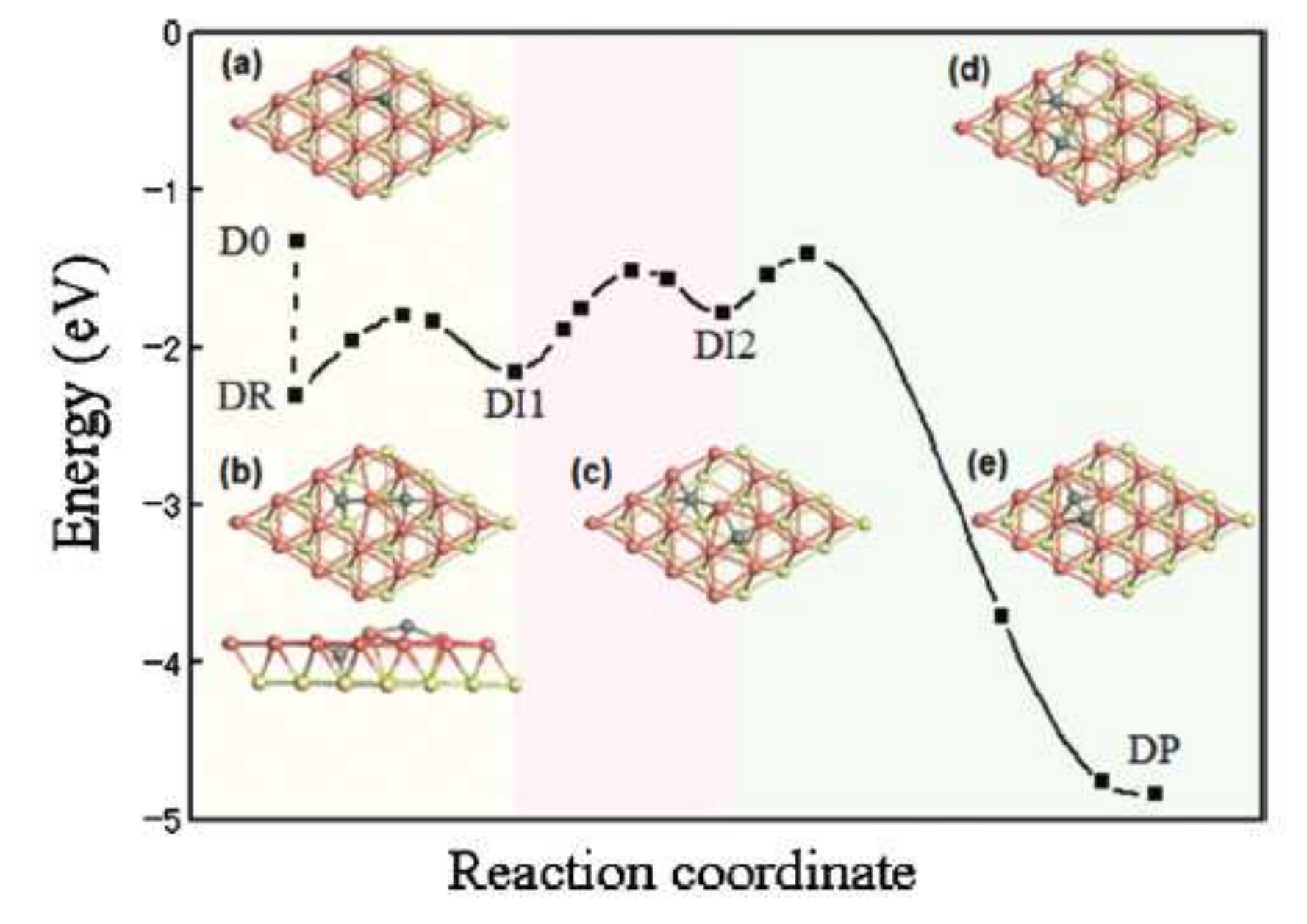}
\par\end{centering}

\caption{\textcolor{black}{Calculated dimerization process path (a)-(e) showing
formation of a dimer (e) on the Cu(111) surface from two monomers
sharing the same Cu atom (a). Several intermediate states were found
along the minimum energy path. {[}Reprinte}d with permission from
\cite{Wu2010}. Copyright (2010), AIP Publishing LLC.{]} \label{fig:Wu-dimerisation}}
\end{figure}

A study by Wu \emph{et al. }\cite{Wu2010}\emph{ }used both DFT and
MD to investigate steps in the reaction sequence describing dimer
formation that had not previously been considered. Their work is based
on the following observation: while it is indeed true that spontaneous
dimerization occurs when two atoms are placed on nearest neighbor
sites, there are intermediate states that must occur before monomers
occupy these sites. These states are derived by considering the beginning
of the dimerization process with two monomers occupying next nearest
neighbor sites, which share a common Cu atom. Relaxation of this state
leads to the formation of novel structures shown in Fig. \ref{fig:Wu-dimerisation}(b)-(d)
which manifest themselves on the potential energy surface as metastable
minima. The change in the activation barrier that occurs by considering
these intermediate states in dimerization and other reaction sequences
could be important in calculating the rates of processes occurring
on the surface.

\subsubsection{Early stages of growth: small C clusters\label{sub:Early-Stages-of}}

The phenomenological model described in Section \ref{sub:Rate-equation-analysis}
indicates that the nucleation of graphene islands could begin with
the collision of a number of small clusters and indeed, several experimental
papers reported relatively small carbon precursors present on the
surfaces of Rh(111) \cite{Wang} and Ir(111) \cite{Lacovig}. Theoretical
efforts have mostly relied on using DFT to examine the stability of
various structures on different surfaces. In some studies thermodynamic
models of nucleation are used to try and speculate on the minimum
sizes of clusters that would be stable on a surface at a specific
temperature and supersaturation. We now review a selection of papers
that look at this aspect of growth theoretically.

The transformation of carbidic species to free-standing graphene has
been explored by Lacovig \emph{et al. } \cite{Lacovig}. They\emph{
}compared the temperature dependence of surface core level shifts
between $10^{-6}$ and $10^{-9}$ mbar at 300 K and subsequent annealing
to 820, 970 and 1270 K, see Fig. \ref{fig:Lacovig}. These spectra
allowed the authors to deduce that three different carbon species,
referred to as C$_{A}$, C$_{B}$ and C$_{C}$, each interacting differently
with the Ir surface, are present during growth. Further, the change
in spectra with increasing temperature indicates that the number of
these C species is temperature dependent. To explain these observations
the authors performed a number of DFT calculations, relaxing carbon
clusters of different sizes on an iridium surface after a simulated
annealing procedure (this is a way to approach a global energy minimum
in a system, i.e. the most energetically favorable structure: MD simulations
are run starting at rather high $T$ with subsequent slow cooling
down to zero). The structures they obtained were distinctly ``dome-like.''
The interactions between substrate and cluster were limited to carbon
atoms found at the cluster edge, and a relatively large distance between
central atoms of the clusters and the surface was found. This distance
was seen to increase with the size of carbon clusters. Comparison
of the surface core level shift (SCLS) obtained from simulations with
those obtained experimentally allowed the authors to identify the
three carbon species as belonging to cluster edges (C$_{A}$ and C$_{C}$)
and to the central carbon atom (C$_{B}$). The temperature dependence
of the SCLS was then explained as follows: high temperature ripening
processes occurring on Ir \cite{Coraux} cause the average size of
carbon cluster to increase. Increasing cluster size causes the relative
number of edge sites C$_{A}$ and C$_{C}$ to decrease while the amplitude
of the peak in the C $1s$ spectrum associated with the central C$_{B}$
species increases towards that found in graphene.

\begin{figure}
\begin{centering}
\includegraphics[height=6cm]{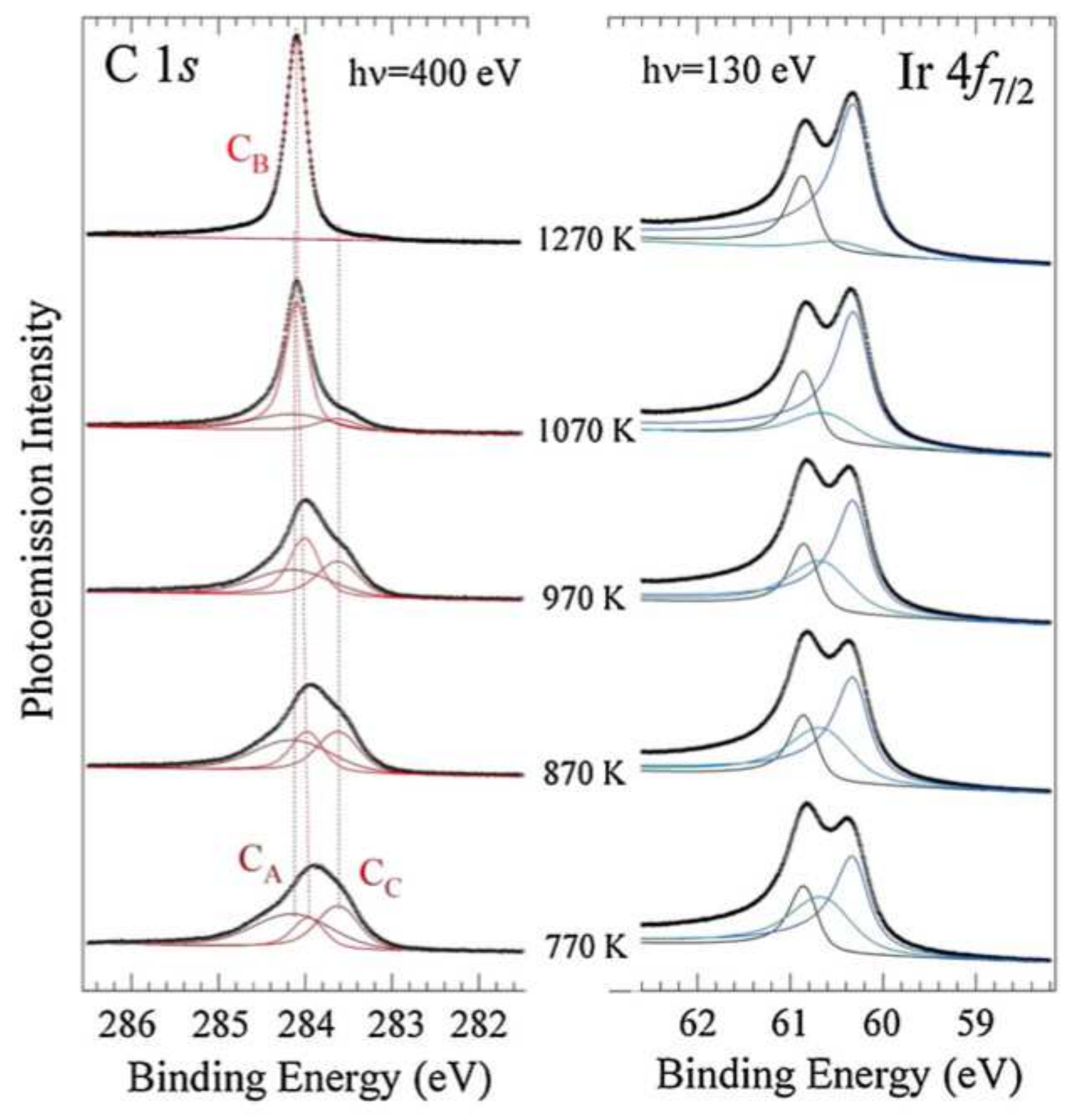}
\par\end{centering}

\caption{C 1s (left) and Ir 4$f_{7/2}$ (right) spectra after annealing at
different temperatures of the Ir(111) surface saturated with C$_{2}$H$_{4}$
at 300 K. The spectra were measured at 300 K. The different components
represent inequivalent C and Ir atoms. {[}Reprinted with permission
from \cite{Lacovig}. Copyright (2009) by the American Physical Society. {]} \label{fig:Lacovig}}
\end{figure}

Similar carbon clusters have been identified on Rh(111) \cite{Wang},
Ru(0001) \cite{Cui2011}, and Ir(111) \cite{Coraux}. In a study by
Yuan \emph{et al.} \cite{Yuan} different sized carbon clusters C$_{N}$
with $N=16,\ldots,26$ on the Rh(111), Ru(0001), Ni(111) and Cu(111)
surfaces were explored using \emph{ab initio} DFT calculations. The
formation energies of these clusters were calculated as a function
of $N$ by determining the difference between the energy of the supported
clusters and the energies of $N$ carbon atoms in the free standing
graphene (i.e. presumably using Eq. (\ref{eq:cluster-form-energy})
with $\mu_{C}$ being the energy of a C atom in free-standing graphene).
The results of these calculations are shown in Fig. \ref{fig:Yuan1}(a).

\begin{figure}
\centering{}\includegraphics[scale=0.4]{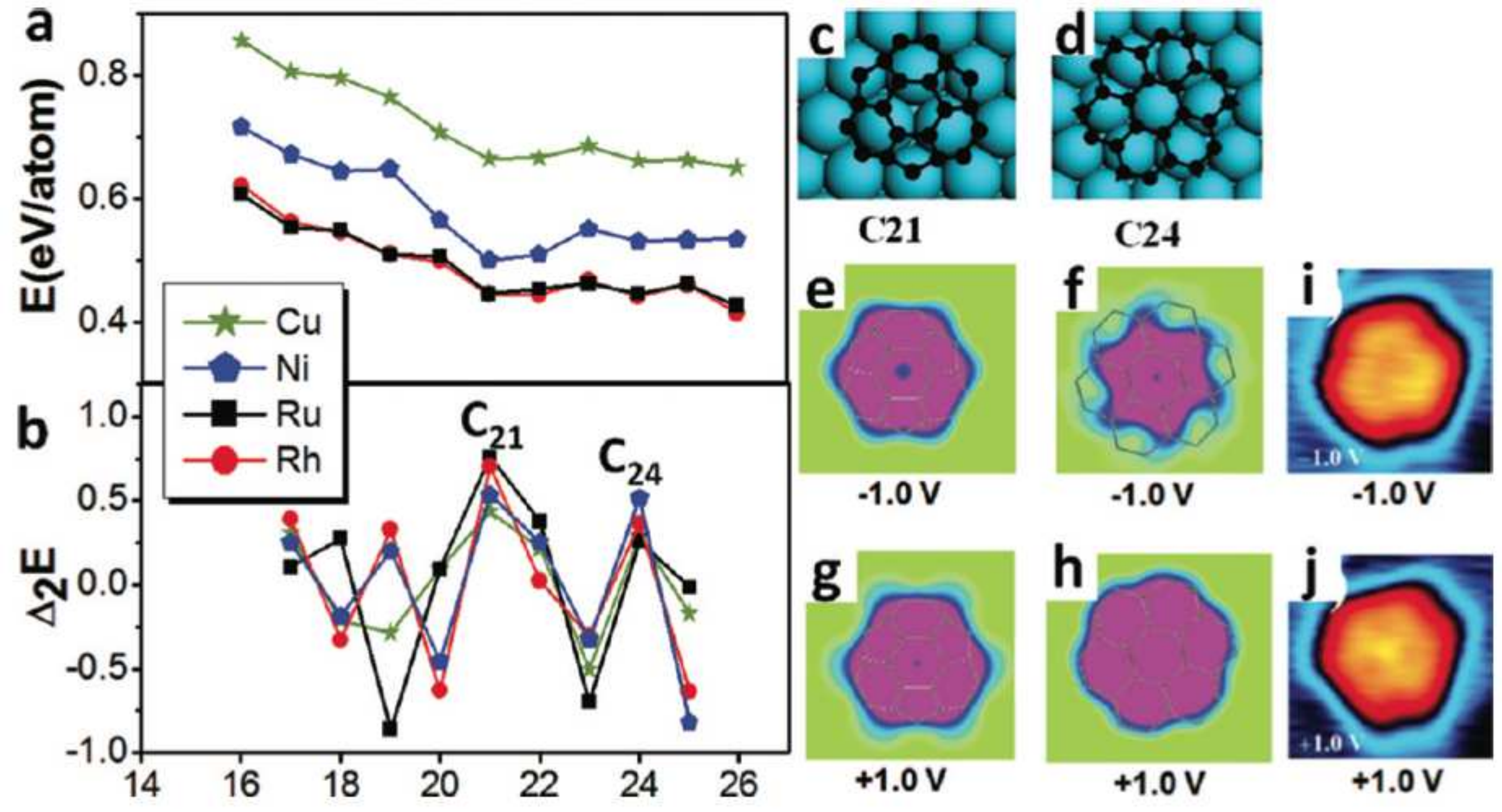}\caption{(a) The formation energies of carbon clusters C$_{N}$ on Cu(111),
Ni(111), Ru(0001) and Rh(111) with $N=16,\ldots,26$ and (b) the second
derivative of the energy with respect to $N$. Top view images show
(c) the C$_{21}$ and (d) the C$_{24}$ clusters on a Rh(111) surface.
Calculated STM images of the (e,g) C$_{21}$ and (f,h) C$_{24}$ clusters
for negative and positive bias voltages, and the corresponding experimental
STM images of the most abundant carbon cluster (i,j). {[}Reprinted
with permission from \cite{Yuan}. Copyright (2012) American Chemical Society{]} \label{fig:Yuan1}}
\end{figure}

There is a clear valley for all the surfaces around size $N=21$.
For the second derivative of the formation energy, $\Delta_{2}E$,
shown in Fig. \ref{fig:Yuan1}(b), there is also a maximum for C$_{21}$
and C$_{24}$. Therefore it is deduced, as previously shown in experiments
\cite{Cui2011,Coraux,Wang}, that both these cluster sizes are stable
on the surfaces. This is in part attributed to their high symmetry
and the tendency of the graphene edges to stand at some tilting angle
to the metal surface; the latter explains the dome-like shape of most
studied compact clusters. Fig. \ref{fig:Yuan1}(c) and (d) show the
DFT relaxed models of the C$_{21}$ and C$_{24}$ clusters, respectively,
on the Rh(111) surface. Taking into account the substrate, the C$_{21}$
cluster has C$_{3\nu}$ symmetry, whereas the C$_{24}$ cluster is
rotated by $\sim15$\textdegree{} from the high symmetry position
and therefore has a lower C$_{3}$ symmetry. A comparison of calculated
STM images of the clusters (using the Tersoff-Hamann approximation
\cite{Tersoff1985}) with the experimental images, all shown in
Fig. \ref{fig:Yuan1}(e-h) and (i,j), respectively, clearly demonstrates
that the C$_{21}$ cluster must be the dominant species on the Rh(111)
surface as only its image matches well the experimental one at both
bias voltages. To explain the high stability of this particular ``magic''
cluster size, the authors note that its formation energy, as seen
in Fig. \ref{fig:Yuan1}(a), is the smallest amongst all smaller clusters
($N<21$) and, at the same time, it is also smaller than that of the
next $N=22$ cluster. More or less the same is true for the C$_{24}$
cluster, however, as it is larger, the C$_{21}$ would be formed first
and hence must be in relative abundance.

To understand formation of graphene islands from those clusters and
estimate their lifetime, the authors studied their coalescence using
a NEB like technique. The transition path involved several steps with
the highest energy barrier $\Delta E$ during the dimerization process,
shown in Fig. \ref{fig:Yuan2}(a), calculated as 3.01 eV. From this
the lifetime of the C$_{21}$ cluster was estimated using a simplified
transition-state theory \cite{Nitzan-book} expression: $\tau=(h/k_{B}T)\exp(\Delta E/k_{B}T)$,
where $h$ is the Planck constant. In Fig. \ref{fig:Yuan2}(b) it
can be seen that in the temperature range of 930 - 985 K the lifetime
is 100 - 1000 s. This agrees with the experimental observations that
the coalescence of clusters into graphene islands occurs at $\sim$870
K on Rh(111) \cite{Wang}, at 1000-1100 K on Ru(0001) \cite{Cui2011}
and at 970 K on Ir(111) \cite{Coroux2008b}.

\begin{figure}
\centering{}\includegraphics[scale=0.4]{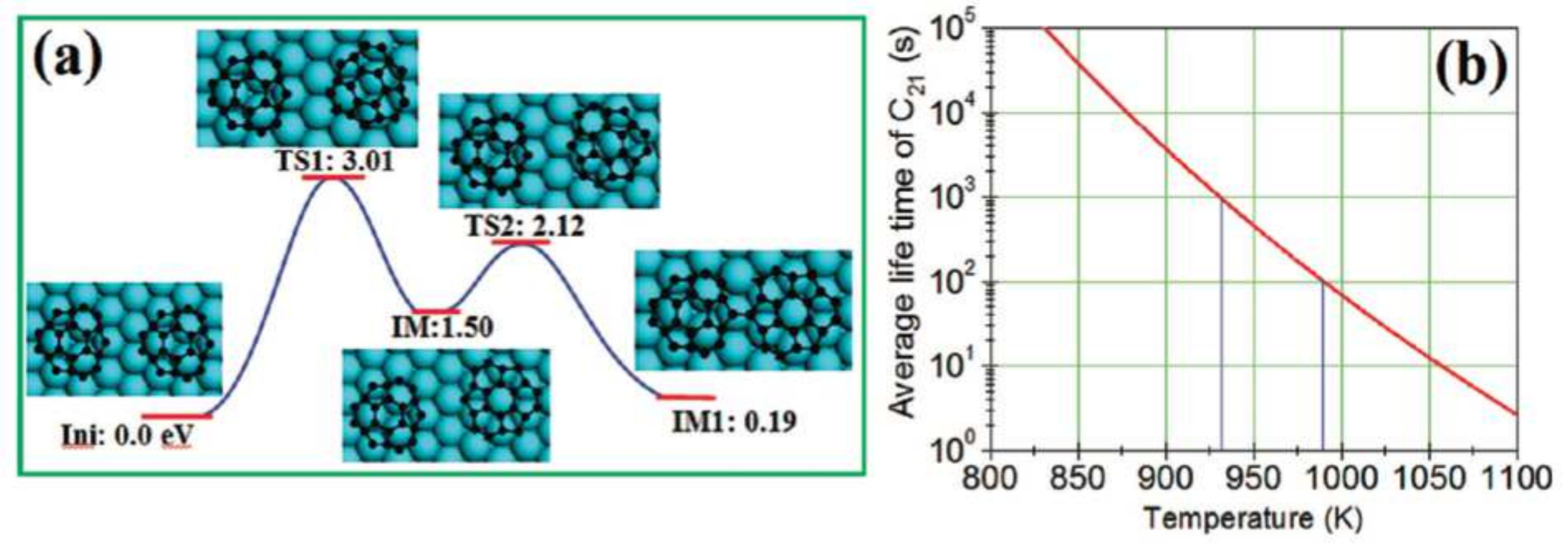}\caption{(a) The kinetic process of the dimerization of two C$_{21}$ clusters
on a Rh(111) surface and (b) the average lifetime of C$_{21}$ on
Rh(111) as calculated from the highest energy barrier in (a). {[}Reprinted
with permission from \cite{Yuan}. Copyright (2012) American Chemical Society.{]} \label{fig:Yuan2}}
\end{figure}

The high stability of these clusters is suggested to be due to the
strong binding of their edge atoms to the surface \cite{Lacovig,Wang}.
The side views of the clusters structures are shown in Fig. \ref{fig:Yuan3}.
The strong edge binding causes the islands to be dome-shaped which
elevates their height above the surface when compared with the coronene
molecule (which is basically a hexagonal C$_{24}$ cluster in which
all border C atoms are terminated by hydrogen atoms). The C$_{21}$
cluster was calculated to be $\sim$1.2 $\textrm{\AA}$ higher off
the surface than coronene and the C$_{24}$ cluster is 0.3 $\textrm{\AA}$
higher. Overall it is concluded that formation of these clusters is
not ideal for producing high quality graphene. This is because their
lack of mobility is suggested to lead to the formation of a high concentration
of small graphene islands. When these coalesce grain boundaries will
form and the graphene quality will be reduced.

\begin{figure}
\centering{}\includegraphics[scale=0.4]{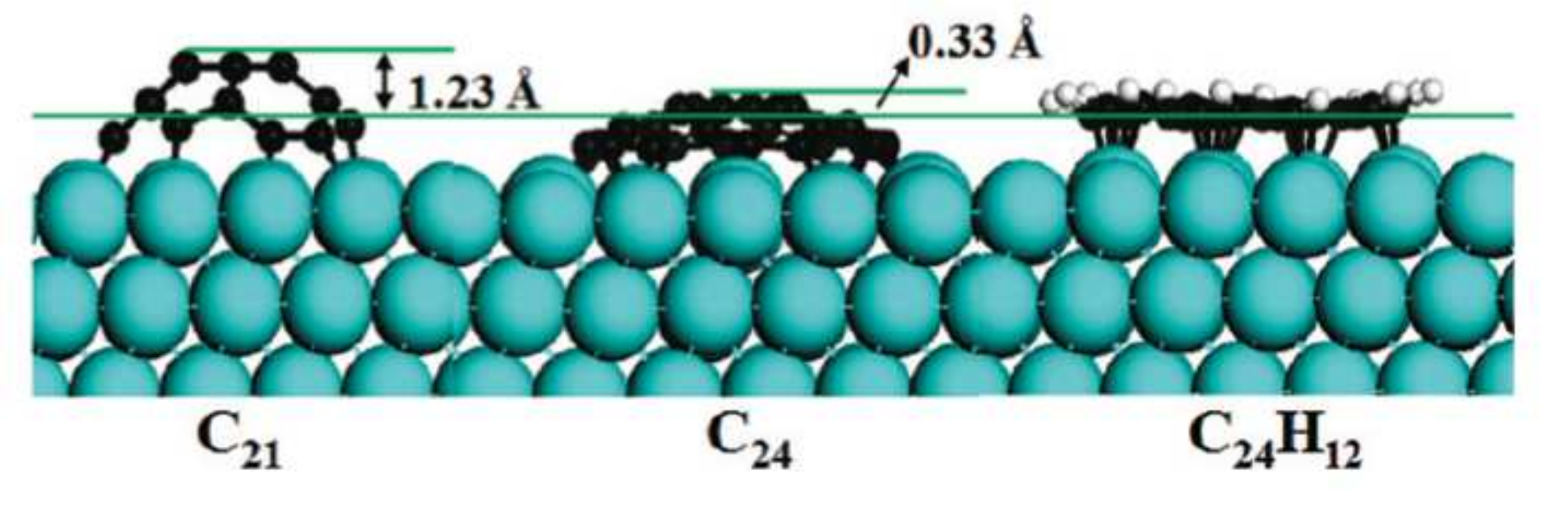}\caption{Side view images of the structure of the C$_{21}$ and C$_{24}$ clusters
and coronene molecule C$_{24}$H$_{12}$ on the Rh(111) surface. The
heights of the clusters with respect to the coronene molecule are
also shown. {[}Reprinted with permission from \cite{Yuan}. Copyright (2012) American Chemical Society.{]} \label{fig:Yuan3}}
\end{figure}

In the work of Wu \emph{et al.} \cite{Wu} the relative stability
of small carbon species (clusters) on the Ir(111) surface was calculated
with DFT. The stabilities of different clusters were compared by calculating
their ``potential energy''
\begin{equation}
E_{p}=\left(E_{C/Ir}-E_{Ir}\right)/N\;,\label{eq:Wu-cluster-pot-en}
\end{equation}
which is in fact the formation energy (per C atom) calculated relative
to the energy of an isolated cluster C$_{N}$ (i.e. in the gas phase).
Here $E_{C/Ir}$ is the total energy of the cluster of $N$ atoms
on the Ir surface, and $E_{Ir}$ is the energy of the surface. A stability
comparison was performed in order to determine how different carbon
species C$_{N}$ with $N=1,\ldots,10$ are likely to evolve on the
surface. The most stable site for carbon monomers to adsorb onto the
surface was calculated to be the hollow hcp site, followed by the
fcc site. For small carbon clusters the most stable formed structure
is usually a chain structure, formed between two carbon atoms that
both occupy an hcp site. Compact structures can also be formed, which
are often dome-shaped. These are less stable than chains when the
cluster size is small, but the compact structure stability increases
with size. Results from the calculations, shown in Fig. \ref{fig:Wu1},
demonstrate that C monomers are more stable on the Ir(111) surface
terraces than larger C clusters (the black line) and should therefore
occur more frequently. Also, clusters of 5 atoms, C$_{5}$, were also
identified as relatively stable on terraces meaning that their role
in growth is expected to be significant. This is in accord with observations
made in the LEEM study of Loginova \emph{et al.} \cite{Loginova08,Loginova09}.
The authors also considered C clusters placed near Ir(322) (red) and
Ir(332) (green) steps and found them on average to be more stable
than the same clusters on the terrace, Fig. \ref{fig:Wu1}. Interestingly,
the dimer at step edges is more stable by about 0.2 eV than a monomer.
These calculations suggest that C monomers and clusters migrating
on the surface would eventually end up at step edges; in particular,
migrating monomers may preferentially form dimers at steps. This conclusion
is also supported by the calculated diffusion barriers for $C_{N}$
species ($N=1,2,3$) using the NEB method which were found to be in
the range of 0.4-0.8 eV; these are thought to be perfectly accessible
at the elevated temperatures used for graphene growth. Interestingly,
some of the clusters diffuse atom-by-atom (e.g. C$_{2}$), but a concerted
motion is the preferred mechanism for three-atom clusters; chains
diffuse in a more complex manner by moving parts of their ``bodies''
at a time. Higher barriers were found for incorporation of a C atom
into a cluster (1.42 eV) or a chain (0.86 eV), and a combination of
two neighboring monomers required a 1.44 eV barrier to be overcome
in order to coalesce. Generally then, diffusion of clusters and monomers
is easier than their coalescence.

\begin{figure}
\begin{centering}
\includegraphics[height=4cm]{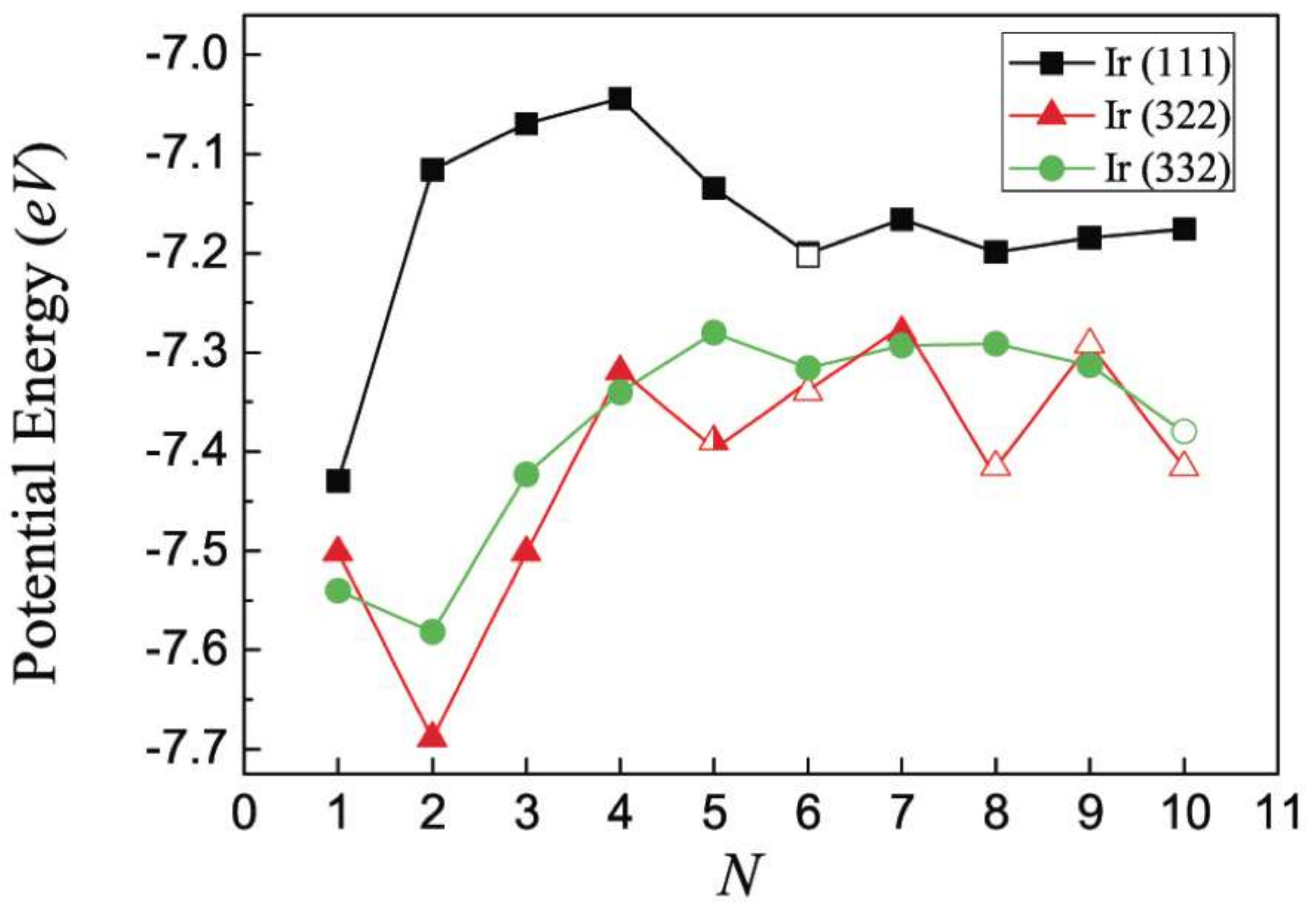}
\par\end{centering}

\caption{The calculated potential energy, Eq. (\ref{eq:Wu-cluster-pot-en}),
for different carbon species C$_{N}$ on the Ir(111) surface and at
the Ir(322) and Ir(332) steps. {[}Reprinted with permission from
\cite{Wu}. Copyright (2012) American Chemical Society.{]}\label{fig:Wu1}}
\end{figure}

The formation energies of different optimized carbon clusters on Ni(111)
were calculated using DFT in \cite{Gao11}. The formation energy was
defined using Eq. (\ref{eq:cluster-form-energy}) with $\mu_{C}$
being taken to be the energy of a C atom in a free standing graphene.
The authors examined a number of different cluster structures that
can be formed from a certain number of carbon atoms, $N=1,\ldots,24$.
Various cluster types with the same $N$ were systematically explored
including chains, rings and $sp^{2}$-networks (the closest analog
of the graphene-like structure) on the terrace of the Ni(111) surface
with their formation energies shown in Fig. \ref{fig:Gao-energies}
as a function of $N$. It was shown that, as expected, the presence
of a transition metal surface causes the energy of formation of linear
chains to be greatly reduced, largely due to the saturation of the
dangling bonds present at the free ends of the linear chains. Surprisingly,
if graphene-like clusters (with the most hexagons) are the most favorable
in the gas phase, on the surface these structures are less favorable
then ones which have one to three pentagons. The authors explain this
by the interaction with the metal which is sensitive to the number
of edge atoms of the clusters. It follows from Fig. \ref{fig:Gao-energies}
that $sp^{2}$ networks for $N\preceq12$ are less favorable than
chains; however, this situation changes for larger $N$ values and
the $sp^{2}$ network structures become the most energetically favorable
as one would expect.

\begin{figure}
\begin{centering}
\includegraphics[height=5cm]{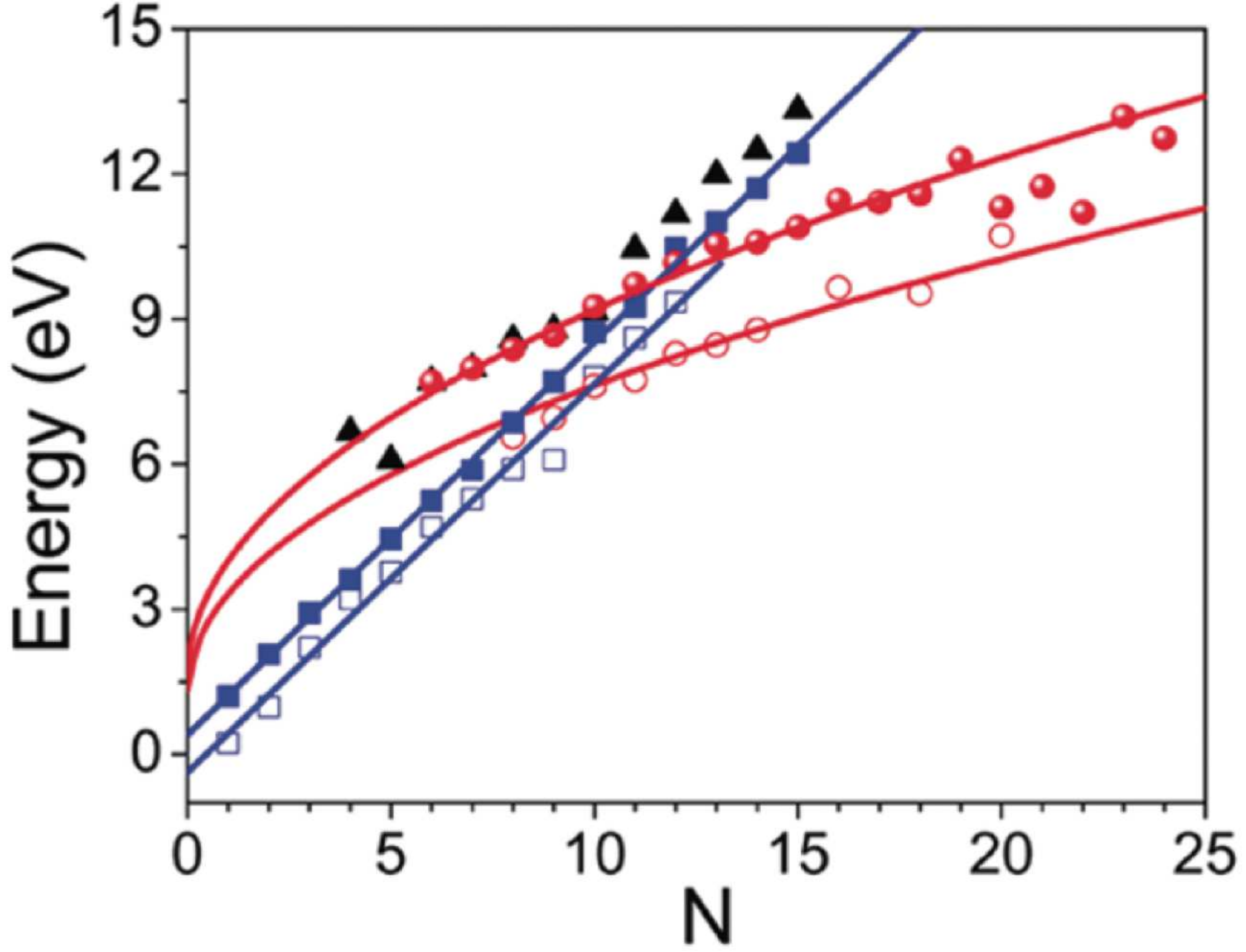}
\par\end{centering}

\caption{Calculated formation energies of $N$ atom clusters on the terrace
(solid symbols) and near the step (empty symbols) of the Ni(111) surface.
Squares, triangles, and circles correspond to C chains, rings and
$sp^{2}$ networks, respectively. Lines correspond to the fitting
to analytical expressions. {[}Reprinted with permission from \cite{Gao11}. Copyright (2011) American Chemical Society{]}.
\label{fig:Gao-energies}}
\end{figure}

Similar results were obtained on Cu(111) by Wesep \emph{et al}. \cite{wesep:171105}\emph{.}
They also used DFT to compare the formation energies (stabilities)
of a number of different carbon clusters on the terraces and like
\cite{Gao11} found that for $N<13$ the most stable structures were
chains of carbon atoms. The details of the structure of the clusters
they studied on the surface ar\textcolor{black}{e shown in Fig. \ref{fig:Wesep}.
The }chains of atoms can be seen to ``arch'' away from the surface,
with the height of the middle atoms in the chain increasing with $N$.
Wesep \emph{et al. }\cite{wesep:171105} also examined the energy
required to break a six member ring using the climbing image NEB (CI-NEB)
method, and found a surprisingly low breakup energy of 0.66 eV. Thus,
at the early stages of growth chains with $N<13$ will be the dominant
species on Cu(111). However, because removal of a C atom from islands
larger than $N=10$ would require at least two bonds to be broken,
these clusters become more long-lived since it gets increasingly more
difficult to break them up; correspondingly, the surface will likely
end up being dominated by islands of $N=10$ or $13$ atoms.

\begin{figure}
\centering{}\includegraphics[scale=0.5]{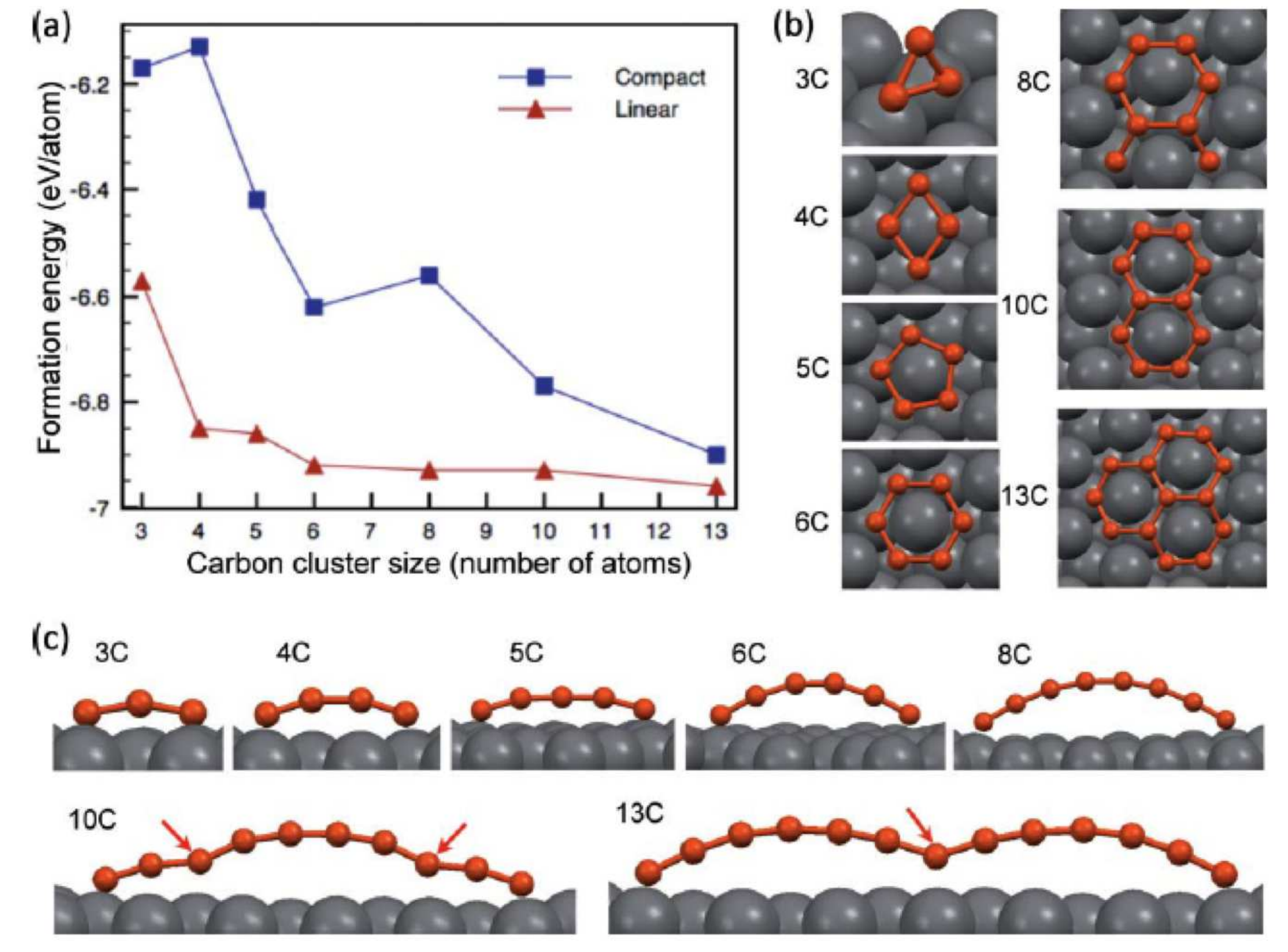}\caption{(a) The formation energies of the most stable ring and linear chain
structures on Cu(111) with the corresponding structures shown in (b)
and (c). {[}Reprinted with permission from \cite{wesep:171105}. Copyright (2011), AIP Publishing LLC.{]}
\label{fig:Wesep}}
\end{figure}

The authors in \cite{wesep:171105} ignored steps in their consideration
of the nucleation of graphene on Cu. Although this assumption was
motivated by a\textcolor{red}{{} }theoretical study suggesting that
nucleation on terraces should be dominant on Cu(111), recent experimental
observations suggest that this might not be entirely corr\textcolor{black}{ect
\cite{Nie}. Ga}o \emph{et al. }\cite{Gao11} on the other hand repeated
their calculations for structures on terraces, discussed above, but
allowed carbon atoms to bond to a (110) Ni step edge. It was found
that the formation energies $E_{N}$ of $N$ atom clusters close to
the step edge decreased (by as much as 2 eV for $N>12$), see Fig.
\ref{fig:Gao-energies}, and it was also observed that for $N\succeq12$
$sp^{2}$ network structures (those which are closest to that in graphene)
become the most stable.

The calculated values for the formation energies of different clusters
allowed the authors to apply the ideas of classical nucleation theory
to estimate the nucleation barrier, $\Delta F^{*}=\Delta F(N^{*})$,
and critical nucleus size, $N^{*}$, for different values of $\Delta\mu$,
the difference between the C atom chemical potentials in the carbon
cluster and in the carbon feedstock. The $\Delta F^{*}$ and $N^{*}$
were obtained by considering the maximum of the free energy function

\begin{equation}
\Delta F=E_{N}-N\Delta\mu\label{eq: Gao}
\end{equation}
with respect to $N$. Comparing this with equation (\ref{eq:nuc}),
notice that the entropic contributions to the right hand side are
ignored: this is a common approximation employed when considering
solid structures. The quantity $E_{N}$ should correspond to the energetic
part of the excess free energy of cluster formation. The calculated
dependences, both for the case of the clusters on the terrace and
near the step edge, are shown in Fig. \ref{fig:Gao11-nucleus-size-and-G}.
One can see that across almost the whole range of $\Delta\mu$ values
studied, the nucleation barrier and the critical nucleus size are
smaller at the step edge than on the terrace. Next, there is a region
of $\Delta\mu$ where the nucleation barrier decreases linearly, but
the critical cluster size is fixed at $N^{*}=12$ and $10$ for the
terrace and step, respectively. For $\Delta\mu>0.81$ eV the barrier
drops to 0.2 eV, while $N^{*}\simeq1$, implying that nucleation may
occur from practically any cluster size, and hence the process of
growth will be controlled by the rates of C deposition and diffusion
on the surface. The authors believe, however, that this regime of
rather high $\Delta\mu$ is less likely to arise during the CVD growth
of graphene. In the other extreme of small $\Delta\mu$ values the
barriers and critical nucleation sizes are so large that it is concluded
that nucleation is unlikely to occur in this supersaturation region.
Consequently, the nucleation rate, calculated from the classical nucleation
t\textcolor{black}{heory \cite{Kalikmanov-book2012} (Section \ref{sub:Nucleation-Theory})
as}
\begin{equation}
R_{nucl}=R_{0}\exp\left(-\Delta F^{*}/k_{B}T\right)\;,\label{eq:nulceation-rate}
\end{equation}
rapidly increases with $\Delta\mu$, demonstrating extremely sensitive
behavior to both $T$ and $\Delta\mu$, with the nucleation rates
at step edges being consistently larger than those on the terrace.
At the same time, the authors warn that these results may not necessarily
mean that nucleation starts at steps in all regions of the parameter
space as the effective area at steps is much smaller than that at
terraces. For instance, nucleation may be preferable at terraces for
large values of $\Delta\mu$. Since in the parameter regions of large
nucleation rates many nuclei would start forming at the same time,
leading over time to coalescing graphene flakes with grain boundaries
(which are not desirable for a high quality graphene), the regimes
of relatively low $R_{nucl}$ are preferable. The authors conclude
that this regime is achieved at lower $T$ and with $\Delta\mu$ in
the region of 0.3-0.5 eV.

\begin{figure}
\begin{centering}
\includegraphics[height=5cm]{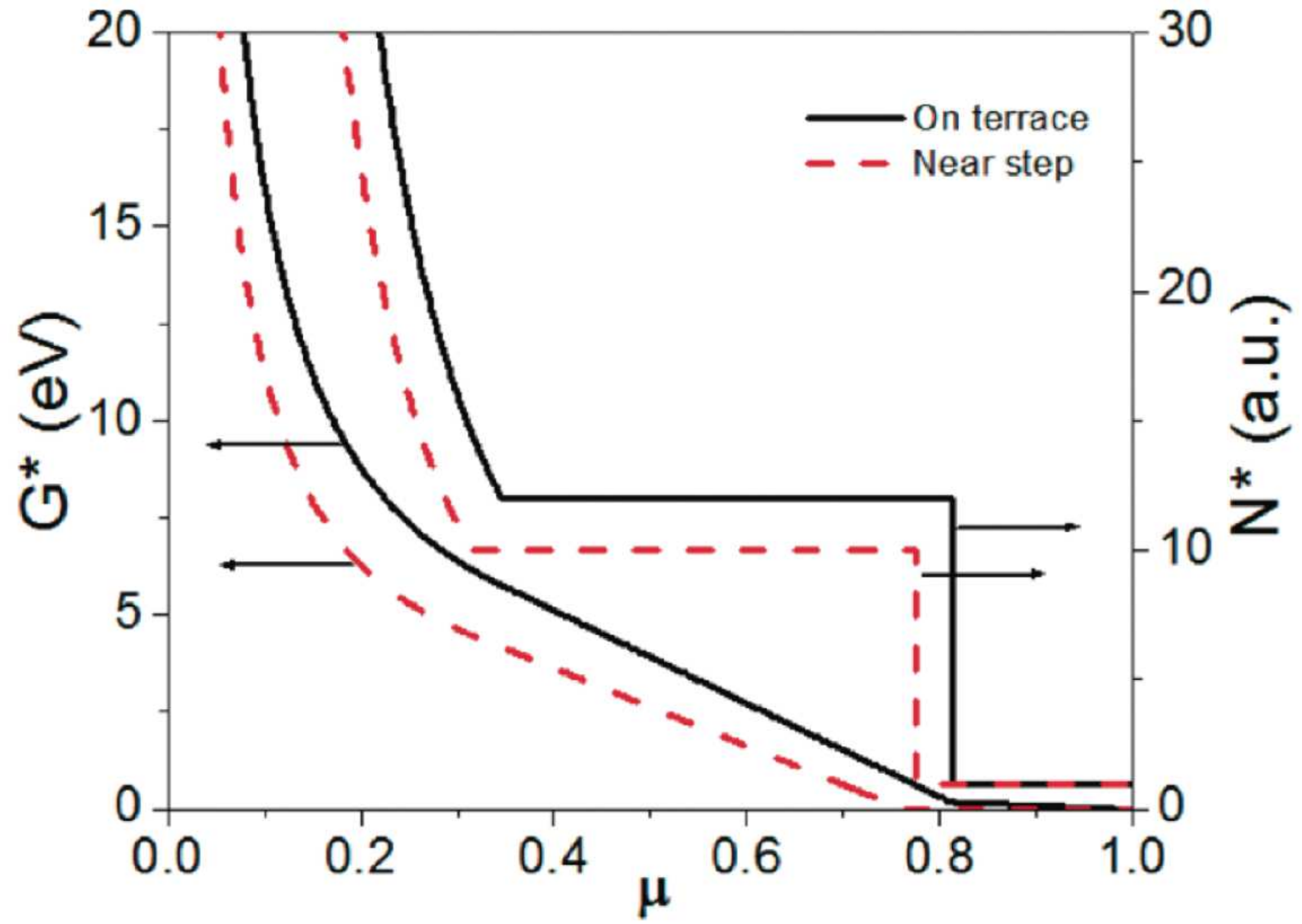}
\par\end{centering}

\caption{Calculated nucleus size, $N^{*}$, and nucleation barrier, here denoted
$G^{*}$, as functions of the difference $\Delta\mu$ of the carbon
chemical potential in clusters and in the C source. {[}Reprinted
with permission from \cite{Gao11}. Copyright (2011) American Chemical Society.{]}\emph{\label{fig:Gao11-nucleus-size-and-G}}}
\end{figure}

\begin{figure}
\centering{}\includegraphics[height=8cm]{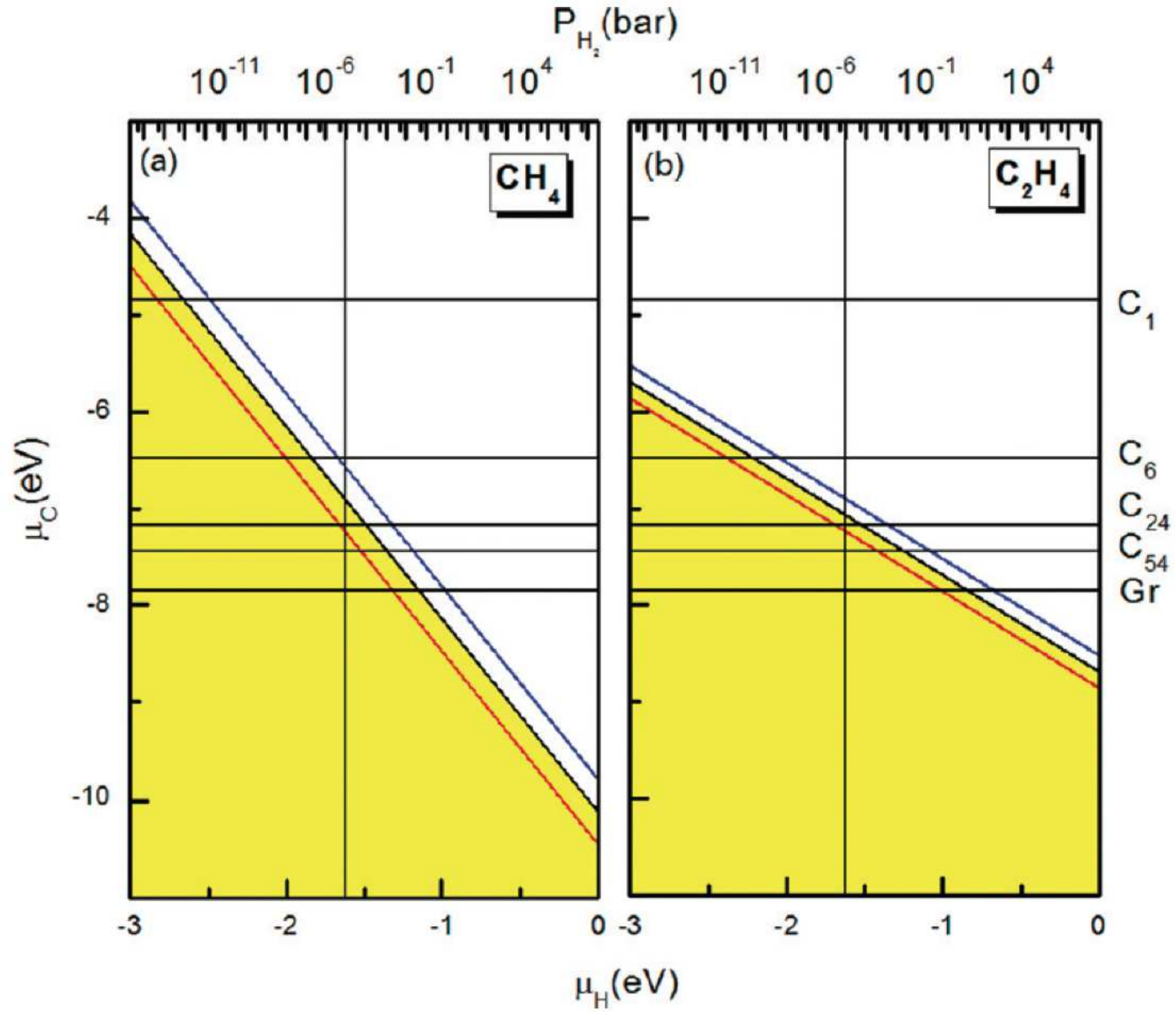}\caption{The relationship between the chemical potential of C and H for CVD
growth on the Cu(111) surface at 1300 K with (a) methane and (b) ethene
precursors. Different ratios of the partial pressures of CH$_{4}$
and H$_{2}$, $\chi$ = 1, 20 and 1/20, are indicated by black, blue
and red straight lines, respectively. A typical partial pressure of
hydrogen gas in shown by a vertical line in each case. Horizontal
lines indicate calculated chemical potentials of carbon atoms in clusters
C$_{N}$ with $N=1,6,24,54$; the carbon chemical potential of graphene
on Cu(111) surface is also indicated as ``Gr''. {[}Reprinted with
the permission of \cite{Zhangdoi:10.1021/jp2006827}. Copyright (2011) American Chemical Society.{]} \label{fig:Zhang-relationship-between}}
\end{figure}

In a similar spirit, in order to understand better the driving force
of graphene nucleation on the Cu(111) surface, thermodynamic considerations
have been proposed in \cite{Zhangdoi:10.1021/jp2006827} based on
the fact that both a carbon feedstock and the hydrogen gas are added
simultaneously when growing graphene on Cu foils \cite{Zhanchengdoi:10.1021/nn200854p}.
In the case of methane the chemical potential of C atoms in the methane
gas above the surface (which is in equilibrium with H$_{2}$ gas as
well) is given by (in eV)
\[
\mu_{C}=-2\mu_{H}-10.152+0.112\ln\chi\;,
\]
where $\mu_{H}$ is the chemical potential of H atoms and $\chi=P_{CH_{4}}/P_{H_{2}}$
is the ratio of the partial pressures of the two gases. Both chemical
potentials are functions of their partial pressures and temperature.
On the other hand, one can calculate the chemical potentials $\mu_{C_{N}}$
of various carbon clusters C$_{N}$ on the surface ($N\geq1$) including
carbon atoms ($N=1$), and compare them with $\mu_{C}$ in the gas
phase. If $\mu_{C_{N}}>\mu_{C}$ for the given ratio of pressures
$\chi$ and temperature $T$, then the cluster C$_{N}$ is deemed
unstable and would decompose with subsequent reaction of C atoms with
hydrogen to form methane, which desorbs from the surface into the
gas phase. If however $\mu_{C_{N}}<\mu_{C}$, then the cluster C$_{N}$
must be thermodynamically stable on the surface. One can see from
the yellow area in Fig. \ref{fig:Zhang-relationship-between}(a) (shown
for $\chi=1$) that atomic carbon is indeed highly unfavorable at
the typical pressures (the vertical line); however, the chemical potential
of a carbon atom in a cluster C$_{6}$ (calculated in a usual way
and with respect to an isolated C atom in vacuum) becomes very close
to $\mu_{C}$. Increasing the number of carbon atoms in the adsorbed
clusters results in the chemical potentials of their C atoms getting
more negative, indicating their eventual stability on the surface
starting from some critical size.

Similar consideration has been made for ethene as a feedstock as well,
in which case
\[
\mu_{C}=-\mu_{H}-8.692+0.056\ln\chi\;,
\]
with the corresponding stability diagram shown in Fig. \ref{fig:Zhang-relationship-between}(b).
In this case it is noticed that the critical value $\mu_{C}$ of the
carbon chemical potential is less sensitive to the ratio of partial
pressures $\chi$.

Controlling the nucleation size of carbon clusters is essential in
growing graphene of sufficient quality and size. The considerations
above demonstrate that reducing the partial pressure of the feedstock
for a given hydrogen pressure (smaller $\chi$ ) increases the rate
of nucleation, but lowers the nucleation density. This creates higher
quality graphene.

\subsubsection{Atomistic Attachment Processes\label{sub:Atomistic}}

Various carbon species that might serve as nuclei in the early stages
of graphene growth have been discussed in the previous section. Once
the graphene nucleus is formed, further growth will depend, in particular,
on the structure of the graphene flake boundary (edge). Hence, determination
of the most stable edge structures is a vital step in modeling graphene
growth. A DFT investigation into the most stable edge structures on
Ru(0001) \cite{Wei} found that the zigzag (ZZ) edge is the most stable
whereas on Cu(111) \cite{Shu} the armchair (AC) structure passivated
by Cu atoms from the bulk was shown to be preferred. Both edge terminations
are illustrated in Fig. \ref{fig:ZZ-and-AC-edges}. It is essential
to note the importance of the substrate in stabilizing particular
edge configurations: the pentagon-heptagon termination being the most
energetically favorable edge structure in vacuum is usually the least
favorable on most metal surfaces used for graphene growth \cite{Artyukhov}.
DFT studies were also conducted into the attachment of different species
onto the edge of a graphene flake on Cu(111) \cite{Shu}, Ir(111)
\cite{Wu} and Ni(111) \cite{Artyukhov}.

\begin{figure}
\begin{centering}
\includegraphics[height=4cm]{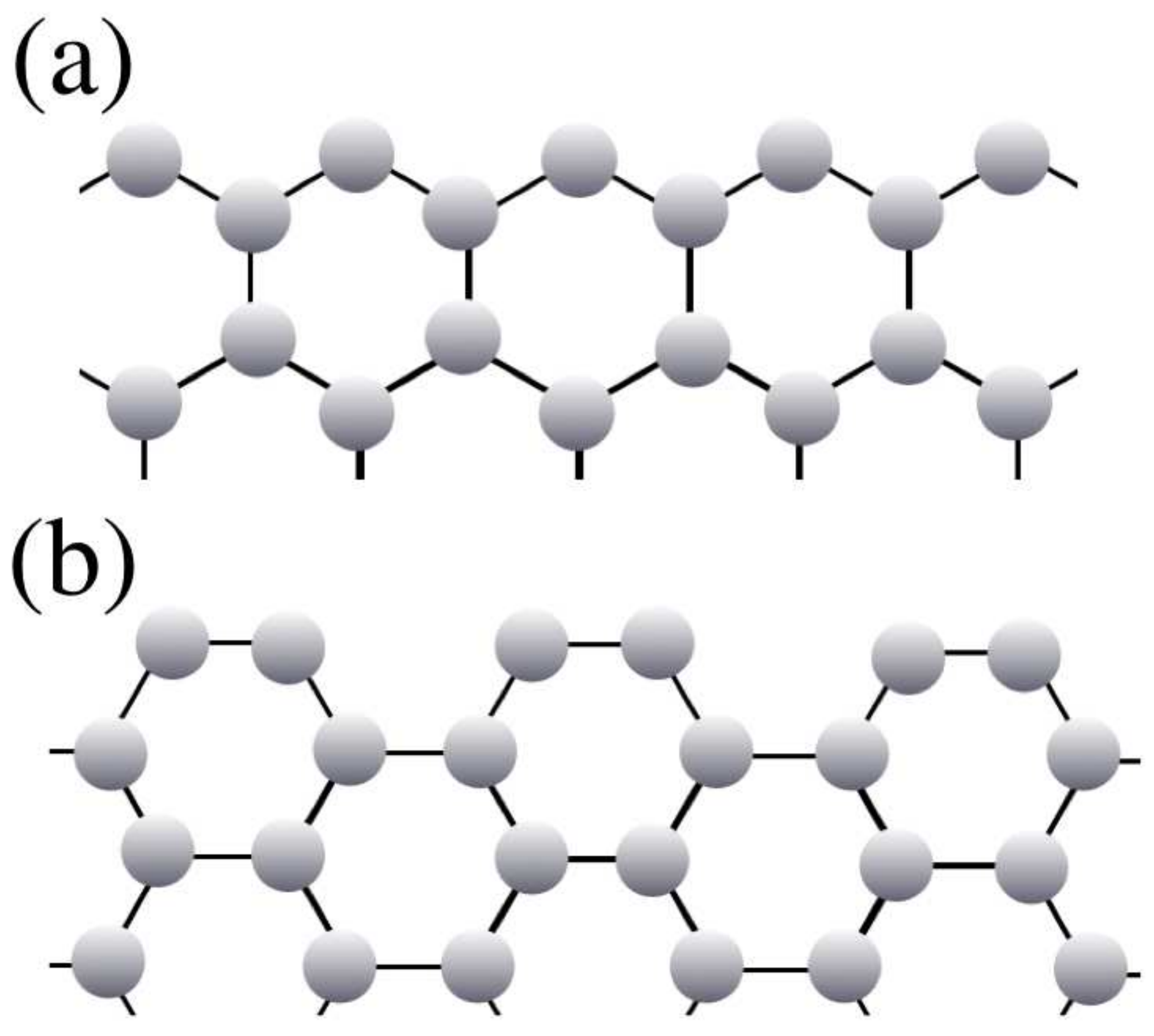}
\par\end{centering}

\caption{(a) Zigzag (ZZ) and (b) armchair (AC) edges of a graphene ribbon.\label{fig:ZZ-and-AC-edges}}
\end{figure}

We shall start by mentioning recent calculations by Haghighatpanah
\emph{et al.} \cite{Haghi} in which the catalyst-free graphene ZZ
and AC edges (i.e. free standing) and attachment of C atoms to them
were considered. The authors used both DFT and tight binding (TB)
models to describe the attachment energetics. A complete hexagon is
formed after adding 3 atoms to the ZZ and 2 to the AC edges. After
each C atom is added a geometry optimization follows. The formation
energy for each of the structures is calculated from

\begin{equation}
E_{f}=E_{G/C}-E_{G}-N\mu_{C}\;,\label{eq:Haghighatpanah_Ef-1}
\end{equation}
where $E_{G/C}$ is the energy of the graphene ribbon with $N$ carbon
atoms attached to it, $E_{G}$ is the energy of the isolated graphene
ribbon and $\mu_{C}$ here denotes the energy of a single C atom.

For both the DFT and TB methods the energy decreases (becomes more
negative) as C atoms are added, with the new atoms finally forming
a stable low-energy hexagon structure in each case; obviously, the
energy is gained every time a C atom is attached to the edge. These
preliminary calculations were done to test the TB model parametrization
intended for subsequent MC simulations of growth (unthinkable with
DFT) which we shall discuss later on in Section \ref{sub:kMC}. The
comparison of the two methods shows that in the DFT calculations the
energy is lower as the C-C bonds are determined to be stronger. In
addition different structures are predicted following the addition
of the first C atom with the two methods: the TB method yields a dangling
bond, whereas for DFT ring structures are formed on both the ZZ and
AC edges. However, the addition of further C atoms goes along the
same path for both methods.

\begin{figure}
\begin{centering}
\includegraphics[height=6cm]{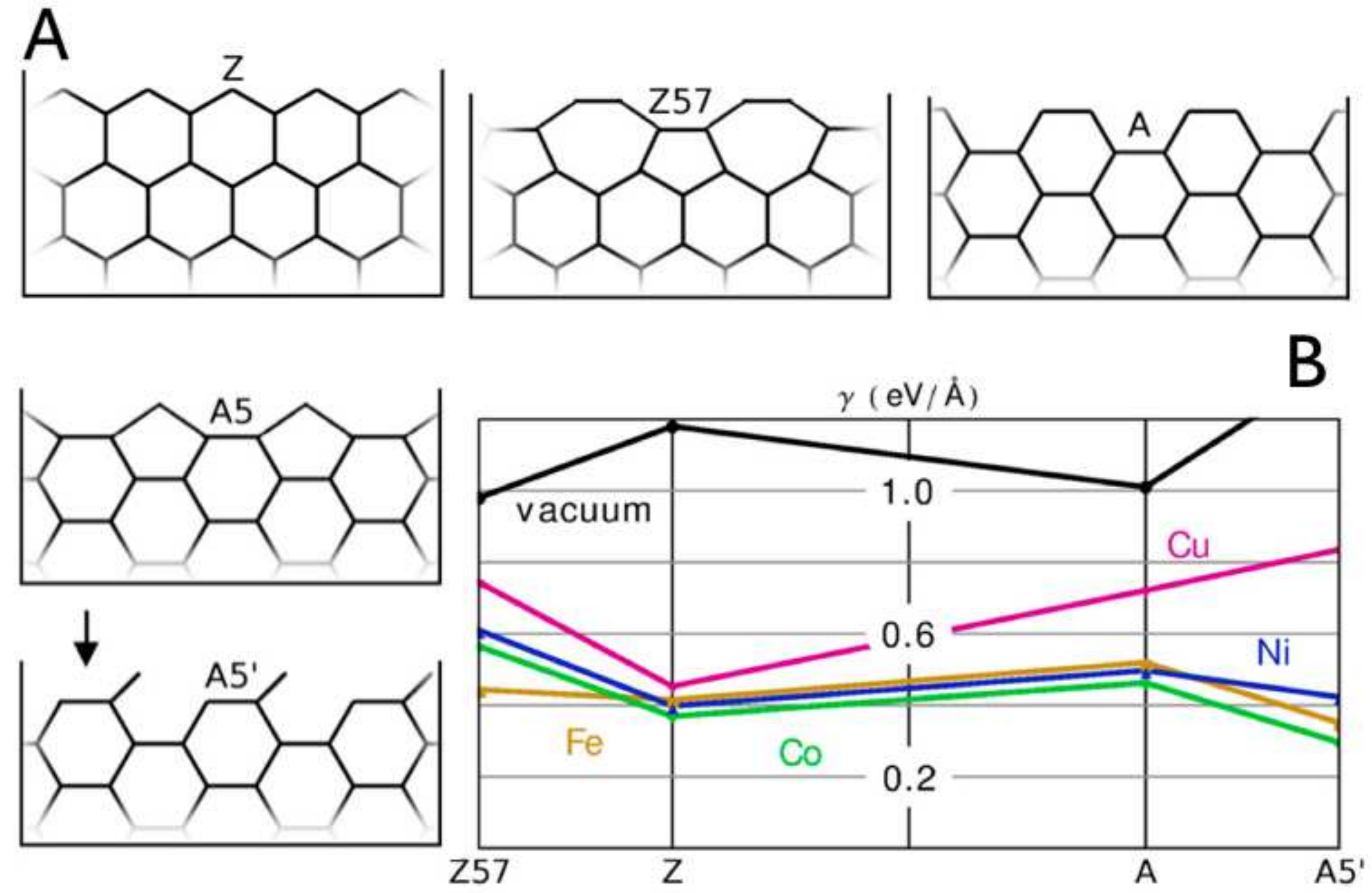}
\par\end{centering}

\caption{(A) Schematics of graphene edge structures: Z (zigzag), Z57 (pentagon-heptagon),
A (armchair), A5 (armchair-pentagon) and A5' (open armchair). (B)
Diagram showing DFT-calculated edge energies in the vacuum and on
several metals (per unit length). {[}Reproduced with permission from
\cite{Artyukhov}.{]} \label{fig:Arthyukhov_fig1} }
\end{figure}

We shall now move on to discuss theoretical studies in w\textcolor{black}{hich
the effect of the metal substrate was taken into account. A number
of experiments have observed the ZZ edge to be the predominant edge
structure on polycrystalline Cu \cite{Yu2011,Jauregui2011,Tian11}.
Several attempts have been made to explain these observations. Artyukhov
}\textcolor{black}{\emph{et al. }}\textcolor{black}{\cite{Artyukhov}
considered the thermodynamics of different edge structures. They used
DFT }calculations in which graphene ribbon was considered with one side attached
to a step edge of a metal and another free. This would be a very realistic
configuration of a growing graphene flake nucleated at a step edge.
Several basic terminations of the free edge were investigated as shown
in Fig. \ref{fig:Arthyukhov_fig1}. It is seen that the pentagon-heptagon
termination Z57, being the most stable in vacuum, becomes the least
stable on most metals. Similarly, interaction with the metals makes
the AC edge (indicated as A edge in Fig. \ref{fig:Arthyukhov_fig1})
less favorable than the ZZ one (Z in Fig. \ref{fig:Arthyukhov_fig1}).
Surprisingly, they also find that the open AC edge (denoted A5'),
which is highly unfavorable in vacuum due to unsaturated dangling
bonds, becomes more stable than the AC edge for most metals except
Cu.

\begin{figure}
\begin{centering}
\includegraphics[height=9cm]{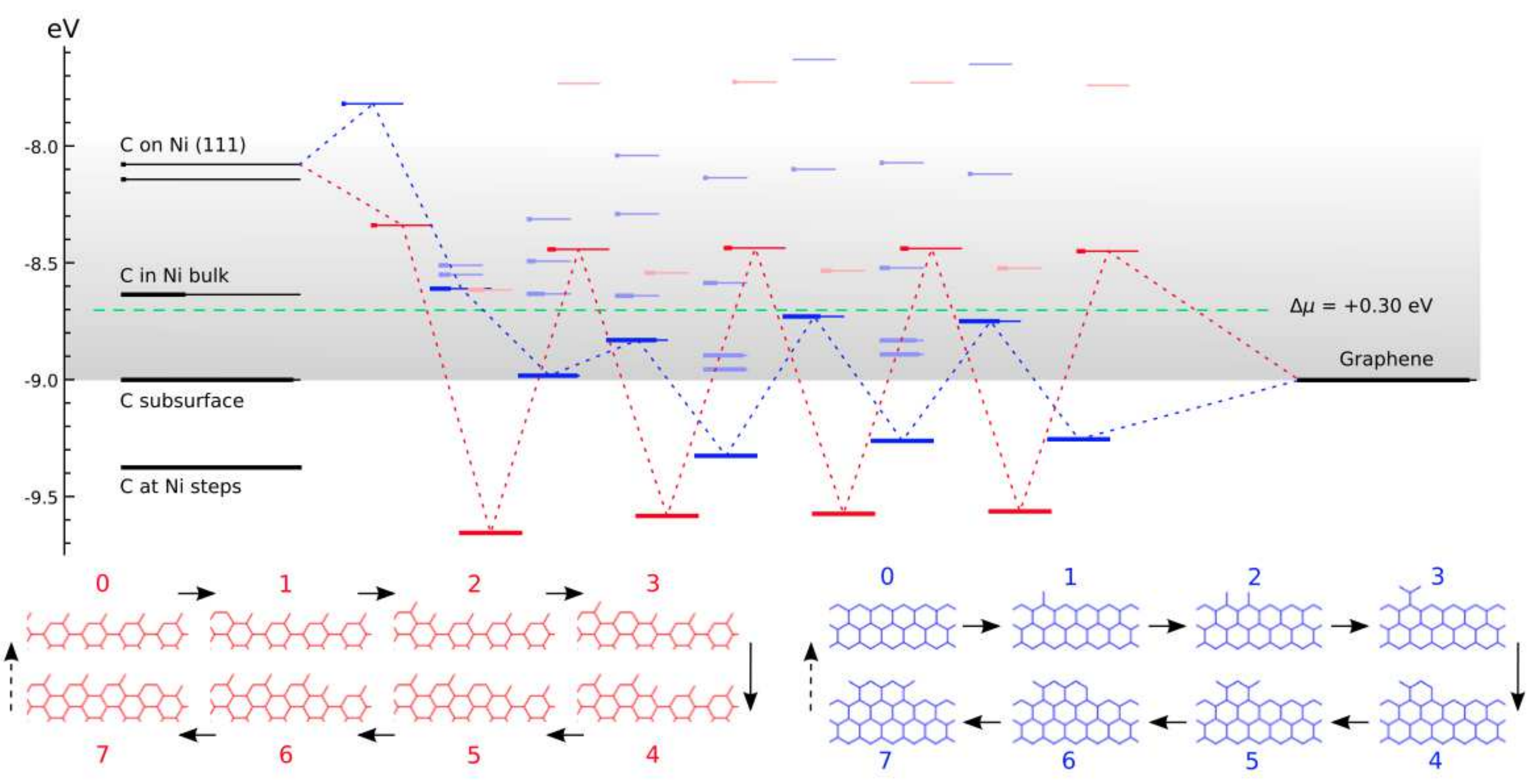}
\par\end{centering}

\caption{Energy diagram showing states of different structures formed on Ni(111)
surface formed by adding (from left to right) up to 7 C atoms to either
ZZ (blue) or A5' (red) graphene ribbon edge. The most favorable structures
after each addition are shown at the bottom. {[}Reproduced with permission
from \cite{Artyukhov}.{]} \label{fig:Arthyukhov_fig2}}
\end{figure}

Further, to understand the thermodynamics of edge growth, up to 7
atoms were added one-by-one to the ZZ or AC edges, and in each case
multiple possible geometries were investigated. Each structure is
shown in Fig. \ref{fig:Arthyukhov_fig2} by a short horizontal line
of either blue (attachment to ZZ edge) or red (to open AC) color. In the
left part of the diagram states of a single C atom on the surface
or in the bulk are shown, while on the right the chemical potential
$\mu_{C}$ of C atoms in graphene is displayed. The horizontal green
dashed line represents the position of the chemical potential of C
atoms in the feedstock. The most favorable states in each case after
adding a C atom are connected by dashed lines with the schematics
at the bottom illustrating each structure. Although the corresponding
barriers for the C atom additions were not calculated, these results
still provide an interesting insight into the steps along the growth
path. The growth of the ZZ edge involves an energy barrier for the
addition of the first atom (leading to a noticeable nucleation barrier),
and after that further additions become more favorable. As is seen
from the blue structures at the bottom of Fig. \ref{fig:Arthyukhov_fig2},
the newly formed hexagon serves as the nucleus for further growth
which proceeds after that along the edge in both directions rather
easily by two kinks propagating in both directions along the step
edge. For the A5' edge (red) the first addition is favorable, but
each subsequent addition requires energy intake. A slightly different
propagating kink mechanism is at work here as well, as shown by a
series of red structures at the bottom of Fig. \ref{fig:Arthyukhov_fig2}.
Using the calculated energies, directional growth velocities were
then calculated which allowed the authors to work out the shapes of
growing graphene islands. Since overall the nucleation barriers were
estimated to be lower for the A5' edge than for the ZZ one, the A5'
edges grow much faster with slow ZZ edges lagging behind, leading
to predominantly ZZ edged hexagon shapes for the graphene islands
in agreement with experiments. The energy state diagram also suggests
that adding C$_{2}$ clusters instead of monomers would be much more
favorable and result in larger growth rates as this would avoid the
calculated oscillations of the energy levels.

Another interesting observation made in \cite{Artyukhov} on the basis
of their Fig. \ref{fig:Arthyukhov_fig2} concerns the dependence of
the growth mechanism on the C atoms feedstock chemical potential $\mu_{C}$
($\Delta\mu$ in the Figure). The derived sequence of states described
above and based on $\mu_{C}=$0.3 eV does not lead to defects during
growth (the structures at the bottom of the Figure). However, higher
$\mu_{C}$ values (a higher position of the dashed green line) would
result in other structures being involved which would inevitably result
in defective growth.

\begin{figure}
\begin{centering}
\includegraphics[height=7cm]{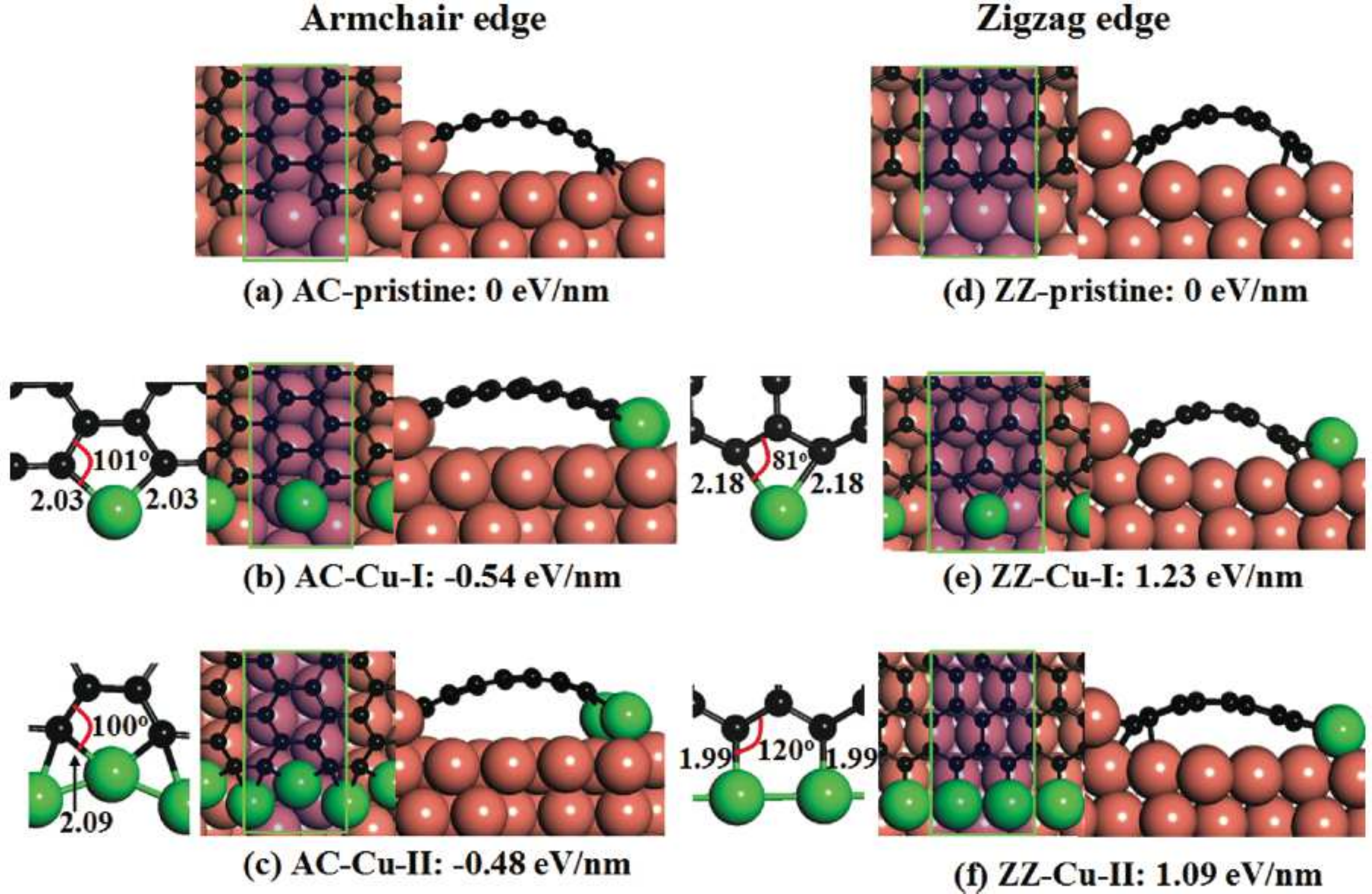}
\par\end{centering}

\caption{DFT calculated geometries of the AC (left panels) and ZZ (right panels)
structures: pristine (upper row), with a single passivated Cu atom
per cell (middle row) and attached Cu chain (bottom row). Both top
and side views are shown together with the corresponding formation
energies per unit length under the pictures. C atoms are shown in
black, other colors are used for various Cu atoms. {[}Reprinted with
permission from \cite{Shu}. Copyright (2012) American Chemical Society.{]} \label{fig:Shu-GNR-models}}
\end{figure}

In a similar attempt to explain observations of the predominantly
ZZ edges observed in growing isles, Shu \emph{et al. }\cite{Shu}
carried out a comprehensive DFT study ranging from calculating stabilities
of several edge terminations on the Cu(111) surface, their ability
to incorporate additional C adatoms and the C$_{2}$ cluster, and
finally estimating the growth rate of different edges and correspondingly
the island shapes and their dominant termination during growth. In
their calculations they attached either a ZZ or AC graphene nanoribbon
(GNR) to a step on the metal surface, in a similar fashion to the
study in \cite{Artyukhov}. In addition to considering the pristine
termination of the GNR at the free end attached to the metal, however,
either individual Cu atoms or their chain (Cu passivation) were added
as well in some calculations, see Fig. \ref{fig:Shu-GNR-models}.
The stabilization (formation) energy was calculated using the formula:
\begin{equation}
E_{f}=\left(E_{tot}-E_{GNR}-E_{M}-N_{Cu}\mu_{Cu}\right)/L\;,\label{eq:Shu-formation-energy}
\end{equation}
where $E_{tot}$, $E_{GNR}$ and $E_{M}$ are the DFT energies of
the total system (metal and GNR), individual GNR and of the metal,
respectively, while $\mu_{Cu}$ is the Cu atom chemical potential
(at $T=0$) taken from the DFT energy of the Cu bulk (the last term
with $\mu_{Cu}$ is only needed when extra Cu atoms are added
to passivate the GNR). The formation energy $E_{f}$ is calculated
per unit length $L$ of the unit cell along the Cu step direction.
The very important results they find were that the stability of the
AC termination of GNR is greatly enhanced by passivating the edge
with individual Cu atoms (in spite of the obvious energy penalty of
$\mu_{Cu}$ required to bring additional Cu atoms from the bulk).
At the same time, the pristine ZZ termination was found to be more
favorable than the passivated ZZ edges. Next, incorporation of single
C atoms and its smallest cluster C$_{2}$ to the GNR was considered
with several edge structures using a set of NEB calculations. Note,
as was mentioned above, that two C atoms need to be added to the AC
structure to form a complete hexagon nucleus, while in the case of
the ZZ edge three atoms are required. It was found that attachment
of two single C atoms to the passivated AC edge involved a much lower
barrier than attachment to the pristine AC edge; in either case the
whole process is highly exothermic. On the other hand, attachment
of three C atoms (one by one) to the ZZ edge requires considerably
larger barriers. Qualitatively the same result is obtained when a
dimer cluster C$_{2}$ was added instead. Hence, their results indicate
that, in agreement with conclusions drawn in the study by Artyukhov
\emph{et al.} \cite{Artyukhov} (see above), the growth rate of AC
edges must be much higher than that of ZZ edges, which agrees with
the experimental observation of mostly ZZ terminated graphene islands
during their growth. Therefore, the authors use the calculated energy
barriers $\Delta E$ in expressions derived in \cite{Liu10} to obtain
the rates of propagation of the two edges:
\[
R\propto\exp\left(-\Delta E/k_{B}T\right)\;.
\]
Since the whole attachment process was found to consist of several
steps, $\Delta E$ from the threshold step (with the largest barrier)
was considered in calculating the rate. On the basis of this model
it was shown that the high-growth-rate AC edge quickly disappears,
while the slowly growing ZZ edge dominates the graphene edges, in
agreement with experiment and similar to the conclusion made in \cite{Artyukhov}
as we have already mentioned. Also, the mechanism of growth of the
ZZ edge was proposed to be the kink propagation mechanism, similar
to that found in \cite{Artyukhov}. It was concluded that Cu passivation
of the graphene AC edges plays an essential role in its growth on
the Cu(111) surface.

In the work of Wu \emph{et al.} \cite{Wu} the attachment of carbon
clusters to the ZZ edge of the R0 phase of graphene, fixed with one
edge to the Ir(111) surface, was considered by adding different carbon
species to the other (free) edge of the graphene ribbon and calculating
their formation energies. This was then used to determine which C$_{N}$
species are favorable for the attachment. The formation energy $E_{f}$
per carbon atom for a particular attachment process of a carbon cluster
of $N$ atoms to the graphene ribbon placed on the surface was calculated
(relative to the binding energy of a single C atom on the terrace)
from

\begin{equation}
E_{f}=(E_{C+R/Ir}-E_{R/Ir}-E_{C/Ir}+E_{Ir})/N\;,
\end{equation}
where $E_{C+R/Ir}$ is the total energy of the ribbon after the C
cluster is attached, $E{}_{R/Ir}$ and $E{}_{C/Ir}$ are the energies
of the ribbon and the cluster before the attachment, respectively
(both on the surface); the energy of the surface, $E_{Ir}$, was added
due to double counting of it in the preceding two energy terms. The
formation energy is dependent on the metal sites that the additional
carbon atoms occupy. For a carbon monomer to attach and occupy a hollow
hcp site $E{}_{f}$ was calculated as -0.86 eV (with a barrier of
0.75 eV) and therefore a C atom attachment is energetically favorable.
However carbon monomer attachment in a top site is unfavorable with
$E_{f}=$0.58 eV, Fig. \ref{fig:Wu2}(a), i.e. the C monomers will
detach faster from this site than they can attach. Instead the growth
relies on the attachment of clusters. Dimer and trimer attachment,
Fig. \ref{fig:Wu2}(b,c), is unfavorable, while the C$_{4}$, C$_{5}$,
C$_{6}$ carbon clusters each have a formation energy of -0.36 eV,
Fig. \ref{fig:Wu2}(d-f), and hence provide the driving mechanism
for the edge attachment. This suggests a possible explanation for
the 5 atom cluster growth mechanism that causes the non-linear growth
observed in the Loginova \emph{et al.} experiments \cite{Loginova08,Loginova2009b}.

\begin{figure}
\centering{}\includegraphics[scale=0.55]{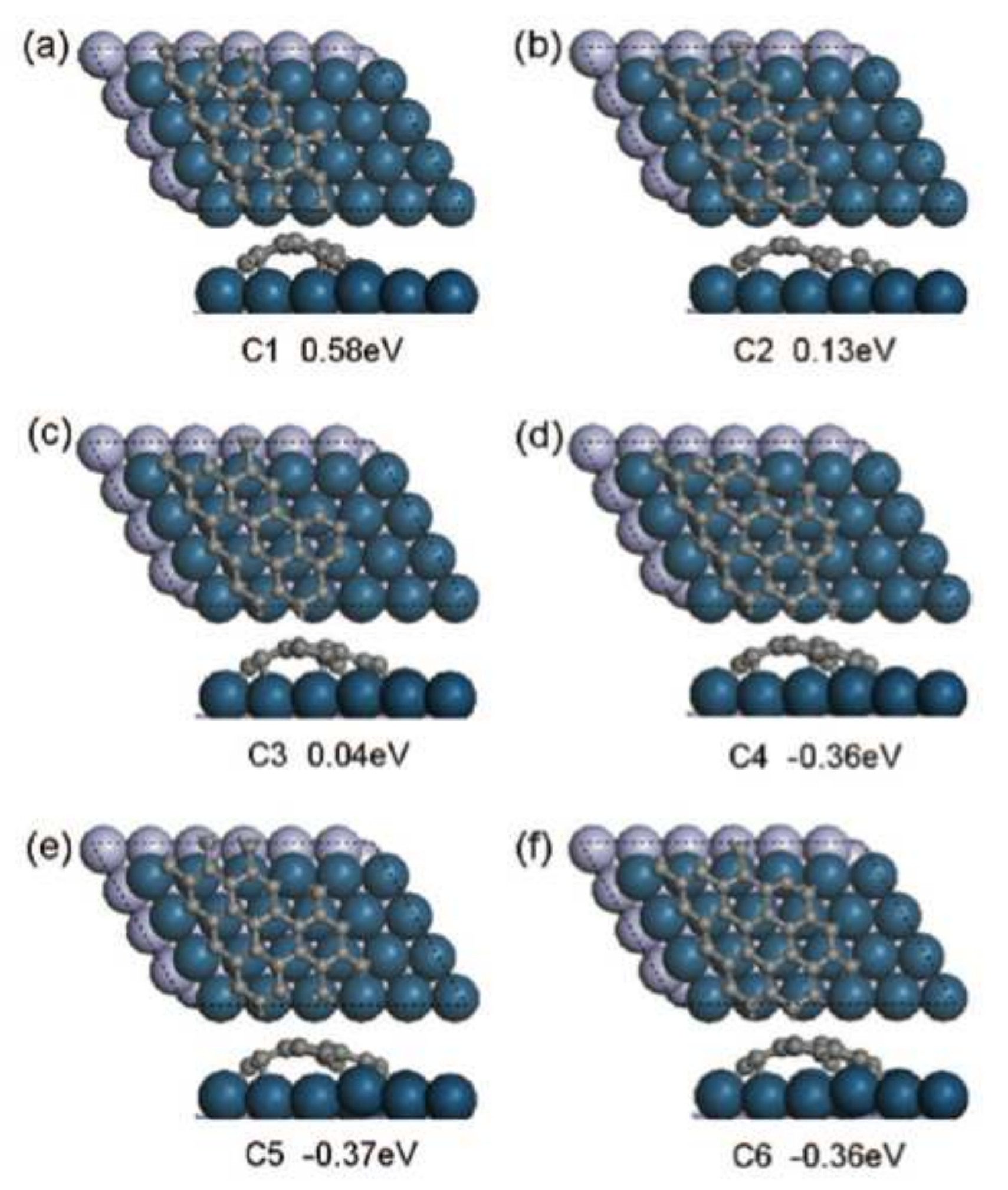}\caption{Selection of optimized structures of carbon clusters attached to the
ZZ edge and their calculated formation energies $E_{f}$. {[}Reprinted
with permission from \cite{Wu}. Copyright (2012) American Chemical Society.{]} \label{fig:Wu2} }
\end{figure}

In addition to the growth of the Ir(111) R0 phase, the R30 phase was
also investigated using DFT \cite{Wu} which is important for understanding
the effect that the orientation of the graphene island with respect
to the substrate metal has on the growth. Since different edge atoms
are not equivalent, the monomer attachment may be either energetically
favorable or not. However, the growth can continue via attachment
of clusters. In particular, C$_{2}$ and C$_{3}$ clusters can be
attached to the ZZ edge. Since the number of C$_{2}$ clusters is
the largest, their attachment would determine the growth of the ZZ
edge. Compared to the R0 phase which relied on larger clusters, the
R30 phase would then be growing faster since the concentration of
C$_{2}$ clusters is very high. It is worth noting however that the
edges of graphene on Ir(111) are not actually purely ZZ or AC, be
they attached or not to a substrate edge: see \cite{Diaye2008} and
a more recent work \cite{Phark2012}.

\subsection{Role of defects\label{sub:Role-of-defects}}

During graphene growth defects in the structure are formed. The impact
on the properties of graphene is regarded either as negative, e.g.
a strong reduction in its carrier mobility \cite{Yu2011}, or positive,
e.g. a possible route to turning on magnetic moments in an otherwise
diamagnetic material \cite{Yazyev2010}. These defects mainly include
grain boundaries which are formed where different domains meet during
growth and also point defects such as vacancies and other local imperfections
of the pristine hexagonal network. The grain boundaries are created
when the growth has begun at different nucleation sites. As these
regions merge together during growth it leads to grain boundaries
between the different crystalline domains. Point defects on the other
hand could be due to a vacancy or impurity atoms becoming trapped
in the structure during growth: their healing requires energy barriers
to overcome and hence takes much longer then the time needed for new
C atoms to attach. Consequently, graphene flakes grow with defects
which heal on a much longer timescale. Point
defects may also be the result of defects in the underlying metal
surface (e.g. steps, kinks) which imprint on the graphene structure
stimulating its growth away from the perfect C atom arrangement.

It is possible to reduce the amount of grain boundaries by carefully
controlling CVD growth conditions so that graphene nucleation does
not occur at many different sites. This was achieved e.g. by Yan \emph{et
al.} for graphene on a copper foil \cite{Yan2012}. The graphene was
grown using very low pressure CVD of methane with a low flow rate.
Large domains of up to few mm were observed. For these large domains,
transferred to SiO$_{2}$/Si, it was shown that the carrier
mobility reaches values above $\sim$10000 cm$^{2}$V$^{-1}$s$^{-1}$.
Such values approach those obtained with exfoliated graphene (in much
smaller domains), which were also obtained with CVD graphene in single-crystal
domains of few 100 microns in size, transferred to SiO$_{2}$/Si
with the help of a dry process \cite{Petrone2012}.

Point defects may also influence the electronic transport properties
of graphene prepared on substrates, as do point defects induced in
exfoliated graphene \cite{Chen2009}. We note that, to our knowledge,
the influence of defects on the conduction properties of a SiC graphene
has not been much studied.

\subsubsection{Defects in free standing graphene\label{sub:Defects-in-free-standing}}

Even though we are interested mostly in the growth of epitaxial graphene,
consideration of defects in the free standing graphene is more than
an academic exercise. This is not only because the free standing graphene
seems to be the simplest system to consider; it is also that this
rather simple model system may already suggest possible defective
structures in the real situation of graphene on the metal substrate,
even though one would expect that the metal might significantly affect
the defective structures and introduce further complexity, not least
due to the moir\'{e} pattern.

For free standing graphene many types of structural defect have been
investigated. A detailed review of theoretical studies into these
has been recently published \cite{Banhartdoi:10.1021/nn102598m},
so that only a brief account of them will be given here. The types
of point defects considered so far are single vacancies (SV), double
vacancies (DV), interstitial carbon atoms, impurities or Stone-Wales
defects. The structures of some of these defects are shown in Fig.
\ref{fig:Relaxed-structures-of-defects}. Stone Wales defects do not
involve the removal of any carbon atoms, but are caused by rotation
of a carbon bond that results in two pentagons and two heptagons being
formed. Single vacancies result from the removal of one C atom. These
can form a 5-9 structure (the notation used corresponds to the fact
that instead of two hexagons, a pentagon (5 vertices) and an adjacent
nonagon (9 vertices) are formed), where a C-C bond forms between two
of the dangling bonds that remain. For double vacancies there are
three different structures that are proposed. These include the 5-8-5
structure, where the dangling bonds form two pentagons and an octagon;
the 555-777 structure, which consists of three pentagons and three
heptagons; and its reconstruction, the 5555-6-7777 defect \cite{KotakoskiPhysRevLett.106.105505,KotakoskiPhysRevB.83.245420}.

In order to understand the stabilities of different point defects
and hence assess the feasibility of their formation, knowledge of
their formation energies is useful. The calculated formation energies
of these defects are presented in Table \ref{tab:The-various-defects}.
These calculations were performed using DFT, with the exception of
\cite{LeePhysRevLett.95.205501} where tight binding molecular dynamics
simulations were used. The formation energies of both SV and DV were
found to be around 7 eV \cite{JiePhysRevB.80.033407,LiPhysRevB.72.184109,ElbarbaryPhysRevB.68.144107}.
For SV this energy is high as there is a dangling bond left in the
structure which is unfavorable. As DVs have two missing atoms, the
energy per removed atom is lower than that for the SV and therefore
thermodynamically the DVs will be more likely to form. Multiple vacancies
may also be formed in graphene; however, it seems that it must be
more favorable to have an even number of missing atoms rather than
an odd number. This is because for the removal of an odd number of
carbon atoms there will be a remaining dangling bond.

Once defects are present they can migrate across the graphene surface.
The mobility of the various defects is determined by calculating their
diffusion barriers. For mobile defects it is possible that they will
migrate to the graphene edge where there is a greater chance for the
defect to disappear and hence for the whole structure to heal. Moreover,
two defects may merge together to create a larger defect and defects
can reconstruct themselves by rotating particular bonds. Calculated
diffusion barriers for the defects are given in Table \ref{tab:The-various-defects}.

\begin{figure}
\begin{centering}
\includegraphics[height=7cm]{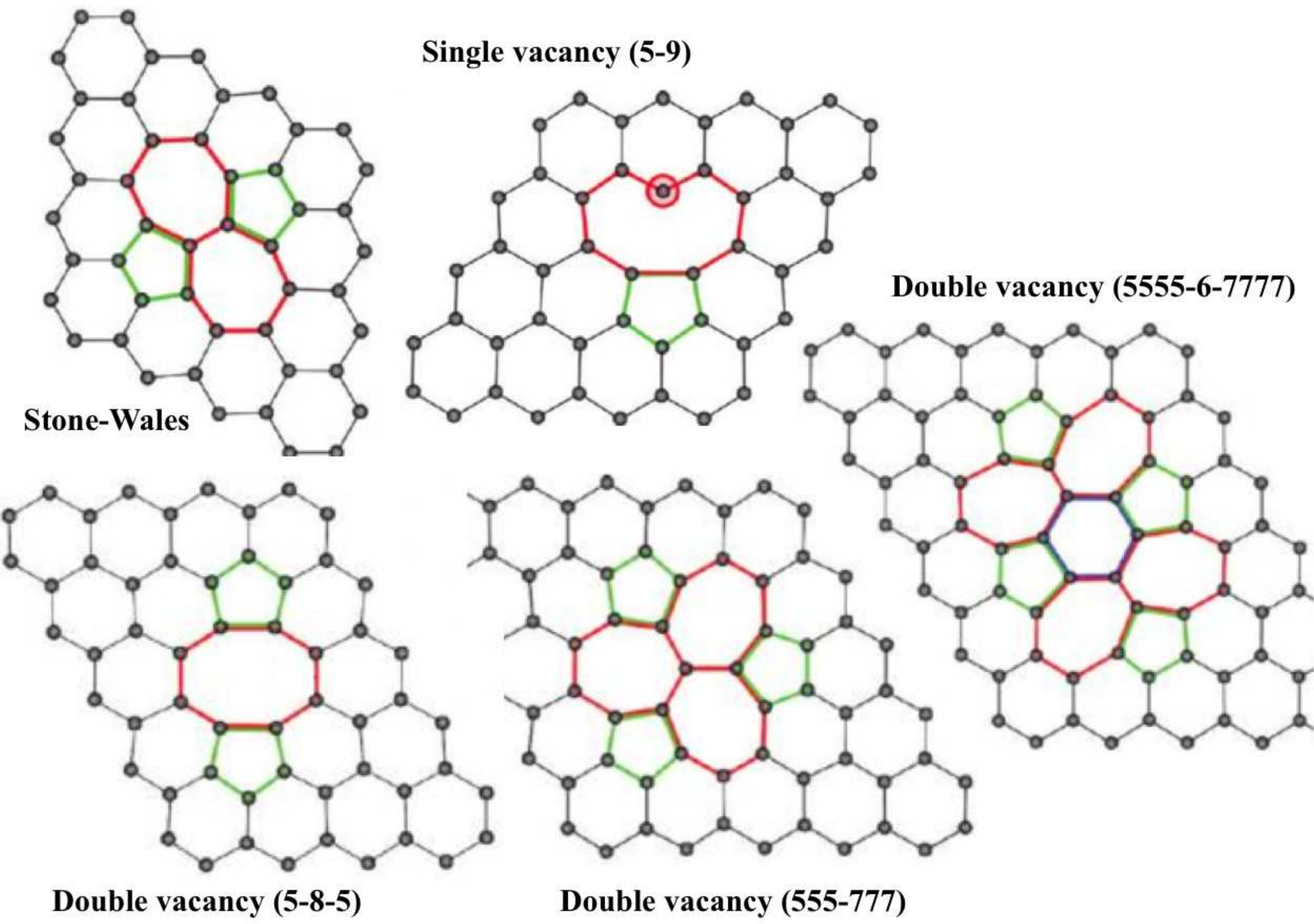}
\par\end{centering}

\caption{Relaxed structures of some of the point defects in free standing graphene.
{[}Reprinted with permission from \cite{Banhartdoi:10.1021/nn102598m}. Copyright (2011) American Chemical Society.{]}\label{fig:Relaxed-structures-of-defects}}
\end{figure}

It is seen from the table that SVs have a much lower diffusion barrier
than DVs, and are therefore expected to be highly mobile, which can
lead to the formation of DVs when two single ones coalesce. This was
investigated by Lee \emph{et al. }who found that at 3000 K two SVs
can diffuse and coalesce to form 5-8-5 DV \cite{LeePhysRevLett.95.205501}.
These vacancies can then reconstruct into a 555-777 vacancy by undergoing
a Stone-Wales type transformation where a bond rotates by 90 degrees.
The 555-777 vacancy was found to be more stable than the 5-8-5 vacancy.
DVs have been judged to be less mobile but could possibly migrate
via atom jumps \cite{ElbarbaryPhysRevB.68.144107} or switching between
configurations \cite{KotakoskiPhysRevLett.106.105505}.

\begin{center}
\begin{table}
\centering{}%
\begin{tabular}{|c|c|c|c|}
\hline
Defect  & %
\begin{tabular}{c}
Formation\tabularnewline
Energy {[}eV{]}\tabularnewline
\end{tabular} & %
\begin{tabular}{c}
Diffusion \tabularnewline
Barrier {[}eV{]}\tabularnewline
\end{tabular} & Reference\tabularnewline
\hline
\hline
\textbf{}%
\begin{tabular}{c}
Stone-Wales\tabularnewline
\textbf{55-77}\tabularnewline
\end{tabular} & 4.5-5.3 & 10 & \cite{LiPhysRevB.72.184109,JiePhysRevB.80.033407}\tabularnewline
\hline
\textbf{}%
\begin{tabular}{c}
Single Vacancy\tabularnewline
\textbf{5-9}\tabularnewline
\end{tabular} & 7.3-7.5 & 1.2-1.4 & \cite{ElbarbaryPhysRevB.68.144107}\tabularnewline
\hline
\textbf{}%
\begin{tabular}{c}
Double Vacancy\tabularnewline
 \textbf{5-8-5}\tabularnewline
\end{tabular} & 7.2-7.9 & 7 & \cite{Krasheninnikov2006132,ElbarbaryPhysRevB.68.144107}\tabularnewline
\hline
\textbf{}%
\begin{tabular}{c}
Double Vacancy\tabularnewline
\textbf{555-777}\tabularnewline
\end{tabular} & 6.4-7.5 & 6 & \cite{LeePhysRevLett.95.205501,CretuPhysRevLett.105.196102}\tabularnewline
\hline
\textbf{}%
\begin{tabular}{c}
Double Vacancy\tabularnewline
\textbf{5555-6-7777}\tabularnewline
\end{tabular} & 7 & 6 & \cite{KotakoskiPhysRevLett.106.105505,KotakoskiPhysRevB.83.245420}\tabularnewline
\hline
\end{tabular}\caption{Various defects found in free standing graphene and their formation
energies and diffusion barriers calculated using DFT. {[}Adapted from
\cite{Banhartdoi:10.1021/nn102598m}.{]} \label{tab:The-various-defects}}
\end{table}

\par\end{center}

\subsubsection{Defects in graphene on transition metal surfaces\label{sub:Defects-in-graphene-on-TM}}

When graphene is grown with CVD or TPG, it is supported by a transition
metal surface. In this case the structure, formation and mobility
of any graphene point defects must be affected by the metal. Recently,
theoretical work started to appear focused on determining defect properties
on different growth surfaces in an attempt to find how to limit the
formation of defects during graphene growth.

The formation of vacancy defects in graphene on the Cu(111), Ni(111)
and Co(0001) surfaces has been studied using DFT calculations \cite{Wangdoi:10.1021/ja312687a}.
As mentioned above, for graphene grown on a surface different types
of defects are possible due to the ability of the metal atoms to be
incorporated into them. For instance, as is shown in Fig. \ref{fig:Optimized-structures-defects-on-TM},
in the case of SVs the effect of the missing atom can be reduced by
lifting a metal atom from the substrate (this defect is denoted 3DBs
in the Figure and is also known as 5-9 defect in the free-standing
graphene, see Fig. \ref{fig:Relaxed-structures-of-defects}), or replacing it with
 an extra metal atom which would interact strongly with the defect
(M@3DBs). Note that there are always free metal atoms on the surface
diffusing across terraces between steps \cite{Pai:1996vo}; their
concentration depends on the surface roughness and their mobility
on temperature. The surface atom attaches to the three dangling carbon
bonds left after C atom removal thereby forming a new type of the
SV de\textcolor{black}{fect. For DVs there are the analogues of the
5-8-5 and 555-777 structures found in free-standing graphene (}Fig.
\ref{fig:Relaxed-structures-of-defects})\textcolor{black}{; however,
structures where a metal atom passivates the four dangling bonds of
the defect have also been proposed. As wit}h SV, two possibilities
have been considered \cite{Wangdoi:10.1021/ja312687a}: either the
metal atom comes from the substrate relaxing upwards (4DBs) or it
is a free surface atom migrating around the surface and eventually
being trapped by the defect (M@4DBs), see Fig. \ref{fig:Optimized-structures-defects-on-TM}.

\begin{figure}
\centering{}\includegraphics[scale=0.5]{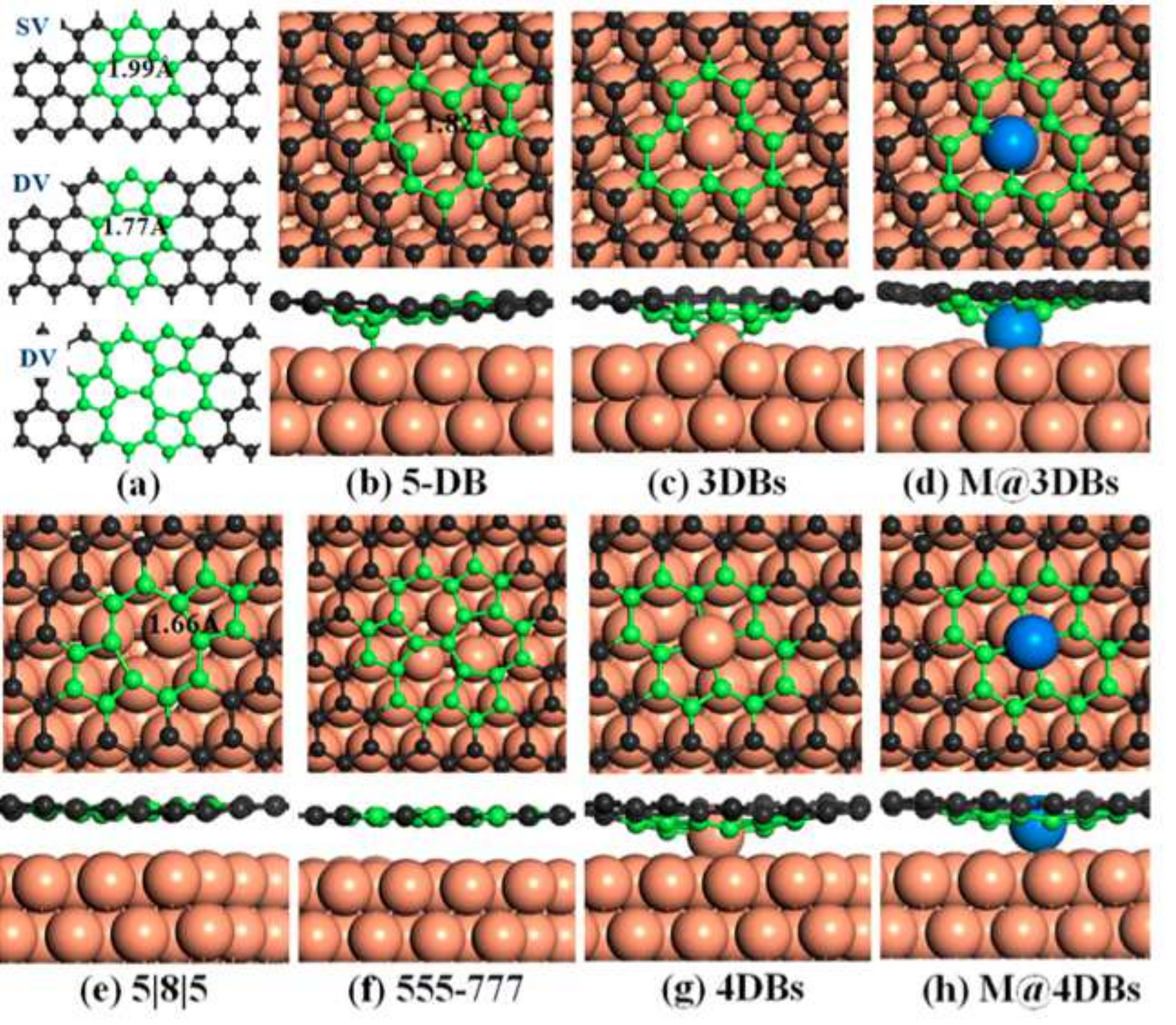}\caption{(a) Optimized structures of a SV (denoted 5-DB) and DVs (5-8-5 and
555-777) in free-standing graphene. The SV structures on
the Cu(111) surface: (b) 5-DBs, (c) 3DBs (5-9) and (d) M@3DBs, as
well as the DV structures (e) 5-8-5, (f) 555-777, (g) 4DBs and (h)
M@4DBs. The carbon atoms around the defect and the extra metal adatom
in the M@3DBs and M@4DBs structures are colored green and blue, respectively.
{[}Reprinted with permission from \cite{Wangdoi:10.1021/ja312687a}. Copyright (2013) American Chemical Society.{]}
\label{fig:Optimized-structures-defects-on-TM}}
\end{figure}

It is essential to know the formation energies of the point defects
as these indicate how feasible is their formation and hence their
likely concentrations. In \cite{Wangdoi:10.1021/ja312687a} the formation
energies were calculated using the C chemical potential from free-standing
graphene, while the chemical potential of the metal atom (needed when
an extra metal atom is involved as in the cases of M@3DBs and M@4DBs)
was taken from the metal bulk energy. The DFT calculated formation
energies of the vacancies are shown in Fig. \ref{fig:The-formation-energies}.
Of the SV structures the 3DB has the lowest formation energy for all
surfaces. This is deduced to be due to the passivation of the double
bonds by the metal atom. The M@3DBs structure has a higher energy
possibly because of a steric effect as there is limited space for
the extra metal atom. The vacancies formed on the Cu(111) surface
have a higher formation energy compared to the other two surfaces,
Ni(111) and Co(0001), and should therefore be a better catalyst for
graphene growth with an expected lower defect concentration. Importantly,
compared to the free-standing graphene the formation energies of SVs
on a substrate are significantly lower. This is because the dangling
bonds left behind around the vacancy after removal of the C atom bind
to the metal which reduces the system energy.

\begin{figure}
\centering{}\includegraphics[height=6cm]{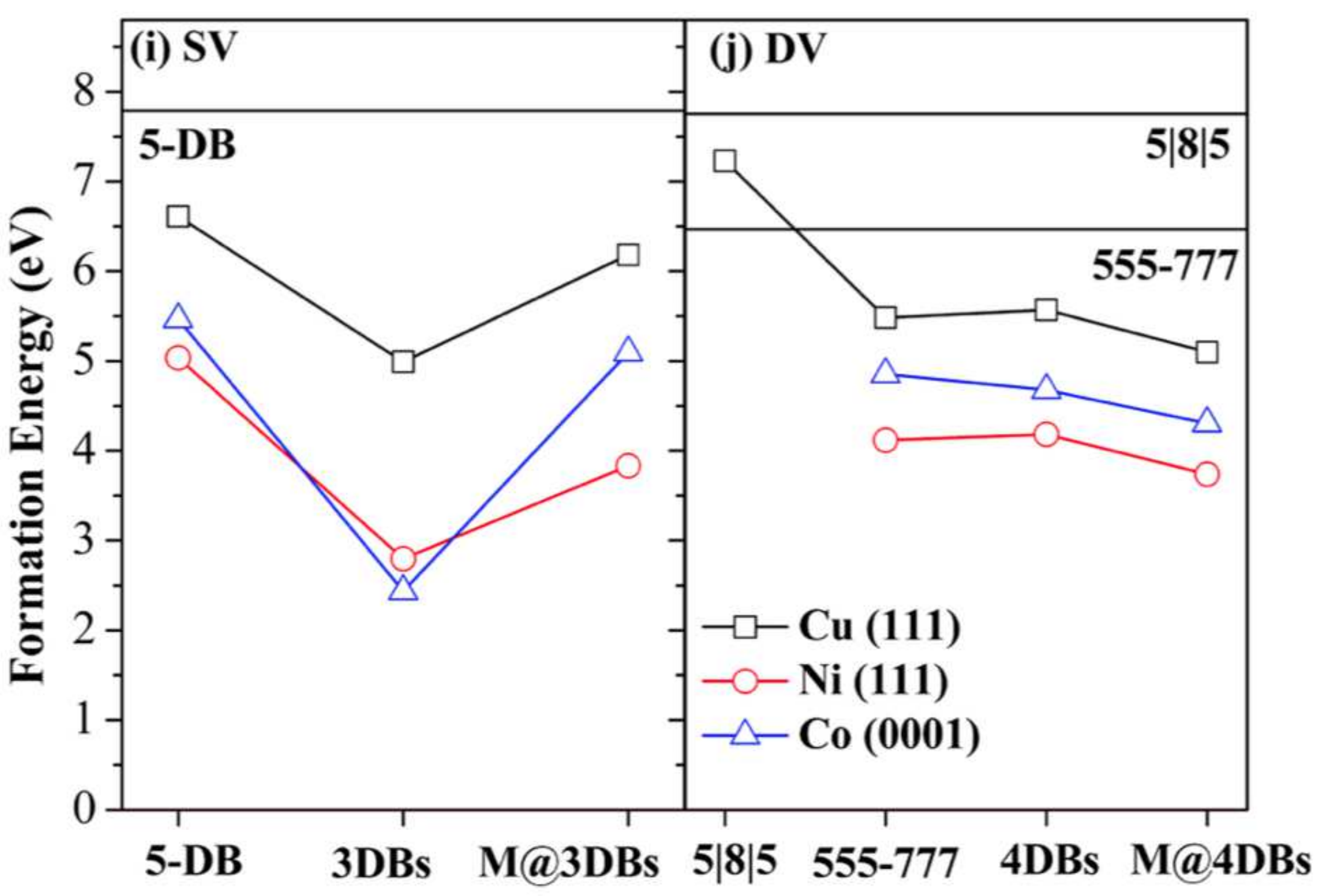}\caption{Formation energies of the various single and double vacancy structures
in free standing graphene (black horizontal lines) and on the Cu(111),
Ni(111) and Co(0001) surfaces. {[}Reprinted with permission from
\cite{Wangdoi:10.1021/ja312687a}. Copyright (2013) American Chemical Society.{]} \label{fig:The-formation-energies}}
\end{figure}

For DVs the calculated formation energies are also smaller than those
for vacancies in free-standing graphene. The 5-8-5 structure was found
to be stable only on the Cu(111) surface, since on the other two surfaces
the strength of the graphene-metal interaction causes the C-C bonds
to break, leading to the creation of the 4DBs structure instead. For
the 555-777 structure on the Cu(111), Co(0001) and Ni(111) surfaces
there is a drop in the formation energy of 0.92, 1.68 and 2.33 eV,
respectively, compared to unsupported graphene. From this the authors
concluded that the strength of the graphene-metal interaction increases
in the order of Cu(111) < Co(0001) < Ni(111). The 4DBs and M@4DBs
structures were found to be either similar to or even lower in formation
energy than the 555-777 structure which shows the significance of
the participation of the metal atoms in defect formation.

The diffusion of these vacancies on the surfaces was also calculated
in \cite{Wangdoi:10.1021/ja312687a} using the NEB method, the results
of which are shown in Fig. \ref{fig:The-diffusion-barriers}(a-b)
for the two most stable structures. Also included are the diffusion
pathways for M@3DBs SV and M@4DBs DV on Cu(111). For SVs the diffusion
barriers are higher than for free standing graphene. This is likely
to be due to the interaction of the metal surface with the vacancy.
The diffusion pathway depicted in Fig. \ref{fig:The-diffusion-barriers}(c)
involves the detachment of the metal atom from the graphene, then
switching of the carbon atoms and finally a different metal atom is
lifted from the surface to fill the vacancy. For the diffusion pathway
of the M@4DBs DV found by the authors the metal and carbon atoms switch
places. The diffusion barriers on each of the substrates are higher
than for SVs, and cannot be overcome at the growth temperature (1300
K). Therefore it is concluded that DVs must be immobile. Also note
that the diffusion barriers for the SV across these three metals increase
in the order G < Cu < Ni < Co, while for the DV they increase in the
opposite direction. Moreover, either order is different from that
established from the stability of the 555-777 DV defect which was
mentioned above. Therefore, care is needed in establishing relative
strengths of the graphene-metal interaction on the basis of graphene
defects.

\begin{figure}
\centering{}\includegraphics[height=7cm]{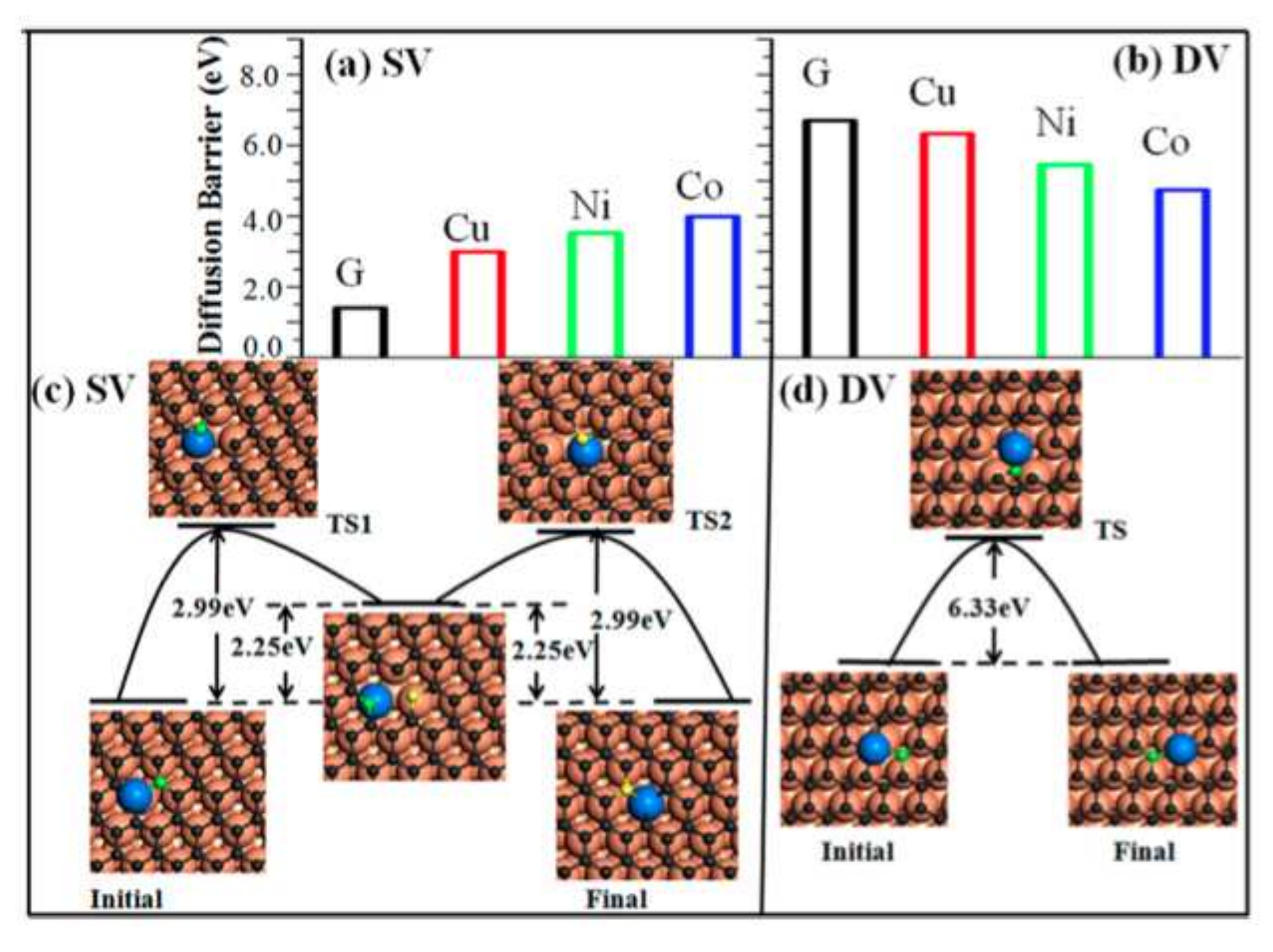}\caption{The diffusion barriers of (a) the M@3DBs single vacancy and (b) the
M@4DBs double vacancy on the Cu(111), Co(0001) and Ni(111) surfaces.
The corresponding values for the SV ad DV defects in the free standing
graphene are also given for comparison (as G). The diffusion pathways
for each of these vacancies on the Cu(111) surface are shown in (c)
and (d), respectively. {[}Reprinted with permission from \cite{Wangdoi:10.1021/ja312687a}. Copyright (2013) American Chemical Society.{]}
\label{fig:The-diffusion-barriers}}
\end{figure}

Other than the work described above there has been very little theoretical
work on defects in graphene on transition metal surfaces. The structure
of a SV in graphene on the Pt(111) surface has been calculated in
\cite{UgedaPhysRevLett.107.116803} using DFT with the van der Waals
interaction included. It was found that the relaxed structure which
most closely matches STM results is the 5-9 structure shown in Fig.
\ref{fig:(a)-The-relaxed}. It is seen that two carbon atoms of the
defect displace strongly towards the surface.

\begin{figure}
\centering{}\includegraphics[height=3cm]{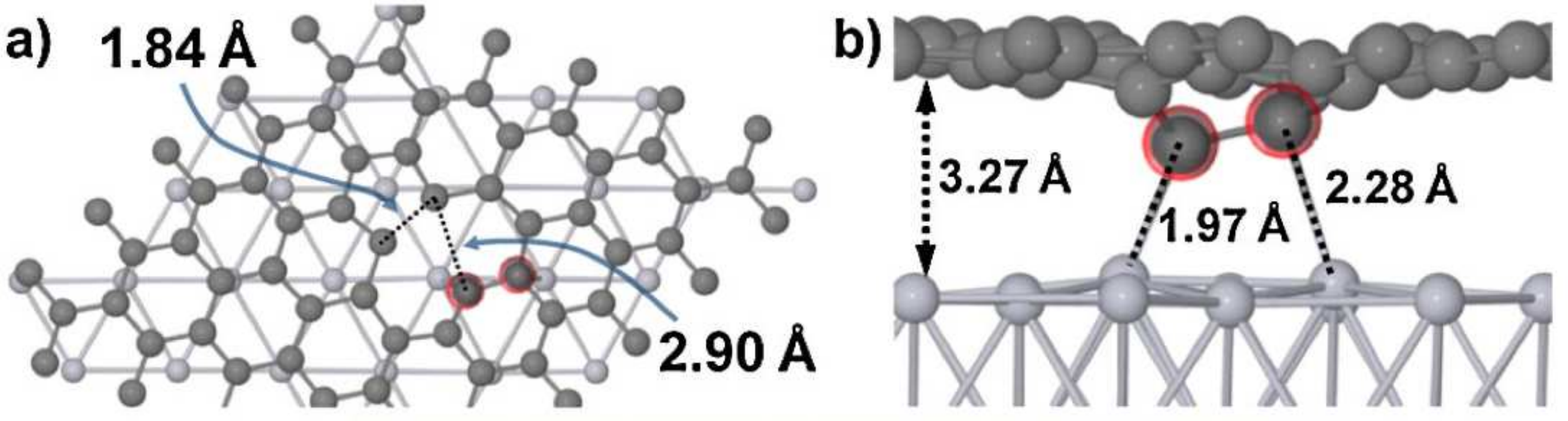}\caption{The top (a) and side (b) views of the relaxed SV in a 6 $\times$
6 unit cell (4 cells of the 3 $\times$ 3 graphene Pt(111) moir\'{e}).
The two carbon atoms that move towards the surface are highlighted.
{[}Reprinted with permission from \cite{UgedaPhysRevLett.107.116803}. Copyright (2011) by the American Physical Society.{]}
\label{fig:(a)-The-relaxed}}
\end{figure}

There has also been work on determining how graphene defects might
heal once they have formed on transition metal surfaces. The healing
pathway of a Stone-Wales defect on a Ni(111) surface was calculated
by Jacobson \emph{et al. }\cite{Jacobsondoi:10.1021/jz2015007}. From
this it was determined that compared to free-standing graphene the
Stone-Wales defect is stabilized by the Ni(111) surface and has a
formation energy 1.0 eV lower. Furthermore the barrier to healing
the defect on Ni(111) is calculated as 2.88 eV, which is lower than
the 4.10 eV required for healing in free-standing graphene. This is
due to the presence of the surface breaking the symmetry of the transition
state, which proceeds by the rotation of a C-C bond as in the free
standing case. In the intermediate state the carbon atoms are stabilized
by the surface and the barrier of the reconstruction is reduced.

The removal of metal atoms at the edge of a graphene front was investigated
in \cite{Wangdoi:10.1021/ja312687a}. If these remain in the graphene
they would become embedded in the graphene as it grows, forming M@4DBs
defects. To remove the metal atoms two carbon atoms are required to
diffuse to the edge and then the suggested process occurs in two steps.
First, one carbon atom must attach onto the metal atom to form a hexagon,
and then the next carbon atom replaces the metal to complete the graphene
edge. The whole process is depicted in Fig. \ref{fig:The-healing-process}
for graphene on the Cu(111) surface. The DFT calculated barriers in
this case are 1.86 eV for the first transition state and 0.33 eV for
the second transition state. By comparison of this process on the
Ni(111) and Cu(111) surfaces it was deduced that healing in this manner
has a higher rate on the latter surface for the commonly used growth
conditions (with $T=1300$ K and a growth rate of 10-100 nm/s).

\begin{figure}
\centering{}\includegraphics[height=7cm]{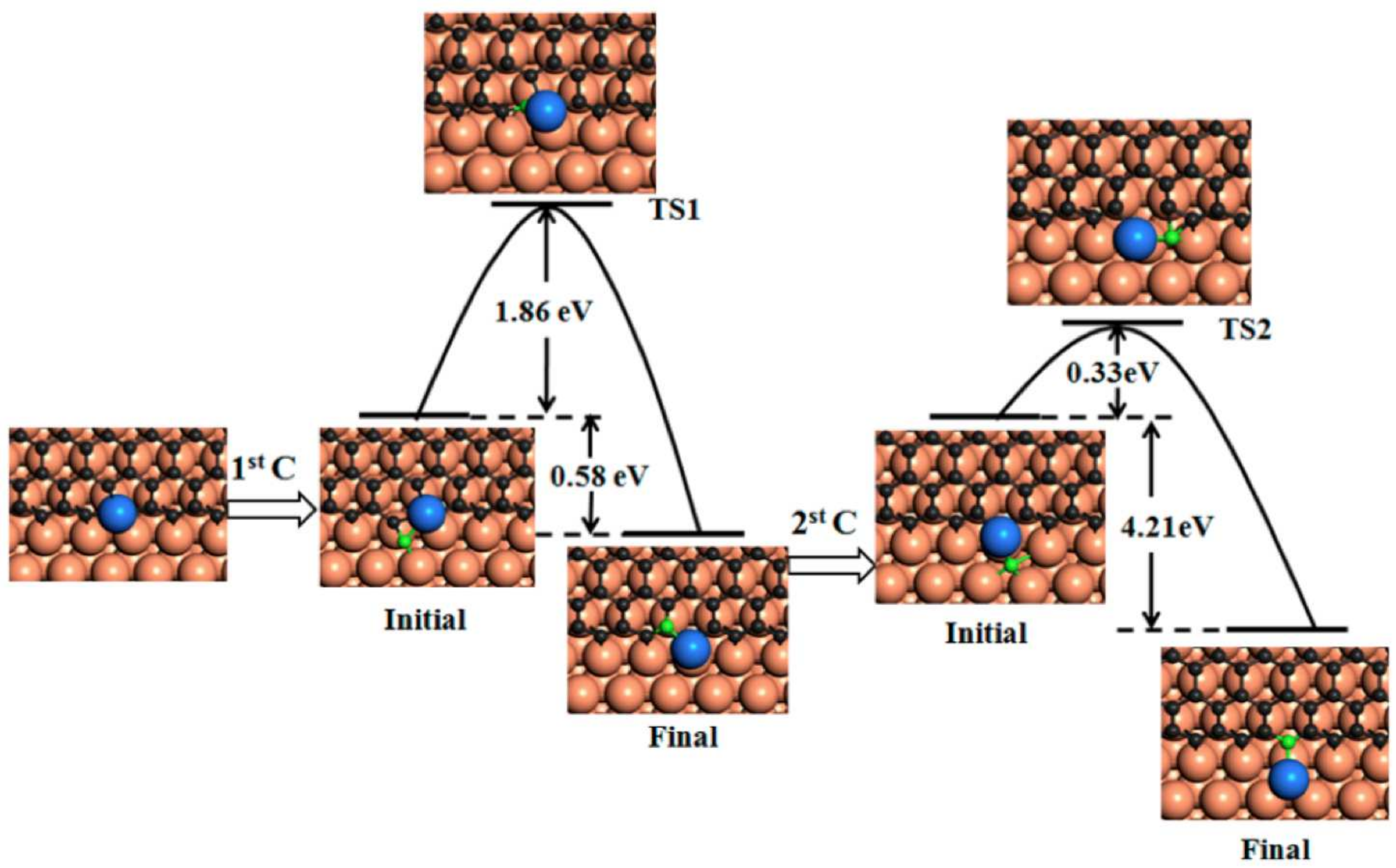}\caption{The healing process for an M@4DBs defect on a graphene edge on the
Cu(111) surface. The barriers for each of the transition states are
indicated. {[}Reprinted with permission from \cite{Wangdoi:10.1021/ja312687a}. Copyright (2013) American Chemical Society.{]}
\label{fig:The-healing-process}}
\end{figure}

\subsection{Attempts to simulate the growth\label{sub:kMC}}

Above we reviewed studies of monomers, dimers and various larger C
clusters on several metal surfaces used for growing graphene. Not
only stabilization energies of these C species were calculated (usually
relative to the energies the C atoms would have in the graphene feedstock
serving as their source), but also mechanisms of their migration across
the surface were studied, as well as (in some limited specific cases)
association and dissociation energies. It is widely agreed that C
clusters tend to attach to metal steps where they are more stable,
and hence nucleation of graphene may well start mostly at the steps,
i.e. heterogeneously. Further growth would require attachment of monomers,
dimers and other C species to the formed nucleus. If the C species
are added extremely slowly, the growing film would have sufficient
time to ``heal'' itself leading eventually to a reduction in free
energy of the growing structure; however, depending on temperature
and the supersaturation $\Delta\mu$, the influx of C atoms may be
so fast that only partial healing is possible leading to a defective
graphene structure (see also section \ref{sub:Defects-in-graphene-on-TM}).
That is why it is paramount to model the growth kinetics in real time
under different growth conditions in order to derive the best experimental
procedure leading to large graphene sheets with a low concentration
of defects.

In this Section several simulations of graphene growth found in the
literature will be reviewed. This direction of research in application
to graphene growth is in its infancy mostly due to the sheer complexity
of the system; however, in the near future we should expect more simulations
of this kind.

In the KMC simulations by Wu\emph{ et al.} \cite{Wu} addition of
carbon species to the edges of graphene islands on the Ir(111) surface
was modeled based on first principles calculations performed for the
adsorption of different carbon species on the surface, their diffusion
and attachment to the graphene edges, as has already been described
in Section \ref{sub:Early-Stages-of}. This allowed the authors to
determine the dependence of the growth rate on the C monomer concentration.
For this study Wu \emph{et al.} developed a multi-scale ``growth-front-focused''
KMC model. This allowed only the moving growth front to be considered
in the actual simulation, and also the authors implemented a special
procedure which takes into account the vastly different carbon species
densities (and hence their fluxes) that vary over many orders of magnitude.
In their model a honeycomb lattice was divided into a graphene region,
a growth front, a diffusion layer and the far field, all shown in
Fig.\ref{fig:Wu3}(a). In the far field thermal equilibrium is assumed
between differently sized C$_{N}$ species, so that densities of the
species with $N>2$ (dimers, trimers, etc., up to hexamers) were directly
related to the monomer density $C_{1}$ via the $NC{}_{1}\rightleftharpoons C_{N}$
equilibrium condition. From the far field the various carbon species
can diffuse across the diffusion layer with a flux that is calculated
from the known species concentration in the far field region and the
corresponding diffusion barriers calculated from first principles.
Then at the growth front the attachment and detachment of the C$_{N}$
species to and from the graphene edge are considered explicitly using
the KMC method. The attachment and detachment rates for a particular
species were determined from first principles calculations (section
\ref{sub:Atomistic}), which took into account the effect of the underlying
Ir(111) surface. To overcome the problem of vastly different species
fluxes, the authors proposed a scheme whereby at every single time
only one species contributes to the growth until the growth gets stuck.
At this point a different species with lower flux is ``switched on''
and this process is continued until the growth stops again. The
simulation starts from monomers which are the only ones contributing
to the flux in the diffusion layer region, then it is switched to
dimers, then trimers, etc. Finally, after the last species (C$_{6}$)
has been considered, attention is returned into the monomers again.
Although this procedure is very artificial and is likely to be far
from being realistic, this is undoubtedly an interesting attempt to
tackle this vastly complicated problem of many species contributing
in their own way to the growth under realistic conditions when their
concentrations differ by many orders of magnitude. This clearly demonstrates
the sheer complexity of modeling the growth of graphene at the atomic
level which still needs to be properly addressed.

The nucleation stage of the growth process was not considered explicitly
by Wu\emph{ et al.} \cite{Wu}; it was assumed that the initial nucleus
of the graphene ribbon was already formed, most likely at a step edge.
This approach therefore can only be applied to the steady-state growth
stage of the graphene islands. In particular, it can be used, and
this is exactly what has been done by Wu\emph{ et al.} \cite{Wu},
to determine the steady-state rate, $r$, of graphene growth.

The results from these KMC simulations are shown in Fig. \ref{fig:Wu3}(b).
It was found that the growth rate $r$ can be fitted to an equation
of the form

\begin{equation}
r=aC_{1}^{n}+d\;,\label{eq:nonlinear-growth-rate-Wu-KMC}
\end{equation}
where $C_{1}$ is the monomer concentration and the constant $n$
= 5.25. This agrees very closely with the experimental results from
the LEEM studies of Loginova \emph{et al.} \cite{Loginova08} where
it was found that the growth rate has a nearly quintic dependence
on the monomer concentration. Unfortunately, no detailed analysis
has been made in \cite{Wu} concerning the origin of this highly non-linear
dependence; in particular, it is not clear whether this result can
be reproduced, at least qualitatively, by just accepting clusters
$C_{N}$ with $N$ values of just 1 (monomers) and 5 (quintets), or
whether the existence of other species is essential. Recall (section
7.1) that in the phenomenological rate equation-based
study by Zangwill and Vvedensky \cite{Vvedensky} where the non-linear
growth was also explained, only monomers and C$_{5}$ clusters were
accounted for. In addition, only reactions between $C_{N}$ and $C_{1}$
species were considered by Wu\emph{ et al.} \cite{Wu} and since the
decomposition of $C_{N}$ clusters does not happen completely in one
step, but rather gradually, producing smaller clusters along the way,
it is not obvious how this assumption may have affected the calculated
kinetics.

\begin{figure}
\centering{}\includegraphics[height=6cm]{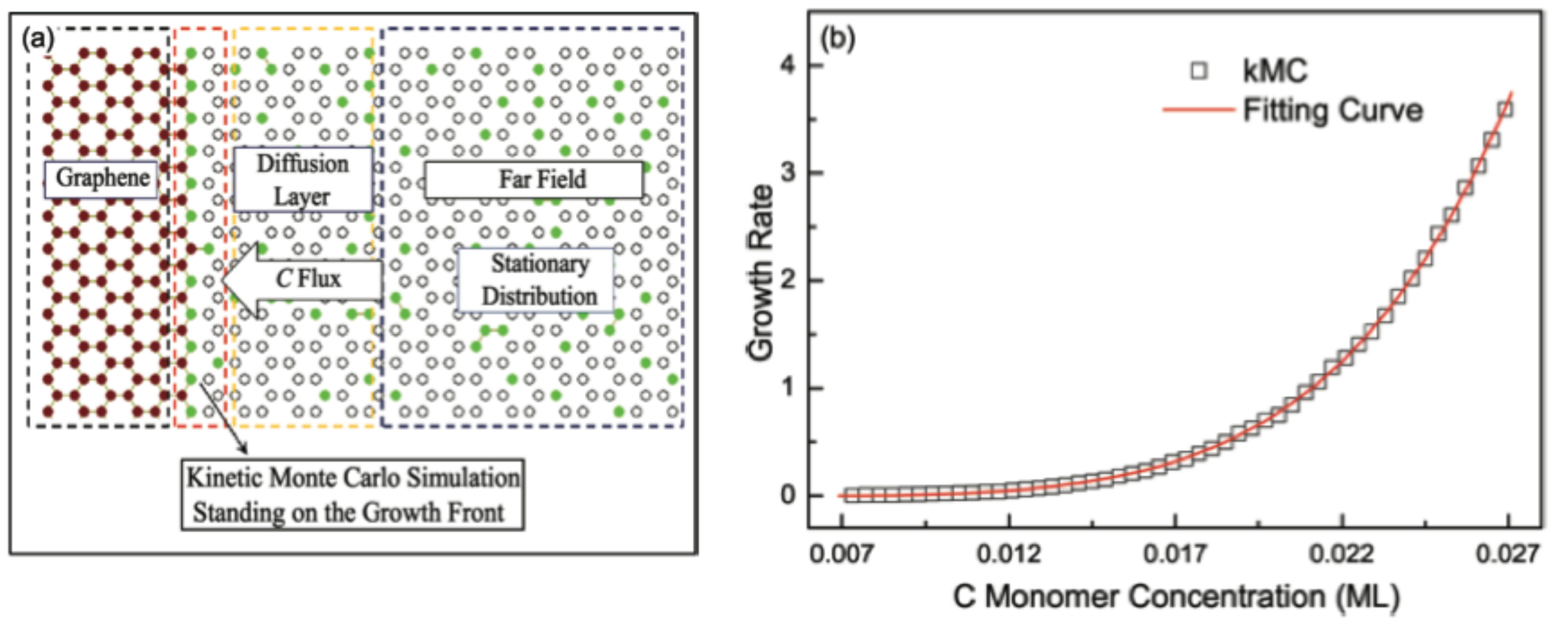}\caption{(a) The KMC model of graphene growth based on a division of the whole
system into four regions (from left to right): already grown graphene
ribbon, growth front where new species are able to attach/detach to/from
the ribbon edge, diffusion layer providing influx of C$_{N}$ species
($N=1,\ldots,6$), and the far field region where thermal equilibrium
between C$_{N}$ species is assumed; (b) the calculated dependence
of the growth rate on the C monomer concentration showing the fifth
power growth rate by the curve fitted to the KMC calculated growth
rate (boxes). {[}Reprinted with permission from \cite{Wu}. Copyright (2012) American Chemical Society.{]} \label{fig:Wu3}}
\end{figure}

Haghighatpanah \emph{et al.} used a Grand Canonical Monte Carlo (GCMC)
method to determine how various factors, such as temperature $T$
and chemical potential $\mu$ of the C feedstock, may affect the growth
of graphene both in vacuum and on the Ni(111) surface \cite{Haghi}.
As calculations of this type would be prohibitively expensive with
DFT, a much computationally cheaper TB model was used instead whose
parameters were fitted to the DFT results on a number of reference
systems (as reported in section \ref{sub:Atomistic}), and then used
as a basis for their GCMC simulations. As C atoms are added to the
structure, a certain equilibration consisting of a relatively large
number of displacement moves was performed. Then, the average numbers
of C atoms in hexagons $N_{H}$, non-hexagonal rings $N_{R}$, and
strings $N_{S}$ were determined and used to assess the quality of
the growing graphene. From this the ideal temperature and chemical
potential were determined for producing the highest quality graphene
where $N_{H}$ is large and $N_{R}$ and $N_{S}$ are small.

First, the most efficient parameters for the GCMC calculations were
investigated using a free standing graphene sheet (no surface). It
was noticed that at 1000 K defects were formed that were not easily
healed and became embedded in the structure, whereas at 2000 K healing
of defects was reduced due to the rapid motion of the C atoms. Therefore
1500 K was used as the growth temperature for production runs.

For all chemical potentials $N_{H}$, $N_{R}$ and $N_{S}$ increase
with the number of added carbon atoms. For $\mu_{C}$= -8.5 eV most
carbon atoms attach at the end of chains as this is the most energetically
favorable site. As a result the increase in $N_{H}$ is lower than
for the cases with higher chemical potentials where attachment is
also likely at sites which are energetically less favorable (at hexagonal
and non-hexagonal sites). Since fewer carbon atoms are added to the
non-hexagonal rings in the early stages of growth for the chemical
potential of -8.0 eV, this value of $\mu_{C}$ was used in further
calculations. The changes in $N_{H}$, $N_{R}$ and $N_{S}$ were
also investigated for different numbers of equilibration steps (denoted
ND) computed between C atom additions, and the results show that larger
ND values allow better healing of newly formed defective structures
at the edges; however, once a defect is embedded inside the growing
island, it stabilizes and its healing is found to be greatly suppressed.

\begin{figure}
\begin{centering}
\includegraphics[height=10cm]{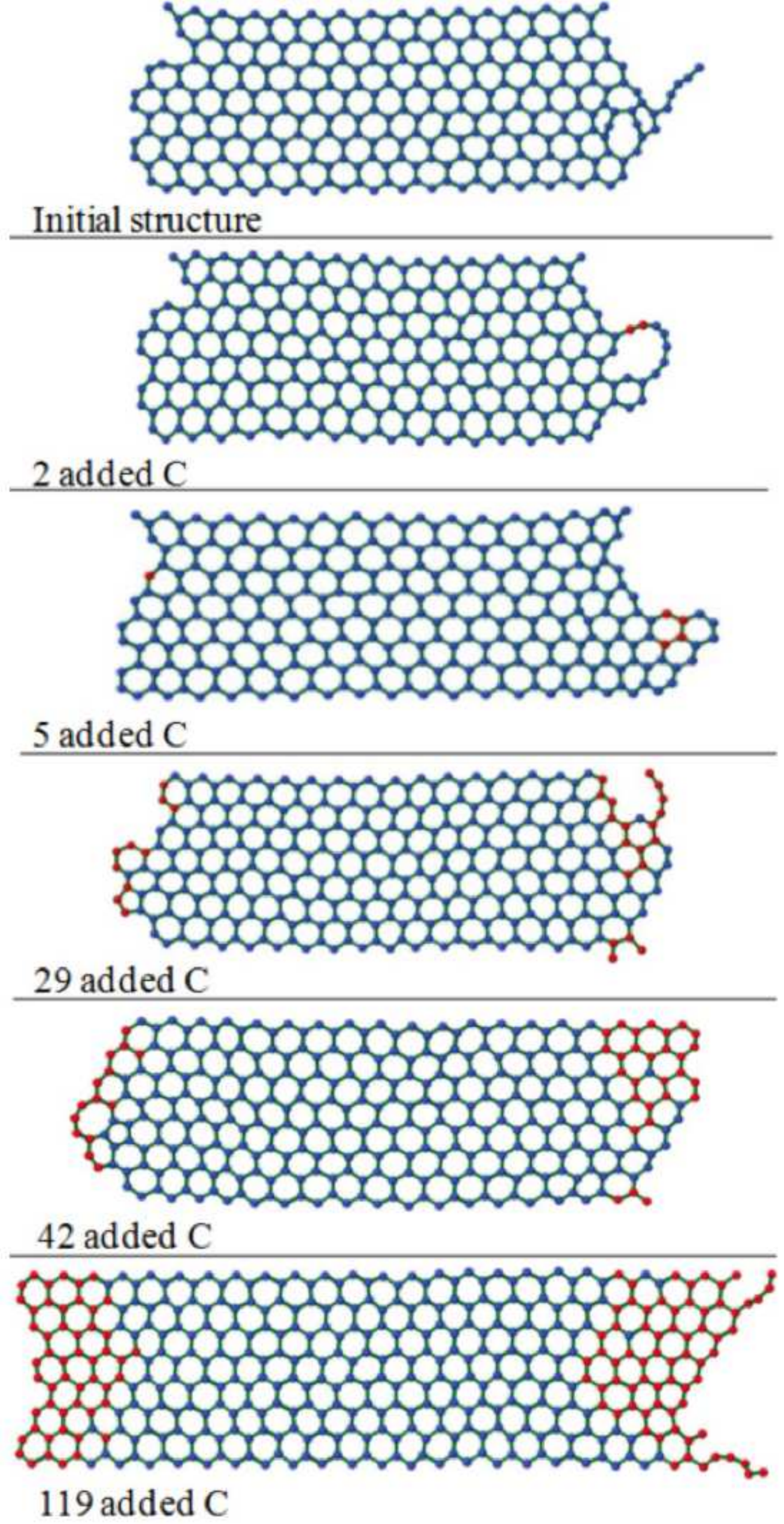}
\par\end{centering}

\caption{Snapshots of growth of free standing graphene during MC simulations
starting from the initial structure (the top panel) as more C atoms
are added. While the initial C atoms are shown in blue, newly added
areas are colored in red. {[}Reprinted with permission from \cite{Haghi}. Copyright (2012) by the American Physical Society.{]}
\label{fig:growth-free-Haghigh1}}
\end{figure}

\begin{figure}
\begin{centering}
\includegraphics[height=9cm]{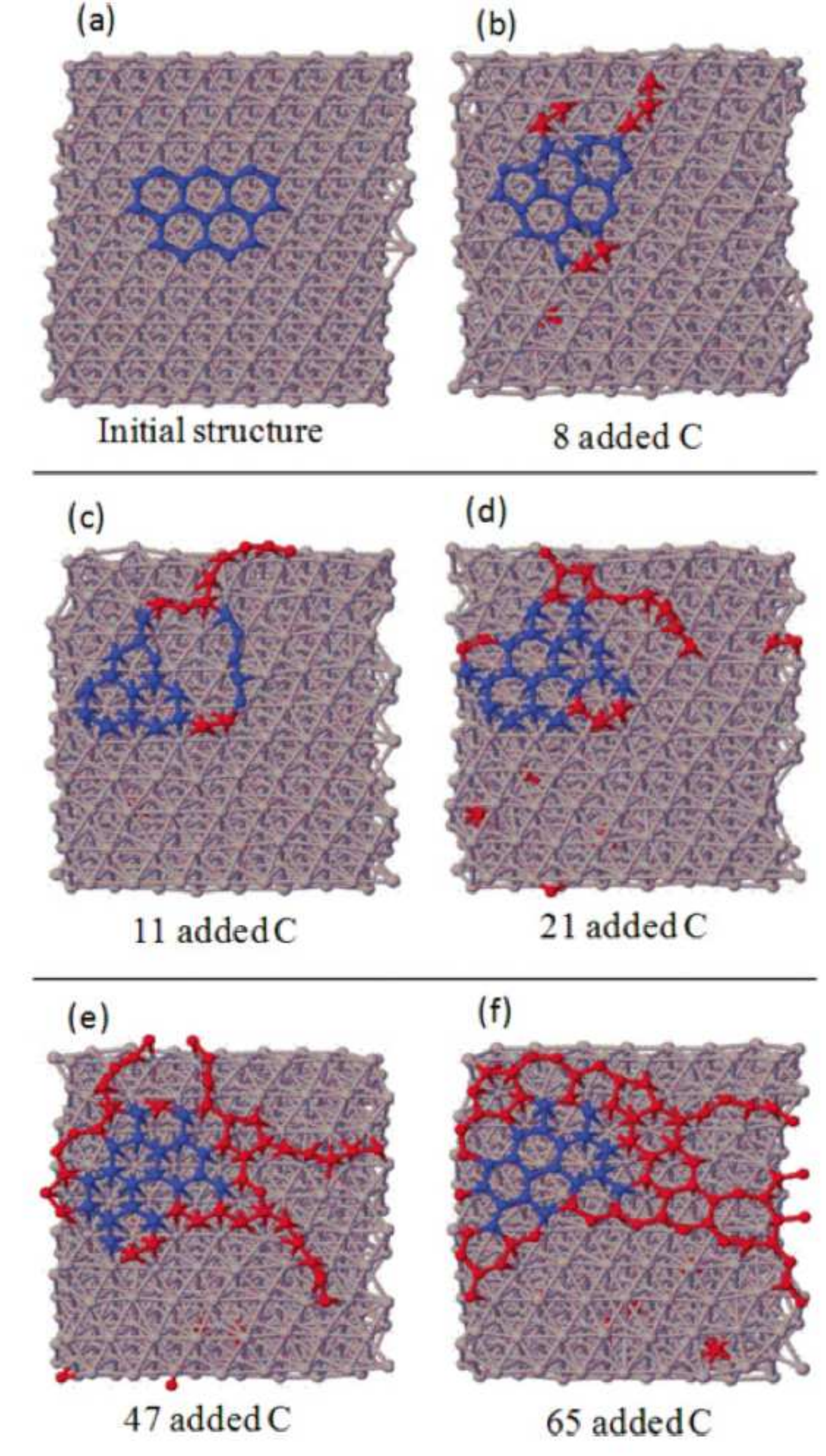}
\par\end{centering}

\caption{Selected snapshots of a graphene island grown on the Ni(111) surface.
Blue and red colors correspond to initial and newly added C atoms.
{[}Reprinted with permission from \cite{Haghi}. Copyright (2012) by the American Physical Society.{]} \label{fig:growth-on-Ni-Highhigh}}
\end{figure}

Using $T=1500$ K, $\mu_{C}$ = -8.0 eV and ND $=3\times10^{5}$ the
growth of graphene was simulated first without any surface present,
Fig. \ref{fig:growth-free-Haghigh1}, and then on the Ni(111) surface,
Fig. \ref{fig:growth-on-Ni-Highhigh} (Note that the temperature used
is to be rescaled downward when comparing theory and experiment since
the melting temperature of Ni with this model is about 15\% higher
than measured \cite{Los-Pellenq-2010}.) The graphene produced with
these conditions was optimized to grow without forming many strings
and non-hexagonal rings, and this was successful for the free standing
graphene. The situation is more complex, however, when calculations
were performed on the Ni surface as exemplified in Fig. \ref{fig:growth-on-Ni-Highhigh}.
It was found that C atoms may go to subsurface sites, form chains
on the surface, and existing hexagons may rupture and then reconnect
as more C atoms are added. It was also found that flakes become relatively
more stable on the surface than in free space once they have grown
in size: the defect-free area in flakes (with the same initial nucleus
size and shape) on average is larger on the Ni surface. The two cases
are not directly comparable as different nuclei were used in each
case, shown in Figs. \ref{fig:growth-free-Haghigh1} and \ref{fig:growth-on-Ni-Highhigh}:
the initial 5 hexagon graphene flake used in the simulation on Ni
is much smaller than the graphene sheet used for the free standing
graphene with no surface present. Still, a general conclusion was
made, based on many such calculations, that the surface stabilizes
the formation of graphene.

There are some indications that the calculations presented above may
be incomplete. Indeed, after 25 atoms have been added $N_{S}$ reaches
around 6-8 atoms and then does not increase any further as more atoms
are added. This is suggested to be because the surface stabilizes
the formation of strings due to the strength of the Ni-C bond. It
can also be noted that $N_{H}$ and $N_{M}$ increase with the addition
of C atoms. Over time it is presumed that these values will saturate
as the surface becomes covered in carbon, however, longer simulations
are needed to determine whether this is the case. Comparing the number
of C atoms in hexagons for the growth on Ni(111) and without a surface
suggests that graphene on the Ni surface is less stable.

Finally, Haghighatpanah \emph{et al.} \cite{Haghi} tested the ability
of a defected graphene sheet to heal under simulated annealing. For
this the growth of two defected graphene sheets was simulated starting
from one created with a very high chemical potential that gave 3D
defects, Fig. \ref{fig:HH5}(a), and another one grown using a very
low chemical potential which resulted in long carbon strings, Fig.
\ref{fig:HH5}(b). To simulate the annealing process a temperature
of 1500 K and ND $\thickapprox2\times10^{8}$ were used. The figure
shows that the 3D and embedded defects do not heal easily when annealed.
However chain and edge defects are more susceptible to healing and
can normally form hexagons. Defect free hexagons formed during the
annealing (shown in red) can become separated by defects that appear
similar to domain boundaries. These are difficult to heal by annealing.
Therefore it is suggested that in order to produce high quality graphene
the initial growth conditions should be controlled rather than relying
on annealing as a postprocessing procedure to remove defects, especially
as the metal substrate stabilizes defects.

\begin{figure}
\centering{}\includegraphics[height=6cm]{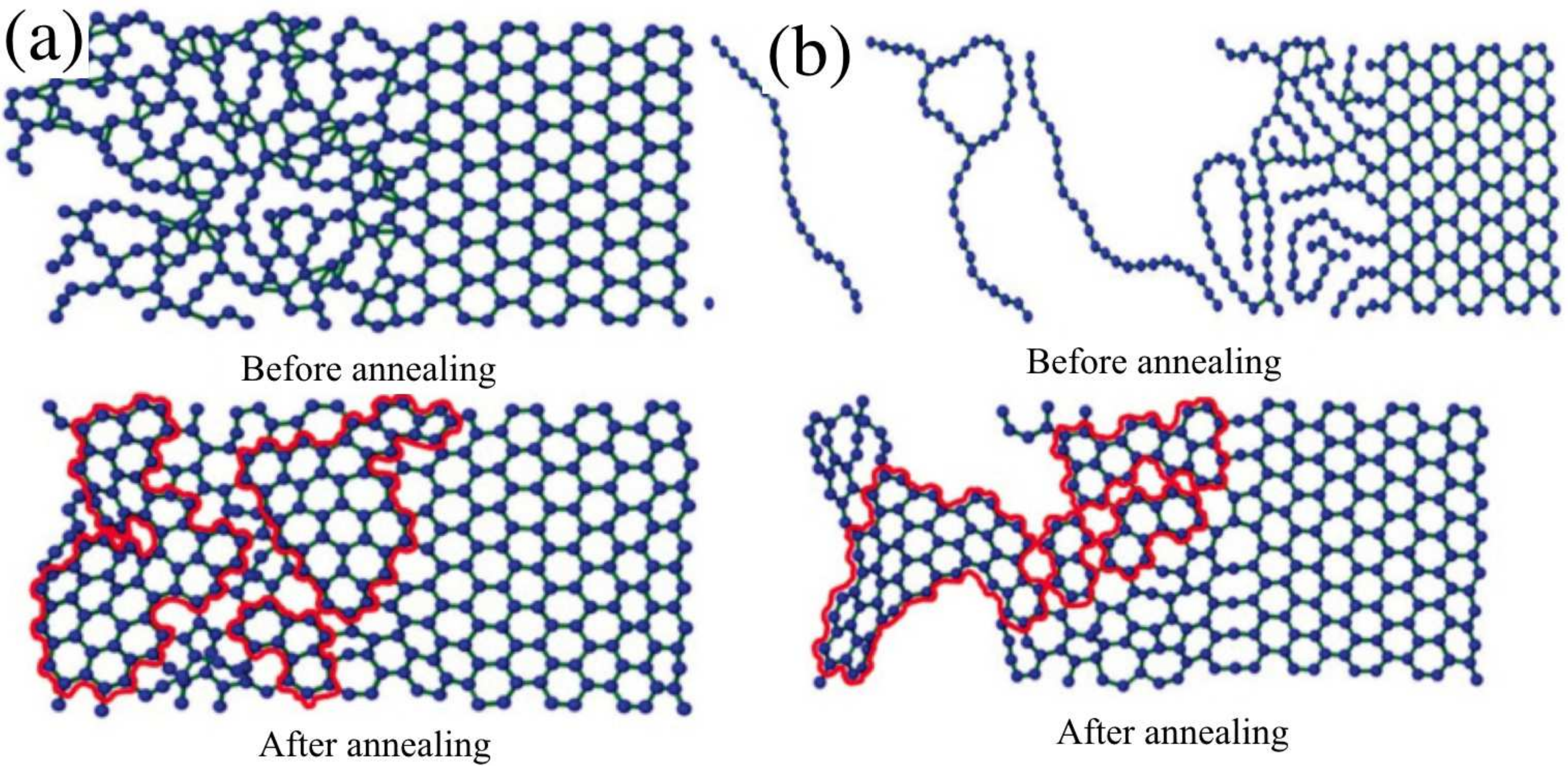}\caption{Defective graphene structures before and after thermal annealing at
1500 K for ND $\thickapprox2\times10^{8}$. The top panel structure
in (a) is formed using a high chemical potential, whereas in (b) a
low chemical potential was used. {[}Reprinted with permission from
\cite{Haghi}. Copyright (2012) by the American Physical Society.{]} \label{fig:HH5}}
\end{figure}

Although the results of the simulations like this one are undoubtedly
useful, one has to be careful in interpreting them, as was noted in
Section \ref{sub:Simulations-of-dynamics-methods}, as realistic simulations
of growth. Indeed, everything depends on the nature of ``moves''
performed when equilibrating the system between C atom additions,
where the C atoms were added, etc. In our view the GCMC simulations
are to be used with care; however, KMC based simulations are capable
of simulating the growth, though the latter are prone to yielding
incorrect results if some of the essential elementary processes are
missing or their barriers were under- or overestimated.

In GCMC simulations based on a TB model a wide range of temperature
and C atom chemical potential $\mu_{C}$ was studied in order to understand
the favorable conditions for formation of a graphene-like layer on
the surface. In these simulations the MC moves corresponded only to
C atoms being added and/or removed from the surface. Examples of
the results of the simulations for 1000 K for five values of $\mu_{C}$
are shown in Fig. \ref{fig:Snapshots-of-GCMC-Amara-2006} (the less
negative value corresponds to a higher ``pressure'' or ``concentration''
of C atoms in the feedstock with which C atoms on the surface are
in equilibrium). At large negative values of $\mu_{C}$ carbon atoms
penetrate inside the Ni. As the value of $\mu_{C}$ is increased (gets
less negative), first the appearance of C chains creeping on the surface
is observed ($\mu_{C}=-5.75$ eV), and then at $\mu_{C}=-5.0$ eV
a graphene-like structure on the surface emerges. Note that the structure
is defective which is most likely due to the fact that the simulation
was not run for sufficiently long time, i.e. the system was not fully
equilibrated. It is noted by the authors that the graphene-like structure
appears only after chains start to ``collide'' and three-coordinated
C atoms appear serving as nucleation sites for the carbon monolayer.
This requires a sufficiently high value of the C chemical potential
($\mu_{C}\succeq-5.0$ eV) and a certain temperature window up to
and around 1000 K. For instance, at 1500 K a carbidic phase was seen
formed.

\begin{figure}
\begin{centering}
\includegraphics[height=4cm]{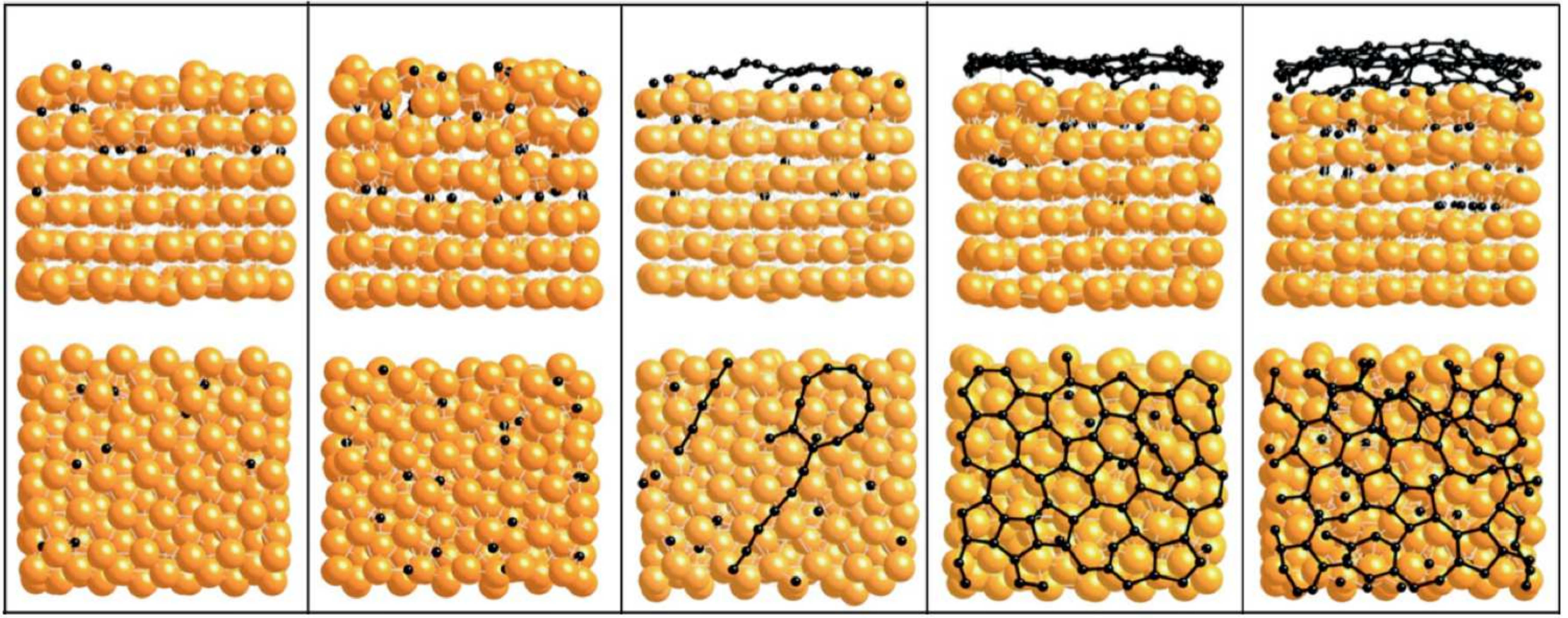}
\par\end{centering}

\caption{Snapshots of GCMC simulations of a Ni(111) surface in thermodynamic
equilibrium at 1000 K with the C atom system at chemical potentials
$\mu_{C}$ (from left to right) equal to -6.5, -6.0, -5.75, -5.0 and
-4.5 eV. Lower panels: top view; upper panels: side view. {[}Reprinted
with permission from \cite{Amara:2006hd}. Copyright (2006) by the American Physical Society.{]}\label{fig:Snapshots-of-GCMC-Amara-2006}}
\end{figure}

A recent study \cite{meca13} based on the phase-field method has
modelled the epitaxial growth of graphene islands on copper by CVD
of methane (CH$_{4}$) in the presence of H$_{2}$. The calculations
were based on a standard phase-field formulation of the motion of
island boundaries, but with anisotropies in the kinetic coefficients,
step energies, and diffusion, with the latter included to account
for the crystallinity of the substrate. The calculations assumed an
effective adatom deposition flux to avoid the complications of the
reaction sequence of a polyatomic precursor. Experiments were carried
out on several orientations of copper, such as (100), (310), (221)
and (111), and the comparisons with phase-field calculations presented
as an island morphological ``phase diagram'' shown in Fig.~\ref{fig-last-DDV-phase}.

\begin{figure}[h!]
\centering{}\includegraphics[height=6cm]{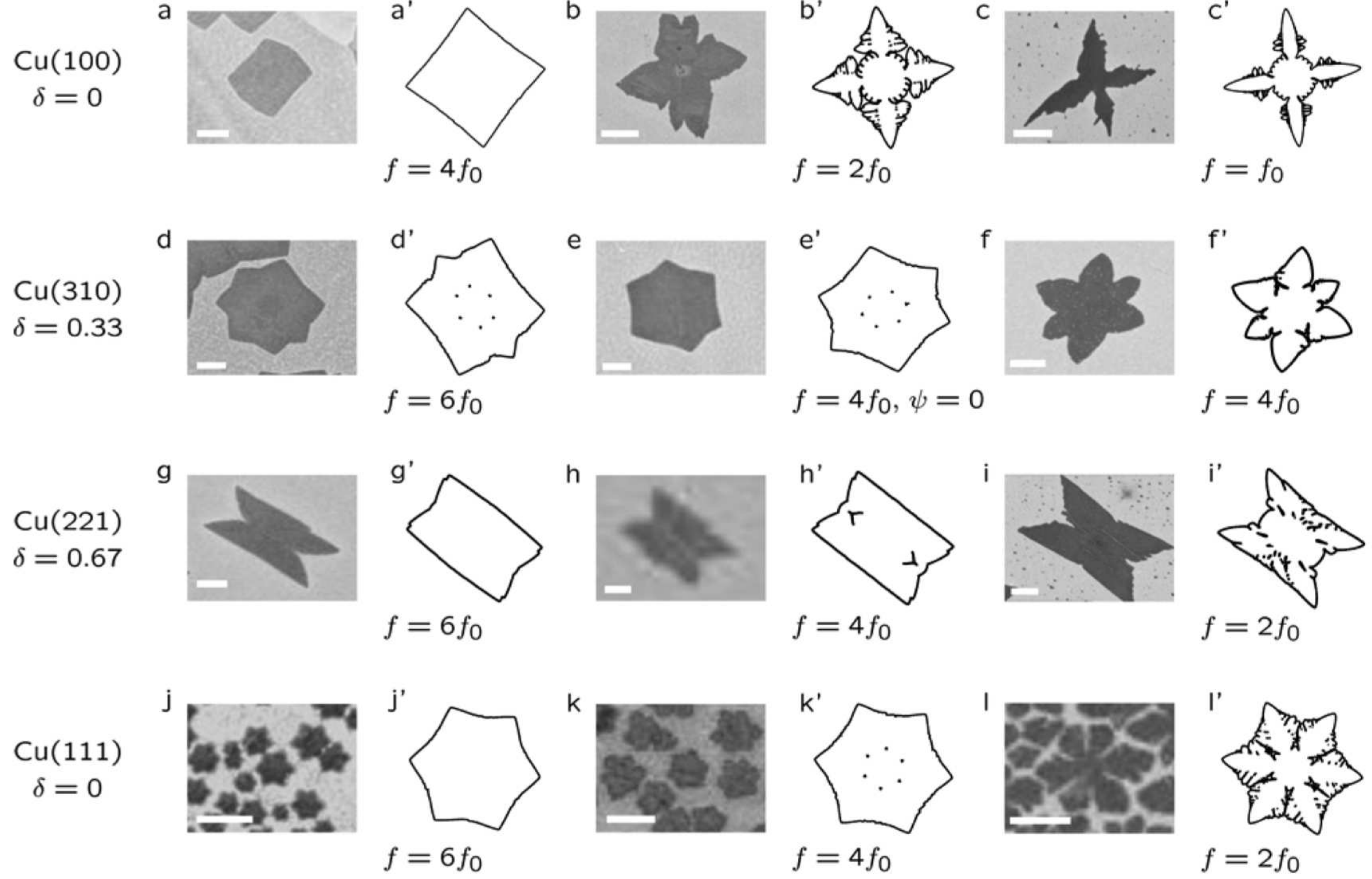} \caption{Shapes of graphene islands on copper facets with different exposure
times and partial pressures of C$_{2}$H$_{4}$ and H$_{2}$ at $T=$ 1303 K.
(a), (b), (c):~Cu(100) after 6, 4, and 2 min exposure with partial
methane/hydrogen pressure ratios of 3$\times$10$^{-4}$, 0.028, and
0.81, respectively. Scale bars are 10, 4, and 2~$\mu$m, respectively.
(d), (e), (f):~Cu(310) after 20, 10, and 2 min exposure with partial
pressure ratios of 3$\times$10$^{-4}$, 7.7$\times$10$^{-3}$, and
0.028, respectively. Scale bars are 10, 10, and 4~$\mu$m, respectively.
(g), (h), (i):~Cu(221) after 7, 6, and 3 min exposure with partial
pressure ratios of 1.6$\times$10$^{-5}$, 0.028, and 0.035, respectively.
Scale bars are 4, 4, and 2~$\mu$m, respectively. (j), (k), (l):~Cu(111)
after 7, 4, and 3 min exposure with partial pressure ratios of of
3$\times$10$^{-6}$, 1.6$\times$10$^{-5}$, and 0.026, respectively.
Scale bars are 4, 2, and 2~$\mu$m, respectively. Shapes from simulations
are presented next to the corresponding experimental images. {[}Reprinted
with permission from \cite{meca13}. Copyright (2013) American Chemical Society.{]}\label{fig-last-DDV-phase} }
\end{figure}

The main features of this phase diagram are as follows. The graphene
nuclei on Cu(100) begin as squares, but evolve into shapes with four
approximately symmetric dendritic extensions. The phase-field calculations
used six-fold symmetry for the kinetic and edge anisotropies and an
isotropic diffusion tensor. The nuclei on Cu(111) show hexagonal symmetry
with dendritic features which develop with time. The phase-field calculations
used six-fold symmetry for the kinetic and edge anisotropies with
isotropic diffusion. The nuclei on Cu(310) show a polygonal shape
initially, which evolves into a distorted hexagon before developing
cusps at the island boundary. A six-fold symmetry and an asymmetric
diffusion constant was used for the phase-field calculations. Finally,
the nuclei in Cu(221) exhibit approximately rectangular shapes with
shallow depressions in the center of the shorter sides, which evolve
into a ``butterfly'' shape. The phase-field calculations used a
four-fold symmetry with an asymmetric diffusion constant.

These comprehensive phase-field calculations show that the main features
of the complex morphologies of graphene islands on various copper
surfaces result from the competing effects of the anisotropy of the
kinetics of the growing graphene edge and the crystallographic orientation
of the substrate through the anisotropy of surface diffusion. The
value of performing experiments in tandem with theoretical calculations
is evident, as the growth conditions (surface orientation, flux, temperature)
can be varied systematically. This study provides a platform for simulations
with atomistic spatial resolution, as was done some years ago for
semiconductors \cite{itoh98,kratzer02}.

\section{Simulations of graphene growth from silicon carbide \label{sec:SiC}}

In spite of the fact that graphene growth from SiC is experimentally
well established (Section \ref{sec:SiC}), the corresponding theoretical
effort is lagging behind. This is partially explained by the challenges
this system presents, related, in the first instance, to the complexity
of the SiC surface itself as was explained in Section \ref{sec:SiC}
and the way graphene grows. Correspondingly, atomistic processes associated
with  graphene growth on the SiC surface are much less understood.

\subsection{Explorations into growth of graphene on SiC\label{sub:explorations-of-growth-SiC}}

The previously described studies give useful information about the
structure of the buffer and first graphene layers on SiC, but they
do not really consider processes which might be relevant for the graphene
growth. Several studies have been conducted to date in which some
of the relevant processes were considered. In all these studies it
was still assumed that the carbon for buffer and first layer graphene
growth comes predominantly from the sublimation of SiC at step edges
and also to a lesser degree from the sublimation of point defects
on terraces.

\begin{figure}
\begin{centering}
\includegraphics[height=10cm]{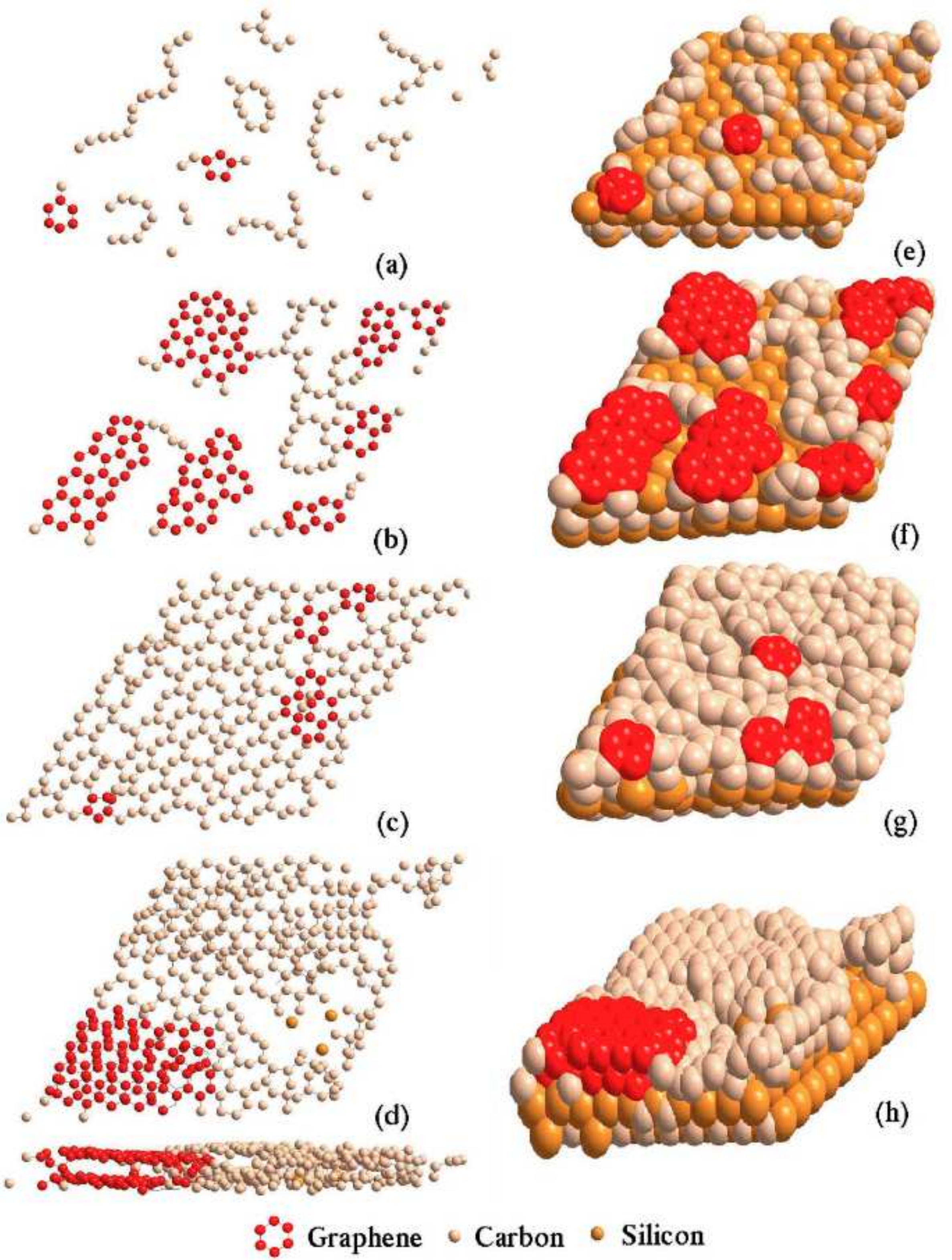}
\par\end{centering}

\caption{Typical calculated geometries of carbon atoms on top of the Si-terminated
SiC surface for various C coverages achieved by removing Si atoms
from a different number of SiC layers: (a,e) one; (b,f) two; (c,g)
three; (d,h) four. The annealing temperature is 2000 K. Graphene-like
clusters with hexagonal ordering are shown in red. The topmost layers
of carbon atoms are shown in the left panels, while the SiC layer
below is seen in the corresponding right panels. {[}Reprinted with
permission from \cite{Tang08}. Copyright (2008), AIP Publishing LLC.{]}\label{fig:Tang-1-4-layers-structures}}
\end{figure}

Tang \emph{et al.} \cite{Tang08} and\textcolor{red}{{} }Jakse \emph{et
al.} \cite{Jakse} have used different versions of Tersoff's EP \cite{Tersoff-potentials-1989,Tersoff-potentials-1994}
based MD simulations to examine the transformation of carbon layers
on top of SiC into graphene, and particularly the effect of annealing
temperature and carbon coverage on the graphitization temperature
which is the lowest temperature at which the graphene growth kicks
in. This was accomplished by removing a number of Si layers from the
top of 6H-SiC to provide a source of carbon atoms and then annealing
to different temperatures and examining the structures that were subsequently
formed after cooling down. In Ref. \cite{Tang08}, using Tersoff potentials,
it was found that a single layer of graphene clusters can only be
produced after annealing a 6H-SiC surface with two Si layers removed
and when the annealing temperature is greater than 1500 K, see Fig.
\ref{fig:Tang-1-4-layers-structures}(b\textcolor{black}{,f); gra}phene
did not form however if a single layer of Si atoms was removed from
the 6H-SiC surface in which case only chains and ring-like structures
were formed, Fig. \ref{fig:Tang-1-4-layers-structures}(a,e). This
result supports an assumption that an intermediate buffer layer between
the graphene sheet and the SiC surface is formed during the graphene
growth \cite{Hibino2010,Virojanadara10} . When three layers were
removed, small graphene clusters surrounded by amorphous C structures
were seen, Fig. \ref{fig:Tang-1-4-layers-structures}(c,g), while
a graphene bilayer started to form when four layers of Si atoms were
removed, Fig. \ref{fig:Tang-1-4-layers-structures}(d,h).

Different results were obtained in \cite{Jakse}: using a modified
Tersoff potential \cite{Erhart-mod-Tersoff-pot} it was found that
graphene formed after annealing to between 1200 and 1260 K even when
a single Si layer was removed. Furthermore, Jakse \emph{et al.} \cite{Jakse}
showed that using the original Tersoff potential parametrization gave
a graphitization temperature that is about 100 K higher. The authors
used a very similar methodology to that applied in\textcolor{black}{{}
\cite{Tang08}, so the qualitative difference in the results must
be attributed to subtle details of the simulations (e.g. the thermostat
used, different annealing protocol, etc.). How}ever, while both these
simulations see graphene forming at temperatures similar to those
observed in experiment, there is still no connection to observed growth
processes. For example, there is no reference to nucleation at step
edges or the observed step flow growth mode on SiC. Also, the way
in which Si atoms are removed prior to MD simulations is artificial:
in reality Si atoms are evaporated from SiC gradually at the same
time as the buffer and graphene layers are formed. In spite of these
drawbacks, these simulations have a potential for considering large
cells in real time and hence provide a motivation for theorists to
consider EP as a viable tool in further explorations of nucleation
and growth on SiC.

A number of experimental observations have suggested that the nucleation
and growth of graphene is influenced by SiC steps \cite{Hibino2010,Virojanadara10,Hanon}.
Kageshima \emph{et al. }\cite{Kageshima11,Kageshima12}\emph{ }performed
a series of DFT studies based on the geometry optimization of specific
structures in order to uncover the main processes responsible for
growth on  SiC and the role of SiC steps. Two types of numerical
experiments were performed inspired by the GCMC approach: (i) sublimation
simulations in which, starting from some SiC surface structure, Si
atoms were \emph{removed} one-by-one and the structure thus obtained
was relaxed, and (ii) aggregation simulations in which a C atom excess
was created by \emph{adding} C atoms to the surface one-by-one instead.
In the case of each particular structure formed in this way the most
energetically favorable geometry was sought, and its formation energy
was calculated. For instance, in the case of a structure obtained
by removing $n$ Si atoms, the formation energy was calculated via
\[
E_{f}(n)=E_{tot}(n)-E_{substr}+n\mu_{Si}\;,
\]
where $E_{tot}(n)$ is the total energy of the system with $n$ removed
Si atoms, $E_{substr}$ is the initial energy of the system, and $\mu_{Si}=E_{SiC}-\mu_{C}$
is the chemical potential of Si atoms, related (assuming thermodynamic
equilibrium) to the bulk energy of the SiC unit cell and the chemical
potential of C atoms in the free-standing graphene. The authors also
stressed that $\mu_{Si}$ can be controlled by Si gas pressure and
temperature, assuming again a thermodynamic equilibrium between the
Si atoms on the surface during growth and in the gas phase after their
sublimation. Then, differences between formation energies, $\Delta E(n)=E_{f}(n)-E_{f}(n-1)$,
were used to understand the energetics of adding (removing) C (Si)
atoms to (from) the surface. The authors called these differences
``growth barriers'', however, this may be misleading as these are
only differences in energies between two stable structures; the actual
energy barriers must be larger.

\begin{figure}
\begin{centering}
\includegraphics[height=6cm]{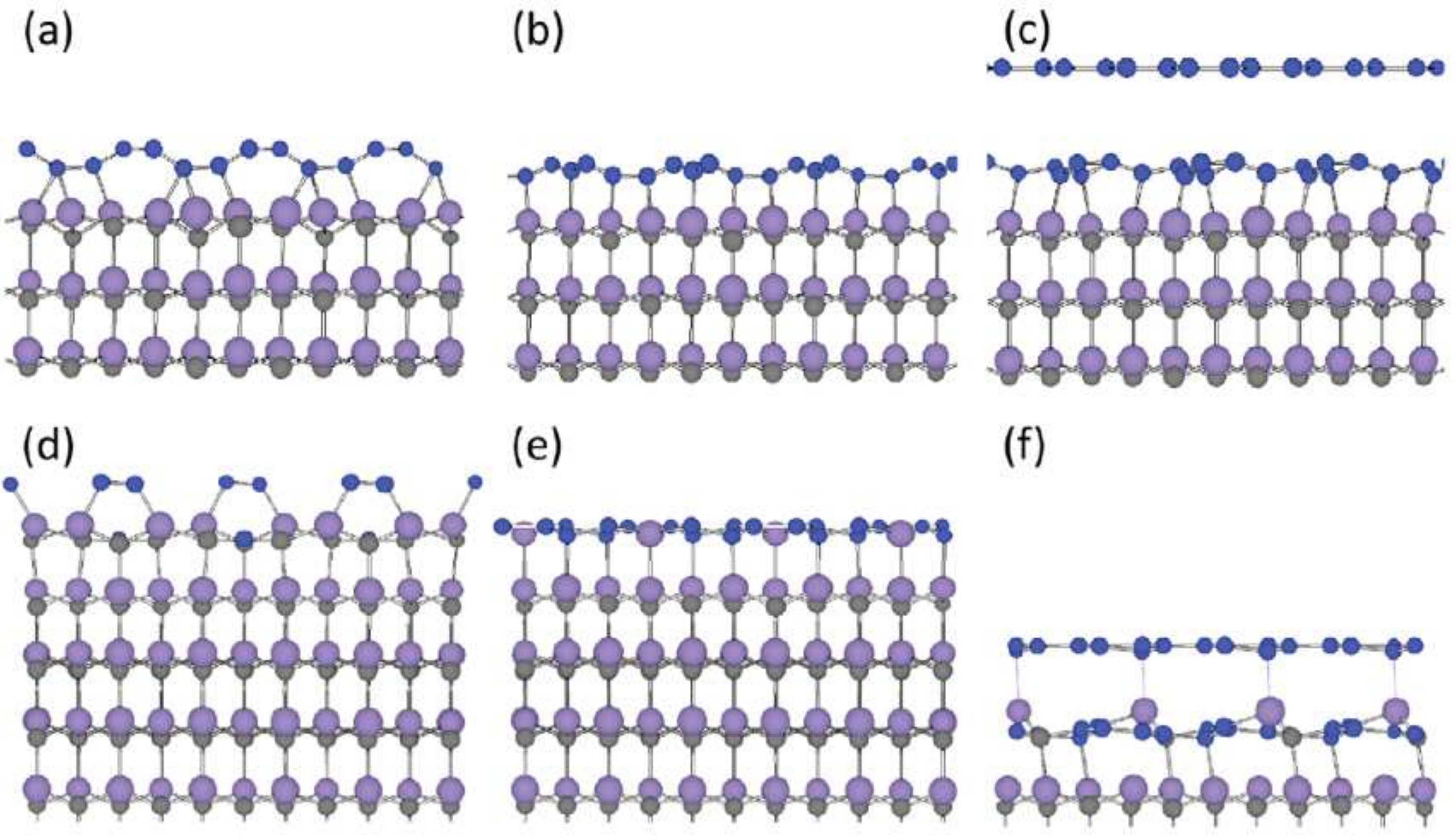}
\par\end{centering}

\caption{DFT relaxed geometries after adding C atoms (top panels) or removing
Si atoms (bottom panels). The number of excess C atoms is: 4 (a,d),
5 (b,e) and 15 (c,f). Large and small circles correspond to Si and
C atoms, respectively. {[}Reproduced from \cite{Kageshima12} by permission of IOP Publishing. All rights reserved.{]}\label{fig:Kageshima_comparision}}
\end{figure}

In Fig. \ref{fig:Kageshima_comparision} the two types of calculations
are compared. It is seen that in both cases a graphene layer is formed.
In the cases (b) and (f) this is the buffer layer as it is still bound
to the SiC; however, a proper graphene layer is seen to be formed
in (c) after adding sufficient numbers of C atoms to the simulation
cell, with the layer underneath serving as the buffer. In spite of
some similarities in both cases, there are important differences.
In particular, there is a considerable reconstruction of the surface
layers in the sublimation case (for instance, Si vacancies may trap
C atoms), and also the formation energies in the addition case are
generally lower than in the sublimation case, i.e. it follows from
these calculations that the sublimation route is generally harder.

\begin{figure}
\begin{centering}
\includegraphics[height=6cm]{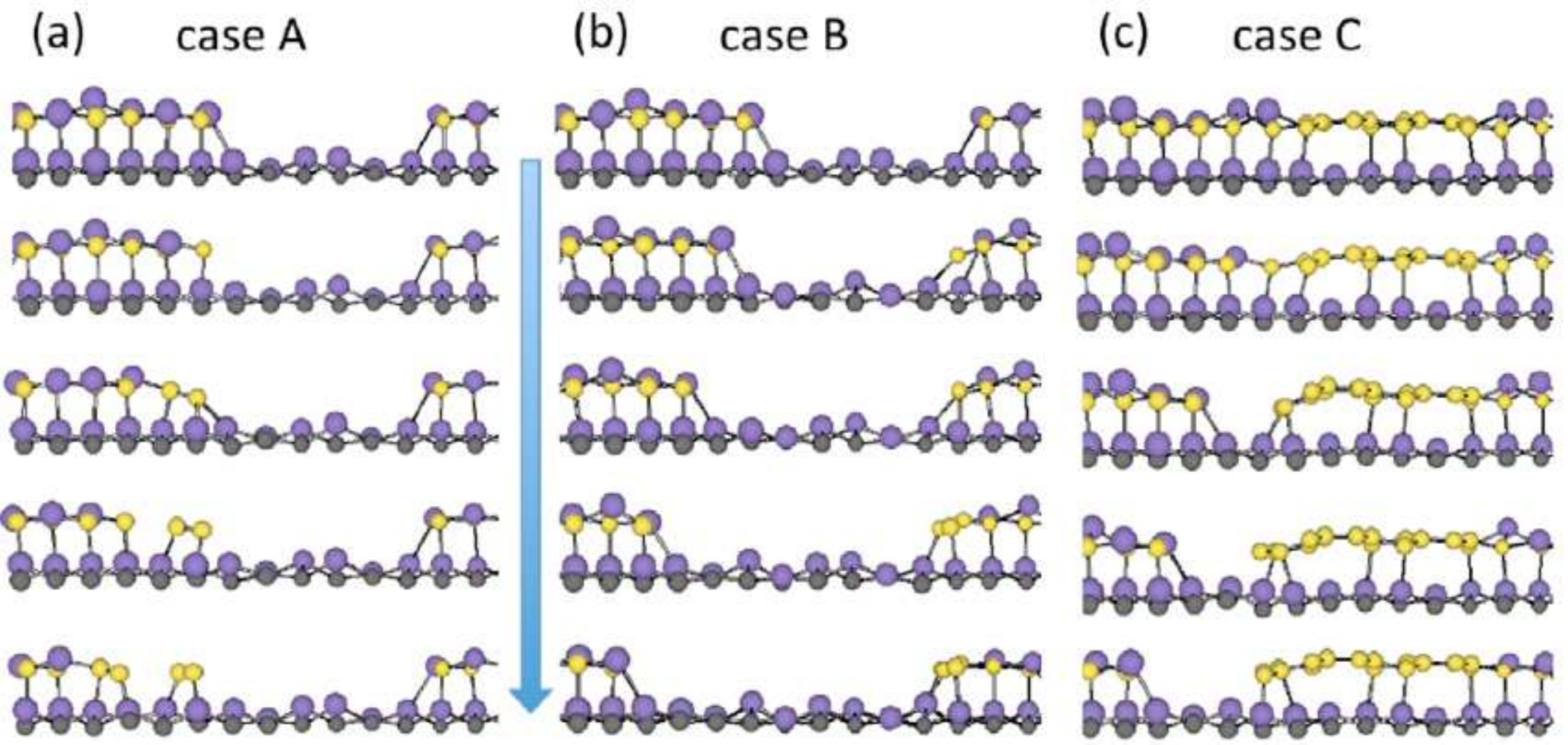}
\par\end{centering}

\caption{DFT relaxed geometries related to different possible routes of graphene
growth near the steps. (a) Si atoms are removed from one of the steps;
(b) while Si atoms are removed, the excess C atoms are moved across
to the other step and form C aggregates; (c) a graphene ribbon is
prepared, and excess C atoms are moved to its edge near the opposite
step. {[}Reproduced from \cite{Kageshima12} by permission of IOP Publishing. All rights reserved.{]}\label{fig:Kageshima_steps}}
\end{figure}

In order to investigate the role of SiC steps in supplying excess
C atoms to the surface, Kageshima \emph{et al.} \cite{Kageshima12}
also simulated several possible growth scenarios at the steps using
again a series of simple relaxation calculations as shown in Fig.
\ref{fig:Kageshima_steps}. Three cases were considered. In the first
two, when Si atoms were removed, then either no additional steps were
taken (case A), or excess C atoms were moved to the opposite step
(case B). In the third case a graphene ribbon was prepared near one
step and the excess C atoms were taken to its edge near the other
step (case C). In case A only C dimers are formed; in case B a C aggregate
forms serving as a nucleus for graphene growth. Finally, in  case
C the migrating C atoms help to increase the width of the ribbon.
Formation energy calculations showed that the C atoms preferentially
aggregate at step edges from where graphene starts to grow. This process
is assisted by C atoms attaching to the edges of the graphene islands.
Based on these observations a growth mechanisms was suggested \cite{Kageshima12}
whereby carbon atoms released by one step aggregate at adjacent steps,
and eventually grow to form the buffer layer. This mechanism obviously
requires the diffusion of C atoms on SiC which has not been investigated.
However, the authors raised a number of important issues which must
also be addressed, for example, how Si atoms that desorb under the
buffer layer are able to escape and what effect this may have on th\textcolor{black}{e
growth. Also, as pointed out in \cite{WuSiC}, the rea}dsorption of
Si to reform SiC could be another important process, as yet not considered,
in the growth of graphene from SiC in a gaseous environment.

\subsection{Simulations of growth dynamics \label{sub:Simulations-of-dynamics-SiC}}

In order to understand the growth process of graphene on SiC Ming\emph{
}and Zangwill \cite{Ming1} studied the growth using both rate equations
and one-dimensional KMC simulations. In the latter case it was suggested
how the energy barriers for the various growth processes involved,
such as nucleation at steps and growth of already nucleated graphene
sheet, i.e. propagation of both the first and second layers, could
be calculated from comparison with experiment. Unlike graphene growth
on transition metals, first principles calculations of the barriers
are not yet available as the structure of the buffer layer found on
SiC is not yet fully understood; moreover, the steps on this layer
have an unknown structure. Therefore, simulations of kinetics of growth
based on fully first principles calculations of the barriers and processes
have not yet been done to our knowledge.

In the KMC modeling \cite{Ming1} triple-bilayer (half unit cell)
steps formed on a high-index (vicinal) SiC surface consisting of long
terraces separated by steps, were considered. These steps act as nucleation
sites for graphene growth, which propagates by the sublimation of
Si atoms from a unit of the SiC step and its subsequent replacement
by a unit area of graphene, because of mass conservation. In this
model the buffer layer is assumed to be reconstructed from the SiC
terrace immediately upon its exposure to the surface. The processes
involved in this model for the growth are shown schematically in Fig.
\ref{fig:The-kinetic-process-nanofacets}. In (a) an empty step is
shown (green). With the nucleation rate of $r_{nuc}=v_{0}\exp(-E_{nuc}/k_{B}T)$,
where $E_{nuc}$ is the corresponding barrier and $v_{0}\thickapprox10^{12}$
s$^{-1}$ the attempt frequency, the embryonic graphene sheet appears
at the step as shown in (b). The effective nucleation barrier includes
the combined effects of the Si atoms sublimating, the C atoms recrystallizing
and the graphene growing at the step edge. The next step in the growth
process is the propagation of the nucleated graphene sheet leading
to an increase in the width of the graphene strip. This occurs via
SiC step (red) sublimation, Fig. \ref{fig:The-kinetic-process-nanofacets}(c),
when the step unit disappears and in its place a graphene strip is
produced at a rate $r_{prop}=v_{0}\exp(-E_{prop}/k_{B}T)$ with $E_{prop}$
being the corresponding effective energy barrier. As the strip propagates,
it may either run into an empty SiC step as shown in (d), or another
graphene strip on the upper terrace, demonstrated in (f). In the first
case the strip can ``climb over'' the upper terrace at the assumed
rate $r_{prop}$ and continue to propagate, as shown in (e); in the
latter case the two strips coalesce, with the same assumed rate $r_{prop}$,
see the transition from (f) to (g). Both of these processes cause
a step edge to be covered by a graphene layer. After the SiC step
is covered by the single layer of graphene, a second layer of graphene
can nucleate underneath the graphene layer as schematically shown
in (h), at the same rate $r_{nuc}$ as used for the first layer nucleation.
The second layer strip will then grow, see (i), at the rate of $r_{prop}$
or at the slower rate of $r'_{prop}$ related to a larger barrier
$E_{prop}^{\prime}$. The slower propagation rate of the second layer
sandwiched between the first layer and the SiC surface is expected
to be due to the fact that it must be more difficult to sublimate
Si atoms from the SiC with the first graphene layer already present
above. Because of the same reason, an assumption of the same nucleation
rate for the second layer, used in this study, must also lead to a
significant overestimation of the rate.

\begin{figure}
\centering{}\includegraphics[height=6cm]{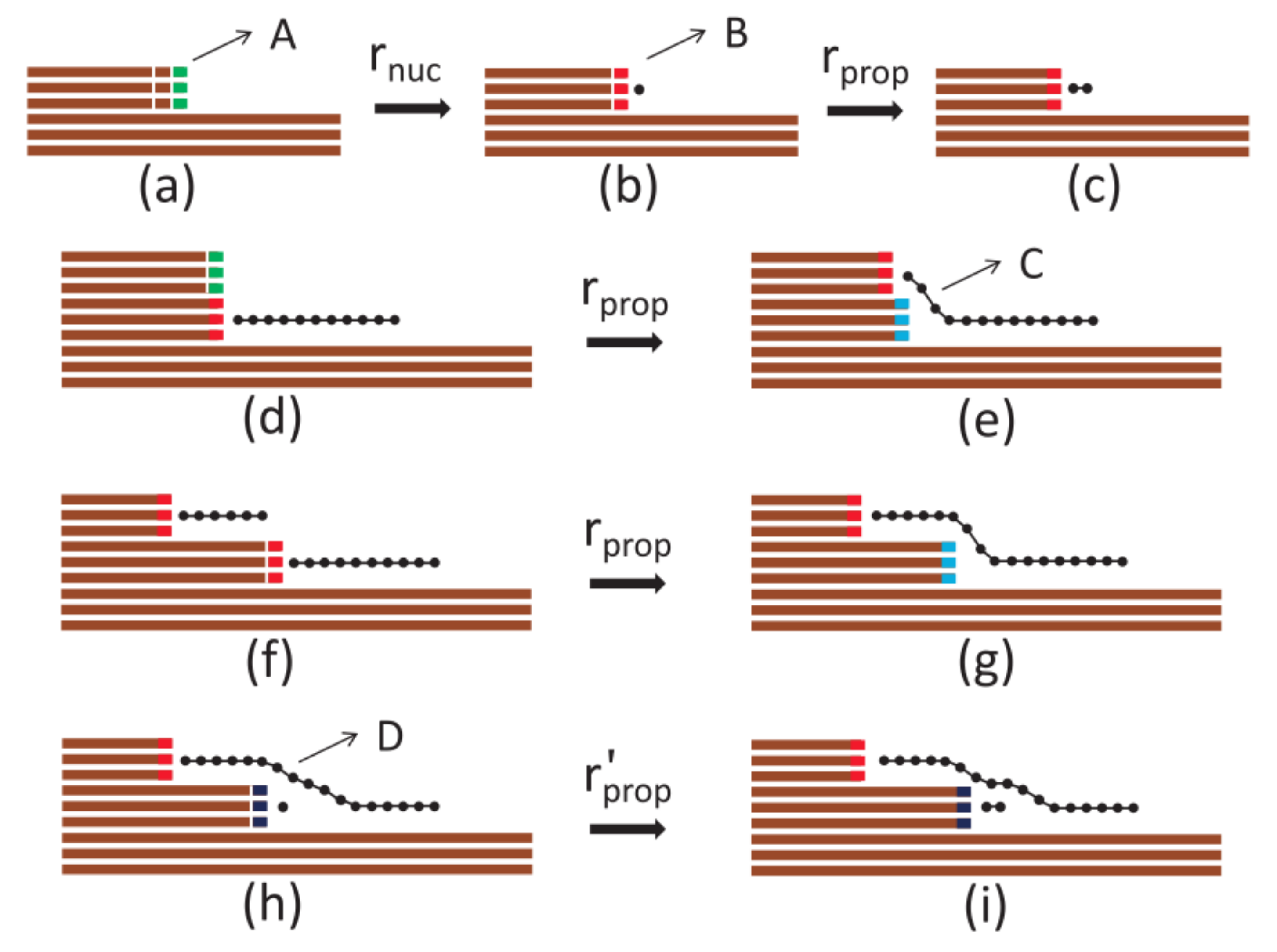}\caption{The kinetic processes for graphene growth on SiC used in the simulation
in \cite{Ming1}. In the initial step structure (a) the Si atoms in
the green cell are extracted, leaving a nucleus of the graphene cell
formed (b), which then propagates to the left (c) by consuming the
next SiC cell in the same manner. When the propagating graphene consumes
the whole terrace (d), the graphene sheet propagates up carpeting
the step (e). Alternatively, nucleation may happen at the step of
the upper terrace (f), in which case the two graphene sheets may coalesce
(g) upon further propagation. Once the step is fully covered with
the graphene sheet as in (e) or (g), the step below is freed and the
nucleation of the second layer underneath the first one may then start
(h) and subsequently propagate further (i). {[}Reprinted with permission
from \cite{Ming1}. Copyright (2011) by the American Physical Society.{]} \label{fig:The-kinetic-process}}
\end{figure}

To model growth on the SiC surface, simulations consisting of 5000
steps were made, and up to 10$^{6}$ independent runs were averaged
to accumulate enough statistics. Initially the graphene coverage of
the surface was investigated as a function of the energy barrier difference
$\Delta E=E_{nuc}-E_{prop}$. For this the dependence of the graphene
coverage of the first layer $\Theta_{1}$ on the total coverage $\Theta$
as well as the vicinal angle $\phi$ were considered. It was found
that for increasing $\Delta E$ the coverage of layer one, $\Theta_{1}$,
decreases. This is suggested to be because when $\Delta E$ is large
(and the nucleation barrier is much larger than the propagation barrier),
the nucleation of new strips is less likely than the propagation of
existing ones. In this case ``climb over'' steps are proposed to
occur more frequently and second layer nucleation happens sooner.
Since the second layer will have a longer time to grow, it will end
up being larger, hence the first layer coverage decreases. As $\Delta E$
is further increased the probability of nucleation of the second layer
is also reduced and the coverage of both layers becomes balanced.
At $\Delta E=0$ the step nucleation happens almost simultaneously,
and then growing strips coalesce when sufficient overall coverage
is achieved. The dynamics of growth was also found to depend on the
difference $\Delta E^{\prime}=E_{prop}^{\prime}-E_{prop}$ of the
propagation barriers between the second and first layers: as one would
expect, the first layer coverage increases substantially with the
increase of $\Delta E^{\prime}$ as the propagation of the first layer
is preferable. It is then suggested that the results of the simulations
for both layers as a function of the total coverage can be used to
extract the three parameters of the model ($E_{nuc}$, $E_{prop}$
and $E_{prop}^{\prime}$) from corresponding experimental dependencies.

It was also proposed that there is another way to extracting barriers,
based on comparing directly observed on the surface, and simulated,
distributions of strips with respect to their width. The comparison
of experimental LEEM images of the graphene growth in Fig. \ref{fig:LEEM-images-of}
with simulated images using the KMC method shows that for increasing
$\Delta E$ the number of strips decreases and the width of single
strips becomes larger and the strips cover many steps. The best agreement
with experiment is achieved for $\Delta E=0$. The calculated strip
width distributions demonstrate further details of the first and second
layer growth. For instance, a uniform distribution upon the increase
in the nucleation energy barrier $E_{nuc}$ is observed, and this
behavior is explained by the authors due to the fact that the ``climb
over'' and coalescence processes occur at step edges during growth
very rapidly.

\begin{figure}
\centering{}\includegraphics[height=6cm]{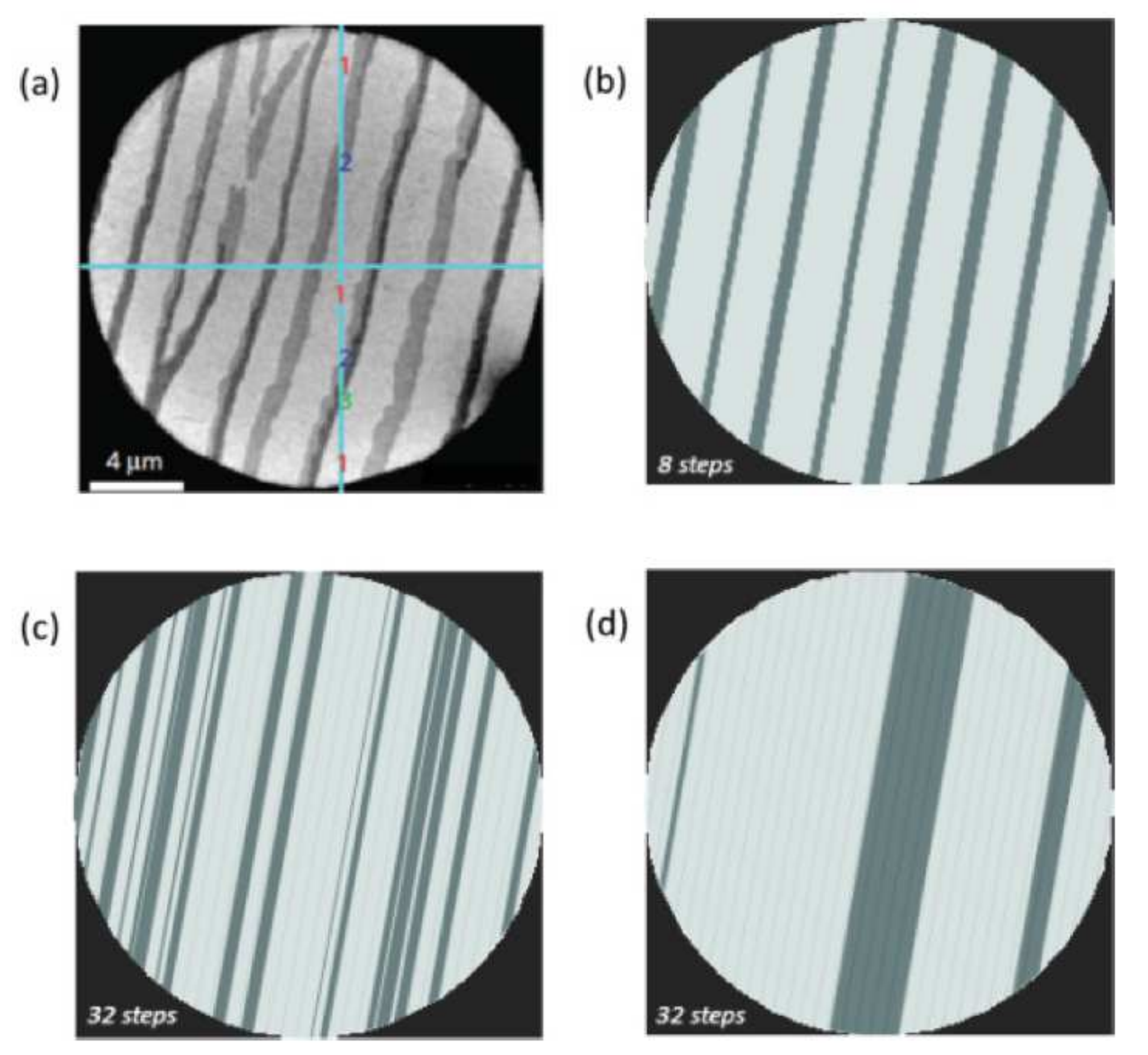}\caption{LEEM image (a) fr\textcolor{black}{om \cite{Emtsev:2009wd} of g}raphene
grown on SiC is compared with the simulated ones using KMC employing:
$\Delta E/k_{B}T$ = 0 (b), 5.8 (c) and 11.6 (d). The darker areas
indicate that more than one graphene layer is present. The total coverage
is 0.25, the vicinal angle is 0.9\textdegree{} and $\Delta E^{\prime}$
was assumed to be zero. {[}Reprinted with permission from \cite{Ming1}. Copyright (2011) by the American Physical Society.{]}
\label{fig:LEEM-images-of}}
\end{figure}

This paper also presents a mean-field rate equation analysis for the
growth mechanism described in Fig. \ref{fig:The-kinetic-process}
\cite{Ming1}. For this, four different types of steps are defined,
identified in Fig. \ref{fig:The-kinetic-process} as: (A) a bare SiC
step (green); (B) a step which has a layer of graphene segment attached
to it (red); (C) a step covered by a layer of graphene (blue), and
(D) a step which has a second layer of graphene segment attached (purple).
The proposed rate equations describe how the different steps are lost
and gained. Steps (A) are lost by first layer nucleation events and
climb-over processes. (B) steps are gained from nucleation events
and lost from coalescence with other strips, whereas (C) steps are
lost from second layer nucleation events and gained from both climb-over
processes and the coalescence of steps. Finally, (D) steps are only
created from the nucleation of the second graphene layer, i.e. they
require type (C) steps to proceed. These rate equations provide good
comparison with the KMC simulation results.

In further work by the same authors \cite{Ming2} graphene growth
on a different type of SiC surface which contains nano-facets was
studied using a similar method. In this paper nano-facets are described
as several closely and evenly spaced triple bilayer steps on the SiC
surface. These facets form when vicinal SiC with triple bilayer steps
is heated to the graphitization temperature. The steps become bunched
together on the surface, which is suggested to occur in order to minimize
the surface free energy. Fig. \ref{fig:The-kinetic-process-nanofacets}(a)
shows a nano-facet. The initial angle that the nano-facet makes with
respect to the basal plane depends on the orientation of the substrate
and the growth conditions. It is also dependent on the width and the
spacing between subsequent steps.

\begin{figure}
\centering{}\includegraphics[height=5cm]{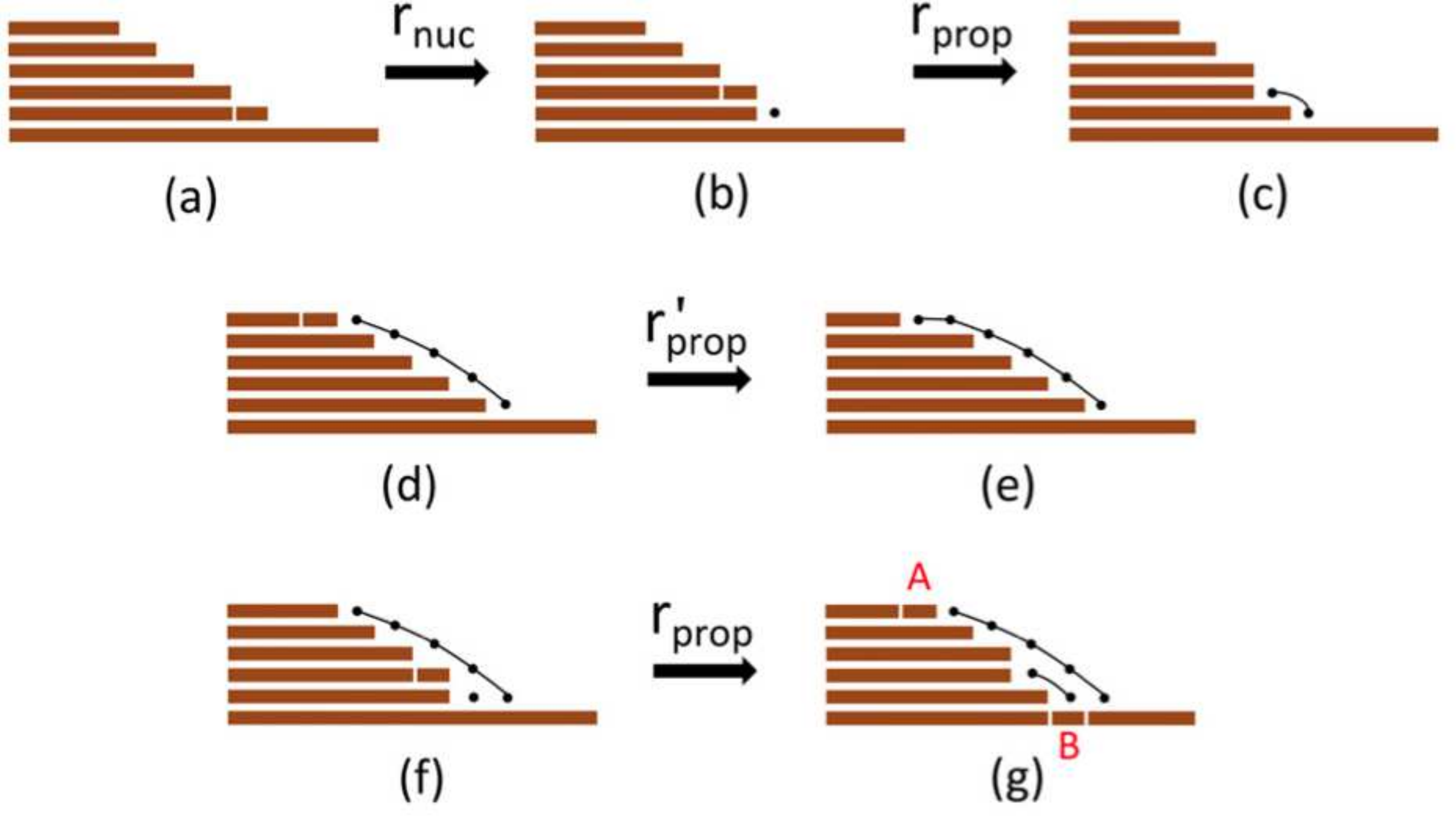}\caption{The kinetic processes accounted for in the KMC model for the growth
of graphene on a nano-facetted surface \cite{Ming2}: (a) initial
structure of the nano-facet; (b) a graphene layer nucleates at the
bottom terrace with the rate $r_{nuc}$ and then (c) propagates up
the step with the rate $r_{prop}$ until (d) it reaches the upper
terrace where (e) the growth proceeds with a reduced rate $r_{prop}^{\prime}<r_{prop}$;
once the step is fully covered with the carpet of graphene first layer,
the second layer underneath it nucleates (f) and start propagating
up the step (g) with the rate $r_{prop}$. A and B are the step units
at the edges of the top and the bottom terraces used for determining
the fracture angle $\theta$ indicating the slope of the nano-facetted
step. {[}Reproduced from \cite{Ming2} by permission of IOP Publishing. All rights reserved.{]} \label{fig:The-kinetic-process-nanofacets}}
\end{figure}

For growth on the nano-facet it is assumed that the bottom layer nucleates
first with the rate $r_{nuc}=v_{0}\exp(-E_{nuc}/k_{B}T)$, as shown
in Fig. \ref{fig:The-kinetic-process-nanofacets}(a,b). Then the growth
continues up the steps of the facet with the propagation rate $r_{prop}=v_{0}\exp(-E_{prop}/k_{B}T)$.
As a complete coverage of the facet is almost always observed in experiments,
the propagation barrier $E_{prop}$ was set equal to zero, assuming
that the propagation is the fastest kinetic process. Once the propagating
strip reaches the top layer it will continue growing on the (0001)
plane with the growth rate $r{}_{prop}^{\prime}=\nu_{0}\exp(-E{}_{prop}^{\prime}/k_{B}T)$
which is assumed to be lower than the propagation rate up the step.
This was concluded on the basis of analyzing experimental data on
growth on multifaceted surfaces when it was noticed that many layers
of graphene grow on the nano-facet before growth occurs on the (0001)
plane; this seems to be a reasonable assumption since sublimation
of Si atoms should be easier at the steps than on terraces. Once the
first layer has grown over a step, as with the case of the vicinal
surfaces \cite{Ming1} (see above), the second layer can nucleate
underneath, Fig. \ref{fig:The-formation-process} (f), and continue
growing in the same way as the first layer, see (g).

Fig. \ref{fig:The-formation-process} shows the formation process
schematically where the fracturing of the nano-facet occurs after
the first layer has started to propagate on the upper terrace shown
in (b). Since its further propagation on the terrace would be slower
than the propagation up the facets, the second layer quickly forms
and propagates up as demonstrated in (c). This is followed by the
formation and climbing (d) of the third layer and the corresponding
fracturing of the original nano-facets at the top (e) by the second
and third layers.

\begin{figure}
\centering{}\includegraphics[scale=0.3]{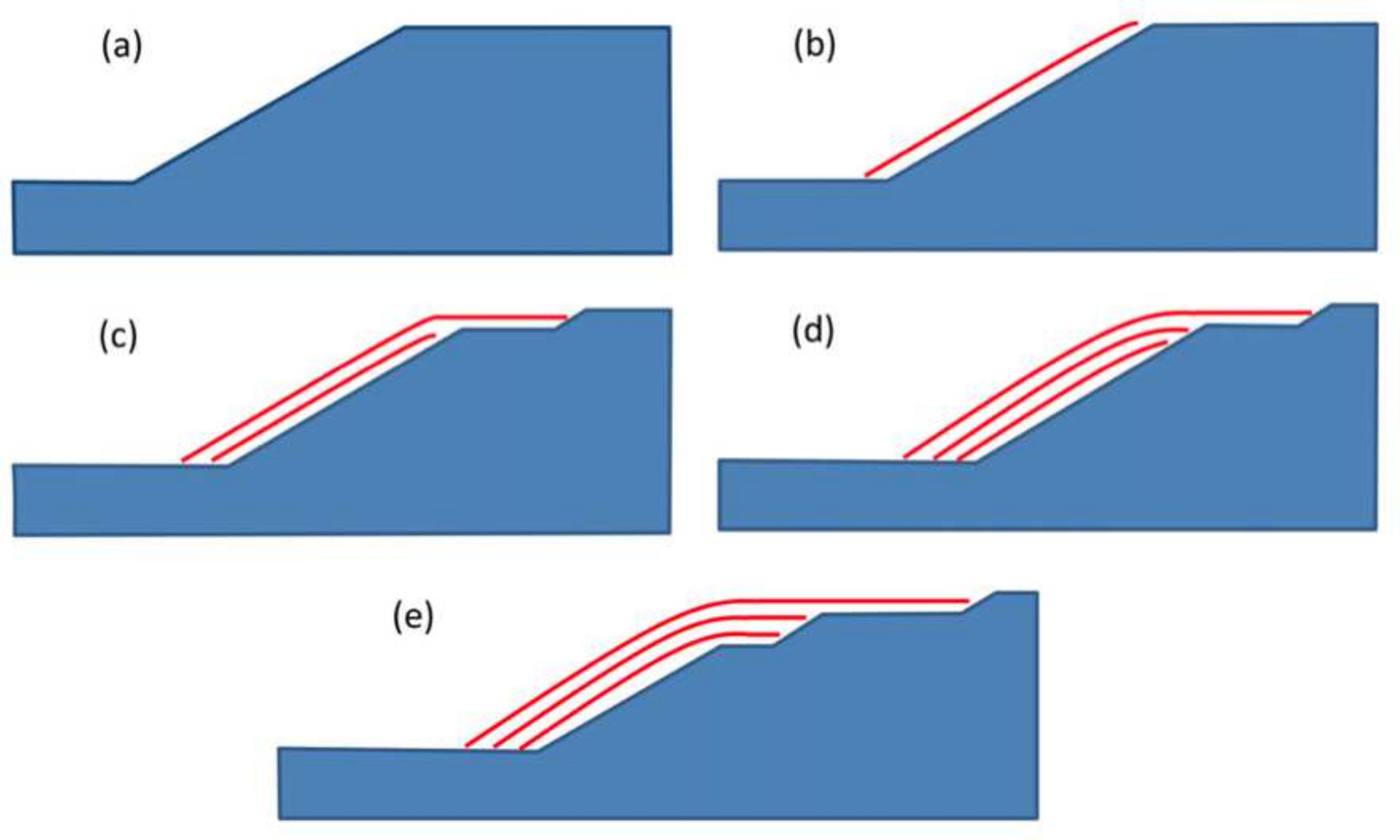}\caption{A possible formation process of graphene multilayers on a fractured
nano-facet. {[}Reproduced from \cite{Ming2} by permission of IOP Publishing. All rights reserved.{]} \label{fig:The-formation-process}}
\end{figure}

The results of the KMC simulations compare well with experimental
TEM images of graphene growth as seen in Fig. \ref{fig:Images-of-graphene}.
Both show graphene growth beginning sharply on the bottom terrace
of the facet and being continuous along the nano-facet to the basal
plane. The morphology of the surface imaged by TEM in Fig. \ref{fig:Images-of-graphene}(b)
is the same as seen in Fig. \ref{fig:The-formation-process}(e), where
two small nano-facets have been fractured from the original facet,
one with a height of two triple bilayers and the other with a height
of one triple bilayer.

\begin{figure}
\centering{}\includegraphics[scale=0.3]{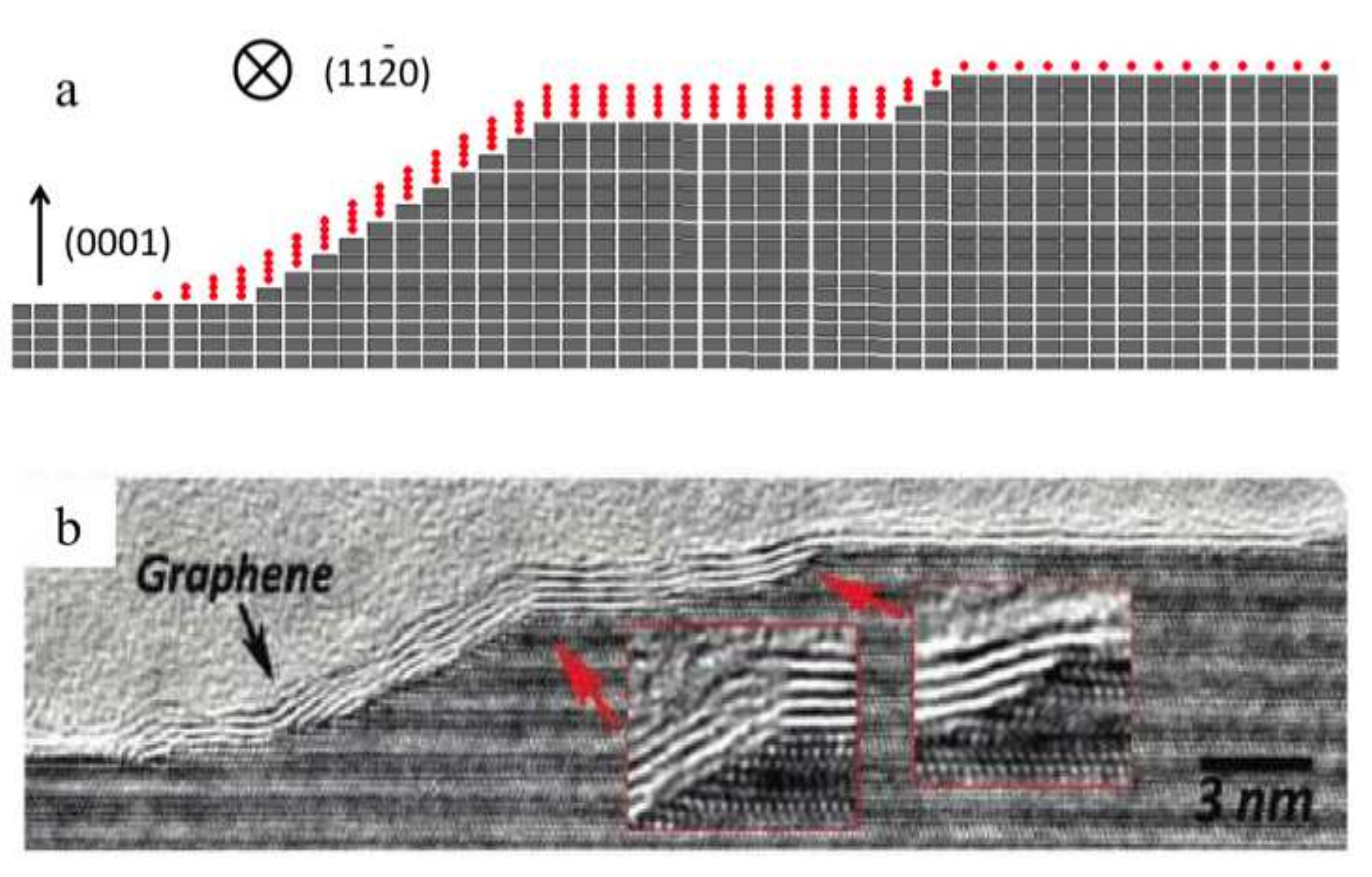}\caption{Images of graphene growth on a nano-facet from (a) a KMC simulation
with total surface coverage $\Theta$ = 0.8, $T=1800$ K, $E_{nuc}/k_{B}T$
= 7.7, and $E{}_{prop}^{\prime}/k_{B}T$ being either 3.9 or 7.1 for
the top and the lower layers, respectively, and (b) a TEM image adapted
from Robinson \emph{et al.} \cite{Robinson}. {[}Reproduced from \cite{Ming2} by permission of IOP Publishing. All rights reserved.{]} \label{fig:Images-of-graphene}}
\end{figure}

In order to quantify the fracturing of the nano-facet during the graphene
growth in their KMC simulations, Ming and Zangwill \cite{Ming2} introduced
the fracture angle $\theta$ characterizing the slope of the nano-facetted
step with respect to the basal plane direction. Various distributions
of the angle were obtained for different values of the difference
between the nucleation and terrace propagation barriers $\Delta E_{f}=E_{nuc}-E{}_{prop}^{\prime}$:
for small $\Delta E_{f}$ the distribution is Gaussian, but at larger
values it becomes a power-law. It was suggested that by comparing
the calculated angle distributions with experimental observations
of the fracture angle, $\Delta E_{f}$ can be determined.

\subsection{Finger-like growth: phenomenological models\label{sub:Finger-like-growth}}

We shall finish this review of the literature by discussing the use
of phenomenological models developed to explore the growth of graphene
on SiC. The obvious advantage of such a treatment is that these works
often suggest relevant atomistic processes which may then be modeled
using \emph{ab initio }methods. The interesting morphological characteristics
of graphene grown on SiC by heating to 1873 K and
then exposed to Si at a pressure of 1$\times10^{-5}$ torr take the
form of finger-like structures as seen in Fig. \ref{fig:AFM-images-of fingers}.
These observations motivated Borovikov and Zangwill \cite{Borovikov2009}
to use a continuum model to model the growth of graphene islands on
SiC in order to understand the mechanism of growth of these peculiar
features. Though the model was motivated by growth under specific
conditions, because the same finger-like structures seen in Fig. \ref{fig:AFM-images-of fingers}
have also been observed under UHV conditions, this could suggest that
similar growth processes occur for different conditions.

\begin{figure}
\begin{centering}
\includegraphics[height=4cm]{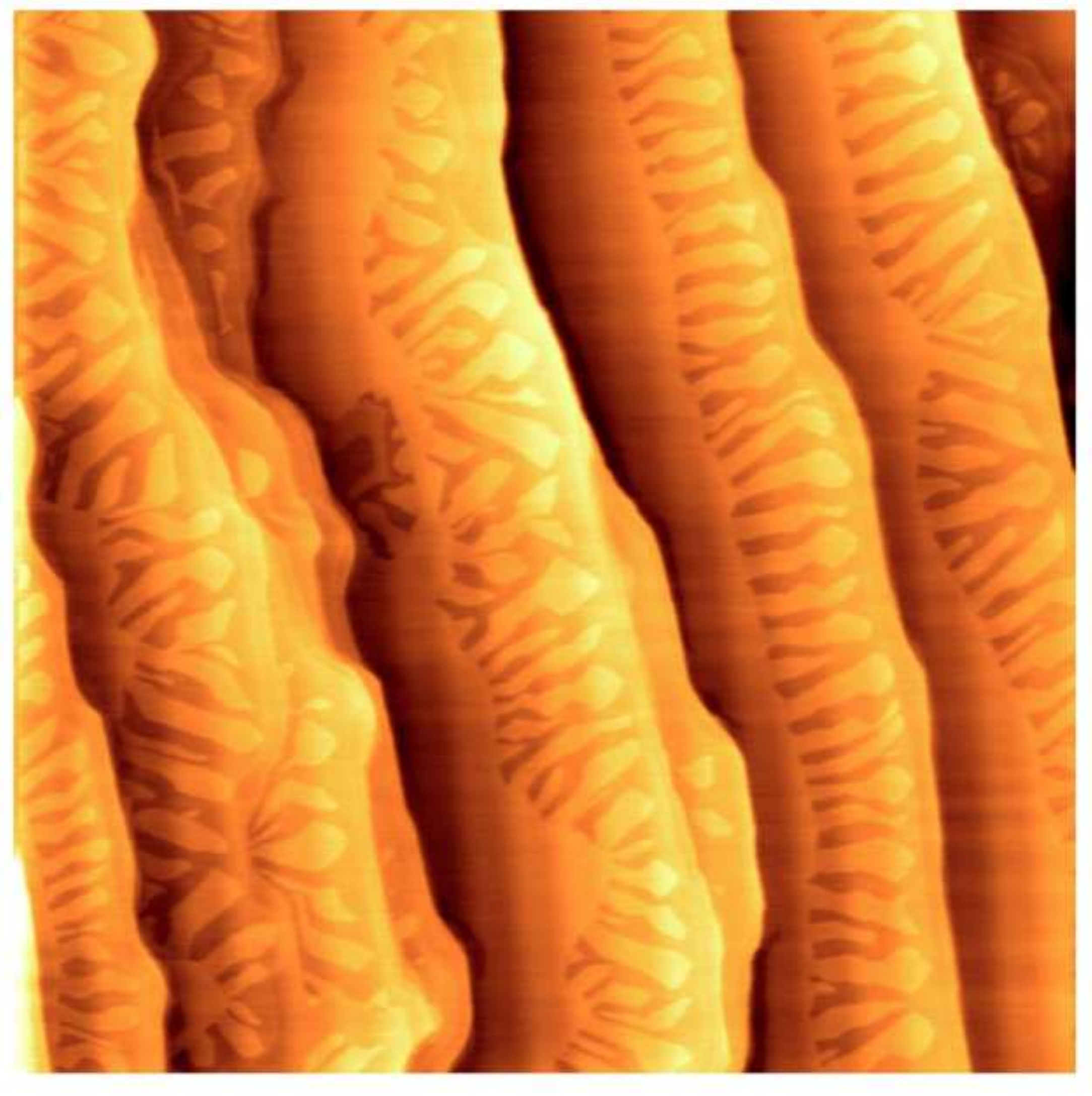}
\par\end{centering}

\caption{AFM images of graphene growth on SiC at 1873 K and with
a background pressure of Si of 1$\times10^{-5}$ torr. The field of
view is 10 $\mu\mbox{m.}$ {[}Reprinted with permission from \cite{Borovikov2009}. Copyright (2009) by the American Physical Society.{]}\label{fig:AFM-images-of fingers}}
\end{figure}

Graphene is assumed to form when the SiC layer which lies directly
below the buffer layer decomposes. The released carbon then recrystallizes
underneath the step to form a new buffer layer with the old buffer
layer forming quasi-free-standing graphene. The authors assume that
the decomposition of SiC step edges is dependent on the local curvature
of the growth front. They then write down a continuum equation for
the evolution of step height, $h(x,t),$

\begin{equation}
\frac{\partial h(x,t)}{\partial t}=-V-aV\frac{\partial^{2}h(x,t)}{\partial x^{2}}+\sigma\Gamma\frac{\partial^{2}h(x,t)}{\partial x^{2}}-\sigma D\frac{\partial^{4}h(x,t)}{\partial x^{4}}\;,\label{eq:Continuum-Borovikov}
\end{equation}
where the velocity of the growth front is $V$, $a$ is the lattice
constant of SiC, $\sigma=a^{3}\gamma/k_{B}T$ ($\gamma$ is the SiC
step stiffness), $\Gamma=\nu\mbox{exp\ensuremath{(-E_{1}/K_{B}T})}$
is the mean rate at which atomic species detach from a straight SiC
step and $D=a^{2}\nu\mbox{exp(\ensuremath{-E_{2}/K_{B}T\mbox{)}}}$
is the edge diffusion constant. The second term in Eq. (\ref{eq:Continuum-Borovikov})
describes the change in the rate of step edge evaporation due to curvature.
Thus, perturbations to a flat growth front which produce a concave
(convex) region have a higher (lower) rate of decomposition. The authors
also suggest a physical origin for this term whereby the release of
heat by the recrystallization of carbon atoms into a new buffer layer
increases the local temperature and drives further decomposition.
The exact meaning of the last two terms in the above equations can
be found in \cite{Borovikov2009}, but essentially the third term
describes detachment of carbon atoms and eventual reattachment after
diffusion along a terrace, and the fourth term models the diffusion
of adatoms along a step edge.

The authors perform a linear stability analysis of Eq. (\ref{eq:Continuum-Borovikov})
by assuming a solution of the form,

\[
h(x,t)=-Vt+\varepsilon(t)\mbox{sin\ensuremath{(2\pi x/\lambda)}}\;,
\]
where $\varepsilon(t)$ is the amplitude of a perturbation with wavelength
$\lambda$. Growth or decay of this amplitude (and therefore the step
edge) occurs exponentially when $\lambda$ is greater or less than
some critical wavelength. The wavelength for which growth occurs the
fastest is given by

\[
\lambda_{m}=\sqrt{\frac{8\pi^{2}\sigma D}{aV-\sigma\Gamma}}
\]
and its temperature ``phase diagram'' is shown in Fig. \ref{fig:wavelengths}.
Using estimates for the various parameters occurring in Eq. (\ref{eq:Continuum-Borovikov})
the authors plot this critical wavelength as a function of temperature
for a range of pressures of SiC and thus obtain regions for which
curvature effects drive instabilities at the growth front, producing
the observed finger-like features during growth. Their analysis produces
two temperatures of interest, $T_{G}$ and $T_{S}$, both indicated
in the Figure. Below $T_{G}$ no graphene growth occurs and above
$T_{S}$ the growth front is stable so that for $T>T_{S}$ no finger-like
structures are expected to grow. For the unstable growth, the values
they obtain for $\lambda_{m}$ agree well with the separation between
fingers found in the experiments described in Fig. \ref{fig:AFM-images-of fingers}.
Further, experimental observations of growth at higher pressures see
only straight steps which agrees with the results shown in Fig. \ref{fig:wavelengths}.
Thus, the results obtained using continuum modeling of the growth
process on SiC agree semi-quantitatively with experimental observations.

\begin{figure}
\centering{}\includegraphics[scale=0.5]{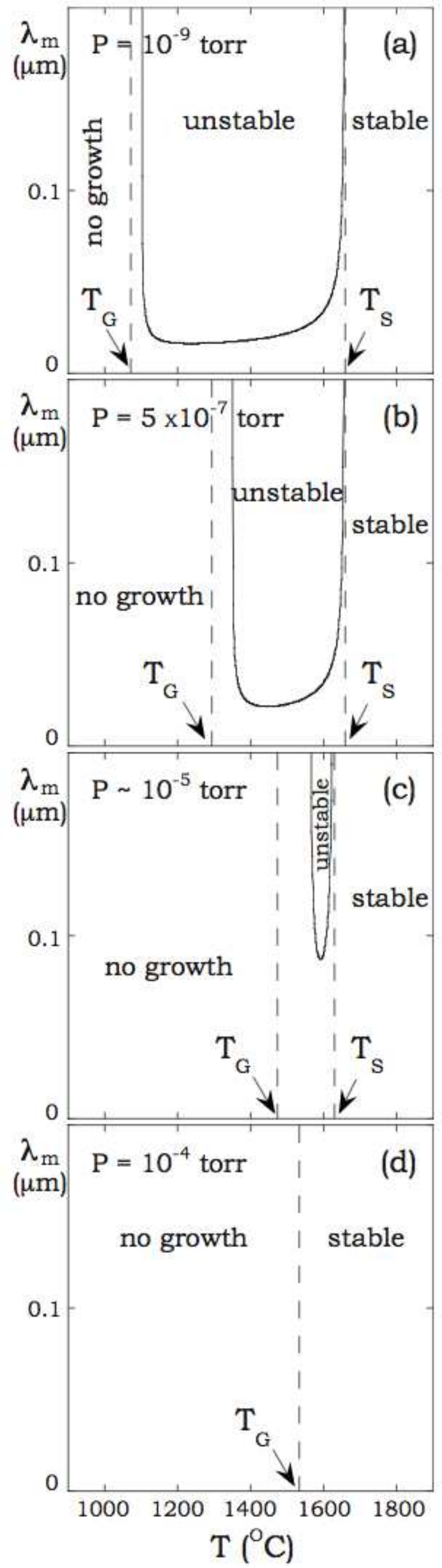}\caption{Temperature and pressure dependence of the fastest growing wavelength
obtained from the analysis of the linear stability of Eq. (\ref{eq:Continuum-Borovikov}).
{[}Reprinted with permission from \cite{Borovikov2009}. Copyright (2009) by the American Physical Society.{]} \label{fig:wavelengths}}
\end{figure}

\section{Discussion\label{sec:Discussion}}

Though much progress has been made since 2006 in understanding and
controlling the growth of graphene on metals and SiC at experimental
level, it is worth briefly pointing out that there are still unaddressed
or unanswered questions that suggest future directions of research
and new architectures based on graphene. First, only a limited region
of the growth parameter space has been explored in a systematic manner.
Systematic studies with varying growth temperature, carbon precursor
pressure, or inert gas back-pressure, for instance, are needed in
order to understand better the extent to which the structure (e.g.
strain, defect density, number of layers) of graphene can be controlled.
Such studies are usually time-consuming if performed with \emph{ex
situ }approaches, but may be considerably speeded up with the help
of \emph{in operando} analysis. This poses another problem, that of
implementing \emph{in operando} analysis in constrained growth environments,
especially at high temperatures (above 1800 K) and pressures (above
10$^{-6}$ mbar).

The case of graphene growth on SiC (e.g., how can a high quality buffer
layer be prepared?) and on Cu foils (e.g., how do substrate defects
impact graphene nucleation and structure?) under close-to-ambient-pressure
reactors is a telling example. Probes exploiting light scattering
seem here the best candidates. Overall, by extending the range of
conditions accessible to \emph{in operando} analysis and by more systematic
exploration of the growth parameter space, one may expect, for example,
new graphene growth modes (e.g. spiral, step-flow) to be unveiled.
Most growth studies are motivated by the prospect of preparing high
quality, defect-free graphene.

Obviously defect densities have been indeed much decreased in graphene
samples on both metals and SiC, still it remains difficult to eliminate
completely some kinds of defects, such as wrinkles, which actually
do not form upon growth but during cooling down. Although strategies
for preventing the formation of such defects remain to be imagined,
one may expect that strain and substrate engineering may provide interesting
angles of attack. In-plane rotations between successive layers in
multilayers, which can also be viewed as defects, offer stimulating
horizons for manipulating charge carrier transport in graphene. This
example illustrates that defect engineering, rather than defect minimization,
is also a stimulating avenue of research. Other examples include periodic
modulations of strain and chemical doping (by impurities replacing
C atoms for instance) or patterned growth of graphene. Only a few
research studies to date report the bottom-up demonstration of such
advanced graphene structures, mostly because few possibilities have
been explored.

On the theoretical side, there is still much to be done to understand
fully the growth mechanisms on different substrates and via different
growth methods. Modeling the process of decomposition of hydrocarbons
has only just started. In Fig. \ref{fig:processes} we showed as an
example all essential processes which happen in the CVD growth of
graphene when ethene molecules E are deposited on the hot metal surface:
upon impact with the surface the molecules decompose into intermediate
species E$_{d}$ (C$_{x}$H$_{y}$ in this case) via a complex chain
of interconnected dehydrogenation, isomerization and hydrogenation
reactions. The original molecules E and the decomposed species E$_{d}$
may diffuse on the surface, but their mobility is yet to be understood.
Eventually, carbon monomers M and dimers D are produced which must
be quite mobile. Because of the strong C-C bonds as compared to much
weaker C-H bonds, one may expect the dimer concentration to be
much larger than that of monomers. Colliding carbon species form larger
clusters C which eventually nucleate islands G which grow if the flux
of adsorbing species at their boundary exceeds the desorption flux.
Island growth depends on whether they have reached their critical
size. The H atoms, initially adsorbed on the surface after the dehydrogenation
reactions, will freely diffuse across the surface and may either attach
to the species E and E$_{d}$ of the feedstock thereby transforming
them, or may instead collide with other H atoms in which case H$_{2}$
molecules are formed. The latter can easily desorb, and this must
be the main mechanism for the hydrogen to leave the surface as H atoms
are relatively strongly bound to the surface as compared to the hydrogen
molecules. Correct understanding of the multitude of these reactions
for various hydrocarbon molecules and hence the distribution of their
products (C clusters) on the surface during growth is necessary for
correctly modeling growth kinetics (e.g. Avrami vs. Gompertzian type
of kinetics \cite{Celebi:2013cc}). It also seems plausible that the
various reactions sketched above and their rates (and hence their
significance in the global picture of growth kinetics) should depend
on the particular hydrocarbon feedstock used. For instance, in the
case of ethene one would expect abundance of carbon dimers as compared
to monomers, whereas in the case of methane feedstock one might think
that the opposite would be true.

Another important point affecting kinetics must be the spatial distribution
of all species on the surface during the growth and their diffusion,
as the reactions can only happen when the species collide. Clusters
of various sizes and shapes may be formed and their mobility and ability
to serve as nuclei for further growth would vary.

The corresponding theoretical simulations and experimental data that
would help in understanding all these processes are either sketchy
or simply missing.

The first steps in addressing the above issues from the theoretical
side could be done by performing detailed \emph{ab initio} based MD
simulations in which various hydrocarbon molecules are made to impinge
upon the surface and their decomposition is studied as a function
of their kinetic energies. Note that care should be taken in considering
correctly the dissipation of energy into the surface from the impinging
molecules in these MD simulations: equilibrium thermostats which are
widely applied in this kind of problems may not actually be appropriate
(see e.g. discussion in \cite{my-SBC-3}). The subsequent processes
such as diffusion, further decomposition, cluster formation and nucleation
and growth, can be investigated, e.g. with KMC based methods.

Modeling TPG growth is somewhat similar to that of the CVD as we described
above, but there are important differences since one would expect
that the impact of the hydrocarbon molecules with the surface is not
significant in the decomposition of the molecules since deposition
is performed when the surface is kept at room temperature. Instead,
one starts with intact adsorbed molecules on top of the surface, then
the surface is gradually heated up, the molecules become mobile, collide
with each other and dehydrogenation reactions commence. Therefore,
the distribution of various C$_{x}$H$_{y}$ species on the surface
during nucleation and growth may in this case be different to the
CVD case, thereby modifying the kinetics. Consequently, MD simulations
in the case of TPG may be unnecessary and what is required is detailed
KMC-like modeling based on transition rates calculated from first
principles simulations.

Further, kinetic modeling of nucleation events in real time, apart
from a number of simulations made using the GCMC method within a TB
Hamiltonian, are practically non-existent, and only a single KMC simulation
of steady-state growth exists to date. An understanding of nucleation
would enable one to rationalize the formation of rotational domains
and discover how this can be prevented, if desired. Better understanding
of the diffusion of clusters of various sizes and shapes, including
large ones to explain the Smoluchowski ripening mechanism observed
in many systems, will help in learning their detailed role in nucleation
and growth. More also needs to be done on the formation of defects
during growth, their mobility and the mechanism of their ``healing''
(e.g. by diffusion to the edges, coalescence or by decomposition of
hydrocarbon molecules evaporated onto the surface). There are a number
of simulations of particular aspects for the given system, but no
systematic simulations even for a single growth method and a single
transition metal has been done.

The situation with SiC is even worse because of the sheer complexity
of this system: our understanding of the growth mechanism is at a
rudimentary stage, and there are only a few simulations in which Si
sublimation was modeled properly. However, without modeling sublimation
in detail it is not possible to understand the formation of the buffer
layer or assess, for instance, whether one needs a high graphitization
temperature in order to promote both Si sublimation and C-clustering,
or only the sublimation process.

Challenges posed by this very complicated system clearly show limitations
of the current toolkit of computational methods, and hence the development
of cheaper techniques retaining the precision of the most expensive
methods is required so that larger systems can be studied. Problems
in modeling realistic kinetics of growth of graphene call for applications
of adaptive KMC based techniques \cite{Adaptive-KMC-2008}; however,
at the moment for practical calculations these can only be combined
with TB or EP methods; using standard DFT methods to explore the PES
is still prohibitively expensive.

\section*{Note added in proof}

Since this review went to press, we have become aware of several articles that bear directly on the nucleation and growth
of graphene. Kim et al. \cite{Kim2012} examined the growth of graphene on copper by using the chemical vapor deposition (CVD) of
methane (CH$_4$).

Methane is deposited onto the Cu surface, decomposes, and the reaction products desorb, except for the carbon. This leads
to an effective flux of carbon atoms and a supersaturation at the surface, that is, a concentration of carbon adatoms which
exceeds that at equilibrium. The concentration continues to rise linearly until, at a critical supersaturation, graphene nuclei
form and begin to grow. This supersaturation then drops until the nucleation ceases, and the carbon adatom concentration
reaches a steady state, where the deposited atoms are incorporated into the growing domains. The resulting morphology
of the graphene layer depends on the temperature and exposure time to the methane. The temperature ranges from 720 to
1000$^\circ$C and the exposure time from a flash exposure to 30 min. Many small nucleation centres are discernible at the lower
temperatures, but there are fewer and larger such centres at increasing temperatures. This is typical of growth dominated
by diffusion. Even at the highest temperatures, however, the nuclei do not coalesce to form a continuous film (with grain
boundaries). The light regions between the dark domains is the underlying copper substrate. We see from these samples
that continuous films are obtained only for temperatures of about 1000$^\circ$C and exposures of 30 min. To understand the
mechanisms behind these results, the number of nuclei as a function of temperature taken from SEM images at the flash
exposure were plotted as the natural logarithm of the density of nuclei as a function of $1/T$. There resulted two straight
lines: one at low-temperatures ($< 850 ^\circ$C) and the other at high-temperatures ($> 850 ^\circ$C). This is suggestive of different
mechanisms operating in the two temperature regimes and the authors identify these as the attachment-dominated regime
at low temperatures, and the desorption-dominated regime at high temperatures. At low temperatures, desorption is
negligible, so the formation and growth of clusters is dominated by the diffusion of carbon adatoms on the surface and
their capture at the perimeters of the nuclei. At higher temperatures, desorption becomes activated (this has a much higher
activation barrier than diffusion), so the nucleation rate is dominated by desorption.

Nemec et al. \cite{Nemec2013} used van der Waals corrected Perdew-Burke-Ernzerhof density functional theory to calculate surface
phase diagrams of the precursor phase, monolayer, and bilayer graphene on the Si face of 3C-SiC(111). These calculations
enable several observations related to the growth of graphene on these surfaces to be rationalized. The background
pressures of Si and C are low in UHV and not well-defined, which leads to widely-varying experimental conditions and
inferior morphologies. Maintaining a controlled Si partial pressure at a constant high temperature appears to be the most important factor in producing homogeneous graphene layers. The main conclusion from this study is that near-equilibrum
conditions may provide the best opportunities for the expitaxial growth of graphene on SiC.

\section*{Acknowledgement}
J.P. de B. was supported through a studentship in the Centre for Doctoral Training on the Theory and Simulation of
Materials at Imperial College London funded by EPSRC under grant number EP/G036888/1, while H.T. was supported by
EPSRC under grant number EP/K502868/1.


\begin{thebibliography}{381}
\expandafter\ifx\csname natexlab\endcsname\relax\def\natexlab#1{#1}\fi
\providecommand{\url}[1]{\texttt{#1}}
\providecommand{\href}[2]{#2}
\providecommand{\path}[1]{#1}
\providecommand{\DOIprefix}{doi:}
\providecommand{\ArXivprefix}{arXiv:}
\providecommand{\URLprefix}{URL: }
\providecommand{\Pubmedprefix}{pmid:}
\providecommand{\doi}[1]{\href{http://dx.doi.org/#1}{\path{#1}}}
\providecommand{\Pubmed}[1]{\href{pmid:#1}{\path{#1}}}
\providecommand{\bibinfo}[2]{#2}
\ifx\xfnm\relax \def\xfnm[#1]{\unskip,\space#1}\fi
\bibitem[{Slonczewski and Weiss(1958)}]{weiss58}
\bibinfo{author}{J.~C. Slonczewski}, \bibinfo{author}{P.~R. Weiss},
\newblock \bibinfo{title}{Band structure of graphite},
\newblock \bibinfo{journal}{Phys. Rev.} \bibinfo{volume}{109}
  (\bibinfo{year}{1958}) \bibinfo{pages}{272--279}.
\bibitem[{Castro~Neto et~al.(2006)Castro~Neto, Guinea, and
  Peres}]{C-transformation-source}
\bibinfo{author}{A.~H. Castro~Neto}, \bibinfo{author}{F.~Guinea},
  \bibinfo{author}{N.~M.~R. Peres},
\newblock \bibinfo{title}{Drawing conclusions from graphene},
\newblock \bibinfo{journal}{Physics World} \bibinfo{volume}{19}
  (\bibinfo{year}{2006}) \bibinfo{pages}{33}.
\bibitem[{Novoselov et~al.(2004)Novoselov, Geim, Morozov, Jiang, Zhang,
  Dubonos, Grigorieva, and Firsov}]{novoselov04}
\bibinfo{author}{K.~S. Novoselov}, \bibinfo{author}{A.~K. Geim},
  \bibinfo{author}{S.~V. Morozov}, \bibinfo{author}{D.~Jiang},
  \bibinfo{author}{Y.~Zhang}, \bibinfo{author}{S.~V. Dubonos},
  \bibinfo{author}{I.~V. Grigorieva}, \bibinfo{author}{A.~A. Firsov},
\newblock \bibinfo{title}{Electric field effect in atomically thin carbon
  films},
\newblock \bibinfo{journal}{Science} \bibinfo{volume}{306}
  (\bibinfo{year}{2004}) \bibinfo{pages}{666--669}.
\bibitem[{Novoselov et~al.(2005{\natexlab{a}})Novoselov, Jiang, Schedin, Booth,
  Khotkevich, Morozov, and Geim}]{novoselov05a}
\bibinfo{author}{K.~S. Novoselov}, \bibinfo{author}{D.~Jiang},
  \bibinfo{author}{F.~Schedin}, \bibinfo{author}{T.~J. Booth},
  \bibinfo{author}{V.~V. Khotkevich}, \bibinfo{author}{S.~V. Morozov},
  \bibinfo{author}{A.~K. Geim},
\newblock \bibinfo{title}{Two-dimensional atomic crystals},
\newblock \bibinfo{journal}{Proceedings of the National Academy of Sciences of
  the United States of America} \bibinfo{volume}{102}
  (\bibinfo{year}{2005}{\natexlab{a}}) \bibinfo{pages}{10451--10453}.
\bibitem[{Novoselov et~al.(2005{\natexlab{b}})Novoselov, Geim, Morozov, Jiang,
  Katsnelson, Grigorieva, Dubonos, and Firsov}]{novoselov05b}
\bibinfo{author}{K.~S. Novoselov}, \bibinfo{author}{A.~K. Geim},
  \bibinfo{author}{S.~V. Morozov}, \bibinfo{author}{D.~Jiang},
  \bibinfo{author}{M.~I. Katsnelson}, \bibinfo{author}{I.~V. Grigorieva},
  \bibinfo{author}{S.~V. Dubonos}, \bibinfo{author}{A.~A. Firsov},
\newblock \bibinfo{title}{Two-dimensional gas of massless {D}irac fermions in
  graphene},
\newblock \bibinfo{journal}{Nature} \bibinfo{volume}{438}
  (\bibinfo{year}{2005}{\natexlab{b}}) \bibinfo{pages}{197--200}.
\bibitem[{Berger et~al.(2004)Berger, Song, Li, Li, Ogbazghi, Feng, Dai,
  Marchenkov, Conrad, First, and de~Heer}]{berger04}
\bibinfo{author}{C.~Berger}, \bibinfo{author}{Z.~Song},
  \bibinfo{author}{T.~Li}, \bibinfo{author}{X.~Li}, \bibinfo{author}{A.~Y.
  Ogbazghi}, \bibinfo{author}{R.~Feng}, \bibinfo{author}{Z.~Dai},
  \bibinfo{author}{A.~N. Marchenkov}, \bibinfo{author}{E.~H. Conrad},
  \bibinfo{author}{P.~N. First}, \bibinfo{author}{W.~A. de~Heer},
\newblock \bibinfo{title}{Ultrathin epitaxial graphite: 2d electron gas
  properties and a route toward graphene-based nanoelectronics},
\newblock \bibinfo{journal}{J. Phys. Chem. {B}} \bibinfo{volume}{108}
  (\bibinfo{year}{2004}) \bibinfo{pages}{19912--19916}.
\bibitem[{Zhang et~al.(2005)Zhang, Tan, Stormer, and Kim}]{zhang05}
\bibinfo{author}{Y.~Zhang}, \bibinfo{author}{Y.~Tan}, \bibinfo{author}{H.~L.
  Stormer}, \bibinfo{author}{P.~Kim},
\newblock \bibinfo{title}{Experimental observation of the quantum {H}all effect
  and {B}erry's phase in graphene},
\newblock \bibinfo{journal}{Nature} \bibinfo{volume}{438}
  (\bibinfo{year}{2005}) \bibinfo{pages}{201--204}.
\bibitem[{Neto et~al.(2009)Neto, Guinea, Peres, Novoselov, and Geim}]{neto09}
\bibinfo{author}{A.~H.~C. Neto}, \bibinfo{author}{F.~Guinea},
  \bibinfo{author}{N.~M.~R. Peres}, \bibinfo{author}{K.~S. Novoselov},
  \bibinfo{author}{A.~K. Geim},
\newblock \bibinfo{title}{The electronic properties of graphene},
\newblock \bibinfo{journal}{Rev. Mod. Phys.} \bibinfo{volume}{81}
  (\bibinfo{year}{2009}) \bibinfo{pages}{109--162}.
\bibitem[{Soldano et~al.(2010)Soldano, Mahmood, and Dujardin}]{soldano10}
\bibinfo{author}{C.~Soldano}, \bibinfo{author}{A.~Mahmood},
  \bibinfo{author}{E.~Dujardin},
\newblock \bibinfo{title}{Production, properties and potential of graphene},
\newblock \bibinfo{journal}{Carbon} \bibinfo{volume}{48} (\bibinfo{year}{2010})
  \bibinfo{pages}{2127--2150}.
\bibitem[{Avouris(2010)}]{avouris10}
\bibinfo{author}{P.~Avouris},
\newblock \bibinfo{title}{Graphene: Electronic and photonic properties and
  devices},
\newblock \bibinfo{journal}{Nano Lett.} \bibinfo{volume}{10}
  (\bibinfo{year}{2010}) \bibinfo{pages}{4285--4294}.
\bibitem[{Bonaccorso et~al.(2010)Bonaccorso, Sun, Hasan, and
  Ferrari}]{bonaccorso10}
\bibinfo{author}{F.~Bonaccorso}, \bibinfo{author}{Z.~Sun},
  \bibinfo{author}{T.~Hasan}, \bibinfo{author}{A.~C. Ferrari},
\newblock \bibinfo{title}{Graphene photonics and optoelectronics},
\newblock \bibinfo{journal}{Nature Photon.} \bibinfo{volume}{4}
  (\bibinfo{year}{2010}) \bibinfo{pages}{611--622}.
\bibitem[{Sarma et~al.(2011)Sarma, Adam, Hwang, and Rossi}]{dassarma11}
\bibinfo{author}{S.~D. Sarma}, \bibinfo{author}{S.~Adam},
  \bibinfo{author}{E.~H. Hwang}, \bibinfo{author}{E.~Rossi},
\newblock \bibinfo{title}{Electronic transport in two-dimensional graphene},
\newblock \bibinfo{journal}{Rev. Mod. Phys.} \bibinfo{volume}{83}
  (\bibinfo{year}{2011}) \bibinfo{pages}{407--470}.
\bibitem[{Sun et~al.(2011)Sun, Wu, and Shi}]{sun11}
\bibinfo{author}{Y.~Sun}, \bibinfo{author}{Q.~Wu}, \bibinfo{author}{G.~Shi},
\newblock \bibinfo{title}{Graphene based new energy materials},
\newblock \bibinfo{journal}{Energy and Environmental Science}
  \bibinfo{volume}{4} (\bibinfo{year}{2011}) \bibinfo{pages}{1113--1132}.
\bibitem[{Novoselov et~al.(2012)Novoselov, Fal'ko, Colombo, Gellert, Chwab, and
  Kim}]{Review_applications_2012}
\bibinfo{author}{K.~S. Novoselov}, \bibinfo{author}{V.~I. Fal'ko},
  \bibinfo{author}{L.~Colombo}, \bibinfo{author}{P.~R. Gellert},
  \bibinfo{author}{M.~G. Chwab}, \bibinfo{author}{K.~Kim},
\newblock \bibinfo{title}{A roadmap for graphene},
\newblock \bibinfo{journal}{Nature} \bibinfo{volume}{490}
  (\bibinfo{year}{2012}) \bibinfo{pages}{192}.
\bibitem[{Hudson(2011)}]{bbc}
\bibinfo{author}{A.~Hudson}, \bibinfo{title}{Is graphene a miracle material?},
  \bibinfo{year}{2011}. \URLprefix
  \url{http://news.bbc.co.uk/1/hi/programmes/click_online/9491789.stm}.
\bibitem[{Wallace(1947)}]{wallace47}
\bibinfo{author}{P.~R. Wallace},
\newblock \bibinfo{title}{The band theory of graphite},
\newblock \bibinfo{journal}{Phys. Rev.} \bibinfo{volume}{71}
  (\bibinfo{year}{1947}) \bibinfo{pages}{622--634}.
\bibitem[{Kusminskiy et~al.(2009)Kusminskiy, Campbell, and Neto}]{viola09}
\bibinfo{author}{S.~V. Kusminskiy}, \bibinfo{author}{D.~K. Campbell},
  \bibinfo{author}{A.~H.~C. Neto},
\newblock \bibinfo{title}{Lenosky's energy and the phonon dispersion of
  graphene},
\newblock \bibinfo{journal}{Phys. Rev. B} \bibinfo{volume}{80}
  (\bibinfo{year}{2009}) \bibinfo{pages}{035401}.
\bibitem[{Lee et~al.(2008)Lee, Wei, Kysar, and Hone}]{lee08}
\bibinfo{author}{C.~Lee}, \bibinfo{author}{X.~Wei}, \bibinfo{author}{J.~W.
  Kysar}, \bibinfo{author}{J.~Hone},
\newblock \bibinfo{title}{Measurement of the elastic properties and intrinsic
  strength of monolayer graphene},
\newblock \bibinfo{journal}{Science} \bibinfo{volume}{321}
  (\bibinfo{year}{2008}) \bibinfo{pages}{385--388}.
\bibitem[{Park and Ruoff(2009)}]{park09}
\bibinfo{author}{S.~Park}, \bibinfo{author}{R.~S. Ruoff},
\newblock \bibinfo{title}{Chemical methods for the production of graphenes},
\newblock \bibinfo{journal}{Nature Nanotech.} \bibinfo{volume}{4}
  (\bibinfo{year}{2009}) \bibinfo{pages}{217--224}.
\bibitem[{Shao et~al.(2009)Shao, Wang, Wu, Liu, Aksay, and Lin}]{shao09}
\bibinfo{author}{Y.~Shao}, \bibinfo{author}{J.~Wang}, \bibinfo{author}{H.~Wu},
  \bibinfo{author}{J.~Liu}, \bibinfo{author}{I.~A. Aksay},
  \bibinfo{author}{Y.~Lin},
\newblock \bibinfo{title}{Graphene based electrochemical sensors and
  biosensors: A review},
\newblock \bibinfo{journal}{Electroanalysis} \bibinfo{volume}{22}
  (\bibinfo{year}{2009}) \bibinfo{pages}{1027--1036}.
\bibitem[{Choi et~al.(2010)Choi, Lahiri, Seelaboyina, and Kang}]{choi10}
\bibinfo{author}{W.~Choi}, \bibinfo{author}{I.~Lahiri},
  \bibinfo{author}{R.~Seelaboyina}, \bibinfo{author}{Y.~S. Kang},
\newblock \bibinfo{title}{Synthesis of graphene and its applications: a
  review},
\newblock \bibinfo{journal}{Critical Reviews in Solid State and Materials
  Sciences} \bibinfo{volume}{35} (\bibinfo{year}{2010})
  \bibinfo{pages}{52--71}.
\bibitem[{Dreyer et~al.(2010)Dreyer, Ruoff, and Bielawski}]{dreyer10}
\bibinfo{author}{D.~R. Dreyer}, \bibinfo{author}{R.~S. Ruoff},
  \bibinfo{author}{C.~W. Bielawski},
\newblock \bibinfo{title}{From conception to realization: An historial account
  of graphene and some perspectives for its future},
\newblock \bibinfo{journal}{Angew. Chemie Intern. Ed.} \bibinfo{volume}{49}
  (\bibinfo{year}{2010}) \bibinfo{pages}{9336--9344}.
\bibitem[{First et~al.(2010)First, de~Heer, Seyller, Berger, Stroscio, and
  Moon}]{first10}
\bibinfo{author}{P.~N. First}, \bibinfo{author}{W.~A. de~Heer},
  \bibinfo{author}{T.~Seyller}, \bibinfo{author}{C.~Berger},
  \bibinfo{author}{J.~A. Stroscio}, \bibinfo{author}{J.-S. Moon},
\newblock \bibinfo{title}{Epitaxial graphenes on silicon carbide},
\newblock \bibinfo{journal}{MRS Bulletin} \bibinfo{volume}{35}
  (\bibinfo{year}{2010}) \bibinfo{pages}{296--305}.
\bibitem[{Riedl et~al.(2010)Riedl, Coletti, and Starke}]{riedl10}
\bibinfo{author}{C.~Riedl}, \bibinfo{author}{C.~Coletti},
  \bibinfo{author}{U.~Starke},
\newblock \bibinfo{title}{Structural and electronic properties of epitaxial
  graphene on sic(0001): a review of growth, characterization, transfer doping
  and hydrogen intercalation},
\newblock \bibinfo{journal}{J. of Physics D: Appl. Phys.} \bibinfo{volume}{43}
  (\bibinfo{year}{2010}) \bibinfo{pages}{374009}.
\bibitem[{Schwierz(2010)}]{schwierz10}
\bibinfo{author}{F.~Schwierz},
\newblock \bibinfo{title}{Graphene transistors},
\newblock \bibinfo{journal}{Nature Nanotech.} \bibinfo{volume}{5}
  (\bibinfo{year}{2010}) \bibinfo{pages}{487--496}.
\bibitem[{Balandin(2011)}]{balandin11}
\bibinfo{author}{A.~A. Balandin},
\newblock \bibinfo{title}{Thermal properties of graphene and nanostructured
  carbon materials},
\newblock \bibinfo{journal}{Nature Mater.} \bibinfo{volume}{10}
  (\bibinfo{year}{2011}) \bibinfo{pages}{569--581}.
\bibitem[{Neto and Novoselov(2011)}]{neto11}
\bibinfo{author}{A.~H.~C. Neto}, \bibinfo{author}{K.~Novoselov},
\newblock \bibinfo{title}{New directions in science and technology:
  two-dimensional crystals},
\newblock \bibinfo{journal}{Rep. Prog. Phys.} \bibinfo{volume}{74}
  (\bibinfo{year}{2011}) \bibinfo{pages}{082501}.
\bibitem[{Pumera(2011)}]{pumera11}
\bibinfo{author}{M.~Pumera},
\newblock \bibinfo{title}{Graphene in biosensing},
\newblock \bibinfo{journal}{Mater. Today} \bibinfo{volume}{14}
  (\bibinfo{year}{2011}) \bibinfo{pages}{308--315}.
\bibitem[{Singh et~al.(2011)Singh, Joung, Zhai, Das~Saiful, Khondaker, and
  Seal}]{singh11}
\bibinfo{author}{V.~Singh}, \bibinfo{author}{D.~Joung},
  \bibinfo{author}{L.~Zhai}, \bibinfo{author}{S.~Das~Saiful},
  \bibinfo{author}{I.~Khondaker}, \bibinfo{author}{S.~Seal},
\newblock \bibinfo{title}{Graphene based materials: Past, present and future},
\newblock \bibinfo{journal}{Progr. Mater. Sci.} \bibinfo{volume}{56}
  (\bibinfo{year}{2011}) \bibinfo{pages}{1178--1271}.
\bibitem[{Young and Kim(2011)}]{young11}
\bibinfo{author}{A.~F. Young}, \bibinfo{author}{P.~Kim},
\newblock \bibinfo{title}{Electronic transport in graphene heterostructures},
\newblock \bibinfo{journal}{Ann. Rev. Cond. Matter Phys.} \bibinfo{volume}{2}
  (\bibinfo{year}{2011}) \bibinfo{pages}{101--120}.
\bibitem[{Batzill(2012)}]{batzill12}
\bibinfo{author}{M.~Batzill},
\newblock \bibinfo{title}{The surface science of graphene: Metal interfaces,
  {CVD} synthesis, nanoribbons, chemical modifications, and defects},
\newblock \bibinfo{journal}{Surf. Sci. Rep.} \bibinfo{volume}{67}
  (\bibinfo{year}{2012}) \bibinfo{pages}{83--115}.
\bibitem[{Bonaccorso et~al.(2012)Bonaccorso, Lombardo, Hasan, Sun, Colombo, and
  Ferrari}]{bonaccorso12}
\bibinfo{author}{F.~Bonaccorso}, \bibinfo{author}{A.~Lombardo},
  \bibinfo{author}{T.~Hasan}, \bibinfo{author}{Z.~Sun},
  \bibinfo{author}{L.~Colombo}, \bibinfo{author}{A.~C. Ferrari},
\newblock \bibinfo{title}{Production and processing of graphene and 2d
  crystals},
\newblock \bibinfo{journal}{Mater. Today} \bibinfo{volume}{15}
  (\bibinfo{year}{2012}) \bibinfo{pages}{564--589}.
\bibitem[{Shen et~al.(2012)Shen, Zhang, Liu, and Zhang}]{shen12}
\bibinfo{author}{H.~Shen}, \bibinfo{author}{L.~Zhang},
  \bibinfo{author}{M.~Liu}, \bibinfo{author}{Z.~Zhang},
\newblock \bibinfo{title}{Biomedical applications of graphene},
\newblock \bibinfo{journal}{Theranostics} \bibinfo{volume}{2}
  (\bibinfo{year}{2012}) \bibinfo{pages}{283--294}.
\bibitem[{Young et~al.(2012)Young, Kinlocha, Gonga, and Novoselov}]{young12}
\bibinfo{author}{R.~J. Young}, \bibinfo{author}{I.~A. Kinlocha},
  \bibinfo{author}{L.~Gonga}, \bibinfo{author}{K.~S. Novoselov},
\newblock \bibinfo{title}{The mechanics of graphene nanocomposites: A review},
\newblock \bibinfo{journal}{Compos. Sci. Techn.} \bibinfo{volume}{72}
  (\bibinfo{year}{2012}) \bibinfo{pages}{1459--1476}.
\bibitem[{Zhang et~al.(2013)Zhang, Zhang, and Zhou}]{zhang13}
\bibinfo{author}{Y.~Zhang}, \bibinfo{author}{L.~Zhang},
  \bibinfo{author}{C.~Zhou},
\newblock \bibinfo{title}{Review of chemical vapor deposition of graphene and
  related applications},
\newblock \bibinfo{journal}{Acc. Chem. Res.} \bibinfo{volume}{46}
  (\bibinfo{year}{2013}) \bibinfo{pages}{2329--2339}.
\bibitem[{Ford(2004)}]{Ford2004}
\bibinfo{author}{I.~J. Ford},
\newblock \bibinfo{title}{Statistical mechanics of nucleation: a review},
\newblock \bibinfo{journal}{Proc. Instn Mech. Engnrs Part C: J. Mech. Eng.
  Sci.} \bibinfo{volume}{218} (\bibinfo{year}{2004}) \bibinfo{pages}{883--899}.
\bibitem[{Kalikmanov(2012)}]{Kalikmanov-book2012}
\bibinfo{author}{V.~I. Kalikmanov}, \bibinfo{title}{Nucleation Theory},
  \bibinfo{publisher}{Springer}, \bibinfo{year}{2012}.
\bibitem[{Ford(2013)}]{Ford-book2013}
\bibinfo{author}{I.~J. Ford}, \bibinfo{title}{Statistical Physics: an entropic
  approach}, \bibinfo{publisher}{Wiley}, \bibinfo{year}{2013}.
\bibitem[{Ford(1997)}]{ford97}
\bibinfo{author}{I.~J. Ford},
\newblock \bibinfo{title}{Nucleation theorems, the statistical mechanics of
  molecular clusters, and a revision of classical nucleation theory},
\newblock \bibinfo{journal}{Phys. Rev. E} \bibinfo{volume}{56}
  (\bibinfo{year}{1997}) \bibinfo{pages}{5615--5629}.
\bibitem[{Harris and Ford(2003)}]{harris03}
\bibinfo{author}{S.~A. Harris}, \bibinfo{author}{I.~J. Ford},
\newblock \bibinfo{title}{A dynamical definition of quasibound molecular
  clusters},
\newblock \bibinfo{journal}{J. Chem. Phys.} \bibinfo{volume}{118}
  (\bibinfo{year}{2003}) \bibinfo{pages}{9216}.
\bibitem[{Smoluchowski(1906)}]{Smoluchowski06}
\bibinfo{author}{M.~Smoluchowski},
\newblock \bibinfo{journal}{Ann. Physik. (Leipzig)} \bibinfo{volume}{21}
  (\bibinfo{year}{1906}) \bibinfo{pages}{756}.
\bibitem[{Vehkam{\"a}ki(2006)}]{Vehkamaki-book2006}
\bibinfo{author}{H.~Vehkam{\"a}ki}, \bibinfo{title}{Classical Nucleation Theory
  in Multicomponent Systems}, \bibinfo{publisher}{Springer},
  \bibinfo{year}{2006}.
\bibitem[{Binnig et~al.(1982)Binnig, Rohrer, Gerber, and Weibel}]{Binnig1982}
\bibinfo{author}{G.~Binnig}, \bibinfo{author}{H.~Rohrer},
  \bibinfo{author}{C.~Gerber}, \bibinfo{author}{E.~Weibel},
\newblock \bibinfo{title}{Surface studies by scanning tunneling microscopy},
\newblock \bibinfo{journal}{Phys. Rev. Lett.} \bibinfo{volume}{49}
  (\bibinfo{year}{1982}) \bibinfo{pages}{57--61}.
\bibitem[{Hansma and Tersoff(1987)}]{Hansma1987}
\bibinfo{author}{P.~K. Hansma}, \bibinfo{author}{J.~Tersoff},
\newblock \bibinfo{title}{Scanning tunneling microscopy},
\newblock \bibinfo{journal}{J. Appl. Phys.} \bibinfo{volume}{61}
  (\bibinfo{year}{1987}) \bibinfo{pages}{R1--R24}.
\bibitem[{Tersoff and Hamann(1985)}]{Tersoff1985}
\bibinfo{author}{J.~Tersoff}, \bibinfo{author}{D.~R. Hamann},
\newblock \bibinfo{title}{Theory of the scanning tunneling microscope},
\newblock \bibinfo{journal}{Phys. Rev. B} \bibinfo{volume}{31}
  (\bibinfo{year}{1985}) \bibinfo{pages}{805}.
\bibitem[{Rost et~al.(2005)Rost, Crama, Schakel, Van~Tol, van Velzen-Williams,
  Overgauw, Ter~Horst, Dekker, Okhuijsen, Seynen, Vijftigschild, Han, Katan,
  Schoots, Schumm, van Loo, Oosterkamp, and Frenken}]{Rost2005}
\bibinfo{author}{M.~J. Rost}, \bibinfo{author}{L.~Crama},
  \bibinfo{author}{P.~Schakel}, \bibinfo{author}{E.~Van~Tol},
  \bibinfo{author}{G.~B. E.~M. van Velzen-Williams}, \bibinfo{author}{C.~F.
  Overgauw}, \bibinfo{author}{H.~Ter~Horst}, \bibinfo{author}{H.~Dekker},
  \bibinfo{author}{B.~Okhuijsen}, \bibinfo{author}{M.~Seynen},
  \bibinfo{author}{A.~Vijftigschild}, \bibinfo{author}{P.~Han},
  \bibinfo{author}{A.~J. Katan}, \bibinfo{author}{K.~Schoots},
  \bibinfo{author}{R.~Schumm}, \bibinfo{author}{W.~van Loo},
  \bibinfo{author}{T.~H. Oosterkamp}, \bibinfo{author}{J.~W.~M. Frenken},
\newblock \bibinfo{title}{Scanning probe microscopes go video rate and beyond},
\newblock \bibinfo{journal}{Rev. Sci. Instr.} \bibinfo{volume}{76}
  (\bibinfo{year}{2005}) \bibinfo{pages}{053710--053710}.
\bibitem[{Hoogeman et~al.(1998)Hoogeman, van Loon, Loos, Ficke, de~Haas,
  van~der Linden, Zeijlemaker, Kuipers, Chang, Klik, and
  Frenken}]{Hoogeman1998}
\bibinfo{author}{M.~S. Hoogeman}, \bibinfo{author}{D.~G. van Loon},
  \bibinfo{author}{R.~W.~M. Loos}, \bibinfo{author}{H.~G. Ficke},
  \bibinfo{author}{E.~de~Haas}, \bibinfo{author}{J.~J. van~der Linden},
  \bibinfo{author}{H.~Zeijlemaker}, \bibinfo{author}{L.~Kuipers},
  \bibinfo{author}{M.~F. Chang}, \bibinfo{author}{M.~A.~J. Klik},
  \bibinfo{author}{J.~W.~M. Frenken},
\newblock \bibinfo{title}{Design and performance of a programmable-temperature
  scanning tunneling microscope},
\newblock \bibinfo{journal}{Rev. Sci. Instr.} \bibinfo{volume}{69}
  (\bibinfo{year}{1998}) \bibinfo{pages}{2072--2080}.
\bibitem[{Land et~al.(1992)Land, Michely, Behm, Hemminger, and
  Comsa}]{Land1992}
\bibinfo{author}{T.~A. Land}, \bibinfo{author}{T.~Michely},
  \bibinfo{author}{R.~J. Behm}, \bibinfo{author}{J.~C. Hemminger},
  \bibinfo{author}{G.~Comsa},
\newblock \bibinfo{title}{{STM investigation of single layer graphite
  structures produced on Pt(111) by hydrocarbon decomposition}},
\newblock \bibinfo{journal}{Surf. Sci.} \bibinfo{volume}{264}
  (\bibinfo{year}{1992}) \bibinfo{pages}{261--270}.
\bibitem[{Mallet et~al.(2007)Mallet, Varchon, Naud, Magaud, Berger, and
  Veuillen}]{Mallet2007}
\bibinfo{author}{P.~Mallet}, \bibinfo{author}{F.~Varchon},
  \bibinfo{author}{C.~Naud}, \bibinfo{author}{L.~Magaud},
  \bibinfo{author}{C.~Berger}, \bibinfo{author}{J.-Y. Veuillen},
\newblock \bibinfo{title}{Electron states of mono- and bilayer graphene on
  {SiC} probed by scanning-tunneling microscopy},
\newblock \bibinfo{journal}{Phys. Rev. B} \bibinfo{volume}{76}
  (\bibinfo{year}{2007}) \bibinfo{pages}{041403(R)}.
\bibitem[{N'Diaye et~al.(2006)N'Diaye, Bleikamp, Feibelman, and
  Michely}]{Diaye2006}
\bibinfo{author}{A.~T. N'Diaye}, \bibinfo{author}{S.~Bleikamp},
  \bibinfo{author}{P.~J. Feibelman}, \bibinfo{author}{T.~Michely},
\newblock \bibinfo{title}{{Two-Dimensional Ir Cluster Lattice on a Graphene
  moir\'{e} on Ir(111)}},
\newblock \bibinfo{journal}{Phys. Rev. Lett.} \bibinfo{volume}{97}
  (\bibinfo{year}{2006}) \bibinfo{pages}{215501}.
\bibitem[{Marchini et~al.(2007)Marchini, G\"unther, and
  Wintterlin}]{Marchini2007}
\bibinfo{author}{S.~Marchini}, \bibinfo{author}{S.~G\"unther},
  \bibinfo{author}{J.~Wintterlin},
\newblock \bibinfo{title}{{Scanning tunneling microscopy of graphene on
  Ru(0001)}},
\newblock \bibinfo{journal}{Phys. Rev. B} \bibinfo{volume}{76}
  (\bibinfo{year}{2007}) \bibinfo{pages}{075429}.
\bibitem[{Riedl et~al.(2007)Riedl, Starke, Bernhardt, Franke, and
  Heinz}]{Riedl2007}
\bibinfo{author}{C.~Riedl}, \bibinfo{author}{U.~Starke},
  \bibinfo{author}{J.~Bernhardt}, \bibinfo{author}{M.~Franke},
  \bibinfo{author}{K.~Heinz},
\newblock \bibinfo{title}{Structural properties of the {graphene-SiC(0001)}
  interface as a key for the preparation of homogeneous large-terrace graphene
  surfaces},
\newblock \bibinfo{journal}{Phys. Rev. B} \bibinfo{volume}{76}
  (\bibinfo{year}{2007}) \bibinfo{pages}{245406}.
\bibitem[{Varchon et~al.(2008)Varchon, Mallet, Veuillen, and
  Magaud}]{Varchon08}
\bibinfo{author}{F.~Varchon}, \bibinfo{author}{P.~Mallet},
  \bibinfo{author}{J.-Y. Veuillen}, \bibinfo{author}{L.~Magaud},
\newblock \bibinfo{title}{{Ripples in epitaxial graphene on the Si-terminated
  SiC(0001) surface}},
\newblock \bibinfo{journal}{Phys. Rev. B} \bibinfo{volume}{77}
  (\bibinfo{year}{2008}) \bibinfo{pages}{235412}.
\bibitem[{Loginova et~al.(2008)Loginova, Bartelt, Feibelman, and
  McCarty}]{Loginova08}
\bibinfo{author}{E.~Loginova}, \bibinfo{author}{N.~C. Bartelt},
  \bibinfo{author}{P.~J. Feibelman}, \bibinfo{author}{K.~F. McCarty},
\newblock \bibinfo{title}{Evidence for growth by {C} cluster attachment},
\newblock \bibinfo{journal}{New J. Phys.} \bibinfo{volume}{10}
  (\bibinfo{year}{2008}) \bibinfo{pages}{093026}.
\bibitem[{Hiebel et~al.(2008)Hiebel, Mallet, Varchon, Magaud, and
  Veuillen}]{Hiebel2008}
\bibinfo{author}{F.~Hiebel}, \bibinfo{author}{P.~Mallet},
  \bibinfo{author}{F.~Varchon}, \bibinfo{author}{L.~Magaud},
  \bibinfo{author}{J.~Y. Veuillen},
\newblock \bibinfo{title}{{Graphene-substrate interaction on 6H-SiC
  (000$\bar{1}$): A scanning tunneling microscopy study}},
\newblock \bibinfo{journal}{Phys. Rev. B} \bibinfo{volume}{78}
  (\bibinfo{year}{2008}) \bibinfo{pages}{153412}.
\bibitem[{Otero et~al.(2010)Otero, Gonzalez, Pinardi, Merino, Gardonio, Lizzit,
  Blanco-Rey, Van~de Ruit, Flipse, M{\'e}ndez et~al.}]{Otero2010}
\bibinfo{author}{G.~Otero}, \bibinfo{author}{C.~Gonzalez},
  \bibinfo{author}{A.~L. Pinardi}, \bibinfo{author}{P.~Merino},
  \bibinfo{author}{S.~Gardonio}, \bibinfo{author}{S.~Lizzit},
  \bibinfo{author}{M.~Blanco-Rey}, \bibinfo{author}{K.~Van~de Ruit},
  \bibinfo{author}{C.~F.~J. Flipse}, \bibinfo{author}{J.~M{\'e}ndez}, et~al.,
\newblock \bibinfo{title}{{Ordered vacancy network induced by the growth of
  epitaxial graphene on Pt(111)}},
\newblock \bibinfo{journal}{Phys. Rev. Lett.} \bibinfo{volume}{105}
  (\bibinfo{year}{2010}) \bibinfo{pages}{216102}.
\bibitem[{Coraux et~al.(2008)Coraux, N`Diaye, Busse, and Michely}]{Coroux2008b}
\bibinfo{author}{J.~Coraux}, \bibinfo{author}{A.~T. N`Diaye},
  \bibinfo{author}{C.~Busse}, \bibinfo{author}{T.~Michely},
\newblock \bibinfo{title}{{Structural Coherency of Graphene on Ir(111)}},
\newblock \bibinfo{journal}{Nano Lett.} \bibinfo{volume}{8}
  (\bibinfo{year}{2008}) \bibinfo{pages}{565--570}.
\bibitem[{Sutter et~al.(2009)Sutter, Sadowski, and Sutter}]{Sutter2009}
\bibinfo{author}{P.~Sutter}, \bibinfo{author}{J.~T. Sadowski},
  \bibinfo{author}{E.~Sutter},
\newblock \bibinfo{title}{{Graphene on Pt(111): Growth and substrate
  interaction}},
\newblock \bibinfo{journal}{Phys. Rev. B} \bibinfo{volume}{80}
  (\bibinfo{year}{2009}) \bibinfo{pages}{245411}.
\bibitem[{Biedermann et~al.(2009)Biedermann, Bolen, Capano, Zemlyanov, and
  Reifenberger}]{Biedermann2009}
\bibinfo{author}{L.~B. Biedermann}, \bibinfo{author}{M.~L. Bolen},
  \bibinfo{author}{M.~A. Capano}, \bibinfo{author}{D.~Zemlyanov},
  \bibinfo{author}{R.~G. Reifenberger},
\newblock \bibinfo{title}{Insights into few-layer epitaxial graphene growth on
  4h$\text{-SiC}(000\overline{1})$ substrates from {STM} studies},
\newblock \bibinfo{journal}{Phys. Rev. B} \bibinfo{volume}{79}
  (\bibinfo{year}{2009}) \bibinfo{pages}{125411}.
\bibitem[{Levy et~al.(2010)Levy, Burke, Meaker, Panlasigui, Zettl, Guinea,
  Neto, and Crommie}]{Levy2010}
\bibinfo{author}{N.~Levy}, \bibinfo{author}{S.~A. Burke},
  \bibinfo{author}{K.~L. Meaker}, \bibinfo{author}{M.~Panlasigui},
  \bibinfo{author}{A.~Zettl}, \bibinfo{author}{F.~Guinea},
  \bibinfo{author}{A.~H.~C. Neto}, \bibinfo{author}{M.~F. Crommie},
\newblock \bibinfo{title}{Strain-induced pseudo--magnetic fields greater than
  300 tesla in graphene nanobubbles},
\newblock \bibinfo{journal}{Science} \bibinfo{volume}{329}
  (\bibinfo{year}{2010}) \bibinfo{pages}{544--547}.
\bibitem[{Coraux et~al.(2009)Coraux, N'Diaye, Engler, Busse, Wall, Buckanie,
  zu~Heringdorf, van Gastel, Poelsema, and Michely}]{Coraux}
\bibinfo{author}{J.~Coraux}, \bibinfo{author}{A.~T. N'Diaye},
  \bibinfo{author}{M.~Engler}, \bibinfo{author}{C.~Busse},
  \bibinfo{author}{D.~Wall}, \bibinfo{author}{N.~Buckanie},
  \bibinfo{author}{F.~J.~M. zu~Heringdorf}, \bibinfo{author}{R.~van Gastel},
  \bibinfo{author}{B.~Poelsema}, \bibinfo{author}{T.~Michely},
\newblock \bibinfo{title}{Growth of graphene on {I}r(111)},
\newblock \bibinfo{journal}{New J. Phys.} \bibinfo{volume}{11}
  (\bibinfo{year}{2009}) \bibinfo{pages}{023006}.
\bibitem[{Kwon et~al.(2009)Kwon, Ciobanu, Petrova, Shenoy, Bare$\~n$o, Gambin,
  Petrov, and Kodambaka}]{Kwon_NL_2009}
\bibinfo{author}{S.-Y. Kwon}, \bibinfo{author}{C.~V. Ciobanu},
  \bibinfo{author}{V.~Petrova}, \bibinfo{author}{V.~B. Shenoy},
  \bibinfo{author}{J.~Bare$\~n$o}, \bibinfo{author}{V.~Gambin},
  \bibinfo{author}{I.~Petrov}, \bibinfo{author}{S.~Kodambaka},
\newblock \bibinfo{title}{Growth of semiconducting graphene on palladium},
\newblock \bibinfo{journal}{Nano Lett.} \bibinfo{volume}{9}
  (\bibinfo{year}{2009}) \bibinfo{pages}{3985}.
\bibitem[{Dong et~al.(2012)Dong, van Baarle, Rost, and Frenken}]{Dong}
\bibinfo{author}{G.~C. Dong}, \bibinfo{author}{D.~W. van Baarle},
  \bibinfo{author}{M.~J. Rost}, \bibinfo{author}{J.~W.~M. Frenken},
\newblock \bibinfo{title}{Graphene formation on metal surfaces investigated by
  in-situ scanning tunneling microscopy},
\newblock \bibinfo{journal}{New J. Phys.} \bibinfo{volume}{14}
  (\bibinfo{year}{2012}) \bibinfo{pages}{053033}.
\bibitem[{Binnig et~al.(1986)Binnig, Quate, and Gerber}]{Binnig1986}
\bibinfo{author}{G.~Binnig}, \bibinfo{author}{C.~F. Quate},
  \bibinfo{author}{C.~Gerber},
\newblock \bibinfo{title}{Atomic force microscope},
\newblock \bibinfo{journal}{Phys. Rev. Lett.} \bibinfo{volume}{56}
  (\bibinfo{year}{1986}) \bibinfo{pages}{930--933}.
\bibitem[{Giessibl and Binnig(1992)}]{Giessibl1992}
\bibinfo{author}{F.~J. Giessibl}, \bibinfo{author}{G.~Binnig},
\newblock \bibinfo{title}{{True atomic resolution on KBr with a low-temperature
  atomic force microscope in ultrahigh vacuum}},
\newblock \bibinfo{journal}{Ultramicroscopy} \bibinfo{volume}{42}
  (\bibinfo{year}{1992}) \bibinfo{pages}{281--286}.
\bibitem[{Meyer(1992)}]{Meyer1992}
\bibinfo{author}{E.~Meyer},
\newblock \bibinfo{title}{Atomic force microscopy},
\newblock \bibinfo{journal}{Progr. Surf. Sci.} \bibinfo{volume}{41}
  (\bibinfo{year}{1992}) \bibinfo{pages}{3--49}.
\bibitem[{Giessibl(2003)}]{Giessibl2003}
\bibinfo{author}{F.~J. Giessibl},
\newblock \bibinfo{title}{Advances in atomic force microscopy},
\newblock \bibinfo{journal}{Rev. Mod. Phys.} \bibinfo{volume}{75}
  (\bibinfo{year}{2003}) \bibinfo{pages}{949}.
\bibitem[{Hibino et~al.(2010)Hibino, Kageshima, and Nagase}]{Hibino2010}
\bibinfo{author}{H.~Hibino}, \bibinfo{author}{H.~Kageshima},
  \bibinfo{author}{M.~Nagase},
\newblock \bibinfo{title}{Epitaxial few-layer graphene: towards single crystal
  growth},
\newblock \bibinfo{journal}{J. Phys. D: Appl. Phys.} \bibinfo{volume}{43}
  (\bibinfo{year}{2010}) \bibinfo{pages}{374005}.
\bibitem[{Boneschanscher et~al.(2012)Boneschanscher, van~der Lit, Sun, Swart,
  Liljeroth, and Vanmaekelbergh}]{Boneschanscher2012}
\bibinfo{author}{M.~P. Boneschanscher}, \bibinfo{author}{J.~van~der Lit},
  \bibinfo{author}{Z.~Sun}, \bibinfo{author}{I.~Swart},
  \bibinfo{author}{P.~Liljeroth}, \bibinfo{author}{D.~Vanmaekelbergh},
\newblock \bibinfo{title}{Quantitative atomic resolution force imaging on
  epitaxial graphene with reactive and nonreactive {AFM} probes},
\newblock \bibinfo{journal}{ACS Nano} \bibinfo{volume}{6}
  (\bibinfo{year}{2012}) \bibinfo{pages}{10216--10221}.
\bibitem[{Novoselov et~al.(2004)Novoselov, Geim, Morozov, Jiang, Zhang,
  Dubonos, Grigorieva, and Firsov}]{Novoselov2004}
\bibinfo{author}{K.~S. Novoselov}, \bibinfo{author}{A.~K. Geim},
  \bibinfo{author}{S.~V. Morozov}, \bibinfo{author}{D.~Jiang},
  \bibinfo{author}{Y.~Zhang}, \bibinfo{author}{S.~V. Dubonos},
  \bibinfo{author}{I.~V. Grigorieva}, \bibinfo{author}{A.~A. Firsov},
\newblock \bibinfo{title}{Electric field effect in atomically thin carbon
  films},
\newblock \bibinfo{journal}{Science} \bibinfo{volume}{306}
  (\bibinfo{year}{2004}) \bibinfo{pages}{666--669}.
\bibitem[{Berger et~al.(2006)Berger, Song, Li, Wu, Brown, Naud, Mayou, Li,
  Hass, Marchenkov, Conrad, First, and de~Heer}]{Berger2006}
\bibinfo{author}{C.~Berger}, \bibinfo{author}{Z.~Song},
  \bibinfo{author}{X.~Li}, \bibinfo{author}{X.~Wu}, \bibinfo{author}{N.~Brown},
  \bibinfo{author}{C.~Naud}, \bibinfo{author}{D.~Mayou},
  \bibinfo{author}{T.~Li}, \bibinfo{author}{J.~Hass}, \bibinfo{author}{A.~N.
  Marchenkov}, \bibinfo{author}{E.~H. Conrad}, \bibinfo{author}{P.~N. First},
  \bibinfo{author}{W.~A. de~Heer},
\newblock \bibinfo{title}{Electronic confinement and coherence in patterned
  epitaxial graphene},
\newblock \bibinfo{journal}{Science} \bibinfo{volume}{312}
  (\bibinfo{year}{2006}) \bibinfo{pages}{1191--1196}.
\bibitem[{Reina et~al.(2008)Reina, Jia, Ho, Nezich, Son, Bulovic, Dresselhaus,
  and Kong}]{Reina2008}
\bibinfo{author}{A.~Reina}, \bibinfo{author}{X.~Jia}, \bibinfo{author}{J.~Ho},
  \bibinfo{author}{D.~Nezich}, \bibinfo{author}{H.~Son},
  \bibinfo{author}{V.~Bulovic}, \bibinfo{author}{M.~S. Dresselhaus},
  \bibinfo{author}{J.~Kong},
\newblock \bibinfo{title}{Large area, few-layer graphene films on arbitrary
  substrates by chemical vapor deposition},
\newblock \bibinfo{journal}{Nano Lett.} \bibinfo{volume}{9}
  (\bibinfo{year}{2008}) \bibinfo{pages}{30--35}.
\bibitem[{Bolen et~al.(2009)Bolen, Harrison, Biedermann, and
  Capano}]{Bolen2009}
\bibinfo{author}{M.~L. Bolen}, \bibinfo{author}{S.~E. Harrison},
  \bibinfo{author}{L.~B. Biedermann}, \bibinfo{author}{M.~A. Capano},
\newblock \bibinfo{title}{{Graphene formation mechanisms on 4H-SiC(0001)}},
\newblock \bibinfo{journal}{Phys. Rev. B} \bibinfo{volume}{80}
  (\bibinfo{year}{2009}) \bibinfo{pages}{115433}.
\bibitem[{Hannon and Tromp(2008)}]{Hannon2008}
\bibinfo{author}{J.~B. Hannon}, \bibinfo{author}{R.~M. Tromp},
\newblock \bibinfo{title}{Pit formation during graphene synthesis on
  {SiC}(0001): In situ electron microscopy},
\newblock \bibinfo{journal}{Phys. Rev. B} \bibinfo{volume}{77}
  (\bibinfo{year}{2008}) \bibinfo{pages}{241204}.
\bibitem[{Sun et~al.(2011)Sun, H\"{a}m\"{a}l\"{a}inen, Sainio, Lahtinen,
  Vanmaekelbergh, and Liljeroth}]{Sun2011}
\bibinfo{author}{Z.~Sun}, \bibinfo{author}{S.~K. H\"{a}m\"{a}l\"{a}inen},
  \bibinfo{author}{J.~Sainio}, \bibinfo{author}{J.~Lahtinen},
  \bibinfo{author}{D.~Vanmaekelbergh}, \bibinfo{author}{P.~Liljeroth},
\newblock \bibinfo{title}{Topographic and electronic contrast of the graphene
  moir\'{e} on {I}r(111) probed by scanning tunneling microscopy and noncontact
  atomic force microscopy},
\newblock \bibinfo{journal}{Phys. Rev. B} \bibinfo{volume}{83}
  (\bibinfo{year}{2011}) \bibinfo{pages}{081415}.
\bibitem[{Voloshina et~al.(2013)Voloshina, Fertitta, Garhofer, Mittendorfer,
  Fonin, Thissen, and Dedkov}]{Voloshina2013}
\bibinfo{author}{E.~N. Voloshina}, \bibinfo{author}{E.~Fertitta},
  \bibinfo{author}{A.~Garhofer}, \bibinfo{author}{F.~Mittendorfer},
  \bibinfo{author}{M.~Fonin}, \bibinfo{author}{A.~Thissen},
  \bibinfo{author}{Y.~S. Dedkov},
\newblock \bibinfo{title}{{Electronic structure and imaging contrast of
  graphene moir\'{e} on metals}},
\newblock \bibinfo{journal}{Nature Sci. Rep.} \bibinfo{volume}{3}
  (\bibinfo{year}{2013}) \bibinfo{pages}{1072}.
\bibitem[{Ruska(1980)}]{Ruska1980}
\bibinfo{author}{E.~Ruska}, \bibinfo{title}{The early development of electron
  lenses and electron microscopy}, \bibinfo{publisher}{Hirzel},
  \bibinfo{year}{1980}.
\bibitem[{Haider et~al.(1998)Haider, Uhlemann, Schwan, Rose, Kabius, and
  Urban}]{Haider1998}
\bibinfo{author}{M.~Haider}, \bibinfo{author}{S.~Uhlemann},
  \bibinfo{author}{E.~Schwan}, \bibinfo{author}{H.~Rose},
  \bibinfo{author}{B.~Kabius}, \bibinfo{author}{K.~Urban},
\newblock \bibinfo{title}{Electron microscopy image enhanced},
\newblock \bibinfo{journal}{Nature} \bibinfo{volume}{392}
  (\bibinfo{year}{1998}) \bibinfo{pages}{768--769}.
\bibitem[{Batson et~al.(2002)Batson, Dellby, and Krivanek}]{Batson2002}
\bibinfo{author}{P.~E. Batson}, \bibinfo{author}{N.~Dellby},
  \bibinfo{author}{O.~L. Krivanek},
\newblock \bibinfo{title}{Sub-{\aa}ngstrom resolution using aberration
  corrected electron optics},
\newblock \bibinfo{journal}{Nature} \bibinfo{volume}{418}
  (\bibinfo{year}{2002}) \bibinfo{pages}{617--620}.
\bibitem[{Hansen et~al.(2002)Hansen, Wagner, Helveg, Rostrup-Nielsen, Clausen,
  and Tops{\o}e}]{Hansen2002}
\bibinfo{author}{P.~L. Hansen}, \bibinfo{author}{J.~B. Wagner},
  \bibinfo{author}{S.~Helveg}, \bibinfo{author}{J.~R. Rostrup-Nielsen},
  \bibinfo{author}{B.~S. Clausen}, \bibinfo{author}{H.~Tops{\o}e},
\newblock \bibinfo{title}{Atom-resolved imaging of dynamic shape changes in
  supported copper nanocrystals},
\newblock \bibinfo{journal}{Science} \bibinfo{volume}{295}
  (\bibinfo{year}{2002}) \bibinfo{pages}{2053--2055}.
\bibitem[{Rodriguez-Manzo et~al.(2011)Rodriguez-Manzo, Pham-Huu, and
  Banhart}]{RodriguezManzo2011}
\bibinfo{author}{J.~A. Rodriguez-Manzo}, \bibinfo{author}{C.~Pham-Huu},
  \bibinfo{author}{F.~Banhart},
\newblock \bibinfo{title}{Graphene growth by a metal-catalyzed solid-state
  transformation of amorphous carbon},
\newblock \bibinfo{journal}{ACS Nano} \bibinfo{volume}{5}
  (\bibinfo{year}{2011}) \bibinfo{pages}{1529--1534}.
\bibitem[{Smith and Luzzi(2001)}]{Smith2001}
\bibinfo{author}{B.~W. Smith}, \bibinfo{author}{D.~E. Luzzi},
\newblock \bibinfo{title}{Electron irradiation effects in single wall carbon
  nanotubes},
\newblock \bibinfo{journal}{J. Appl. Phys.} \bibinfo{volume}{90}
  (\bibinfo{year}{2001}) \bibinfo{pages}{3509--3515}.
\bibitem[{Irving and Walker(1967)}]{Irving1967}
\bibinfo{author}{S.~M. Irving}, \bibinfo{author}{P.~L. Walker},
\newblock \bibinfo{title}{Interaction of evaporated carbon with heated metal
  substrates},
\newblock \bibinfo{journal}{Carbon} \bibinfo{volume}{5} (\bibinfo{year}{1967})
  \bibinfo{pages}{399--402}.
\bibitem[{Turner and Bauer(1966)}]{Turner1966}
\bibinfo{author}{G.~Turner}, \bibinfo{author}{E.~Bauer},
\newblock \bibinfo{title}{An ultrahigh vacuum electron microscope and its
  application to work function studies},
\newblock \bibinfo{journal}{Electron Microscopy} \bibinfo{volume}{1}
  (\bibinfo{year}{1966}) \bibinfo{pages}{163}.
\bibitem[{Telieps and Bauer(1985{\natexlab{a}})}]{Telieps1985a}
\bibinfo{author}{W.~Telieps}, \bibinfo{author}{E.~Bauer},
\newblock \bibinfo{title}{An analytical reflection and emission {UHV} surface
  electron microscope},
\newblock \bibinfo{journal}{Ultramicroscopy} \bibinfo{volume}{17}
  (\bibinfo{year}{1985}{\natexlab{a}}) \bibinfo{pages}{57}.
\bibitem[{Telieps and Bauer(1985{\natexlab{b}})}]{Telieps1985b}
\bibinfo{author}{W.~Telieps}, \bibinfo{author}{E.~Bauer},
\newblock \bibinfo{title}{The (7 $\times$ 7) $\leftrightarrow$ (1 $\times$ 1)
  phase transition on {S}i(111)},
\newblock \bibinfo{journal}{Surf. Sci.} \bibinfo{volume}{162}
  (\bibinfo{year}{1985}{\natexlab{b}}) \bibinfo{pages}{163}.
\bibitem[{Bauer(2012)}]{Bauer2011}
\bibinfo{author}{E.~Bauer},
\newblock \bibinfo{title}{{LEEM and UHV-PEEM: A retrospective}},
\newblock \bibinfo{journal}{Ultramicroscopy} \bibinfo{volume}{119}
  (\bibinfo{year}{2012}) \bibinfo{pages}{18}.
\bibitem[{Tromp et~al.(2010)Tromp, Hannon, Ellis, Wan, Berghaus, and
  Schaff}]{Tromp2010}
\bibinfo{author}{R.~M. Tromp}, \bibinfo{author}{J.~B. Hannon},
  \bibinfo{author}{A.~W. Ellis}, \bibinfo{author}{W.~Wan},
  \bibinfo{author}{A.~Berghaus}, \bibinfo{author}{O.~Schaff},
\newblock \bibinfo{title}{A new aberration-corrected, energy-filtered
  {LEEM}/{PEEM} instrument. {I}. {P}rinciples and design},
\newblock \bibinfo{journal}{Ultramicroscopy} \bibinfo{volume}{110}
  (\bibinfo{year}{2010}) \bibinfo{pages}{852--861}.
\bibitem[{Schmidt et~al.(1998)Schmidt, Heun, Slezak, Diaz, Prince, Lilienkamp,
  and Bauer}]{Schmidt1998}
\bibinfo{author}{T.~Schmidt}, \bibinfo{author}{S.~Heun},
  \bibinfo{author}{J.~Slezak}, \bibinfo{author}{J.~Diaz},
  \bibinfo{author}{K.~C. Prince}, \bibinfo{author}{G.~Lilienkamp},
  \bibinfo{author}{E.~Bauer},
\newblock \bibinfo{title}{{SPELEEM: Combining LEEM and spectroscopic imaging}},
\newblock \bibinfo{journal}{Surf. Rev. Lett.} \bibinfo{volume}{5}
  (\bibinfo{year}{1998}) \bibinfo{pages}{1287}.
\bibitem[{Ohta et~al.(2008)Ohta, Gabaly, Bostwick, McChesney, Emtsev, Schmid,
  Seyller, Horn, and Rotenberg}]{Ohta2008}
\bibinfo{author}{T.~Ohta}, \bibinfo{author}{F.~E. Gabaly},
  \bibinfo{author}{A.~Bostwick}, \bibinfo{author}{J.~L. McChesney},
  \bibinfo{author}{K.~V. Emtsev}, \bibinfo{author}{A.~K. Schmid},
  \bibinfo{author}{T.~Seyller}, \bibinfo{author}{K.~Horn},
  \bibinfo{author}{E.~Rotenberg},
\newblock \bibinfo{title}{Morphology of graphene thin film growth on
  {SiC(0001)}},
\newblock \bibinfo{journal}{New J. Phys.} \bibinfo{volume}{10}
  (\bibinfo{year}{2008}) \bibinfo{pages}{023034}.
\bibitem[{Sutter et~al.(2008)Sutter, Flege, and Sutter}]{Sutter08}
\bibinfo{author}{P.~W. Sutter}, \bibinfo{author}{J.~Flege},
  \bibinfo{author}{E.~A. Sutter},
\newblock \bibinfo{title}{Epitaxial graphene on ruthenium},
\newblock \bibinfo{journal}{Nature Mater.} \bibinfo{volume}{7}
  (\bibinfo{year}{2008}) \bibinfo{pages}{406--411}.
\bibitem[{Loginova et~al.(2009)Loginova, Bartelt, Feibelman, and
  McCarty}]{Loginova09}
\bibinfo{author}{E.~Loginova}, \bibinfo{author}{N.~C. Bartelt},
  \bibinfo{author}{P.~J. Feibelman}, \bibinfo{author}{K.~F. McCarty},
\newblock \bibinfo{title}{Factors influencing graphene growth on metal
  surfaces},
\newblock \bibinfo{journal}{New J. Phys} \bibinfo{volume}{11}
  (\bibinfo{year}{2009}) \bibinfo{pages}{063046}.
\bibitem[{van Gastel et~al.(2009)van Gastel, N'Diaye, Wall, Coraux, Busse,
  Buckanie, zu~Heringdorf, von Hoegen, Michely, and Poelsema}]{vanGastel2009}
\bibinfo{author}{R.~van Gastel}, \bibinfo{author}{A.~T. N'Diaye},
  \bibinfo{author}{D.~Wall}, \bibinfo{author}{J.~Coraux},
  \bibinfo{author}{C.~Busse}, \bibinfo{author}{N.~M. Buckanie},
  \bibinfo{author}{F.-J.~M. zu~Heringdorf}, \bibinfo{author}{M.~H. von Hoegen},
  \bibinfo{author}{T.~Michely}, \bibinfo{author}{B.~Poelsema},
\newblock \bibinfo{title}{Selecting a single orientation for millimeter sized
  graphene sheets},
\newblock \bibinfo{journal}{Appl. Phys. Lett.} \bibinfo{volume}{95}
  (\bibinfo{year}{2009}) \bibinfo{pages}{121901}.
\bibitem[{Lagally(1985)}]{Lagally1985}
\bibinfo{author}{M.~G. Lagally}, \bibinfo{title}{Methods in Experimental
  Physics}, volume~\bibinfo{volume}{22}, \bibinfo{publisher}{Elsevier},
  \bibinfo{year}{1985}, pp. \bibinfo{pages}{237--298}.
\bibitem[{Van~Hove et~al.(1986)Van~Hove, Weinberg, and Chan}]{vanHove1986}
\bibinfo{author}{M.~A. Van~Hove}, \bibinfo{author}{W.~H. Weinberg},
  \bibinfo{author}{C.~Chan}, \bibinfo{title}{Low-energy electron diffraction},
  \bibinfo{publisher}{Springer-Verlag Berlin}, \bibinfo{year}{1986}.
\bibitem[{Scheithauer et~al.(1986)Scheithauer, Meyer, and
  Henzler}]{Sheithauer1986}
\bibinfo{author}{U.~Scheithauer}, \bibinfo{author}{G.~Meyer},
  \bibinfo{author}{M.~Henzler},
\newblock \bibinfo{title}{A new {LEED} instrument for quantitative spot profile
  analysis},
\newblock \bibinfo{journal}{Surf. Sci.} \bibinfo{volume}{178}
  (\bibinfo{year}{1986}) \bibinfo{pages}{441}.
\bibitem[{Blanc et~al.(2013)Blanc, Jean, Krasheninnikov, Renaud, and
  Coraux}]{Blanc-Coraux-2013}
\bibinfo{author}{N.~Blanc}, \bibinfo{author}{F.~Jean}, \bibinfo{author}{A.~V.
  Krasheninnikov}, \bibinfo{author}{G.~Renaud}, \bibinfo{author}{J.~Coraux},
\newblock \bibinfo{title}{Strains induced by point defects in graphene on a
  metal},
\newblock \bibinfo{journal}{Phys. Rev. Lett.} \bibinfo{volume}{111}
  (\bibinfo{year}{2013}) \bibinfo{pages}{085501}.
\bibitem[{van Bommel et~al.(1975)van Bommel, Crombeen, and van
  Tooren}]{vanBommel1975}
\bibinfo{author}{A.~J. van Bommel}, \bibinfo{author}{J.~E. Crombeen},
  \bibinfo{author}{A.~van Tooren},
\newblock \bibinfo{title}{{LEED} and {A}uger electron observations of the
  {SiC}(0001) surface},
\newblock \bibinfo{journal}{Surf. Sci.} \bibinfo{volume}{48}
  (\bibinfo{year}{1975}) \bibinfo{pages}{463}.
\bibitem[{Hattab et~al.(2011)Hattab, N'Diaye, Wall, Jnawali, Coraux, Busse, van
  Gastel, Poelsema, Michely, zu~Heringdorf, and von Hoegen}]{Hattab2011}
\bibinfo{author}{H.~Hattab}, \bibinfo{author}{A.~T. N'Diaye},
  \bibinfo{author}{D.~Wall}, \bibinfo{author}{G.~Jnawali},
  \bibinfo{author}{J.~Coraux}, \bibinfo{author}{C.~Busse},
  \bibinfo{author}{R.~van Gastel}, \bibinfo{author}{B.~Poelsema},
  \bibinfo{author}{T.~Michely}, \bibinfo{author}{F.-J.~M. zu~Heringdorf},
  \bibinfo{author}{M.~H. von Hoegen},
\newblock \bibinfo{title}{{Growth temperature dependent graphene alignment on
  Ir(111)}},
\newblock \bibinfo{journal}{Appl. Phys. Lett.} \bibinfo{volume}{98}
  (\bibinfo{year}{2011}) \bibinfo{pages}{141903}.
\bibitem[{Vo-Van et~al.(2011)Vo-Van, Kimouche, Reserbat-Plantey, Fruchart,
  Bayle-Guillemaud, Bendiab, and Coraux}]{VoVan2011}
\bibinfo{author}{C.~Vo-Van}, \bibinfo{author}{A.~Kimouche},
  \bibinfo{author}{A.~Reserbat-Plantey}, \bibinfo{author}{O.~Fruchart},
  \bibinfo{author}{P.~Bayle-Guillemaud}, \bibinfo{author}{N.~Bendiab},
  \bibinfo{author}{J.~Coraux},
\newblock \bibinfo{title}{Epitaxial graphene prepared by chemical vapor
  deposition on single crystal thin iridium films on sapphire},
\newblock \bibinfo{journal}{Appl. Phys. Lett.} \bibinfo{volume}{98}
  (\bibinfo{year}{2011}) \bibinfo{pages}{181903}.
\bibitem[{Forbeaux et~al.(2008)Forbeaux, Themlin, and Debever}]{Forbeaux1998}
\bibinfo{author}{I.~Forbeaux}, \bibinfo{author}{J.-M. Themlin},
  \bibinfo{author}{J.-M. Debever},
\newblock \bibinfo{title}{Heteroepitaxial graphite on 6{H}-{SiC}(0001):
  Interface formation through conduction-band electronic structure},
\newblock \bibinfo{journal}{Phys. Rev. B} \bibinfo{volume}{58}
  (\bibinfo{year}{2008}) \bibinfo{pages}{16396}.
\bibitem[{Nieuwenhuys et~al.(1976)Nieuwenhuys, Hagen, Rovida, and
  Somorjai}]{Nieuwenhuys-TPD-1976}
\bibinfo{author}{B.~E. Nieuwenhuys}, \bibinfo{author}{D.~I. Hagen},
  \bibinfo{author}{G.~Rovida}, \bibinfo{author}{G.~A. Somorjai},
\newblock \bibinfo{title}{{LEED, AES} and thermal desorption studies of
  chemisorbed hydrogen and hydrocarbons ({C$_2$H$_2$, C$_2$H$_4$, C$_6$H$_6$,
  C$_6$H$_{12}$}) on the (111) and stepped [6(111) $\times$ (100)] iridium
  crystal surfaces; comparison with platinum},
\newblock \bibinfo{journal}{Surf. Sci.} \bibinfo{volume}{155}
  (\bibinfo{year}{1976}) \bibinfo{pages}{59}.
\bibitem[{Gamo et~al.(1997)Gamo, Nagashima, Wakabayashi, Terai, and
  Oshima}]{Gamo1997}
\bibinfo{author}{Y.~Gamo}, \bibinfo{author}{A.~Nagashima},
  \bibinfo{author}{M.~Wakabayashi}, \bibinfo{author}{M.~Terai},
  \bibinfo{author}{C.~Oshima},
\newblock \bibinfo{title}{Atomic structure of monolayer graphite formed on
  {N}i(111)},
\newblock \bibinfo{journal}{Surf. Sci.} \bibinfo{volume}{374}
  (\bibinfo{year}{1997}) \bibinfo{pages}{61}.
\bibitem[{Moritz et~al.(2010)Moritz, Wang, Bocquet, Brugger, Greber,
  Wintterlin, and G{\"u}nther}]{Moritz}
\bibinfo{author}{W.~Moritz}, \bibinfo{author}{B.~Wang},
  \bibinfo{author}{M.~Bocquet}, \bibinfo{author}{T.~Brugger},
  \bibinfo{author}{T.~Greber}, \bibinfo{author}{J.~Wintterlin},
  \bibinfo{author}{S.~G{\"u}nther},
\newblock \bibinfo{title}{{Structure Determination of the Coincidence Phase of
  Graphene on Ru(0001)}},
\newblock \bibinfo{journal}{Phys. Rev. Lett.} \bibinfo{volume}{104}
  (\bibinfo{year}{2010}) \bibinfo{pages}{136102}.
\bibitem[{H\"{a}m\"{a}l\"{a}inen et~al.(2014)H\"{a}m\"{a}l\"{a}inen,
  Boneschanscher, Jacobse, Swart, Pussi, Moritz, Lahtinen, Liljeroth, and
  Sainio}]{Hamalaninen-2014}
\bibinfo{author}{S.~K. H\"{a}m\"{a}l\"{a}inen}, \bibinfo{author}{M.~P.
  Boneschanscher}, \bibinfo{author}{P.~H. Jacobse}, \bibinfo{author}{I.~Swart},
  \bibinfo{author}{K.~Pussi}, \bibinfo{author}{W.~Moritz},
  \bibinfo{author}{J.~Lahtinen}, \bibinfo{author}{P.~Liljeroth},
  \bibinfo{author}{J.~Sainio},
\newblock \bibinfo{title}{The structure and local variations of the graphene
  moir\'{e} on {I}r(111).},
\newblock \bibinfo{journal}{submitted}  (\bibinfo{year}{2014}).
\bibitem[{Loginova et~al.(2009)Loginova, Nie, Th\"urmer, Bartelt, and
  McCarty}]{Loginova2009b}
\bibinfo{author}{E.~Loginova}, \bibinfo{author}{S.~Nie},
  \bibinfo{author}{K.~Th\"urmer}, \bibinfo{author}{N.~C. Bartelt},
  \bibinfo{author}{K.~F. McCarty},
\newblock \bibinfo{title}{{Defects of graphene on Ir(111): Rotational domains
  and ridges}},
\newblock \bibinfo{journal}{Phys. Rev. B} \bibinfo{volume}{80}
  (\bibinfo{year}{2009}) \bibinfo{pages}{085430}.
\bibitem[{Moreau et~al.(2010)Moreau, Godey, Ferrer, Vignaud, Wallart, Avila,
  Asensio, Bournel, and Gallet}]{Moreau2010}
\bibinfo{author}{E.~Moreau}, \bibinfo{author}{S.~Godey}, \bibinfo{author}{F.~J.
  Ferrer}, \bibinfo{author}{D.~Vignaud}, \bibinfo{author}{X.~Wallart},
  \bibinfo{author}{J.~Avila}, \bibinfo{author}{M.~C. Asensio},
  \bibinfo{author}{F.~Bournel}, \bibinfo{author}{J.-J. Gallet},
\newblock \bibinfo{title}{{Graphene growth by molecular beam epitaxy on the
  carbon-face of SiC}},
\newblock \bibinfo{journal}{Appl. Phys. Lett.} \bibinfo{volume}{97}
  (\bibinfo{year}{2010}) \bibinfo{pages}{241907}.
\bibitem[{Vo-Van et~al.(2010)Vo-Van, Kassir-Bodon, Yang, Coraux, Vogel,
  Pizzini, Bayle-Guillemaud, Chshiev, Ranno, Guisset, David, Salvador, and
  Fruchart}]{VoVan2010}
\bibinfo{author}{C.~Vo-Van}, \bibinfo{author}{Z.~Kassir-Bodon},
  \bibinfo{author}{Yang}, \bibinfo{author}{J.~Coraux},
  \bibinfo{author}{J.~Vogel}, \bibinfo{author}{S.~Pizzini},
  \bibinfo{author}{P.~Bayle-Guillemaud}, \bibinfo{author}{M.~Chshiev},
  \bibinfo{author}{L.~Ranno}, \bibinfo{author}{V.~Guisset},
  \bibinfo{author}{P.~David}, \bibinfo{author}{V.~Salvador},
  \bibinfo{author}{O.~Fruchart},
\newblock \bibinfo{title}{Ultrathin epitaxial cobalt films on graphene for
  spintronic investigations and applications},
\newblock \bibinfo{journal}{New J. Phys.} \bibinfo{volume}{12}
  (\bibinfo{year}{2010}) \bibinfo{pages}{103040}.
\bibitem[{Von~Ardenne(1938)}]{vonArdenne1938}
\bibinfo{author}{M.~Von~Ardenne},
\newblock \bibinfo{title}{Das elektronen-rastermikroskop},
\newblock \bibinfo{journal}{Zeitschrift f{\"u}r Physik} \bibinfo{volume}{109}
  (\bibinfo{year}{1938}) \bibinfo{pages}{553--572}.
\bibitem[{Biswas and Nath(1983)}]{Biswas1983}
\bibinfo{author}{D.~R. Biswas}, \bibinfo{author}{D.~K. Nath},
\newblock \bibinfo{title}{Direct observation of sintering of {VAD} soot
  particles by hot stage scanning electron microscopy},
\newblock \bibinfo{journal}{J. Mater. Sci. Lett.} \bibinfo{volume}{2}
  (\bibinfo{year}{1983}) \bibinfo{pages}{245--248}.
\bibitem[{Erhart et~al.(1984)Erhart, Wang, and Rapp}]{Erhart1984}
\bibinfo{author}{H.~Erhart}, \bibinfo{author}{R.~Wang}, \bibinfo{author}{R.~A.
  Rapp},
\newblock \bibinfo{title}{In situ {SEM} study of the high-temperature oxidation
  of an {Fe-Mn-Al-Si} alloy},
\newblock \bibinfo{journal}{Oxidation of metals} \bibinfo{volume}{21}
  (\bibinfo{year}{1984}) \bibinfo{pages}{81--88}.
\bibitem[{Vlassiouk et~al.(2011)Vlassiouk, Regmi, Fulvio, Dai, Datskos, Eres,
  and Smirnov}]{Vlassiouk2011}
\bibinfo{author}{I.~Vlassiouk}, \bibinfo{author}{M.~Regmi},
  \bibinfo{author}{P.~Fulvio}, \bibinfo{author}{S.~Dai},
  \bibinfo{author}{P.~Datskos}, \bibinfo{author}{G.~Eres},
  \bibinfo{author}{S.~Smirnov},
\newblock \bibinfo{title}{Role of hydrogen in chemical vapor deposition growth
  of large single-crystal graphene},
\newblock \bibinfo{journal}{ACS Nano} \bibinfo{volume}{5}
  (\bibinfo{year}{2011}) \bibinfo{pages}{6069}.
\bibitem[{Kim et~al.(2009)Kim, Zhao, Jang, Lee, Kim, Kim, Ahn, Kim, Choi, and
  Hong}]{Kim2009}
\bibinfo{author}{K.~S. Kim}, \bibinfo{author}{Y.~Zhao},
  \bibinfo{author}{H.~Jang}, \bibinfo{author}{S.~Y. Lee},
  \bibinfo{author}{J.~M. Kim}, \bibinfo{author}{K.~S. Kim},
  \bibinfo{author}{J.-H. Ahn}, \bibinfo{author}{P.~Kim}, \bibinfo{author}{J.-Y.
  Choi}, \bibinfo{author}{B.~H. Hong},
\newblock \bibinfo{title}{Large-scale pattern growth of graphene films for
  stretchable transparent electrodes},
\newblock \bibinfo{journal}{Nature} \bibinfo{volume}{457}
  (\bibinfo{year}{2009}) \bibinfo{pages}{706}.
\bibitem[{Li et~al.(2009)Li, Cai, An, Kim, Nah, Yang, Piner, Velamakanni, Jung,
  Tutuc, Banerjee, Colombo, and Ruoff}]{Li2009}
\bibinfo{author}{X.~Li}, \bibinfo{author}{W.~Cai}, \bibinfo{author}{J.~An},
  \bibinfo{author}{S.~Kim}, \bibinfo{author}{J.~Nah},
  \bibinfo{author}{D.~Yang}, \bibinfo{author}{R.~Piner},
  \bibinfo{author}{A.~Velamakanni}, \bibinfo{author}{I.~Jung},
  \bibinfo{author}{E.~Tutuc}, \bibinfo{author}{S.~K. Banerjee},
  \bibinfo{author}{L.~Colombo}, \bibinfo{author}{R.~S. Ruoff},
\newblock \bibinfo{title}{Large-area synthesis of high-quality and uniform
  graphene films on copper foils},
\newblock \bibinfo{journal}{Science} \bibinfo{volume}{324}
  (\bibinfo{year}{2009}) \bibinfo{pages}{1312--1314}.
\bibitem[{Chae et~al.(2009)Chae, G\"{u}nes, Kim, Kim, Han, Kim, Shin, Yoon,
  Choi, Park, Yang, Pribat, and Lee}]{Chae2009}
\bibinfo{author}{S.~J. Chae}, \bibinfo{author}{F.~G\"{u}nes},
  \bibinfo{author}{K.~K. Kim}, \bibinfo{author}{E.~S. Kim},
  \bibinfo{author}{G.~H. Han}, \bibinfo{author}{S.~M. Kim},
  \bibinfo{author}{H.-J. Shin}, \bibinfo{author}{S.-M. Yoon},
  \bibinfo{author}{J.-Y. Choi}, \bibinfo{author}{M.~H. Park},
  \bibinfo{author}{C.~W. Yang}, \bibinfo{author}{D.~Pribat},
  \bibinfo{author}{Y.~H. Lee},
\newblock \bibinfo{title}{Synthesis of large-area graphene layers on
  poly-nickel substrate by chemical vapor deposition: Wrinkle formation},
\newblock \bibinfo{journal}{Adv. Mater.} \bibinfo{volume}{21}
  (\bibinfo{year}{2009}) \bibinfo{pages}{2328}.
\bibitem[{Wood et~al.(2011)Wood, Schmucker, Lyons, Pop, and Lyding}]{Wood2011}
\bibinfo{author}{J.~D. Wood}, \bibinfo{author}{S.~W. Schmucker},
  \bibinfo{author}{A.~S. Lyons}, \bibinfo{author}{E.~Pop},
  \bibinfo{author}{J.~W. Lyding},
\newblock \bibinfo{title}{Effects of polycrystalline {C}u substrate on graphene
  growth by chemical vapor deposition},
\newblock \bibinfo{journal}{Nano Lett.} \bibinfo{volume}{11}
  (\bibinfo{year}{2011}) \bibinfo{pages}{4547--4554}.
\bibitem[{Subramanian(2006)}]{Subramanian2006}
\bibinfo{author}{S.~Subramanian}, \bibinfo{title}{In situ high temperature
  environmental scanning electron microscopic investigation of sintering
  behavior in barium titanate}, Ph.D. thesis, University of Cincinnati,
  \bibinfo{year}{2006}.
\bibitem[{Li et~al.(2009)Li, Cai, Colombo, and Ruoff}]{Li2009b}
\bibinfo{author}{X.~Li}, \bibinfo{author}{W.~Cai},
  \bibinfo{author}{L.~Colombo}, \bibinfo{author}{R.~S. Ruoff},
\newblock \bibinfo{title}{Evolution of graphene growth on {N}i and {C}u by
  carbon isotope labeling},
\newblock \bibinfo{journal}{Nano Lett.} \bibinfo{volume}{9}
  (\bibinfo{year}{2009}) \bibinfo{pages}{4268}.
\bibitem[{Ferrari et~al.(2006)Ferrari, Meyer, Scardaci, Casiraghi, Lazzeri,
  Mauri, Piscanec, Jiang, Novoselov, Roth, and Geim}]{Ferrari2006}
\bibinfo{author}{A.~C. Ferrari}, \bibinfo{author}{J.~C. Meyer},
  \bibinfo{author}{V.~Scardaci}, \bibinfo{author}{C.~Casiraghi},
  \bibinfo{author}{M.~Lazzeri}, \bibinfo{author}{F.~Mauri},
  \bibinfo{author}{S.~Piscanec}, \bibinfo{author}{D.~Jiang},
  \bibinfo{author}{K.~S. Novoselov}, \bibinfo{author}{S.~Roth},
  \bibinfo{author}{A.~K. Geim},
\newblock \bibinfo{title}{Raman spectrum of graphene and graphene layers},
\newblock \bibinfo{journal}{Phys. Rev. Lett.} \bibinfo{volume}{97}
  (\bibinfo{year}{2006}) \bibinfo{pages}{187401}.
\bibitem[{Ferrari(2007)}]{Ferrari2007}
\bibinfo{author}{A.~C. Ferrari},
\newblock \bibinfo{title}{Raman spectroscopy of graphene and graphite:
  Disorder, electron-phonon coupling, doping and nonadiabatic effects},
\newblock \bibinfo{journal}{Solid State Comm.} \bibinfo{volume}{143}
  (\bibinfo{year}{2007}) \bibinfo{pages}{47}.
\bibitem[{Malard et~al.(2009)Malard, Pimenta, Dresselhaus, and
  Dresselhaus}]{Malard2009}
\bibinfo{author}{L.~M. Malard}, \bibinfo{author}{M.~A. Pimenta},
  \bibinfo{author}{G.~Dresselhaus}, \bibinfo{author}{M.~Dresselhaus},
\newblock \bibinfo{title}{Raman spectroscopy of graphene},
\newblock \bibinfo{journal}{Phys. Reports} \bibinfo{volume}{473}
  (\bibinfo{year}{2009}) \bibinfo{pages}{51}.
\bibitem[{Tan et~al.(2012)Tan, Han, Zhao, Wu, Chang, Wang, Wang, Bonini,
  Marzari, Savini, Lombardo, and Ferrari}]{Tan2012}
\bibinfo{author}{P.~H. Tan}, \bibinfo{author}{W.~P. Han},
  \bibinfo{author}{W.~J. Zhao}, \bibinfo{author}{Z.~H. Wu},
  \bibinfo{author}{K.~Chang}, \bibinfo{author}{H.~Wang}, \bibinfo{author}{Y.~F.
  Wang}, \bibinfo{author}{N.~Bonini}, \bibinfo{author}{N.~Marzari},
  \bibinfo{author}{G.~Savini}, \bibinfo{author}{A.~Lombardo},
  \bibinfo{author}{A.~C. Ferrari},
\newblock \bibinfo{title}{The shear mode of multilayer graphene},
\newblock \bibinfo{journal}{Nature Mater.} \bibinfo{volume}{11}
  (\bibinfo{year}{2012}) \bibinfo{pages}{294}.
\bibitem[{Casiraghi(2009)}]{Casiraghi2009}
\bibinfo{author}{C.~Casiraghi},
\newblock \bibinfo{title}{{Doping dependence of the Raman peaks intensity of
  graphene close to the Dirac point}},
\newblock \bibinfo{journal}{Phys. Rev. B} \bibinfo{volume}{80}
  (\bibinfo{year}{2009}) \bibinfo{pages}{233407}.
\bibitem[{Righi et~al.(2011)Righi, Costa, Chacham, Fantini, Venezuela,
  Magnuson, Colombo, Bacsa, Ruoff, and Pimenta}]{Righi2011}
\bibinfo{author}{A.~Righi}, \bibinfo{author}{S.~D. Costa},
  \bibinfo{author}{H.~Chacham}, \bibinfo{author}{C.~Fantini},
  \bibinfo{author}{P.~Venezuela}, \bibinfo{author}{C.~Magnuson},
  \bibinfo{author}{L.~Colombo}, \bibinfo{author}{W.~S. Bacsa},
  \bibinfo{author}{R.~S. Ruoff}, \bibinfo{author}{M.~A. Pimenta},
\newblock \bibinfo{title}{Graphene moir\'{e} patterns observed by umklapp
  double-resonance raman scattering},
\newblock \bibinfo{journal}{Phys. Rev. B} \bibinfo{volume}{84}
  (\bibinfo{year}{2011}) \bibinfo{pages}{241409}.
\bibitem[{Mohiuddin et~al.(2009)Mohiuddin, Lombardo, Nair, Bonetti, Savini,
  Jalil, Bonini, Basko, Galiotis, Marzari, Novoselov, Geim, and
  Ferrari}]{Mohiuddin-2009}
\bibinfo{author}{T.~M.~G. Mohiuddin}, \bibinfo{author}{A.~Lombardo},
  \bibinfo{author}{R.~R. Nair}, \bibinfo{author}{A.~Bonetti},
  \bibinfo{author}{G.~Savini}, \bibinfo{author}{R.~Jalil},
  \bibinfo{author}{N.~Bonini}, \bibinfo{author}{D.~M. Basko},
  \bibinfo{author}{C.~Galiotis}, \bibinfo{author}{N.~Marzari},
  \bibinfo{author}{K.~S. Novoselov}, \bibinfo{author}{A.~K. Geim},
  \bibinfo{author}{A.~C. Ferrari},
\newblock \bibinfo{title}{Uniaxial strain in graphene by raman spectroscopy: G
  peak splitting, {G}r{\"u}neisen parameters, and sample orientation},
\newblock \bibinfo{journal}{Phys. Rev. B} \bibinfo{volume}{79}
  (\bibinfo{year}{2009}) \bibinfo{pages}{205433}.
\bibitem[{Das et~al.(2008)Das, Pisana, Chakraborty, Piscanec, Saha, Waghmare,
  Novoselov, Krishnamurthy, Geim, Ferrari, and Sood}]{Das-Pisana-2008}
\bibinfo{author}{A.~Das}, \bibinfo{author}{S.~Pisana},
  \bibinfo{author}{B.~Chakraborty}, \bibinfo{author}{S.~Piscanec},
  \bibinfo{author}{S.~K. Saha}, \bibinfo{author}{U.~V. Waghmare},
  \bibinfo{author}{K.~S. Novoselov}, \bibinfo{author}{H.~R. Krishnamurthy},
  \bibinfo{author}{A.~K. Geim}, \bibinfo{author}{A.~C. Ferrari},
  \bibinfo{author}{A.~K. Sood},
\newblock \bibinfo{title}{Monitoring dopants by raman scattering on an
  electrochemically top-gated graphene transistor},
\newblock \bibinfo{journal}{Nature Nanotech.} \bibinfo{volume}{3}
  (\bibinfo{year}{2008}) \bibinfo{pages}{210--215}.
\bibitem[{Marra et~al.(1979)Marra, Eisenberger, and Cho}]{Marra1979}
\bibinfo{author}{W.~Marra}, \bibinfo{author}{P.~Eisenberger},
  \bibinfo{author}{A.~Cho},
\newblock \bibinfo{title}{{X-ray total-external-reflection--Bragg diffraction:
  A structural study of the GaAs-Al interface}},
\newblock \bibinfo{journal}{J. Appl. Phys.} \bibinfo{volume}{50}
  (\bibinfo{year}{1979}) \bibinfo{pages}{6927--6933}.
\bibitem[{Robinson(1986)}]{Robinson1986}
\bibinfo{author}{I.~K. Robinson},
\newblock \bibinfo{title}{Crystal truncation rods and surface roughness},
\newblock \bibinfo{journal}{Phys. Rev. B} \bibinfo{volume}{33}
  (\bibinfo{year}{1986}) \bibinfo{pages}{3830}.
\bibitem[{Feidenhans'l(1989)}]{Feidenhans'l1989}
\bibinfo{author}{R.~Feidenhans'l},
\newblock \bibinfo{title}{Surface structure determination by {X}-ray
  diffraction},
\newblock \bibinfo{journal}{Surf. Sci. Rep.} \bibinfo{volume}{10}
  (\bibinfo{year}{1989}) \bibinfo{pages}{105}.
\bibitem[{Hass et~al.(2007)Hass, Feng, Mill\'{a}n-Otoya, Li, Sprinkle, First,
  de~Heer, Conrad, and Berger}]{Hass2007}
\bibinfo{author}{J.~Hass}, \bibinfo{author}{R.~Feng}, \bibinfo{author}{J.~E.
  Mill\'{a}n-Otoya}, \bibinfo{author}{X.~Li}, \bibinfo{author}{M.~Sprinkle},
  \bibinfo{author}{P.~N. First}, \bibinfo{author}{W.~A. de~Heer},
  \bibinfo{author}{E.~H. Conrad}, \bibinfo{author}{C.~Berger},
\newblock \bibinfo{title}{Structural properties of the multilayer
  graphene/4{H}-{SiC}(000$\bar{1}$) system as determined by surface x-ray
  diffraction},
\newblock \bibinfo{journal}{Phys. Rev. B} \bibinfo{volume}{75}
  (\bibinfo{year}{2007}) \bibinfo{pages}{214109}.
\bibitem[{Martoccia et~al.(2008)Martoccia, Willmott, Brugger, Bj{\"o}rck,
  G{\"u}nther, Schlep{\"u}tz, Cervellino, Pauli, Patterson, Marchini,
  Wintterlin, Moritz, and Greber}]{Martoccia}
\bibinfo{author}{D.~Martoccia}, \bibinfo{author}{P.~R. Willmott},
  \bibinfo{author}{T.~Brugger}, \bibinfo{author}{M.~Bj{\"o}rck},
  \bibinfo{author}{S.~G{\"u}nther}, \bibinfo{author}{C.~M. Schlep{\"u}tz},
  \bibinfo{author}{A.~Cervellino}, \bibinfo{author}{S.~A. Pauli},
  \bibinfo{author}{B.~D. Patterson}, \bibinfo{author}{S.~Marchini},
  \bibinfo{author}{J.~Wintterlin}, \bibinfo{author}{W.~Moritz},
  \bibinfo{author}{T.~Greber},
\newblock \bibinfo{title}{{Graphene on Ru(0001): A 25 x 25 Supercell}},
\newblock \bibinfo{journal}{Phys. Rev. Lett.} \bibinfo{volume}{101}
  (\bibinfo{year}{2008}) \bibinfo{pages}{126102}.
\bibitem[{Blanc et~al.(2012)Blanc, Coraux, Vo-Van, N'Diaye, Geaymond, and
  Renaud}]{Blanc2012}
\bibinfo{author}{N.~Blanc}, \bibinfo{author}{J.~Coraux},
  \bibinfo{author}{C.~Vo-Van}, \bibinfo{author}{A.~T. N'Diaye},
  \bibinfo{author}{O.~Geaymond}, \bibinfo{author}{G.~Renaud},
\newblock \bibinfo{title}{Local deformations and incommensurability of
  high-quality epitaxial graphene on a weakly interacting transition metal},
\newblock \bibinfo{journal}{Phys. Rev. B} \bibinfo{volume}{86}
  (\bibinfo{year}{2012}) \bibinfo{pages}{235439}.
\bibitem[{Jean et~al.(2013)Jean, Zhou, Blanc, Felici, Coraux, and
  Renaud}]{Jean-Zhou-2013}
\bibinfo{author}{F.~Jean}, \bibinfo{author}{T.~Zhou},
  \bibinfo{author}{N.~Blanc}, \bibinfo{author}{R.~Felici},
  \bibinfo{author}{J.~Coraux}, \bibinfo{author}{G.~Renaud},
\newblock \bibinfo{title}{Effect of preparation on the commensurabilities and
  thermal expansion of graphene on {I}r(111) between 10 and 1300 {K}},
\newblock \bibinfo{journal}{Phys. Rev. B} \bibinfo{volume}{88}
  (\bibinfo{year}{2013}) \bibinfo{pages}{165406}.
\bibitem[{Siegbahn and Edvarson(1956)}]{Siegbahn1956}
\bibinfo{author}{K.~Siegbahn}, \bibinfo{author}{K.~I. Edvarson},
\newblock \bibinfo{title}{$\beta$-ray spectroscopy in the precision range of 1
  : 10$^5$},
\newblock \bibinfo{journal}{Nucl. Phys.} \bibinfo{volume}{1}
  (\bibinfo{year}{1956}) \bibinfo{pages}{137--159}.
\bibitem[{Turner and Jobory(1962)}]{Turner1962}
\bibinfo{author}{D.~W. Turner}, \bibinfo{author}{M.~I. Jobory},
\newblock \bibinfo{title}{Determination of ionization potentials by
  photoelectron energy measurement},
\newblock \bibinfo{journal}{J. Chem. Phys.} \bibinfo{volume}{37}
  (\bibinfo{year}{1962}) \bibinfo{pages}{3007}.
\bibitem[{Nagashima et~al.(1994)Nagashima, Tejima, and Oshima}]{Nagashima1994}
\bibinfo{author}{A.~Nagashima}, \bibinfo{author}{N.~Tejima},
  \bibinfo{author}{C.~Oshima},
\newblock \bibinfo{title}{Electronic states of the pristine and
  alkali-metal-intercalated monolayer graphite/{N}i(111) systems},
\newblock \bibinfo{journal}{Phys. Rev. B} \bibinfo{volume}{50}
  (\bibinfo{year}{1994}) \bibinfo{pages}{17487--17495}.
\bibitem[{Rollings et~al.(2006)Rollings, Gweon, Zhou, Mun, McChesney, Hussain,
  Fedorov, First, de~Heer, and Lanzara}]{Rollings2006}
\bibinfo{author}{E.~Rollings}, \bibinfo{author}{G.-H. Gweon},
  \bibinfo{author}{S.~Y. Zhou}, \bibinfo{author}{B.~S. Mun},
  \bibinfo{author}{J.~L. McChesney}, \bibinfo{author}{B.~S. Hussain},
  \bibinfo{author}{A.~V. Fedorov}, \bibinfo{author}{P.~N. First},
  \bibinfo{author}{W.~A. de~Heer}, \bibinfo{author}{A.~Lanzara},
\newblock \bibinfo{title}{Synthesis and characterization of atomically thin
  graphite films on a silicon carbide substrate},
\newblock \bibinfo{journal}{J. Phys. Chem. Sol.} \bibinfo{volume}{67}
  (\bibinfo{year}{2006}) \bibinfo{pages}{2172--2177}.
\bibitem[{Preobrajenski et~al.(2008)Preobrajenski, Ng, Vinogradov, and
  M\aa{}rtensson}]{Preobrajenski2008}
\bibinfo{author}{A.~B. Preobrajenski}, \bibinfo{author}{M.~L. Ng},
  \bibinfo{author}{A.~S. Vinogradov}, \bibinfo{author}{N.~M\aa{}rtensson},
\newblock \bibinfo{title}{Controlling graphene corrugation on
  lattice-mismatched substrates},
\newblock \bibinfo{journal}{Phys. Rev. B} \bibinfo{volume}{78}
  (\bibinfo{year}{2008}) \bibinfo{pages}{073401}.
\bibitem[{Lacovig et~al.(2009)Lacovig, Pozzo, Alfe, Vilmercati, Baraldi, and
  Lizzit}]{Lacovig}
\bibinfo{author}{P.~Lacovig}, \bibinfo{author}{M.~Pozzo},
  \bibinfo{author}{D.~Alfe}, \bibinfo{author}{P.~Vilmercati},
  \bibinfo{author}{A.~Baraldi}, \bibinfo{author}{S.~Lizzit},
\newblock \bibinfo{title}{Growth of dome-shaped carbon nanoislands on {I}r
  (111): the intermediate between carbidic clusters and quasi-free-standing
  graphene},
\newblock \bibinfo{journal}{Phys. Rev. Lett.} \bibinfo{volume}{103}
  (\bibinfo{year}{2009}) \bibinfo{pages}{166101}.
\bibitem[{Lizzit and Baraldi(2010)}]{Lizzit201068}
\bibinfo{author}{S.~Lizzit}, \bibinfo{author}{A.~Baraldi},
\newblock \bibinfo{title}{{High-resolution fast X-ray photoelectron
  spectroscopy study of ethylene interaction with Ir(111): From chemisorption
  to dissociation and graphene formation}},
\newblock \bibinfo{journal}{Catalysis Today} \bibinfo{volume}{154}
  (\bibinfo{year}{2010}) \bibinfo{pages}{68 -- 74}.
\bibitem[{Miniussi et~al.(2011)Miniussi, Pozzo, Baraldi, Vesselli, Zhan,
  Comelli, Mente{\c{s}}, Ni{\~n}o, Locatelli, Lizzit et~al.}]{Miniussi2011}
\bibinfo{author}{E.~Miniussi}, \bibinfo{author}{M.~Pozzo},
  \bibinfo{author}{A.~Baraldi}, \bibinfo{author}{E.~Vesselli},
  \bibinfo{author}{R.~R. Zhan}, \bibinfo{author}{G.~Comelli},
  \bibinfo{author}{T.~O. Mente{\c{s}}}, \bibinfo{author}{M.~A. Ni{\~n}o},
  \bibinfo{author}{A.~Locatelli}, \bibinfo{author}{S.~Lizzit}, et~al.,
\newblock \bibinfo{title}{{Thermal stability of corrugated epitaxial graphene
  grown on Re (0001)}},
\newblock \bibinfo{journal}{Phys. Rev. Lett.} \bibinfo{volume}{106}
  (\bibinfo{year}{2011}) \bibinfo{pages}{216101}.
\bibitem[{Koningsberger and Prins(1988)}]{koningsberger1988}
\bibinfo{author}{D.~Koningsberger}, \bibinfo{author}{R.~Prins},
  \bibinfo{title}{X-ray absorption: principles, applications, techniques of
  EXAFS, SEXAFS, and XANES}, \bibinfo{publisher}{John Wiley and Sons, New York,
  NY}, \bibinfo{year}{1988}.
\bibitem[{Aristov et~al.(2010)Aristov, Urbanik, Kummer, Vyalikh, Molodtsova,
  Preobrajenski, Zakharov, Hess, H{\"a}nke, B{\"u}chner, Vobornik, Fujii,
  Panaccione, Ossipyan, and Knupfer}]{aristov2010}
\bibinfo{author}{V.~Y. Aristov}, \bibinfo{author}{G.~Urbanik},
  \bibinfo{author}{K.~Kummer}, \bibinfo{author}{D.~V. Vyalikh},
  \bibinfo{author}{O.~V. Molodtsova}, \bibinfo{author}{A.~B. Preobrajenski},
  \bibinfo{author}{A.~A. Zakharov}, \bibinfo{author}{C.~Hess},
  \bibinfo{author}{T.~H{\"a}nke}, \bibinfo{author}{B.~B{\"u}chner},
  \bibinfo{author}{I.~Vobornik}, \bibinfo{author}{J.~Fujii},
  \bibinfo{author}{G.~Panaccione}, \bibinfo{author}{Y.~A. Ossipyan},
  \bibinfo{author}{M.~Knupfer},
\newblock \bibinfo{title}{Graphene synthesis on cubic {SiC/Si} wafers.
  perspectives for mass production of graphene-based electronic devices},
\newblock \bibinfo{journal}{Nano Lett.} \bibinfo{volume}{10}
  (\bibinfo{year}{2010}) \bibinfo{pages}{992--995}.
\bibitem[{Br\"{u}hwiler et~al.(1995)Br\"{u}hwiler, Maxwell, Puglia, Nilsson,
  Andersson, and M{\aa}rtensson}]{bruhwiler1995}
\bibinfo{author}{P.~A. Br\"{u}hwiler}, \bibinfo{author}{A.~J. Maxwell},
  \bibinfo{author}{C.~Puglia}, \bibinfo{author}{A.~Nilsson},
  \bibinfo{author}{S.~Andersson}, \bibinfo{author}{N.~M{\aa}rtensson},
\newblock \bibinfo{title}{$\pi$* and $\sigma$* excitons in c 1 s absorption of
  graphite},
\newblock \bibinfo{journal}{Phys. Rev. Lett.} \bibinfo{volume}{74}
  (\bibinfo{year}{1995}) \bibinfo{pages}{614}.
\bibitem[{Fischer et~al.(1991)Fischer, Wentzcovitch, Carr, Continenza, and
  Freeman}]{fischer1991}
\bibinfo{author}{D.~A. Fischer}, \bibinfo{author}{R.~M. Wentzcovitch},
  \bibinfo{author}{R.~G. Carr}, \bibinfo{author}{A.~Continenza},
  \bibinfo{author}{A.~J. Freeman},
\newblock \bibinfo{title}{Graphitic interlayer states: A carbon {K} near-edge
  {X}-ray-absorption fine-structure study},
\newblock \bibinfo{journal}{Phys. Rev. B} \bibinfo{volume}{44}
  (\bibinfo{year}{1991}) \bibinfo{pages}{1427}.
\bibitem[{Lee et~al.(2010)Lee, Park, Jaye, Fischer, Yu, Wu, Liu, Bao, Pei,
  Smith, Lysaght, and Banerjee}]{lee2010}
\bibinfo{author}{V.~Lee}, \bibinfo{author}{C.~Park}, \bibinfo{author}{C.~Jaye},
  \bibinfo{author}{D.~A. Fischer}, \bibinfo{author}{Q.~Yu},
  \bibinfo{author}{W.~Wu}, \bibinfo{author}{Z.~Liu}, \bibinfo{author}{J.~Bao},
  \bibinfo{author}{S.-S. Pei}, \bibinfo{author}{C.~Smith},
  \bibinfo{author}{P.~Lysaght}, \bibinfo{author}{S.~Banerjee},
\newblock \bibinfo{title}{Substrate hybridization and rippling of graphene
  evidenced by near-edge {X}-ray absorption fine structure spectroscopy},
\newblock \bibinfo{journal}{L. Phys. Chem. Lett.} \bibinfo{volume}{1}
  (\bibinfo{year}{2010}) \bibinfo{pages}{1247--1253}.
\bibitem[{Batterman(1964)}]{Batteran1964}
\bibinfo{author}{B.~W. Batterman},
\newblock \bibinfo{title}{Effect of dynamical diffraction in {X}-ray
  fluorescence scattering},
\newblock \bibinfo{journal}{Phys. Rev. B} \bibinfo{volume}{133}
  (\bibinfo{year}{1964}) \bibinfo{pages}{A759}.
\bibitem[{Zegenhagen(1993)}]{Zegenhagen1993}
\bibinfo{author}{J.~Zegenhagen},
\newblock \bibinfo{title}{Surface structure determination with {X}-ray standing
  waves},
\newblock \bibinfo{journal}{Surf. Sci. Rep.} \bibinfo{volume}{18}
  (\bibinfo{year}{1993}) \bibinfo{pages}{202}.
\bibitem[{Woodruff(2005)}]{Woodruff2005}
\bibinfo{author}{D.~P. Woodruff},
\newblock \bibinfo{title}{{Surface structure determination using X-ray standing
  waves}},
\newblock \bibinfo{journal}{Rep. Progr. Phys.} \bibinfo{volume}{68}
  (\bibinfo{year}{2005}) \bibinfo{pages}{743}.
  
\bibitem[{Busse et~al.(2011)Busse, Lazi{\'c}, Djemour, Coraux, Gerber,
  Atodiresei, Caciuc, Brako, N'Diaye, Bl{\"u}gel, Zegenhagen, and
  Michely}]{Busse2011}
\bibinfo{author}{C.~Busse}, \bibinfo{author}{P.~Lazi{\'c}},
  \bibinfo{author}{R.~Djemour}, \bibinfo{author}{J.~Coraux},
  \bibinfo{author}{T.~Gerber}, \bibinfo{author}{N.~Atodiresei},
  \bibinfo{author}{V.~Caciuc}, \bibinfo{author}{R.~Brako},
  \bibinfo{author}{A.~T. N'Diaye}, \bibinfo{author}{S.~Bl{\"u}gel},
  \bibinfo{author}{J.~Zegenhagen}, \bibinfo{author}{T.~Michely},
\newblock \bibinfo{title}{Graphene on {I}r(111): Physisorption with chemical
  modulation},
\newblock \bibinfo{journal}{Phys. Rev. Lett.} \bibinfo{volume}{107}
  (\bibinfo{year}{2011}) \bibinfo{pages}{036101}.
  
\bibitem[{Runte et~al.(2014)Runte, Lazi{\'c}, Chi Vo-Van, Coraux, Zegenhagen, and Busse}]{Runte2014}
\bibinfo{author}{S.~Runte}, \bibinfo{author}{P.~Lazi{\'c}},
  \bibinfo{author}{C.~vo-Van}, \bibinfo{author}{J.~Coraux},
  \bibinfo{author}{J.~Zegenhagen}, \bibinfo{author}{C.~Busse},
\newblock \bibinfo{title}{Graphene buckles under stress: an X-ray standing wave and scanning tunneling microscopy study},
\newblock \bibinfo{journal}{Phys. Rev. B} \bibinfo{volume}{89}
  (\bibinfo{year}{2014}) \bibinfo{pages}{155427}.
  
\bibitem[{Payne et~al.(1992)Payne, Teter, Allan, Arias, and
  Joannopoulos}]{Payne-DFT}
\bibinfo{author}{M.~C. Payne}, \bibinfo{author}{M.~P. Teter},
  \bibinfo{author}{D.~C. Allan}, \bibinfo{author}{T.~A. Arias},
  \bibinfo{author}{J.~D. Joannopoulos},
\newblock \bibinfo{title}{Iterative minimization techniques for abinitio
  total-energy calculations - molecular-dynamics and conjugate gradients},
\newblock \bibinfo{journal}{Rev. Mod. Phys.} \bibinfo{volume}{64}
  (\bibinfo{year}{1992}) \bibinfo{pages}{1045--1097}.
\bibitem[{Stich(2007)}]{Stich:2007vt}
\bibinfo{author}{I.~Stich},
\newblock \bibinfo{title}{Computer simuations for the nano-scale},
\newblock \bibinfo{journal}{Acta Physica Slovaca} \bibinfo{volume}{57}
  (\bibinfo{year}{2007}) \bibinfo{pages}{1}.
\bibitem[{Kantorovich(2004)}]{LK-book-2004}
\bibinfo{author}{L.~Kantorovich}, \bibinfo{title}{Quantum Theory of the Solid
  State: An Introduction}, \bibinfo{publisher}{Kluwer}, \bibinfo{year}{2004}.
\bibitem[{Allen and Tildesley(1987)}]{Allen-Tildesley-simul-book}
\bibinfo{author}{M.~P. Allen}, \bibinfo{author}{D.~J. Tildesley},
  \bibinfo{title}{Computer Simulations of Liquids}, \bibinfo{publisher}{Oxford
  University Press}, \bibinfo{year}{1987}.
\bibitem[{Parr and Yang(1989)}]{Parr-Yang-DFT-book-1989}
\bibinfo{author}{R.~G. Parr}, \bibinfo{author}{W.~Yang},
  \bibinfo{title}{Density-Functional Theory of Atoms and Molecules},
  \bibinfo{publisher}{Oxford Univ. Press}, \bibinfo{year}{1989}.
\bibitem[{Martin(2008)}]{Martin-DFT-book-2008}
\bibinfo{author}{R.~Martin}, \bibinfo{title}{Electronic structure. Basic Theory
  and Practical Methods}, \bibinfo{publisher}{Cambridge}, \bibinfo{year}{2008}.
\bibitem[{Frenkel and Smit(2002)}]{Frenkel-Smit-simul-book}
\bibinfo{author}{D.~Frenkel}, \bibinfo{author}{B.~Smit},
  \bibinfo{title}{Understanding Molecular Simulation. From Algorithms to
  Applications}, \bibinfo{publisher}{Aced. Press}, \bibinfo{year}{2002}.
\bibitem[{Hohenberg and Kohn(1964)}]{Hohenberg-Kohn}
\bibinfo{author}{P.~Hohenberg}, \bibinfo{author}{W.~Kohn},
\newblock \bibinfo{title}{Inhomogeneous electron gas},
\newblock \bibinfo{journal}{Phys. Rev.} \bibinfo{volume}{136}
  (\bibinfo{year}{1964}) \bibinfo{pages}{B864--871}.
\bibitem[{Kohn and Sham(1965)}]{Kohn-Sham}
\bibinfo{author}{W.~Kohn}, \bibinfo{author}{L.~J. Sham},
\newblock \bibinfo{title}{Self-consistent equations including exchange and
  correlation effects},
\newblock \bibinfo{journal}{Phys. Rev.} \bibinfo{volume}{140}
  (\bibinfo{year}{1965}) \bibinfo{pages}{A1133--A1138}.
\bibitem[{Baroni et~al.(2001)Baroni, de~Gironcoli, Dal~Corso, and
  Giannozzi}]{Baroni:2001tn}
\bibinfo{author}{S.~Baroni}, \bibinfo{author}{S.~de~Gironcoli},
  \bibinfo{author}{A.~Dal~Corso}, \bibinfo{author}{P.~Giannozzi},
\newblock \bibinfo{title}{{Phonons and related crystal properties from
  density-functional perturbation theory}},
\newblock \bibinfo{journal}{Rev. Mod. Phys.} \bibinfo{volume}{73}
  (\bibinfo{year}{2001}) \bibinfo{pages}{515--562}.
\bibitem[{Boys and Bernardi(1970)}]{Boys:1970BSSE}
\bibinfo{author}{S.~F. Boys}, \bibinfo{author}{F.~Bernardi},
\newblock \bibinfo{title}{The calculation of small molecular interactions by
  the differences of separate total energies. {S}ome procedures with reduced
  errors},
\newblock \bibinfo{journal}{Molecular Physics} \bibinfo{volume}{19}
  (\bibinfo{year}{1970}) \bibinfo{pages}{553}.
\bibitem[{Wales(2003)}]{Wales-PES-book}
\bibinfo{author}{D.~Wales}, \bibinfo{title}{Energy Landscapes: Applications to
  Clusters, Biomolecules and Glasses}, \bibinfo{publisher}{Cambridge Molecular
  Science}, \bibinfo{year}{2003}.
\bibitem[{Wales and Bogdan(2006)}]{Wales:2006tn}
\bibinfo{author}{D.~J. Wales}, \bibinfo{author}{T.~V. Bogdan},
\newblock \bibinfo{title}{{Potential energy and free energy landscapes}},
\newblock \bibinfo{journal}{J. Phys. Chem. B} \bibinfo{volume}{110}
  (\bibinfo{year}{2006}) \bibinfo{pages}{20765--20776}.
\bibitem[{Goedecker et~al.(2005)Goedecker, Hellmann, and
  Lenosky}]{Goedecker:2005vj}
\bibinfo{author}{S.~Goedecker}, \bibinfo{author}{W.~Hellmann},
  \bibinfo{author}{T.~Lenosky},
\newblock \bibinfo{title}{{Global minimum determination of the Born-Oppenheimer
  surface within density functional theory}},
\newblock \bibinfo{journal}{Phys. Rev. Lett.} \bibinfo{volume}{95}
  (\bibinfo{year}{2005}) \bibinfo{pages}{55501}.
\bibitem[{Zhao et~al.(2010)Zhao, Moskaleva, Aleksandrov, Basaran, and
  R{\"o}sch}]{Zhao:2010gq}
\bibinfo{author}{Z.-J. Zhao}, \bibinfo{author}{L.~V. Moskaleva},
  \bibinfo{author}{H.~A. Aleksandrov}, \bibinfo{author}{D.~Basaran},
  \bibinfo{author}{N.~R{\"o}sch},
\newblock \bibinfo{title}{{Ethylidyne Formation from Ethylene over Pt(111): A
  Mechanistic Study from First-Principle Calculations}},
\newblock \bibinfo{journal}{J. Phys. Chem. C} \bibinfo{volume}{114}
  (\bibinfo{year}{2010}) \bibinfo{pages}{12190--12201}.
\bibitem[{Henkelman and J{\'o}nsson(2001)}]{Henkelman:2001fg}
\bibinfo{author}{G.~Henkelman}, \bibinfo{author}{H.~J{\'o}nsson},
\newblock \bibinfo{title}{{Theoretical Calculations of Dissociative Adsorption
  of CH$_{4}$ on an Ir(111) Surface}},
\newblock \bibinfo{journal}{Phys. Rev. Lett.} \bibinfo{volume}{86}
  (\bibinfo{year}{2001}) \bibinfo{pages}{664--667}.
\bibitem[{Goedecker(1999)}]{Goedecker-1999}
\bibinfo{author}{S.~Goedecker},
\newblock \bibinfo{title}{Linear scaling electronic structure methods},
\newblock \bibinfo{journal}{Rev. Mod. Phys.} \bibinfo{volume}{71}
  (\bibinfo{year}{1999}) \bibinfo{pages}{1085--1123}.
\bibitem[{Bowler et~al.(2002)Bowler, Miyazaki, and Gillan}]{Bowler-Gillan-2002}
\bibinfo{author}{D.~R. Bowler}, \bibinfo{author}{T.~Miyazaki},
  \bibinfo{author}{M.~J. Gillan},
\newblock \bibinfo{title}{Recent progress in linear scaling {\em ab initio}
  electronic structure techniques},
\newblock \bibinfo{journal}{J. Phys. Condens. Matter} \bibinfo{volume}{14}
  (\bibinfo{year}{2002}) \bibinfo{pages}{2781--2798}.
\bibitem[{Bowler et~al.(2008)Bowler, Fattebert, Gillan, Haynes, and
  Skylaris}]{O-N-workshop-JPCM-2008}
\bibinfo{author}{D.~R. Bowler}, \bibinfo{author}{J.-L. Fattebert},
  \bibinfo{author}{M.~J. Gillan}, \bibinfo{author}{P.~D. Haynes},
  \bibinfo{author}{C.-K. Skylaris},
\newblock \bibinfo{title}{Introductory remarks: Linear scaling methods},
\newblock \bibinfo{journal}{J. Phys. Cond. Matter} \bibinfo{volume}{20}
  (\bibinfo{year}{2008}) \bibinfo{pages}{290301}.
\bibitem[{Skylaris et~al.(2005)Skylaris, Haynes, Mostofi, and Payne}]{ONETEP}
\bibinfo{author}{C.~Skylaris}, \bibinfo{author}{P.~D. Haynes},
  \bibinfo{author}{A.~A. Mostofi}, \bibinfo{author}{M.~C. Payne},
\newblock \bibinfo{title}{{Introducing ONETEP: Linear-scaling density
  functional simulations on parallel computers}},
\newblock \bibinfo{journal}{J. Chem. Phys.} \bibinfo{volume}{122}
  (\bibinfo{year}{2005}) \bibinfo{pages}{084119}.
\bibitem[{Soler et~al.(2002)Soler, Artacho, Gale, Garc{\'i}a, Junquera,
  Ordej{\'o}n, and S{\'a}nchez-Portal}]{SIESTA}
\bibinfo{author}{J.~M. Soler}, \bibinfo{author}{E.~Artacho},
  \bibinfo{author}{J.~D. Gale}, \bibinfo{author}{A.~Garc{\'i}a},
  \bibinfo{author}{J.~Junquera}, \bibinfo{author}{P.~Ordej{\'o}n},
  \bibinfo{author}{D.~S{\'a}nchez-Portal},
\newblock \bibinfo{title}{{The SIESTA method for {\it ab initio} order-N
  materials simulation}},
\newblock \bibinfo{journal}{J. Phys.: Condens. Matter} \bibinfo{volume}{14}
  (\bibinfo{year}{2002}) \bibinfo{pages}{2745--2779}.
\bibitem[{Ceperley and Alder(1980)}]{LDA}
\bibinfo{author}{D.~M. Ceperley}, \bibinfo{author}{B.~J. Alder},
\newblock \bibinfo{title}{ground state of the electronic gas by a stochastic
  method},
\newblock \bibinfo{journal}{Phys. Rev. Lett.} \bibinfo{volume}{45}
  (\bibinfo{year}{1980}) \bibinfo{pages}{566--569}.
\bibitem[{Perdew et~al.(1996)Perdew, Burke, and Ernzerhof}]{Perdew:1996PBE}
\bibinfo{author}{J.~Perdew}, \bibinfo{author}{K.~Burke},
  \bibinfo{author}{M.~Ernzerhof},
\newblock \bibinfo{title}{Generalized gradient approximation made simple},
\newblock \bibinfo{journal}{Phys. Rev. Lett.} \bibinfo{volume}{77}
  (\bibinfo{year}{1996}) \bibinfo{pages}{3865}.
\bibitem[{Perdew et~al.(1998)Perdew, Burke, and Ernzerhof}]{Perdew:1998PBE}
\bibinfo{author}{J.~Perdew}, \bibinfo{author}{K.~Burke},
  \bibinfo{author}{M.~Ernzerhof},
\newblock \bibinfo{title}{Perdew, burke, and ernzerhof reply},
\newblock \bibinfo{journal}{Phys. Rev. Lett.} \bibinfo{volume}{80}
  (\bibinfo{year}{1998}) \bibinfo{pages}{891}.
\bibitem[{Bj\'{o}rkman et~al.(2012{\natexlab{a}})Bj\'{o}rkman, Gulans,
  Krasheninnikov, and Nieminen}]{Bjorkman-JPCM-2012}
\bibinfo{author}{T.~Bj\'{o}rkman}, \bibinfo{author}{A.~Gulans},
  \bibinfo{author}{A.~V. Krasheninnikov}, \bibinfo{author}{R.~M. Nieminen},
\newblock \bibinfo{title}{Are we van der waals ready?},
\newblock \bibinfo{journal}{J. Phys.: Cond. Matter} \bibinfo{volume}{24}
  (\bibinfo{year}{2012}{\natexlab{a}}) \bibinfo{pages}{424218}.
\bibitem[{Bj\'{o}rkman et~al.(2012{\natexlab{b}})Bj\'{o}rkman, Gulans,
  Krasheninnikov, and Nieminen}]{Bjorkman-PRL-2012}
\bibinfo{author}{T.~Bj\'{o}rkman}, \bibinfo{author}{A.~Gulans},
  \bibinfo{author}{A.~V. Krasheninnikov}, \bibinfo{author}{R.~M. Nieminen},
\newblock \bibinfo{title}{van der waals bonding in layered compounds from
  advanced density-functional first-principles calculations},
\newblock \bibinfo{journal}{Phys. Rev. Lett.} \bibinfo{volume}{108}
  (\bibinfo{year}{2012}{\natexlab{b}}) \bibinfo{pages}{235502}.
\bibitem[{Grimme(2004)}]{Grimme-JComptChem-2004}
\bibinfo{author}{S.~Grimme},
\newblock \bibinfo{title}{Accurate description of van der waals complexes by
  density functional theory including empirical corrections},
\newblock \bibinfo{journal}{J. Comp. Chem.} \bibinfo{volume}{25}
  (\bibinfo{year}{2004}) \bibinfo{pages}{1463--1473}.
\bibitem[{Grimme(2006)}]{Grimme-JCC-2006}
\bibinfo{author}{S.~Grimme},
\newblock \bibinfo{title}{Semiempirical gga-type density functional constructed
  with a long-range dispersion correction},
\newblock \bibinfo{journal}{J. Comp. Chem.} \bibinfo{volume}{27}
  (\bibinfo{year}{2006}) \bibinfo{pages}{1787--1799}.
\bibitem[{Grimme(2011)}]{Grimme-vdW}
\bibinfo{author}{S.~Grimme},
\newblock \bibinfo{title}{Density functional theory with {L}ondon dispersion
  corrections},
\newblock \bibinfo{journal}{Comput. Mol. Sci.} \bibinfo{volume}{1}
  (\bibinfo{year}{2011}) \bibinfo{pages}{211--228}.
\bibitem[{Dion et~al.(2004)Dion, Rydberg, Schr{\"o}der, Langreth, and
  Lundqvist}]{Dion-Rydberg-Schroder-Langreth-Lundqvist-PRL-2004}
\bibinfo{author}{M.~Dion}, \bibinfo{author}{H.~Rydberg},
  \bibinfo{author}{E.~Schr{\"o}der}, \bibinfo{author}{D.~C. Langreth},
  \bibinfo{author}{B.~I. Lundqvist},
\newblock \bibinfo{title}{Van der waals density functional for general
  geometries},
\newblock \bibinfo{journal}{Phys. Rev. Lett.} \bibinfo{volume}{92}
  (\bibinfo{year}{2004}) \bibinfo{pages}{246401}.
\bibitem[{Lee et~al.(2010)Lee, Murray, Kong, Lundqvist, and Langreth}]{vdW-DF2}
\bibinfo{author}{K.~Lee}, \bibinfo{author}{E.~D. Murray},
  \bibinfo{author}{L.~Kong}, \bibinfo{author}{B.~I. Lundqvist},
  \bibinfo{author}{D.~C. Langreth},
\newblock \bibinfo{title}{Higher-accuracy van der waals density functional},
\newblock \bibinfo{journal}{Phys. Rev. B} \bibinfo{volume}{82}
  (\bibinfo{year}{2010}) \bibinfo{pages}{081101(R)}.
\bibitem[{Langreth et~al.(2009)Langreth, Lundqvist, Chakarova-K{\" a}ck,
  Cooper, Dion, Hyldgaard, Kelkkanen, Kleis, Kong, Li, Moses, Murray, Puzder,
  Rydberg, Schr{\" o}der, and Thonhauser}]{Langreth-review}
\bibinfo{author}{D.~C. Langreth}, \bibinfo{author}{B.~I. Lundqvist},
  \bibinfo{author}{S.~D. Chakarova-K{\" a}ck}, \bibinfo{author}{V.~R. Cooper},
  \bibinfo{author}{M.~Dion}, \bibinfo{author}{P.~Hyldgaard},
  \bibinfo{author}{A.~Kelkkanen}, \bibinfo{author}{J.~Kleis},
  \bibinfo{author}{L.~Kong}, \bibinfo{author}{S.~Li}, \bibinfo{author}{P.~G.
  Moses}, \bibinfo{author}{E.~Murray}, \bibinfo{author}{A.~Puzder},
  \bibinfo{author}{H.~Rydberg}, \bibinfo{author}{E.~Schr{\" o}der},
  \bibinfo{author}{T.~Thonhauser},
\newblock \bibinfo{title}{A density functional for sparse matter},
\newblock \bibinfo{journal}{J. Phys. Condens. Matter.} \bibinfo{volume}{21}
  (\bibinfo{year}{2009}) \bibinfo{pages}{084203}.
\bibitem[{Bj\'{o}rkman(2012)}]{VV10-PRB-2012}
\bibinfo{author}{T.~Bj\'{o}rkman},
\newblock \bibinfo{title}{van der waals density functional for solids},
\newblock \bibinfo{journal}{Phys. Rev. B} \bibinfo{volume}{86}
  (\bibinfo{year}{2012}) \bibinfo{pages}{165109}.
\bibitem[{Vydrov and Voorhis(2010)}]{Vydrov-VV10-2010}
\bibinfo{author}{O.~A. Vydrov}, \bibinfo{author}{T.~V. Voorhis},
\newblock \bibinfo{title}{Nonlocal van der waals density functional: The
  simpler the better},
\newblock \bibinfo{journal}{J. Chem. Phys.} \bibinfo{volume}{133}
  (\bibinfo{year}{2010}) \bibinfo{pages}{244103}.
\bibitem[{Klime\^{s} et~al.(2010)Klime\^{s}, Bowler, and
  Michaelides}]{Klimes-JPCM-2010}
\bibinfo{author}{J.~Klime\^{s}}, \bibinfo{author}{D.~R. Bowler},
  \bibinfo{author}{A.~Michaelides},
\newblock \bibinfo{title}{Chemical accuracy for the van der waals density
  functional},
\newblock \bibinfo{journal}{J. Phys.: Cond. Matter} \bibinfo{volume}{22}
  (\bibinfo{year}{2010}) \bibinfo{pages}{022201}.
\bibitem[{Tkatchenko et~al.(2012)Tkatchenko, DiStasio~Jr, Car, and
  Scheffler}]{Tkatchenko:2012fm}
\bibinfo{author}{A.~Tkatchenko}, \bibinfo{author}{R.~A. DiStasio~Jr},
  \bibinfo{author}{R.~Car}, \bibinfo{author}{M.~Scheffler},
\newblock \bibinfo{title}{{Accurate and efficient method for many-body van der
  Waals interactions}},
\newblock \bibinfo{journal}{Phys. Rev. Lett.} \bibinfo{volume}{108}
  (\bibinfo{year}{2012}) \bibinfo{pages}{236402}.
\bibitem[{Tkatchenko et~al.(2013)Tkatchenko, Ambrosetti, and
  DiStasio}]{Tkatchenko-JCP-2013}
\bibinfo{author}{A.~Tkatchenko}, \bibinfo{author}{A.~Ambrosetti},
  \bibinfo{author}{R.~A. DiStasio},
\newblock \bibinfo{title}{Interatomic methods for the dispersion energy derived
  from the adiabatic connection fluctuation-dissipation theorem},
\newblock \bibinfo{journal}{J. Chem. Phys.} \bibinfo{volume}{138}
  (\bibinfo{year}{2013}) \bibinfo{pages}{074106}.
\bibitem[{Bu\^{c}ko et~al.(2013)Bu\^{c}ko, Leb\'{e}gue, Hafner, and
  \'{A}ngy\'{a}n}]{Bucko-PRB-2013}
\bibinfo{author}{T.~Bu\^{c}ko}, \bibinfo{author}{S.~Leb\'{e}gue},
  \bibinfo{author}{J.~Hafner}, \bibinfo{author}{J.~G. \'{A}ngy\'{a}n},
\newblock \bibinfo{title}{Tkatchenko-scheffler van der waals correction method
  with and without self-consistent screening applied to solids},
\newblock \bibinfo{journal}{Phys. Rev. B} \bibinfo{volume}{87}
  (\bibinfo{year}{2013}) \bibinfo{pages}{064110}.
\bibitem[{Langreth and Perdew(1977)}]{ACFDT-Langreth-Perdew-1977}
\bibinfo{author}{D.~C. Langreth}, \bibinfo{author}{J.~Perdew},
\newblock \bibinfo{title}{Exchange correlation energy of a metallic surface:
  wave-vector analysis},
\newblock \bibinfo{journal}{Phys. Rev. B} \bibinfo{volume}{15}
  (\bibinfo{year}{1977}) \bibinfo{pages}{2884--2901}.
\bibitem[{Harl and Kresse(2009)}]{Harl-Kresse-PRL-2009}
\bibinfo{author}{J.~Harl}, \bibinfo{author}{G.~Kresse},
\newblock \bibinfo{title}{Accurate bulk properties from approximate many-body
  techniques},
\newblock \bibinfo{journal}{Phys. Rev. Lett.} \bibinfo{volume}{103}
  (\bibinfo{year}{2009}) \bibinfo{pages}{056401}.
\bibitem[{Bamidele et~al.(2013)Bamidele, Brndiar, Gulans, Kantorovich, and
  \^{S}tich}]{Bamidele-Stich-JCTC-2013}
\bibinfo{author}{J.~Bamidele}, \bibinfo{author}{J.~Brndiar},
  \bibinfo{author}{A.~Gulans}, \bibinfo{author}{L.~Kantorovich},
  \bibinfo{author}{I.~\^{S}tich},
\newblock \bibinfo{title}{Importance of van der waals stabilization in strongly
  chemically bonded surfaces: Cu(110):o},
\newblock \bibinfo{journal}{J. Chem. Theory Comp.} \bibinfo{volume}{9}
  (\bibinfo{year}{2013}) \bibinfo{pages}{5578--5584}.
\bibitem[{Jones(1924)}]{jones24}
\bibinfo{author}{J.~E. Jones},
\newblock \bibinfo{title}{On the determination of molecular fields. {II} from
  the equation of state of a gas},
\newblock \bibinfo{journal}{Proc. Roy. Soc. A} \bibinfo{volume}{106}
  (\bibinfo{year}{1924}) \bibinfo{pages}{463--477}.
\bibitem[{Morse(1929)}]{morse29}
\bibinfo{author}{P.~M. Morse},
\newblock \bibinfo{title}{{Diatomic molecules according to the wave mechanics.
  II. Vibrational levels}},
\newblock \bibinfo{journal}{Phys. Rev.} \bibinfo{volume}{34}
  (\bibinfo{year}{1929}) \bibinfo{pages}{57--64}.
\bibitem[{Carlsson(1990)}]{carlsson90}
\bibinfo{author}{A.~E. Carlsson}, \bibinfo{title}{Solid State Physics: Advances
  in Research and Applications}, volume~\bibinfo{volume}{43},
  \bibinfo{publisher}{Academic Press}, \bibinfo{year}{1990}.
\bibitem[{Stillinger and Weber(1985)}]{stillinger85}
\bibinfo{author}{F.~H. Stillinger}, \bibinfo{author}{T.~A. Weber},
\newblock \bibinfo{title}{Computer simulation of local order in condensed
  phases of silicon},
\newblock \bibinfo{journal}{Phys. Rev. B} \bibinfo{volume}{31}
  (\bibinfo{year}{1985}) \bibinfo{pages}{5262--5271}.
\bibitem[{Tersoff(1986)}]{tersoff86}
\bibinfo{author}{J.~Tersoff},
\newblock \bibinfo{title}{New empirical model for the structural properties of
  silicon},
\newblock \bibinfo{journal}{Phys. Rev. Lett.} \bibinfo{volume}{56}
  (\bibinfo{year}{1986}) \bibinfo{pages}{632--635}.
\bibitem[{Tersoff(1988)}]{tersoff88}
\bibinfo{author}{J.~Tersoff},
\newblock \bibinfo{title}{New empirical approach for the structure and energy
  of covalent systems},
\newblock \bibinfo{journal}{Phys. Rev. B} \bibinfo{volume}{37}
  (\bibinfo{year}{1988}) \bibinfo{pages}{6991--7000}.
\bibitem[{Brenner(1990)}]{brenner90}
\bibinfo{author}{D.~W. Brenner},
\newblock \bibinfo{title}{Empirical potential for hydrocarbons for use in
  simulating the chemical vapor deposition of diamond films},
\newblock \bibinfo{journal}{Phys. Rev. B} \bibinfo{volume}{42}
  (\bibinfo{year}{1990}) \bibinfo{pages}{9458--9471}.
\bibitem[{Daw and Baskes(1983)}]{daw83}
\bibinfo{author}{M.~S. Daw}, \bibinfo{author}{M.~I. Baskes},
\newblock \bibinfo{title}{Semiempirical, quantum mechanical calculation of
  hydrogen embrittlement in metals},
\newblock \bibinfo{journal}{Phys. Rev. Lett.} \bibinfo{volume}{50}
  (\bibinfo{year}{1983}) \bibinfo{pages}{1285--1288}.
\bibitem[{Daw and Baskes(1984)}]{daw84}
\bibinfo{author}{M.~S. Daw}, \bibinfo{author}{M.~I. Baskes},
\newblock \bibinfo{title}{Embedded-atom method: Derivation and application to
  impurities, surfaces, and other defects in metals},
\newblock \bibinfo{journal}{Phys. Rev. B} \bibinfo{volume}{29}
  (\bibinfo{year}{1984}) \bibinfo{pages}{6443--6453}.
\bibitem[{Jacobsen et~al.(1987)Jacobsen, Norskov, and Puska}]{jacobsen87}
\bibinfo{author}{K.~W. Jacobsen}, \bibinfo{author}{J.~K. Norskov},
  \bibinfo{author}{M.~J. Puska},
\newblock \bibinfo{title}{Interatomic interactions in the effective-medium
  theory},
\newblock \bibinfo{journal}{Phys. Rev. B} \bibinfo{volume}{35}
  (\bibinfo{year}{1987}) \bibinfo{pages}{7423--7442}.
\bibitem[{Finnis and Sinclair(1984)}]{finnis84}
\bibinfo{author}{M.~W. Finnis}, \bibinfo{author}{J.~E. Sinclair},
\newblock \bibinfo{title}{A simple empirical $n$-body potential for transition
  metals},
\newblock \bibinfo{journal}{Phil. Mag. A} \bibinfo{volume}{50}
  (\bibinfo{year}{1984}) \bibinfo{pages}{45--55}.
\bibitem[{Haghighatpanah and Borjesson(2012)}]{Haghi}
\bibinfo{author}{S.~Haghighatpanah}, \bibinfo{author}{A.~Borjesson},
\newblock \bibinfo{title}{Computational studies of graphene growth mechanisms},
\newblock \bibinfo{journal}{Phys. Rev. B.} \bibinfo{volume}{85}
  (\bibinfo{year}{2012}) \bibinfo{pages}{205448}.
\bibitem[{Amara et~al.(2009)Amara, Roussel, Bichara, Gaspard, and
  Ducastelle}]{Amara:2009hn}
\bibinfo{author}{H.~Amara}, \bibinfo{author}{J.~M. Roussel},
  \bibinfo{author}{C.~Bichara}, \bibinfo{author}{J.~P. Gaspard},
  \bibinfo{author}{F.~Ducastelle},
\newblock \bibinfo{title}{{Tight-binding potential for atomistic simulations of
  carbon interacting with transition metals: Application to the Ni-C system}},
\newblock \bibinfo{journal}{Phys. Rev. B} \bibinfo{volume}{79}
  (\bibinfo{year}{2009}) \bibinfo{pages}{014109}.
\bibitem[{Gao et~al.(2011)Gao, Yip, Zhao, Yakobson, and Ding}]{Gao11}
\bibinfo{author}{J.~Gao}, \bibinfo{author}{J.~Yip}, \bibinfo{author}{J.~Zhao},
  \bibinfo{author}{B.~I. Yakobson}, \bibinfo{author}{F.~Ding},
\newblock \bibinfo{title}{Graphene nucleation on transition metal surface:
  Structure transformation and role of the metal step edge},
\newblock \bibinfo{journal}{J. Am. Chem. Soc.} \bibinfo{volume}{133}
  (\bibinfo{year}{2011}) \bibinfo{pages}{5009--5015}.
\bibitem[{Wesep et~al.(2011)Wesep, Chen, Zhu, and Zhang}]{wesep:171105}
\bibinfo{author}{R.~G.~V. Wesep}, \bibinfo{author}{H.~Chen},
  \bibinfo{author}{W.~Zhu}, \bibinfo{author}{Z.~Zhang},
\newblock \bibinfo{title}{Communication: Stable carbon nanoarches in the
  initial stages of epitaxial growth of graphene on {C}u(111)},
\newblock \bibinfo{journal}{J. Chem. Phys.} \bibinfo{volume}{134}
  (\bibinfo{year}{2011}) \bibinfo{pages}{171105}.
\bibitem[{Riikonen et~al.(2012)Riikonen, Krasheninnikov, Halonen, and
  Nieminen}]{Riikonen}
\bibinfo{author}{S.~Riikonen}, \bibinfo{author}{A.~V. Krasheninnikov},
  \bibinfo{author}{L.~Halonen}, \bibinfo{author}{R.~M. Nieminen},
\newblock \bibinfo{title}{The role of stable and mobile carbon adspecies in
  copper-promoted graphene growth},
\newblock \bibinfo{journal}{J. Phys. Chem. {C}} \bibinfo{volume}{116}
  (\bibinfo{year}{2012}) \bibinfo{pages}{5802--5809}.
\bibitem[{Zhang et~al.(2011)Zhang, Wu, Li, and
  Yang}]{Zhangdoi:10.1021/jp2006827}
\bibinfo{author}{W.~Zhang}, \bibinfo{author}{P.~Wu}, \bibinfo{author}{Z.~Li},
  \bibinfo{author}{J.~Yang},
\newblock \bibinfo{title}{First-principles thermodynamics of graphene growth on
  {C}u surfaces},
\newblock \bibinfo{journal}{J. Phys. Chem. C} \bibinfo{volume}{115}
  (\bibinfo{year}{2011}) \bibinfo{pages}{17782--17787}.
\bibitem[{Jonsson et~al.(1999)Jonsson, Mills, and Jacobsen}]{Jonsson:1999tf}
\bibinfo{author}{H.~Jonsson}, \bibinfo{author}{G.~Mills},
  \bibinfo{author}{K.~W. Jacobsen},
\newblock \bibinfo{title}{{Nudged elastic band method for finding minimum
  energy paths of transitions}},
\newblock \bibinfo{journal}{Classical and Quantum Dynamics in Condensed Phase
  Simulations}  (\bibinfo{year}{1999}) \bibinfo{pages}{385}.
\bibitem[{Henkelman and J{\'o}nsson(2000)}]{Henkelman:2000tl}
\bibinfo{author}{G.~Henkelman}, \bibinfo{author}{H.~J{\'o}nsson},
\newblock \bibinfo{title}{{Improved tangent estimate in the nudged elastic band
  method for finding minimum energy paths and saddle points}},
\newblock \bibinfo{journal}{J. Chem. Phys.} \bibinfo{volume}{113}
  (\bibinfo{year}{2000}) \bibinfo{pages}{9978}.
\bibitem[{Henkelman et~al.(2000)Henkelman, Uberuaga, and
  J{\'o}nsson}]{Henkelman:2000wb}
\bibinfo{author}{G.~Henkelman}, \bibinfo{author}{B.~P. Uberuaga},
  \bibinfo{author}{H.~J{\'o}nsson},
\newblock \bibinfo{title}{{A climbing image nudged elastic band method for
  finding saddle points and minimum energy paths}},
\newblock \bibinfo{journal}{J. Chem. Phys.} \bibinfo{volume}{113}
  (\bibinfo{year}{2000}) \bibinfo{pages}{9901}.
\bibitem[{Voter et~al.(2002)Voter, Montalenti, and Germann}]{voter02}
\bibinfo{author}{A.~F. Voter}, \bibinfo{author}{F.~Montalenti},
  \bibinfo{author}{T.~C. Germann},
\newblock \bibinfo{title}{Extending the time scale in atomistic simulation of
  materials},
\newblock \bibinfo{journal}{Annu. Rev. Mater. Res.} \bibinfo{volume}{32}
  (\bibinfo{year}{2002}) \bibinfo{pages}{321--346}.
\bibitem[{Landau and Binder(2000)}]{landau00}
\bibinfo{author}{D.~P. Landau}, \bibinfo{author}{K.~Binder}, \bibinfo{title}{A
  Guide to Monte Carlo Simulations in Statistical Physics},
  \bibinfo{publisher}{Cambridge University Press},
  \bibinfo{address}{Cambridge}, \bibinfo{year}{2000}.
\bibitem[{Kang and Weinberg(1989)}]{kang89}
\bibinfo{author}{H.~C. Kang}, \bibinfo{author}{W.~H. Weinberg},
\newblock \bibinfo{title}{Dynamic {M}onte {C}arlo with a proper energy barrier:
  Surface diffusion and two-dimensional domain ordering},
\newblock \bibinfo{journal}{J. Chem. Phys.} \bibinfo{volume}{90}
  (\bibinfo{year}{1989}) \bibinfo{pages}{2824--2830}.
\bibitem[{Fichthorn and Weinberg(1991)}]{fichthorn91}
\bibinfo{author}{K.~A. Fichthorn}, \bibinfo{author}{W.~H. Weinberg},
\newblock \bibinfo{title}{Theoretical foundations of dynamical {M}onte {C}arlo
  simulations},
\newblock \bibinfo{journal}{J. Chem. Phys.} \bibinfo{volume}{95}
  (\bibinfo{year}{1991}) \bibinfo{pages}{1090--1096}.
\bibitem[{Gillespie(1976)}]{Gillespie-1976}
\bibinfo{author}{D.~T. Gillespie},
\newblock \bibinfo{title}{A general method for numerically simulating the
  stochastic time evolution of coupled chemical reactions},
\newblock \bibinfo{journal}{J. Comp. Physics} \bibinfo{volume}{22}
  (\bibinfo{year}{1976}) \bibinfo{pages}{403--434}.
\bibitem[{Reuter et~al.(2005)Reuter, Stampfl, and
  Scheffler}]{Reuter-Stampfl-Scheffler-2005}
\bibinfo{author}{K.~Reuter}, \bibinfo{author}{C.~Stampfl},
  \bibinfo{author}{M.~Scheffler}, \bibinfo{title}{Ab initio atomistic
  thermodynamic and stastical mechanics of surface properties and functions},
  volume~\bibinfo{volume}{A} of \textit{\bibinfo{series}{Handbook of Material
  Modeling}}, \bibinfo{publisher}{Springer}, \bibinfo{year}{2005}, p.
  \bibinfo{pages}{149}.
\bibitem[{Gillespie(2007)}]{Gillespie-KMC}
\bibinfo{author}{D.~T. Gillespie},
\newblock \bibinfo{title}{Stochastic simulation of chemical kinetics},
\newblock \bibinfo{journal}{Annu. Rev. Phys. Chem.} \bibinfo{volume}{58}
  (\bibinfo{year}{2007}) \bibinfo{pages}{35}.
\bibitem[{Voter(2005)}]{Voter-review-2005}
\bibinfo{author}{A.~F. Voter}, \bibinfo{title}{Radiation effects in solids},
  Handbook of Material Modeling, Part A. Methods, \bibinfo{publisher}{Springer,
  NATO Publishing Unit, Dordrecht, The Netherlands}, \bibinfo{year}{2005}.
\bibitem[{Kampen(1981)}]{vankampen81}
\bibinfo{author}{N.~G.~V. Kampen}, \bibinfo{title}{Stochastic Processes in
  Physics and Chemistry}, \bibinfo{publisher}{North-Holland},
  \bibinfo{address}{Amsterdam}, \bibinfo{year}{1981}.
\bibitem[{Nitzan(2006)}]{Nitzan-book}
\bibinfo{author}{A.~Nitzan}, \bibinfo{title}{Chemical Dynamics in Condensed
  Phases: \ Relaxation, Transfer, and Reactions in Condensed Molecular
  Systems}, \bibinfo{publisher}{Oxford Univ. Press}, \bibinfo{year}{2006}.
\bibitem[{Zangwill(1988)}]{zangwill88}
\bibinfo{author}{A.~Zangwill}, \bibinfo{title}{Physics at Surfaces},
  \bibinfo{publisher}{Cambridge University Press},
  \bibinfo{address}{Cambridge}, \bibinfo{year}{1988}.
\bibitem[{Wang and Seebauer(2001)}]{wang01}
\bibinfo{author}{Z.~Wang}, \bibinfo{author}{E.~G. Seebauer},
\newblock \bibinfo{title}{Estimating pre-exponential factors for desorption
  from semiconductors: Consequences for {\it a priori} process modelling},
\newblock \bibinfo{journal}{Appl. Surf. Sci.} \bibinfo{volume}{181}
  (\bibinfo{year}{2001}) \bibinfo{pages}{111--120}.
\bibitem[{Amara et~al.(2006)Amara, Bichara, and Ducastelle}]{Amara:2006hd}
\bibinfo{author}{H.~Amara}, \bibinfo{author}{C.~Bichara},
  \bibinfo{author}{F.~Ducastelle},
\newblock \bibinfo{title}{{Formation of carbon nanostructures on nickel
  surfaces: A tight-binding grand canonical Monte Carlo study}},
\newblock \bibinfo{journal}{Phys. Rev. B} \bibinfo{volume}{73}
  (\bibinfo{year}{2006}) \bibinfo{pages}{113404}.
\bibitem[{Burton et~al.(1951)Burton, Cabrera, and Frank}]{burton51}
\bibinfo{author}{W.~K. Burton}, \bibinfo{author}{N.~Cabrera},
  \bibinfo{author}{F.~C. Frank},
\newblock \bibinfo{title}{The growth of crystals and the equilibrium structure
  of their surfaces},
\newblock \bibinfo{journal}{Phil. Trans. R. Soc. Lond. A} \bibinfo{volume}{243}
  (\bibinfo{year}{1951}) \bibinfo{pages}{299--358}.
\bibitem[{Langer(1986)}]{langer86}
\bibinfo{author}{J.~S. Langer}, \bibinfo{title}{Directions in Condensed Matter
  Physics}, \bibinfo{publisher}{World Scientific}, \bibinfo{year}{1986}, pp.
  \bibinfo{pages}{164--186}.
\bibitem[{Sekerka(2004)}]{sekerka04}
\bibinfo{author}{R.~F. Sekerka},
\newblock \bibinfo{title}{Morphology: From sharp interface to phase field
  models},
\newblock \bibinfo{journal}{Journal of Crystal Growth} \bibinfo{volume}{264}
  (\bibinfo{year}{2004}) \bibinfo{pages}{530--540}.
\bibitem[{Kassner et~al.(2001)Kassner, Misbah, M{\"u}ller, Kappey, and
  Kohlert}]{kassner01}
\bibinfo{author}{K.~Kassner}, \bibinfo{author}{C.~Misbah},
  \bibinfo{author}{J.~M{\"u}ller}, \bibinfo{author}{J.~Kappey},
  \bibinfo{author}{P.~Kohlert},
\newblock \bibinfo{title}{Phase-field modeling of stress-induced
  instabilities},
\newblock \bibinfo{journal}{Phys. Rev. E} \bibinfo{volume}{63}
  (\bibinfo{year}{2001}) \bibinfo{pages}{036117}.
\bibitem[{Liu and Metiu(1994)}]{liu94}
\bibinfo{author}{F.~Liu}, \bibinfo{author}{H.~Metiu},
\newblock \bibinfo{title}{Stability and kinetics of step motion on crystal
  surfaces},
\newblock \bibinfo{journal}{Phy. Rev. E} \bibinfo{volume}{49}
  (\bibinfo{year}{1994}) \bibinfo{pages}{2601--2616}.
\bibitem[{Karma and Plapp(1998)}]{karma98}
\bibinfo{author}{A.~Karma}, \bibinfo{author}{M.~Plapp},
\newblock \bibinfo{title}{Spiral surface growth without desorption},
\newblock \bibinfo{journal}{Phys. Rev. Lett.} \bibinfo{volume}{81}
  (\bibinfo{year}{1998}) \bibinfo{pages}{4444--4447}.
\bibitem[{R\"{a}tz and Voigt(2004)}]{ratz04}
\bibinfo{author}{A.~R\"{a}tz}, \bibinfo{author}{A.~Voigt},
\newblock \bibinfo{title}{Various phase-field approximations for epitaxial
  growth},
\newblock \bibinfo{journal}{Journal of Crystal Growth} \bibinfo{volume}{266}
  (\bibinfo{year}{2004}) \bibinfo{pages}{278--282}.
\bibitem[{Castro(2003)}]{castro03}
\bibinfo{author}{M.~Castro},
\newblock \bibinfo{title}{Phase-field approach to heterogeneous nucleation},
\newblock \bibinfo{journal}{Phy. Rev. B} \bibinfo{volume}{67}
  (\bibinfo{year}{2003}) \bibinfo{pages}{035412}.
\bibitem[{Yu and Liu(2004)}]{yu04}
\bibinfo{author}{Y.~Yu}, \bibinfo{author}{B.~Liu},
\newblock \bibinfo{title}{Phase-field model of island growth in epitaxy},
\newblock \bibinfo{journal}{Phys. Rev. E} \bibinfo{volume}{69}
  (\bibinfo{year}{2004}) \bibinfo{pages}{021601}.
\bibitem[{Ming and Zangwill(2010)}]{ming10}
\bibinfo{author}{F.~Ming}, \bibinfo{author}{A.~Zangwill},
\newblock \bibinfo{title}{Phase field modeling of submonolayer epitaxial
  growth},
\newblock \bibinfo{journal}{Phys. Rev. B} \bibinfo{volume}{81}
  (\bibinfo{year}{2010}) \bibinfo{pages}{235431}.
\bibitem[{Kim and Lu(2004)}]{kim04}
\bibinfo{author}{D.~Kim}, \bibinfo{author}{W.~Lu},
\newblock \bibinfo{title}{Self-organized nanostructures in multi-phase
  epilayers},
\newblock \bibinfo{journal}{Nanotechnology} \bibinfo{volume}{15}
  (\bibinfo{year}{2004}) \bibinfo{pages}{667--674}.
\bibitem[{Venables et~al.(1984)Venables, Spiller, and
  Hanb\"{u}cken}]{venables84}
\bibinfo{author}{J.~A. Venables}, \bibinfo{author}{G.~D.~T. Spiller},
  \bibinfo{author}{M.~Hanb\"{u}cken},
\newblock \bibinfo{title}{Nucleation and growth of thin films},
\newblock \bibinfo{journal}{Rep. Prog. Phys.} \bibinfo{volume}{47}
  (\bibinfo{year}{1984}) \bibinfo{pages}{399--458}.
\bibitem[{Bales and Chrzan(1994)}]{bales94}
\bibinfo{author}{G.~S. Bales}, \bibinfo{author}{D.~C. Chrzan},
\newblock \bibinfo{title}{Dynamics of irreversible island growth during
  submonolayer epitaxy},
\newblock \bibinfo{journal}{Phys. Rev. B} \bibinfo{volume}{50}
  (\bibinfo{year}{1994}) \bibinfo{pages}{6057--6067}.
\bibitem[{Bartelt and Evans(1996)}]{bartelt96}
\bibinfo{author}{M.~C. Bartelt}, \bibinfo{author}{J.~W. Evans},
\newblock \bibinfo{title}{Exact island-size distributions for submonolayer
  deposition: Influence of correlations between island size and separation},
\newblock \bibinfo{journal}{Phys. Rev. B} \bibinfo{volume}{54}
  (\bibinfo{year}{1996}) \bibinfo{pages}{R17359--R17362}.
\bibitem[{Amar et~al.(2001)Amar, Popescu, and Family}]{amar01}
\bibinfo{author}{J.~G. Amar}, \bibinfo{author}{M.~N. Popescu},
  \bibinfo{author}{F.~Family},
\newblock \bibinfo{title}{Rate-equation approach to island capture zones and
  size distributions in epitaxial growth},
\newblock \bibinfo{journal}{Phys. Rev. Lett} \bibinfo{volume}{86}
  (\bibinfo{year}{2001}) \bibinfo{pages}{3092--3095}.
\bibitem[{Bartelt et~al.(1999)Bartelt, Stoldt, Jenks, Thiel, and
  Evans}]{bartelt99}
\bibinfo{author}{M.~C. Bartelt}, \bibinfo{author}{C.~R. Stoldt},
  \bibinfo{author}{C.~J. Jenks}, \bibinfo{author}{P.~A. Thiel},
  \bibinfo{author}{J.~W. Evans},
\newblock \bibinfo{title}{{Adatom capture by arrays of two-dimensional Ag
  islands on Ag(100)}},
\newblock \bibinfo{journal}{Phys. Rev. B} \bibinfo{volume}{59}
  (\bibinfo{year}{1999}) \bibinfo{pages}{3125--3134}.
\bibitem[{Wintterlin and Bocquet(2009)}]{Wintterlin}
\bibinfo{author}{J.~Wintterlin}, \bibinfo{author}{M.~L. Bocquet},
\newblock \bibinfo{title}{Graphene on metal surfaces},
\newblock \bibinfo{journal}{Surf. Sci.} \bibinfo{volume}{603}
  (\bibinfo{year}{2009}) \bibinfo{pages}{1841--1852}.
\bibitem[{Ago et~al.(2010)Ago, Ito, Mizuta, Yoshida, Hu, Orofeo, Tsuji, Ikeda,
  and Mizuno}]{Ago2010}
\bibinfo{author}{H.~Ago}, \bibinfo{author}{Y.~Ito},
  \bibinfo{author}{N.~Mizuta}, \bibinfo{author}{K.~Yoshida},
  \bibinfo{author}{B.~Hu}, \bibinfo{author}{C.~M. Orofeo},
  \bibinfo{author}{M.~Tsuji}, \bibinfo{author}{K.~Ikeda},
  \bibinfo{author}{S.~Mizuno},
\newblock \bibinfo{title}{Epitaxial chemical vapor deposition growth of
  single-layer graphene over cobalt film crystallized on sapphire},
\newblock \bibinfo{journal}{ACS Nano} \bibinfo{volume}{4}
  (\bibinfo{year}{2010}) \bibinfo{pages}{7407--7414}.
\bibitem[{Iwasaki et~al.(2011)Iwasaki, Park, Konuma, Lee, Smet, and
  Starke}]{Iwasaki2010}
\bibinfo{author}{T.~Iwasaki}, \bibinfo{author}{H.~J. Park},
  \bibinfo{author}{M.~Konuma}, \bibinfo{author}{D.~S. Lee},
  \bibinfo{author}{J.~H. Smet}, \bibinfo{author}{U.~Starke},
\newblock \bibinfo{title}{{Long-range ordered single-crystal graphene on
  high-quality heteroepitaxial Ni thin films grown on MgO(111)}},
\newblock \bibinfo{journal}{Nano Lett.} \bibinfo{volume}{11}
  (\bibinfo{year}{2011}) \bibinfo{pages}{79--84}.
\bibitem[{Reddy et~al.(2011)Reddy, Gledhill, Chen, Drexler, and
  Padture}]{Reddy201}
\bibinfo{author}{K.~M. Reddy}, \bibinfo{author}{A.~D. Gledhill},
  \bibinfo{author}{C.~Chen}, \bibinfo{author}{J.~M. Drexler},
  \bibinfo{author}{N.~P. Padture},
\newblock \bibinfo{title}{{High quality, transferrable graphene grown on single
  crystal Cu(111) thin films on basal-plane sapphire}},
\newblock \bibinfo{journal}{Appl. Phys. Lett.} \bibinfo{volume}{98}
  (\bibinfo{year}{2011}) \bibinfo{pages}{113117}.
\bibitem[{Sutter et~al.(2010)Sutter, Albrecht, and Sutter}]{Sutter2010}
\bibinfo{author}{P.~W. Sutter}, \bibinfo{author}{P.~M. Albrecht},
  \bibinfo{author}{E.~A. Sutter},
\newblock \bibinfo{title}{{Graphene growth on epitaxial Ru thin films on
  sapphire}},
\newblock \bibinfo{journal}{Appl. Phys. Lett.} \bibinfo{volume}{97}
  (\bibinfo{year}{2010}) \bibinfo{pages}{213101}.
\bibitem[{Tonnoir et~al.(2013)Tonnoir, Kimouche, Coraux, Magaud, Delsol,
  Gilles, and Chapelier}]{Tonnoir2013}
\bibinfo{author}{C.~Tonnoir}, \bibinfo{author}{A.~Kimouche},
  \bibinfo{author}{J.~Coraux}, \bibinfo{author}{L.~Magaud},
  \bibinfo{author}{B.~Delsol}, \bibinfo{author}{B.~Gilles},
  \bibinfo{author}{C.~Chapelier},
\newblock \bibinfo{title}{Induced superconductivity in graphene grown on
  rhenium},
\newblock \bibinfo{journal}{Phys. Rev. Lett.} \bibinfo{volume}{111}
  (\bibinfo{year}{2013}) \bibinfo{pages}{246805}.
\bibitem[{Al-Temimy et~al.(2009)Al-Temimy, Riedl, and Starke}]{AlTemimy2009}
\bibinfo{author}{A.~Al-Temimy}, \bibinfo{author}{C.~Riedl},
  \bibinfo{author}{U.~Starke},
\newblock \bibinfo{title}{{Low temperature growth of epitaxial graphene on SiC
  induced by carbon evaporation}},
\newblock \bibinfo{journal}{Appl. Phys. Lett.} \bibinfo{volume}{95}
  (\bibinfo{year}{2009}) \bibinfo{pages}{231907}.
\bibitem[{Michon et~al.(2010)Michon, V{\'e}zian, Ouerghi, Zielinski, Chassagne,
  and Portail}]{Michon2010}
\bibinfo{author}{A.~Michon}, \bibinfo{author}{S.~V{\'e}zian},
  \bibinfo{author}{A.~Ouerghi}, \bibinfo{author}{M.~Zielinski},
  \bibinfo{author}{T.~Chassagne}, \bibinfo{author}{M.~Portail},
\newblock \bibinfo{title}{{Direct growth of few-layer graphene on 6H-SiC and
  3C-SiC/Si via propane chemical vapor deposition}},
\newblock \bibinfo{journal}{Appl. Phys. Lett.} \bibinfo{volume}{97}
  (\bibinfo{year}{2010}) \bibinfo{pages}{171909}.
\bibitem[{Hwang et~al.(2010)Hwang, Shields, Thomas, Shivaraman, Hao, Kim, Woll,
  Tompa, and Spencer}]{Hwang2010}
\bibinfo{author}{J.~Hwang}, \bibinfo{author}{V.~B. Shields},
  \bibinfo{author}{C.~I. Thomas}, \bibinfo{author}{S.~Shivaraman},
  \bibinfo{author}{D.~Hao}, \bibinfo{author}{M.~Kim}, \bibinfo{author}{A.~R.
  Woll}, \bibinfo{author}{G.~S. Tompa}, \bibinfo{author}{M.~G. Spencer},
\newblock \bibinfo{title}{{Epitaxial growth of graphitic carbon on C-face SiC
  and sapphire by chemical vapor deposition (CVD)}},
\newblock \bibinfo{journal}{J. Cryst. Growth} \bibinfo{volume}{312}
  (\bibinfo{year}{2010}) \bibinfo{pages}{3219--3224}.
\bibitem[{Wang et~al.(2011)Wang, Ma, Caffio, Schaub, and Li}]{Wang}
\bibinfo{author}{B.~Wang}, \bibinfo{author}{X.~Ma},
  \bibinfo{author}{M.~Caffio}, \bibinfo{author}{R.~Schaub},
  \bibinfo{author}{W.-X. Li},
\newblock \bibinfo{title}{Size-selective carbon nanoclusters as precursors to
  the growth of epitaxial graphene},
\newblock \bibinfo{journal}{Nano Lett.} \bibinfo{volume}{11}
  (\bibinfo{year}{2011}) \bibinfo{pages}{424--430}.
\bibitem[{Hwang et~al.(2011)Hwang, Yoo, Kim, Seo, Yu, and Biro}]{Hwang11}
\bibinfo{author}{C.~Hwang}, \bibinfo{author}{K.~Yoo}, \bibinfo{author}{S.~J.
  Kim}, \bibinfo{author}{E.~K. Seo}, \bibinfo{author}{H.~Yu},
  \bibinfo{author}{L.~P. Biro},
\newblock \bibinfo{title}{Initial stage of graphene growth on a {Cu}
  substrate},
\newblock \bibinfo{journal}{J. Phys. Chem. C} \bibinfo{volume}{115}
  (\bibinfo{year}{2011}) \bibinfo{pages}{22369--22374}.
\bibitem[{Gao et~al.(2010)Gao, Guest, and Guisinger}]{Gao10}
\bibinfo{author}{L.~Gao}, \bibinfo{author}{J.~R. Guest}, \bibinfo{author}{N.~P.
  Guisinger},
\newblock \bibinfo{title}{Epitaxial graphene on {Cu(111)}},
\newblock \bibinfo{journal}{Nano Lett.} \bibinfo{volume}{10}
  (\bibinfo{year}{2010}) \bibinfo{pages}{3512--3516}.
\bibitem[{Yamamoto et~al.(1992)Yamamoto, Fukushima, Osaka, and
  Oshima}]{Yamamoto1992}
\bibinfo{author}{K.~Yamamoto}, \bibinfo{author}{M.~Fukushima},
  \bibinfo{author}{T.~Osaka}, \bibinfo{author}{C.~Oshima},
\newblock \bibinfo{title}{Charge-transfer mechanism for the (monolayer
  graphite)/{N}i(111) system},
\newblock \bibinfo{journal}{Phys. Rev. B} \bibinfo{volume}{45}
  (\bibinfo{year}{1992}) \bibinfo{pages}{11358}.
\bibitem[{Lahiri et~al.(2010)Lahiri, Miller, Adamska, Oleynik, and
  Batzill}]{Lahiri2010}
\bibinfo{author}{J.~Lahiri}, \bibinfo{author}{T.~Miller},
  \bibinfo{author}{L.~Adamska}, \bibinfo{author}{I.~I. Oleynik},
  \bibinfo{author}{M.~Batzill},
\newblock \bibinfo{title}{Graphene growth on {Ni}(111) by transformation of a
  surface carbide},
\newblock \bibinfo{journal}{Nano Lett.} \bibinfo{volume}{11}
  (\bibinfo{year}{2010}) \bibinfo{pages}{518--522}.
\bibitem[{Eom et~al.(2009)Eom, Prezzi, Rim, Zhou, Lefenfeld, Xiao, Nuckolls,
  Hybertsen, Heinz, and Flynn}]{Eom2009}
\bibinfo{author}{D.~Eom}, \bibinfo{author}{D.~Prezzi}, \bibinfo{author}{K.~T.
  Rim}, \bibinfo{author}{H.~Zhou}, \bibinfo{author}{M.~Lefenfeld},
  \bibinfo{author}{S.~Xiao}, \bibinfo{author}{C.~Nuckolls},
  \bibinfo{author}{M.~S. Hybertsen}, \bibinfo{author}{T.~F. Heinz},
  \bibinfo{author}{G.~W. Flynn},
\newblock \bibinfo{title}{Structure and electronic properties of graphene
  nanoislands on {C}o(0001)},
\newblock \bibinfo{journal}{Nano Lett.} \bibinfo{volume}{9}
  (\bibinfo{year}{2009}) \bibinfo{pages}{2844--2848}.
\bibitem[{Rader et~al.(2009)Rader, Varykhalov, S\'{a}nchez-Barriga, Marchenko,
  Rybkin, and Shikin}]{Rader2009}
\bibinfo{author}{O.~Rader}, \bibinfo{author}{A.~Varykhalov},
  \bibinfo{author}{J.~S\'{a}nchez-Barriga}, \bibinfo{author}{O.~Marchenko},
  \bibinfo{author}{A.~Rybkin}, \bibinfo{author}{A.~M. Shikin},
\newblock \bibinfo{title}{Is there a Rashba effect in graphene on 3d
  ferromagnets?},
\newblock \bibinfo{journal}{Phys. Rev. Lett.} \bibinfo{volume}{102}
  (\bibinfo{year}{2009}) \bibinfo{pages}{057602}.
\bibitem[{Vinogradov et~al.(2012)Vinogradov, Zakharov, Kocevski, Rusz, Simonov,
  Eriksson, Mikkelsen, Lundgren, Vinogradov, M{\aa}rtensson, and
  Preobrajenski}]{Vinogradov2012}
\bibinfo{author}{N.~A. Vinogradov}, \bibinfo{author}{A.~A. Zakharov},
  \bibinfo{author}{V.~Kocevski}, \bibinfo{author}{J.~Rusz},
  \bibinfo{author}{K.~A. Simonov}, \bibinfo{author}{O.~Eriksson},
  \bibinfo{author}{A.~Mikkelsen}, \bibinfo{author}{E.~Lundgren},
  \bibinfo{author}{A.~S. Vinogradov}, \bibinfo{author}{N.~M{\aa}rtensson},
  \bibinfo{author}{A.~B. Preobrajenski},
\newblock \bibinfo{title}{Formation and structure of graphene waves on {Fe}
  (110)},
\newblock \bibinfo{journal}{Phys. Rev. Lett.} \bibinfo{volume}{109}
  (\bibinfo{year}{2012}) \bibinfo{pages}{026101}.
\bibitem[{Nie et~al.(2012)Nie, Bartelt, Wofford, Dubon, McCarty, and
  Th\"{u}rmer}]{Nie2012}
\bibinfo{author}{S.~Nie}, \bibinfo{author}{N.~C. Bartelt},
  \bibinfo{author}{J.~M. Wofford}, \bibinfo{author}{O.~D. Dubon},
  \bibinfo{author}{K.~F. McCarty}, \bibinfo{author}{K.~Th\"{u}rmer},
\newblock \bibinfo{title}{Scanning tunneling microscopy study of graphen on
  {A}u(111): growth mechanisms and substrate interactions},
\newblock \bibinfo{journal}{Phys. Rev. B} \bibinfo{volume}{85}
  (\bibinfo{year}{2012}) \bibinfo{pages}{205406}.
\bibitem[{Gall et~al.(1987)Gall, Mikhailov, Rutkov, and Tontegode}]{Gall1987}
\bibinfo{author}{N.~R. Gall}, \bibinfo{author}{S.~R. Mikhailov},
  \bibinfo{author}{E.~V. Rutkov}, \bibinfo{author}{A.~Y. Tontegode},
\newblock \bibinfo{title}{Carbon interaction with the rhenium surface},
\newblock \bibinfo{journal}{Surf. Sci.} \bibinfo{volume}{191}
  (\bibinfo{year}{1987}) \bibinfo{pages}{187--202}.
\bibitem[{Voloshina et~al.(2012)Voloshina, Dedkov, Torbr{\"u}gge, Thissen, and
  Fonin}]{Voloshina}
\bibinfo{author}{E.~N. Voloshina}, \bibinfo{author}{Y.~S. Dedkov},
  \bibinfo{author}{S.~Torbr{\"u}gge}, \bibinfo{author}{A.~Thissen},
  \bibinfo{author}{M.~Fonin},
\newblock \bibinfo{title}{Graphene on {R}h(111): {S}canning tunneling and
  atomic force microscopies studies},
\newblock \bibinfo{journal}{Appl. Phys. Lett.} \bibinfo{volume}{100}
  (\bibinfo{year}{2012}) \bibinfo{pages}{241606}.
\bibitem[{Wofford et~al.(2010)Wofford, Nie, McCarty, Bartelt, and
  Dubon}]{Wofford}
\bibinfo{author}{J.~M. Wofford}, \bibinfo{author}{S.~Nie},
  \bibinfo{author}{K.~F. McCarty}, \bibinfo{author}{N.~C. Bartelt},
  \bibinfo{author}{O.~D. Dubon},
\newblock \bibinfo{title}{Graphene islands on {Cu} foils: The interplay between
  shape, orientation, and defects},
\newblock \bibinfo{journal}{Nano Lett.} \bibinfo{volume}{10}
  (\bibinfo{year}{2010}) \bibinfo{pages}{4890--4896}.
\bibitem[{Li et~al.(2010)Li, Magnuson, Venugopal, An, Suk, Han, Borysiak, Cai,
  Velamakanni, Zhu, Fu, Vogel, Voelkl, Colombo, and Ruoff}]{Li10}
\bibinfo{author}{X.~Li}, \bibinfo{author}{C.~W. Magnuson},
  \bibinfo{author}{A.~Venugopal}, \bibinfo{author}{J.~An},
  \bibinfo{author}{J.~W. Suk}, \bibinfo{author}{B.~Han},
  \bibinfo{author}{M.~Borysiak}, \bibinfo{author}{W.~Cai},
  \bibinfo{author}{A.~Velamakanni}, \bibinfo{author}{Y.~Zhu},
  \bibinfo{author}{L.~Fu}, \bibinfo{author}{E.~M. Vogel},
  \bibinfo{author}{E.~Voelkl}, \bibinfo{author}{L.~Colombo},
  \bibinfo{author}{R.~S. Ruoff},
\newblock \bibinfo{title}{Graphene films with large domain size by a two-step
  chemical vapor deposition process},
\newblock \bibinfo{journal}{Nano Lett.} \bibinfo{volume}{10}
  (\bibinfo{year}{2010}) \bibinfo{pages}{4328--4334}.
\bibitem[{Vo-Van et~al.(2011)Vo-Van, Kimouche, Reserbat-Plantey, Fruchart,
  Bayle-Guillemaud, Bendiab, and Coraux}]{Chi}
\bibinfo{author}{C.~Vo-Van}, \bibinfo{author}{A.~Kimouche},
  \bibinfo{author}{A.~Reserbat-Plantey}, \bibinfo{author}{O.~Fruchart},
  \bibinfo{author}{P.~Bayle-Guillemaud}, \bibinfo{author}{N.~Bendiab},
  \bibinfo{author}{J.~Coraux},
\newblock \bibinfo{title}{Epitaxial graphene prepared by chemical vapor
  deposition on single crystal thin iridium films on sapphire},
\newblock \bibinfo{journal}{Appl. Phys. Lett.} \bibinfo{volume}{98}
  (\bibinfo{year}{2011}) \bibinfo{pages}{181903}.
\bibitem[{Reddy et~al.(2011)Reddy, Gledhill, Chen, Drexler, and
  Padture}]{Kongara}
\bibinfo{author}{K.~M. Reddy}, \bibinfo{author}{A.~D. Gledhill},
  \bibinfo{author}{C.-H. Chen}, \bibinfo{author}{J.~M. Drexler},
  \bibinfo{author}{N.~P. Padture},
\newblock \bibinfo{title}{High quality, transferrable graphene grown on single
  crystal {C}u(111) thin films on basal-plane sapphire},
\newblock \bibinfo{journal}{Appl. Phys. Lett.} \bibinfo{volume}{98}
  (\bibinfo{year}{2011}) \bibinfo{pages}{113117}.
\bibitem[{Nie et~al.(2011)Nie, Wofford, Bartelt, Dubon, and McCarty}]{Nie}
\bibinfo{author}{S.~Nie}, \bibinfo{author}{J.~M. Wofford},
  \bibinfo{author}{N.~C. Bartelt}, \bibinfo{author}{O.~D. Dubon},
  \bibinfo{author}{K.~F. McCarty},
\newblock \bibinfo{title}{{Origin of the mosaicity in graphene grown on
  Cu(111)}},
\newblock \bibinfo{journal}{Phys. Rev. B} \bibinfo{volume}{84}
  (\bibinfo{year}{2011}) \bibinfo{pages}{155425}.
\bibitem[{Celebi et~al.(2013)Celebi, Cole, Choi, Wyczisk, Legagneux,
  Rupesinghe, Robertson, Teo, and Park}]{Celebi:2013cc}
\bibinfo{author}{K.~Celebi}, \bibinfo{author}{M.~T. Cole},
  \bibinfo{author}{J.~W. Choi}, \bibinfo{author}{F.~Wyczisk},
  \bibinfo{author}{P.~Legagneux}, \bibinfo{author}{N.~Rupesinghe},
  \bibinfo{author}{J.~Robertson}, \bibinfo{author}{K.~B.~K. Teo},
  \bibinfo{author}{H.~G. Park},
\newblock \bibinfo{title}{{Evolutionary Kinetics of Graphene Formation on
  Copper}},
\newblock \bibinfo{journal}{Nano Lett.} \bibinfo{volume}{13}
  (\bibinfo{year}{2013}) \bibinfo{pages}{967--974}.
\bibitem[{Wang et~al.(2010)Wang, G\"{u}nther, Wintterlin, and
  Bocquet}]{Gunther}
\bibinfo{author}{B.~Wang}, \bibinfo{author}{S.~G\"{u}nther},
  \bibinfo{author}{J.~Wintterlin}, \bibinfo{author}{M.-L. Bocquet},
\newblock \bibinfo{title}{{Periodicity, work function and reactivity of
  graphene on Ru(0001) from first principles}},
\newblock \bibinfo{journal}{New J. Phys.} \bibinfo{volume}{12}
  (\bibinfo{year}{2010}) \bibinfo{pages}{043041}.
\bibitem[{G\"{u}nther et~al.(2011)G\"{u}nther, D{\"a}nhardt, Wang, Bocquet,
  Schmitt, and Wintterlin}]{Gunther:2011jq}
\bibinfo{author}{S.~G\"{u}nther}, \bibinfo{author}{S.~D{\"a}nhardt},
  \bibinfo{author}{B.~Wang}, \bibinfo{author}{M.~L. Bocquet},
  \bibinfo{author}{S.~Schmitt}, \bibinfo{author}{J.~Wintterlin},
\newblock \bibinfo{title}{{Single Terrace Growth of Graphene on a Metal
  Surface}},
\newblock \bibinfo{journal}{Nano Lett.} \bibinfo{volume}{11}
  (\bibinfo{year}{2011}) \bibinfo{pages}{1895--1900}.
\bibitem[{Li et~al.(2011)Li, Magnuson, Venugopal, Tromp, Hannon, Vogel,
  Colombo, and Ruoff}]{Li2012}
\bibinfo{author}{X.~Li}, \bibinfo{author}{C.~W. Magnuson},
  \bibinfo{author}{A.~Venugopal}, \bibinfo{author}{R.~M. Tromp},
  \bibinfo{author}{J.~B. Hannon}, \bibinfo{author}{E.~M. Vogel},
  \bibinfo{author}{L.~Colombo}, \bibinfo{author}{R.~S. Ruoff},
\newblock \bibinfo{title}{Large-area graphene single crystals grown by
  low-pressure chemical vapor deposition of methane on copper},
\newblock \bibinfo{journal}{J. Am. Chem. Soc.} \bibinfo{volume}{133}
  (\bibinfo{year}{2011}) \bibinfo{pages}{2816--2819}.
\bibitem[{Yan et~al.(2012)Yan, Lin, Peng, Sun, Zhu, Li, Xiang, Samuel,
  Kittrell, and Tour}]{Yan2012}
\bibinfo{author}{Z.~Yan}, \bibinfo{author}{J.~Lin}, \bibinfo{author}{Z.~Peng},
  \bibinfo{author}{Z.~Sun}, \bibinfo{author}{Y.~Zhu}, \bibinfo{author}{L.~Li},
  \bibinfo{author}{C.~Xiang}, \bibinfo{author}{E.~L. Samuel},
  \bibinfo{author}{C.~Kittrell}, \bibinfo{author}{J.~M. Tour},
\newblock \bibinfo{title}{Toward the synthesis of wafer-scale single-crystal
  graphene on copper foils},
\newblock \bibinfo{journal}{ACS Nano} \bibinfo{volume}{6}
  (\bibinfo{year}{2012}) \bibinfo{pages}{9110}.
\bibitem[{Skrdla(2012)}]{Skrdla:2012jt}
\bibinfo{author}{P.~J. Skrdla},
\newblock \bibinfo{title}{{Roles of Nucleation, Denucleation, Coarsening, and
  Aggregation Kinetics in Nanoparticle Preparations and Neurological Disease}},
\newblock \bibinfo{journal}{Langmuir} \bibinfo{volume}{28}
  (\bibinfo{year}{2012}) \bibinfo{pages}{4842--4857}.
\bibitem[{Addou et~al.(2012)Addou, Dahal, Sutter, and Batzill}]{Addou}
\bibinfo{author}{R.~Addou}, \bibinfo{author}{A.~Dahal},
  \bibinfo{author}{P.~Sutter}, \bibinfo{author}{M.~Batzill},
\newblock \bibinfo{title}{Monolayer graphene growth on {N}i(111) by low
  temperature chemical vapour deposition},
\newblock \bibinfo{journal}{Appl. Phys. Lett.} \bibinfo{volume}{100}
  (\bibinfo{year}{2012}) \bibinfo{pages}{021601}.
\bibitem[{Huang et~al.(2012)Huang, Xu, Que, Pan, Gao, Pan, Guo, Wang, Du, and
  Gao}]{Huang2012}
\bibinfo{author}{L.~Huang}, \bibinfo{author}{W.-Y. Xu}, \bibinfo{author}{Y.-D.
  Que}, \bibinfo{author}{Y.~Pan}, \bibinfo{author}{M.~Gao},
  \bibinfo{author}{L.-D. Pan}, \bibinfo{author}{H.-M. Guo},
  \bibinfo{author}{Y.-L. Wang}, \bibinfo{author}{S.-X. Du},
  \bibinfo{author}{H.-J. Gao},
\newblock \bibinfo{title}{The influence of annealing temperature on the
  morphology of graphene islands},
\newblock \bibinfo{journal}{Chin. Phys. B} \bibinfo{volume}{21}
  (\bibinfo{year}{2012}) \bibinfo{pages}{088102}.
\bibitem[{Cui et~al.(2011)Cui, Fu, Zhang, and Bao}]{Cui2011}
\bibinfo{author}{Y.~Cui}, \bibinfo{author}{Q.~Fu}, \bibinfo{author}{H.~Zhang},
  \bibinfo{author}{X.~Bao},
\newblock \bibinfo{title}{{Formation of identical-size graphene nanoclusters on
  Ru(0001)}},
\newblock \bibinfo{journal}{Chem. Commun.} \bibinfo{volume}{47}
  (\bibinfo{year}{2011}) \bibinfo{pages}{1470--1472}.
\bibitem[{Voorhees(1985)}]{Ostwaldripening}
\bibinfo{author}{P.~W. Voorhees},
\newblock \bibinfo{title}{{The theory of Ostwald ripening}},
\newblock \bibinfo{journal}{J. Stat. Phys.} \bibinfo{volume}{38}
  (\bibinfo{year}{1985}) \bibinfo{pages}{231--252}.
\bibitem[{Stoldt et~al.(1999)Stoldt, Jenks, Thiel, Cadilhe, and
  Evans}]{Schmoluchowskiripening}
\bibinfo{author}{C.~R. Stoldt}, \bibinfo{author}{C.~J. Jenks},
  \bibinfo{author}{P.~A. Thiel}, \bibinfo{author}{A.~M. Cadilhe},
  \bibinfo{author}{J.~W. Evans},
\newblock \bibinfo{title}{{Smoluchowski ripening of Ag islands on Ag(100)}},
\newblock \bibinfo{journal}{J. Chem. Phys.} \bibinfo{volume}{111}
  (\bibinfo{year}{1999}) \bibinfo{pages}{5157--5166}.
\bibitem[{Chen et~al.(2010)Chen, Zhu, and Zhanh}]{Chen}
\bibinfo{author}{H.~Chen}, \bibinfo{author}{W.~Zhu},
  \bibinfo{author}{Z.~Zhanh},
\newblock \bibinfo{title}{Contrasting behaviour of carbon nucleation in the
  initial stages of graphene epitaxial growth on stepped metal surfaces},
\newblock \bibinfo{journal}{Phys. Rev. Lett.} \bibinfo{volume}{104}
  (\bibinfo{year}{2010}) \bibinfo{pages}{186101}.
\bibitem[{Fuhrmann et~al.(2005)Fuhrmann, Kinne, Tr{\"a}nkenschuh, Papp, Zhu,
  Denecke, and Steinr{\"u}ck}]{Fuhrmann:2005jr}
\bibinfo{author}{T.~Fuhrmann}, \bibinfo{author}{M.~Kinne},
  \bibinfo{author}{B.~Tr{\"a}nkenschuh}, \bibinfo{author}{C.~Papp},
  \bibinfo{author}{J.~F. Zhu}, \bibinfo{author}{R.~Denecke},
  \bibinfo{author}{H.-P. Steinr{\"u}ck},
\newblock \bibinfo{title}{{Activated adsorption of methane on Pt(111) - an {\em
  in situ} XPS study}},
\newblock \bibinfo{journal}{New J. Phys.} \bibinfo{volume}{7}
  (\bibinfo{year}{2005}) \bibinfo{pages}{107}.
\bibitem[{Andersin et~al.(2009)Andersin, Lopez, and Honkala}]{Andersin:2009kd}
\bibinfo{author}{J.~Andersin}, \bibinfo{author}{N.~Lopez},
  \bibinfo{author}{K.~Honkala},
\newblock \bibinfo{title}{{DFT Study on the Complex Reaction Networks in the
  Conversion of Ethylene to Ethylidyne on Flat and Stepped Pd}},
\newblock \bibinfo{journal}{J. Phys. Chem. C} \bibinfo{volume}{113}
  (\bibinfo{year}{2009}) \bibinfo{pages}{8278--8286}.
\bibitem[{Li et~al.(2011)Li, Wu, Wang, Fan, Zhang, Zhai, Zeng, Li, Yang, and
  Hou}]{Zhanchengdoi:10.1021/nn200854p}
\bibinfo{author}{Z.~Li}, \bibinfo{author}{P.~Wu}, \bibinfo{author}{C.~Wang},
  \bibinfo{author}{X.~Fan}, \bibinfo{author}{W.~Zhang},
  \bibinfo{author}{X.~Zhai}, \bibinfo{author}{C.~Zeng},
  \bibinfo{author}{Z.~Li}, \bibinfo{author}{J.~Yang}, \bibinfo{author}{J.~Hou},
\newblock \bibinfo{title}{Low-temperature growth of graphene by chemical vapor
  deposition using solid and liquid carbon sources},
\newblock \bibinfo{journal}{ACS Nano} \bibinfo{volume}{5}
  (\bibinfo{year}{2011}) \bibinfo{pages}{3385--3390}.
\bibitem[{Zhang et~al.(2010)Zhang, Gomez, Ishikawa, Madaria, Ryu, Wang,
  Badmaev, and Zhou}]{Zhang10}
\bibinfo{author}{Y.~Zhang}, \bibinfo{author}{L.~Gomez}, \bibinfo{author}{F.~N.
  Ishikawa}, \bibinfo{author}{A.~Madaria}, \bibinfo{author}{K.~Ryu},
  \bibinfo{author}{C.~Wang}, \bibinfo{author}{A.~Badmaev},
  \bibinfo{author}{C.~Zhou},
\newblock \bibinfo{title}{Comparison of graphene growth on single-crystalline
  and polycrystalline {Ni} by chemical vapor deposition},
\newblock \bibinfo{journal}{J.Phys. Chem. Lett.} \bibinfo{volume}{1}
  (\bibinfo{year}{2010}) \bibinfo{pages}{3101--3107}.
\bibitem[{McCarty et~al.(2009)McCarty, Feibelman, Loginova, and
  Bartelt}]{McCarty}
\bibinfo{author}{K.~F. McCarty}, \bibinfo{author}{F.~J. Feibelman},
  \bibinfo{author}{E.~Loginova}, \bibinfo{author}{N.~C. Bartelt},
\newblock \bibinfo{title}{Kinetics and thermodynamics of carbon segregation and
  graphene growth on {Ru(0001)}},
\newblock \bibinfo{journal}{Carbon} \bibinfo{volume}{47} (\bibinfo{year}{2009})
  \bibinfo{pages}{1806--1813}.
\bibitem[{Nie et~al.(2011)Nie, Walter, Bartelt, Starodub, Bostwick, Rotenberg,
  and McCarty}]{Nie2011}
\bibinfo{author}{S.~Nie}, \bibinfo{author}{A.~L. Walter},
  \bibinfo{author}{N.~C. Bartelt}, \bibinfo{author}{E.~Starodub},
  \bibinfo{author}{A.~Bostwick}, \bibinfo{author}{E.~Rotenberg},
  \bibinfo{author}{K.~F. McCarty},
\newblock \bibinfo{title}{Growth from below: Graphene bilayers on {Ir}(111)},
\newblock \bibinfo{journal}{ACS Nano} \bibinfo{volume}{5}
  (\bibinfo{year}{2011}) \bibinfo{pages}{2298--2306}.
\bibitem[{Su et~al.(2011)Su, Lu, Wu, Li, Liu, Zhang, Lin, Juang, Zhong, Chen,
  and Li}]{Su2011}
\bibinfo{author}{C.~Su}, \bibinfo{author}{A.~Lu}, \bibinfo{author}{C.~Wu},
  \bibinfo{author}{Y.~Li}, \bibinfo{author}{K.~Liu},
  \bibinfo{author}{W.~Zhang}, \bibinfo{author}{S.~Lin},
  \bibinfo{author}{Z.~Juang}, \bibinfo{author}{Y.~Zhong},
  \bibinfo{author}{F.~Chen}, \bibinfo{author}{L.~Li},
\newblock \bibinfo{title}{Direct formation of wafer scale graphene thin layers
  on insulating substrates by chemical vapor deposition},
\newblock \bibinfo{journal}{Nano Lett.} \bibinfo{volume}{11}
  (\bibinfo{year}{2011}) \bibinfo{pages}{3612--3616}.
\bibitem[{Odahara et~al.(2011)Odahara, Otani, Oshima, Suzuki, Yasue, and
  Koshikawa}]{Odahara}
\bibinfo{author}{G.~Odahara}, \bibinfo{author}{S.~Otani},
  \bibinfo{author}{C.~Oshima}, \bibinfo{author}{M.~Suzuki},
  \bibinfo{author}{T.~Yasue}, \bibinfo{author}{T.~Koshikawa},
\newblock \bibinfo{title}{In-situ observation of graphene growth on {N}i(111)},
\newblock \bibinfo{journal}{Surf. Sci.} \bibinfo{volume}{605}
  (\bibinfo{year}{2011}) \bibinfo{pages}{1095 -- 1098}.
\bibitem[{Lahiri et~al.(2011)Lahiri, Miller, Ross, Adamska, Oleynik, and
  Batzill}]{Lahiri11}
\bibinfo{author}{J.~Lahiri}, \bibinfo{author}{T.~S. Miller},
  \bibinfo{author}{A.~J. Ross}, \bibinfo{author}{L.~Adamska},
  \bibinfo{author}{I.~I. Oleynik}, \bibinfo{author}{M.~Batzill},
\newblock \bibinfo{title}{Graphene growth and stability at nickel surfaces},
\newblock \bibinfo{journal}{New J. Phys.} \bibinfo{volume}{13}
  (\bibinfo{year}{2011}) \bibinfo{pages}{025001}.
\bibitem[{de~Heer et~al.(2007)de~Heer, Berger, Wu, First, Conrad, Li, Li,
  Sprinkle, Hass, Sadowski, Potemski, and Martinez}]{deHeer2007}
\bibinfo{author}{W.~A. de~Heer}, \bibinfo{author}{C.~Berger},
  \bibinfo{author}{X.~Wu}, \bibinfo{author}{P.~N. First},
  \bibinfo{author}{E.~H. Conrad}, \bibinfo{author}{X.~Li},
  \bibinfo{author}{T.~Li}, \bibinfo{author}{M.~Sprinkle},
  \bibinfo{author}{J.~Hass}, \bibinfo{author}{M.~L. Sadowski},
  \bibinfo{author}{M.~Potemski}, \bibinfo{author}{G.~Martinez},
\newblock \bibinfo{title}{Epitaxial graphene},
\newblock \bibinfo{journal}{Solid State Commun.} \bibinfo{volume}{143}
  (\bibinfo{year}{2007}) \bibinfo{pages}{92}.
\bibitem[{Virojanadara et~al.(2008)Virojanadara, Syv\"{a}jarvi, Yakimova,
  Johansson, Zakharov, and Balasubramanian}]{Virojanadar2008}
\bibinfo{author}{C.~Virojanadara}, \bibinfo{author}{M.~Syv\"{a}jarvi},
  \bibinfo{author}{R.~Yakimova}, \bibinfo{author}{L.~I. Johansson},
  \bibinfo{author}{A.~A. Zakharov}, \bibinfo{author}{T.~Balasubramanian},
\newblock \bibinfo{title}{{Homogeneous large-area graphene layer growth on
  6H-SiC(0001)}},
\newblock \bibinfo{journal}{Phys. Rev. B} \bibinfo{volume}{78}
  (\bibinfo{year}{2008}) \bibinfo{pages}{245403}.
\bibitem[{Tromp and Hannon(2009)}]{Tromp2009}
\bibinfo{author}{R.~M. Tromp}, \bibinfo{author}{J.~B. Hannon},
\newblock \bibinfo{title}{{Thermodynamics and Kinetics of Graphene Growth on
  SiC(0001)}},
\newblock \bibinfo{journal}{Phys. Rev. Lett.} \bibinfo{volume}{102}
  (\bibinfo{year}{2009}) \bibinfo{pages}{106104}.
\bibitem[{Emtsev et~al.(2009)Emtsev, Bostwick, Horn, Jobst, Kellogg, Ley,
  McChesney, Ohta, Reshanov, and R{\"o}hrl}]{Emtsev:2009wd}
\bibinfo{author}{K.~V. Emtsev}, \bibinfo{author}{A.~Bostwick},
  \bibinfo{author}{K.~Horn}, \bibinfo{author}{J.~Jobst}, \bibinfo{author}{G.~L.
  Kellogg}, \bibinfo{author}{L.~Ley}, \bibinfo{author}{J.~L. McChesney},
  \bibinfo{author}{T.~Ohta}, \bibinfo{author}{S.~A. Reshanov},
  \bibinfo{author}{J.~R{\"o}hrl},
\newblock \bibinfo{title}{{Towards wafer-size graphene layers by atmospheric
  pressure graphitization of silicon carbide}},
\newblock \bibinfo{journal}{Nat. Mater.} \bibinfo{volume}{8}
  (\bibinfo{year}{2009}) \bibinfo{pages}{203--207}.
\bibitem[{Ming(2011)}]{ming2011thesis}
\bibinfo{author}{F.~Ming}, \bibinfo{title}{Theoretical studies of the epitaxial
  growth of graphene}, Ph.D. thesis, Georgia Institute of Technology,
  \bibinfo{year}{2011}.
\bibitem[{Forbeaux et~al.(2000)Forbeaux, Themlin, Charrier, Thibaudau, and
  Debever}]{Forbeaux2000}
\bibinfo{author}{I.~Forbeaux}, \bibinfo{author}{J.-M. Themlin},
  \bibinfo{author}{A.~Charrier}, \bibinfo{author}{F.~Thibaudau},
  \bibinfo{author}{J.-M. Debever},
\newblock \bibinfo{title}{Solid-state graphitization mechanisms of silicon
  carbide {6H-SiC} polar faces},
\newblock \bibinfo{journal}{Appl. Surf. Sci.} \bibinfo{volume}{162-163}
  (\bibinfo{year}{2000}) \bibinfo{pages}{406--412}.
\bibitem[{Muehlhoff et~al.(1986)Muehlhoff, Choyke, Bozack, and John
  T.~Yates}]{muehlhoff:2842}
\bibinfo{author}{L.~Muehlhoff}, \bibinfo{author}{W.~J. Choyke},
  \bibinfo{author}{M.~J. Bozack}, \bibinfo{author}{J.~John T.~Yates},
\newblock \bibinfo{title}{{Comparative electron spectroscopic studies of
  surface segregation on SiC(0001) and SiC(000$\bar{1}$)}},
\newblock \bibinfo{journal}{J. Appl. Phys.} \bibinfo{volume}{60}
  (\bibinfo{year}{1986}) \bibinfo{pages}{2842--2853}.
\bibitem[{Sprinkle et~al.(2010)Sprinkle, Ruan, Hu, Hankinson, Rubio-Roy,
  B.~Zhang1, Berger, and de~Heer}]{Sprinkle2010}
\bibinfo{author}{M.~Sprinkle}, \bibinfo{author}{M.~Ruan},
  \bibinfo{author}{Y.~Hu}, \bibinfo{author}{J.~Hankinson},
  \bibinfo{author}{M.~Rubio-Roy}, \bibinfo{author}{X.~W. B.~Zhang1},
  \bibinfo{author}{C.~Berger}, \bibinfo{author}{W.~A. de~Heer},
\newblock \bibinfo{title}{Scalable templated growth of graphene nanoribbons on
  {SiC}},
\newblock \bibinfo{journal}{Nature Nanotech.} \bibinfo{volume}{5}
  (\bibinfo{year}{2010}) \bibinfo{pages}{727}.
\bibitem[{Li and Tsong(1996)}]{Li1996141}
\bibinfo{author}{L.~Li}, \bibinfo{author}{I.~Tsong},
\newblock \bibinfo{title}{{Atomic structures of 6H-SiC(0001) and (000$\bar{1}$)
  surfaces}},
\newblock \bibinfo{journal}{Surf. Sci.} \bibinfo{volume}{351}
  (\bibinfo{year}{1996}) \bibinfo{pages}{141--148}.
\bibitem[{Kaplan(1989)}]{Kaplan1989111}
\bibinfo{author}{R.~Kaplan},
\newblock \bibinfo{title}{{Surface structure and composition of $\beta$- and
  6H-SiC}},
\newblock \bibinfo{journal}{Surf. Sci.} \bibinfo{volume}{215}
  (\bibinfo{year}{1989}) \bibinfo{pages}{111--134}.
\bibitem[{Ohta et~al.(2010)Ohta, Bartelt, Nie, Th\"{u}rmer, and
  Kellogg}]{Ohta2010}
\bibinfo{author}{T.~Ohta}, \bibinfo{author}{N.~C. Bartelt},
  \bibinfo{author}{S.~Nie}, \bibinfo{author}{K.~Th\"{u}rmer},
  \bibinfo{author}{G.~L. Kellogg},
\newblock \bibinfo{title}{Role of carbon surface diffusion on the growth of
  epitaxial graphene on {SiC}},
\newblock \bibinfo{journal}{Phys. Rev. B} \bibinfo{volume}{81}
  (\bibinfo{year}{2010}) \bibinfo{pages}{121411(R)}.
\bibitem[{Lauffer et~al.(2008)Lauffer, Emtsev, Graupner, Seyller, Ley,
  Reshanov, and Weber}]{Lauffer2008}
\bibinfo{author}{P.~Lauffer}, \bibinfo{author}{K.~V. Emtsev},
  \bibinfo{author}{R.~Graupner}, \bibinfo{author}{T.~Seyller},
  \bibinfo{author}{L.~Ley}, \bibinfo{author}{S.~A. Reshanov},
  \bibinfo{author}{H.~B. Weber},
\newblock \bibinfo{title}{Atomic and electronic structure of few-layer graphene
  on {SiC(0001)} studied with scanning tunneling microscopy and spectroscopy},
\newblock \bibinfo{journal}{Phys. Rev. B} \bibinfo{volume}{77}
  (\bibinfo{year}{2008}) \bibinfo{pages}{155426}.
\bibitem[{Mattausch and Pankratov(2007)}]{Mattausch2007}
\bibinfo{author}{A.~Mattausch}, \bibinfo{author}{O.~Pankratov},
\newblock \bibinfo{title}{\textit{Ab~Initio} study of graphene on {SiC}},
\newblock \bibinfo{journal}{Phys. Rev. Lett.} \bibinfo{volume}{99}
  (\bibinfo{year}{2007}) \bibinfo{pages}{076802}.
\bibitem[{Borovikov and Zangwill(2009)}]{Borovikov2009}
\bibinfo{author}{V.~Borovikov}, \bibinfo{author}{A.~Zangwill},
\newblock \bibinfo{title}{Step-edge instability during epitaxial growth of
  graphene from {SiC(0001)}},
\newblock \bibinfo{journal}{Phys. Rev. B} \bibinfo{volume}{80}
  (\bibinfo{year}{2009}) \bibinfo{pages}{121406}.
\bibitem[{V\'azquez~de Parga et~al.(2008)V\'azquez~de Parga, Calleja, Borca,
  Passeggi, Hinarejos, Guinea, and Miranda}]{Vazquez2008}
\bibinfo{author}{A.~L. V\'azquez~de Parga}, \bibinfo{author}{F.~Calleja},
  \bibinfo{author}{B.~Borca}, \bibinfo{author}{M.~C.~G. Passeggi},
  \bibinfo{author}{J.~J. Hinarejos}, \bibinfo{author}{F.~Guinea},
  \bibinfo{author}{R.~Miranda},
\newblock \bibinfo{title}{Periodically rippled graphene: Growth and spatially
  resolved electronic structure},
\newblock \bibinfo{journal}{Phys. Rev. Lett.} \bibinfo{volume}{100}
  (\bibinfo{year}{2008}) \bibinfo{pages}{056807}.
\bibitem[{Martoccia et~al.(2010)Martoccia, Bj{\"o}rck, Schlep{\"u}tz, Brugger,
  nad B.~D.~Patterson, Greber, and Willmott}]{Martoccia10}
\bibinfo{author}{D.~Martoccia}, \bibinfo{author}{M.~Bj{\"o}rck},
  \bibinfo{author}{C.~M. Schlep{\"u}tz}, \bibinfo{author}{T.~Brugger},
  \bibinfo{author}{S.~A.~P. nad B.~D.~Patterson}, \bibinfo{author}{T.~Greber},
  \bibinfo{author}{P.~R. Willmott},
\newblock \bibinfo{title}{{Graphene on Ru(0001): a corrugated and chiral
  structure}},
\newblock \bibinfo{journal}{New J. Phys.} \bibinfo{volume}{12}
  (\bibinfo{year}{2010}) \bibinfo{pages}{043028}.
\bibitem[{Borca et~al.(2010)Borca, Barja, Garnica, Minniti, Politano,
  Rodriguez-Garcia, Hinarejos, Farias, de~Parga, and Miranda}]{Borca2010}
\bibinfo{author}{B.~Borca}, \bibinfo{author}{S.~Barja},
  \bibinfo{author}{M.~Garnica}, \bibinfo{author}{M.~Minniti},
  \bibinfo{author}{A.~Politano}, \bibinfo{author}{J.~M. Rodriguez-Garcia},
  \bibinfo{author}{J.~J. Hinarejos}, \bibinfo{author}{D.~Farias},
  \bibinfo{author}{A.~L.~V. de~Parga}, \bibinfo{author}{R.~Miranda},
\newblock \bibinfo{title}{{Electronic and geometric corrugation of periodically
  rippled, Self-nanostructured graphene epitaxially grown on Ru(0001)}},
\newblock \bibinfo{journal}{New J. Phys.} \bibinfo{volume}{12}
  (\bibinfo{year}{2010}) \bibinfo{pages}{093018}.
\bibitem[{Wang et~al.(2008)Wang, Bocquet, Marchini, G\"{u}nther, and
  Wintterlin}]{WangRuB801785A}
\bibinfo{author}{B.~Wang}, \bibinfo{author}{M.-L. Bocquet},
  \bibinfo{author}{S.~Marchini}, \bibinfo{author}{S.~G\"{u}nther},
  \bibinfo{author}{J.~Wintterlin},
\newblock \bibinfo{title}{{Chemical origin of a graphene moir\'{e} overlayer on
  Ru(0001)}},
\newblock \bibinfo{journal}{Phys. Chem. Chem. Phys.} \bibinfo{volume}{10}
  (\bibinfo{year}{2008}) \bibinfo{pages}{3530--3534}.
\bibitem[{Meng et~al.(2012)Meng, Wu, Zhang, Li, Du, Wang, and Gao}]{Meng2012}
\bibinfo{author}{L.~Meng}, \bibinfo{author}{R.~Wu}, \bibinfo{author}{L.~Zhang},
  \bibinfo{author}{L.~Li}, \bibinfo{author}{S.~Du}, \bibinfo{author}{Y.~Wang},
  \bibinfo{author}{H.-J. Gao},
\newblock \bibinfo{title}{{Multi-oriented moir\'{e} superstructures of graphene
  on Ir(111): experimental observations and theoretical models}},
\newblock \bibinfo{journal}{J. Phys.: Condens. Matter} \bibinfo{volume}{24}
  (\bibinfo{year}{2012}) \bibinfo{pages}{314214}.
\bibitem[{Gao et~al.(2011)Gao, Pan, Huang, Hu, Zhang, Guo, Du, and
  Gao}]{Gao2011}
\bibinfo{author}{M.~Gao}, \bibinfo{author}{Y.~Pan}, \bibinfo{author}{L.~Huang},
  \bibinfo{author}{H.~Hu}, \bibinfo{author}{L.~Z. Zhang},
  \bibinfo{author}{H.~M. Guo}, \bibinfo{author}{S.~X. Du},
  \bibinfo{author}{H.-J. Gao},
\newblock \bibinfo{title}{{Epitaxial growth and structural property of graphene
  on Pt(111)}},
\newblock \bibinfo{journal}{Appl. Phys. Lett.} \bibinfo{volume}{98}
  (\bibinfo{year}{2011}) \bibinfo{pages}{033101}.
\bibitem[{Merino et~al.(2011)Merino, Svec, Pinardi, Otero, and
  Martin-Gago}]{Merino2011}
\bibinfo{author}{P.~Merino}, \bibinfo{author}{M.~Svec}, \bibinfo{author}{A.~L.
  Pinardi}, \bibinfo{author}{G.~Otero}, \bibinfo{author}{J.~A. Martin-Gago},
\newblock \bibinfo{title}{Strain-driven moir\'{e} superstructures of epitaxial
  graphene on transition metal surfaces},
\newblock \bibinfo{journal}{ACS Nano} \bibinfo{volume}{5}
  (\bibinfo{year}{2011}) \bibinfo{pages}{5627--5634}.
\bibitem[{Rosei et~al.(1983)Rosei, De~Crescenzi, Sette, Quaresima, Savoia, and
  Perfetti}]{rosei1983}
\bibinfo{author}{R.~Rosei}, \bibinfo{author}{M.~De~Crescenzi},
  \bibinfo{author}{F.~Sette}, \bibinfo{author}{C.~Quaresima},
  \bibinfo{author}{A.~Savoia}, \bibinfo{author}{P.~Perfetti},
\newblock \bibinfo{title}{Structure of graphitic carbon on ni (111): A surface
  extended-energy-loss fine-structure study},
\newblock \bibinfo{journal}{Phys. Rev. B} \bibinfo{volume}{28}
  (\bibinfo{year}{1983}) \bibinfo{pages}{1161}.
\bibitem[{Zhao et~al.(2011)Zhao, Kozlov, H\"{o}fert, Gotterbarm, Lorenz,
  Vi{\~n}es, Papp, G\"{o}rling, and Steinr\"{u}ck}]{Zhao2011}
\bibinfo{author}{W.~Zhao}, \bibinfo{author}{S.~M. Kozlov},
  \bibinfo{author}{O.~H\"{o}fert}, \bibinfo{author}{K.~Gotterbarm},
  \bibinfo{author}{M.~P.~A. Lorenz}, \bibinfo{author}{F.~Vi{\~n}es},
  \bibinfo{author}{C.~Papp}, \bibinfo{author}{A.~G\"{o}rling},
  \bibinfo{author}{H.-P. Steinr\"{u}ck},
\newblock \bibinfo{title}{Graphene on {N}i(111): Coexistence of different
  surface structures},
\newblock \bibinfo{journal}{J. Phys. Chem. Lett.} \bibinfo{volume}{2}
  (\bibinfo{year}{2011}) \bibinfo{pages}{759 -- 764}.
\bibitem[{Dahal et~al.(2012)Dahal, Addou, Sutter, and Batzill}]{Dahal}
\bibinfo{author}{A.~Dahal}, \bibinfo{author}{R.~Addou},
  \bibinfo{author}{P.~Sutter}, \bibinfo{author}{M.~Batzill},
\newblock \bibinfo{title}{Graphene monolayer rotation on {N}i(111) facilitates
  bilayer graphene growth},
\newblock \bibinfo{journal}{Appl. Phys. Lett.} \bibinfo{volume}{100}
  (\bibinfo{year}{2012}) \bibinfo{pages}{241602}.
\bibitem[{Murata et~al.(2010)Murata, Starodub, Kappes, Ciobanu, Bartelt,
  McCarty, and Kodambaka}]{Murata2010}
\bibinfo{author}{Y.~Murata}, \bibinfo{author}{E.~Starodub},
  \bibinfo{author}{B.~Kappes}, \bibinfo{author}{C.~Ciobanu},
  \bibinfo{author}{N.~Bartelt}, \bibinfo{author}{K.~McCarty},
  \bibinfo{author}{S.~Kodambaka},
\newblock \bibinfo{title}{Orientation-dependent work function of graphene on pd
  (111)},
\newblock \bibinfo{journal}{Applied Physics Letters} \bibinfo{volume}{97}
  (\bibinfo{year}{2010}) \bibinfo{pages}{143114}.
\bibitem[{Mittendorfer et~al.(2011)Mittendorfer, Garhofer, Redinger, Klime,
  Harl, and Kresse}]{Mittendorfer2011}
\bibinfo{author}{F.~Mittendorfer}, \bibinfo{author}{A.~Garhofer},
  \bibinfo{author}{J.~Redinger}, \bibinfo{author}{J.~Klime},
  \bibinfo{author}{J.~Harl}, \bibinfo{author}{G.~Kresse},
\newblock \bibinfo{title}{Graphene on {N}i(111): Strong interaction and weak
  adsorption},
\newblock \bibinfo{journal}{Phys. Rev. B} \bibinfo{volume}{84}
  (\bibinfo{year}{2011}) \bibinfo{pages}{201401(R)}.
\bibitem[{Olsen et~al.(2011)Olsen, Yan, Mortensen, and
  Thygesen}]{Olsen-PRL-2011}
\bibinfo{author}{T.~Olsen}, \bibinfo{author}{J.~Yan}, \bibinfo{author}{J.~J.
  Mortensen}, \bibinfo{author}{K.~S. Thygesen},
\newblock \bibinfo{title}{Dispersive and covalent interactions between graphene
  and metal surfaces from the random phase approximation},
\newblock \bibinfo{journal}{Phys. Rev. Lett.} \bibinfo{volume}{107}
  (\bibinfo{year}{2011}) \bibinfo{pages}{156401}.
\bibitem[{Zhang et~al.(2009)Zhang, Fu, Cui, Tan, and Bao}]{Zhang2009}
\bibinfo{author}{H.~Zhang}, \bibinfo{author}{Q.~Fu}, \bibinfo{author}{Y.~Cui},
  \bibinfo{author}{D.~Tan}, \bibinfo{author}{X.~Bao},
\newblock \bibinfo{title}{{Growth Mechanism of Graphene on Ru(0001) and O$_2$
  Adsorption on the Graphene/Ru(0001) Surface}},
\newblock \bibinfo{journal}{J. Phys. Chem. C} \bibinfo{volume}{113}
  (\bibinfo{year}{2009}) \bibinfo{pages}{8296--8301}.
\bibitem[{Man and Altman(2011)}]{Man2011}
\bibinfo{author}{K.~L. Man}, \bibinfo{author}{M.~S. Altman},
\newblock \bibinfo{title}{Small-angle lattice rotations in graphene on {Ru}
  (0001)},
\newblock \bibinfo{journal}{Phys. Rev. B} \bibinfo{volume}{84}
  (\bibinfo{year}{2011}) \bibinfo{pages}{235415}.
\bibitem[{Wang et~al.(2010)Wang, Caffio, Bromley, Fru{\"u}chtl, and
  Schaub}]{Wang-Schaub-NanoLett-2010}
\bibinfo{author}{B.~Wang}, \bibinfo{author}{M.~Caffio},
  \bibinfo{author}{C.~Bromley}, \bibinfo{author}{H.~Fru{\"u}chtl},
  \bibinfo{author}{R.~Schaub},
\newblock \bibinfo{title}{Coupling epitaxy, chemical bonding, and work function
  at the local scale in transition metal-supported graphene},
\newblock \bibinfo{journal}{ACS Nano} \bibinfo{volume}{4}
  (\bibinfo{year}{2010}) \bibinfo{pages}{5773}.
\bibitem[{N'Diaye et~al.(2008)N'Diaye, Coraux, Plasa, Busse, and
  Michely}]{Diaye2008}
\bibinfo{author}{A.~T. N'Diaye}, \bibinfo{author}{J.~Coraux},
  \bibinfo{author}{T.~N. Plasa}, \bibinfo{author}{C.~Busse},
  \bibinfo{author}{T.~Michely},
\newblock \bibinfo{title}{{Structure of epitaxial graphene on Ir(111)}},
\newblock \bibinfo{journal}{New J. Phys.} \bibinfo{volume}{10}
  (\bibinfo{year}{2008}) \bibinfo{pages}{043033}.
\bibitem[{Bertoni et~al.(2005)Bertoni, Calmels, Altibelli, and
  Serin}]{Bertoni2005}
\bibinfo{author}{G.~Bertoni}, \bibinfo{author}{L.~Calmels},
  \bibinfo{author}{A.~Altibelli}, \bibinfo{author}{V.~Serin},
\newblock \bibinfo{title}{First-principles calculation of the electronic
  structure and {EELS} spectra at the graphene/{Ni}(111) interface},
\newblock \bibinfo{journal}{Phys. Rev. B} \bibinfo{volume}{71}
  (\bibinfo{year}{2005}) \bibinfo{pages}{075402}.
\bibitem[{Kalibaeva et~al.(2006)Kalibaeva, Vuilleumier, Meloni, Alavi,
  Ciccotti, and Rosei}]{Kalibaeva2006}
\bibinfo{author}{G.~Kalibaeva}, \bibinfo{author}{R.~Vuilleumier},
  \bibinfo{author}{S.~Meloni}, \bibinfo{author}{A.~Alavi},
  \bibinfo{author}{G.~Ciccotti}, \bibinfo{author}{R.~Rosei},
\newblock \bibinfo{title}{Ab initio simulation of carbon clustering on an {N}i
  (111) surface: a model of the poisoning of nickel-based catalysts},
\newblock \bibinfo{journal}{J. Phys. Chem. B} \bibinfo{volume}{110}
  (\bibinfo{year}{2006}) \bibinfo{pages}{3638--3646}.
\bibitem[{Giovannetti et~al.(2008)Giovannetti, Khomyakov, Brocks, Karpan,
  Van~den Brink, and Kelly}]{Giovanetti2008}
\bibinfo{author}{G.~Giovannetti}, \bibinfo{author}{P.~A. Khomyakov},
  \bibinfo{author}{G.~Brocks}, \bibinfo{author}{V.~M. Karpan},
  \bibinfo{author}{J.~Van~den Brink}, \bibinfo{author}{P.~J. Kelly},
\newblock \bibinfo{title}{Doping graphene with metal contacts},
\newblock \bibinfo{journal}{Phys. Rev. Lett.} \bibinfo{volume}{101}
  (\bibinfo{year}{2008}) \bibinfo{pages}{026803}.
\bibitem[{Fuentes-Cabrera et~al.(2008)Fuentes-Cabrera, Baskes, Melechko, and
  Simpson}]{Fuentes2008}
\bibinfo{author}{M.~Fuentes-Cabrera}, \bibinfo{author}{M.~I. Baskes},
  \bibinfo{author}{A.~V. Melechko}, \bibinfo{author}{M.~L. Simpson},
\newblock \bibinfo{title}{Bridge structure for the graphene/{N}i(111) system: a
  first principles study},
\newblock \bibinfo{journal}{Phys. Rev. B} \bibinfo{volume}{77}
  (\bibinfo{year}{2008}) \bibinfo{pages}{035405}.
\bibitem[{Vanin et~al.(2010)Vanin, Mortensen, Kelkkanen, Garcia-Lastra,
  Thygesen, and Jacobsen}]{Vanin2010}
\bibinfo{author}{M.~Vanin}, \bibinfo{author}{J.~J. Mortensen},
  \bibinfo{author}{A.~K. Kelkkanen}, \bibinfo{author}{J.~M. Garcia-Lastra},
  \bibinfo{author}{K.~S. Thygesen}, \bibinfo{author}{K.~W. Jacobsen},
\newblock \bibinfo{title}{Graphene on metals: A van der waals density
  functional study},
\newblock \bibinfo{journal}{Phys. Rev. B} \bibinfo{volume}{81}
  (\bibinfo{year}{2010}) \bibinfo{pages}{081408}.
\bibitem[{Rasool et~al.(2010)Rasool, Song, Allen, Wassei, Kaner, Wang, Weiller,
  and Gimzewski}]{Rasool2010}
\bibinfo{author}{H.~I. Rasool}, \bibinfo{author}{E.~B. Song},
  \bibinfo{author}{M.~J. Allen}, \bibinfo{author}{J.~K. Wassei},
  \bibinfo{author}{R.~B. Kaner}, \bibinfo{author}{K.~L. Wang},
  \bibinfo{author}{B.~H. Weiller}, \bibinfo{author}{J.~K. Gimzewski},
\newblock \bibinfo{title}{Continuity of graphene on polycrystalline copper},
\newblock \bibinfo{journal}{Nano Lett.} \bibinfo{volume}{11}
  (\bibinfo{year}{2010}) \bibinfo{pages}{251--256}.
\bibitem[{Inoue et~al.(2012)Inoue, Kageshima, Kangawa, and
  Kakimoto}]{Inoue:2012ih}
\bibinfo{author}{M.~Inoue}, \bibinfo{author}{H.~Kageshima},
  \bibinfo{author}{Y.~Kangawa}, \bibinfo{author}{K.~Kakimoto},
\newblock \bibinfo{title}{{First-principles calculation of $0^{th}$-layer
  graphene-like growth of C on SiC(0001)}},
\newblock \bibinfo{journal}{Phys. Rev. B} \bibinfo{volume}{86}
  (\bibinfo{year}{2012}) \bibinfo{pages}{085417}.
\bibitem[{Kim et~al.(2008)Kim, Ihm, Choi, and Son}]{Kim}
\bibinfo{author}{S.~Kim}, \bibinfo{author}{J.~Ihm}, \bibinfo{author}{H.~J.
  Choi}, \bibinfo{author}{Y.-W. Son},
\newblock \bibinfo{title}{Origin of anomalous electronic structures of
  epitaxial graphene on silicon carbide},
\newblock \bibinfo{journal}{Phys. Rev. Lett.} \bibinfo{volume}{100}
  (\bibinfo{year}{2008}) \bibinfo{pages}{176802}.
\bibitem[{Varchon et~al.(2007)Varchon, Feng, Hass, Li, Nguyen, Naud, Mallet,
  Veuillen, Berger, Conrad, and Magaud}]{Varchon07}
\bibinfo{author}{F.~Varchon}, \bibinfo{author}{R.~Feng},
  \bibinfo{author}{J.~Hass}, \bibinfo{author}{X.~Li}, \bibinfo{author}{B.~N.
  Nguyen}, \bibinfo{author}{C.~Naud}, \bibinfo{author}{P.~Mallet},
  \bibinfo{author}{J.-Y. Veuillen}, \bibinfo{author}{C.~Berger},
  \bibinfo{author}{E.~H. Conrad}, \bibinfo{author}{L.~Magaud},
\newblock \bibinfo{title}{Electronic structure of epitaxial graphene layers on
  {SiC}: Effect of the substrate},
\newblock \bibinfo{journal}{Phys. Rev. Lett.} \bibinfo{volume}{99}
  (\bibinfo{year}{2007}) \bibinfo{pages}{126805}.
\bibitem[{Lampin et~al.(2010)Lampin, Priester, Krzeminski, and Magaud}]{Lampin}
\bibinfo{author}{E.~Lampin}, \bibinfo{author}{C.~Priester},
  \bibinfo{author}{C.~Krzeminski}, \bibinfo{author}{L.~Magaud},
\newblock \bibinfo{title}{Graphene buffer layer on {Si-terminated SiC} studied
  with an empirical interatomic potential},
\newblock \bibinfo{journal}{J. Appl. Phys.} \bibinfo{volume}{107}
  (\bibinfo{year}{2010}) \bibinfo{pages}{103514}.
\bibitem[{Tang et~al.(2008{\natexlab{a}})Tang, Meng, Sun, Zhang, and
  Zhong}]{Tang}
\bibinfo{author}{C.~Tang}, \bibinfo{author}{L.~Meng}, \bibinfo{author}{L.~Sun},
  \bibinfo{author}{K.~Zhang}, \bibinfo{author}{J.~Zhong},
\newblock \bibinfo{title}{Molecular dynamics study of ripples in graphene
  nanoribbons on {6H-SiC(0001)}: Temperature and size effects},
\newblock \bibinfo{journal}{J. Appl. Phys.} \bibinfo{volume}{104}
  (\bibinfo{year}{2008}{\natexlab{a}}) \bibinfo{pages}{113536}.
\bibitem[{Tang et~al.(2008{\natexlab{b}})Tang, Meng, Xiao, and Zhong}]{Tang08}
\bibinfo{author}{C.~Tang}, \bibinfo{author}{L.~Meng},
  \bibinfo{author}{H.~Xiao}, \bibinfo{author}{J.~Zhong},
\newblock \bibinfo{title}{{Growth of graphene structure on 6H-SiC(0001):
  Molecular dynamics simulation}},
\newblock \bibinfo{journal}{J. Appl. Phys.} \bibinfo{volume}{103}
  (\bibinfo{year}{2008}{\natexlab{b}}) \bibinfo{pages}{063505}.
\bibitem[{Jakse et~al.(2011)Jakse, Arifin, and Lai}]{Jakse}
\bibinfo{author}{N.~Jakse}, \bibinfo{author}{R.~Arifin}, \bibinfo{author}{S.~K.
  Lai},
\newblock \bibinfo{title}{Growth of graphene on 6{H}-{S}i{C} by molecular
  dynamics simulation},
\newblock \bibinfo{journal}{Condens. Matt. Phys.} \bibinfo{volume}{14}
  (\bibinfo{year}{2011}) \bibinfo{pages}{43802}.
\bibitem[{Lucas et~al.(2010)Lucas, Bertolus, and Pizzagalli}]{Lucas}
\bibinfo{author}{G.~Lucas}, \bibinfo{author}{M.~Bertolus},
  \bibinfo{author}{L.~Pizzagalli},
\newblock \bibinfo{title}{An environment-dependent interatomic potential for
  silicon carbide: calculation of bulk properties, high-pressure phases, point
  and extended defects, and amorphous structures},
\newblock \bibinfo{journal}{J. Phys.: Condens. Matter} \bibinfo{volume}{22}
  (\bibinfo{year}{2010}) \bibinfo{pages}{035802}.
\bibitem[{Tersoff(1994)}]{Tersoff-potentials-1994}
\bibinfo{author}{J.~Tersoff},
\newblock \bibinfo{title}{Structural properties of a carbon-nanotube crystal},
\newblock \bibinfo{journal}{Phys. Rev. Lett.} \bibinfo{volume}{73}
  (\bibinfo{year}{1994}) \bibinfo{pages}{676}.
  
\bibitem[{Zangwill and Vvedensky(2011)}]{Vvedensky}
\bibinfo{author}{A.~Zangwill}, \bibinfo{author}{D.~Vvedensky},
\newblock \bibinfo{title}{Novel growth mechanism of epitaxial graphene on
  metals},
\newblock \bibinfo{journal}{Nano Lett.} \bibinfo{volume}{11}
  (\bibinfo{year}{2011}) \bibinfo{pages}{2092--2095}.

\bibitem[{Posthuma(2014)}]{Posthuma14}
\bibinfo{author}{J.~Posthuma de Boer}, \bibinfo{author}{I.~Ford},
\bibinfo{author}{L.~Kantorovich}, \bibinfo{author}{D.~Vvedensky},
\newblock \bibinfo{title}{Optimization algorithm for rate equations with an application to epitaxial graphene},
\newblock \bibinfo{journal}{J. Phys. Cond. Matter} \bibinfo{volume}{26}
  (\bibinfo{year}{2014}) \bibinfo{pages}{185008}.

\bibitem[{Gajewski and Pao(2011)}]{gajewski:064707}
\bibinfo{author}{G.~Gajewski}, \bibinfo{author}{C.~W. Pao},
\newblock \bibinfo{title}{{Ab initio calculations of the reaction pathways for
  methane decomposition over the Cu(111) surface}},
\newblock \bibinfo{journal}{J. Chem. Phys.} \bibinfo{volume}{135}
  (\bibinfo{year}{2011}) \bibinfo{pages}{064707}.
\bibitem[{Aleksandrov et~al.(2012)Aleksandrov, Moskaleva, Zhao, Basaran, Chen,
  Mei, and R{\"o}sch}]{Aleksandrov2012187}
\bibinfo{author}{H.~A. Aleksandrov}, \bibinfo{author}{L.~V. Moskaleva},
  \bibinfo{author}{Z.-J. Zhao}, \bibinfo{author}{D.~Basaran},
  \bibinfo{author}{Z.-X. Chen}, \bibinfo{author}{D.~Mei},
  \bibinfo{author}{N.~R{\"o}sch},
\newblock \bibinfo{title}{{Ethylene conversion to ethylidyne on Pd(111) and
  Pt(111): A first-principles-based kinetic Monte Carlo study}},
\newblock \bibinfo{journal}{J. Catal.} \bibinfo{volume}{285}
  (\bibinfo{year}{2012}) \bibinfo{pages}{187 -- 195}.
\bibitem[{Li et~al.(2010)Li, Guo, Jiang, Zhao, Lu, Zhu, Fu, and
  Shan}]{Li:2010cm}
\bibinfo{author}{M.~Li}, \bibinfo{author}{W.~Guo}, \bibinfo{author}{R.~Jiang},
  \bibinfo{author}{L.~Zhao}, \bibinfo{author}{X.~Lu}, \bibinfo{author}{H.~Zhu},
  \bibinfo{author}{D.~Fu}, \bibinfo{author}{H.~Shan},
\newblock \bibinfo{title}{{Mechanism of the Ethylene Conversion to Ethylidyne
  on Rh(111): A Density Functional Investigation}},
\newblock \bibinfo{journal}{J. Phys. Chem. C} \bibinfo{volume}{114}
  (\bibinfo{year}{2010}) \bibinfo{pages}{8440--8448}.
\bibitem[{Chen and Vlachos(2010)}]{Chen:2010ez}
\bibinfo{author}{Y.~Chen}, \bibinfo{author}{D.~G. Vlachos},
\newblock \bibinfo{title}{{Hydrogenation of Ethylene and Dehydrogenation and
  Hydrogenolysis of Ethane on Pt(111) and Pt(211): A Density Functional Theory
  Study}},
\newblock \bibinfo{journal}{J. Phys. Chem. C} \bibinfo{volume}{114}
  (\bibinfo{year}{2010}) \bibinfo{pages}{4973--4982}.
\bibitem[{Saadi et~al.(2010)Saadi, Abild-Pedersen, Helveg, Sehested, Hinnemann,
  Appel, and Norskov}]{Saadi}
\bibinfo{author}{S.~Saadi}, \bibinfo{author}{F.~Abild-Pedersen},
  \bibinfo{author}{S.~Helveg}, \bibinfo{author}{J.~Sehested},
  \bibinfo{author}{B.~Hinnemann}, \bibinfo{author}{C.~C. Appel},
  \bibinfo{author}{J.~K. Norskov},
\newblock \bibinfo{title}{On the role of metal step-edges in graphene growth},
\newblock \bibinfo{journal}{J. Phys. Chem.} \bibinfo{volume}{114}
  (\bibinfo{year}{2010}) \bibinfo{pages}{11221}.
\bibitem[{Wu et~al.(2010)Wu, Zhang, Li, Yang, and Hou}]{Wu2010}
\bibinfo{author}{P.~Wu}, \bibinfo{author}{W.~Zhang}, \bibinfo{author}{Z.~Li},
  \bibinfo{author}{J.~Yang}, \bibinfo{author}{J.~G. Hou},
\newblock \bibinfo{title}{Communication: Coalescence of carbon atoms on
  {C}u(111) surface: Emergence of a stable bridging-metal structure motif},
\newblock \bibinfo{journal}{J. Chem. Phys.} \bibinfo{volume}{133}
  (\bibinfo{year}{2010}) \bibinfo{pages}{071101}.
\bibitem[{Yuan et~al.(2012)Yuan, Gao, Shu, Zhao, Chen, and Ding}]{Yuan}
\bibinfo{author}{Q.~Yuan}, \bibinfo{author}{J.~Gao}, \bibinfo{author}{H.~Shu},
  \bibinfo{author}{J.~Zhao}, \bibinfo{author}{X.~Chen},
  \bibinfo{author}{F.~Ding},
\newblock \bibinfo{title}{Magic carbon clusters in the chemical vapor
  deposition growth of graphene},
\newblock \bibinfo{journal}{J. Am. Chem. Soc.} \bibinfo{volume}{134}
  (\bibinfo{year}{2012}) \bibinfo{pages}{2970}.
\bibitem[{Wu et~al.(2012)Wu, Jiang, Zhang, Li, Hou, and Yang}]{Wu}
\bibinfo{author}{P.~Wu}, \bibinfo{author}{H.~Jiang},
  \bibinfo{author}{W.~Zhang}, \bibinfo{author}{Z.~Li},
  \bibinfo{author}{Z.~Hou}, \bibinfo{author}{J.~Yang},
\newblock \bibinfo{title}{Lattice mismatch induced nonlinear growth of
  graphene},
\newblock \bibinfo{journal}{J. Am. Chem. Soc.} \bibinfo{volume}{134}
  (\bibinfo{year}{2012}) \bibinfo{pages}{6045--6051}.
\bibitem[{Wei and Wang(2012)}]{Wei}
\bibinfo{author}{D.~Wei}, \bibinfo{author}{F.~Wang},
\newblock \bibinfo{title}{Relative stability of armchair, zigzag and reczag
  graphene edges on the {R}u(0001) surface},
\newblock \bibinfo{journal}{Surf. Sci.} \bibinfo{volume}{606}
  (\bibinfo{year}{2012}) \bibinfo{pages}{485 -- 489}.
\bibitem[{Shu et~al.(2012)Shu, Chen, Tao, and Ding}]{Shu}
\bibinfo{author}{H.~Shu}, \bibinfo{author}{X.~Chen}, \bibinfo{author}{X.~Tao},
  \bibinfo{author}{F.~Ding},
\newblock \bibinfo{title}{Edge structural stability and kinetics of graphene
  chemical vapor deposition growth},
\newblock \bibinfo{journal}{ACS Nano} \bibinfo{volume}{6}
  (\bibinfo{year}{2012}) \bibinfo{pages}{3243--3250}.
\bibitem[{Artyukhov et~al.(2012)Artyukhov, Liu, and Yakobson}]{Artyukhov}
\bibinfo{author}{V.~I. Artyukhov}, \bibinfo{author}{Y.~Liu},
  \bibinfo{author}{B.~I. Yakobson},
\newblock \bibinfo{title}{Equilibrium at the edge and atomistic mechanisms of
  graphene growth},
\newblock \bibinfo{journal}{Proc. Nat. Acad. Sci.} \bibinfo{volume}{109}
  (\bibinfo{year}{2012}) \bibinfo{pages}{235502}.
\bibitem[{Yu et~al.(2011)Yu, Jauregui, Wu, Colby, Tian, Su, Cao, Liu, Pandey,
  Wei, {Chung}, {Peng}, {Guisinger}, {Stach}, {Bao}, {Pei}, and
  {Chen}}]{Yu2011}
\bibinfo{author}{Q.~Yu}, \bibinfo{author}{L.~A. Jauregui},
  \bibinfo{author}{W.~Wu}, \bibinfo{author}{R.~Colby},
  \bibinfo{author}{J.~Tian}, \bibinfo{author}{Z.~Su}, \bibinfo{author}{H.~Cao},
  \bibinfo{author}{Z.~Liu}, \bibinfo{author}{D.~Pandey},
  \bibinfo{author}{D.~Wei}, \bibinfo{author}{T.~F. {Chung}},
  \bibinfo{author}{P.~{Peng}}, \bibinfo{author}{N.~P. {Guisinger}},
  \bibinfo{author}{E.~A. {Stach}}, \bibinfo{author}{J.~{Bao}},
  \bibinfo{author}{S.~S. {Pei}}, \bibinfo{author}{Y.~P. {Chen}},
\newblock \bibinfo{title}{Control and characterization of individual grains and
  grain boundaries in graphene grown by chemical vapour deposition},
\newblock \bibinfo{journal}{Nature Mater.} \bibinfo{volume}{10}
  (\bibinfo{year}{2011}) \bibinfo{pages}{443--449}.
\bibitem[{Jauregui et~al.(2011)Jauregui, Cao, Wu, Yu, and Chen}]{Jauregui2011}
\bibinfo{author}{L.~A. Jauregui}, \bibinfo{author}{H.~Cao},
  \bibinfo{author}{W.~Wu}, \bibinfo{author}{Q.~Yu}, \bibinfo{author}{Y.~P.
  Chen},
\newblock \bibinfo{title}{Electronic properties of grains and grain boundaries
  in graphene grown by chemical vapor deposition},
\newblock \bibinfo{journal}{Solid State Commun} \bibinfo{volume}{151}
  (\bibinfo{year}{2011}) \bibinfo{pages}{1100 -- 1104}.
\bibitem[{Tian et~al.(2011)Tian, Cao, Wu, Yu, and Chen}]{Tian11}
\bibinfo{author}{J.~Tian}, \bibinfo{author}{H.~Cao}, \bibinfo{author}{W.~Wu},
  \bibinfo{author}{Q.~Yu}, \bibinfo{author}{Y.~P. Chen},
\newblock \bibinfo{title}{Direct imaging of graphene edges: Atomic structure
  and electronic scattering},
\newblock \bibinfo{journal}{Nano Lett.} \bibinfo{volume}{11}
  (\bibinfo{year}{2011}) \bibinfo{pages}{3663--3668}.
\bibitem[{Liu et~al.(2010)Liu, Dobrinsky, and Yakobson}]{Liu10}
\bibinfo{author}{Y.~Liu}, \bibinfo{author}{A.~Dobrinsky},
  \bibinfo{author}{B.~I. Yakobson},
\newblock \bibinfo{title}{Graphene edge from armchair to zigzag: The origins of
  nanotube chirality?},
\newblock \bibinfo{journal}{Phys. Rev. Lett.} \bibinfo{volume}{105}
  (\bibinfo{year}{2010}) \bibinfo{pages}{235502}.
\bibitem[{Phark et~al.(2012)Phark, Borme, Vanegas, Corbetta, Sander, and
  Kirschner}]{Phark2012}
\bibinfo{author}{S.~Phark}, \bibinfo{author}{J.~Borme}, \bibinfo{author}{A.~L.
  Vanegas}, \bibinfo{author}{M.~Corbetta}, \bibinfo{author}{D.~Sander},
  \bibinfo{author}{J.~Kirschner},
\newblock \bibinfo{title}{Atomic structure and spectroscopy of graphene edges
  on {Ir}(111)},
\newblock \bibinfo{journal}{Phys. Rev. B} \bibinfo{volume}{86}
  (\bibinfo{year}{2012}) \bibinfo{pages}{045442}.
\bibitem[{Yazyev(2010)}]{Yazyev2010}
\bibinfo{author}{O.~V. Yazyev},
\newblock \bibinfo{title}{Emergence of magnetism in graphene materials and
  nanostructures},
\newblock \bibinfo{journal}{Rep. Progr. Phys.} \bibinfo{volume}{73}
  (\bibinfo{year}{2010}) \bibinfo{pages}{056501}.
\bibitem[{Petrone et~al.(2012)Petrone, Dean, Meric, van~der Zande, Huang, Wang,
  Muller, Shepard, and Hone}]{Petrone2012}
\bibinfo{author}{N.~Petrone}, \bibinfo{author}{C.~R. Dean},
  \bibinfo{author}{I.~Meric}, \bibinfo{author}{A.~M. van~der Zande},
  \bibinfo{author}{P.~Y. Huang}, \bibinfo{author}{L.~Wang},
  \bibinfo{author}{D.~Muller}, \bibinfo{author}{K.~L. Shepard},
  \bibinfo{author}{J.~Hone},
\newblock \bibinfo{title}{Chemical vapor deposition-derived graphene with
  electrical performance of exfoliated graphene},
\newblock \bibinfo{journal}{Nano Lett.} \bibinfo{volume}{12}
  (\bibinfo{year}{2012}) \bibinfo{pages}{2751--2756}.
\bibitem[{Chen et~al.(2009)Chen, Cullen, Jang, Fuhrer, and Williams}]{Chen2009}
\bibinfo{author}{J.~Chen}, \bibinfo{author}{W.~G. Cullen},
  \bibinfo{author}{C.~Jang}, \bibinfo{author}{M.~S. Fuhrer},
  \bibinfo{author}{E.~D. Williams},
\newblock \bibinfo{title}{Defect scattering in graphene},
\newblock \bibinfo{journal}{Phys. Rev. Lett.} \bibinfo{volume}{102}
  (\bibinfo{year}{2009}) \bibinfo{pages}{236805}.
\bibitem[{Banhart et~al.(2011)Banhart, Kotakoski, and
  Krasheninnikov}]{Banhartdoi:10.1021/nn102598m}
\bibinfo{author}{F.~Banhart}, \bibinfo{author}{J.~Kotakoski},
  \bibinfo{author}{A.~V. Krasheninnikov},
\newblock \bibinfo{title}{Structural defects in graphene},
\newblock \bibinfo{journal}{ACS Nano} \bibinfo{volume}{5}
  (\bibinfo{year}{2011}) \bibinfo{pages}{26--41}.
\bibitem[{Kotakoski et~al.(2011{\natexlab{a}})Kotakoski, Krasheninnikov,
  Kaiser, and Meyer}]{KotakoskiPhysRevLett.106.105505}
\bibinfo{author}{J.~Kotakoski}, \bibinfo{author}{A.~V. Krasheninnikov},
  \bibinfo{author}{U.~Kaiser}, \bibinfo{author}{J.~C. Meyer},
\newblock \bibinfo{title}{From point defects in graphene to two-dimensional
  amorphous carbon},
\newblock \bibinfo{journal}{Phys. Rev. Lett.} \bibinfo{volume}{106}
  (\bibinfo{year}{2011}{\natexlab{a}}) \bibinfo{pages}{105505}.
\bibitem[{Kotakoski et~al.(2011{\natexlab{b}})Kotakoski, Meyer, Kurasch,
  Santos-Cottin, Kaiser, and Krasheninnikov}]{KotakoskiPhysRevB.83.245420}
\bibinfo{author}{J.~Kotakoski}, \bibinfo{author}{J.~C. Meyer},
  \bibinfo{author}{S.~Kurasch}, \bibinfo{author}{D.~Santos-Cottin},
  \bibinfo{author}{U.~Kaiser}, \bibinfo{author}{A.~V. Krasheninnikov},
\newblock \bibinfo{title}{Stone-wales-type transformations in carbon
  nanostructures driven by electron irradiation},
\newblock \bibinfo{journal}{Phys. Rev. B} \bibinfo{volume}{83}
  (\bibinfo{year}{2011}{\natexlab{b}}) \bibinfo{pages}{245420}.
\bibitem[{Lee et~al.(2005)Lee, Wang, Yoon, Hwang, Kim, and
  Ho}]{LeePhysRevLett.95.205501}
\bibinfo{author}{G.-D. Lee}, \bibinfo{author}{C.~Z. Wang},
  \bibinfo{author}{E.~Yoon}, \bibinfo{author}{N.-M. Hwang},
  \bibinfo{author}{D.-Y. Kim}, \bibinfo{author}{K.~M. Ho},
\newblock \bibinfo{title}{Diffusion, coalescence, and reconstruction of vacancy
  defects in graphene layers},
\newblock \bibinfo{journal}{Phys. Rev. Lett.} \bibinfo{volume}{95}
  (\bibinfo{year}{2005}) \bibinfo{pages}{205501}.
\bibitem[{Ma et~al.(2009)Ma, Alf{\`e}, Michaelides, and
  Wang}]{JiePhysRevB.80.033407}
\bibinfo{author}{J.~Ma}, \bibinfo{author}{D.~Alf{\`e}},
  \bibinfo{author}{A.~Michaelides}, \bibinfo{author}{E.~Wang},
\newblock \bibinfo{title}{Stone-wales defects in graphene and other planar
  $s{p}^{2}$-bonded materials},
\newblock \bibinfo{journal}{Phys. Rev. B} \bibinfo{volume}{80}
  (\bibinfo{year}{2009}) \bibinfo{pages}{033407}.
\bibitem[{Li et~al.(2005)Li, Reich, and Robertson}]{LiPhysRevB.72.184109}
\bibinfo{author}{L.~Li}, \bibinfo{author}{S.~Reich},
  \bibinfo{author}{J.~Robertson},
\newblock \bibinfo{title}{Defect energies of graphite: Density-functional
  calculations},
\newblock \bibinfo{journal}{Phys. Rev. B} \bibinfo{volume}{72}
  (\bibinfo{year}{2005}) \bibinfo{pages}{184109}.
\bibitem[{El-Barbary et~al.(2003)El-Barbary, Telling, Ewels, Heggie, and
  Briddon}]{ElbarbaryPhysRevB.68.144107}
\bibinfo{author}{A.~A. El-Barbary}, \bibinfo{author}{R.~H. Telling},
  \bibinfo{author}{C.~P. Ewels}, \bibinfo{author}{M.~I. Heggie},
  \bibinfo{author}{P.~R. Briddon},
\newblock \bibinfo{title}{Structure and energetics of the vacancy in graphite},
\newblock \bibinfo{journal}{Phys. Rev. B} \bibinfo{volume}{68}
  (\bibinfo{year}{2003}) \bibinfo{pages}{144107}.
\bibitem[{Krasheninnikov et~al.(2006)Krasheninnikov, Lehtinen, Foster, and
  Nieminen}]{Krasheninnikov2006132}
\bibinfo{author}{A.~V. Krasheninnikov}, \bibinfo{author}{P.~O. Lehtinen},
  \bibinfo{author}{A.~S. Foster}, \bibinfo{author}{R.~M. Nieminen},
\newblock \bibinfo{title}{Bending the rules: Contrasting vacancy energetics and
  migration in graphite and carbon nanotubes},
\newblock \bibinfo{journal}{Chem. Phys. Lett.} \bibinfo{volume}{418}
  (\bibinfo{year}{2006}) \bibinfo{pages}{132 -- 136}.
\bibitem[{Cretu et~al.(2010)Cretu, Krasheninnikov, Rodriguez-Manzo, Sun,
  Nieminen, and Banhart}]{CretuPhysRevLett.105.196102}
\bibinfo{author}{O.~Cretu}, \bibinfo{author}{A.~V. Krasheninnikov},
  \bibinfo{author}{J.~A. Rodriguez-Manzo}, \bibinfo{author}{L.~Sun},
  \bibinfo{author}{R.~M. Nieminen}, \bibinfo{author}{F.~Banhart},
\newblock \bibinfo{title}{Migration and localization of metal atoms on strained
  graphene},
\newblock \bibinfo{journal}{Phys. Rev. Lett.} \bibinfo{volume}{105}
  (\bibinfo{year}{2010}) \bibinfo{pages}{196102}.
\bibitem[{Wang et~al.(2013)Wang, Zhang, Chan, Yan, and
  Ding}]{Wangdoi:10.1021/ja312687a}
\bibinfo{author}{L.~Wang}, \bibinfo{author}{X.~Zhang}, \bibinfo{author}{H.~L.
  Chan}, \bibinfo{author}{F.~Yan}, \bibinfo{author}{F.~Ding},
\newblock \bibinfo{title}{Formation and healing of vacancies in graphene
  chemical vapor deposition {(CVD)} growth},
\newblock \bibinfo{journal}{J. Am. Chem. Soc.} \bibinfo{volume}{135}
  (\bibinfo{year}{2013}) \bibinfo{pages}{4476--4482}.
\bibitem[{Pai et~al.(1996)Pai, Bartelt, and Reutt-Robey}]{Pai:1996vo}
\bibinfo{author}{W.~W. Pai}, \bibinfo{author}{N.~C. Bartelt},
  \bibinfo{author}{J.~E. Reutt-Robey},
\newblock \bibinfo{title}{{Fluctuation kinetics of an isolated {A}g (110)
  step}},
\newblock \bibinfo{journal}{Phys. Rev. B} \bibinfo{volume}{53}
  (\bibinfo{year}{1996}) \bibinfo{pages}{15991}.
\bibitem[{Ugeda et~al.(2011)Ugeda, Fern{\'a}ndez-Torre, Brihuega, Pou,
  Mart{\'i}nez-Galera, P{\'e}rez, and
  G{\'o}mez-Rodr{\'i}guez}]{UgedaPhysRevLett.107.116803}
\bibinfo{author}{M.~M. Ugeda}, \bibinfo{author}{D.~Fern{\'a}ndez-Torre},
  \bibinfo{author}{I.~Brihuega}, \bibinfo{author}{P.~Pou},
  \bibinfo{author}{A.~J. Mart{\'i}nez-Galera}, \bibinfo{author}{R.~P{\'e}rez},
  \bibinfo{author}{J.~M. G{\'o}mez-Rodr{\'i}guez},
\newblock \bibinfo{title}{Point defects on graphene on metals},
\newblock \bibinfo{journal}{Phys. Rev. Lett.} \bibinfo{volume}{107}
  (\bibinfo{year}{2011}) \bibinfo{pages}{116803}.
\bibitem[{Jacobson et~al.(2012)Jacobson, St{\"o}ger, Garhofer, Parkinson,
  Schmid, Caudillo, Mittendorfer, Redinger, and
  Diebold}]{Jacobsondoi:10.1021/jz2015007}
\bibinfo{author}{P.~Jacobson}, \bibinfo{author}{B.~St{\"o}ger},
  \bibinfo{author}{A.~Garhofer}, \bibinfo{author}{G.~S. Parkinson},
  \bibinfo{author}{M.~Schmid}, \bibinfo{author}{R.~Caudillo},
  \bibinfo{author}{F.~Mittendorfer}, \bibinfo{author}{J.~Redinger},
  \bibinfo{author}{U.~Diebold},
\newblock \bibinfo{title}{Disorder and defect healing in graphene on
  {N}i(111)},
\newblock \bibinfo{journal}{J. Phys. Chem. Lett.} \bibinfo{volume}{3}
  (\bibinfo{year}{2012}) \bibinfo{pages}{136--139}.
\bibitem[{Los and Pellenq(2010)}]{Los-Pellenq-2010}
\bibinfo{author}{J.~H. Los}, \bibinfo{author}{R.~J.~M. Pellenq},
\newblock \bibinfo{title}{{Determination of the bulk melting temperature of
  nickel using Monte Carlo simulations: Inaccuracy of extrapolation from
  cluster melting temperatures}},
\newblock \bibinfo{journal}{Phys. Rev. B} \bibinfo{volume}{81}
  (\bibinfo{year}{2010}) \bibinfo{pages}{064112}.
\bibitem[{Meca et~al.(2013)Meca, Lowengrub, Kim, Mattevi, and Shenoy}]{meca13}
\bibinfo{author}{E.~Meca}, \bibinfo{author}{J.~Lowengrub},
  \bibinfo{author}{H.~Kim}, \bibinfo{author}{C.~Mattevi},
  \bibinfo{author}{V.~B. Shenoy},
\newblock \bibinfo{title}{Epitaxial graphene growth and shape dynamics on
  copper: Phase-field modeling and experiments},
\newblock \bibinfo{journal}{Nano Lett.} \bibinfo{volume}{13}
  (\bibinfo{year}{2013}) \bibinfo{pages}{5692--5697}.
\bibitem[{Itoh et~al.(1998)Itoh, Bell, Avery, Jones, Joyce, and
  Vvedensky}]{itoh98}
\bibinfo{author}{M.~Itoh}, \bibinfo{author}{G.~R. Bell}, \bibinfo{author}{A.~R.
  Avery}, \bibinfo{author}{T.~S. Jones}, \bibinfo{author}{B.~A. Joyce},
  \bibinfo{author}{D.~D. Vvedensky},
\newblock \bibinfo{title}{Island nucleation and growth on reconstructed
  {G}a{A}s(001) surfaces},
\newblock \bibinfo{journal}{Phys. Rev. Lett.} \bibinfo{volume}{81}
  (\bibinfo{year}{1998}) \bibinfo{pages}{633--636}.
\bibitem[{Kratzer and Scheffler(2002)}]{kratzer02}
\bibinfo{author}{P.~Kratzer}, \bibinfo{author}{M.~Scheffler},
\newblock \bibinfo{title}{Reaction-limited island nucleation in molecular-beam
  epitaxy of compound semiconductors},
\newblock \bibinfo{journal}{Phys. Rev. Lett.} \bibinfo{volume}{88}
  (\bibinfo{year}{2002}) \bibinfo{pages}{036102}.
\bibitem[{Tersoff(1989)}]{Tersoff-potentials-1989}
\bibinfo{author}{J.~Tersoff},
\newblock \bibinfo{title}{Modeling solid-state chemistry: Interatomic
  potentials for multicomponent systems},
\newblock \bibinfo{journal}{Phys. Rev. B} \bibinfo{volume}{39}
  (\bibinfo{year}{1989}) \bibinfo{pages}{5566}.
\bibitem[{Virojanadara et~al.(2010)Virojanadara, Yakimova, Zakharov, and
  Johansson}]{Virojanadara10}
\bibinfo{author}{C.~Virojanadara}, \bibinfo{author}{R.~Yakimova},
  \bibinfo{author}{A.~A. Zakharov}, \bibinfo{author}{L.~I. Johansson},
\newblock \bibinfo{title}{Large homogeneous mono- bi-layer graphene on
  6{H-SiC}(0001) and buffer layer elimination},
\newblock \bibinfo{journal}{J. Phys. D: Appl. Phys.} \bibinfo{volume}{43}
  (\bibinfo{year}{2010}) \bibinfo{pages}{374010}.
\bibitem[{Erhart and Albe(2005)}]{Erhart-mod-Tersoff-pot}
\bibinfo{author}{P.~Erhart}, \bibinfo{author}{K.~Albe},
\newblock \bibinfo{title}{Analytical potential for atomistic simulations of
  silicon, carbon, and silicon carbide},
\newblock \bibinfo{journal}{Phys. Rev. B} \bibinfo{volume}{71}
  (\bibinfo{year}{2005}) \bibinfo{pages}{035211}.
\bibitem[{Hannon and Tromp(2008)}]{Hanon}
\bibinfo{author}{J.~B. Hannon}, \bibinfo{author}{R.~M. Tromp},
\newblock \bibinfo{title}{Pit formation during graphene synthesis on
  {SiC}(0001):\textit{In situ} electron microscopy},
\newblock \bibinfo{journal}{Phys. Rev. B} \bibinfo{volume}{77}
  (\bibinfo{year}{2008}) \bibinfo{pages}{241404}.
\bibitem[{Kageshima et~al.(2011)Kageshima, Hibino, Yamaguchi, and
  Nagase}]{Kageshima11}
\bibinfo{author}{H.~Kageshima}, \bibinfo{author}{H.~Hibino},
  \bibinfo{author}{H.~Yamaguchi}, \bibinfo{author}{M.~Nagase},
\newblock \bibinfo{title}{Theoretical study on epitaxial graphene growth by
  {S}i sublimation from {SiC(0001)} surface},
\newblock \bibinfo{journal}{Jpn. J. Appl. Phys.} \bibinfo{volume}{50}
  (\bibinfo{year}{2011}) \bibinfo{pages}{095601}.
\bibitem[{Kageshima et~al.(2012)Kageshima, Hibino, and Tanabe}]{Kageshima12}
\bibinfo{author}{H.~Kageshima}, \bibinfo{author}{H.~Hibino},
  \bibinfo{author}{S.~Tanabe},
\newblock \bibinfo{title}{The physics of epitaxial graphene on {SiC(0001)}},
\newblock \bibinfo{journal}{J. Phys.: Condens. Matter} \bibinfo{volume}{24}
  (\bibinfo{year}{2012}) \bibinfo{pages}{314215}.
\bibitem[{Wu(2013)}]{WuSiC}
\bibinfo{author}{X.~Wu}, \bibinfo{title}{Two-Dimensional Carbon - Fundamental
  Properties, Synthesis, Characterization, and Applications},
  \bibinfo{publisher}{Pan Stanford Publishing}, \bibinfo{year}{2013}.
\bibitem[{Ming and Zangwill(2011)}]{Ming1}
\bibinfo{author}{F.~Ming}, \bibinfo{author}{A.~Zangwill},
\newblock \bibinfo{title}{Model for the epitaxial growth of graphene on
  {6H-SiC(0001)}},
\newblock \bibinfo{journal}{Phys. Rev. B} \bibinfo{volume}{84}
  (\bibinfo{year}{2011}) \bibinfo{pages}{115459}.
\bibitem[{{Ming} and {Zangwill}(2012)}]{Ming2}
\bibinfo{author}{F.~{Ming}}, \bibinfo{author}{A.~{Zangwill}},
\newblock \bibinfo{title}{{Model and simulations of the epitaxial growth of
  graphene on non-planar {6H-SiC} surfaces}},
\newblock \bibinfo{journal}{J. Phys. D: Appl. Phys.} \bibinfo{volume}{45}
  (\bibinfo{year}{2012}) \bibinfo{pages}{154007}.
\bibitem[{Robinson et~al.(2010)Robinson, Weng, Trumbull, Cavalero,
  Wetherington, Frantz, LaBella, Hughes, Fanton, and Snyder}]{Robinson}
\bibinfo{author}{J.~Robinson}, \bibinfo{author}{X.~Weng},
  \bibinfo{author}{K.~Trumbull}, \bibinfo{author}{R.~Cavalero},
  \bibinfo{author}{M.~Wetherington}, \bibinfo{author}{E.~Frantz},
  \bibinfo{author}{M.~LaBella}, \bibinfo{author}{Z.~Hughes},
  \bibinfo{author}{M.~Fanton}, \bibinfo{author}{D.~Snyder},
\newblock \bibinfo{title}{{Nucleation of Epitaxial Graphene on SiC(0001)}},
\newblock \bibinfo{journal}{ACS Nano} \bibinfo{volume}{4}
  (\bibinfo{year}{2010}) \bibinfo{pages}{153}.
\bibitem[{Totton et~al.(2010)Totton, Lorenz, Rompotis, Martsinovich, and
  Kantorovich}]{my-SBC-3}
\bibinfo{author}{D.~Totton}, \bibinfo{author}{C.~D. Lorenz},
  \bibinfo{author}{N.~Rompotis}, \bibinfo{author}{N.~Martsinovich},
  \bibinfo{author}{L.~Kantorovich},
\newblock \bibinfo{title}{Temperature control in molecular dynamic simulations
  of non-equilibrium processes},
\newblock \bibinfo{journal}{J. Phys.: Condens. Matter} \bibinfo{volume}{22}
  (\bibinfo{year}{2010}) \bibinfo{pages}{074205}.
  
\bibitem[{Xu and Henkelman(2008)}]{Adaptive-KMC-2008}
\bibinfo{author}{L.~Xu}, \bibinfo{author}{G.~Henkelman},
\newblock \bibinfo{title}{Adaptive kinetic {Monte Carlo} for first-principles
  accelerated dynamics},
\newblock \bibinfo{journal}{J. Chem. Phys.} \bibinfo{volume}{129}
  (\bibinfo{year}{2008}) \bibinfo{pages}{114104}.

\bibitem[{Kim et al(2012)}]{Kim2012}
\bibinfo{author}{H.~Kim}, \bibinfo{author}{C.~Mattevi}, \bibinfo{author}{M.~Reyes Calvo}, 
\bibinfo{author}{J.C.~Oberg}, \bibinfo{author}{L.~Artiglia}, \bibinfo{author}{S.~Agnoli}, 
\bibinfo{author}{C.F.~Hirjibehedin}, \bibinfo{author}{M.~Chhowalla}, \bibinfo{author}{E.~Saiz},
\newblock \bibinfo{title}{Activation energy paths for graphene nucleation and growth on Cu},
\newblock \bibinfo{journal}{ACS Nano} \bibinfo{volume}{6}
  (\bibinfo{year}{2012}) \bibinfo{pages}{3614-3623}.

\bibitem[{Nemec et al(2013)}]{Nemec2013}
\bibinfo{author}{L.~Nemec}, \bibinfo{author}{V.~Bluym}, \bibinfo{author}{P.~Rinjke}, \bibinfo{author}{M.~Scheffler},
\newblock \bibinfo{title}{Thermodynamic equilibrium conditions of graphene films on SiC},
\newblock \bibinfo{journal}{Phys. Rev. Lett.} \bibinfo{volume}{111}
  (\bibinfo{year}{2013}) \bibinfo{pages}{065502}.

\end{thebibliography}



\end{document}